\documentclass[prd,aps,twocolumn,a4paper,showkeys,nofootinbib]{revtex4-1}
\usepackage{amsmath}
\usepackage{amsfonts}
\usepackage{amssymb}	
\usepackage{graphicx}
\usepackage{bm}
\usepackage{color}
\usepackage{commath}
\allowdisplaybreaks

\usepackage{hyperref}
\usepackage{makecell}

\def\GMc2{G M_{\odot} c^{-2}}

\def\F{{\cal F}}
\def\lm{{\ell m}}

\def\lm{{\ell m}}

\def\lm{{\ell m}}

\def\l{{\ell }}

\def\F{{\cal F}}

\def\ha{{\hat{a}}}

\def\TEOBResumS{\texttt{TEOBResumS}}

\def\Teukode{{\texttt{Teukode}}}
\def\RWZ{{\texttt{RWZhyp}}}

\def\Jteuk{{\dot{J}_{\rm teuk}}}
\def\Eteuk{{\dot{E}_{\rm teuk}}}
\def\Fteuk{{\dot{F}_{\rm teuk}}}
\def\JNP{{\dot{J}_{\rm NP}}}
\def\ENP{{\dot{E}_{\rm NP}}}
\def\FNP{{\dot{F}_{\rm NP}}}
\def\JNPold{{\dot{J}_{\rm old}}}
\def\ENPold{{\dot{E}_{\rm old}}}
\def\FNPold{{\dot{F}_{\rm old}}}
\def\JANP{{\dot{J}_{\rm ANP}}}
\def\EANP{{\dot{E}_{\rm ANP}}}
\def\FANP{{\dot{F}_{\rm ANP}}}
\def\Jhlm{{\dot{J}_{h_{\lm}}}}
\def\Ehlm{{\dot{E}_{h_{\lm}}}}
\def\Fhlm{{\dot{F}_{h_{\lm}}}}
\def\HKerr{{\hat{H}^{\rm eq}_{\rm Kerr}}}
\def\FphiNP{{\hat{\F}_\varphi^{\rm NP}}}
\def\FphiANP{{\hat{\F}_\varphi^{\rm ANP}}}
\def\Fphiold{{\hat{\F}_\varphi^{\rm old}}}

\newcommand\fnp[1]{{\hat{f}_{\varphi #1}^{\rm N _{nc}}}}
\def\NP22{{\fnp{,22}}}
\newcommand\av[1]{{\langle #1 \rangle}}
\newcommand\underrel[2]{\mathrel{\mathop{#2}\limits_{#1}}}

\newcommand\be{\begin{equation}}
\newcommand\ee{\end{equation}}

\usepackage{color}
\definecolor{cyan}{rgb}{0,0.9,0.9}
\definecolor{orange}{rgb}{0.9,0.5,0}
\definecolor{magenta}{rgb}{1,0,1}
\definecolor{purple}{rgb}{0.8,0.4,0.8}
\definecolor{gray}{rgb}{0.8242,0.8242,0.8242}
\definecolor{dodgerblue}{rgb}{0.12, 0.56, 1.0}
\definecolor{darkgrey}{rgb}{0.5,0.5,0.5}
\definecolor{darkgreen}{rgb}{0,0.65,0}

\definecolor{colortab1}{rgb}{0.1, 0.1, 1.0}
\definecolor{colortab2}{rgb}{0.9,0,0.1}

\begin{document}

\title{Effective one-body model for extreme-mass-ratio spinning binaries on eccentric equatorial orbits: 
testing radiation reaction and waveform}
\author{Simone \surname{Albanesi}${}^{1,2}$}
\author{Alessandro \surname{Nagar}${}^{2,3}$}
\author{Sebastiano \surname{Bernuzzi}${}^{4}$}
\affiliation{${}^{1}$ Dipartimento di Fisica, Universit\`a di Torino, via P. Giuria 1, 
10125 Torino, Italy}
\affiliation{${}^2$INFN Sezione di Torino, Via P. Giuria 1, 10125 Torino, Italy} 
\affiliation{${}^3$Institut des Hautes Etudes Scientifiques, 91440 Bures-sur-Yvette, France}
\affiliation{${}^4$Theoretisch-Physikalisches Institut, Friedrich-Schiller-Universit{\"a}t 
Jena, 07743, Jena, Germany}  

\begin{abstract}
We provide a systematic analysis of the multipolar gravitational waveform, energy and angular 
momentum fluxes emitted by a  nonspinning test particle orbiting 
a Kerr black hole along equatorial, eccentric orbits.
These quantities are computed by solving numerically the Teukolsky equation in the time domain
using \Teukode ~and are then used to establish the reliability of a recently introduced prescription
to deal with eccentricity-driven effects in the radiation reaction (and waveform) of the 
effective-one-body (EOB) model. 
The prescription relies on the idea of incorporating these effects by replacing
the quasi-circular Newtonian (or leading-order) prefactors in the EOB-factorized multipolar waveform 
(and fluxes) with their generic counterparts. To reliably account for strong-field regimes,
standard factorization and resummation procedures had to be implemented also for the circular 
sector of $\ell=7$ and $\ell=8$ waveform multipoles.
The comparison between numerical and analytical quantities is carried out over a large portion of the 
parameter space, notably for orbits close to the separatrix and with high eccentricities.
The analytical fluxes agree to $\sim 2\%$ with the numerical data for orbits with moderate
eccentricities($e\lesssim 0.3$) and moderate spins ($\ha \lesssim 0.5$),
although this increases up to $\sim 33\%$ for large, positive, black hole spins ($\sim 0.9$) 
and large eccentricities  ($\sim 0.9$). Similar agreement is also found for the waveform. 
For moderate eccentricities, the EOB fluxes can be  used to drive the test-particle dynamics 
through the nonadiabatic transition from eccentric inspiral  to plunge, merger and ringdown.
Over this dynamics, we construct a complete EOB waveform, including merger and ringdown, 
that shows an excellent phasing and amplitude agreement with the numerical one. 
We also show that the same technique can be applied to hyperbolic encounters.
In general, our approach to radiation reaction for eccentric inspirals should be seen as
a first step toward EOB modelization of extreme-mass-ratio-inspirals waveforms for LISA.
\end{abstract}
\date{\today}

\maketitle

\section{Introduction}
\label{sec:introduction}
The gravitational waves (GWs) observed by LIGO and Virgo~\cite{TheVirgo:2014hva,TheLIGOScientific:2014jea}  
are generated by the last stages of the coalescence of compact binary systems with comparable masses~\cite{LIGOScientific:2020ibl}. 
These systems lose energy and angular momentum due to GW emission, that causes the progressive 
circularization of the orbit. Therefore, most of the current gravitational wave templates used for the analysis 
of the signals adopt the quasi-circular approximation. 
Nonetheless, dynamical captures are possible in dense environments, such 
as active galactic nuclei and globular clusters, leading to the formation of binaries with 
nonvanishing eccentricity, as shown in recent population studies~\cite{Mukherjee:2020hnm,Romero-Shaw:2019itr}. 
The dynamical capture scenario is also relevant for the analysis of GW190521~\cite{Abbott:2020tfl,Romero-Shaw:2020thy}.
Other systems where the eccentricity plays a key role are Extreme Mass Ratio 
Inspirals (EMRIs), binaries where a compact stellar object orbits around a supermassive black 
hole. Gravitational waves from EMRIs have characteristic frequencies around the mHz, 
making them one of the most relevant sources for the Laser Interferometer Space Antenna 
(LISA)~\cite{Amaro-Seoane:2018gbb}. For these reasons, the inclusion of eccentricity in 
the current theoretical waveform models has drawn interest in the last few years, leading to 
the realization of many models for {\it bound} configurations with, in general, relatively 
mild eccentricity, such as
{\tt ENIGMA}~\cite{Huerta:2017kez}, {\tt SEOBNRE}
~\cite{Cao:2017ndf,Liu:2019jpg,Liu:2021pkr}, \texttt{NRSur2dq1Ecc}~\cite{Islam:2021mha}, 
{\tt TEOBResumS}~\cite{Chiaramello:2020ehz,Nagar:2020xsk,Nagar:2021gss}
and Refs.~\cite{Tanay:2016zog,Moore:2019xkm,Boetzel:2019nfw,Tiwari:2019jtz,Tiwari:2020hsu}.

The effective one body approach (EOB) currently represents the most complete, reliable
and predictive analytical framework able to deal with inspiraling and coalescing relativistic binaries. 
By design, the EOB approach is superior to standard post-Newtonian techniques because it
structurally incorporates nonperturbative elements (e.g., the existence of a last stable orbit)
that allow for a robust computation of observable quantities, like waveforms and fluxes, 
{\it also} in strong field, i.e. even {\it beyond} the last stable orbit up to merger.
The synergy between the EOB approach and Numerical Relativity (NR) simulations proved
highly successful to provide highly accurate waveform templates for coalescing BBHs and
BNS as observed by LIGO and Virgo. By contrast, systems with large mass ratios like
EMRIs cannot be explored using NR techniques, while they are naturally described within
the EOB approach. In addition: (i) the extreme mass ratio limit plays a pivotal role in the EOB development, 
especially for what concerns waveforms and fluxes, that can be informed/compared with numerical results
~\cite{Nagar:2006xv,Damour:2007xr,Bernuzzi:2010ty,Bernuzzi:2010xj,Bernuzzi:2011aj, Bernuzzi:2012ku, 
Harms:2014dqa,Nagar:2014kha,Harms:2015ixa,Harms:2016ctx,Lukes-Gerakopoulos:2017vkj,Nagar:2019wrt}, 
and (ii) a fairly large amount of work has been dedicated to 
calculate analytically gravitational self-force terms and provide comparisons with numerical 
results~\cite{Damour:2009sm,Barack:2010ny,Akcay:2012ea,Bini:2014ica,Bini:2015mza,Akcay:2015pjz,Barack:2019agd}. 

The EOB relativistic dynamics relies on three building blocks: (i) a Hamiltonian; (ii) a 
prescription for computing 
the radiation reaction; (iii) a prescription for computing the waveform. Typically, points (ii) 
and (iii) are 
interconnected, because the radiation reaction, i.e. the gravitational wave fluxes of energy and angular momentum, 
are obtained by summing together resummed waveform multipoles.
In this respect, Ref.~\cite{Chiaramello:2020ehz} proposed to incorporate noncircular effects
in radiation reaction (and waveform) replacing the quasi-circular Newtonian (pre)factors with
their generic counterparts. Recently~\cite{Nagar:2021gss}, improvement of this approach allowed
to build an EOB eccentric waveform model that is highly faithful with the (tiny) number of eccentric 
NR simulations publicly available. Nonetheless, a systematic understanding and checking of this 
Newtonian-improved quantities is missing, despite the preliminary results shown in 
Ref.~\cite{Chiaramello:2020ehz}. Historically, the systems made by a small black hole of 
mass $\mu$ orbiting a large black hole with mass $M$ such that $\nu\equiv \mu/M\ll 1$ 
proved a useful laboratory to test and verify ideas or methods within the EOB approach 
before adopting them in the comparable-mass case~\cite{Damour:2008gu}.
This practice was in particular followed in Ref.~\cite{Chiaramello:2020ehz}, that
highlighted the excellent agreement between analytical and numerical waveform and fluxes 
in the test-mass case (see Fig.~1 therein, limited to geodesic motion with $e=0.3$).
The purpose of this paper is to systematically extend the analysis of~\cite{Chiaramello:2020ehz} 
up to large values of the eccentricity, also including the black hole spin.
We will thus mainly focus on analytical/numerical flux comparisons of extreme mass ratio binaries:
a test-mass object orbits around a Kerr black hole along eccentric equatorial geodesics. 
The comparisons of the analytical fluxes with the numerical ones obtained from the time-domain (TD) 
code {\Teukode}~\cite{Harms:2014dqa} will provide a reliability-test of the radiation reaction. 
We will also test the reliability of the analytical prescription for the waveform.
Our model should be intended as a first step toward the modelization of EMRIs
within the EOB framework. An physically more faithful description of EMRIs will certainly
need to include high-order noncircular terms beyond the leading order in the radiation reaction
as well as results from Gravitational Self-Force theory within the EOB Hamiltonian (i.e., linear in the 
mass ratio $\nu$ beyond the geodesic dynamics~\cite{Barack:2018yvs}). This aim goes beyond
the purpose of the present work. As a consequence, most of the work presented in this paper
should be considered as an exploratory investigation that will be refined further in the future.
In particular, we shall consider an analytical/numerical agreement for instantaneous eccentric fluxes 
to be good if the fractional differences are around or below the few percents, indicatively  $\lesssim 5\%$. 
For the corresponding averaged fluxes, we expect smaller differences. 
For what concerns eccentric waveforms, we aim at reaching numerical/analytical fractional differences
of a few percents in the amplitude and of a few hundredth of a radian in the phase difference. 
We will show that this accuracy, considered good within our context, is achieved in a large
portion of the parameter space.

The paper is structured as follows. In Sec.~\ref{sec:numsim} we expose the Hamiltonian 
formalism used to describe  the dynamics of a test-particle around a Kerr black hole 
and the numerical methods used to perform the simulations.
In Sec.~\ref{sec:eobmodel}, after a brief introduction of the EOB model, we describe the 
analytical waveform and fluxes in details and the new improvements introduced to the 
circular Post-Newtonian (PN) factors. 
In Sec.~\ref{sec:fluxes}, we study the phenomenology of the fluxes and we 
compare the numerical and analytical fluxes to establish the reliability of the radiation reaction. 
In Sec.~\ref{sec:waves} we provide a comparison between 
numerical and analytical waveforms, also showing the full transition from an eccentric inspiral
to plunge, merger and ringdown, as well as waveforms from dynamical captures.

Through this paper, we will use geometrized units $G=c=1$. Moreover, the time and the 
phase-space variables used in this work are related to the physical ones by $t=T/(GM)$, 
$r=R/(GM)$, $p_{r}=P_{R}/\mu$ and $p_\varphi=P_\varphi/(\mu GM)$. 

\section{Numerical waveforms and fluxes}
\label{sec:numsim}

\subsection{Equatorial dynamics around a Kerr black hole}
\label{sec:Ham_eq}
For a test-particle orbiting in the equatorial plane of a Kerr black hole
of dimensionless spin $\ha\equiv J_{\rm BH}/M^2$ the EOB Hamilton's
equations for spin-aligned objects reduce to~\cite{Damour:2014sva,Harms:2016ctx}
\begin{align}
\dot{r} &=\left(\frac{A}{B}\right)^{1 / 2} \frac{\partial \HKerr }{\partial p_{r_{*}}}  , \\ 
\dot{\varphi} &=\frac{\partial \HKerr}{\partial p_{\varphi}} \equiv \Omega  , \label{eq:freq} \\ 
\dot{p}_{r_*} &=\left(\frac{A}{B}\right)^{1 / 2} \left( \hat{\F}_r -  \frac{\partial \HKerr}{\partial r} \right)  , \\
\dot{p}_\varphi &=\hat{\F}_{\varphi}  .
\end{align}
Here, $\hat{\F}_{r,\varphi} \equiv \F_{r,\varphi} /\nu $ are the radial and angular components 
of the radiation reaction force and $\HKerr$ is the $\mu$-normalized test-particle Hamiltonian~\cite{Damour:2014sva}, 
that reads
\be
\HKerr =  \frac{2\ha p_\varphi}{r r_c^2} + \sqrt{A \del{r} \del{1 + \frac{p_{\varphi}^2}{r_{c}^2}}+p_{r_*}^2} .
\ee
The metric functions $A(r)$ and $B(r)$  read 
\begin{align}
A(r) & = \frac{1+2 u_c}{1+2 u}\left(1- 2 u_c\right) \ , \\
B(r) & = \frac{1}{1-2 u + \ha^2 u^2} \ ,
\end{align}
where $u=1/r$, $u_c = 1/r_c$, and $r_c$ is the centrifugal radius defined as~\cite{Damour:2014sva}
\be
r_c^2=r^2+\ha^2+2\frac{\ha^2}{r} .
\ee
Finally, $p_{r_*}$ is the conjugate momentum of the tortoise coordinate $r_*$, 
defined as $p_{r_*} = \sqrt{A/B}\;p_r$. Being planar, the orbit is fully determined by 
the choice of the eccentricity $e$ and the semilatus rectum $p$. 
Although it is not possible to provide a gauge invariant definition of $(e,p)$,
in the test-particle limit it is natural to define them in analogy with Newtonian mechanics.
For bound orbits, we have
\begin{equation}
\label{eq:ep_definition}
e  = \frac{r_+ - r_-}{r_+ + r_-}, \quad \quad p  = \frac{2 r_+ r_-}{r_+ + r_-} ,
\end{equation}
where $r_\pm $ are the two radial turning points, i.e. the apastron, $r_+$, 
and the periastron, $r_{-}$. Using the definitions of eccentricity and semilatus rectum, 
one finds $r_\pm = p/(1\mp e)$.  
In order to obtain a link between $(e,p)$ and the energy 
and angular momentum $(\hat{E},p_\varphi)$, we analytically solve the two-equations system obtained 
by considering
\begin{equation}
\label{eq:energy}
\hat{E} = \HKerr |_{p_{r_*}=0}  =   \frac{2 \ha p_\varphi}{r r_c^2} + \sqrt{A \del{r} \del{1 + \frac{p_{\varphi}^2}{r_{c}^2}}} 
\end{equation}
evaluated at the two radial turning points $r_\pm$, where $p_{r_*}=0$ by definition.
Then for each pair of initial eccentricity and semilatus rectum $(e_0,p_0)$, 
we obtain the initial energy and angular momentum $(\hat{E}_0,p_{\varphi}^0)$.
Finally, using the convention that the (cyclic) 
azimuthal variable $\varphi$ is set to zero at apastron, we have all the
initial values needed to compute the evolution of the system through Hamilton's equations.
For geodesic motion, i.e. when $\hat{\F}_{\varphi}=\hat{\F}_r=0$, the eccentricity and 
the semilatus rectum (or, equivalently, $\hat{E}$ and $p_\varphi$) are constants of motion. 

It is important to consider that for a given Kerr background, not all the eccentricity-semilatus 
rectum pairs produce stable orbits. In fact, in order to have a bound orbits, the Kerr 
potential (Eq.(2) of~\cite{OShaughnessy:2002tbu}) must have three roots: $(\bar{r},r_-,r_+)$. 
When $r_-=\bar{r}$, the bound motion is only marginally allowed, and when the 
potential has only two roots the particle inevitably  plunges toward the event horizon. 
Therefore, for stable bound orbits we must have $p>p_s$, 
where $p_s$ is known as \textit{separatrix} and depends both on eccentricity and spin. 
The separatrix can be found as a root of~\cite{OShaughnessy:2002tbu,Stein:2019buj}
\begin{align}
& p_s^2 (p_s - 6 - 2 e)^2 + \ha^4 (e-3)^2 (e+1)^2  \\ 
& - 2 \ha^2 (1+e) p_s \left[14 + 2 e^2 + p_s (3 - e)\right] = 0 . \nonumber
\end{align} 
Note that in the Schwarzchild case, the separatrix is simply given by $p_s = 6 + 2e$. 

\subsection{Waveform and fluxes}
From the particle dynamics, we compute waveforms and fluxes at leading order in the mass ratio $\nu$. 
This is done either numerically, solving the Teukolsky or Regge-Wheeler-Zerilli (RWZ) perturbation 
equations~\cite{Regge:1957td,Zerilli:1970se}, or analytically, using suitable resummations
of post-Newtonian results as discussed in Sec.~\ref{sec:eobmodel} below.
To fix conventions, let us remind that the waveform is decomposed in multipoles $h_\lm$
as
\begin{equation}
h_+-{\rm i} h_\times = D_L^{-1} \sum_{\l=2}^{\l_{\rm max}} \sum_{m=-\l}^{\l} h_\lm \,{}_{-2}Y_\lm, 
\end{equation}
where $D_L$ is the luminosity distance and ${}_{-2}Y_\lm$ are the spin-weighted spherical harmonics
with weight $s=-2$. We will often work with the RWZ-normalized 
waveform~\cite{Nagar:2005ea}
$\Psi_{\lm} =h_{\lm}/\sqrt{(\ell+2)(\ell+1)\ell(\ell-1)}$. The energy and angular momentum 
fluxes at future null-infinity, $\dot{E}$ and $\dot{J}$, can be  computed from 
the multipolar waveform as
\begin{subequations}
 \label{eq:fluxes_infty}
\begin{align}
\label{eq:fluxes_infty_E}
\dot{E} &= \;\;\,\frac{1}{16 \pi} \sum_{\l=2}^{\ell_{\rm max}}\sum_{m=-\l}^{\l} | \dot{h}_{\lm} |^2  ,\\
\dot{J} &= -\frac{1}{16 \pi} \sum_{\l=2}^{\ell_{\rm max}} \sum_{m=-\l}^{\l} m \Im \left( \dot{h}_{\lm} h_{\lm}^* \right) \  . \label{eq:fluxes_infty_J}
\end{align}
\end{subequations}
We will fix $\ell_{\rm max}=8$, since the contributions of the higher ones is negligible.
In fact, even the $\ell = 8$ modes have typically contributions $\lesssim 1\%$,  and only in some 
very strong-field regimes their contribution can reach the $2\%$. 
For example, we anticipate that for $(\ha,e,p)=(0.9, 0.9,p_s+0.01)$, 
we get a relative contribution to $\dot{J}$ of  $2.2\%$ (the (8,8) mode alone
contributes to $2.1\%$). The contribution of the subdominant modes to the fluxes will be discussed
in more detail in Sec.~\ref{sec:subdominant}.  

\subsection{Numerical methods and setup}
%
\begin{figure}[t]
	\begin{center}
	\hspace{0.5cm}
	\includegraphics[width=0.20\textwidth]{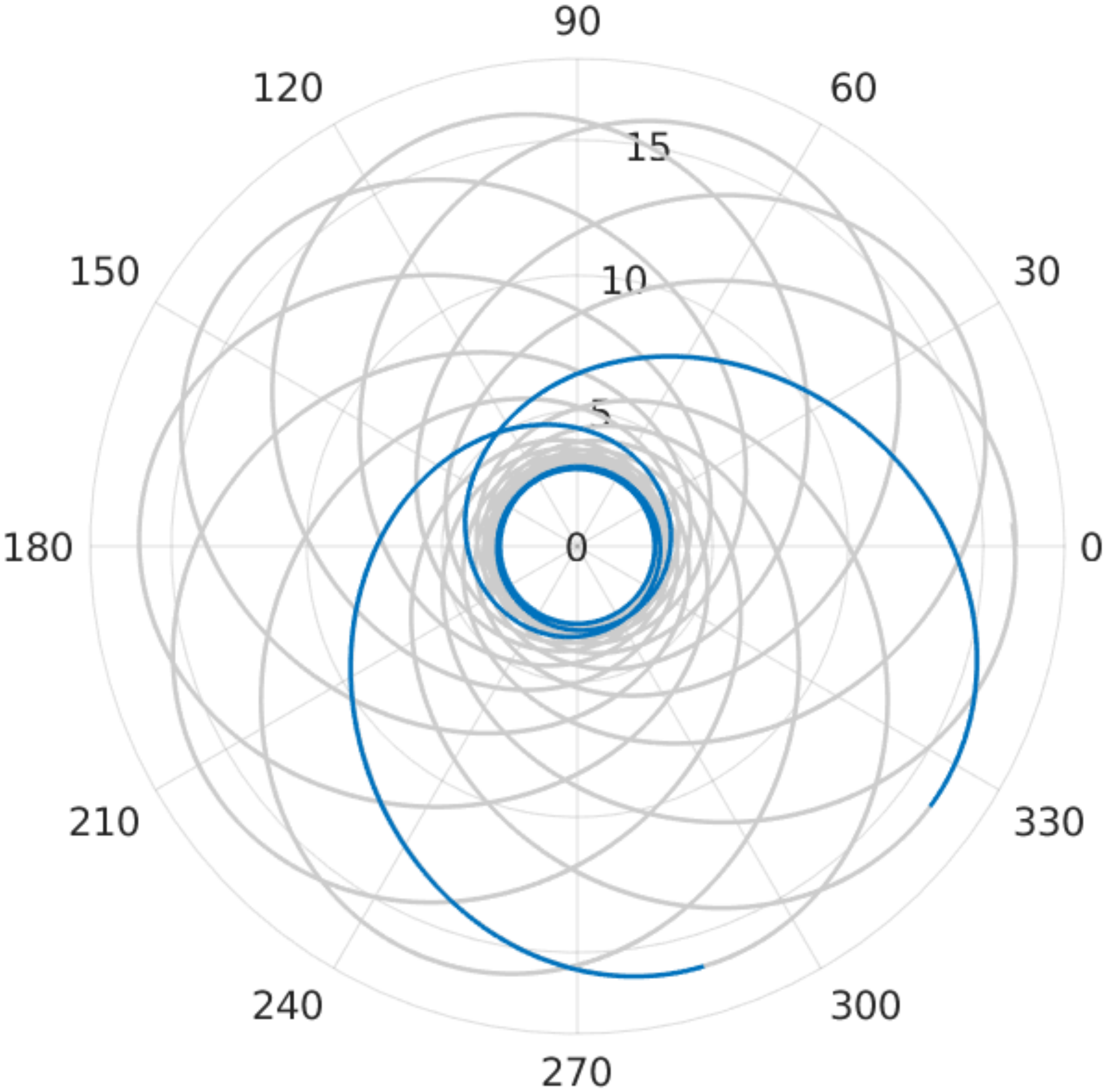}
	\hspace{0.0cm}
	\includegraphics[width=0.23\textwidth]{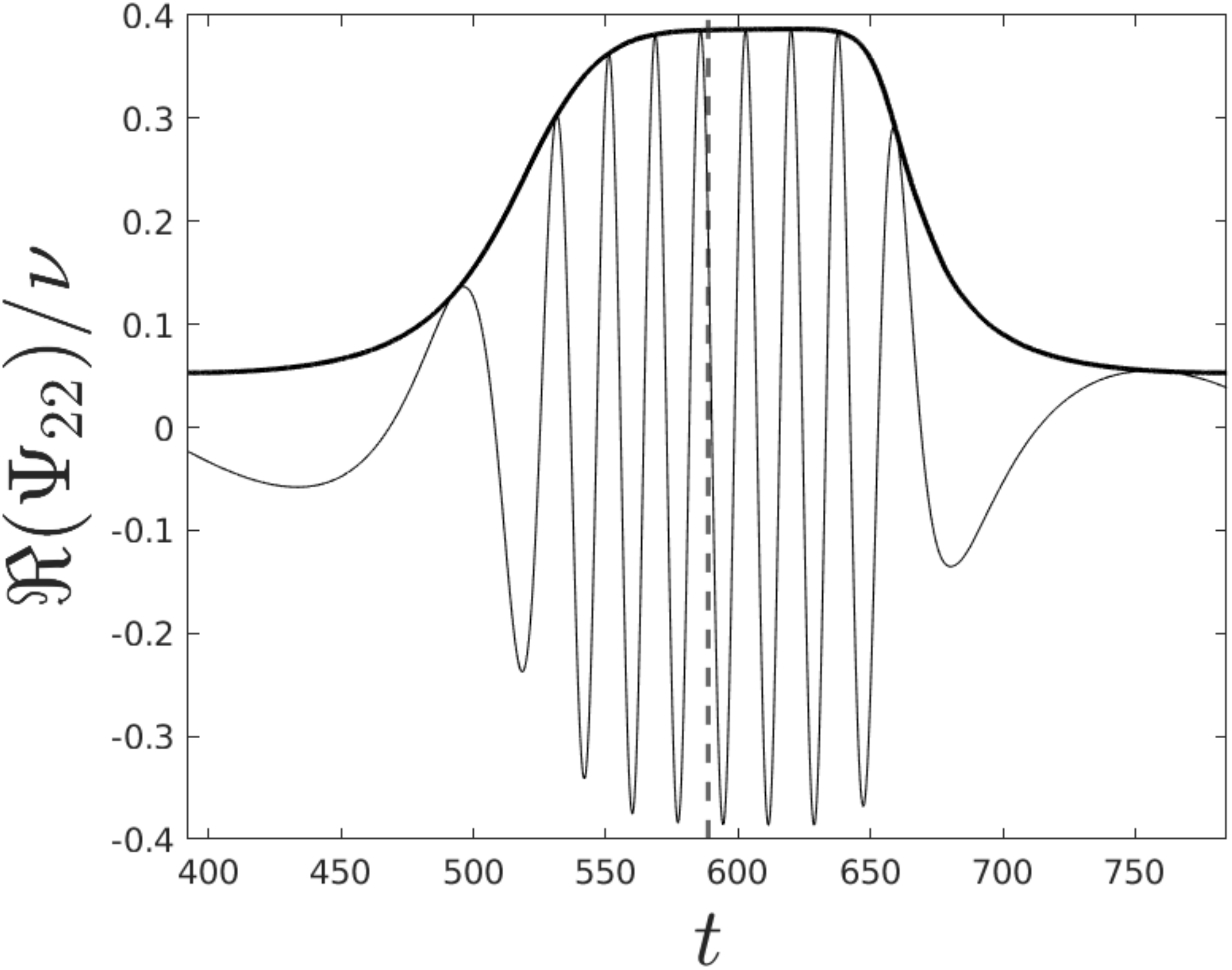} \\
	\includegraphics[width=0.23\textwidth,height=3.1cm]{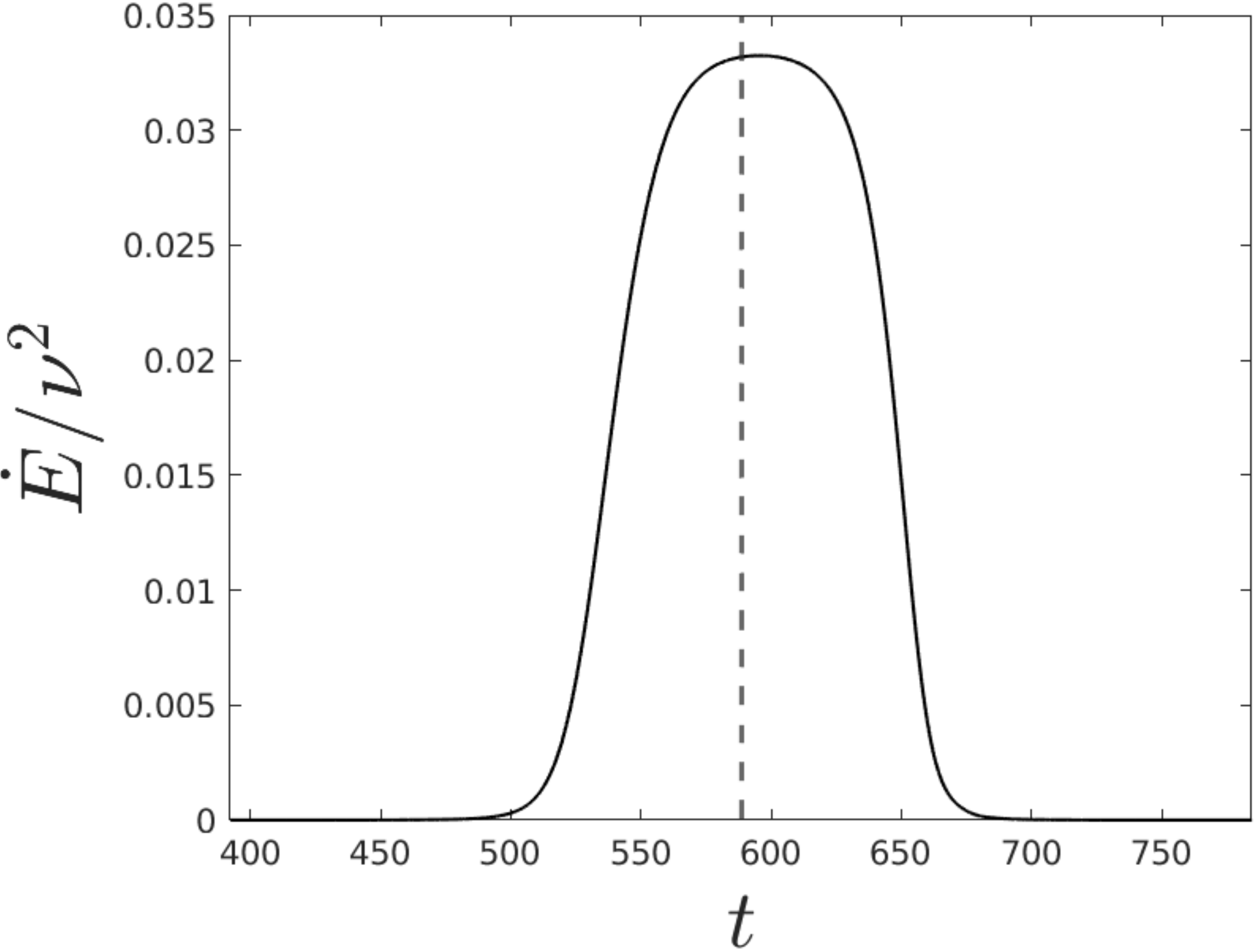}
	\includegraphics[width=0.23\textwidth,height=3.1cm]{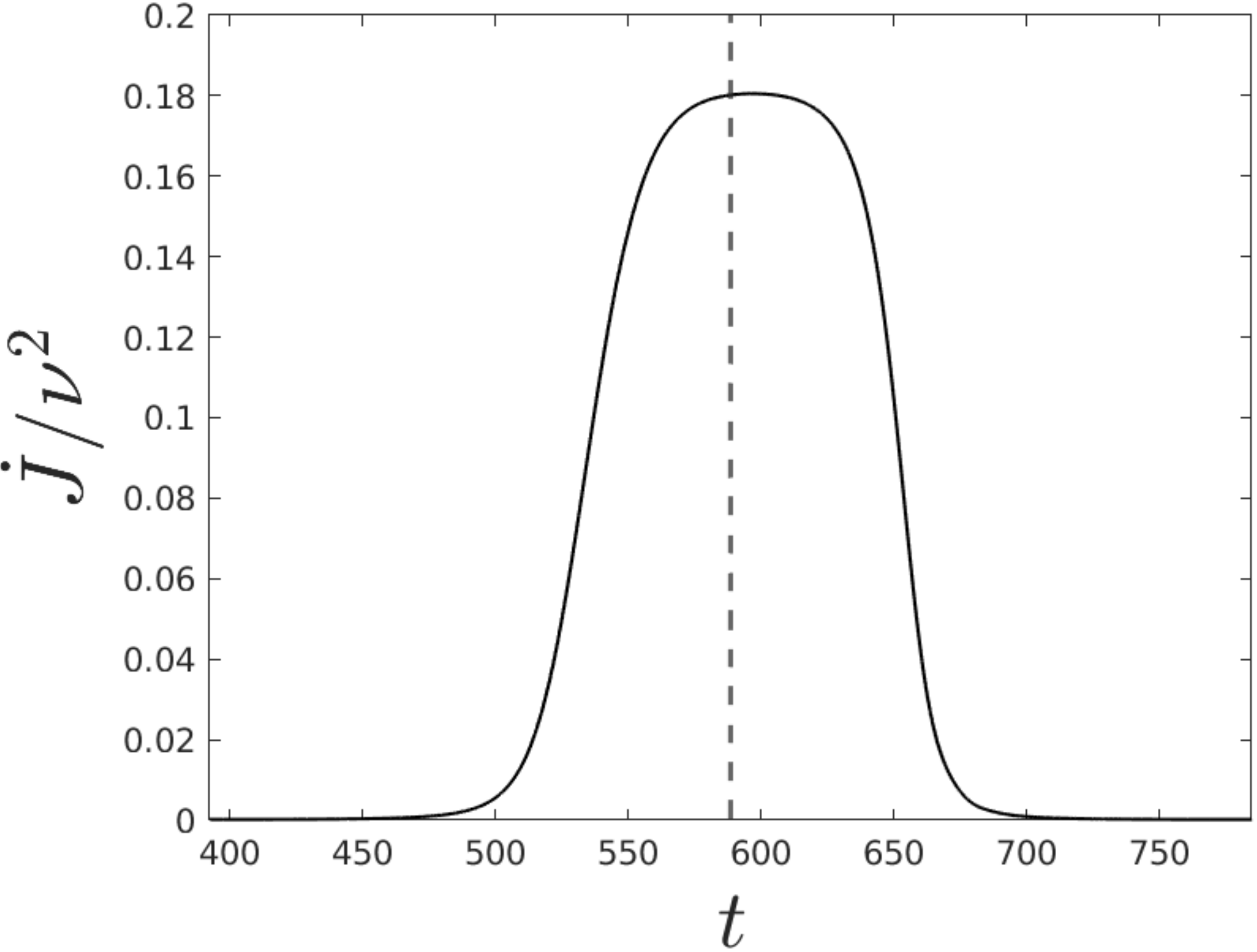} \\
	\caption{\label{fig:fig01} Top panels: a complete radial orbit of the simulation with $\ha=0.6$, 
	$e=0.7$ and $p=p_s+0.01\simeq 4.858$ and the corresponding real part of the dominant mode of the 
	numerical waveform. The dashed vertical line marks the periastron passage. Bottom panels: 
	energy and angular momentum fluxes summed up to $\l_{\rm max}=8$. Note the zoom-whirl 
	behavior and the  periastron precession. Moreover, it is possible to observe the fluxes 
	(and waveform) asymmetry that we will discuss in Sec.~\ref{sec:fluxes}.}
	\end{center}
\end{figure}
To apply black-hole perturbation theory to our waveform (and flux) calculation, we solve, in the time-domain, 
the Regge-Wheeler-Zerilli (RWZ)~\cite{Regge:1957td,Zerilli:1970se} 
or Teukolsky equations in the presence of a point-particle source, that is represented by
a $\delta$-function. Following a practice introduced long ago~\cite{Nagar:2004ns,Nagar:2006xv},
the $\delta$-function is approximated by a narrow Gaussian function.
In the Schwarzschild background case, we solve the RWZ equations in the time-domain using 
the \RWZ~code developed long ago~\cite{Bernuzzi:2010ty, Bernuzzi:2012ku}, that notably employs
the hyperboloidal layer method~\cite{Bernuzzi:2011aj} to extract the waveforms at future null infinity, 
avoiding errors related to the wave extraction at finite radius. 
In the more general case of a rotating black hole, we use {\Teukode}~\cite{Harms:2014dqa} to
solve the Teukolsky equation in the time-domain. In particular, the code uses horizon-penetrating and 
hyperboloidal coordinates that allow for the inclusion of the horizon and the future null infinity 
in the computational domain~\cite{Zenginoglu:2007jw,Zenginoglu:2009hd,Zenginoglu:2010cq}.
The 3+1 equation is decomposed exploiting the axisimmetry of the Kerr spacetime obtaining a 2+1 TD 
equation for each Fourier $m$-mode in the azimuthal direction. Then the wave equation is solved for 
gravitational perturbations, obtaining the Weyl scalar $\Psi_4$, i.e. the contraction of the Weyl scalar 
with a null-tetrad (the Hawking-Hartle tetrad in our case). Each waveform multipole is then obtained
by a double time integration of the Weyl scalar, since at infinity $\ddot{h}_\lm = \Psi_4^\lm$, 
where $\Psi_4^\lm$ is the multipolar decomposition  of $\Psi_4$. Analogously to 
the {\RWZ}~code,  the formal Dirac $\delta$ functions present in the source term are approximated using narrow Gaussian functions.
In our simulations, we use  horizon-penetrating, hyperboloidal coordinates with scri-fixing 
at $S=10$ and resolution of $N_r\times N_\theta = 3600\times 160$, where $(N_r,N_\theta)$
are the number of points in the radial and angular directions respectively.
From the Hamiltonian dynamics for the particle discussed in Sec.~\ref{sec:Ham_eq}, 
we compute the waveforms and the fluxes at infinity using the two numerical 
codes exposed above. {\Teukode} is more general than {\RWZ}, but the computational 
time is also greater. Nonetheless, we will use {\Teukode} also for nonspinning cases, 
with the only exception of few simulations with large semilatera recta ($p=21,31$). 
We will explicitly state when results from {\RWZ} are presented; otherwise, {\Teukode} 
is understood. An example of (zoom-whirl) geodesics dynamics ($\F_r=\F_\varphi=0$) with the 
corresponding $\ell=m=2$ waveform multipole and fluxes is shown in Fig.~\ref{fig:fig01}.

\subsection{Comparisons with previous work}
\label{subsec:comparisons_literature}
\begin{table*}[t]
\centering 
\caption{\label{tab:TeukRWZcheck} Averaged numerical fluxes computed with {\RWZ} and {\Teukode} 
compared with results present in the literature, see discussion in the text.}
\centering
\begin{ruledtabular}
\begin{tabular}{lc|cccccc}  
     & & {\Teukode} (TD) &  {\Teukode} (TD) & {\RWZ} (TD) & Martel (TD) & Barack (GSF) &  Fujita (FD) \\ 
     & & $3600\times 160$& $5400\times 320$ &             &             &              &              \\
\hline
\hline
$p = 7.50478$  & $\av{ \dot{E}} \cdot 10^{4}$ & 3.16885 & 3.16888 & 3.17077 & $3.1770$ & $3.1691$   & $ 3.16899989184$ \\
$e = 0.188917$ & $\av{ \dot{J}} \cdot 10^{3}$ & 5.96731 & 5.96737 & 5.96998 & $5.9329$ & $5.967608$ &  $5.96755215608$ \\
\hline
$p = 8.75455$  & $\av{ \dot{E}} \cdot 10^{4}$ & 2.12276 & 2.12269 & 2.12718 & $2.1484$ & $2.1243$  &  $ 2.12360313326$ \\
$e = 0.764124$ & $\av{ \dot{J}} \cdot 10^{3}$ & 2.77643 & 2.77635 & 2.78077 & $2.7932$ & $2.77746$ &  $2.77735938996$ \\
\end{tabular}
\end{ruledtabular}
\end{table*}
Both {\RWZ} and {\Teukode} have never been used systematically 
for eccentric runs before (see however Ref.~\cite{Chiaramello:2020ehz}), 
therefore we need to test the consistency of our results with published results.
To do so, we compare the averaged fluxes along a radial orbit for both Schwarzschild
and Kerr backgrounds. 

For the nonspinning case, the comparisons are shown in Table~\ref{tab:TeukRWZcheck}, 
where our $\nu$-normalized fluxes averaged over a radial period 
(see Eq.~\eqref{eq:averaged_fluxes} below) 
are compared with the classic TD results of Martel~\cite{Martel:2003jj},
as well as to more recent Gravitation-Self-Force (GSF) calculations of Barack 
and Sago~\cite{Barack:2010tm} and FD results of Fujita~\cite{Fujita:2009us}. 
Note that we sum the multipoles up 
to $\ell_{\rm max}=8$, but in the other works $\ell_{\rm max}$ assumes different values. 
Despite these differences, the agreement with previous computations remains
satisfactory, since the higher multipoles are highly subdominants. 
We can also observe that among the TD codes, {\Teukode} is the most accurate one. 
Note that the difference between resolutions $3600\times 120$ and $5400\times 320$ 
is very small. Since in the second case the computational time increases by 
a factor 6, we will use the former resolution.
We have also compared the instantaneous energy and angular momentum fluxes 
of {\Teukode} and {\RWZ} in the cases $(e,p) = (0.3,9), (0.5, 11)$, $(0.8, 13)$, 
and we have seen that the relative difference between the two numerical fluxes 
reaches its maximum at periastron and it is at most of $0.3\%$.

In the presence of black hole spin, we  have compared the averaged fluxes for $\ha=0.5$ 
and $e=0.5$ obtained using {\Teukode} with the results of Glampedakis and Kennefick~\cite{Glampedakis:2002ya}. 
The comparison, shown in Table~\ref{tab:TeukcheckSpin}, highlights the good agreement, 
with discrepancies below $0.2\%$.
Note however that, as above, we fixed $\l_{\rm max}=8$, while Ref.~\cite{Glampedakis:2002ya} 
sums up to $\l_{\rm max}=10-17$ and this affects the comparison. In fact, in the 
first case reported in Table~\ref{tab:TeukcheckSpin}, the $\l=8$ modes have a relative contribution 
of $\sim 0.25\%$ to the total fluxes, therefore including also the multipoles with $\l>8$ 
would probably improve the agreement. Note also that when the semilatus rectum is increased,
the agreement improves, because  the higher modes become less and less relevant. 
We have also considered a configuration with $\ha=0.9$, $e\simeq 0.3731$ and $p\simeq 12.152$,
calculated both in Ref.~\cite{Shibata:1994xk} and Ref.~\cite{Glampedakis:2002ya}. In this case, 
the contribution of the $\l=8$ modes is only of the order of $2\cdot 10^{-5}$ due to the large 
value of $p$ used, and as a consequence the discrepancy is smaller. 
Nonetheless, comparing the energy and angular momentum fluxes with 
the results of Ref.~\cite{Shibata:1994xk}, we have found, respectively, discrepancies of 
$1.4\%$ and of $0.8\%$, confirming the $\sim 1\%$ disagreement already found by 
 Ref.~\cite{Glampedakis:2002ya}. 
We can thus conclude that our numerical computations of the fluxes along eccentric orbits 
are consistent with all results already present in the literature. Our numerical approach is then
expected to faithfully describe fluxes from eccentric orbits both on Schwarzschild and Kerr spacetimes. 
\begin{table}[t]
\centering 
\caption{\label{tab:TeukcheckSpin} Numerical averaged fluxes in Kerr spacetime. We compare the 
results of {\Teukode} ($\l_{\rm max}=8$) with the results obtained 
by Ref.~\cite{Glampedakis:2002ya} ($\l_{\rm max} = 10-17$). For the last simulation, 
only $\av{ \dot{E}_{GK}}$ is reported in Ref.~\cite{Glampedakis:2002ya}.}
\centering
\begin{ruledtabular}
\begin{tabular}{ccc|cc|cc}  
$\ha$ & $e$ & $p$ & $\av{ \Eteuk }$ & $\av{ \Jteuk }$  &  
$\av{ \dot{E}_{GK}}$ &  $\av{ \dot{J}_{GK}}$\\
&       &     &   $\cdot 10^3$ & $\cdot 10^2$ & $\cdot 10^3$ & $\cdot 10^2$ \\
\hline
\hline
0.5 & 0.5    & 5.1     & 4.20753  & 3.25791   & 4.21594   &   3.26383 \\
0.5 & 0.5    & 5.5     & 2.11538  & 1.89340   & 2.11797   &   1.89546 \\
0.5 & 0.5    & 6.0     & 1.19519  & 1.22870   & 1.19638   &   1.22973 \\
0.9 & 0.3731 & 12.152  & 0.023571 & 0.080743  & 0.023570  &      /    \\
\end{tabular}
\end{ruledtabular}
\end{table}
\subsection{Numerical geodesic simulations}
\label{subsec:144sims}
In order to meaningfully cover the parameter space, we run 144 geodesic simulations with {\Teukode} 
choosing eccentricities $e=(0, 0.1, 0.3,$ $ 0.5, 0.7, 0.9)$ and spins in the 
range $\ha\in[-0.9, 0.9]$, typically $\ha=(0, \pm 0.2, \pm 0.6, \pm 0.9)$. 
For each pair of spin and eccentricity $(\ha,e)$, we 
have chosen three different semilatera recta. The first is $p=p_s+0.01$, 
where $p_s=p_s(e,\ha)$ is the separatrix, while the other two are selected according 
to $p=p_{\rm schw} p_s(e,\ha)/p_s(e,0)$, where $p_{\rm schw}$ is 9 or 13. 
Depending on the value of the semilatus rectum, we will 
refer to the simulations, respectively, as \textit{near}, \textit{intermediate} and \textit{distant}. 
The near simulations exhibit a zoom-whirl behavior, while the others \textit{generally}
have eccentric orbits without whirls at periastron. 
The complete list of the {\Teukode} geodesic simulations with the corresponding averaged 
fluxes can be found in Appendix~\ref{appendix:Fluxes}. Hereafter, when we will report 
a semilatus rectum or a separatrix, we will truncate it to the third decimal. 

\section{Analytical waveforms and fluxes}
\label{sec:eobmodel}
\subsection{Waveform}
\label{sec:eobwave}
Let us turn now to the discussion of the factorized and resummed analytical waveforms 
and fluxes along eccentric orbits. The basic ideas are those introduced  in Ref.~\cite{Damour:2008gu}
for the circular case, that proposed a recipe to factorize and resum the PN-expanded 
waveform multipoles. A simple procedure to generalize it to the case of eccentric orbits 
was introduced in Ref.~\cite{Chiaramello:2020ehz}, that we review here in detail.
Each waveform multipole is factorized as
\begin{equation}
h_{\lm}= h_{\lm}^{(N, \epsilon)} \hat{h}^{(\epsilon)}_{\lm} = h_{\lm}^{(N, \epsilon)} \hat{S}^{(\epsilon)} \hat{h}_\lm^{\rm tail} (\rho_{\lm})^{\l} , \label{eq:eobwave}
\end{equation}
where $\epsilon$ denotes the parity of the multipole ($\epsilon = 0$ if $\l+m$ is even, 
$\epsilon=1$ if $\l+m$ is odd), $h_{\lm}^{(N, \epsilon)}$ is the Newtonian contribution 
and $\hat{h}^{(\epsilon)}_{\lm}$ is the PN correction. The term $\hat{S}^{(\epsilon)}$ 
is the effective-source term, i.e. the energy if $\epsilon=0$ or the Newton-normalized
angular momentum if $\epsilon=1$; $\hat{h}_\lm^{\rm tail}=T_{\lm} e^{i \delta_{\lm}}$ is 
the tail factor and the $\rho_{\lm}$  are the residual amplitude corrections. 
While the $S^{(\epsilon)}$ is general because it is computed along the general dynamics,
the functions $\hat{h}_\lm^{\rm tail}$ and $\rho_\lm$ correspond to those obtained in
the circular case (see Appendix~\ref{appendix:circular_eob} for more details).
Following Ref.~\cite{Chiaramello:2020ehz}, the multipolar waveform is generalized to
generic orbits (including open orbits~\cite{Nagar:2020xsk,Nagar:2021gss}) by simply 
replacing the Newtonian quasi-circular prefactor with its general expression.
We remark that we are considering the waveform at leading order in $\nu$,
therefore we switch off all the subleading $\nu$-dependencies in the factors of 
Eq.~\eqref{eq:eobwave},  except for the leading Newtonian contribution $O(\nu)$.

The Newtonian contribution of the waveform is obtained from the derivatives of the source 
multipoles, explicitly 
\begin{equation}
\label{eq:waveNewt}
h_{\lm}^{(N,\epsilon)} \propto 
\begin{cases} 
I_{\lm}^{(\l)} &  {\rm if} \;\;\; \epsilon=0 \\
S_{\lm}^{(\l)} &  {\rm if} \;\;\; \epsilon=1  \\
\end{cases} ,
\end{equation}
where $I_{\lm}^{(\l)}$ and $S_{\lm}^{(\l)}$ are the $\l$th-derivatives of the mass 
and current source multipoles. For a test-particle orbiting in the equatorial plane of a 
Kerr black hole, they are given by
\begin{equation}
\begin{split}
I_{\lm} & \propto \nu r^{\l} e^{-i m \varphi} ,\\
S_{\lm} & \propto \nu r^{\l+1} \Omega e^{-i m \varphi}.
\end{split}
\end{equation}
For circularized binaries, the derivatives of the radius 
and the orbital frequency are zero, but in the more general case they are not vanishing. 
Therefore the waveform can be generalized~\cite{Chiaramello:2020ehz} (at Newtonian level) 
without neglecting the derivatives of the radius and of the orbital frequency, obtaining 
a general Newtonian contribution 
$h_{\lm}^{(N, \epsilon)_{\rm  tot}}$ that can be separated in circular and noncircular 
factors, respectively $h_\lm^{(N,\epsilon)_c}$ and $h_\lm^{(N,\epsilon)_{nc}}$. Then the 
full multipolar waveform can be written as
\begin{equation}
\label{eq:ecc_wave}
h_{\lm} = h_{\lm}^{(N, \epsilon)_{\rm c}} \hat{h}_{\lm}^{(N, \epsilon)_{\rm  nc}} \hat{h}^{(\epsilon)}_{\lm} ,
\end{equation} 
where
\begin{align}
\hat{h}_{\lm}^{(N, 0)_{\rm nc}} & = \left( h_{\lm}^{(N,0)_{\rm c}} \right)^{-1} I_{\lm}^{(\l)} , \nonumber \\
\hat{h}_{\lm}^{(N, 1)_{\rm nc}} & = \left( h_{\lm}^{(N,1)_{\rm c}} \right)^{-1} S_{\lm}^{(\l)} ,\nonumber\\
h_{\lm}^{(N,\epsilon)_{\rm c}} \;\, & = h_{\lm}^{(N, \epsilon)_{\rm tot}}(r,\Omega, \dot{r}=0, \dot{\Omega}=0,...).\nonumber
\end{align}
For the dominant $(2,2)$ mode, the Newtonian noncircular correction reads 
\begin{equation}
\label{eq:NewtPref22_wave}
\hat{h}_{22}^{(N, 0)_{\rm nc}}  = 1 - \frac{\ddot{r}}{2 r \Omega^2} - \frac{\dot{r}^2}{2 r^2 \Omega^2} + \frac{2 \mathrm{i} \dot{r}}{r \Omega} + \frac{\mathrm{i} \dot{\Omega}}{2 \Omega^2}.
\end{equation}
The explicit noncircular corrections for the most relevant subdominant modes 
can be found in Appendix~\ref{appendix:newtprefs}. 

\subsection{Energy and angular momentum fluxes}
\label{sec:radreac}
The radiation reaction forces $(\F_\varphi, \F_r)$ are obtained requiring the equality between the loss 
of mechanical energy and angular momentum of the system and the energy and angular momentum 
fluxes carried by the GW at infinity, $\dot{E}$ and $\dot{J}$. 
Using the angular and radial components of the radiation reaction
and the equations of motion, the balance equations read~\cite{Bini:2012ji}
\begin{subequations}
\label{eq:enbalance}
\begin{align}
\label{eq:Fr}
\dot{r}\F_r + \Omega \F_\varphi + \dot{E}_{\rm Schott} + \dot{E} = 0 ,\\
\label{eq:Fphi}
\F_\varphi + \dot{J} = 0.
\end{align}
\end{subequations}
where $\dot{E}_{\rm Schott}$ is the time-derivative of the Schott energy, that represents 
the interaction of the source with the local field and its orbital average goes to zero. 
The Schott contribution to the angular momentum, $\dot{J}_{\rm Schott}$, can be gauged away~\cite{Bini:2012ji}. 
In the circular case, $\F_r$ and $\dot{E}_{\rm Schott}$ vanish and we only have the 
angular component of the radiation reaction, that is typically written as
\begin{equation}
\label{eq:circFphi}
\hat{\F}_\varphi = - \frac{32}{5} \nu r_\Omega^4 \Omega^5 \hat{f}, 
\end{equation}
where $\hat{f}$ is Newton-normalized flux function, that incorporates
all the resummed PN corrections, and
\begin{equation}
\label{eq:rOmg}
r_\Omega = r \left( 1 + \ha r^{-3/2} \right)^{2/3}.
\end{equation} 
The form of $\hat{f}$ is reminded in Appendix~\ref{appendix:circular_eob}.
We recall that the analytic expression we use here relies on several
previous works~\cite{Damour:2008gu,Nagar:2016ayt,Messina:2018ghh} 
and, in particular, it uses resummed multipoles up to $\ell=8$.
In Ref.~\cite{Chiaramello:2020ehz}, it was proposed to include noncircular
effects in the angular component of the radiation reaction by means 
of the leading, quadrupolar, noncircular factor $\NP22$, 
\begin{equation}
\label{eq:Fphi_ecc_old}
\hat{\F}_\varphi  = - \frac{32}{5} \nu r_\Omega^4 \Omega^5  \NP22 \hat{f} \equiv \Fphiold,\\
\end{equation}
that reads
\begin{align}
\label{eq:NewtPref22_flux}
\fnp{,22} = 1&+\frac{3 \dot{r}^4}{4 r^4 \Omega ^4}+\frac{3 \dot{r}^3 {\dot{\Omega}}}{4 r^3 \Omega ^5}+\frac{3 \ddot{r}^2}{4 r^2 \Omega ^4}+\frac{3 \ddot{r} \dot{r} {\dot{\Omega}}}{8 r^2 \Omega ^5} \nonumber \\
&-\frac{{r^{(3)}} \dot{r}}{2 r^2 \Omega ^4}+\frac{\dot{r}^2 {\ddot{\Omega}}}{8 r^2 \Omega ^5}+\frac{4 \dot{r}^2}{r^2 \Omega ^2}+\frac{\ddot{r} {\ddot{\Omega}}}{8 r \Omega ^5} \nonumber \\
&-\frac{2 \ddot{r}}{r \Omega ^2}-\frac{{r^{(3)}} {\dot{\Omega}}}{8 r \Omega ^5}+\frac{3 \dot{r} {\dot{\Omega}}}{r \Omega ^3}+\frac{3 {\dot{\Omega}}^2}{4 \Omega ^4}-\frac{{\ddot{\Omega}}}{4 \Omega ^3}  .
\end{align}
This factor is simply the next-to-quasi-circular part of the Newtonian contribution 
to the angular momentum
flux obtained from the Newtonian quadrupolar waveform  $h_{22}^{(N,0)}$. More precisely, one has
\be
\dot{J}^{\rm N}_{22}/\nu \equiv -\dfrac{1}{8\pi}\sum_{m=\pm 2}\Im\left(\dot{h}_{2m}^{(N,0)} h_{2m}^{(N,0)*}\right)= \dfrac{32}{5}\nu r^4 \Omega^5 \NP22,
\ee 
that, after replacing $r$ with $r_\Omega$, changing sign, and considering $\hat{f}$,
finally yields Eq.~\eqref{eq:Fphi_ecc_old}.
Although this choice was used in previous works~\cite{Chiaramello:2020ehz,Nagar:2020xsk} it 
is slightly incorrect, since all subdominant flux multipole are multiplied by the quadrupolar 
noncircular factor. A more consistent approach was proposed in ~\cite{Nagar:2021gss},
where the noncircular correction factor was applied {\it only} to the $\ell=m=2$ multipole.
The angular radiation reaction in this case is written 
as 
\begin{subequations}
\label{eq:Fphi_ecc}
\begin{align}
\label{eq:Fphi_ecc_a}
\FphiNP & =  - \frac{32}{5} \nu r_\Omega^4 \Omega^5  \hat{f}_{{\rm nc}_{22}},\\
\label{eq:Fphi_ecc_b}
\hat{f}_{{\rm nc}_{22}}  & \equiv \hat{F}_{22} \NP22 + \hat{F}_{21} + \sum_{\l\geq 3} \sum_{m=1}^{\l} \hat{F}_{\lm},
\end{align}
\end{subequations}
where $\hat{F}_\lm  = F_\lm /F_{22}^N $ are the Newton-normalized energy fluxes defined 
in Eq.~\eqref{eq:hatf}. In practice, the global factor $\NP22$ 
of Eq.~\eqref{eq:Fphi_ecc_old}
is now a factor of the $\l=m=2$ multipole and we are neglecting the noncircular 
corrections of all the subdominant multipoles. 
A straightforward generalization of Eq.~\eqref{eq:Fphi_ecc} can be obtained 
considering also the noncircular corrections from the subdominant multipoles: 
\begin{equation}
\label{eq:Fphi_ANP}
\begin{split}
\FphiANP = & - \sum_{\l=2}^{6} \sum_{m=1}^{\l} \dot{J}_{\lm}^{(N_{\rm qcirc}, \epsilon)} \fnp{,\lm} |\hat{h}_{\lm}^{(\epsilon)}|^2 \\
 & - \sum_{\l=7}^{\l_{\rm max}} \sum_{m=1}^{\l} \dot{J}_{\lm}^{(N_{\rm qcirc}, \epsilon)} |\hat{h}_{\lm}^{(\epsilon)}|^2,
\end{split}
\end{equation}
where $ J_{\lm}^{(N_{\rm qcirc},\epsilon)} \propto \nu r_\Omega^{2(\l+\epsilon)} \Omega^{2(\l+\epsilon) +1 }$
is the quasi-circular Newtonian contribution to the angular momentum flux\footnote{ For the $\ell=m=2$ mode this is 
$\dot{J}_{22}^{(N_{\rm qcirc}, 0)} = 32/5 \;\nu  r_\Omega^4 \Omega^5$.}, 
$\fnp{,\lm}$ is the Newtonian noncircular factor and $\hat{h}_{\lm}^{(\epsilon)}$ 
is the full PN (circular) correction introduced in Eq.~\eqref{eq:eobwave}. 
For simplicity, we consider noncircular Newtonian prefactors only up to $\l=6$, 
since the others contributions are anyway negligible. The noncircular Newtonian Prefactor 
$\fnp{,\lm}$ for the most relevant subdominant modes can be found in Appendix~\ref{appendix:newtprefs}.

For the radial component of the radiation reaction, $\hat{\F}_r$, we used the 2PN results of
Ref.~\cite{Bini:2012ji} Pad\'e resummed as in ~\cite{Chiaramello:2020ehz}
\begin{equation}
\label{eq:Fr}
\hat{\F}_r = \frac{32}{3} \nu \frac{p_{r_*}}{r^4} P^0_2 [ \hat{f}_r^{\rm  BD} ] ,
\end{equation}
where $P^0_2$ is the $(0,2)$ Pad\'e approximant. 
The explicit expression for the 2PN terms read
\begin{align}
\hat{f}_r^{\rm  BD} =& \hat{f}_{r}^{\rm N} + \hat{f}_{r}^{\rm 1PN} + \hat{f}_{r}^{\rm 2PN}, 
\end{align}
where (see also Ref.~\cite{Nagar:2021gss})
\begin{subequations}
\label{eq:FrPN}
\begin{align}
\hat{f}_{r}^{N} =& -\frac{8}{15} + \frac{56}{5} \frac{p_{\varphi}^{2}}{r}, \\
\hat{f}_{r}^{\rm 1PN} =& -\frac{1228}{105} p_{r_*}^2-\frac{1984}{105} \frac{1}{r}-\frac{124}{105} \frac{p_{r_*}^2 p_{\varphi}^2}{r} \nonumber \\ 
&+\frac{1252}{105} \frac{p_{\varphi}^4}{r^3} -\frac{1696}{35} \frac{p_{\varphi}^2}{r^2},  \\
\hat{f}_{r}^{\rm 2PN} =& \frac{323}{315}p_{r_*}^4+\frac{59554}{2835}{r^2}-\frac{1774}{21} \frac{p_{r_*}^2 p_{\varphi}^2}{r^2}  \nonumber \\
& -\frac{628}{105} \frac{p_{r_*}^2 p_{\varphi}^4}{r^3}-\frac{29438}{315} \frac{p_{\varphi}^2}{r^3} -\frac{461}{315} \frac{p_{r_*}^4 p_{\varphi}^2}{r} \nonumber \\
&+\frac{20666}{315} \frac{p_{r_*}^2}{r}-\frac{3229}{315} \frac{p_{\varphi}^6}{r^5}-\frac{35209}{315} \frac{p_{\varphi}^4}{r^4}. 
\end{align}
\end{subequations}
Finally, for the Schott energy we also follow Ref.~\cite{Chiaramello:2020ehz}
\be
\label{eq:ESchott}
E_{\rm Schott}  = \frac{16}{5} \frac{p_{r_*}}{r^3} P^0_2[E_{\rm Schott}^{\rm c}] P^0_2[E_{\rm Schott}^{\rm nc}]  ,
\ee
where the circular and noncircular parts, taken at 2PN accuracy, are also Pad\'e resummed.
The two contributions explicitly read
\begin{align}
E_{\rm Schott}^{\rm c} =& 1 - \frac{157}{56} \frac{1}{r} -\frac{3421}{756}\frac{1}{r^2},  \\
E_{\rm Schott}^{\rm nc} =& \frac{p_{\varphi}^{2}}{r} -\frac{3}{2} p_{r_*}^2+\frac{2}{21} \frac{1}{r}-\frac{1}{2} \frac{p_{r_*}^2 p_{\varphi}^2}{r}  \nonumber \\
& +\frac{55}{168} \frac{p_{\varphi}^4}{r^3} -\frac{575}{168}
   \frac{p_{\varphi}^2}{r^2} + \frac{5}{8}
   p_{r*}^4  \nonumber \\
   &-\frac{2143}{5292} \frac{1}{r^2}-\frac{61}{48} \frac{p_{r_*}^2 p_{\varphi}^2}{r^2}-\frac{13}{168} \frac{p_{r_*}^2 p_{\varphi}^4}{r^3} \nonumber \\ 
   &+\frac{370189}{84672} \frac{p_{\varphi}^2}{r^3} + \frac{3}{8}
   \frac{p_{r_*}^4 p_{\varphi}^2}{r}-\frac{181}{112} \frac{p_{r*}^2}{r}  \nonumber \\ 
   &-\frac{25}{504} \frac{p_{\varphi}^6}{r^5}-\frac{130223}{28224} \frac{p_{\varphi}^4}{r^4} . 
\end{align}
The functions $(\F_\varphi,\F_r,E_{\rm Schott})$ are then computed along a given 
eccentric (typically geodesic) dynamics and then, using Eqs.~\eqref{eq:enbalance},
eventually yield expressions for the analytical GW fluxes $(\dot{E},\dot{J})$ at infinity.
Given the various possible analytical prescriptions for $\F_\varphi$ that we have discussed
so far, we will consider three different possibilities labeled as follows:
\begin{enumerate}
\item[(i)] $\ENP$, $\JNP$ are computed using $\FphiNP$ from 
Eq.~\eqref{eq:Fphi_ecc}. 
These are the most relevant fluxes for our purposes since they will eventually be our preferred choice
to drive the transition from the eccentric inspiral to plunge, merger and ringdown. Note that 
this choice is the same implemented for the EOB eccentric model of Ref.~\cite{Nagar:2021gss}.
\item[(ii)] $\ENPold$, $\JNPold$ are computed using $\Fphiold$ from Eq.~\eqref{eq:Fphi_ecc_old}, 
i.e. these are fluxes used with the old prescription used in the original EOB eccentric model of 
Ref.~\cite{Chiaramello:2020ehz} as well as in its extension for hyperbolic motion and
dynamical capture~\cite{Nagar:2020xsk}.
\item[(iii)] $\EANP$, $\JANP$ are computed using $\FphiANP$ from Eq.~\eqref{eq:Fphi_ANP}.
\end{enumerate}
We will compare the analytical fluxes with the numerical ones in order to establish 
the strong-field reliability of $(\F_\varphi,\F_r)$. When looking at the instantaneous fluxes,
that provide direct insights on the quality of these functions, the comparison will also involve
$\dot{E}_{\rm Schott}$. By contrast, when considering orbital-averaged fluxes, $E_{\rm Schott}$
does not play any role because its orbital average vanishes, $\av{ \dot{E}_{\rm Schott} } = 0$.

As a final possible analytical choice, one can also compute the fluxes $(\Jhlm,\Ehlm)$
by simply inserting Eq.~\eqref{eq:ecc_wave} into Eqs.~\eqref{eq:fluxes_infty}.  Although
these expressions cannot be conveniently employed to drive the Hamiltonian dynamics,
they can serve as additional consistency check of the waveform. They will be explicitly
discussed in Fig.~\ref{fig:instaFluxes} below.

\subsection{Factorization and resummation of the $\ell=7$ and $\ell=8$ circular amplitudes}
\label{sec:resum_l7_l8}
%
\begin{figure*}[]
  \center
  \includegraphics[width=0.3\textwidth,height=4.0cm]{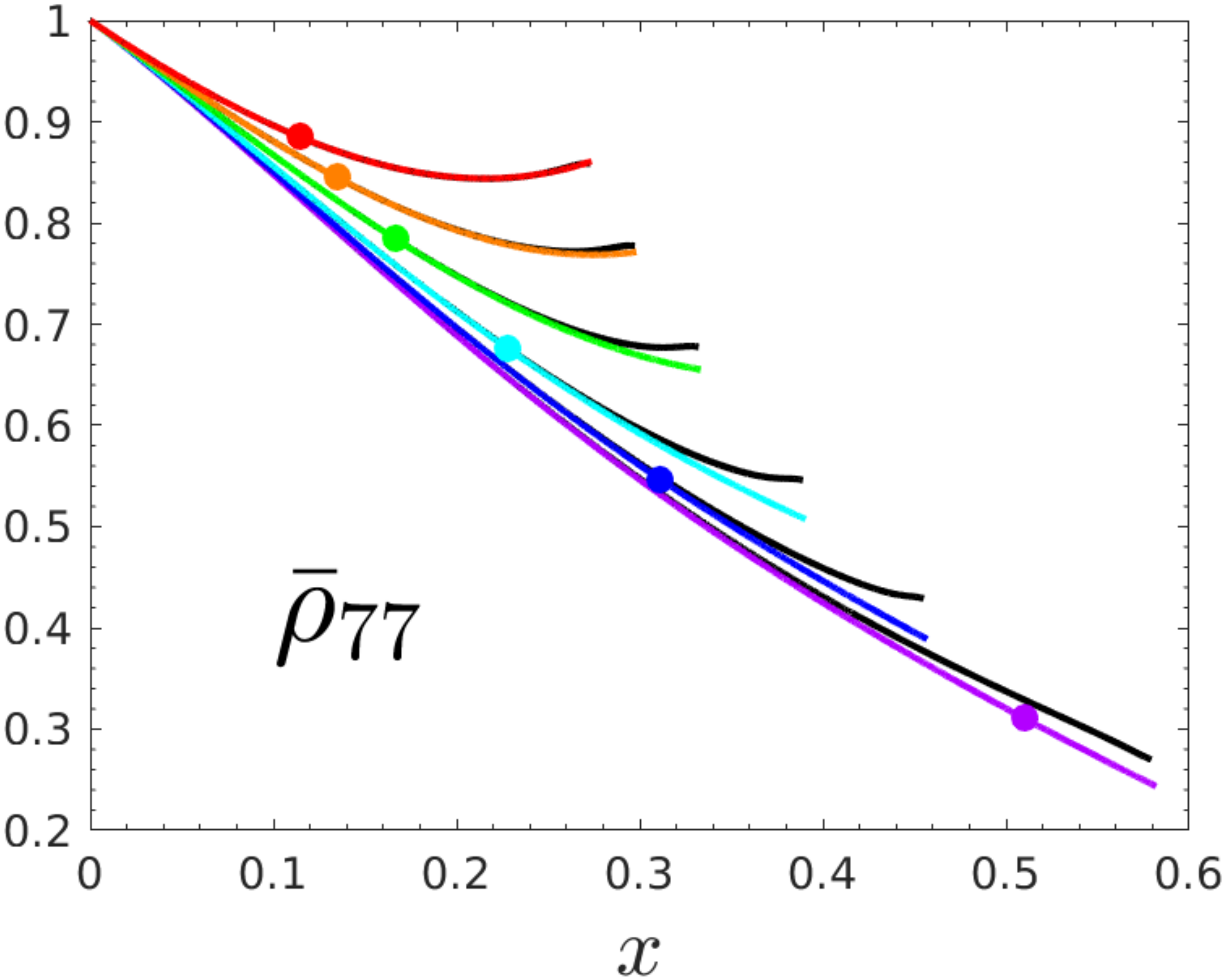} 
  \includegraphics[width=0.3\textwidth,height=4.0cm]{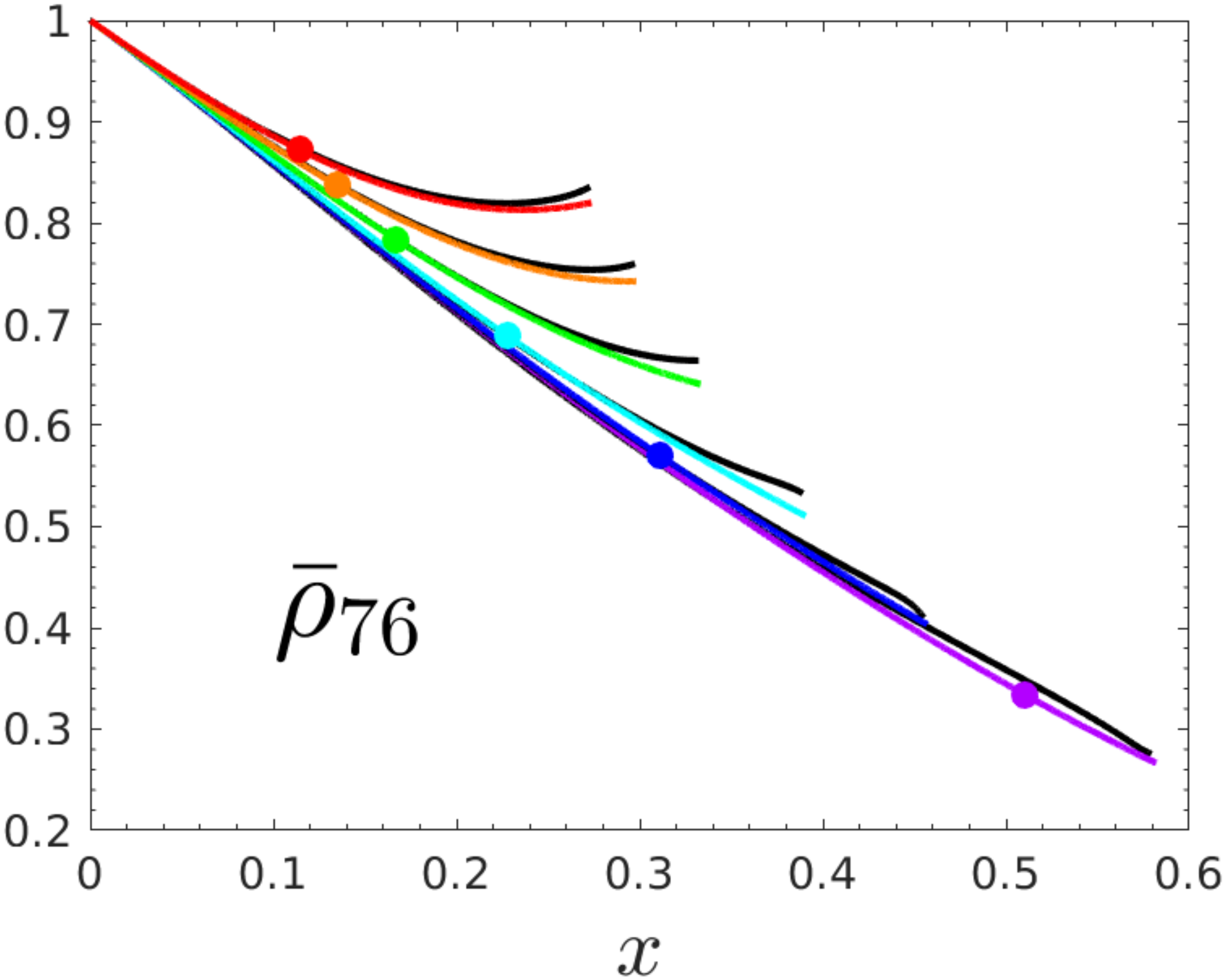} 
  \includegraphics[width=0.3\textwidth,height=4.0cm]{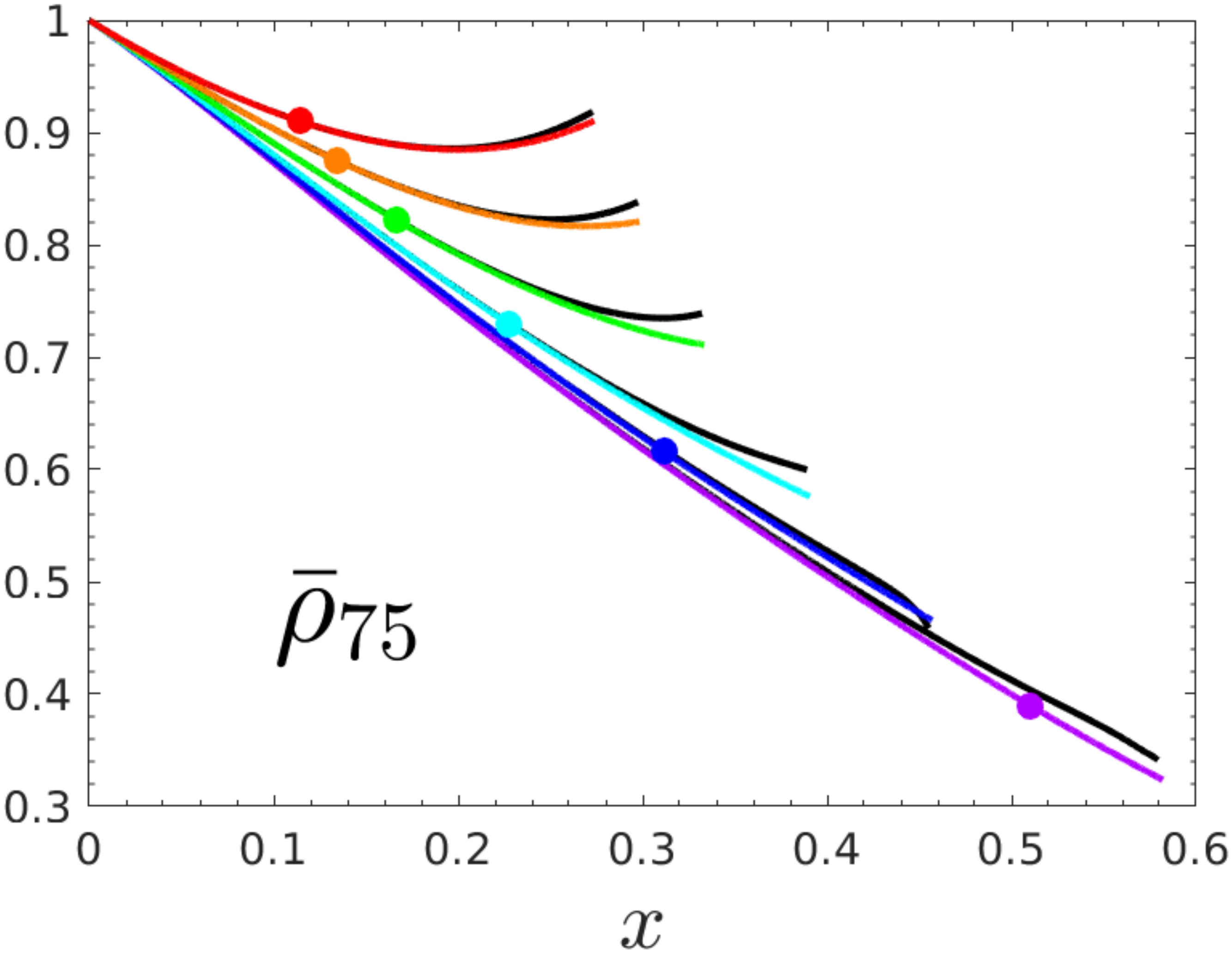} \\
  \includegraphics[width=0.3\textwidth,height=4.0cm]{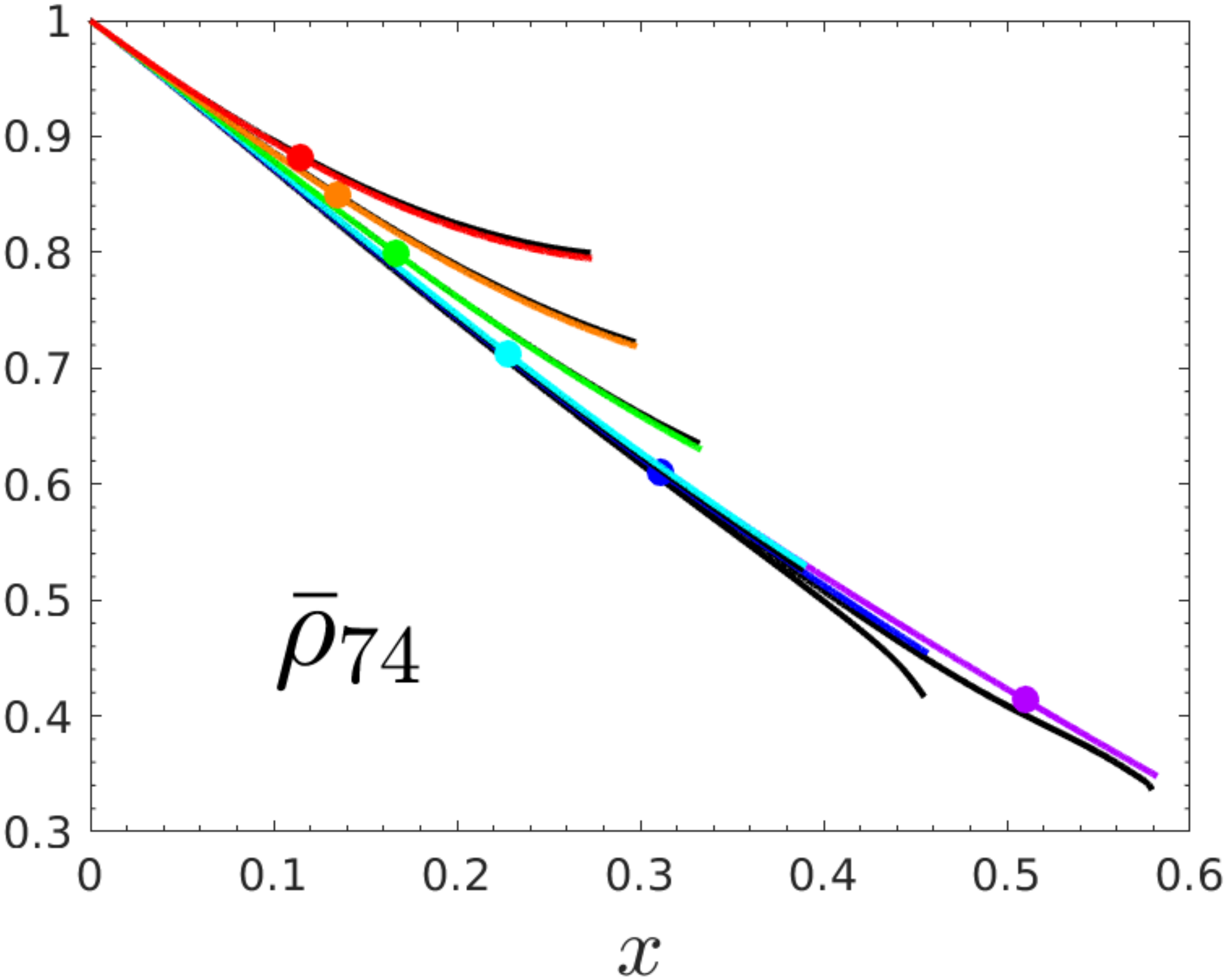} 
  \includegraphics[width=0.3\textwidth,height=4.0cm]{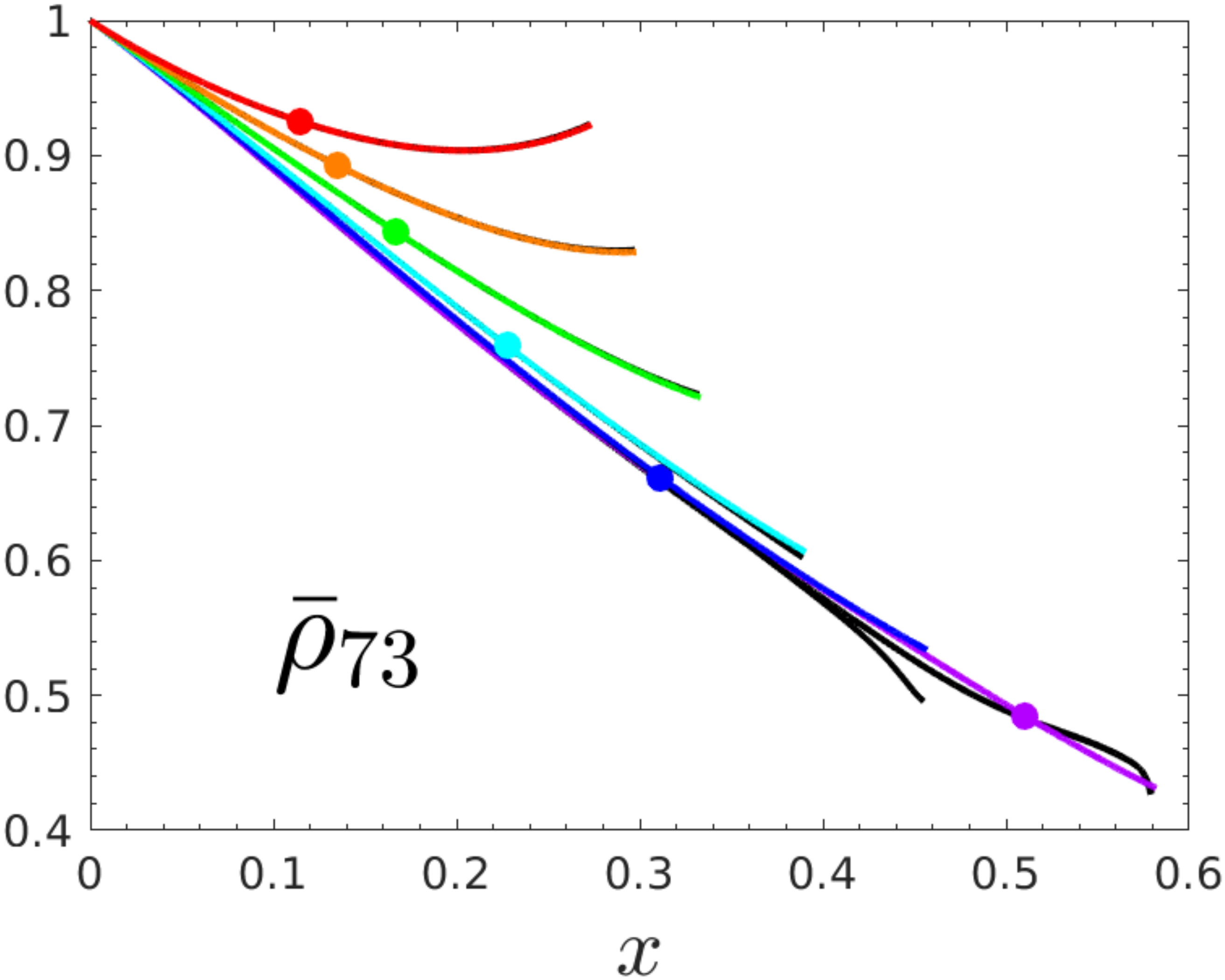} 
  \includegraphics[width=0.3\textwidth,height=4.0cm]{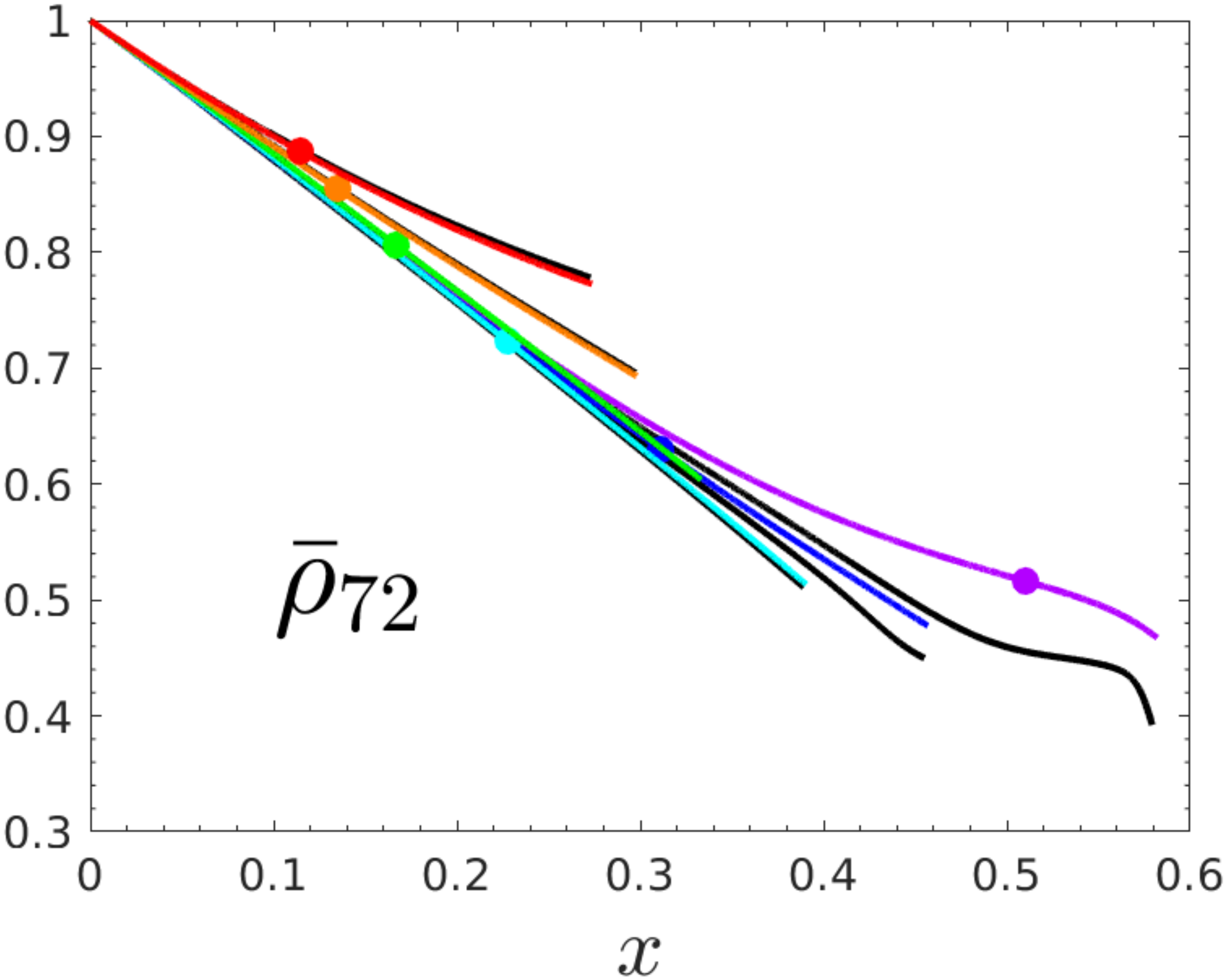} \\
  \includegraphics[width=0.3\textwidth,height=4.0cm]{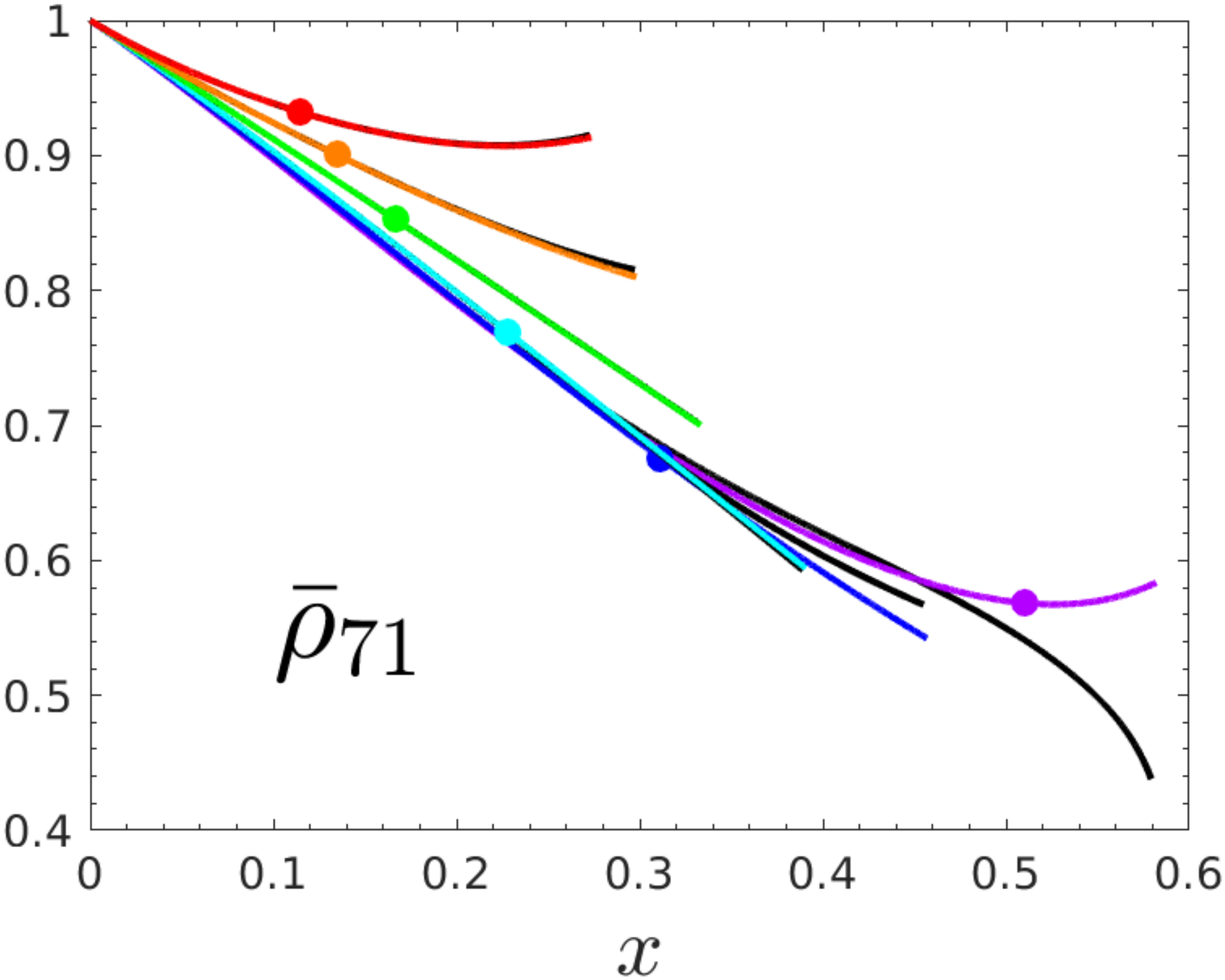} 
  \includegraphics[width=0.3\textwidth,height=4.0cm]{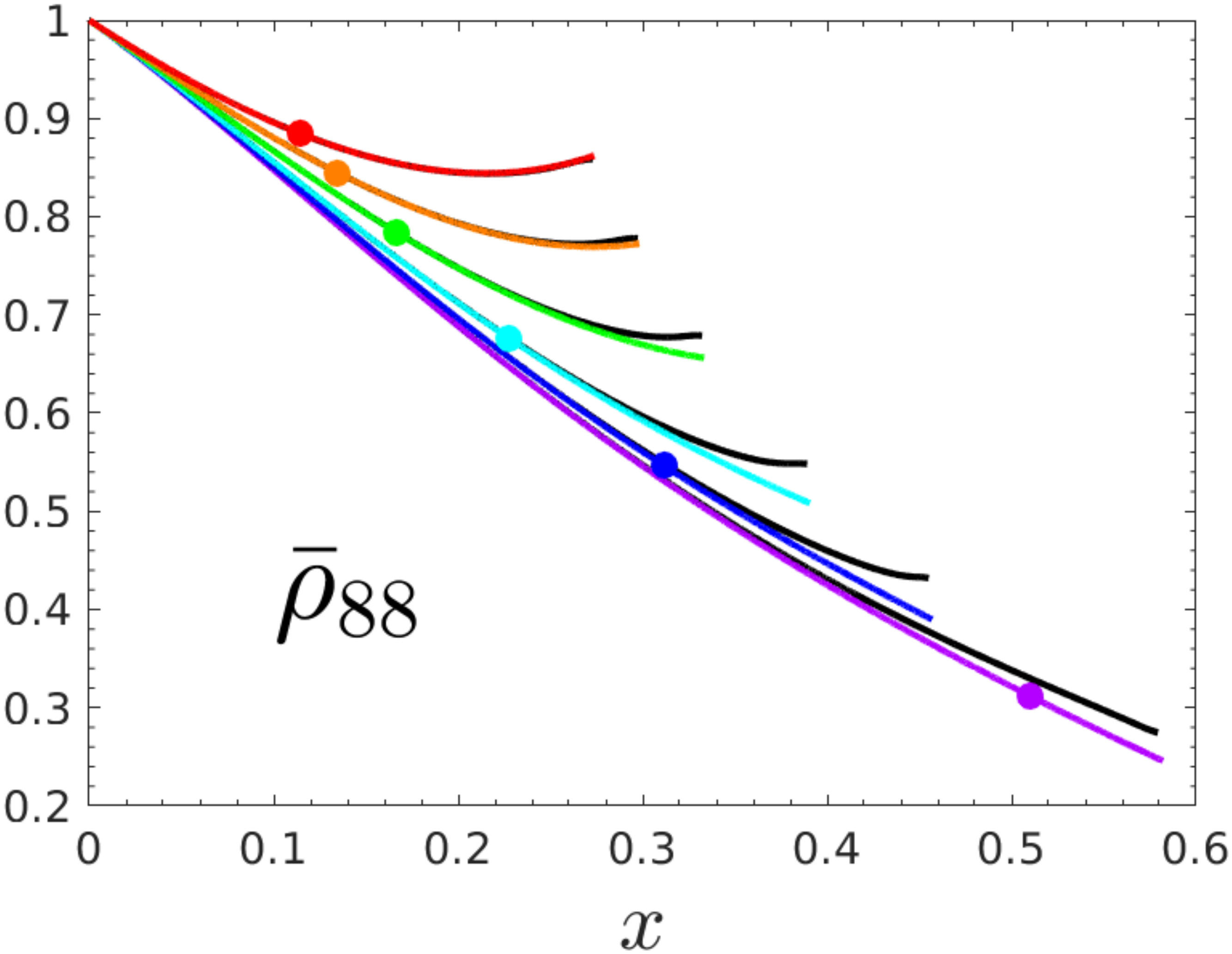} 
  \includegraphics[width=0.3\textwidth,height=4.0cm]{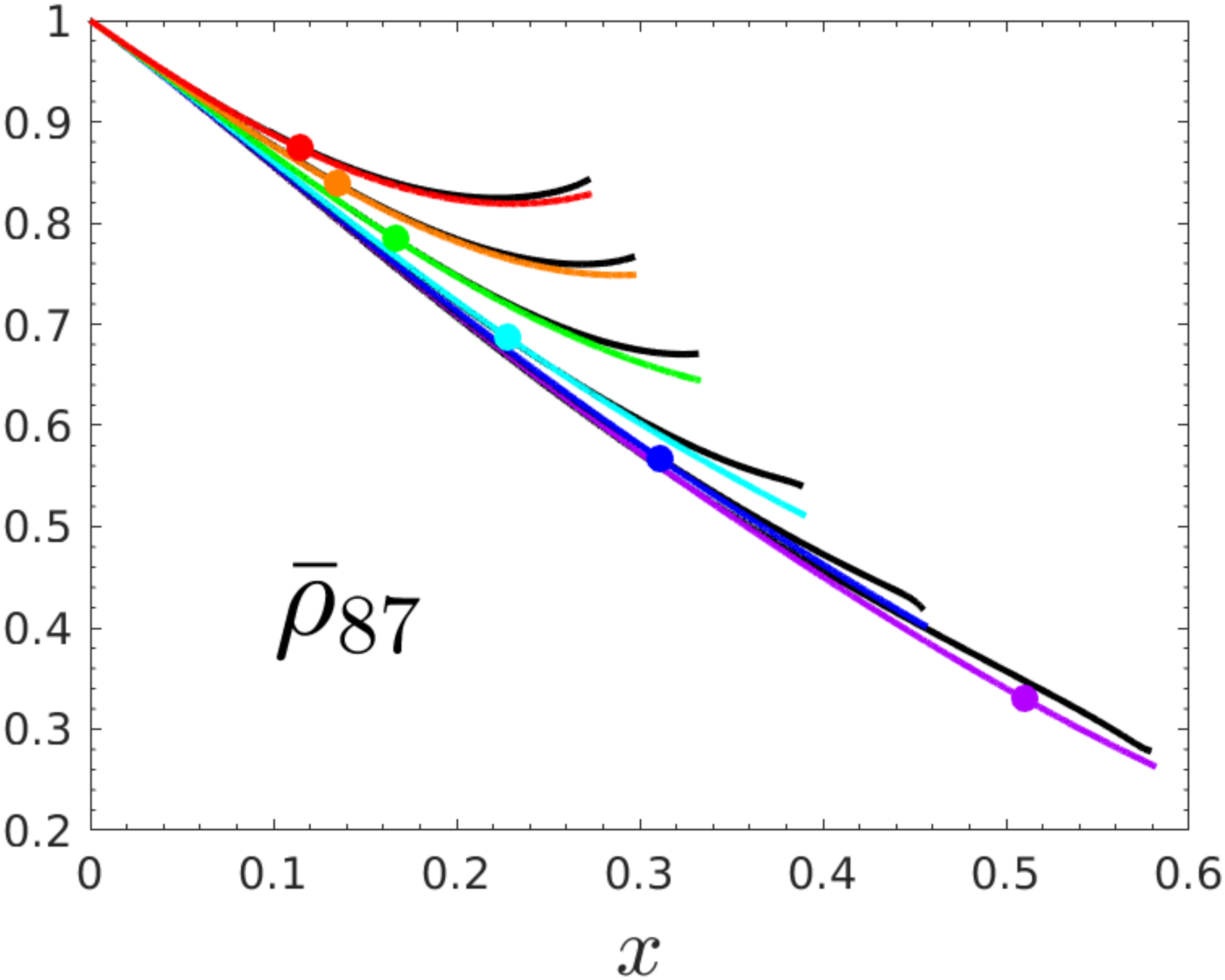} \\ 
  \includegraphics[width=0.3\textwidth,height=4.0cm]{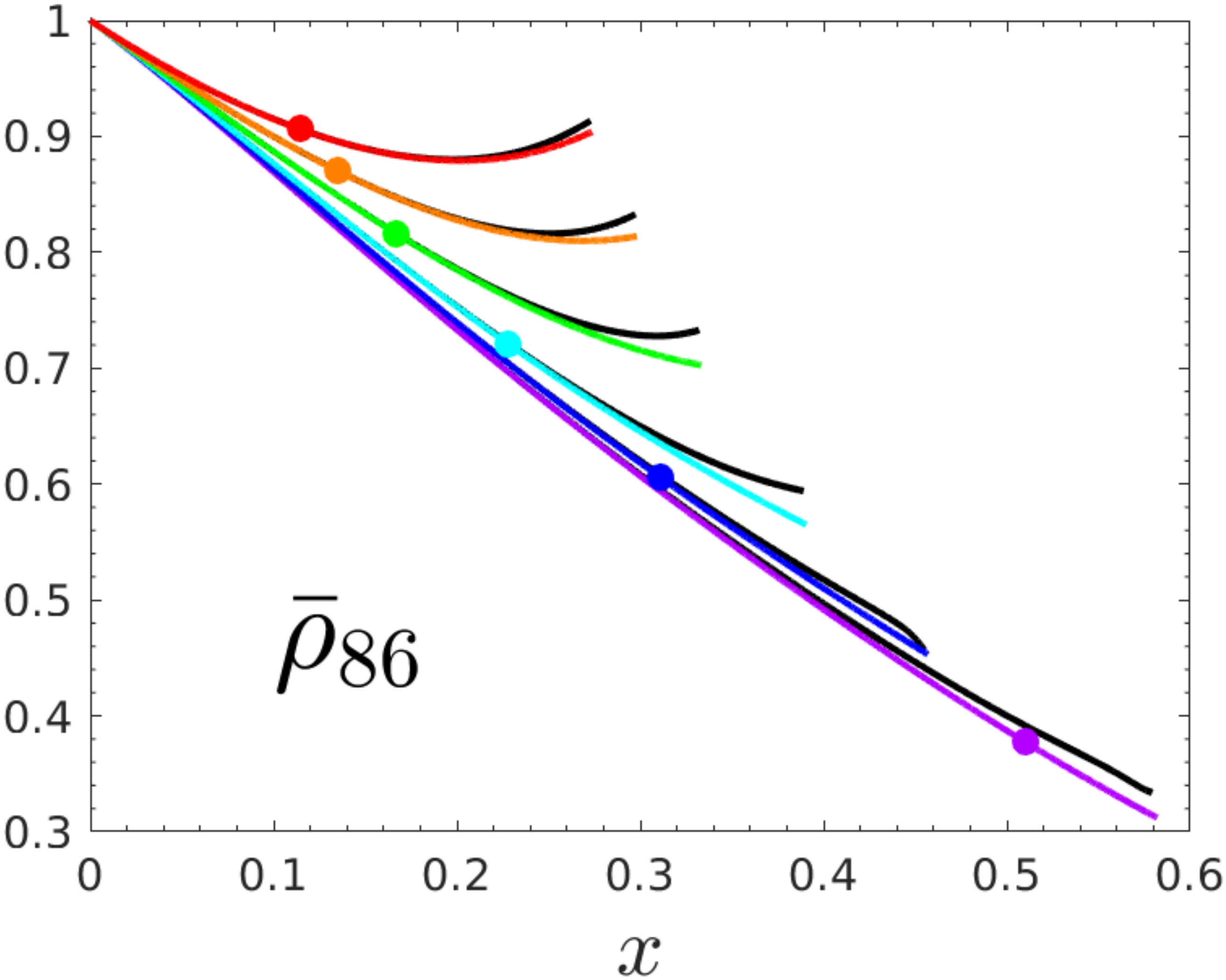} 
  \includegraphics[width=0.3\textwidth,height=4.0cm]{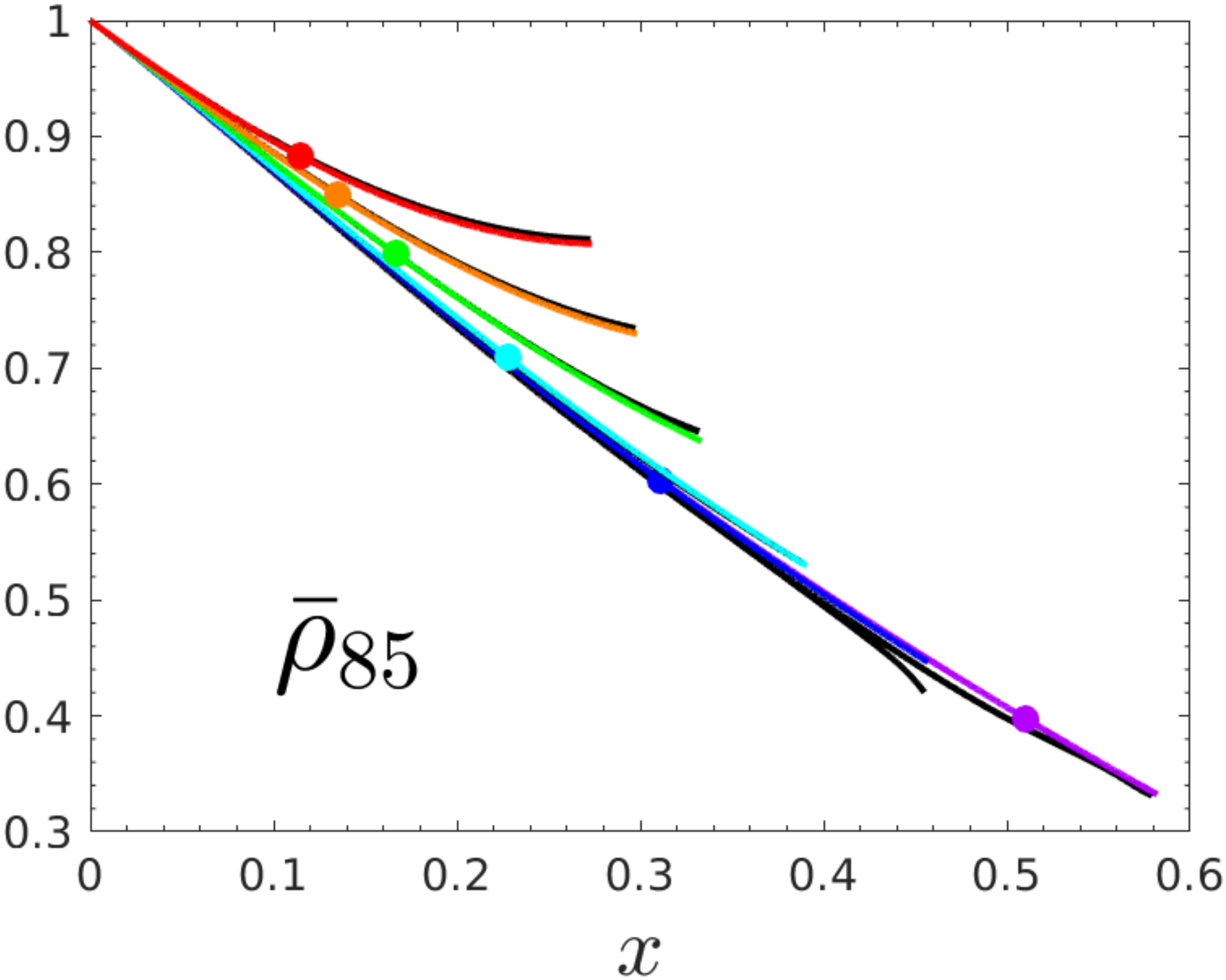} 
  \includegraphics[width=0.3\textwidth,height=4.0cm]{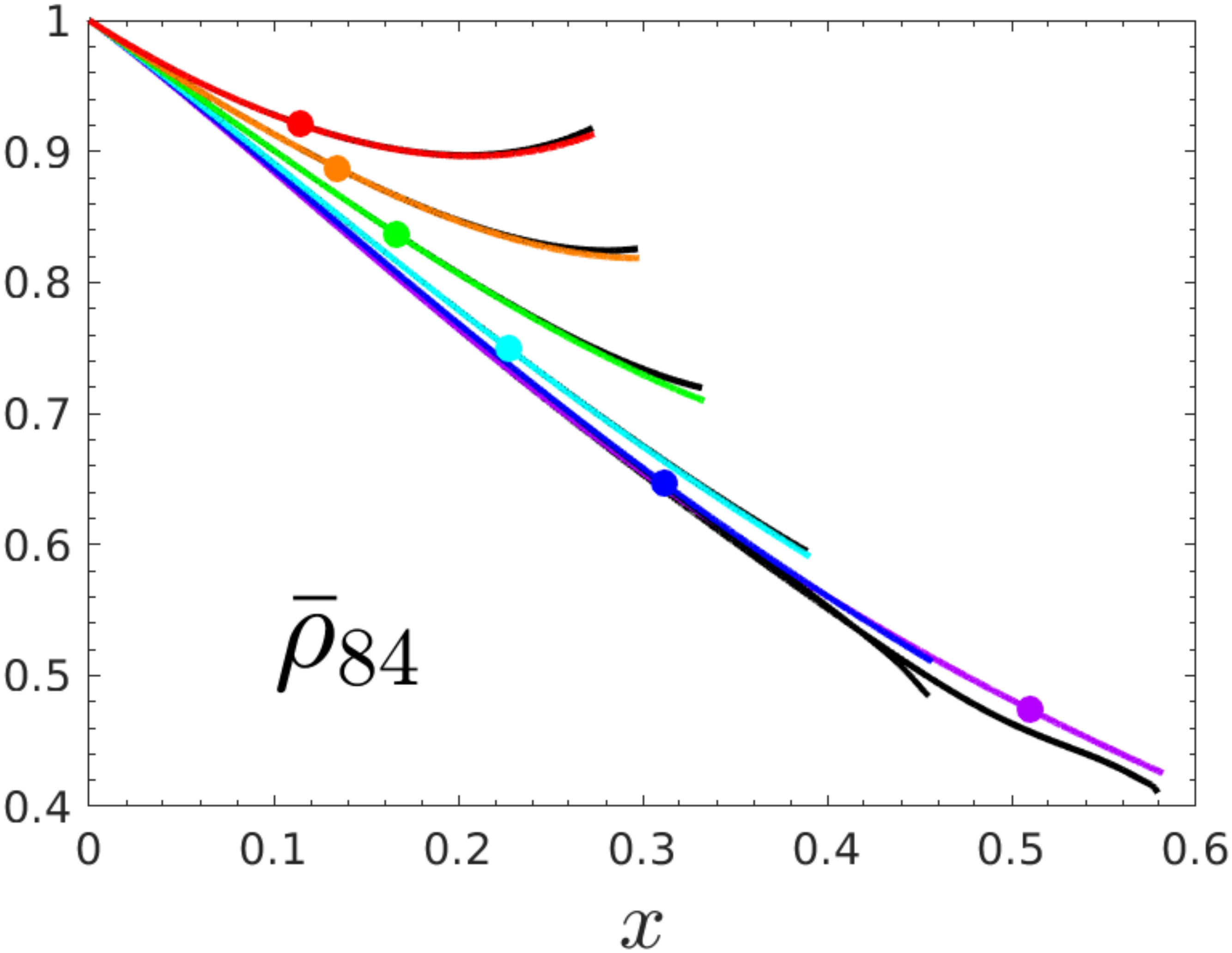} \\
  \includegraphics[width=0.3\textwidth,height=4.0cm]{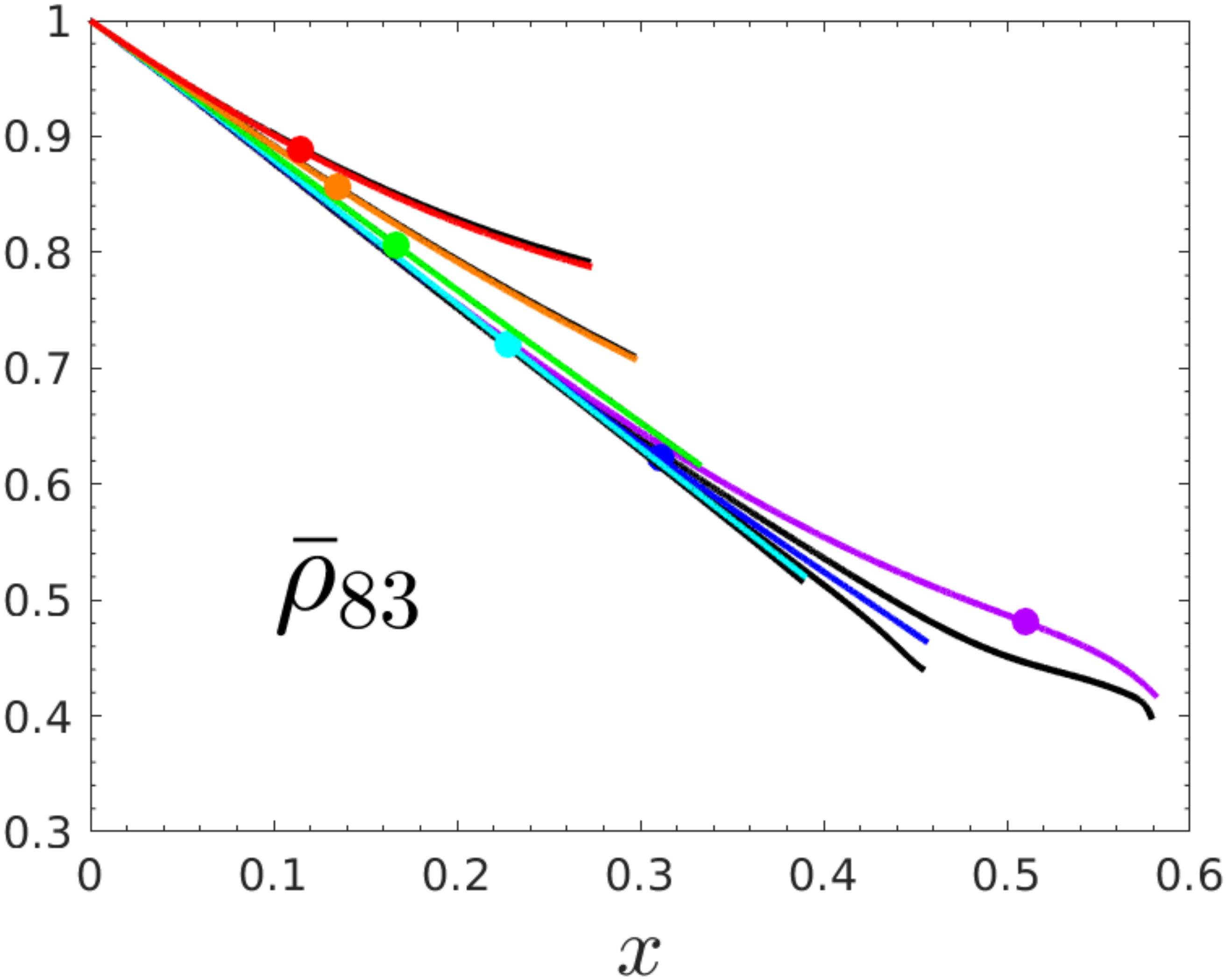} 
  \includegraphics[width=0.3\textwidth,height=4.0cm]{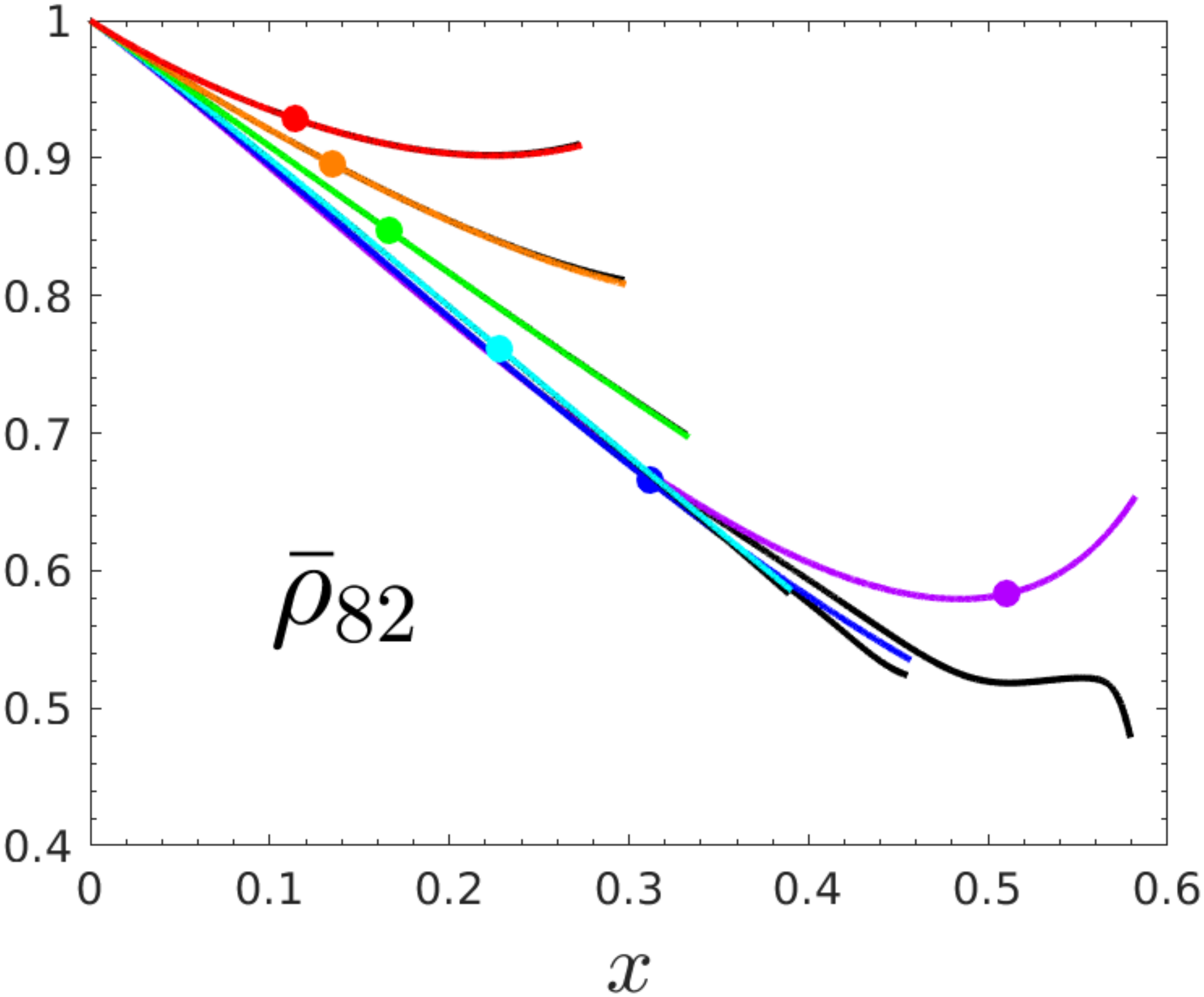} 
  \includegraphics[width=0.3\textwidth,height=4.0cm]{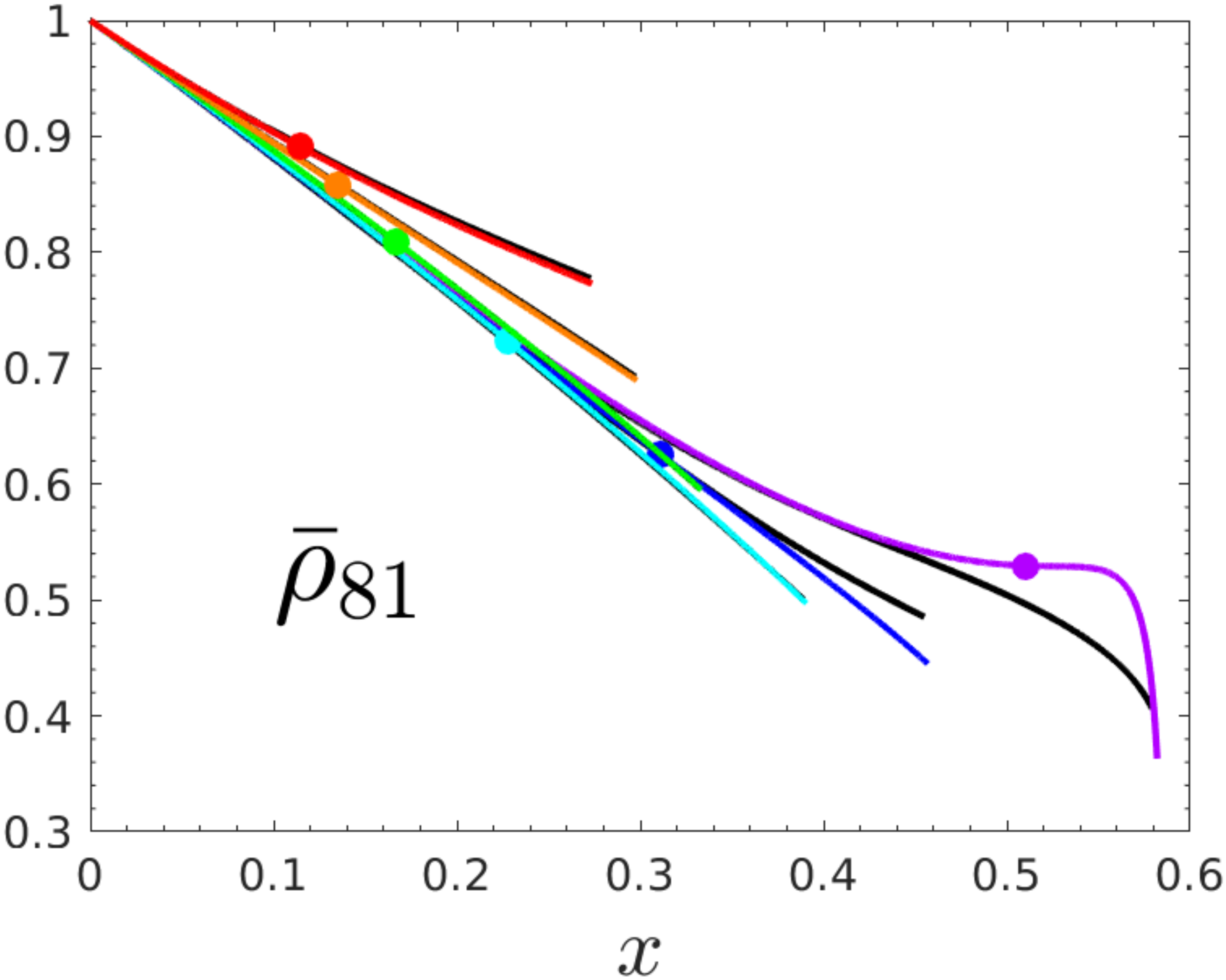} 
  \caption{\label{fig:rho78_iResum} Comparison between the factorized and resummed analytical
    $\bar{\rho}_{\lm}$ (colored lines) and the corresponding numerical (exact) functions
    (black lines) for $\l=7,8$ and for values of the spin parameter 
    $\ha=(-0.99, -0.5, 0, 0.5, 0.8, 0.99)$ (red, orange, green, cyan, blue and purple, respectively). 
    On the $x$-axis we have $x=\Omega^{2/3}$ and the filled circles mark the LSO location. 
    These plots are obtained using 6~PN information for all the modes except for $\bar{\rho}_{73}$ 
    and $\bar{\rho}_{72} $, that employ 8PN information, and for $\bar{\rho}_{71}$ that uses 5PN orders. 
    The Pad\'e approximants used for $\bar{\rho}_\lm^{\rm orb}$ are listed in the second column of
    Table~\ref{tab:rho78_iResum}.}
\end{figure*} 
\begin{table}
   \caption{\label{tab:rho78_iResum} Pad\'e approximants used for the orbital part of the 
   higher multipoles $\ell=7,8$ and  corresponding relative differences 
   $(\rho^{\rm num}_\lm-\bar{\rho}^{\rm analyt}_\lm) / \rho^{\rm num}_\lm$ at LSO for different spin values. 
   The largest differences occur for $\ha=+0.99$.}
   \begin{center}
   \small
   \setlength\tabcolsep{2pt}
\begin{tabular}{ c c | r | r | r | r | r} 
\hline 
\hline 
\multicolumn{1}{c}{$(\ell,m)$} & \multicolumn{1}{c|}{Pad\'e} & \multicolumn{1}{c|}{$-0.99$} & \multicolumn{1}{c|}{$-0.5$} & \multicolumn{1}{c|}{0}   & \multicolumn{1}{c|}{+0.5} &  \multicolumn{1}{c}{+0.99} \\
\hline 
\hline  
$(7,7)$ & $P^{6}_{0}$ & $ -6\cdot 10^{-6} $ & $ 2\cdot 10^{-5} $ & $ 1\cdot 10^{-4} $ & $ 9\cdot 10^{-4} $ & $ 6\cdot 10^{-2} $\\
$(7,6)$ & $P^{4}_{2}$ & $ 2\cdot 10^{-3} $ & $ 1\cdot 10^{-3} $ & $ 2\cdot 10^{-4} $ & $ -1\cdot 10^{-3} $ & $ 4\cdot 10^{-2} $\\
$(7,5)$ & $P^{4}_{2}$ & $ 2\cdot 10^{-5} $ & $ 7\cdot 10^{-5} $ & $ 2\cdot 10^{-4} $ & $ 9\cdot 10^{-4} $ & $ 3\cdot 10^{-2} $\\
$(7,4)$ & $P^{4}_{2}$ & $ 2\cdot 10^{-3} $ & $ 1\cdot 10^{-3} $ & $ 4\cdot 10^{-5} $ & $ -3\cdot 10^{-3} $ & $ -3\cdot 10^{-2} $\\
$(7,3)$ & $P^{6}_{2}$ & $ 5\cdot 10^{-7} $ & $ 1\cdot 10^{-6} $ & $ 2\cdot 10^{-6} $ & $ -6\cdot 10^{-5} $ & $ -4\cdot 10^{-3} $\\
$(7,2)$ & $P^{8}_{0}$ & $ 2\cdot 10^{-3} $ & $ 1\cdot 10^{-3} $ & $ -4\cdot 10^{-7} $ & $ -3\cdot 10^{-3} $ & $ -1\cdot 10^{-1} $\\
$(7,1)$ & $P^{2}_{3}$ & $ 2\cdot 10^{-6} $ & $ 4\cdot 10^{-5} $ & $ 1\cdot 10^{-6} $ & $ -3\cdot 10^{-4} $ & $ -5\cdot 10^{-2} $\\
\hline
$(8,8)$ & $P^{6}_{0}$ & $ -7\cdot 10^{-6} $ & $ 2\cdot 10^{-5} $ & $ 1\cdot 10^{-4} $ & $ 9\cdot 10^{-4} $ & $ 5\cdot 10^{-2} $\\
$(8,7)$ & $P^{4}_{2}$ & $ 2\cdot 10^{-3} $ & $ 1\cdot 10^{-3} $ & $ 3\cdot 10^{-4} $ & $ -9\cdot 10^{-4} $ & $ 5\cdot 10^{-2} $\\
$(8,6)$ & $P^{4}_{2}$ & $ 2\cdot 10^{-5} $ & $ 7\cdot 10^{-5} $ & $ 3\cdot 10^{-4} $ & $ 1\cdot 10^{-3} $ & $ 4\cdot 10^{-2} $\\
$(8,5)$ & $P^{4}_{2}$ & $ 2\cdot 10^{-3} $ & $ 1\cdot 10^{-3} $ & $ 7\cdot 10^{-5} $ & $ -2\cdot 10^{-3} $ & $ -2\cdot 10^{-2} $\\
$(8,4)$ & $P^{4}_{2}$ & $ 9\cdot 10^{-6} $ & $ 3\cdot 10^{-5} $ & $ 9\cdot 10^{-5} $ & $ 2\cdot 10^{-4} $ & $ -4\cdot 10^{-2} $\\
$(8,3)$ & $P^{1}_{5}$ & $ 2\cdot 10^{-3} $ & $ 1\cdot 10^{-3} $ & $ 2\cdot 10^{-6} $ & $ -2\cdot 10^{-3} $ & $ -8\cdot 10^{-2} $\\
$(8,2)$ & $P^{5}_{1}$ & $ 2\cdot 10^{-6} $ & $ 1\cdot 10^{-5} $ & $ 7\cdot 10^{-6} $ & $ -1\cdot 10^{-4} $ & $ -1\cdot 10^{-1} $\\
$(8,1)$ & $P^{3}_{3}$ & $ 2\cdot 10^{-3} $ & $ 1\cdot 10^{-3} $ & $ -1\cdot 10^{-6} $ & $ -2\cdot 10^{-3} $ & $ -7\cdot 10^{-2} $\\
\hline
\hline
\end{tabular}
\end{center}
\end{table}
%
\begin{figure*}
  \center
  \includegraphics[width=0.27\textwidth]{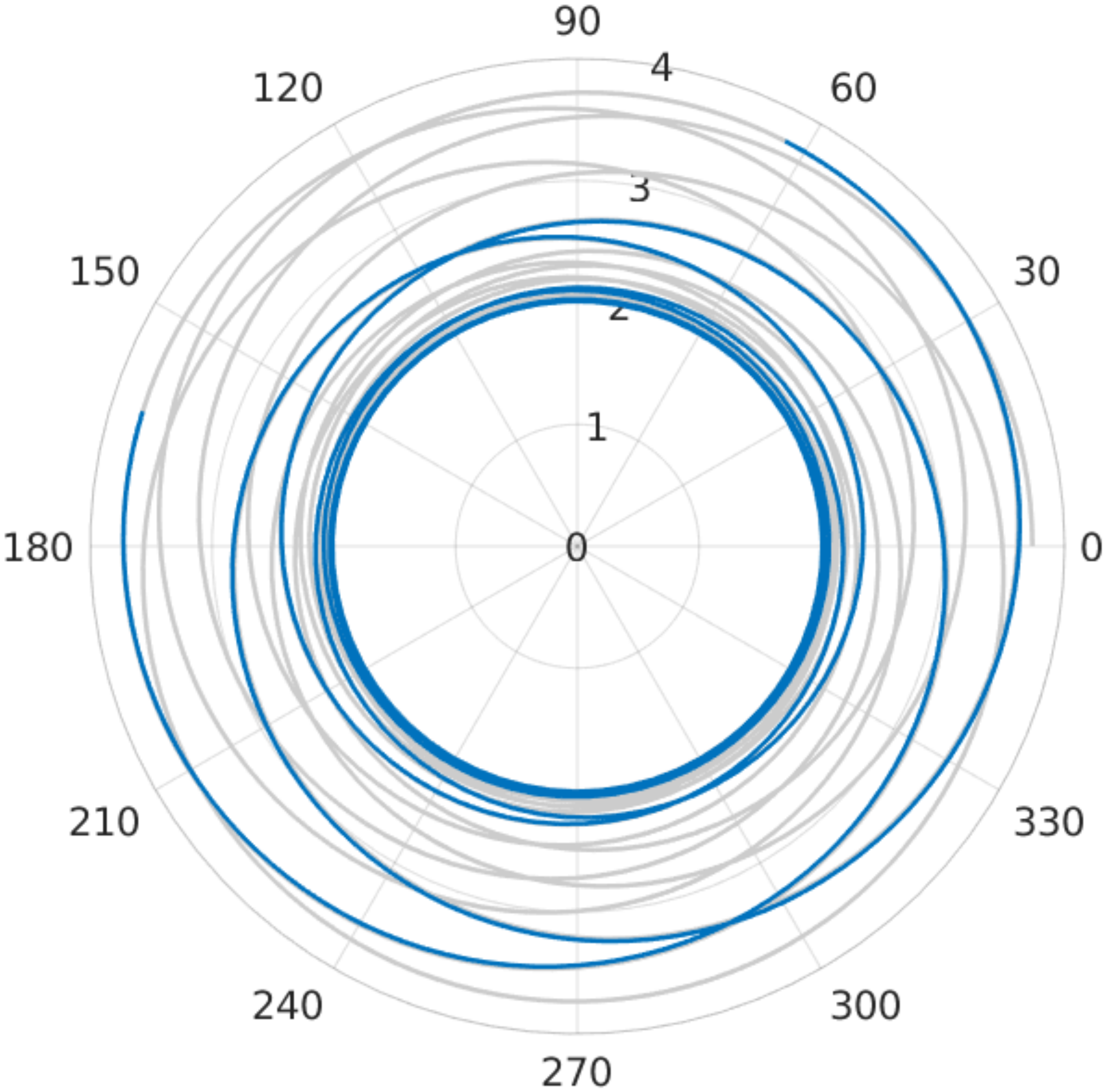}
  \hspace{0.2cm}
  \includegraphics[width=0.31\textwidth,height=4.35cm]{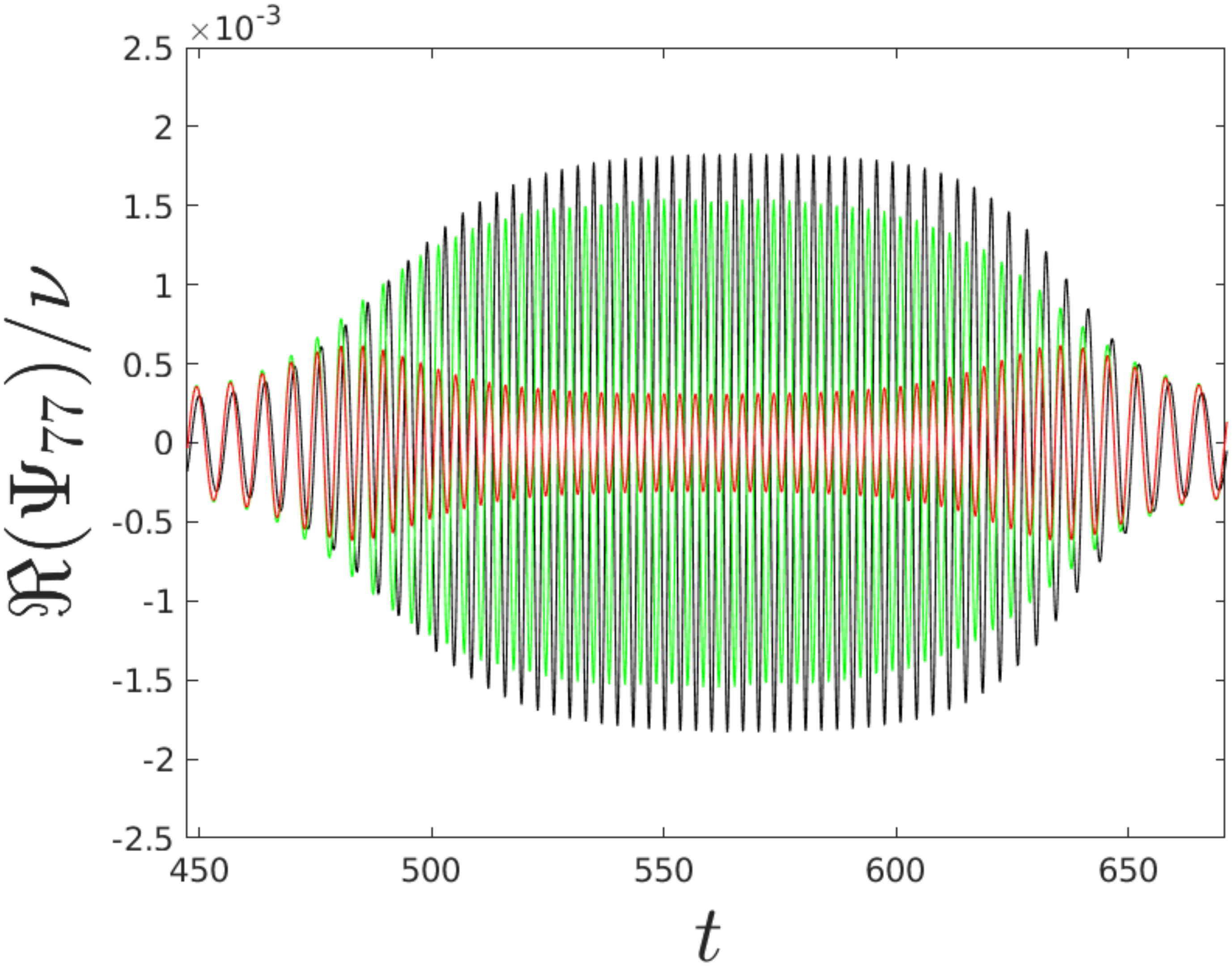} 
  \includegraphics[width=0.31\textwidth,height=4.35cm]{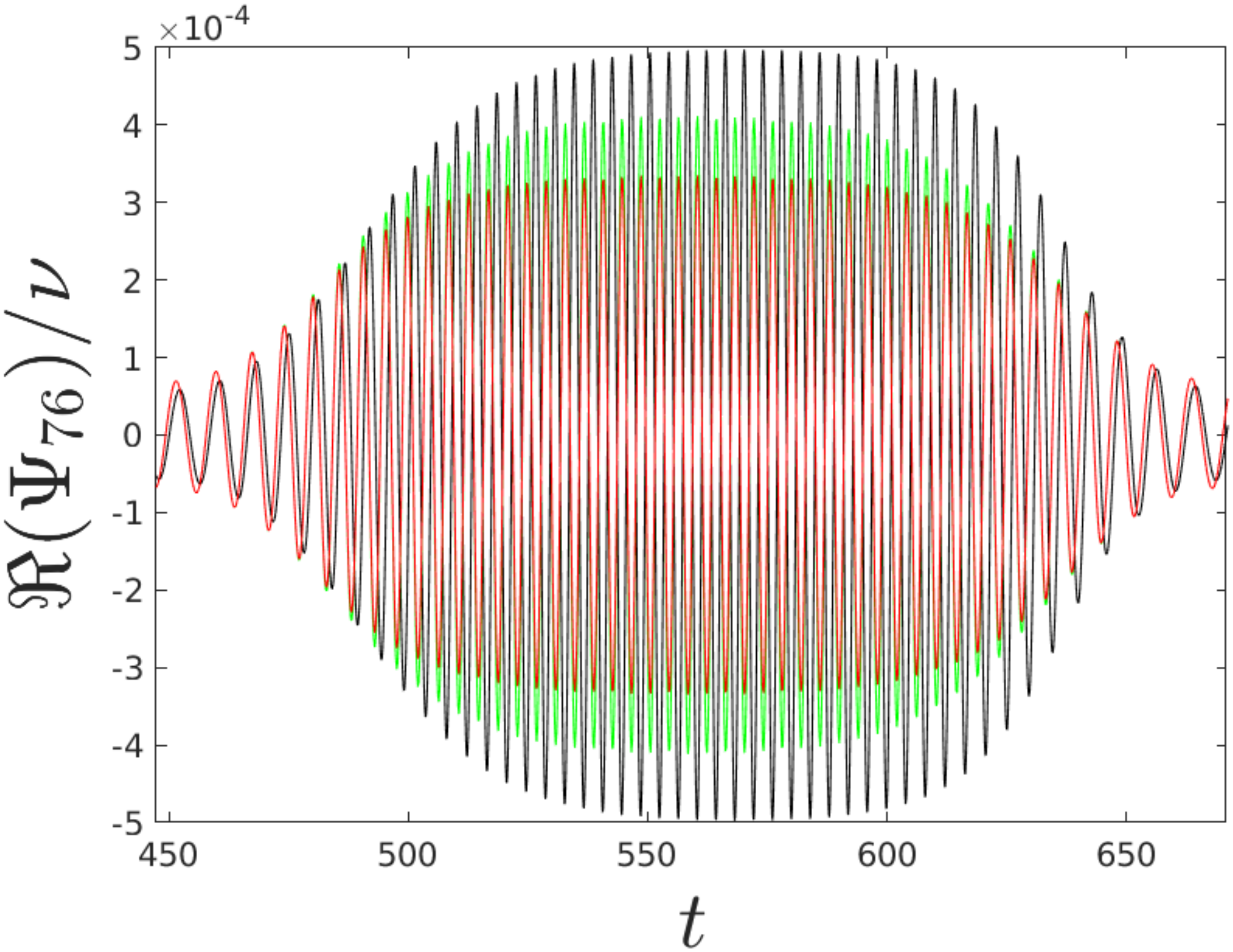} 
  \caption{\label{fig:fig02}High spin and zoom-whirl behavior: left panel,
  trajectory for  $e=0.3$, $\ha=0.9$ and $p=p_s+0.01\simeq 2.615$.
  Middle and right panels: waveform comparison for $(\ell,m)=(7,7)$ and $(7,6)$
  multipoles. Black lines: numerical data. Red lines: analytical waveform
  with the circular $\rho_\lm$ in PN-expanded form at 6~PN. Green lines: analytical waveform 
  with the factorized and resummed $\bar{\rho}_\lm$, always at 6~PN.
  The resummation is essential to obtain a reasonably good agreement between 
  the analytical and numerical waveforms.}
\end{figure*}
The residual amplitude corrections $\rho_{\lm}$ introduced in Eq.~\eqref{eq:eobwave} 
are a crucial building block of any EOB waveform model and need a careful analytical treatment. 
In fact, although they are originally  defined as PN series~\cite{Damour:2008gu}, 
they may need additional resummation procedures to improve their strong-field behavior.
The various PN truncations of the $\rho_\lm$'s, either in the spinning or nonspinning case,
typically oscillate around the function computed numerically solving the Teukolsky equation.
A straight Pad\'e approximant is not the most suitable choice, as pointed out 
in~\cite{Nagar:2016ayt, Messina:2018ghh}. By contrast, Refs.~\cite{Nagar:2016ayt, Messina:2018ghh}
suggested a different resummation scheme that: (i) first factorizes out the orbital, spin-independent,
part and (ii) resums the orbital and spinning factors using, respectively, a Pad\'e approximant
and an inverse Taylor resummation scheme. 
The choice of the  Pad\'e approximant is partly arbitrary and is guided by the comparisons with the 
numerical results. In Refs.~\cite{Nagar:2016ayt, Messina:2018ghh} the resummation scheme 
was applied to all multipoles up to $\ell=6$. For our flux comparisons, especially in 
the presence of zoom-whirl orbits, we found it useful to apply the same scheme to the $\ell=7$ 
and $\ell=8$ modes. Let us recall the main elements of the procedure. As a first step, the orbital 
part is factorized as follows
\begin{equation}
\rho_{\lm}(x) = \rho_{\lm}^{\rm orb}(x) + \rho_{\lm}^{\rm S}(x) = \rho_{\lm}^{\rm orb}(x) \hat{\rho}_{\lm}^{\rm S}(x) ,
\end{equation}
where $\hat{\rho}_{\lm}^{\rm S} = T_n \left[ 1 + \rho_{\lm}^{\rm S}/\rho_{\lm}^{orb} \right]$ 
and $T_n$ indicates the Taylor expansion at the $n$th-PN order. 
Then a Pad\'e approximant $P^i_j$ is used for 
the orbital part and an inverse Taylor scheme for the spin-part:
\begin{equation}
\label{eq:resummedrho}
\begin{split}
\bar{\rho}_{\lm}^{\rm orb}(x) &= P^i_j \left( \rho_{\lm}^{\rm orb}(x) \right) ,\\
\bar{\hat{\rho}}_{\lm}^{\rm S} (x)  &= \left( T_n \left[ \left( \hat{\rho}_{\lm}^{\rm S}(x) \right) ^{-1} \right] \right) ^{-1} .
\end{split}
\end{equation}  
The final result is given by the product of the two resummed factors 
\begin{equation}
\bar{\rho}_{\lm}(x) = \bar{\rho}_{\lm}^{\rm orb}(x) \bar{\rho}_{\lm}^{\rm S}(x) .
\end{equation}

We extend this approach to $\ell=7$ and $\ell=8$ modes using 
the PN series of the relativistic amplitude corrections for a test-particle on
circular orbits around a Kerr black hole~\cite{Fujita:2014eta}. 
We use 6~PN accuracy for almost all the multipoles, 
the only exceptions are the (7,3) and (7,2) modes where we have used 8~PN information, 
and the (7,1) mode where we have resummed the series truncated at 5~PN. 
This choice is motivated by the significantly better agreement with numerical data 
achieved for high spin. However, these modes are highly subdominant and their contribution to the 
fluxes is mostly negligible. In fact, the contribution to the angular momentum flux of the 
modes with $\l=7,8$ and $m\leq 3$ 
all summed together in the case $(\ha,e,p) = (0.9, 0, p_s+0.01)$ is $2\cdot 10^{-7}\%$, 
while in all the other simulations it is even smaller.  In practice, all the relevant information 
in the resummed $\bar{\rho}_\lm$ is at 6~PN. 
We also mention in passing that for extremal cases with 
$\ha > 0.99$, the $\bar{\rho}_{81}$ has a pole in the spin factor, 
so that one is obliged to use the plain PN series. Nonetheless, in this work we consider spins only 
up to  $\ha =0.9$, then we will always use the resummed $\bar{\rho}_\lm$. 
This problem is not present for any of the other modes.
The reliability of the resummation has
been tested using circular frequency-domain data obtained by S.~Hughes kindly
made available to us~\cite{Taracchini:2013wfa}. The comparisons between resummed
analytical expressions and numerical data are shown in Fig.~\ref{fig:rho78_iResum}, 
while the list of Pad\'e used is reported in Table~\ref{tab:rho78_iResum} with the 
corresponding relative differences at the Last Stable Orbit (LSO). We address the 
reader to Ref.~\cite{Messina:2018ghh} for complementary technical details.
The resummation is especially relevant for prograde zoom-whirl orbits around fast spinning 
black holes, as shown in Fig.~\ref{fig:fig02} for the illustrative case $(e,\ha) = (0.3, 0.9)$.
In fact, during the circular whirl the amplitude of the nonresummed $\rho_{77}$ at 6~PN becomes 
unphysically small (red line).
This behavior is mostly solved by the resummation procedure exposed above, 
leading to a better agreement with numerical data (green lines in Fig.~\ref{fig:fig02}). 
Note however that the analytical waveform underestimates the exact result, 
as typical of prograde orbits around fast spinning black holes. 
We will discuss this issue in more details in Sec.~\ref{sec:fluxes}.
Finally, note that in Fig.~\ref{fig:fig02} we are not using any PN correction
for the residual phase of the tail factor, and this leads to a great phase-difference 
between numerical and analytical waveforms during the circular whirl. We will address
this issue in the next subsection.
%
\begin{figure*}[t]
  \center
  \includegraphics[width=0.3\textwidth,height=4cm]{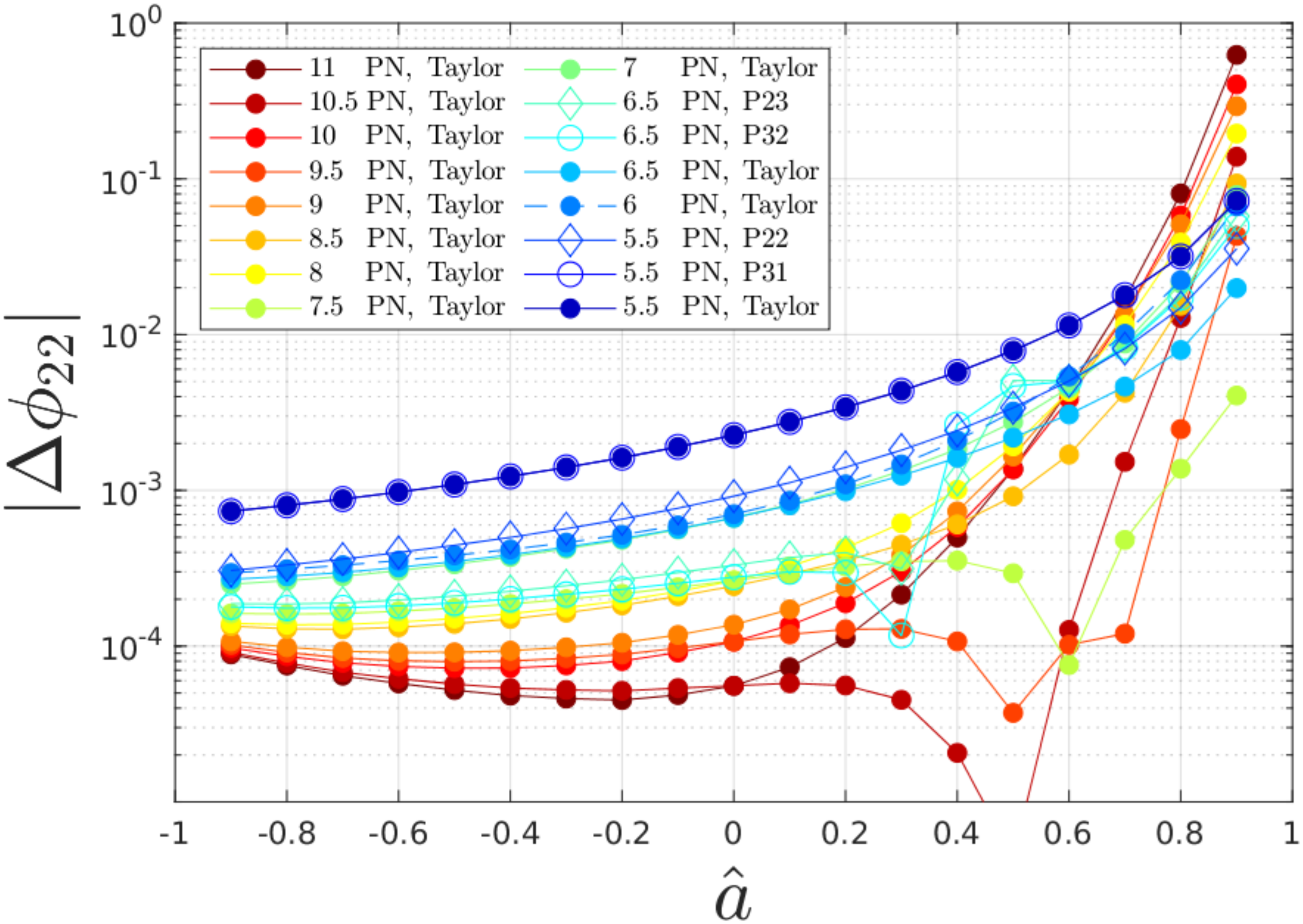} 
  \includegraphics[width=0.3\textwidth,height=4cm]{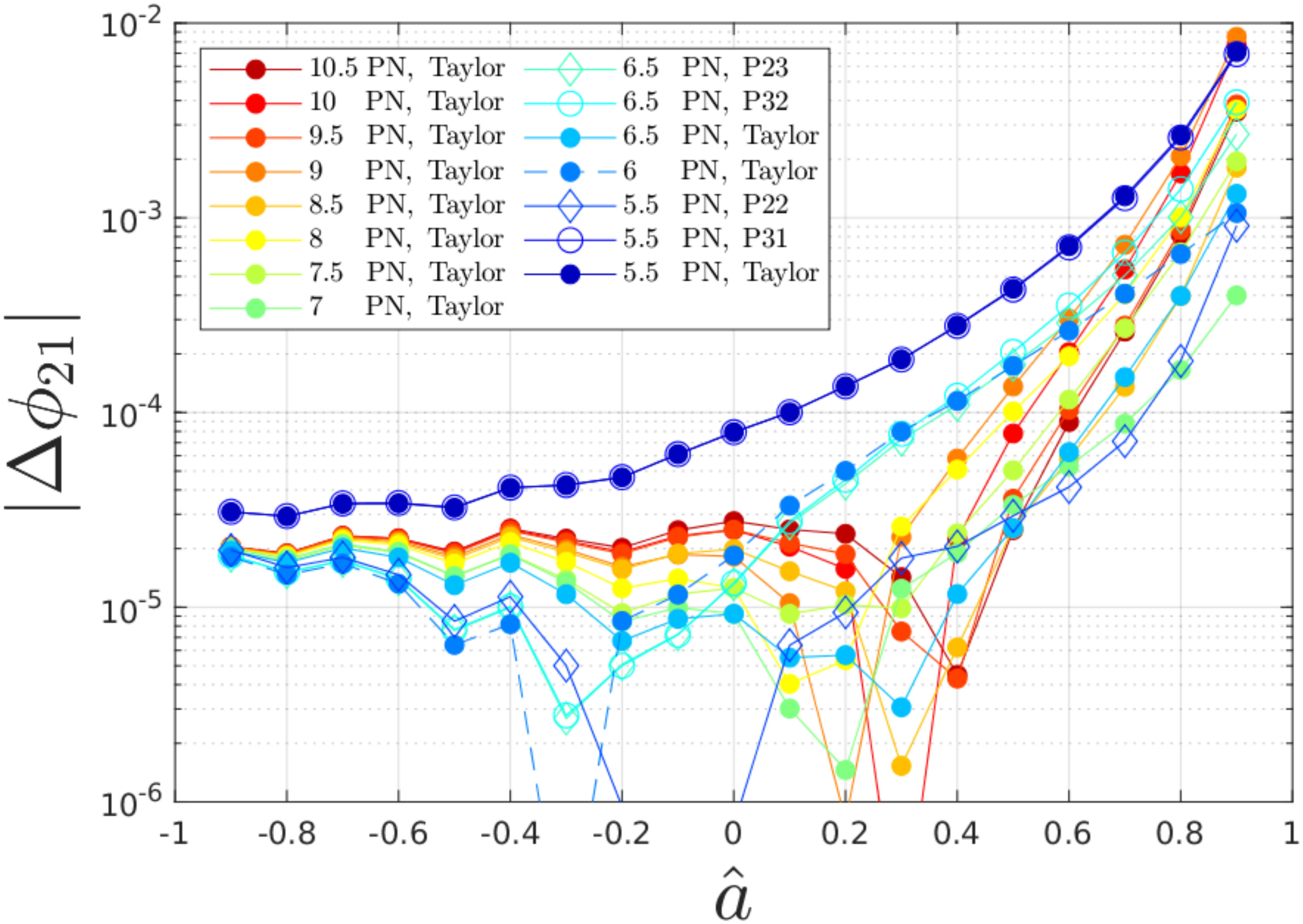} 
  \includegraphics[width=0.3\textwidth,height=4cm]{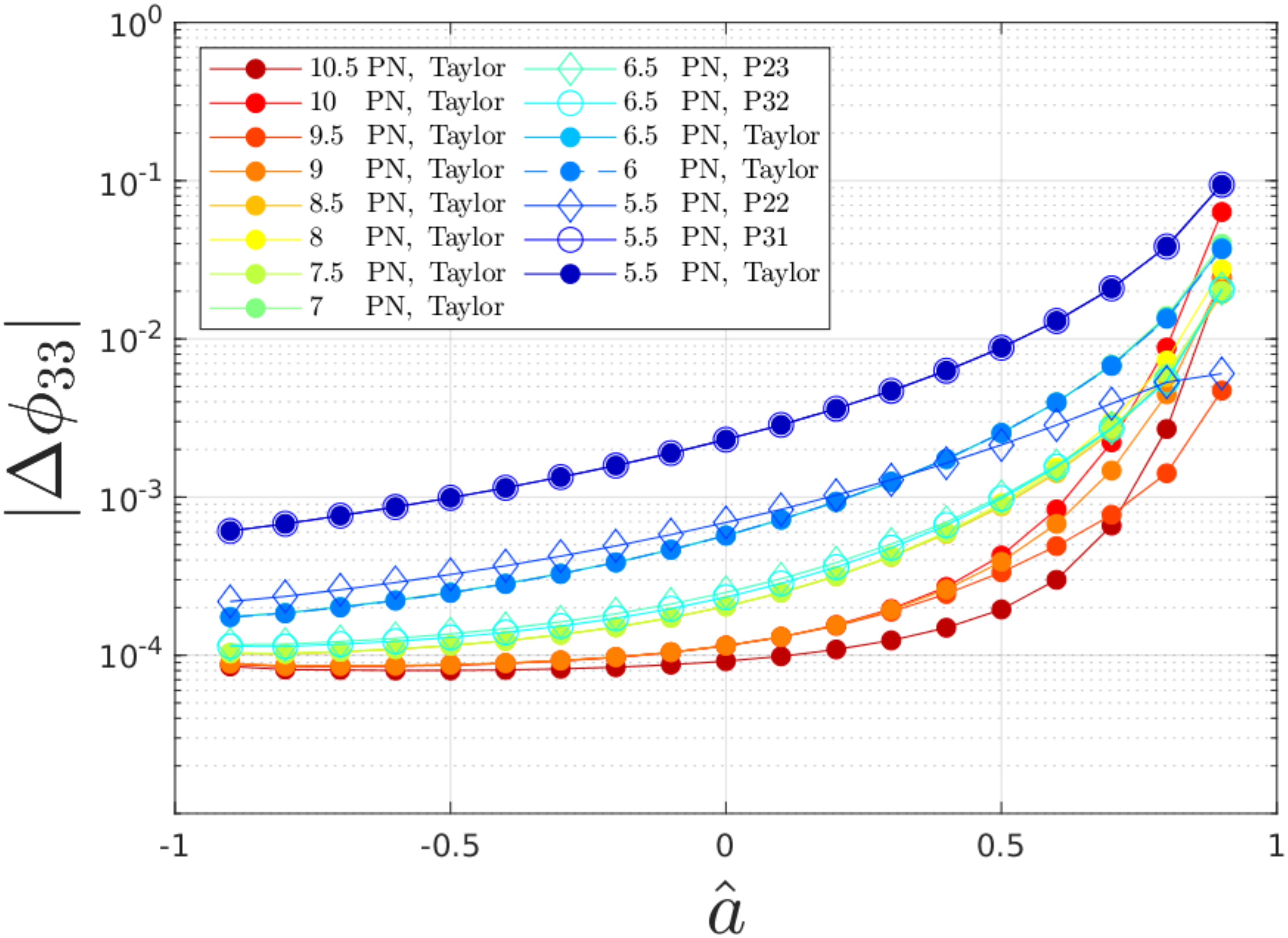} \\
  \includegraphics[width=0.3\textwidth,height=4cm]{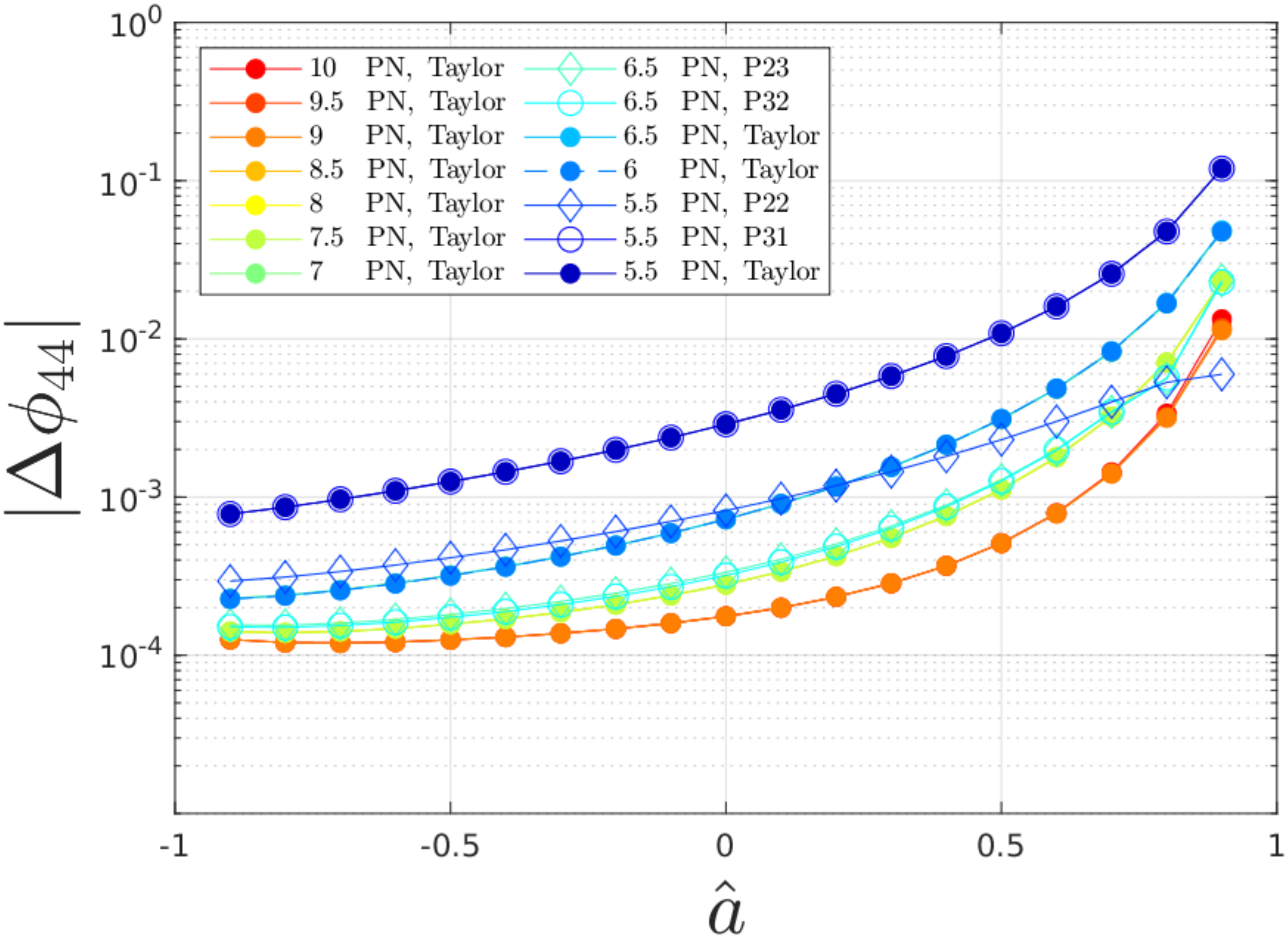}
  \includegraphics[width=0.3\textwidth,height=4cm]{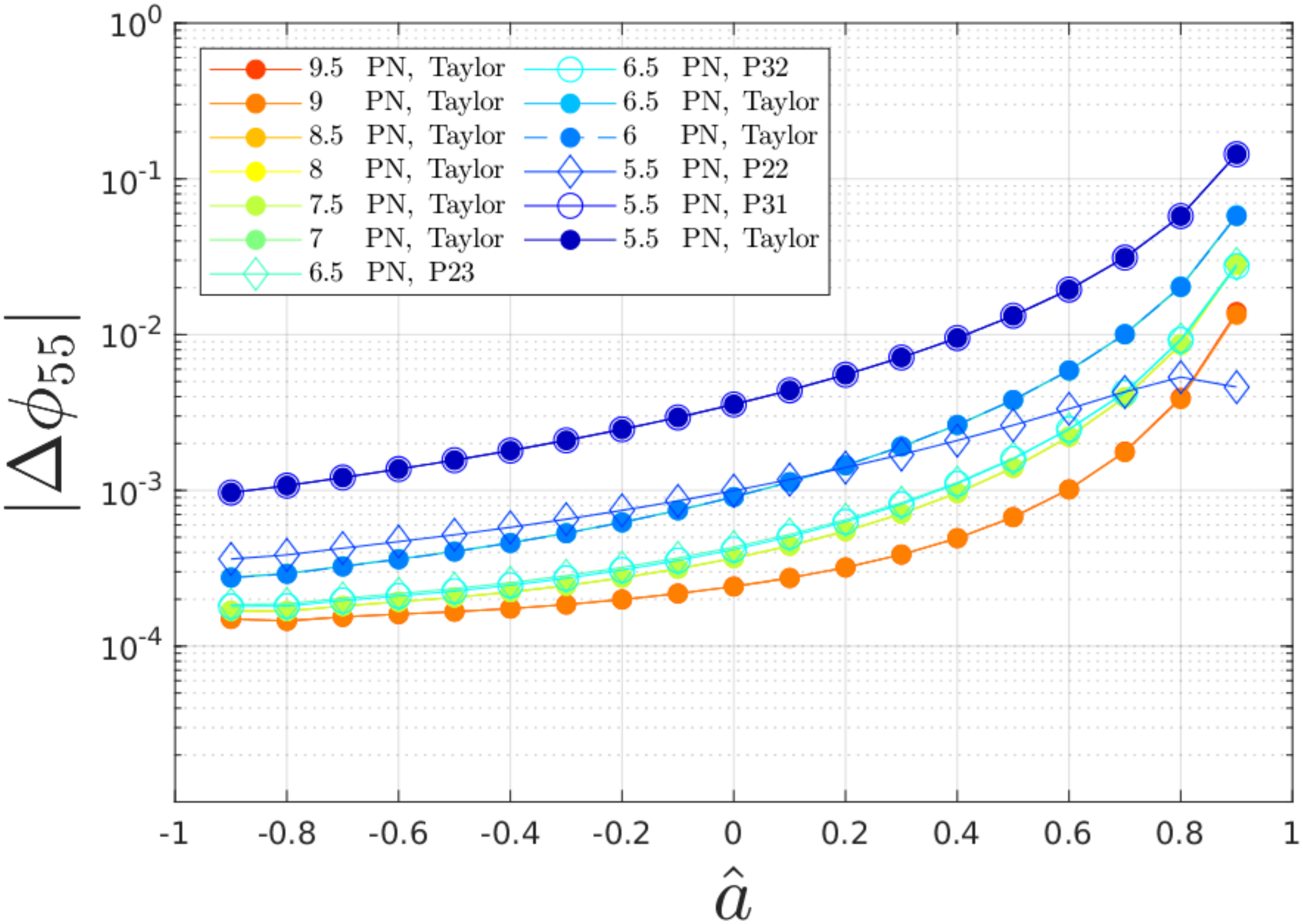}
  \includegraphics[width=0.3\textwidth,height=4cm]{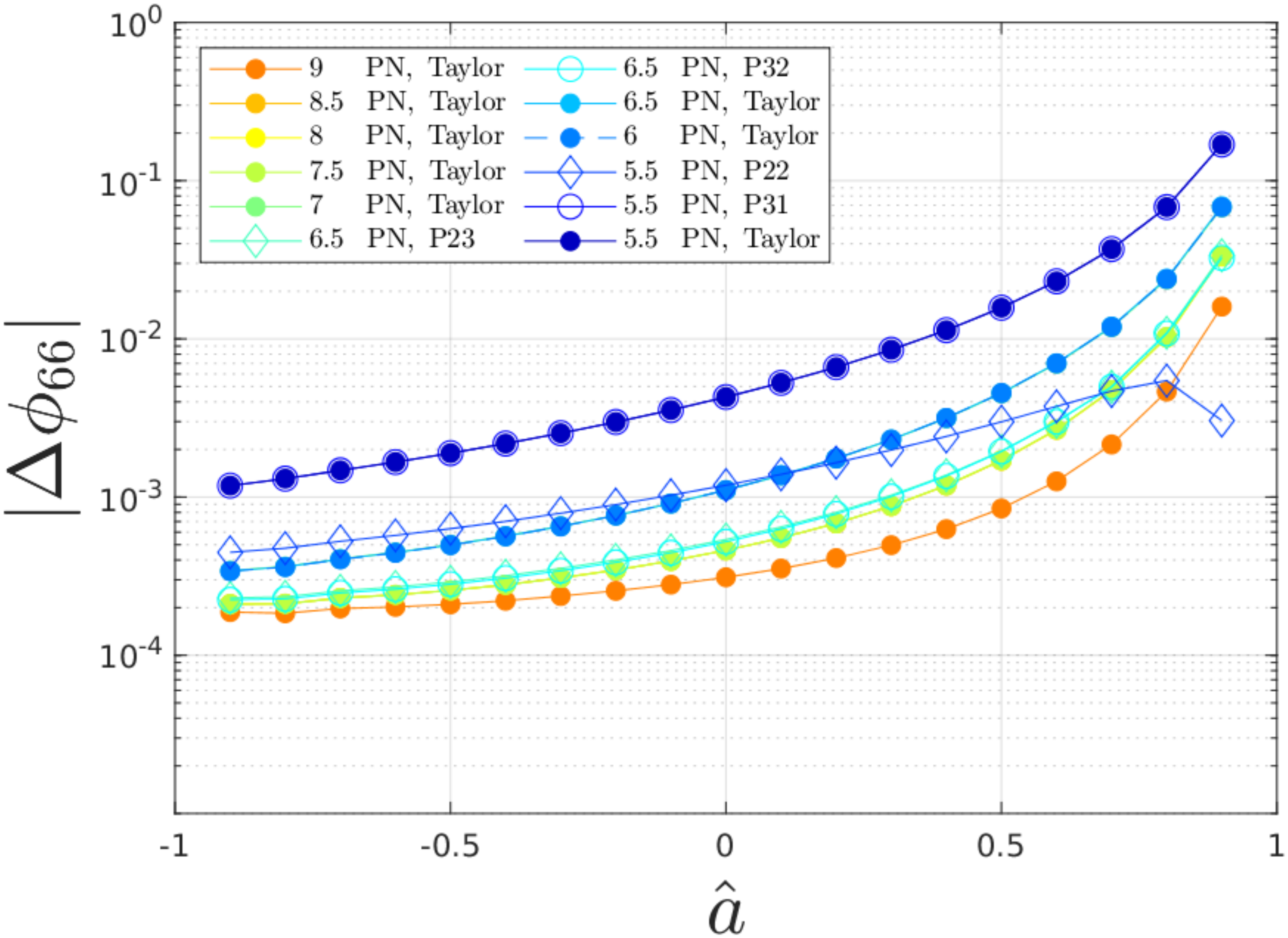} \\
  \caption{\label{fig:delta_PNtest} Phase difference between numerical and analytical waveform
  multipoles for different truncation of the $\delta_\lm$ series 
  (filled markers) and Pad\'e resummations obtained with Eq.~\eqref{eq:deltaPade} (empty markers) 
  for the near circular simulations ($r=r_{\rm LSO}+0.01$).
  Note the different vertical scale for the (2,1) mode.}
\end{figure*}

\subsection{Impact of residual waveform phases $\delta_\lm$}
\label{sec:deltalm}
To improve the circular part of the waveform, we have incorporated high-order 
PN information also in the residual phases $\delta_\lm$ of Eq.~\eqref{eq:eobwave}. 
In particular, we have exploited various PN truncations of the 11~PN series for 
a test-particle on circular orbits around a Kerr black hole~\cite{Fujita:2014eta}.
We have also tested the resummation scheme introduced in~\cite{Damour:2012ky}, where the 
resummed phases are obtained by factorizing the leading order $\delta_\lm^{\rm LO}$ 
contribution and then using a Pad\'e approximant $P^i_j$ for the remaining correcting factor 
$\hat{\delta}_\lm$, explicitly:
\begin{align}
\delta_\lm & = \delta_\lm^{\rm LO} + \delta_\lm^{\rm NLO} + \dots =\delta_\lm^{\rm LO} \hat{\delta}_\lm, \\
\label{eq:deltaPade}
\bar{\delta}_\lm & = \delta_\lm^{\rm LO} P^i_j[\hat{\delta}_\lm] . 
\end{align} 
We report in Fig.~\ref{fig:delta_PNtest} the phase 
differences between numerical and analytical circular waveforms, 
either with the PN-expanded $\delta_\lm$ (at various PN truncations)
or with the resummed $\bar{\delta}_\lm$. 
We consider some of the most relevant modes and only consider 
radii close to the LSO ($r=r_{\rm LSO}+0.01$).
For moderate spins ($\hat{a}\lesssim 0.5$), we see that 
the analytical/numerical phase agreement of the dominant mode is
improved by increasing the PN order of $\delta_{22}$.
Nonetheless, when the spin increases, the series beyond 8~PN 
become unreliable. 
For the subdominant modes, the PN series
at high order are reliable also for $\ha \gtrsim 0.5$, even if for $\delta_{21}$ and 
$\delta_{33}$ some series at lower order give similar (or better) agreements.
 
Applying the resummation scheme of Eq.~\eqref{eq:deltaPade} at 5.5~PN, we see that 
the most suitable choice is the Pad\'e $P^2_2$. The resummation provides a better 
numerical/analytical agreement than the corresponding Taylor-expanded series at 5.5~PN. 
Nonetheless, the Taylor expanded $\delta_{22}$ at 6 or 7.5~PN yields
comparable, or even better, agreement with numerical waveforms. 
For prograde orbits with $\ha\gtrsim 0.3$, the resummation of the $\l=m$ subdominant 
modes provides a more faithful description than the Taylor expanded series at 6.5~PN, 
as shown in Fig.~\ref{fig:delta_PNtest}. For even higher spins ($\ha\gtrsim 0.8$), the resummed
$\bar{\delta}_\lm$ outperform also the series beyond 7~PN.
The hierarchy of the analytical/numerical phase difference is different for the (2,1) mode, 
but in that case all the phase differences are below $0.01$ radians. 
A similar argument holds at larger radii, even if for distant simulation is less
straightforward to evaluate the goodness of the analytical choice, since the comparisons are 
affected by the numerical errors.

Note in passing that, while spurious poles are absent in $\ell=m$ 
and $(2,1)$ modes of the resummed $\delta_\lm$ with Pad\'e $P^2_2$, 
they may occasionally appear in some other subdominant modes. 
Finally, we have also successfully applied the resummation scheme at 6.5~PN accuracy 
as shown in Fig.~\ref{fig:delta_PNtest}. By contrast, when working at 7.5~PN, 
one finds spurious poles even in the $(2,2)$ mulitpole, either with Pad\'e $P^4_2$ 
or $P^3_3$, so that the resummation is not robustly applicable in this case.

Generally speaking, the resummed $\bar{\delta}_\lm$ yield a better phasing agreement
with the numerical results. In particular they are more robust than the high-order PN 
truncations for prograde orbits around fast-spinning black hole. Nonetheless, 
in order to choose a compromise between accuracy and analytical simplicity, 
we have decided to consider the series truncated at 7.5~PN as our preferred choice.  

For noncircular simulations, the $\delta_\lm$ are relevant during the circular whirl of 
zoom-whirl orbits, but are less significant for the other eccentric orbits. 
For higher spins, the analytical choice is more relevant since the separations reached 
are closer to the light ring and thus the PN series are applied in stronger fields.

\section{Fluxes phenomenology and analytical/numerical comparisons}
\label{sec:fluxes}

Now that we have clarified the various analytical structures involved, let us turn to
discussing the comparisons between analytical and numerical fluxes. The final goal of this
procedure is to identify the range of validation of the analytical expressions so that they
can consistently used to drive an eccentric inspiral. We do these comparisons in two complementary
ways. In Sec.~\ref{sec:insta_fluxes} we discuss instantaneous fluxes, while in Sec.~\ref{sec:Flux_averaged}
we discuss orbital averaged fluxes so to easily gain a global picture for all values of $(e,\hat{a})$.

\subsection{Instantaneous fluxes}
\label{sec:insta_fluxes}
\begin{figure}[t]
  \center  
  \includegraphics[width=0.22\textwidth]{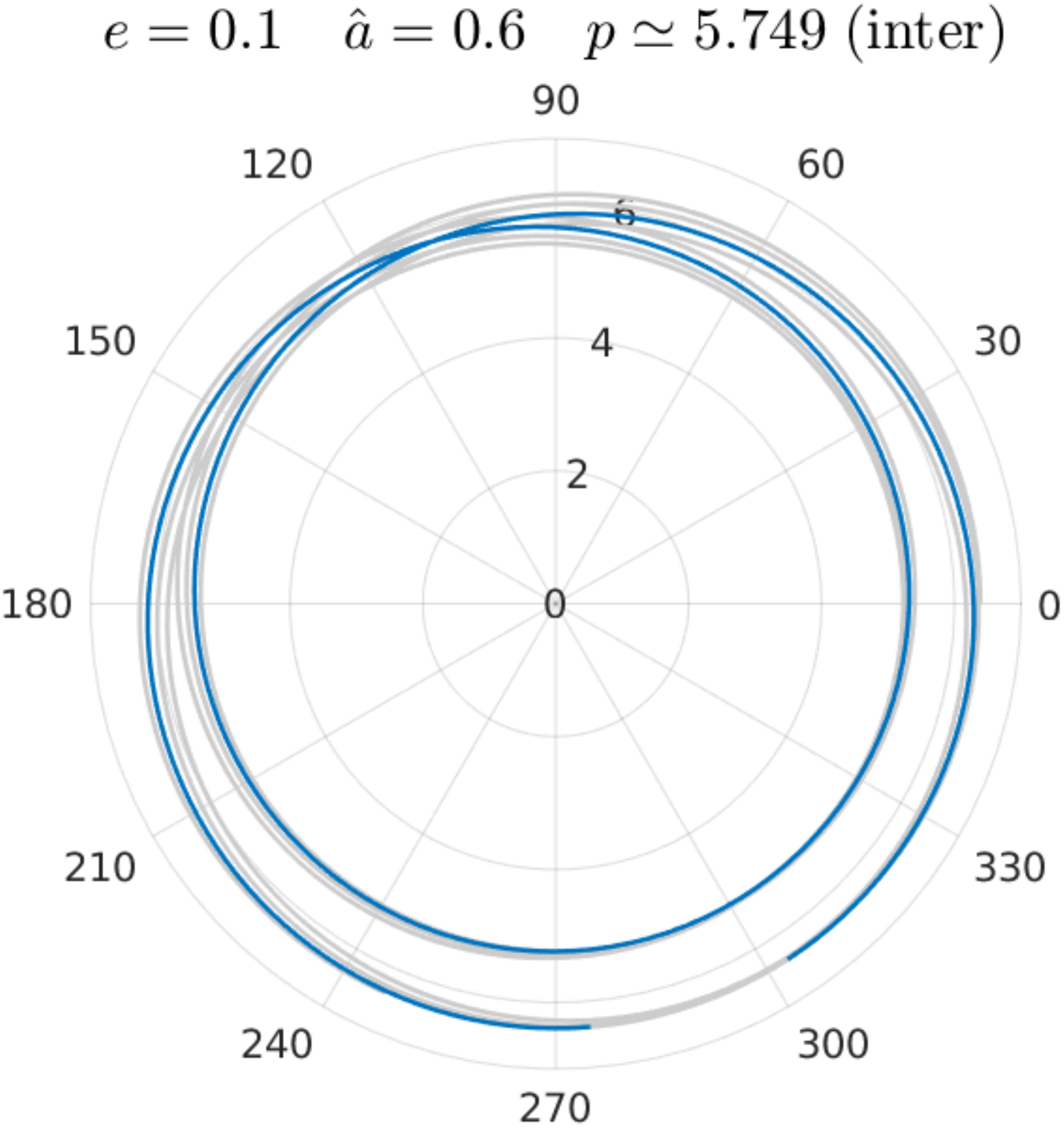}
  \quad
  \includegraphics[width=0.22\textwidth]{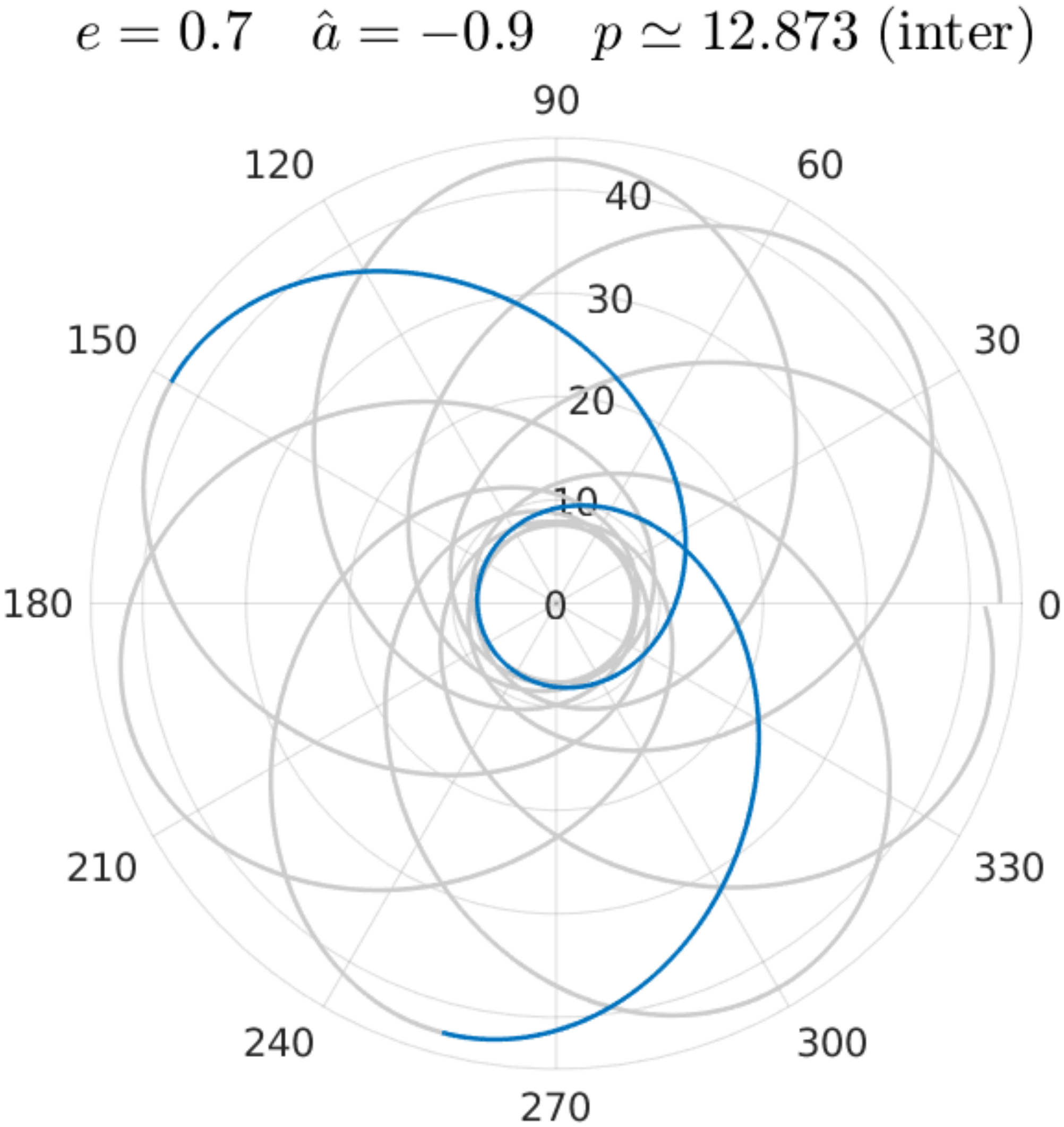} \\
  \vspace{1.0cm}
  \includegraphics[width=0.22\textwidth]{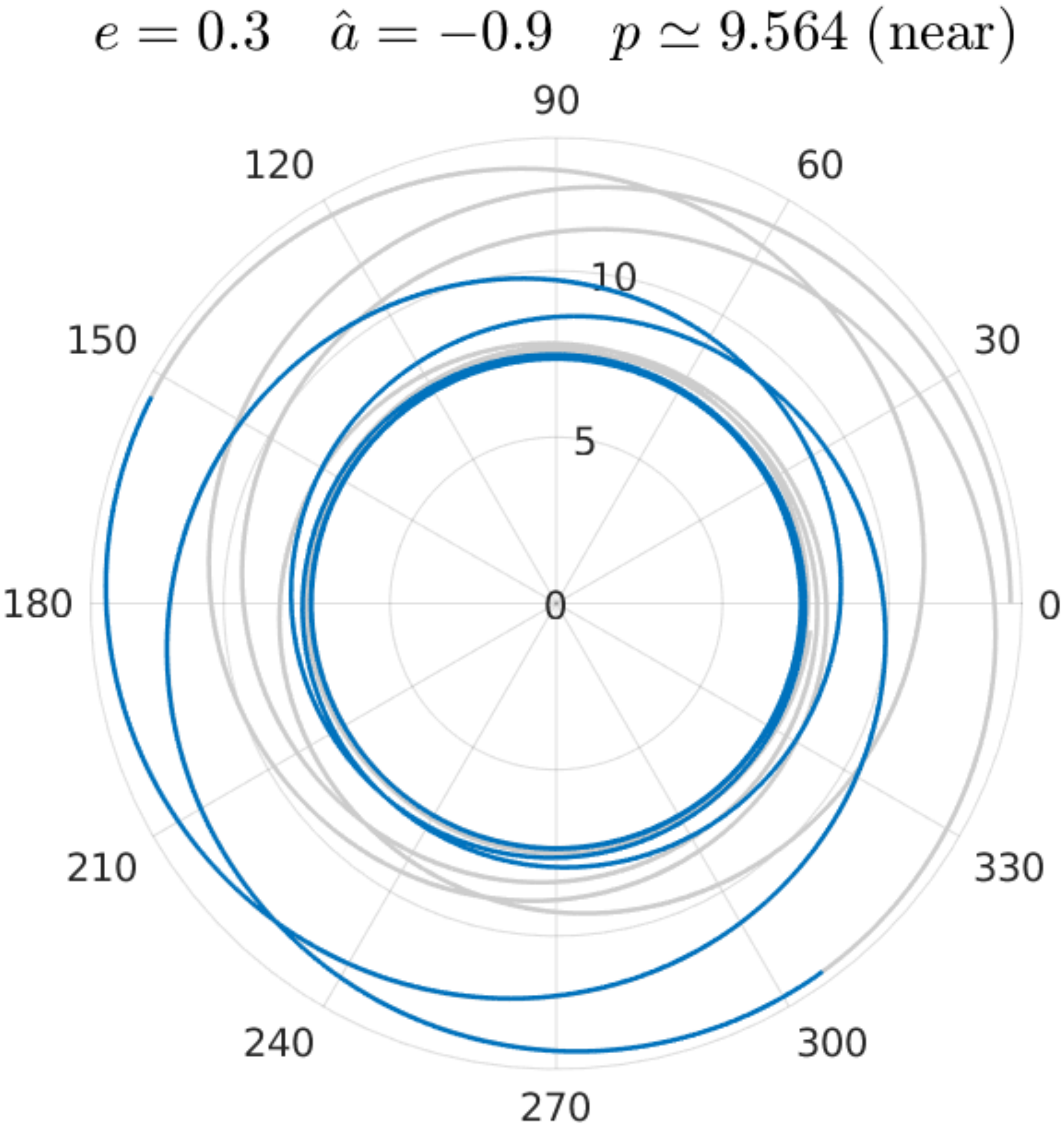} 
  \quad
  \includegraphics[width=0.22\textwidth]{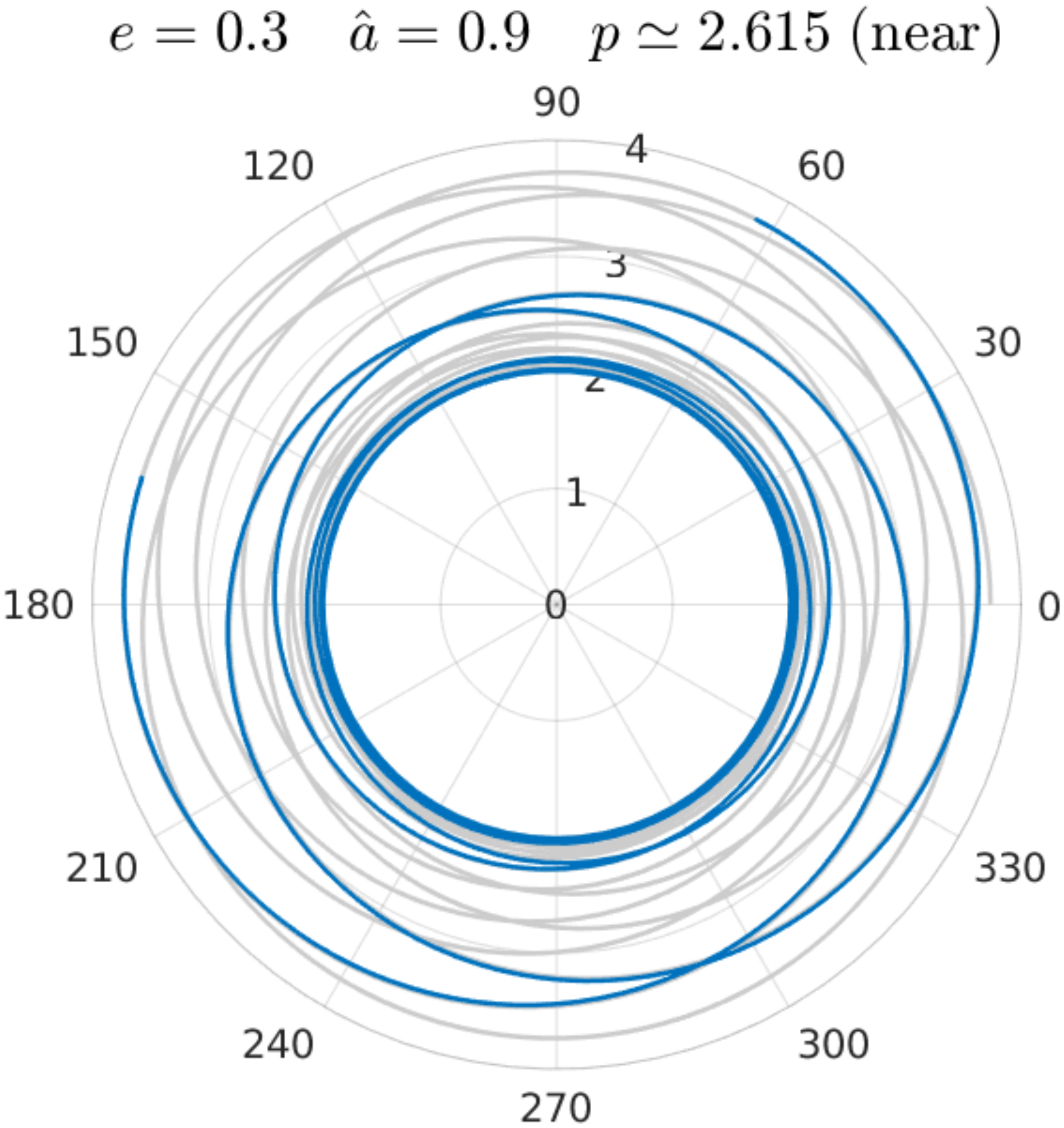} \\
  \vspace{1.0cm}
  \includegraphics[width=0.22\textwidth]{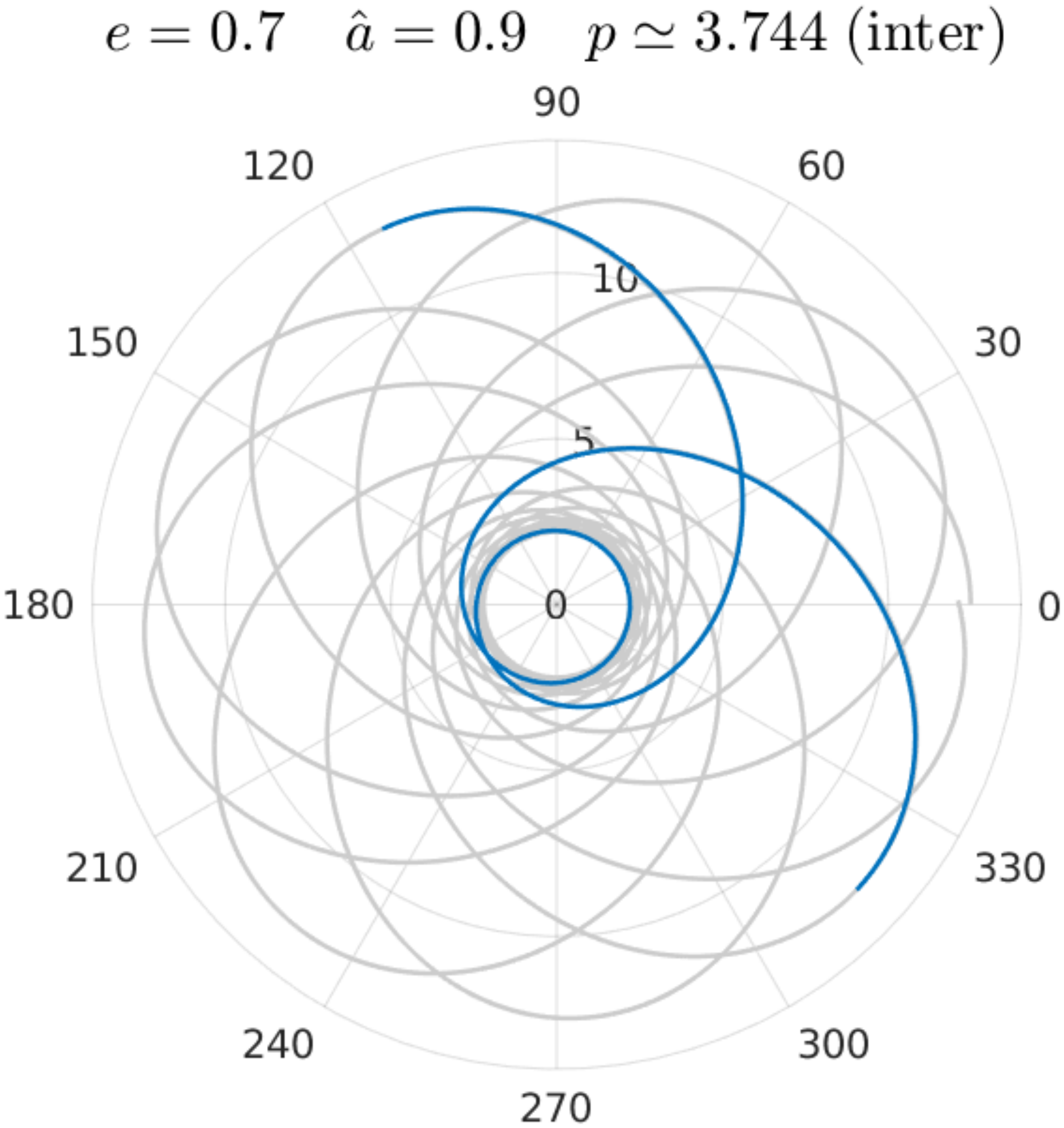}  
  \quad
  \includegraphics[width=0.22\textwidth]{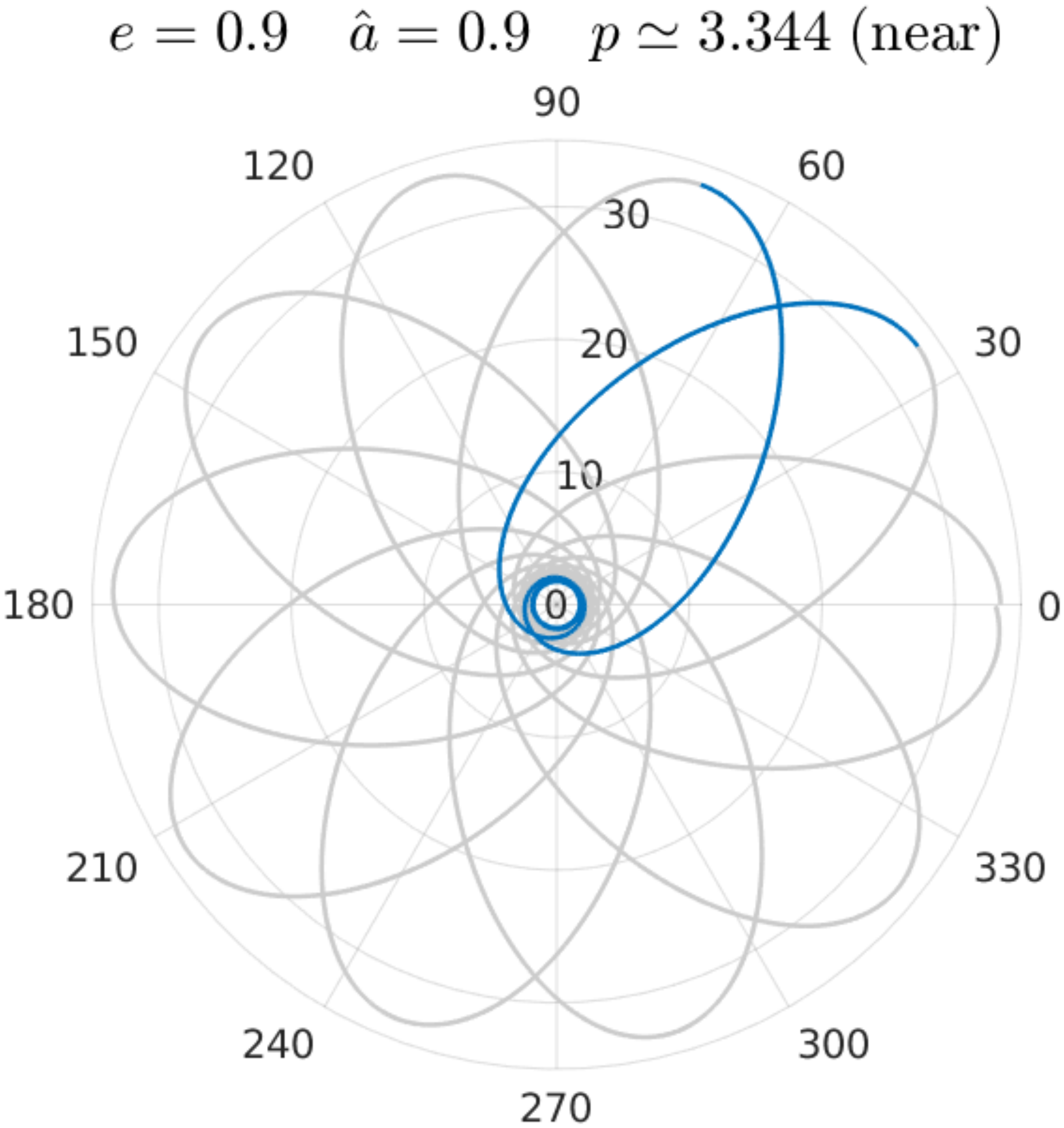}
  \caption{\label{fig:geo_dynamics} Geodesic equatorial orbits for different 
  ($e$, $\ha$, $p$) configurations,
  including the so-called {\it zoom-whirl} behavior (bottom-right panel). We have 
  highlighted one radial orbit for each configuration. The corresponding fluxes and 
  waveforms are shown in Fig.~\ref{fig:instaFluxes} and Fig.~\ref{fig:eobwaves}.}
\end{figure}
\begin{figure*}[t]
  \center  
  \includegraphics[width=0.47\textwidth]{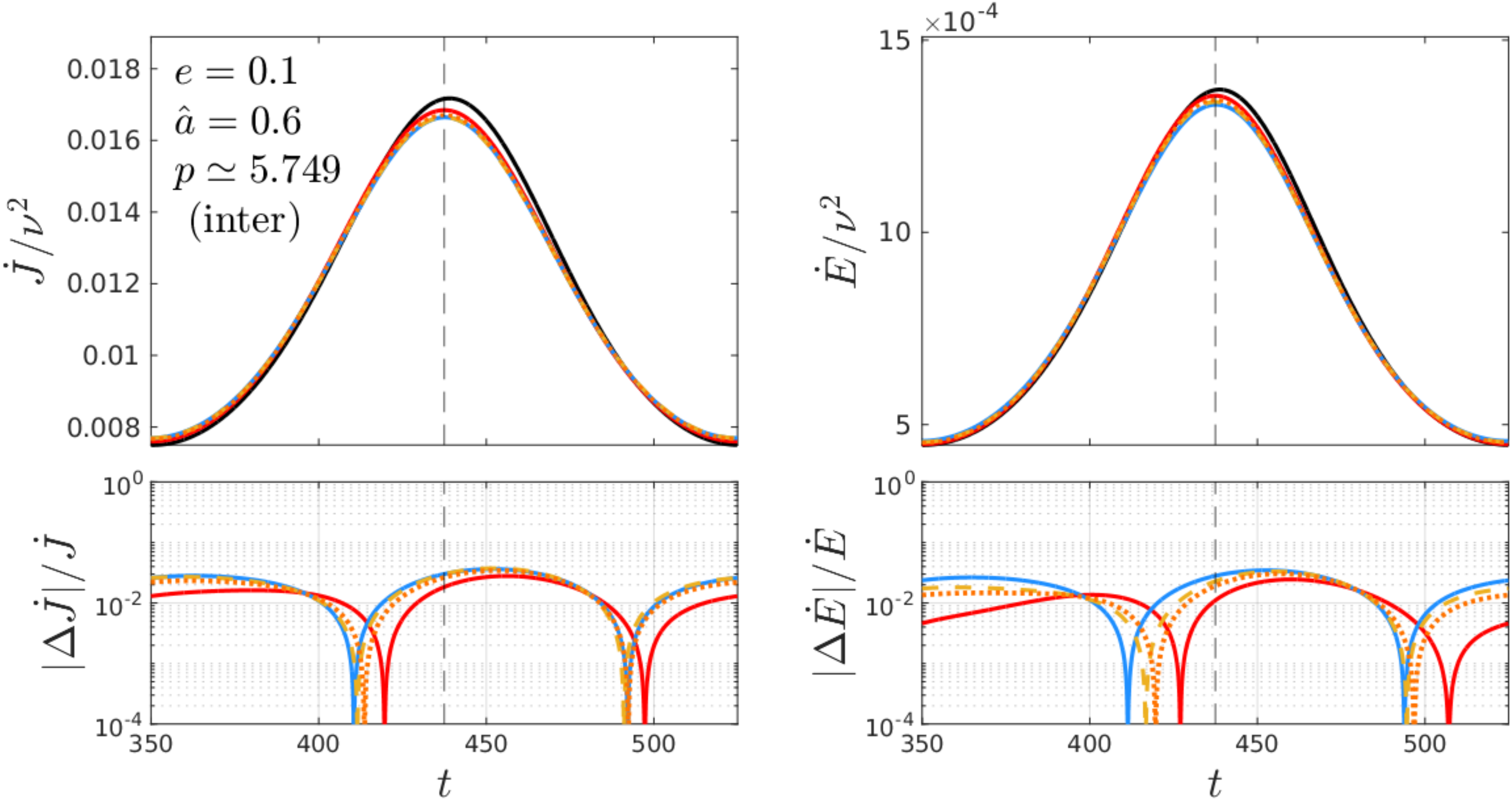}
  \hspace{0.6cm}
  \includegraphics[width=0.47\textwidth]{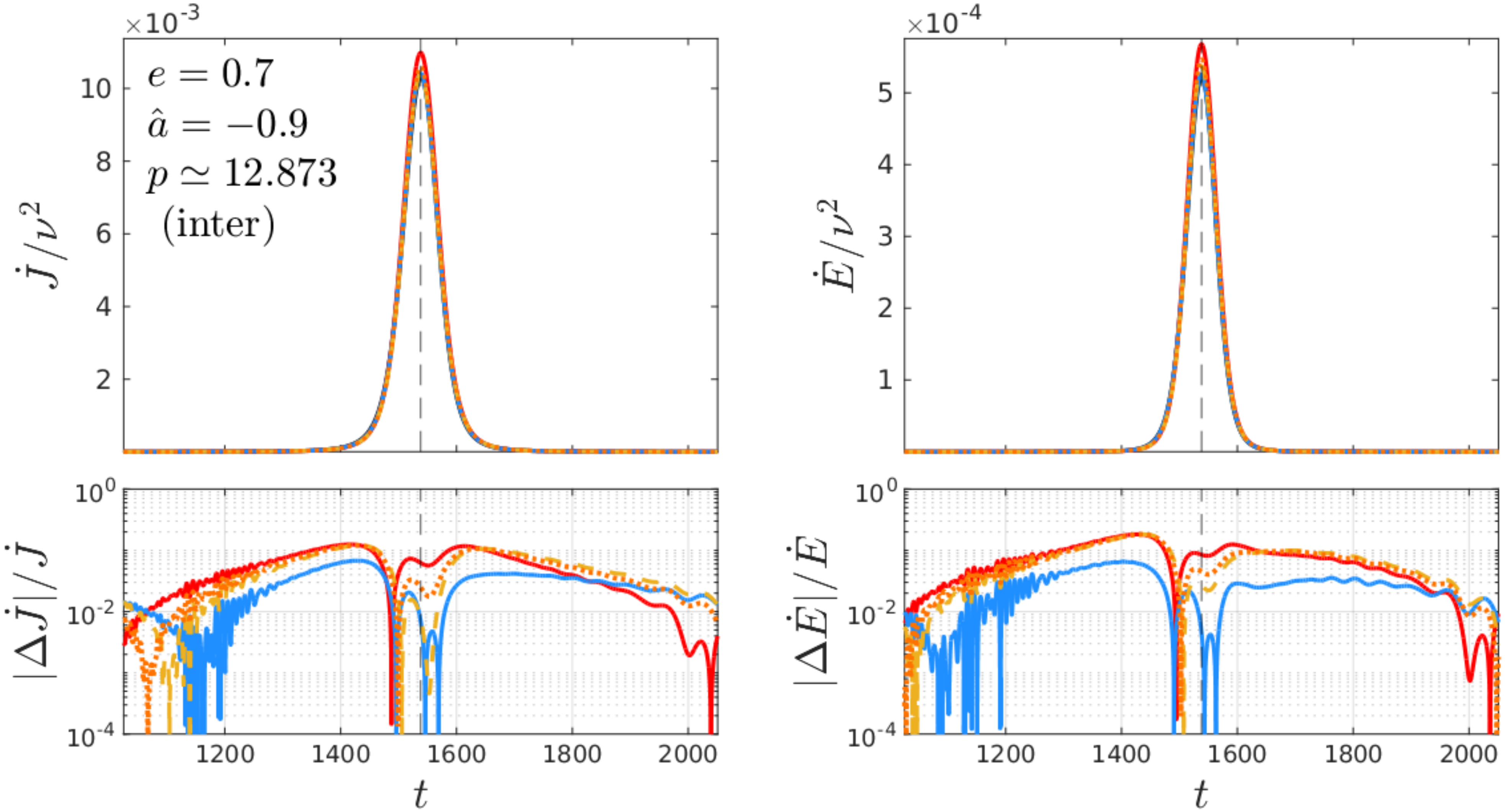} \\
  \vspace{0.6cm}
  \includegraphics[width=0.47\textwidth]{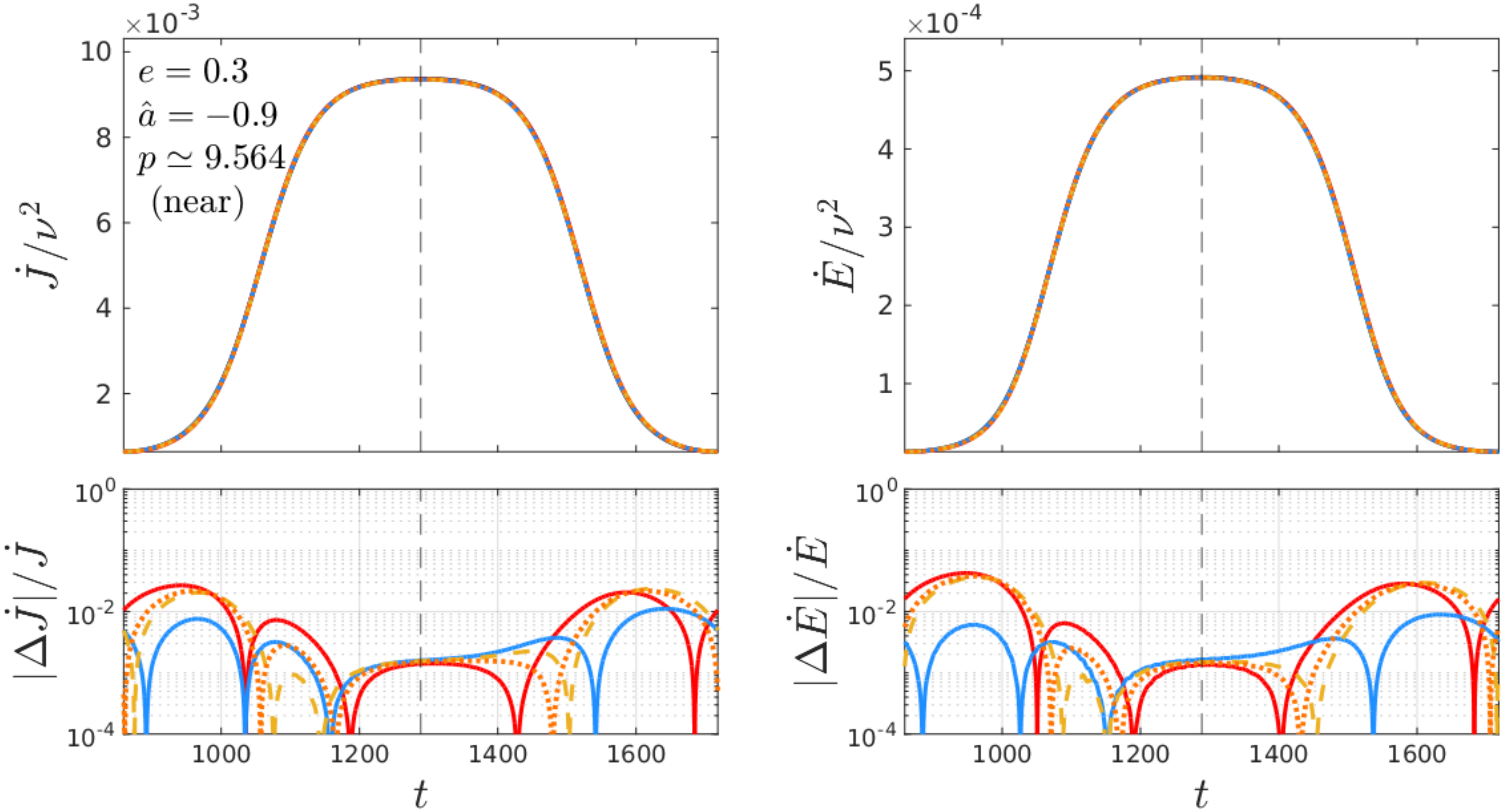} 
  \hspace{0.6cm}
  \includegraphics[width=0.47\textwidth]{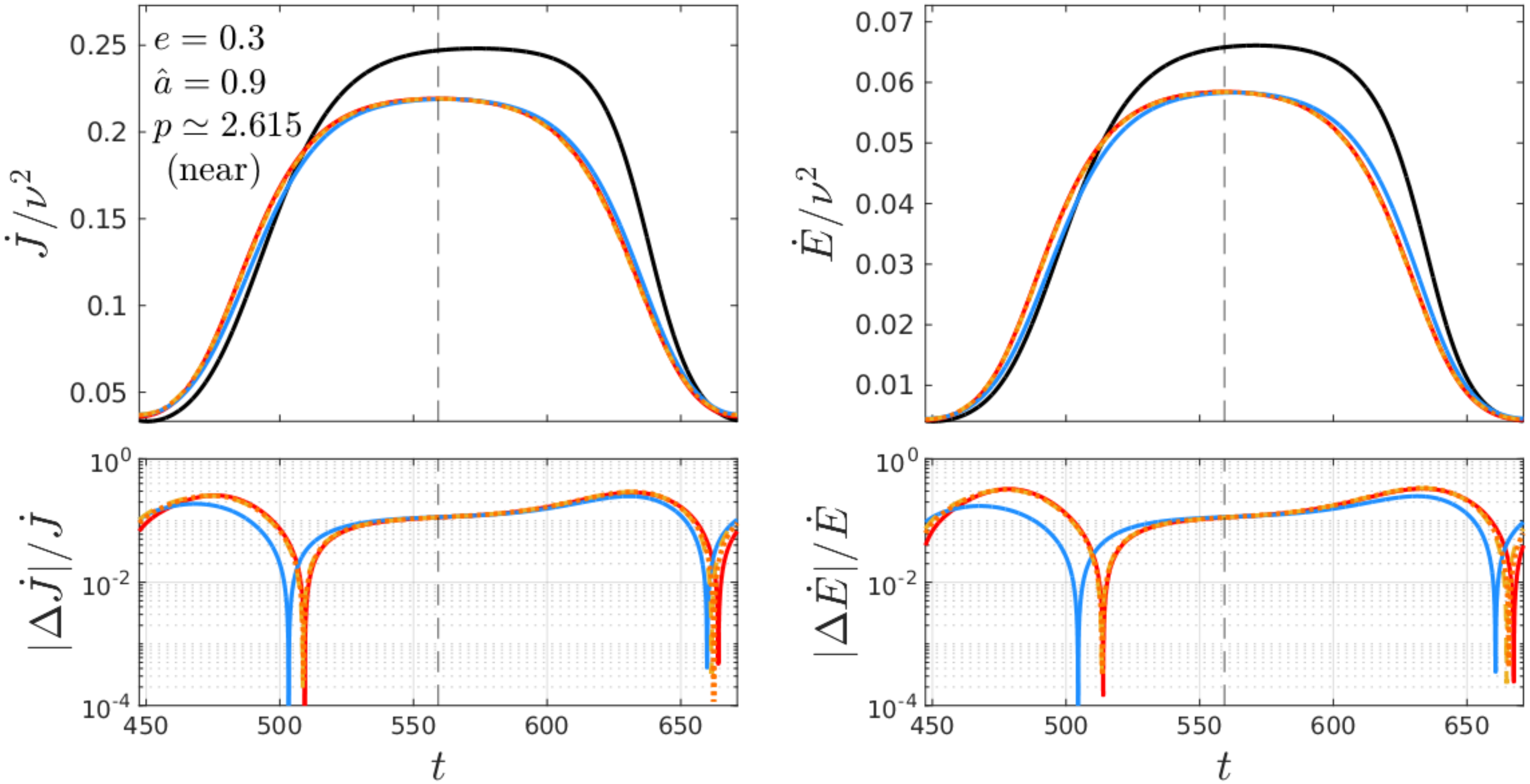} \\
  \vspace{0.6cm}
  \includegraphics[width=0.47\textwidth]{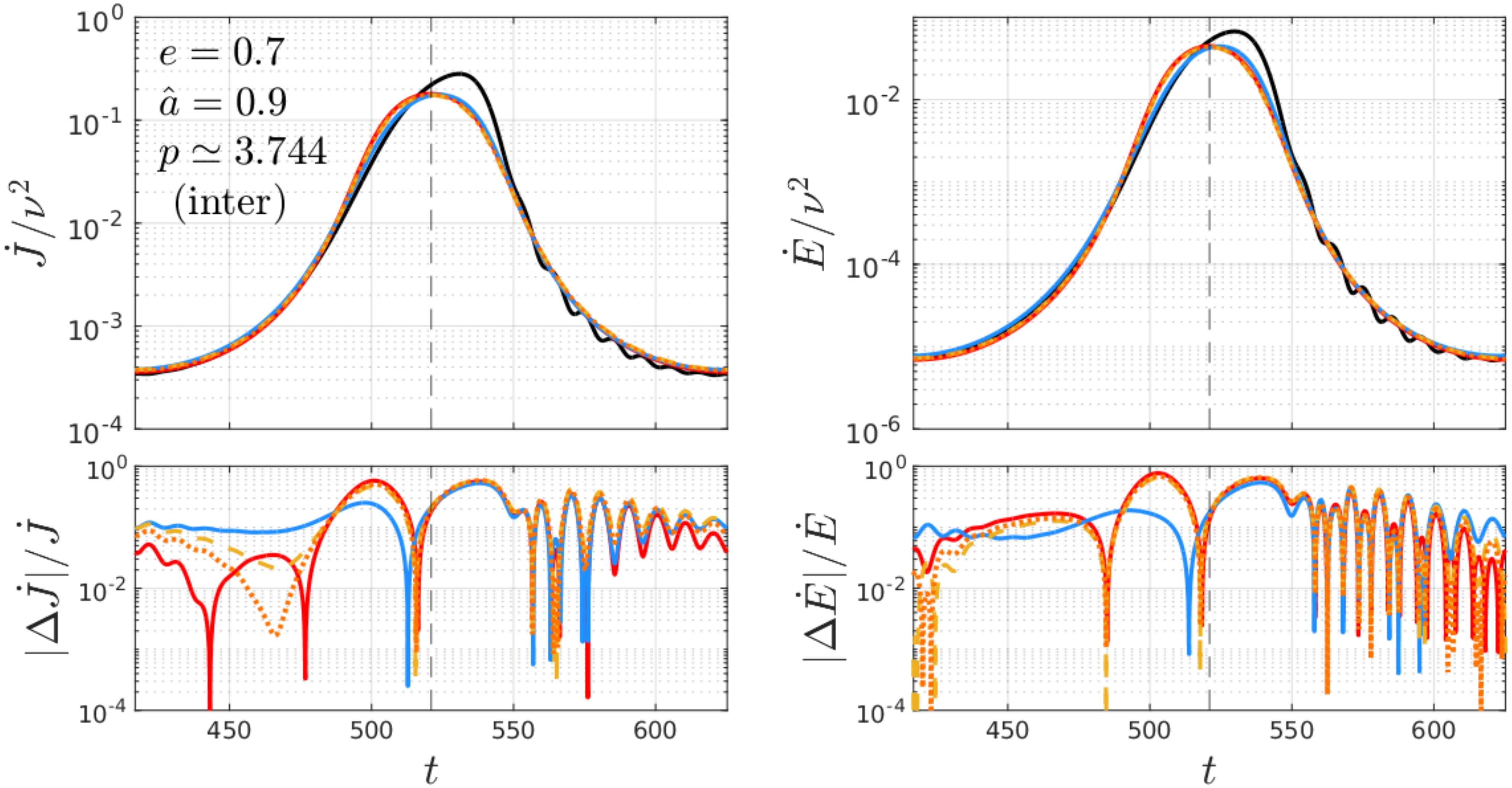} 
  \hspace{0.6cm}
  \includegraphics[width=0.47\textwidth]{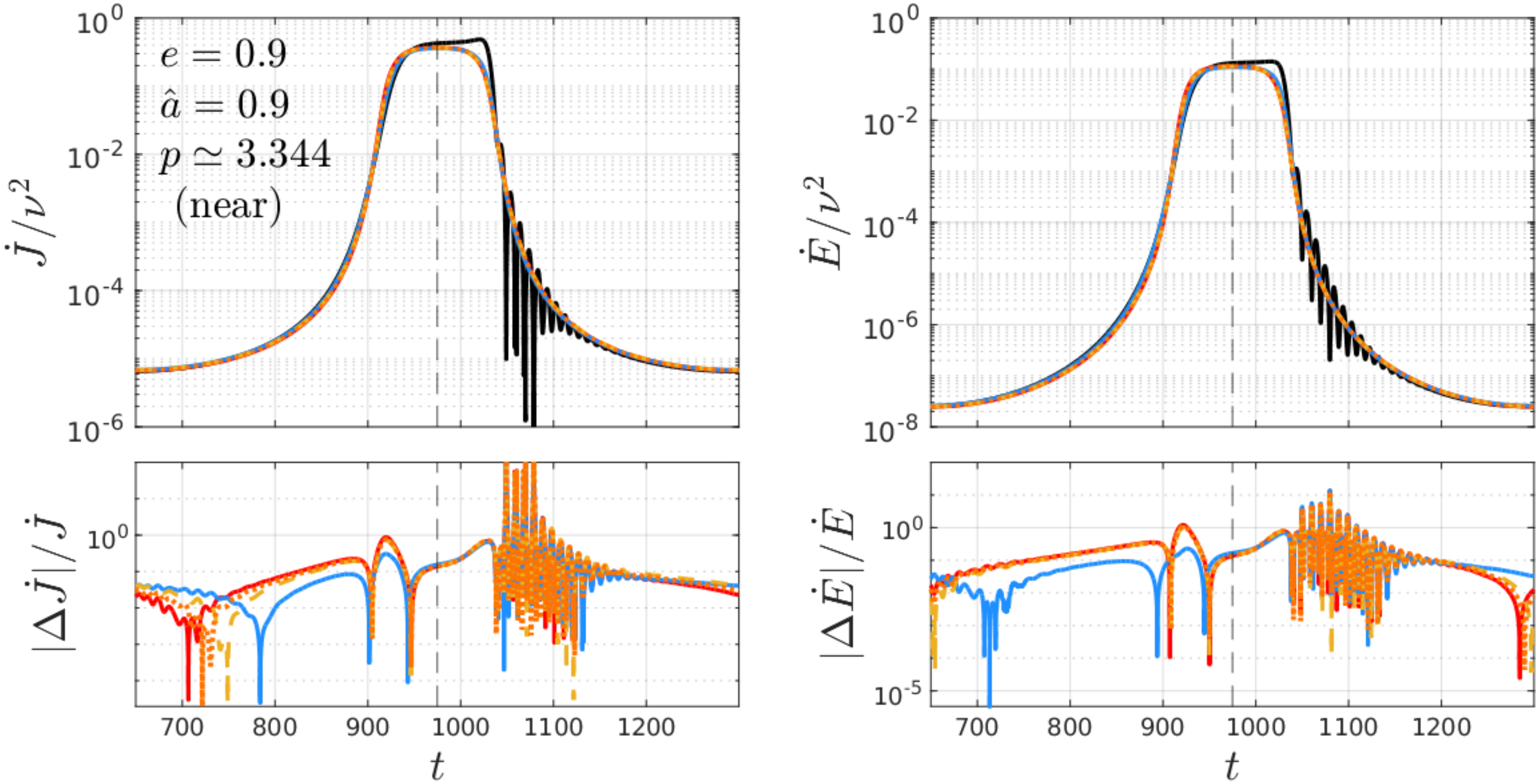}  
  \caption{\label{fig:instaFluxes} Angular momentum, $\dot{J}$, and energy, $\dot{E}$, 
  fluxes at infinity corresponding to the equatorial orbital dynamics shown 
  in Fig.~\ref{fig:geo_dynamics}. 
  The top panel of each plot focuses on a time interval around a periastron passage, that is indicated 
  by a vertical dashed line. The numerical fluxes (black) are contrasted with four different 
  analytical fluxes (in color). $\FNP$ (red), $\FANP$ (dashed orange), 
  $\FNPold$ (dotted orange), $\Fhlm$ (light blue).
  The bottom panel of each plot shows the absolute value of the corresponding relative difference.}
\end{figure*}
The phenomenology of the instantaneous GW fluxes of a particle on an eccentric orbit
around a Kerr black hole depends on the eccentricity, $e$, semilatus rectum, $p$, 
and black hole spin, $\hat{a}$.
To set the stage of our discussion, let us take a sample of illustrative configurations
that survey the different phenomenologies. The selected parameters and trajectories
are shown in Fig.~\ref{fig:geo_dynamics}. We considered both mildly eccentric and highly
eccentric configurations, some of these showing the so-called {\it zoom-whirl} 
behavior~\cite{Glampedakis:2002ya}, involving several revolutions around the central body 
near periastron (see e.g. bottom right panel of the Fig.~\ref{fig:geo_dynamics}).
In particular, an interesting feature arises from the fact that, when a corotating test-particle 
gets close to the light ring at high velocity, the QNMs of  the central black hole can be excited, 
producing high-frequency oscillations (usually addressed as \textit{wiggles}) 
in the radiated GWs at infinity~\cite{10.1143/PTP.72.494}. This phenomenon has been 
recently analyzed in details both using TD and FD codes~\cite{Rifat:2019fkt,Thornburg:2019ukt}. 
Moreover, the QNM excitation leads to strongly asymmetric fluxes with respect to the periastron passage.
The excitations become more relevant when the velocity of the test-particle increases and 
when the periastron of the orbits gets closer to the light ring. Moreover, 
the damping time of the QNMs increases significantly for high spins, therefore these 
excitations are mostly of interest for extremal black holes ($\ha\gtrsim 0.99$).
Nonetheless, QNMs excitations
are present also in less extreme cases, as pointed out in~\cite{Thornburg:2019ukt}. 
These effects are particularly relevant for prograde zoom-whirl orbits with high eccentricity 
around fast spinning black hole since the periastron can get very close to the light ring. 
The energy and angular momentum fluxes corresponding to the trajectories of 
Fig.~\ref{fig:geo_dynamics} are shown in Fig.~\ref{fig:instaFluxes}. 
The main aim of the figure is to compare the  numerical fluxes (in black) with the different 
flavors of the analytical fluxes (in color), but we will comment this in detail below. 
Still, one can clearly see how the QNMs excitations build up when $p$ is changed so to allow 
the system to pass close to the light ring. This occurs either in the case of extreme eccentricity
or in case of zoom-whirl behavior. 

The EOB analytical waveform model is not designed to incorporate QNMs-related
effects. Despite this, one can achieve a reasonable qualitative and quantitative agreement 
also when they are present. For each configuration, Fig.~\ref{fig:instaFluxes} contrast 
the exact, numerical, fluxes with {\it four} different analytical approximations: 
\begin{itemize}
\item[(i)] $\FphiNP$ from Eq.~\eqref{eq:Fphi_ecc} that corresponds to the $\FNP$ fluxes. 
In this case the angular radiation reaction has the noncircular (2,2) Newtonian prefactor 
$\NP22$ in the (2,2) multipole. This is now the standard prescription in state-of-the-art
EOB waveform model~\cite{Nagar:2021gss}.
\item[(ii)] $\Fphiold$ from Eq.~\eqref{eq:Fphi_ecc_old} from which we compute the $\FNPold$ fluxes. 
In $\Fphiold$, the noncircular (2,2) Newtonian prefactor $\NP22$ is a global factor.
\item[(iii)] $\FphiANP$ from Eq.~\eqref{eq:Fphi_ANP}, that provides the $\FANP$ fluxes. 
In this case we consider all the noncircular Newtonian prefactors up to $\l=6$.
\item[(iv)] $\Fhlm$, the fluxes obtained directly plugging the analytical waveform $h_\lm$
in Eq.~\eqref{eq:fluxes_infty} and summing over all multipoles up to $\ell_{\rm max}=8$. 
These quantities are exhibited as blue lines in the plots. 
\end{itemize} 
The various analytical choices offer substantially comparable approximations to
the numerical fluxes. In order to establish a preferred one, we need an efficient way to perform 
analytical/numerical comparisons all over the $(e,p,\hat{a})$ parameter space. 
We do so in the next section studying the behavior of orbital averaged fluxes.

\subsection{Averaged fluxes}
\label{sec:Flux_averaged}
\begin{figure*}[t]
	\center
 	\includegraphics[width=0.32\textwidth,height=3.9cm]{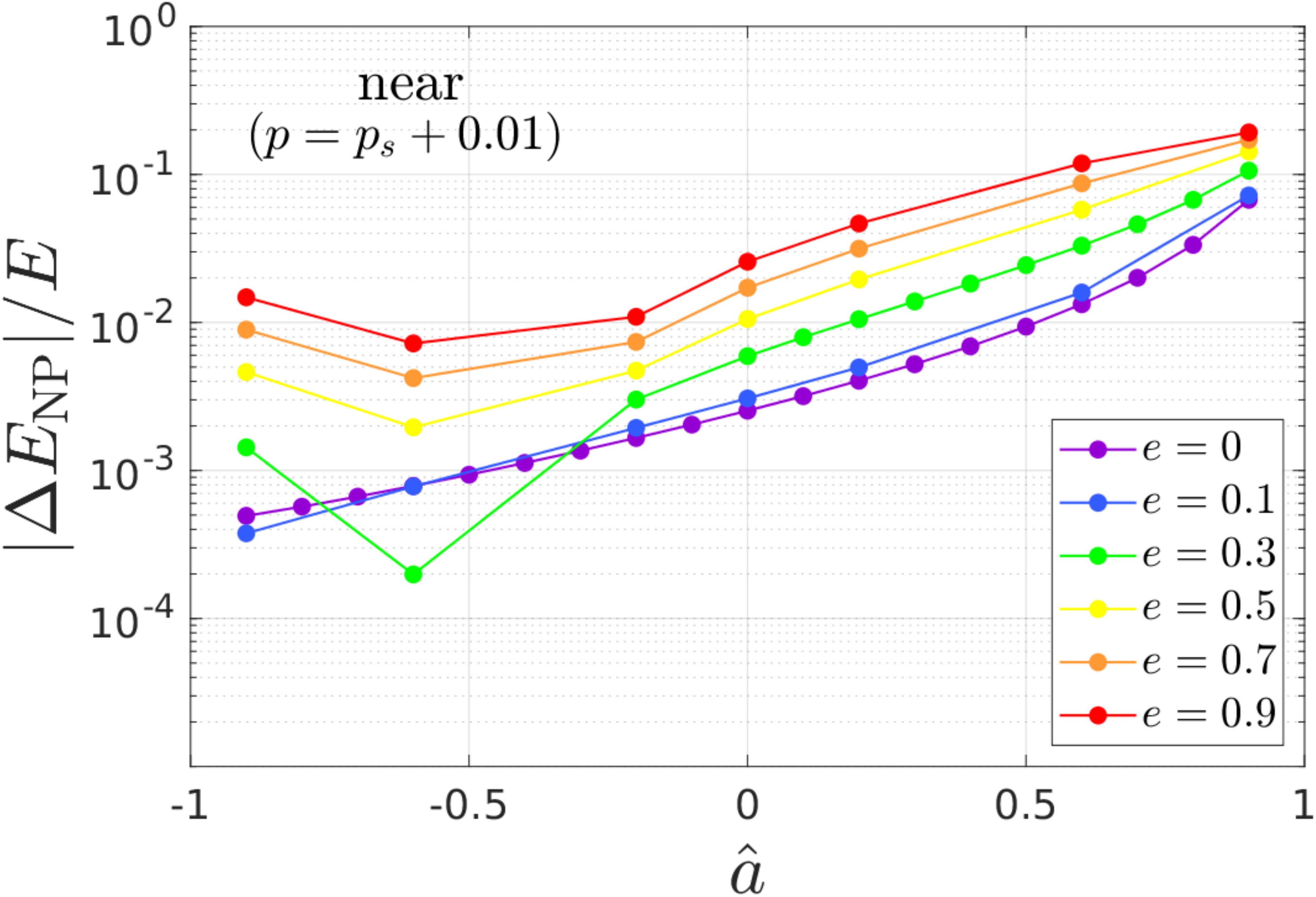}
	\includegraphics[width=0.32\textwidth,height=3.9cm]{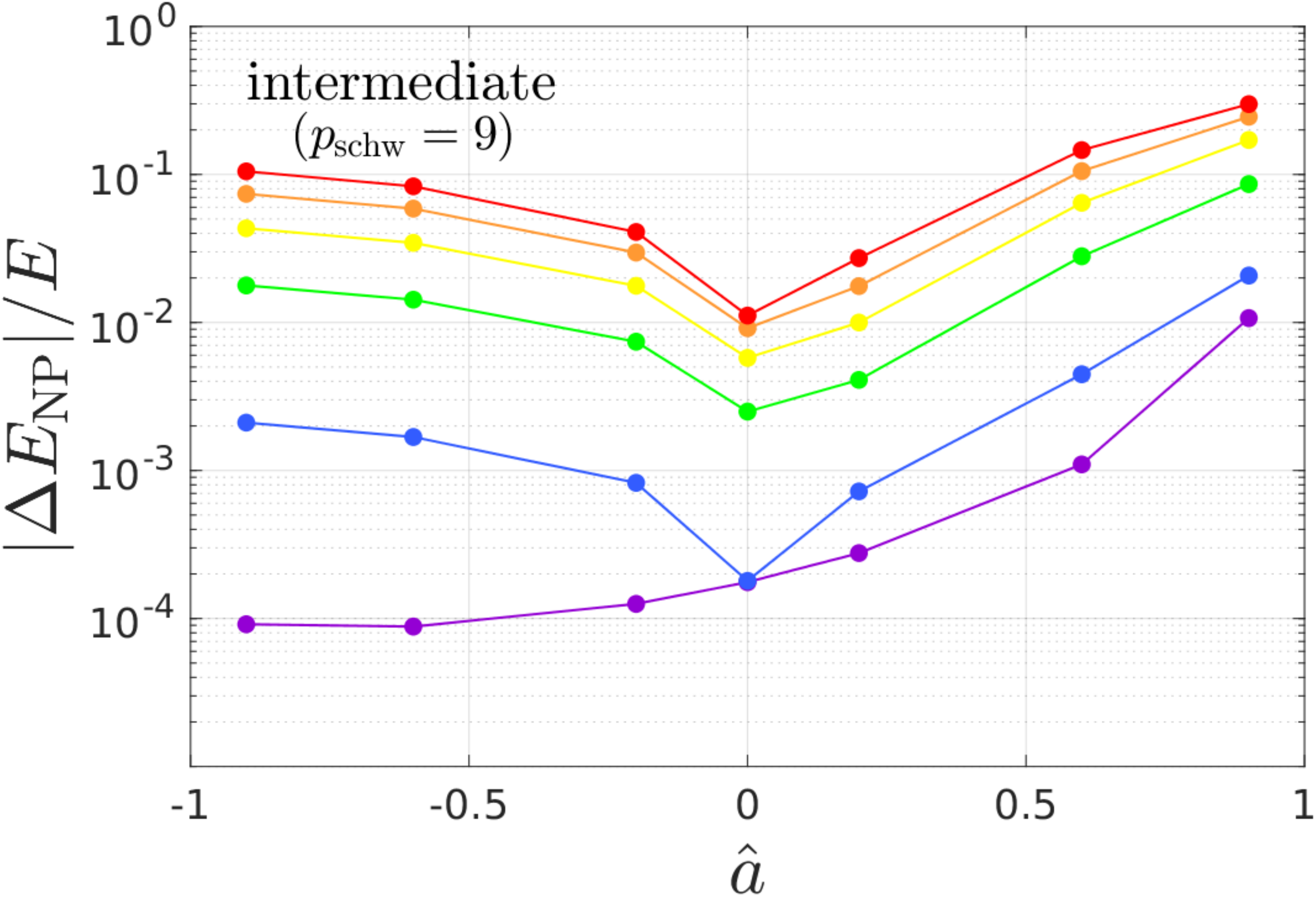}
	\includegraphics[width=0.32\textwidth,height=3.9cm]{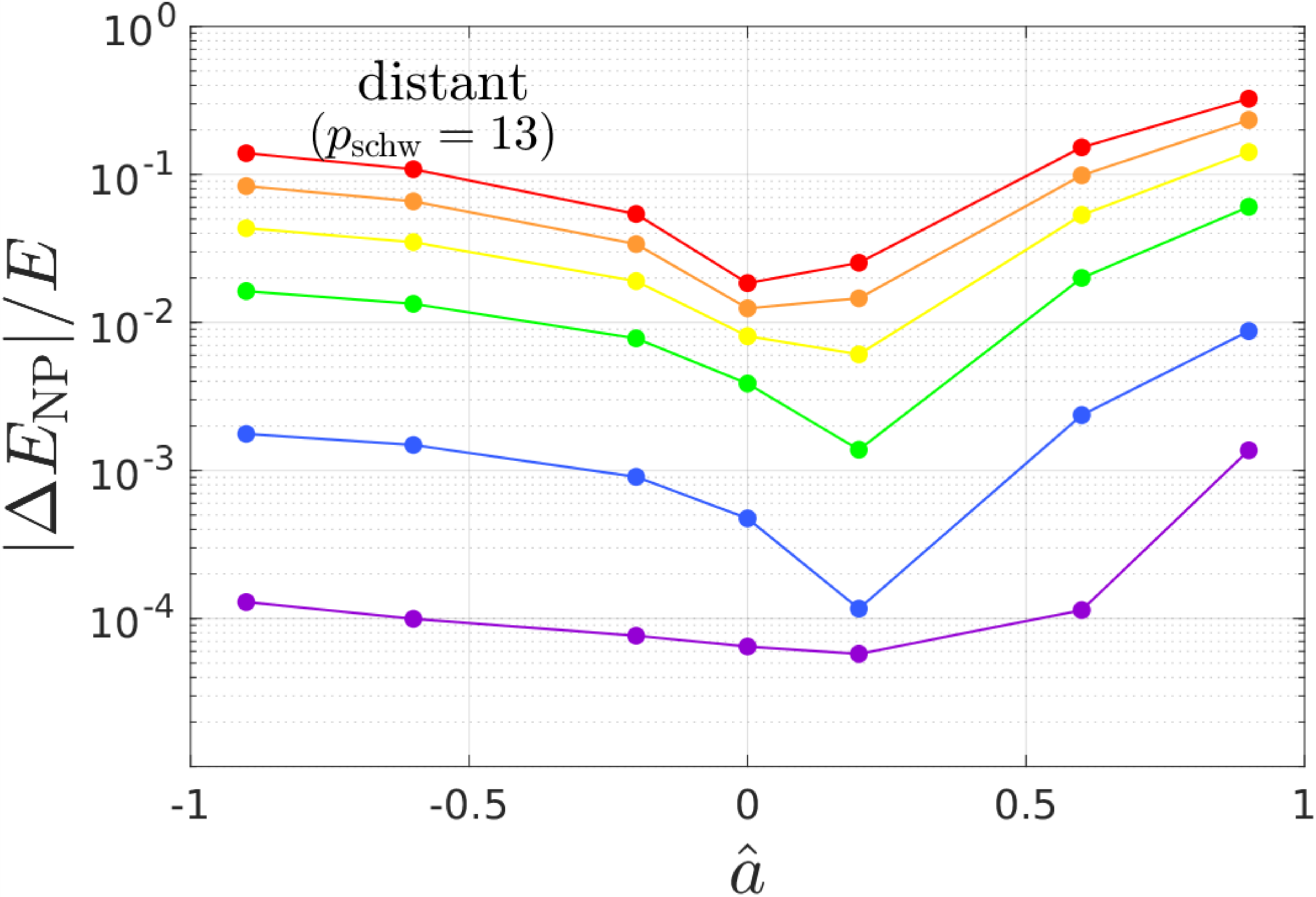} \\ 
	\includegraphics[width=0.32\textwidth,height=3.9cm]{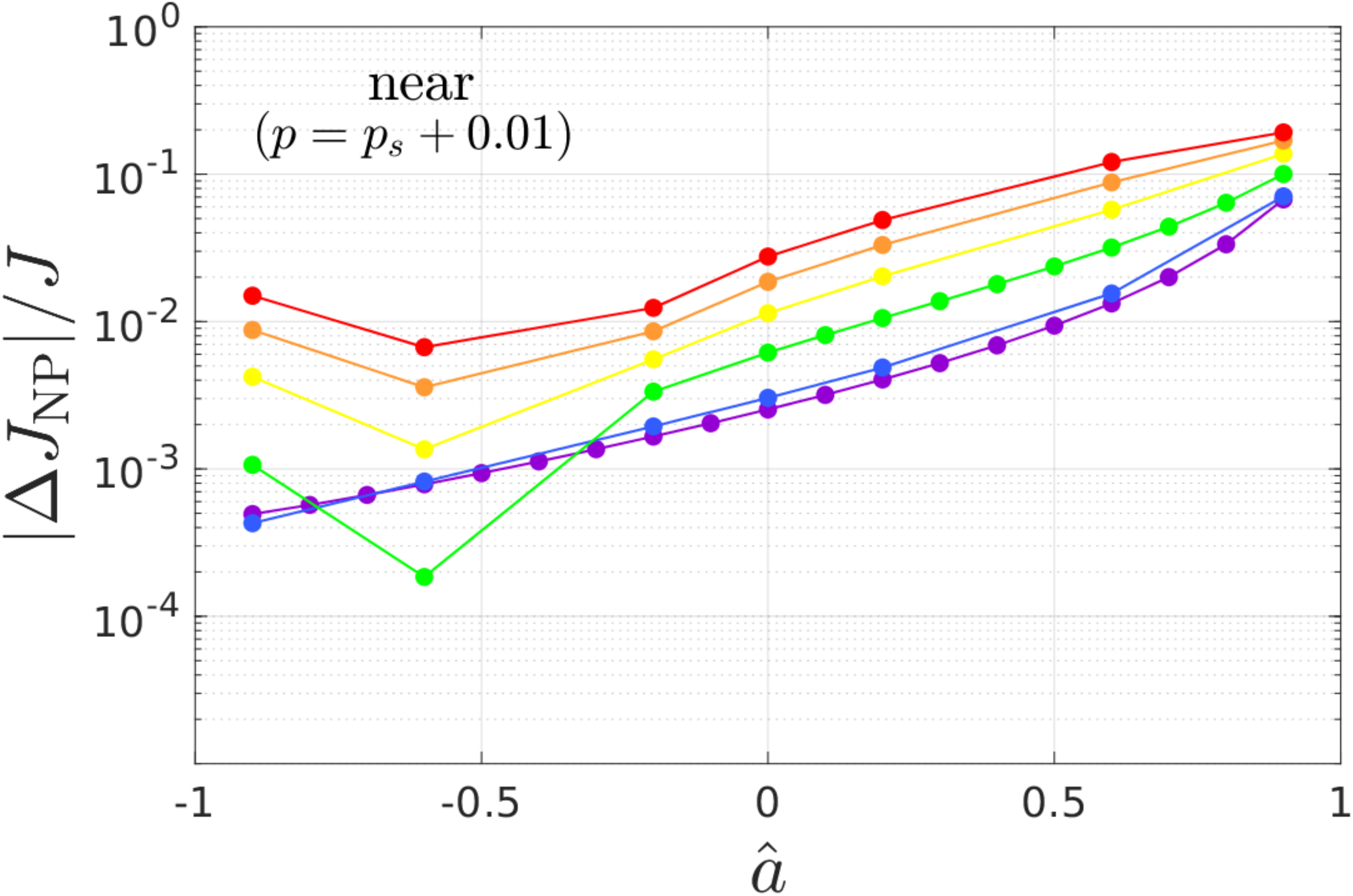}
	\includegraphics[width=0.32\textwidth,height=3.9cm]{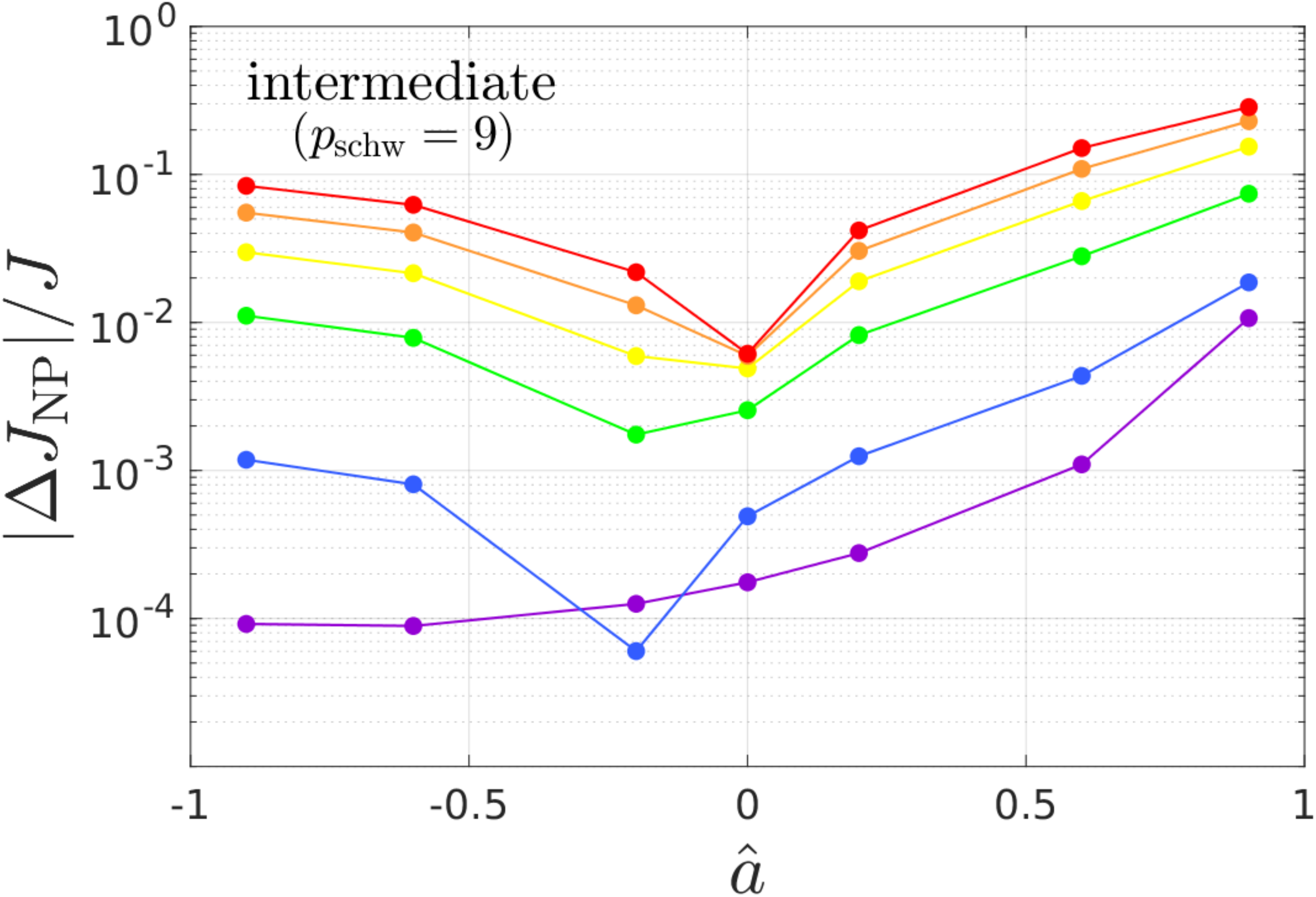}
	\includegraphics[width=0.32\textwidth,height=3.9cm]{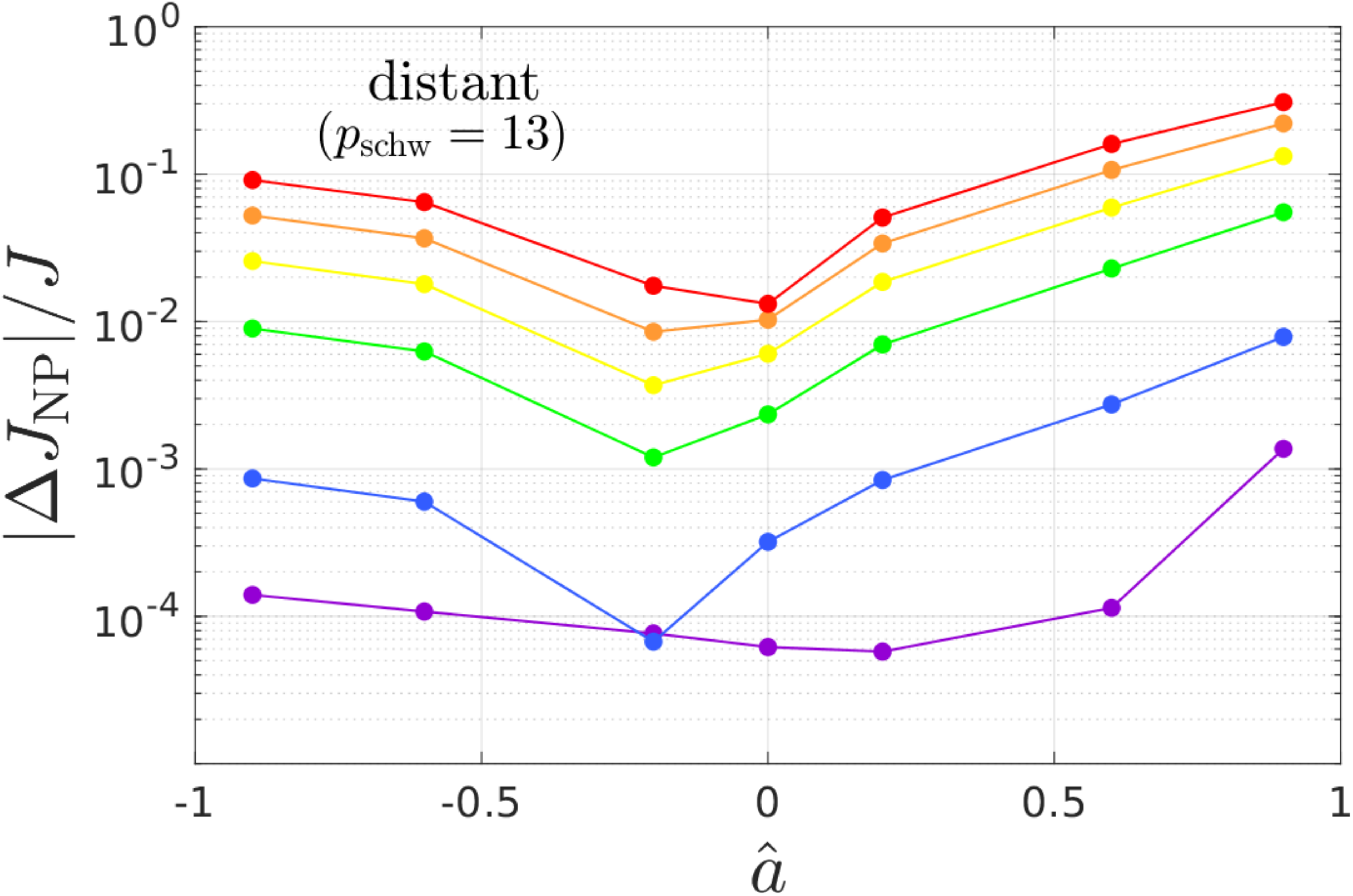} \\
	\caption{\label{fig:reldiff_rainbow_FNP} Relative differences between numerical and analytical
	   averaged fluxes plotted against the spin (absolute value, logarithmic scale). 
	   We consider $\FNP$, the fluxes  with the (2,2) general Newtonian prefactor 
	   of Eq.~\eqref{eq:NewtPref22_flux} in the (2,2) multipole.}
\end{figure*}
\begin{figure*}[t]
	\center
	\includegraphics[width=0.32\textwidth]{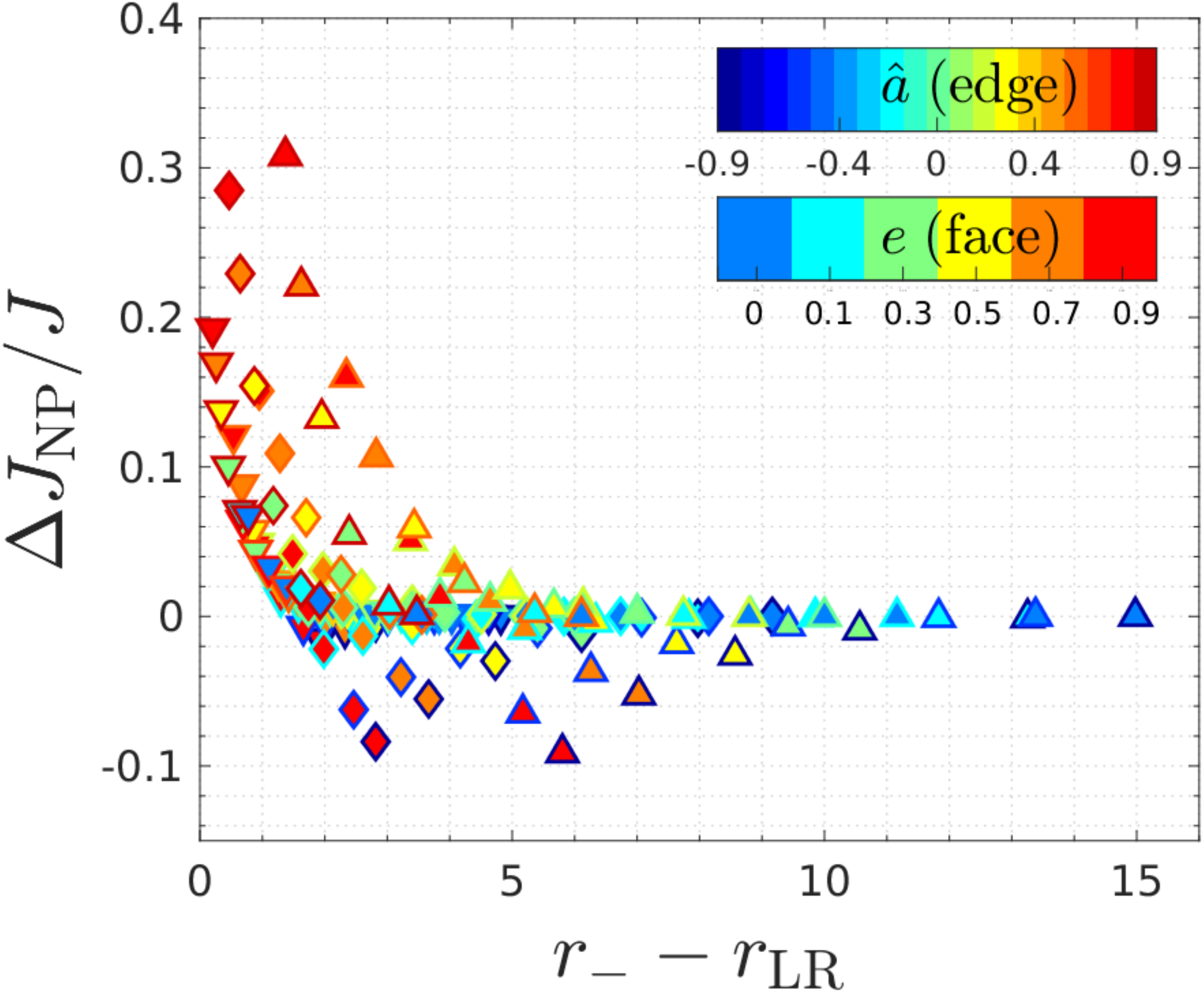}
	\includegraphics[width=0.32\textwidth]{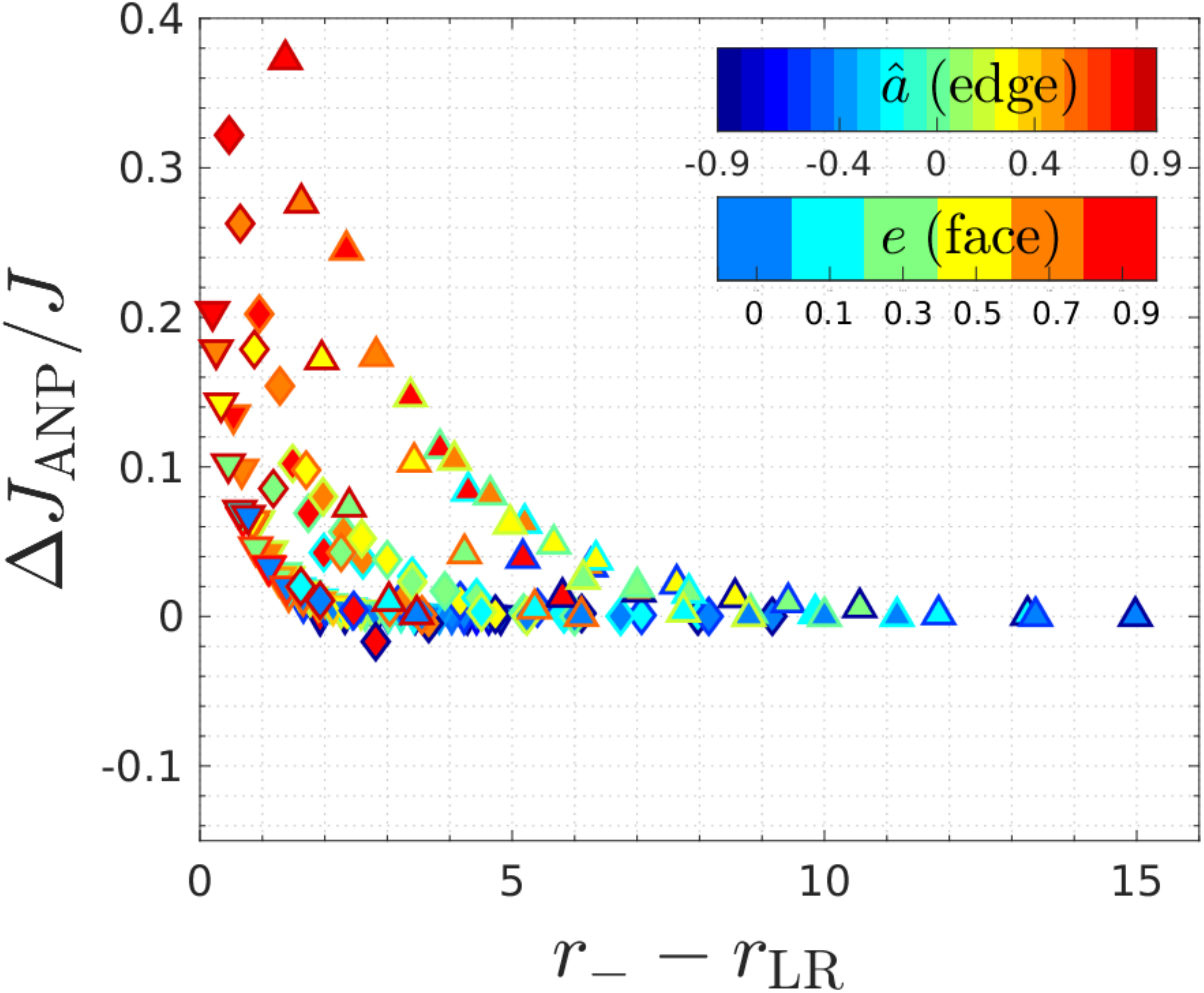}
	\includegraphics[width=0.32\textwidth]{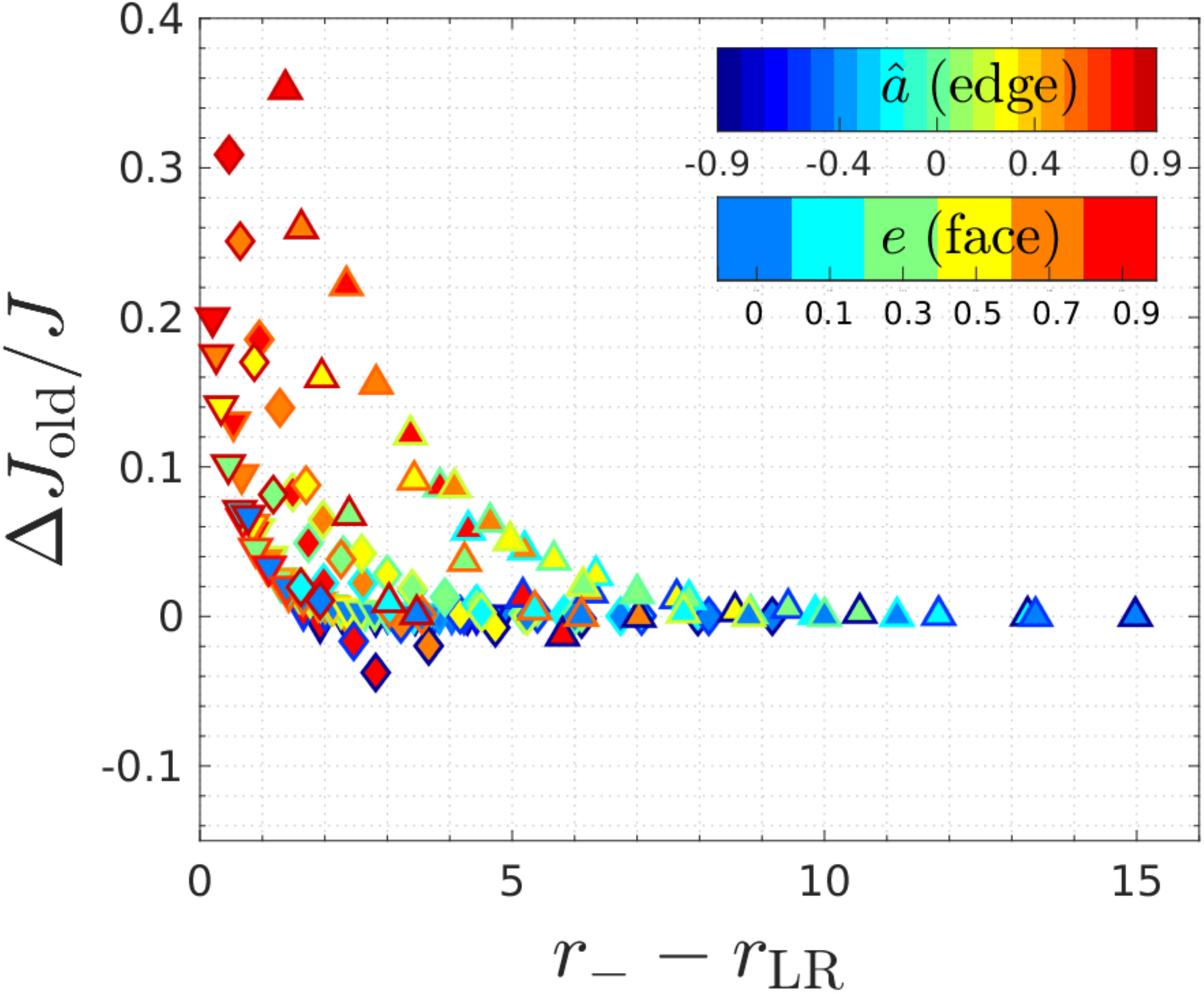}
	\hspace{0.1cm}
	\caption{\label{fig:reldiff_harlequin} Left panel: relative differences
	 $\Delta J_{\rm NP}/J \equiv (\av{\Jteuk} - \av{\JNP})/\av{\Jteuk} $  
	 for the fluxes with the (2,2) general 
	 Newtonian prefactor in the (2,2) multipole plotted against the distance between 
	 the light ring and the periastron. The face color of the markers indicates
	 the eccentricity, while the edge color indicates the spin. 
	 The shape of the markers is related to the rule used for the semilatus rectum: the reverse 
	 triangle indicates the near simulations, the diamond is for the 
	 intermediate simulations and the triangle pointing upward 
	 indicates the distant simulations. Other two panels: same scheme, 
	 but for $J_{\rm ANP}$ and $J_{\rm old}$.
	 See Fig.~\ref{fig:reldiff_harlequin_all_log} for analogous plots 
	 with $\Jhlm$ and also with the energy fluxes.}
\end{figure*}

In order to systematically and efficiently test the various analytical expressions
discussed above, we consider the $\nu$-normalized fluxes averaged over a radial period $T_r$
\begin{equation}
\label{eq:averaged_fluxes}
\av{ \dot{F} } = \frac{1}{T_r} \int_0^{T_r} \frac{\dot{F}}{\nu^2}\, dt,
\end{equation}
and we calculate the relative differences between numerical and analytical fluxes. 
We have considered all the analytical prescriptions
discussed in Sec.~\ref{sec:radreac}: $\FNP$, $\FNPold$, $\FANP$ and $\Fhlm$.
All the values for the relative differences between numerical and analytical averaged 
fluxes can be found in Appendix~\ref{appendix:Fluxes}.

Before analyzing the different analytical prescriptions, we discuss some general features of the 
fluxes over the parameter space. As a priori expected,
the analytical/numerical disagreement grows both with spin and eccentricity,
since the periastron of prograde orbits can get very close to the 
light ring for high spin parameters and high eccentricity. In strong fields, 
the PN series employed in our model lose their reliability, even if strengthened by a proper 
resummation. In particular, from the relative differences between numerical and $\FNP$ fluxes 
reported in Fig.~\ref{fig:reldiff_rainbow_FNP}, it is possible to see that
the spin is a relevant source of disagreement regardless of eccentricity.
On the other hand, the eccentricity is also crucial for the reliability of the fluxes, 
both for the lack of noncircular information in the angular radiation reaction 
beyond the Newtonian order and for 
the approaching of the periastron to stronger fields for higher eccentricities. 
The last issue can be easily 
seen in Fig.~\ref{fig:reldiff_harlequin}, where in the first panel we have plotted 
the same numerical/analytical relative differences of the angular 
momentum fluxes versus the distance between the light ring and the periastron. In the
other two panels of Fig.~\ref{fig:reldiff_harlequin} we have shown the differences for the other
radiation reactions of Eq.~\eqref{eq:Fphi_ANP} and Eq.~\eqref{eq:Fphi_ecc_old}.
The analogous plots for the other analytical fluxes can be found in Appendix~\ref{appendix:Fluxes},
but the general features discussed so far are valid for any of them.

In order to decide which radiation reaction adopt to drive the transition from inspiral to 
plunge (that we shall often address as {\it insplunge} in the following), we focus on 
the three analytical angular momentum fluxes that are strictly linked 
to an analytical prescription of the radiation reaction: 
$\JNP$, $\JNPold$, and $\JANP$ (see Sec.~\ref{sec:radreac} for more details). 
Note that the analytical prescription of $\hat{\F}_r$ is the same in all the radiation reactions
tested in this work, then to analyze them it is sufficient to study the angular momentum fluxes.
The fluxes computed from the EOB waveform (i.e. $\Jhlm$) are less informative since 
they are not related to the radiation reaction and we will not discuss them in details.  

As a first step, we focus on the two averaged analytical fluxes with only the (2,2) 
noncircular Newtonian prefactor $\NP22$, i.e.  $\av{\JNPold}$ and $\av{\JNP}$. 
These fluxes are computed, respectively, using the angular radiation reactions 
$\Fphiold$ from Eq.~\eqref{eq:Fphi_ecc_old} and $\FphiNP$ from Eq.~\eqref{eq:Fphi_ecc}. 
We start by noting that $\FphiNP$ is theoretically more consistent than $\Fphiold$ since
the former includes the (2,2) noncircular Newtonian prefactor 
only in the (2,2) multipole, while in the older prescription the noncircular correction is treated 
as a global factor, and therefore affects even the subdominant modes. 
Nonetheless, for retrograde orbits
$\av{\JNPold}$ has a better numerical/analytical agreement than $\av{\JNP}$. In fact, the latter 
overestimates significantly the numerical fluxes, especially at high eccentricity, as can 
be seen from Fig.~\ref{fig:reldiff_harlequin}. This can be 
explained by noting that the dominant contribution to the averaged fluxes occurs
at periastron, where we have $\NP22 <1$. Then, the lack of noncircular corrections in the subdominant
multipoles of $\av{\JNP}$ leads to an overestimate of the numerical result. 
On the other hand, in $\av{\JNPold}$ the noncircular correction $\NP22$ is a global factor 
and thus artificially reduces the contribution of the subdominant modes, fixing the overestimate.
Nonetheless, this is an artifact and the old radiation reaction is not solid: adding
noncircular information beyond the Newtonian level in the angular radiation reaction
would probably improves $\av{\JNP}$, but not $\av{\JNPold}$. \\
In the Schwarzschild case or for prograde orbits, the new prescription is more accurate. 
In fact, for spins aligned with the orbital angular momentum, the two 
analytical prescriptions generally underestimate 
the numerical fluxes, and therefore a global factor that is smaller than $1$ at periastron 
worsen the numerical/analytical agreement, making the older prescription less accurate. 

Let now consider the averaged fluxes $\av{\JANP}$, that are computed using $\FphiANP$
from Eq.~\eqref{eq:Fphi_ANP} and thus include all the noncircular corrections up to $\l=6$. 
Comparing them with the $\av{\JNP}$ fluxes, we can see that the ANP prescription is 
more reliable for retrograde orbits, but produces bigger relative differences with numerical data
in the nonspinning case and for prograde orbits. 
The reason of this behavior is again related to the fact that at periastron we have $\fnp{,\lm}<1$, 
as discussed above. 

\begin{table}
   \caption{\label{tab:schw_fluxes} Averaged analytical fluxes for test-particle on eccentric 
   orbits around a Schwarzschild black hole compared with numerical results, 
   $ \Delta J_{\rm NP}/J = (\av{ \Jteuk } - \av{ \JNP })/\av{ \Jteuk }$. 
   For $p\leq 13$, the numerical fluxes in this table are obtained with {\Teukode} and are a 
   subset of the ones shown in Appendix~\ref{appendix:Fluxes}. The fluxes with 
    greater semilatera recta are computed with {\RWZ}.} 
   \begin{center}
     \begin{ruledtabular}
\begin{tabular}{ c c | c c r } 
$e$ & $p$ & $\av{ \dot{J}_{\rm num} } $ & $\av{ \JNP } $ & 
\multicolumn{1}{c}{$\Delta J_{\rm NP}/J$} \\
\hline
\hline
$ 0.1 $ & $ 6.21 $ & $ 10.5396 $ & $ 10.5076 $ & $ 3.0\cdot 10^{-3} $ \\ 
$ 0.1 $ & $ 9 $ & $ 0.85442 $ & $ 0.85401 $ & $ 4.9\cdot 10^{-4} $ \\ 
$ 0.1 $ & $ 13 $ & $ 0.30867 $ & $ 0.30857 $ & $ 3.2\cdot 10^{-4} $ \\ 
$ 0.1 $ & $ 21 $ & $ 0.10208 $ & $ 0.10207 $ & $ 7.9\cdot 10^{-5} $ \\ 
$ 0.1 $ & $ 31 $ & $ 0.04445 $ & $ 0.04446 $ & $ -7.0\cdot 10^{-5} $ \\ 
\hline
$ 0.3 $ & $ 6.61 $ & $ 8.73092 $ & $ 8.67715 $ & $ 6.2\cdot 10^{-3} $ \\ 
$ 0.3 $ & $ 9 $ & $ 0.97246 $ & $ 0.96998 $ & $ 2.6\cdot 10^{-3} $ \\ 
$ 0.3 $ & $ 13 $ & $ 0.33935 $ & $ 0.33855 $ & $ 2.3\cdot 10^{-3} $ \\ 
$ 0.3 $ & $ 21 $ & $ 0.11055 $ & $ 0.11048 $ & $ 5.9\cdot 10^{-4} $ \\ 
$ 0.3 $ & $ 31 $ & $ 0.04786 $ & $ 0.04789 $ & $ -6.6\cdot 10^{-4} $ \\ 
\hline
$ 0.5 $ & $ 7.01 $ & $ 9.22978 $ & $ 9.12458 $ & $ 1.1\cdot 10^{-2} $ \\ 
$ 0.5 $ & $ 9 $ & $ 1.23262 $ & $ 1.22659 $ & $ 4.9\cdot 10^{-3} $ \\ 
$ 0.5 $ & $ 13 $ & $ 0.40329 $ & $ 0.40085 $ & $ 6.0\cdot 10^{-3} $ \\ 
$ 0.5 $ & $ 21 $ & $ 0.12784 $ & $ 0.12760 $ & $ 1.9\cdot 10^{-3} $ \\ 
$ 0.5 $ & $ 31 $ & $ 0.05479 $ & $ 0.05488 $ & $ -1.8\cdot 10^{-3} $ \\ 
\hline
$ 0.7 $ & $ 7.41 $ & $ 10.5483 $ & $ 10.3521 $ & $ 1.9\cdot 10^{-2} $ \\ 
$ 0.7 $ & $ 9 $ & $ 1.69513 $ & $ 1.68505 $ & $ 5.9\cdot 10^{-3} $ \\ 
$ 0.7 $ & $ 13 $ & $ 0.50634 $ & $ 0.50113 $ & $ 1.0\cdot 10^{-2} $ \\ 
$ 0.7 $ & $ 21 $ & $ 0.15477 $ & $ 0.15408 $ & $ 4.4\cdot 10^{-3} $ \\ 
$ 0.7 $ & $ 31 $ & $ 0.06542 $ & $ 0.06562 $ & $ -3.1\cdot 10^{-3} $ \\ 
\hline
$ 0.9 $ & $ 7.81 $ & $ 12.5209 $ & $ 12.1752 $ & $ 2.8\cdot 10^{-2} $ \\ 
$ 0.9 $ & $ 9 $ & $ 2.49117 $ & $ 2.47585 $ & $ 6.1\cdot 10^{-3} $ \\ 
$ 0.9 $ & $ 13 $ & $ 0.65853 $ & $ 0.64984 $ & $ 1.3\cdot 10^{-2} $ \\ 
$ 0.9 $ & $ 21 $ & $ 0.19258 $ & $ 0.19098 $ & $ 8.3\cdot 10^{-3} $ \\ 
$ 0.9 $ & $ 31 $ & $ 0.08011 $ & $ 0.08041 $ & $ -3.8\cdot 10^{-3} $ \\
\end{tabular}
\end{ruledtabular}
\end{center}
\end{table}

As anticipated, the fluxes $\av{\JNP}$ are the most accurate in the Schwzarschild case. 
As can be seen from Table~\ref{tab:schw_fluxes},
in the nonspinning case this prescription is highly reliable for every configuration, 
yielding relative differences always below the $3\%$, even for orbits with high eccentricity. 
Instead, for the other analytical prescriptions, 
the relative difference can be even $>7\%$, as shown in Table~\ref{tab:AllFluxesDiff1} 
and Table~\ref{tab:AllFluxesDiff2}. Note that for $p\geq 9$, the accuracy of our analytical model 
is consistent with the averaged eccentric fluxes at 10~PN computed in 
Refs.~\cite{Munna:2020juq, Munna:2020iju}, but the EOB model remains solid even for 
smaller semilatera recta thanks to a robust resummation of the PN series.
 
In conclusion, considering that
(i) the old prescription is not theoretically accurate,
(ii) the radiation reaction with all the multipoles does not drastically 
improves the model for moderate eccentricities and moderate spins
despite being much more complicated,
(iii) the fluxes $\av{\JNP}$ are the most faithful in the nonspinning case, 
we decide to use in our eccentric insplunge simulations the 
angular radiation reaction $\FphiNP$ from Eq.~\eqref{eq:Fphi_ecc}. 

With this choice, the energy and angular momentum fluxes have an agreement of 
few percents for moderate eccentricities ($e\lesssim 0.3$), even if for prograde orbits the spin 
reduces the maximum eccentricity up to which the model has good agreement
with the numerical results.
In the worst case, i.e. the one with $\ha=0.9$, $e=0.9$ and 'distant' semilatus rectum ($p\simeq 5.557$), 
we get relative differences of the $33\%$ and $31\%$ for the energy and angular momentum 
fluxes, respectively (see Table~\ref{tab:simulations_ecc3}). 
Note that the worst case is not a near simulation, since in that case 
the lack of noncircular information is compensated by the zoom-whirl behavior.

\subsection{Subdominant multipoles}
~\label{sec:subdominant}
\begin{figure}
  \center  
  \includegraphics[width=0.48\textwidth]{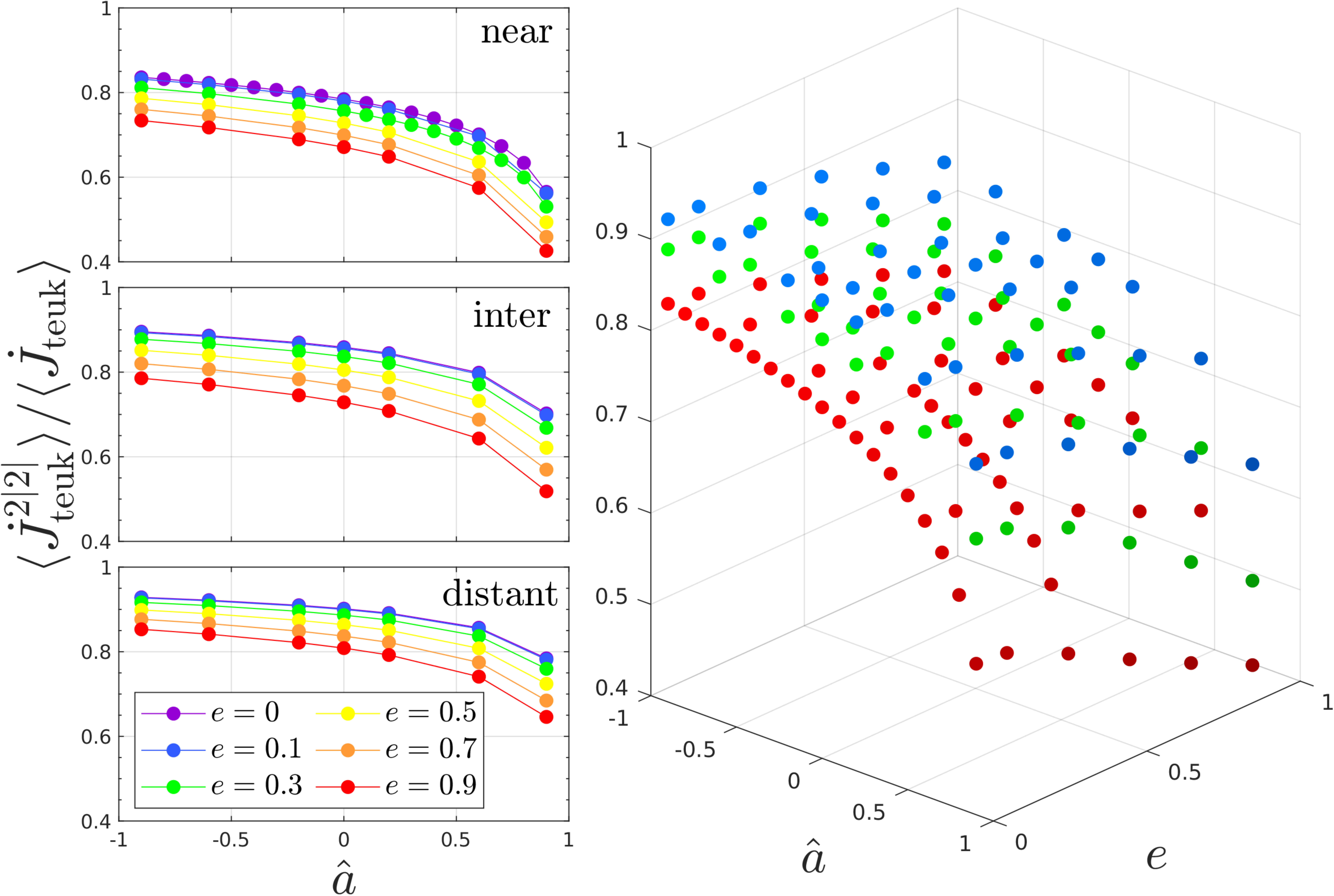}
  \caption{\label{fig:J22_contributions} Relative contributions of the $\l=|m|=2$ modes to 
  the total averaged angular momentum flux over the parameter space. In the 3D plot we 
  show all the near (red), intermediate (green) and distant (blue) simulations, while on the left 
  we have separated the three different classes. }
\end{figure}
Let us finally discuss in some detail the separate contribution of the various subdominant modes
to the angular momentum flux. 
Subdominant modes become more and more relevant for high spins, high 
eccentricities and for semilatera recta near the separatrix as shown 
in Fig.~\ref{fig:J22_contributions}, where we have plotted the  
$\l=|m|=2$ relative contribution to the total averaged angular momentum flux. 
In the distant circular case with $\ha=-0.9$, it provides 
the $92.9\%$ of the total flux; by contrast, for the zoom-whirl orbit with $\ha=0.9$ 
and $e=0.9$, its contribution is only of the $42.6\%$. 
In the former case, the relative contribution of the $\l=8$ 
modes summed together is $3.9\cdot 10^{-5} \%$, while in the latter is $2.2\%$. 
For the complete list of the relative $\l$-contributions to the angular momentum flux, 
see Appendix~\ref{appendix:subdominantJ}. 

The analytical subdominant modes tend to have greater discrepancies with the corresponding
numerical multipoles than the dominant one. In the noncircular case, this result is trivial 
for the fluxes with only the (2,2) noncircular Newtonian prefactor, but holds 
even if we include the noncircular corrections for each multipole.

\section{Multipolar waveform}
\label{sec:waves}

\begin{figure*}
  \center  
  \includegraphics[width=0.22\textwidth,height=3cm]{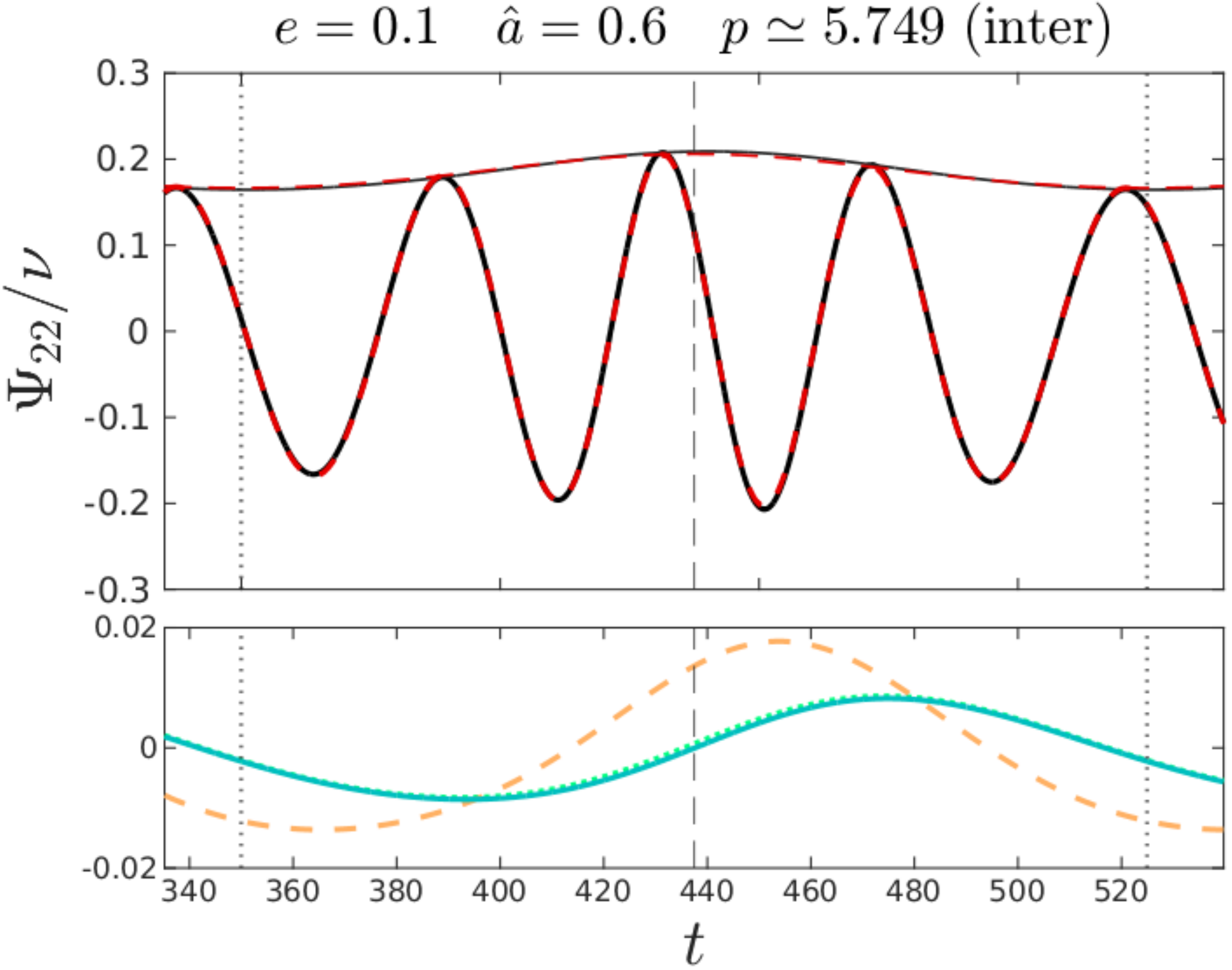}
  \includegraphics[width=0.22\textwidth,height=3cm]{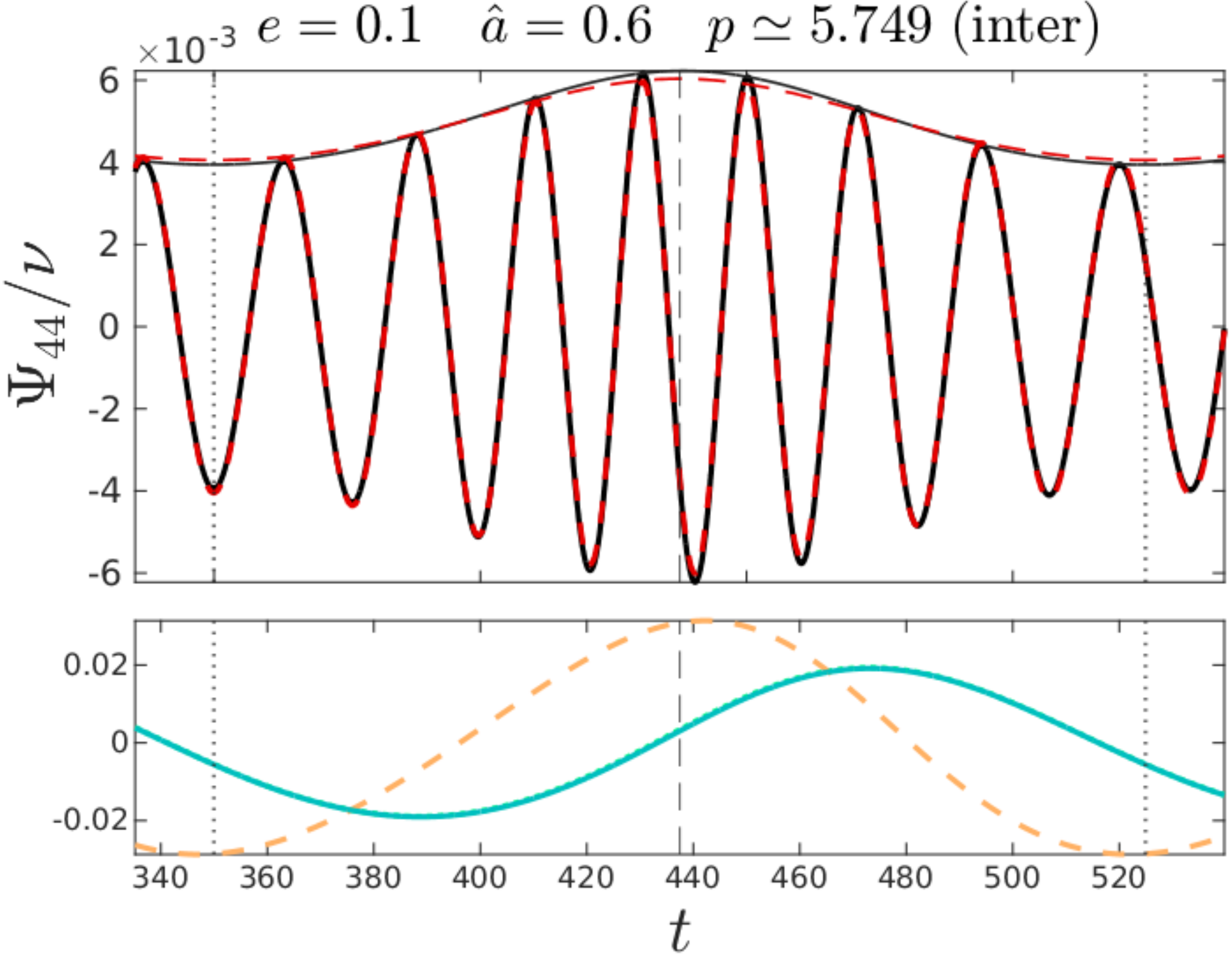}
  \hspace{1.0cm}
  \includegraphics[width=0.22\textwidth,height=3cm]{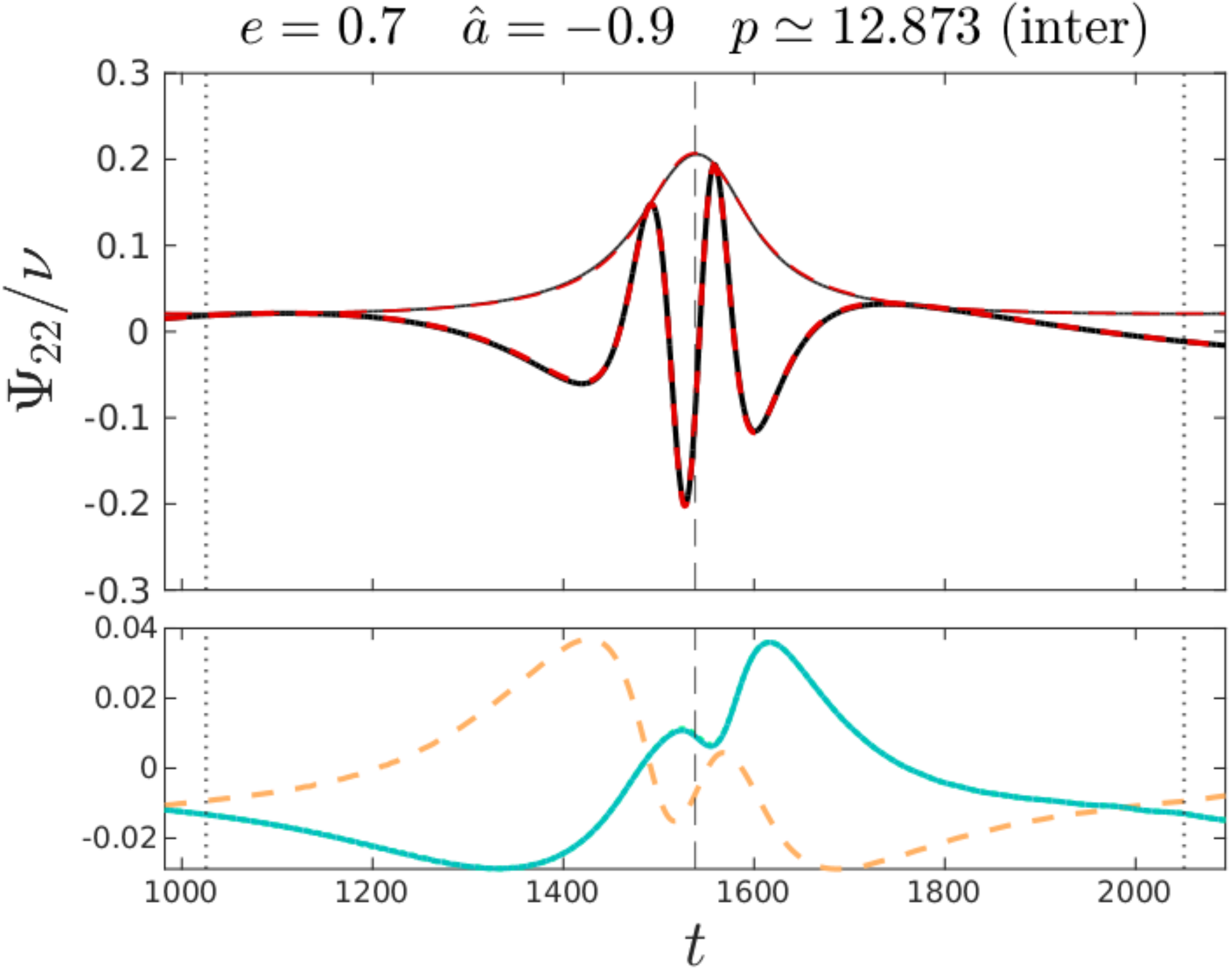}
  \includegraphics[width=0.22\textwidth,height=3cm]{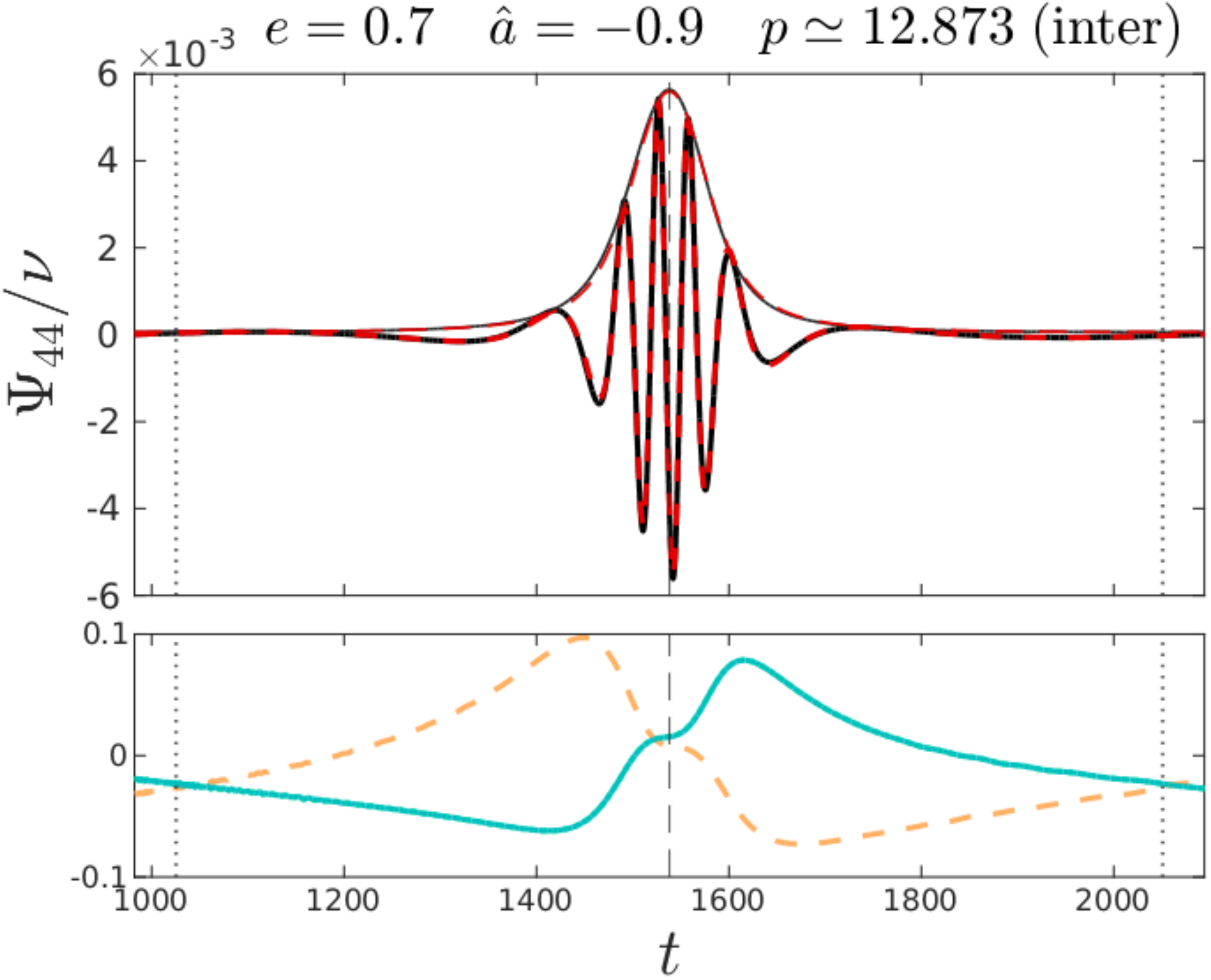} \\
  \vspace{1.0cm}
  \includegraphics[width=0.22\textwidth,height=3cm]{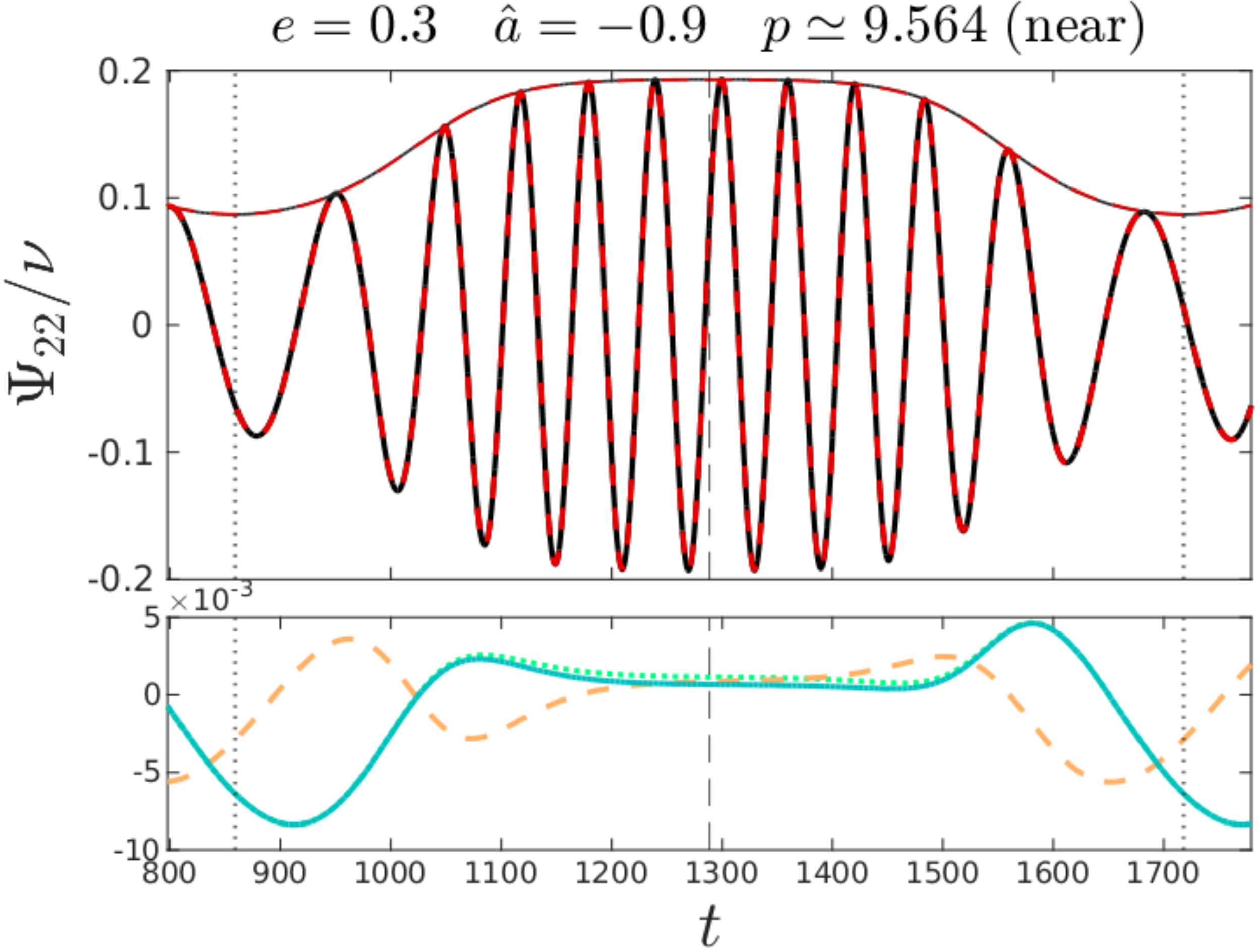} 
  \includegraphics[width=0.22\textwidth,height=3cm]{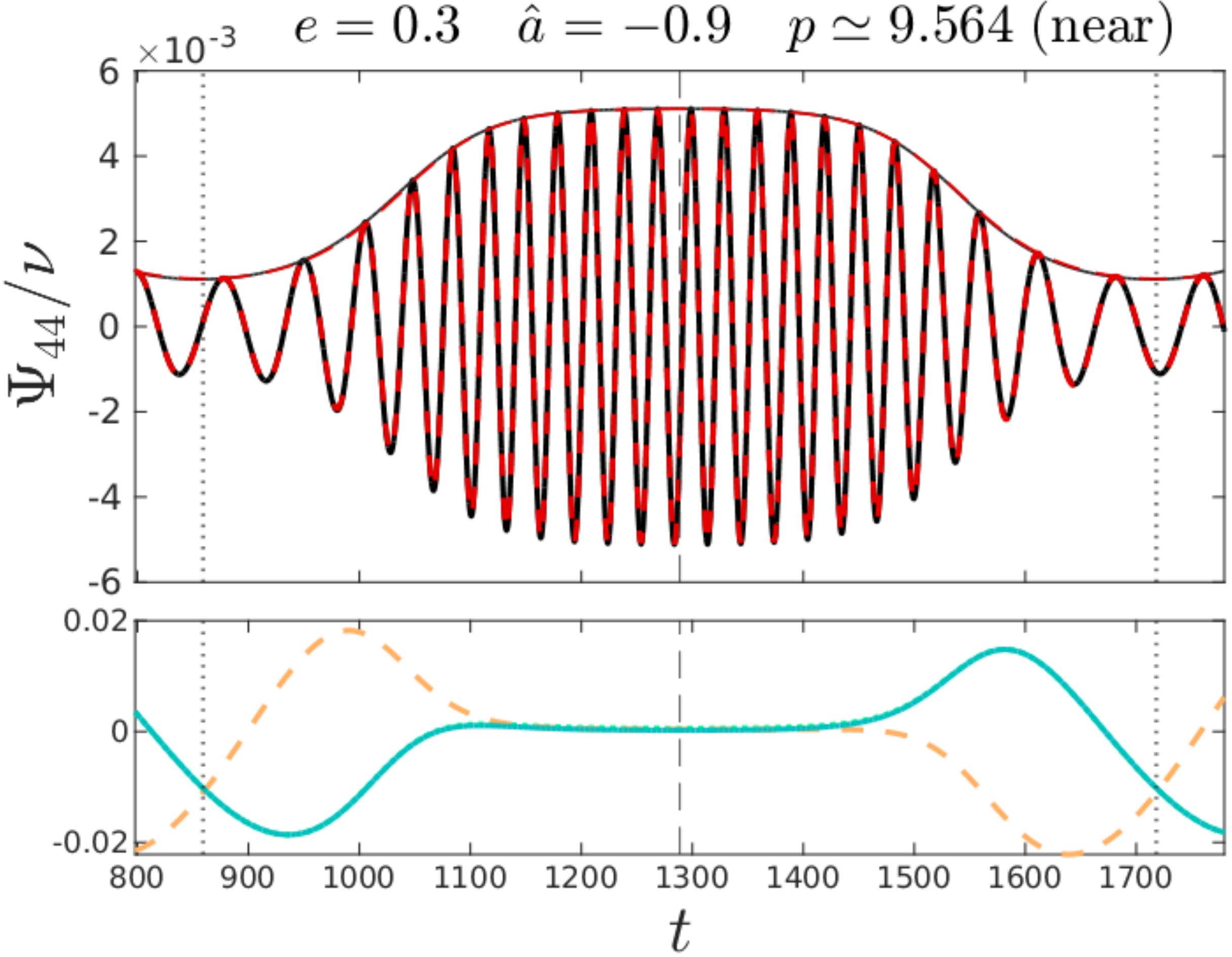} 
  \hspace{1.0cm}
  \includegraphics[width=0.22\textwidth,height=3cm]{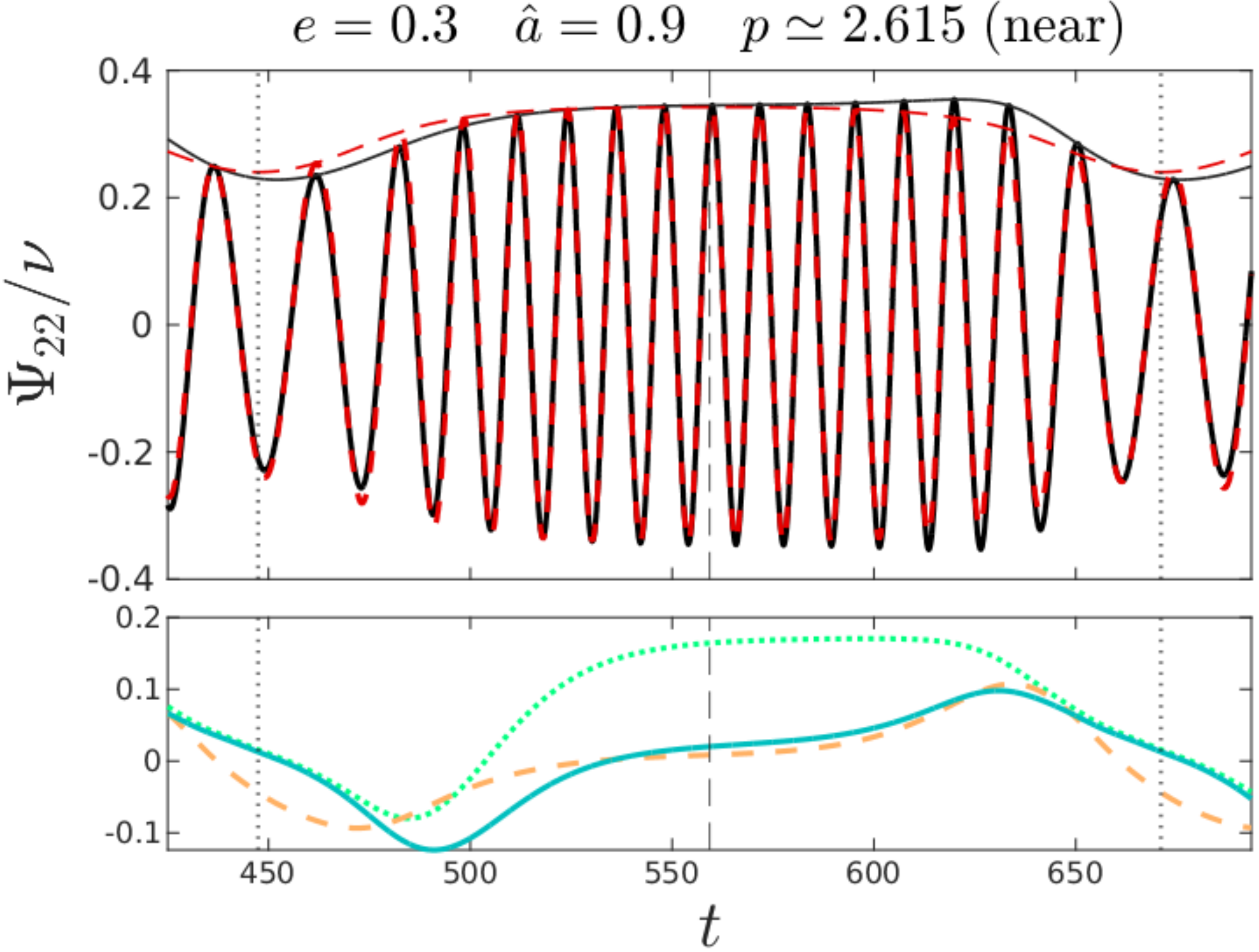} 
  \includegraphics[width=0.22\textwidth,height=3cm]{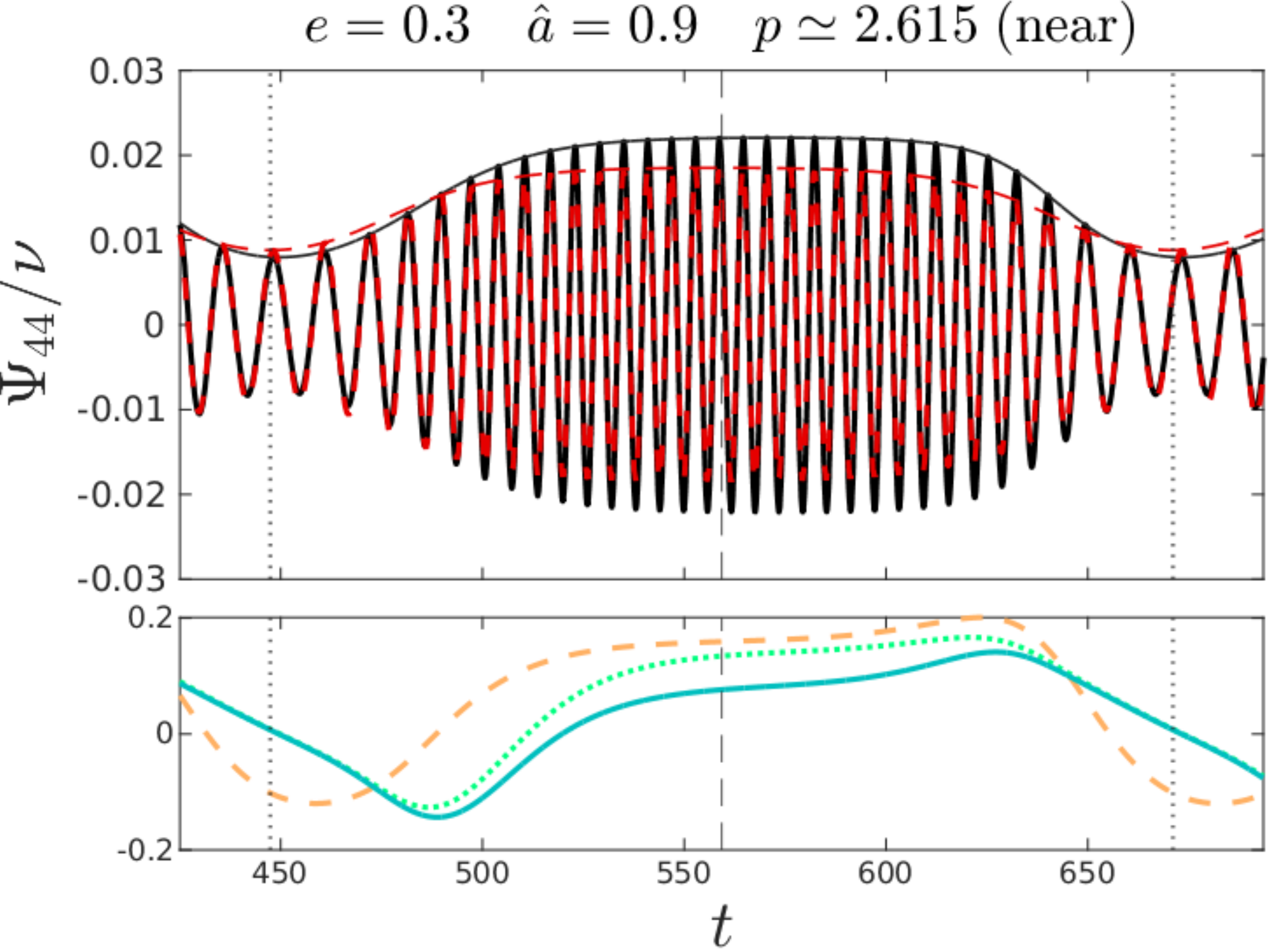} \\
  \vspace{1.0cm}
  \includegraphics[width=0.22\textwidth,height=3cm]{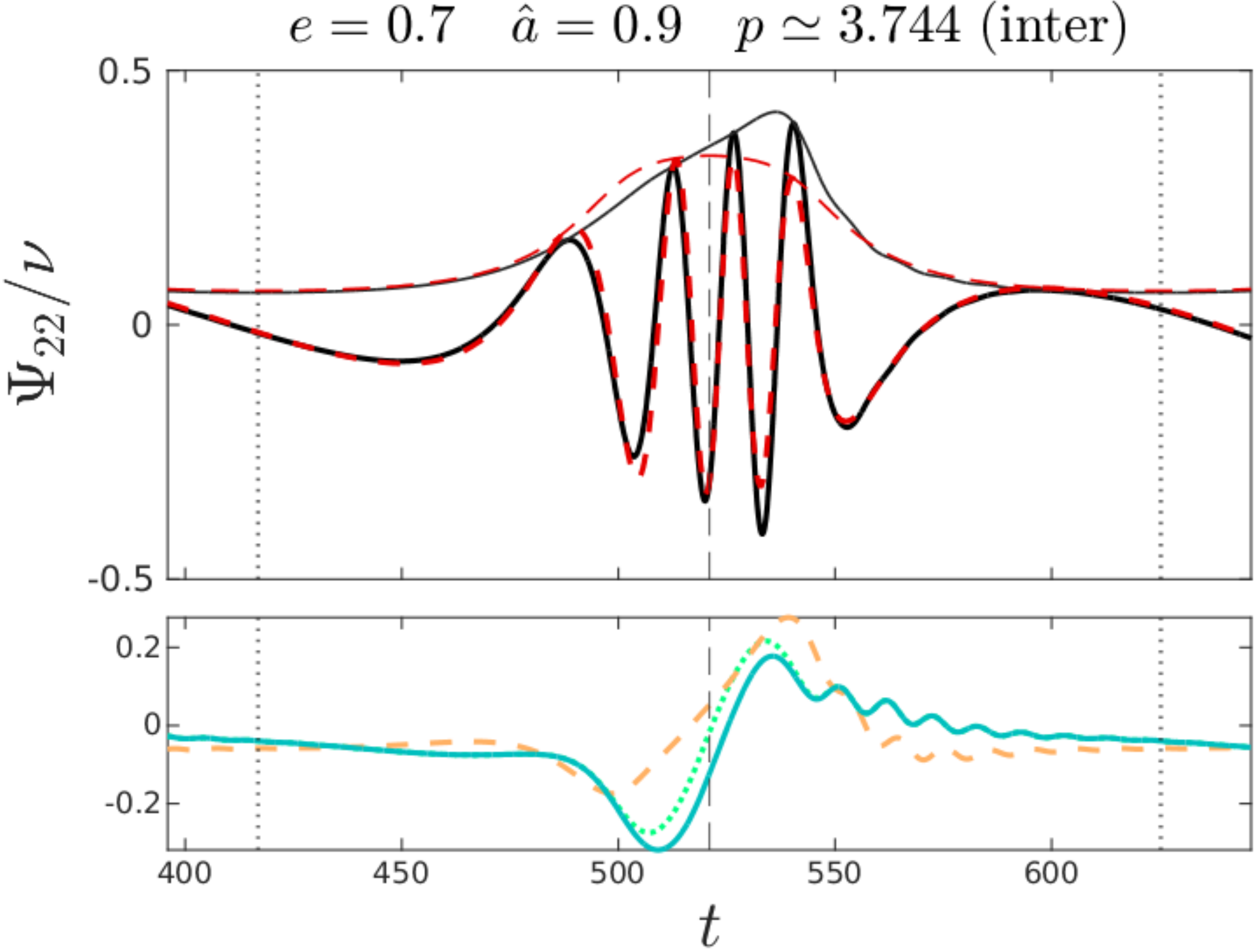}
  \includegraphics[width=0.22\textwidth,height=3cm]{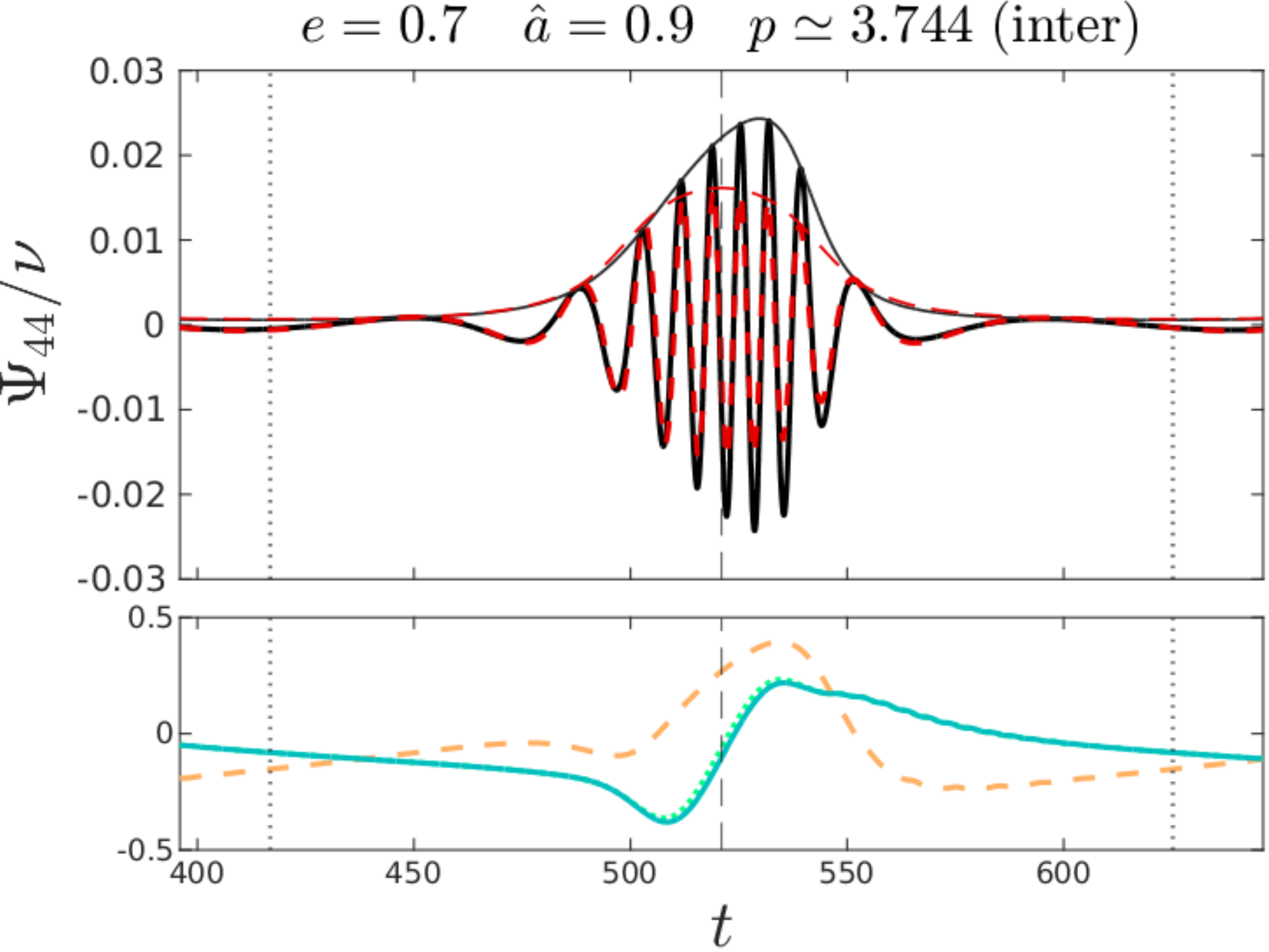}
  \hspace{1.0cm}
  \includegraphics[width=0.22\textwidth,height=3cm]{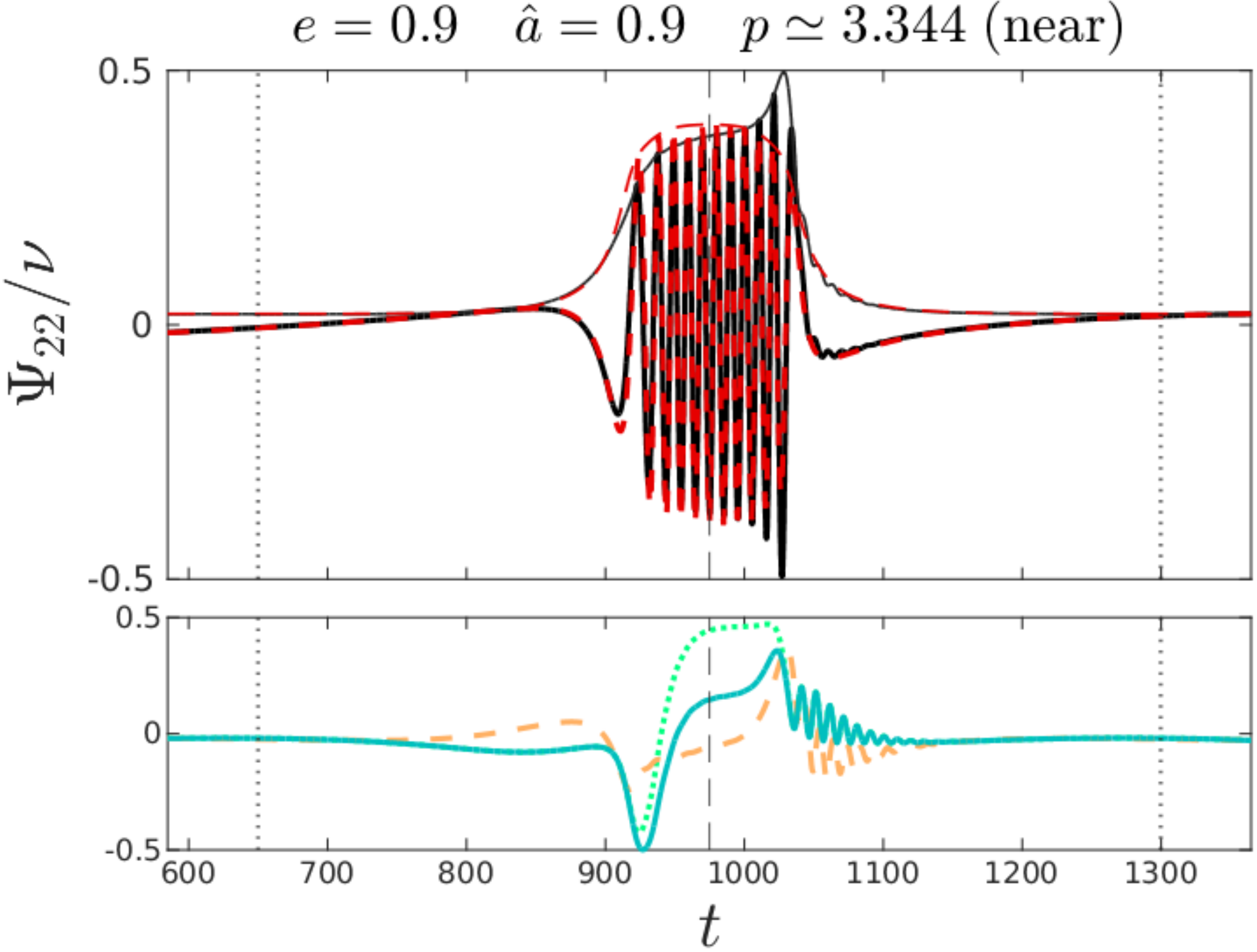}
  \includegraphics[width=0.22\textwidth,height=3cm]{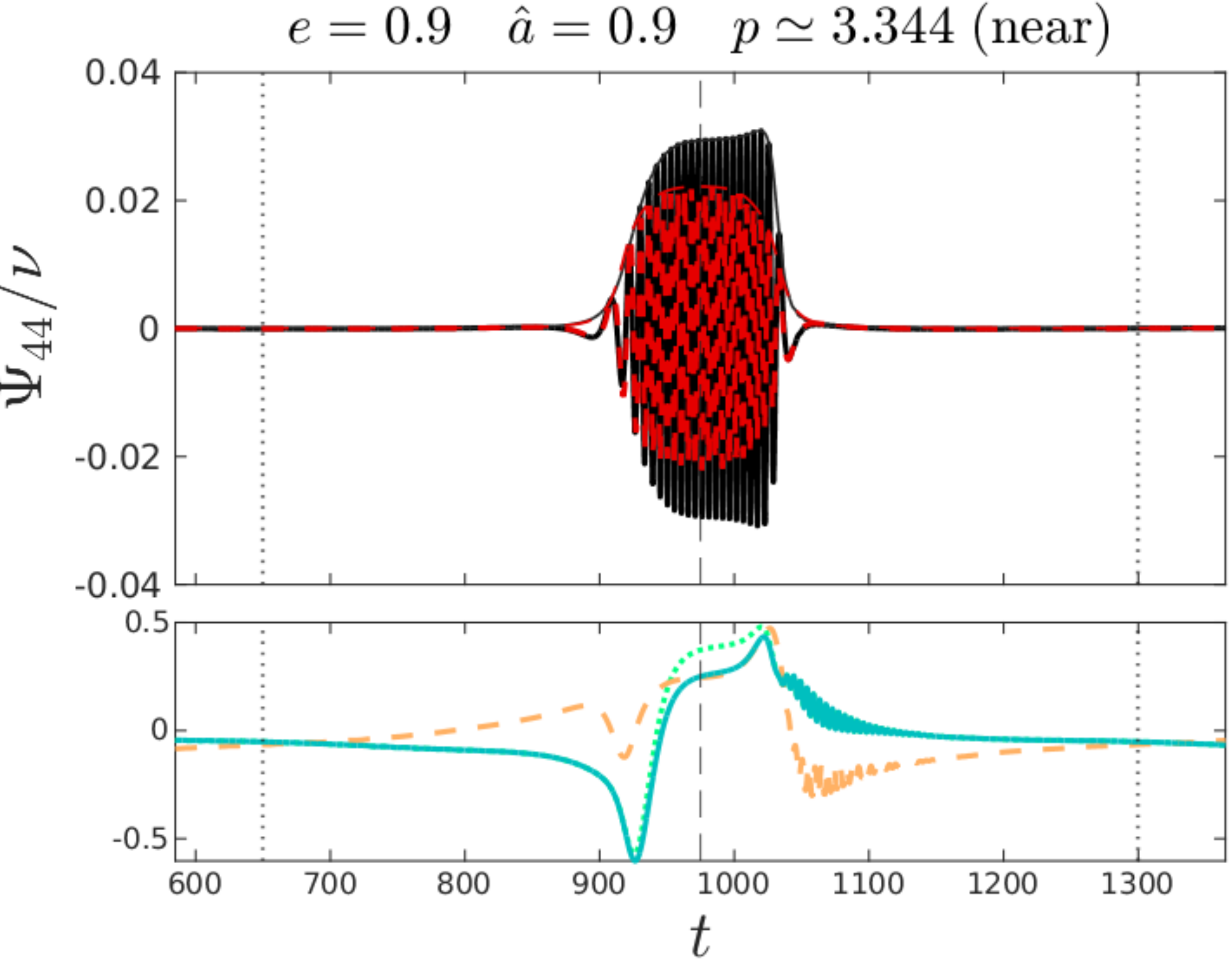}
  \caption{\label{fig:eobwaves} Numerical (black) and EOB (red dashed) $\l=m=2$ and $\l=m=4$ multipoles
  corresponding to the dynamics shown in Fig.~\ref{fig:geo_dynamics}.  
  For each multipole,  we report the relative amplitude difference (dashed orange) 
  and the phase difference (light blue).
  We report also the phase differences for the waveform with 
  the residual tail factor at 6~PN (dotted green). The vertical lines mark the periastron 
  and the apastron.}
\end{figure*}
\subsection{Geodesic, equatorial motion}
Let us turn now to discussing the performance of the analytical multipolar waveform.
Reference~\cite{Chiaramello:2020ehz} already pointed out the fairly good analytical/numerical
agreement that can be obtained only with the prescription of replacing the quasi-circular
Newtonian prefactor by the general one.
A systematic multipole by multipole analysis shows that the relative discrepancies increase 
in the subdominant multipoles, i.e. the $(2,2)$ is in general the most reliable, similarly to what 
discussed for the subdominant contribution to the fluxes. In particular,
in the absence of QNMs excitations,
the relative differences between analytical and numerical amplitudes is generally smaller than
the analogous ones for the fluxes. Figure~\ref{fig:eobwaves} highlights  the analytical/numerical 
agreement for the $\ell=m=2$ and $\ell=m=4$ modes obtained from the six illustrative 
configurations of Fig.~\ref{fig:geo_dynamics}. For two configurations, we explicitly show in 
Appendix~\ref{appendix:subdominant_wave} the comparison for almost all other multipoles, see 
Fig.~\ref{fig:eobwave_manymodes1} and~\ref{fig:eobwave_manymodes2}.
For each configuration we compare, in the top panel, the real part 
of the analytical and numerical waveform
burst emitted around the periastron. For the zoom-whirl configuration this 
corresponds to the many GW cycles
corresponding to the quasi-circular regime. In the bottom panel we show the 
analytical/numerical phase difference
$\Delta\phi=\phi^{\rm analytical}_\lm-\phi^{\rm numerical}_\lm$ and relative 
amplitude difference $|A^{\rm analytical}_\lm-A^{\rm numerical}_\lm|/A^{\rm numerical}_\lm$.

In Fig.~\ref{fig:eobwaves} (as well as~\ref{fig:eobwave_manymodes1} and~\ref{fig:eobwave_manymodes2}),
the residual waveform phases $\delta_\lm$ are kept at 7.5~PN accuracy, 
that is our default choice. Such high-PN accuracy is relevant for
circular or zoom-whirl orbits, but it is less important for other eccentric configurations.
In particular, for the intermediate configurations considered here they are practically equivalent,
except for the high spin case $\ha\simeq 0.9$, since the periastron occurs at small values of
the radial separation. The periastron is, obviously, where the knowledge of high PN information
matters most. To show this fact, we also report in Fig.~\ref{fig:eobwaves} 
the phase  differences obtained with $\delta_\lm$ at 6~PN accuracy
(green dotted online).
Finally, note that the phase difference reaches its minimum near periastron.
This suggests that in the eccentric case this quantity is dominated by the lack
of high-order noncircular information in the waveform beyond the (leading-order) 
Newtonian level. This last sentence could also be justified noting that the discrepancies in the circular 
case are much smaller, as shown in Fig.~\ref{fig:delta_PNtest}.

\subsection{Transition from eccentric inspiral, plunge, merger and ringdown}
\label{subsec:insplunge}
%
\begin{figure}[t]
  \center  
  \includegraphics[width=0.45\textwidth]{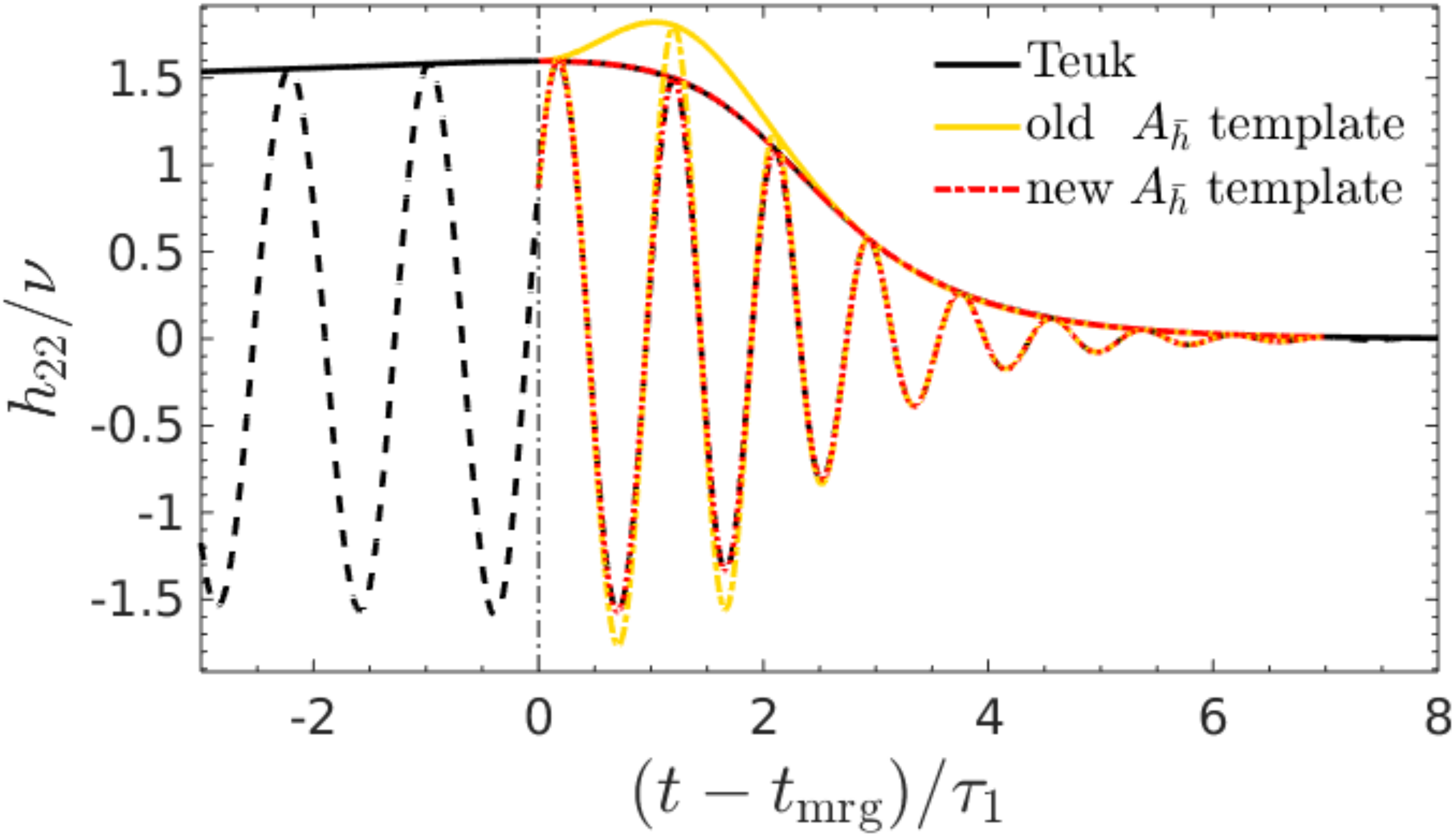} 
  \caption{\label{fig:ringdown_oldA} Ringdown modelization: transition from quasi-circular 
  inspiral to plunge, merger and ringdown with $\nu=10^{-3}$ and $\ha=0.8$. The time-scale is 
  rescaled using $\tau_1 = 1/\alpha_1$.
  The new fitting template for the amplitude, Eq.~\eqref{eq:templateA} accurately reproduces
  the numerical data, while the standard one, proposed in Ref.~\cite{Damour:2014yha}, 
  delivers an unphysical local maximum during ringdown.}
\end{figure}
\begin{figure*}[t]
  \center
  \includegraphics[width=0.32\textwidth,height=7.4cm]{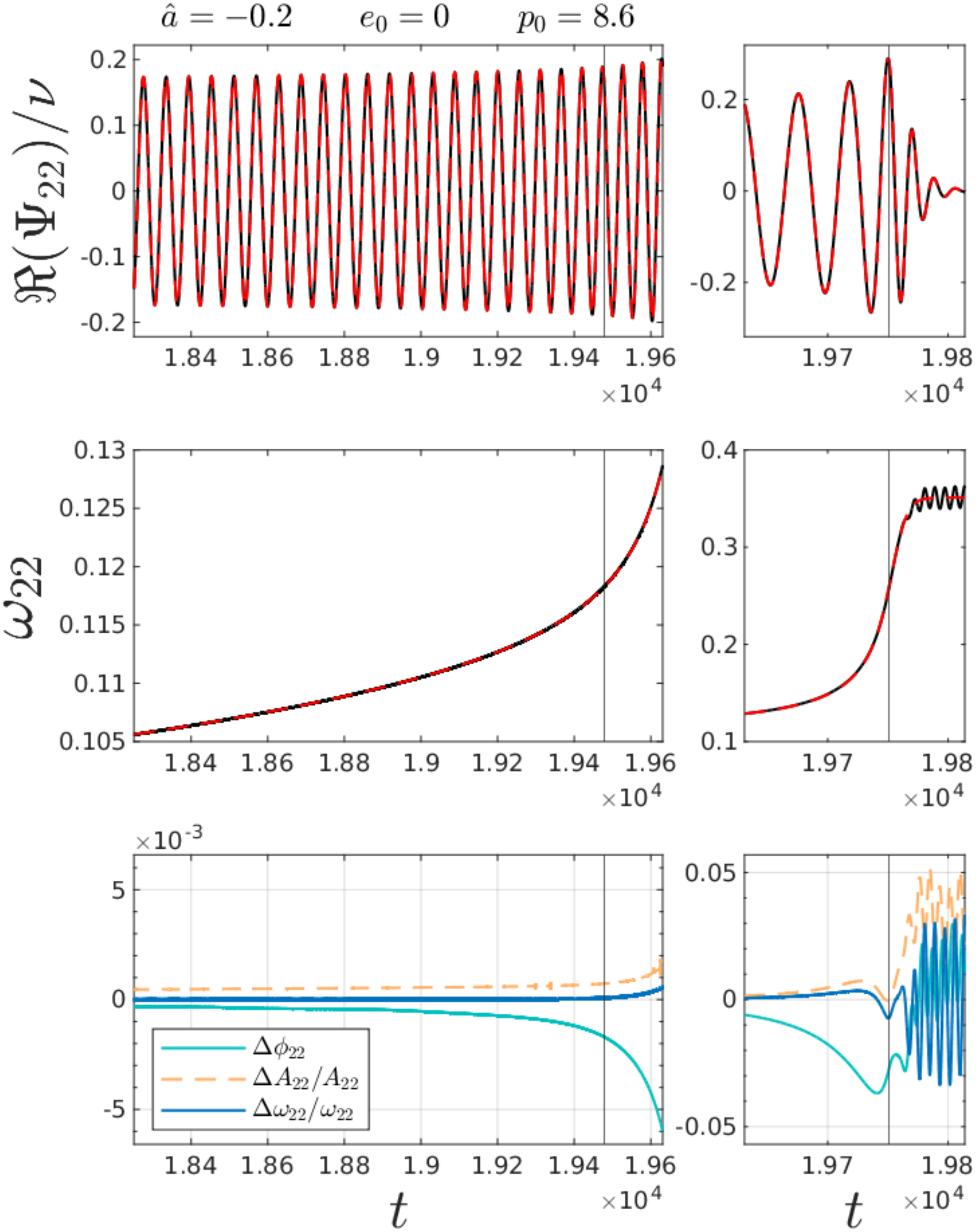}
  \includegraphics[width=0.32\textwidth,height=7.4cm]{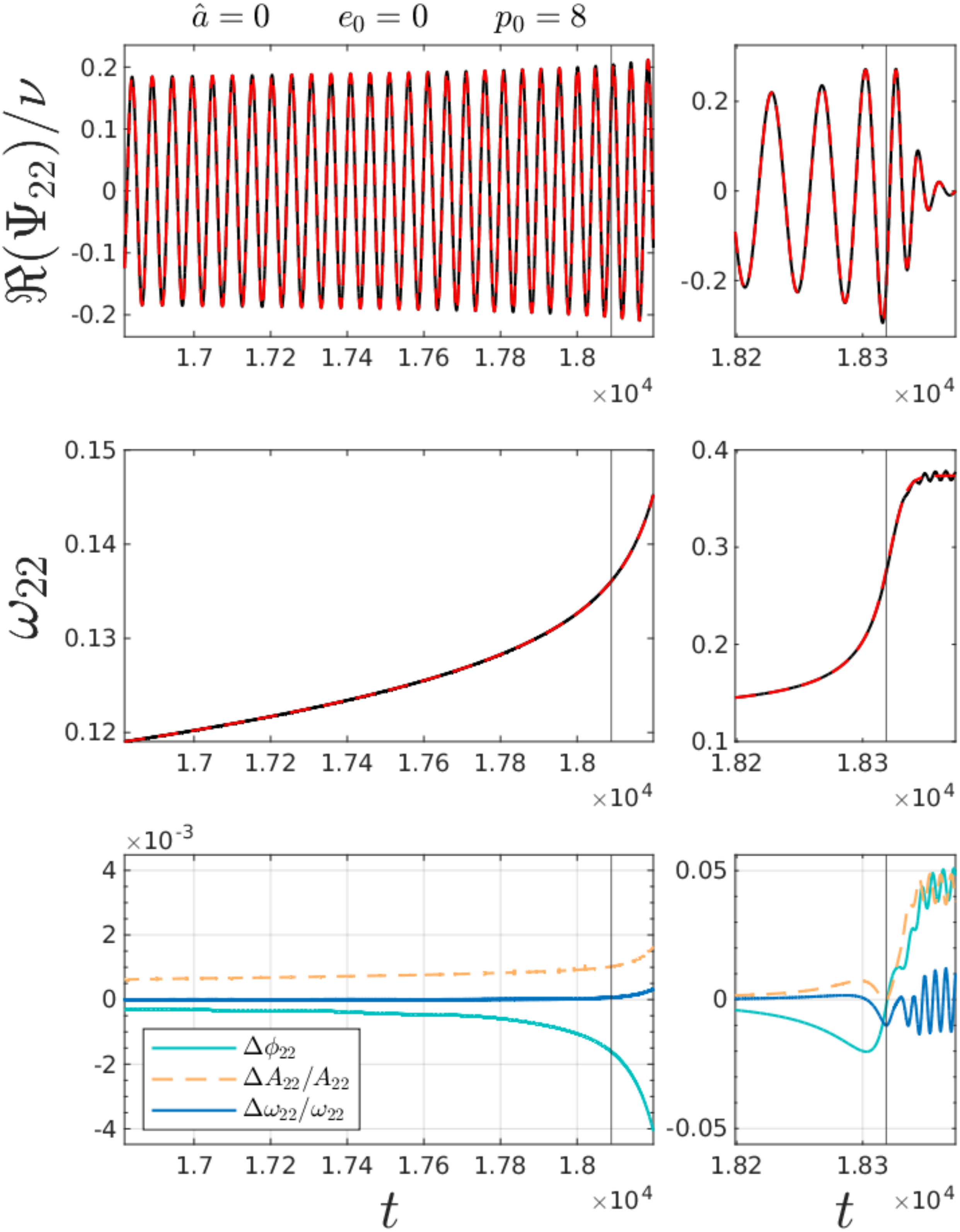}
  \includegraphics[width=0.32\textwidth,height=7.4cm]{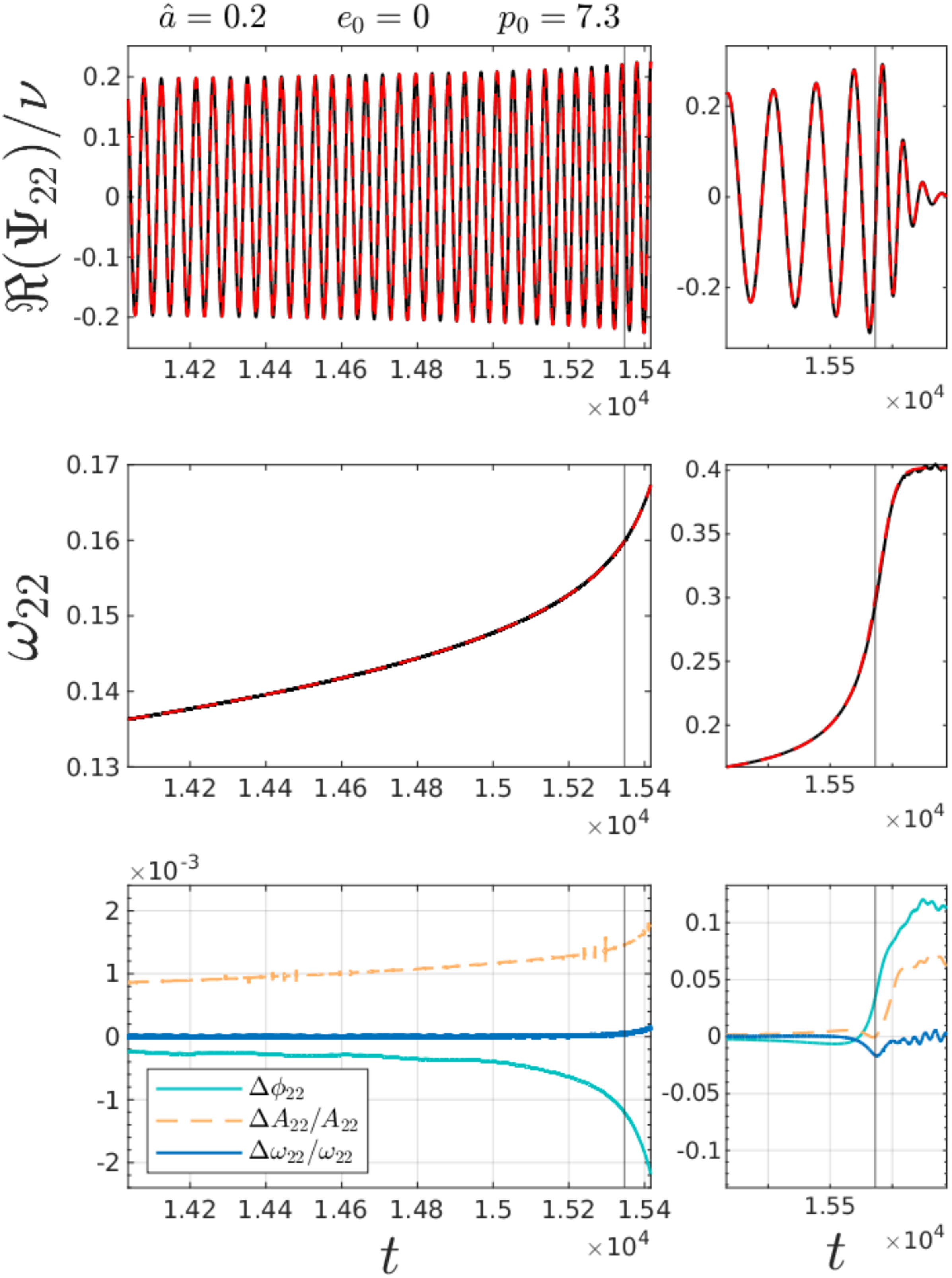} \\
  \caption{\label{fig:insplunge_circ} Comparison between numerical (black) and analytical (red) 
  waveforms for the transition from circular inspiral to plunge, merger and 
  ringdown. We report the comparisons for spins $\ha=0, \pm0.2$. The two vertical lines 
  mark the LSO and the merger time. In the last panel of each comparison, we report the 
  relative difference of the amplitude (orange) and frequency (blue) together with the 
  phase difference (light blue) in radians.
  }
\end{figure*}
Until now we have separately explored the quality of the analytical waveform and fluxes along 
eccentric geodesic orbit. The final aim of this work is to incorporate these two building blocks 
together in order to consistently drive an eccentric inspiral. 
The aim of this section is to discuss the complete resummed analytical waveform, that for 
simplicity we will call EOB waveform, obtained from an eccentric dynamics driven by the
radiation reaction force discussed above. In doing so, we do not limit ourselves to the inspiral,
but we proceed to plunge, merger and ringdown, building a suitable model for this latter.
In this respect, the results we present here, that should be considered preliminary, complement
and generalize previous work within the EOB approach in the extreme-mass-ratio 
limit~\cite{Nagar:2006xv, Damour:2007xr,Yunes:2010zj,Barausse:2011kb}.

In particular, the analytical description of the ringdown stems from the effective model introduced
in Ref.~\cite{Damour:2014yha}, that relies on a certain way of fitting NR waveform data.
In current EOB models informed by NR simulations, and in particular 
\TEOBResumS{}~\cite{Nagar:2019wds,Nagar:2020pcj},
the model is fully informed by spin-aligned NR simulations up to mass ratio $q=18$,
but relied on a limited amount of information coming from test-particle waveform data,
i.e. only the values of amplitude and frequency at merger time. The reason is that the 
fitting template for the amplitude proposed in Ref.~\cite{Damour:2014yha} and used in
\TEOBResumS{}~\cite{Nagar:2019wds,Nagar:2020pcj} is not reliable in the test-particle
limit when the spin of central black hole is large. 
This fact is illustrated in Fig.~\ref{fig:ringdown_oldA}.
The black curves depict the amplitude (solid) and real part (dashed) of a quasi-circular, numerical
waveform corresponding to $\hat{a}=0.8$ and $\nu=10^{-3}$. The yellow curves are the result of
fitting the data\footnote{Over a temporal interval $T=7\tau_1$, where $\tau_1$ is the damping time
of the fundamental quasi-normal mode of the black hole.} with the amplitude template 
of~\cite{Damour:2014yha}. The unphysical peak in the postmerger is a recurrent feature,
that, although practically negligible for small (or negative) values of $\ha$ becomes 
predominant as the black hole spin grows. 
%
\begin{figure*}[t]
  \center
   \includegraphics[width=0.32\textwidth,height=7.4cm]{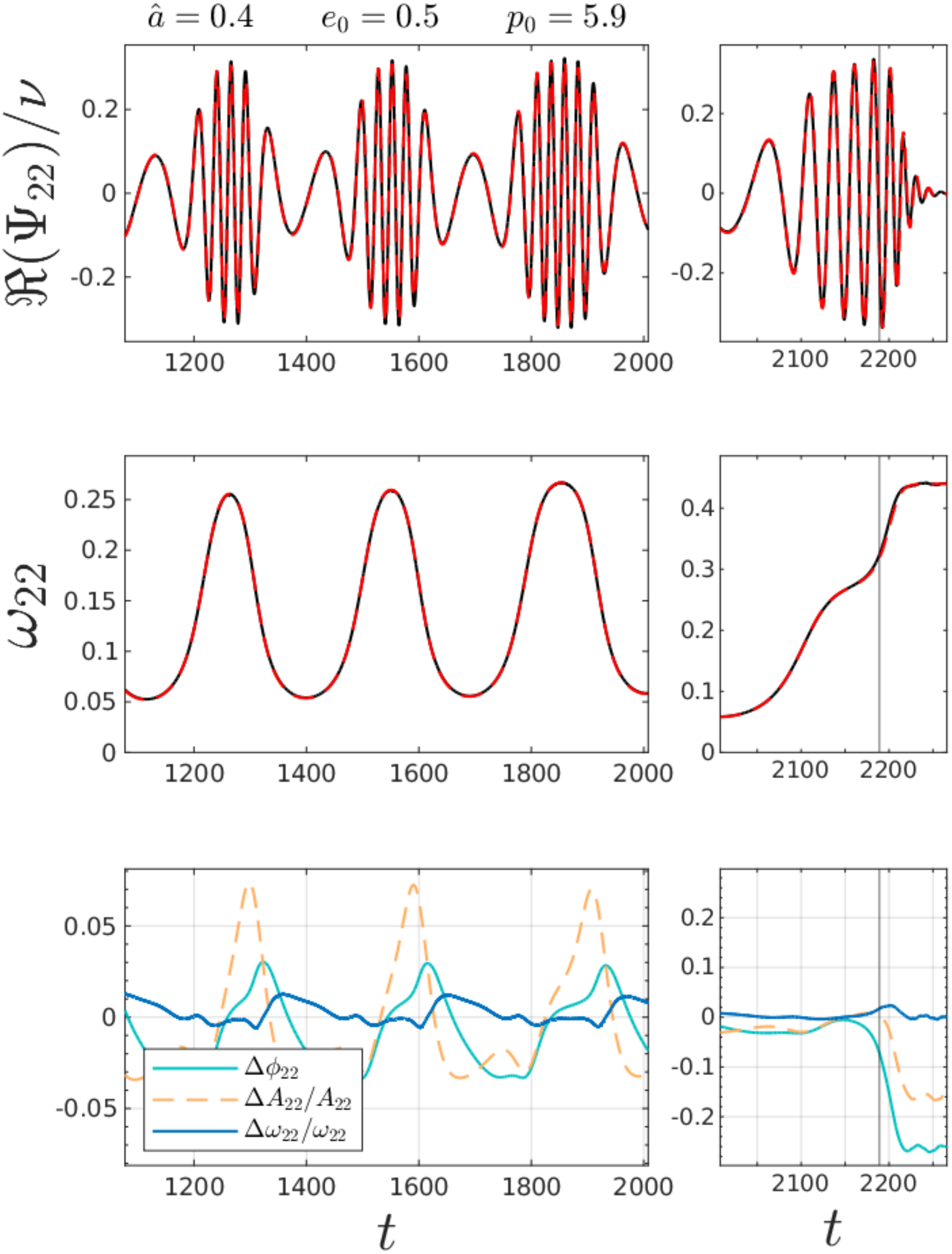}
  \includegraphics[width=0.32\textwidth,height=7.4cm]{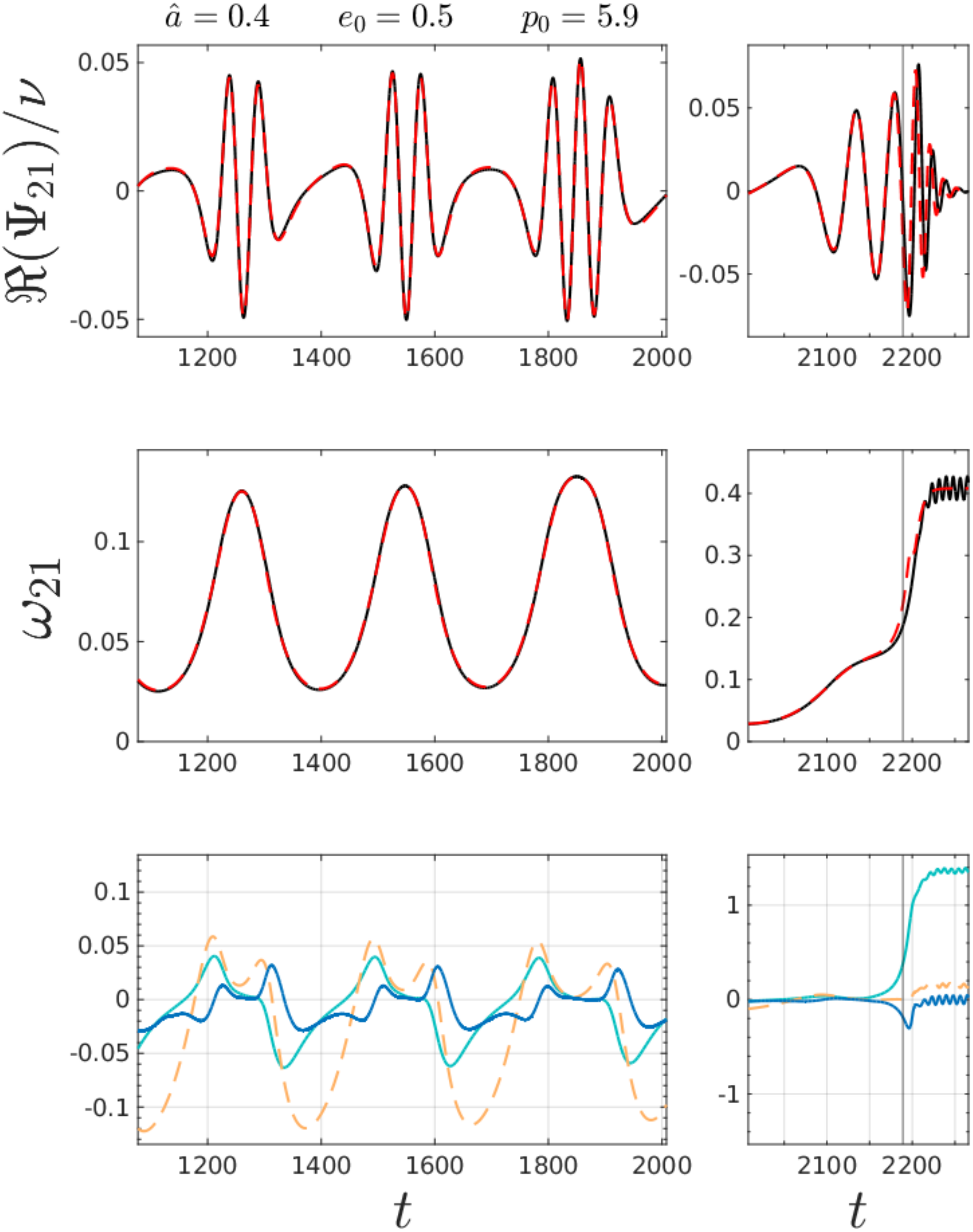}
  \includegraphics[width=0.32\textwidth,height=7.4cm]{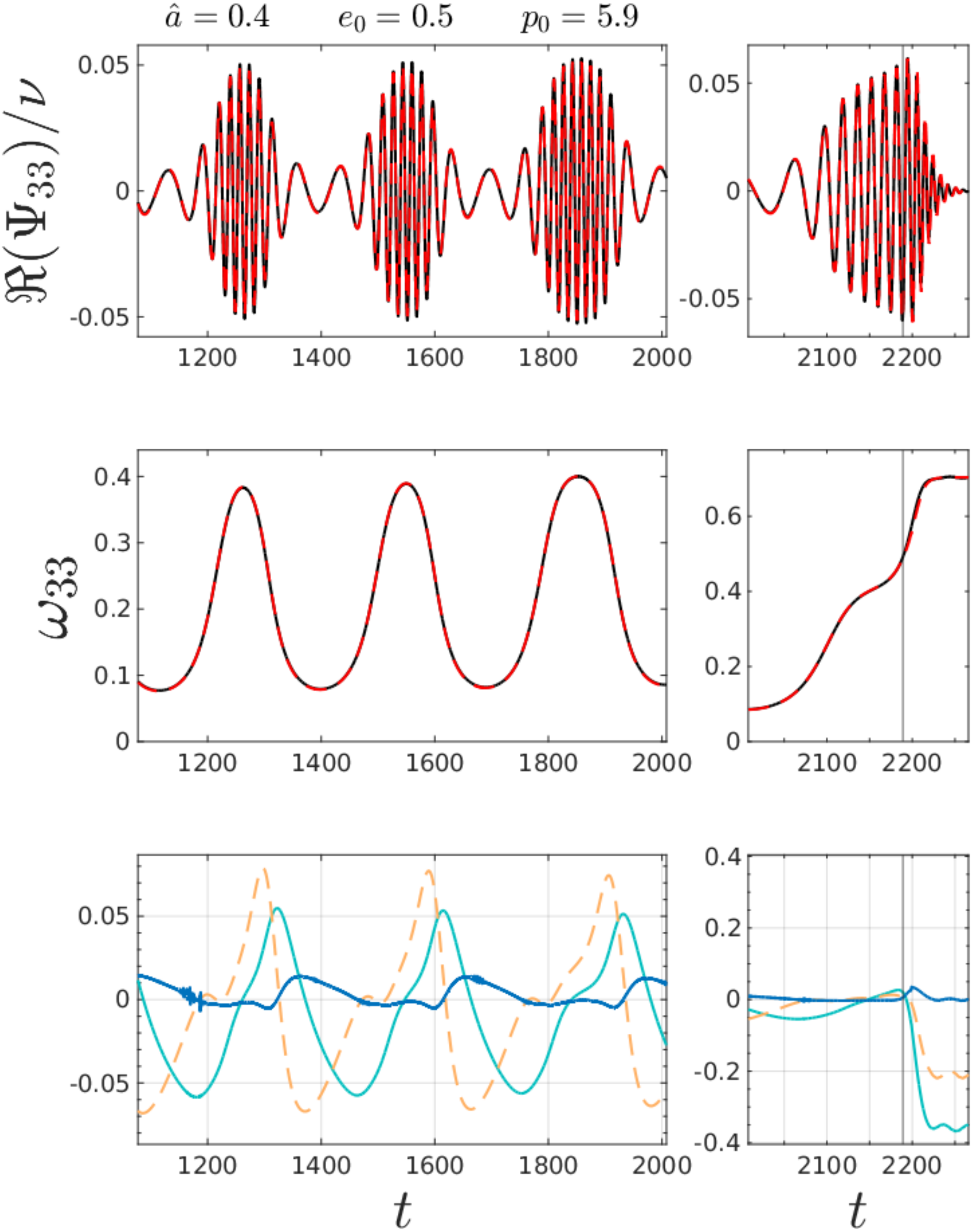} 
  \caption{\label{fig:insplunge_wave} Comparison between numerical (black) and analytical 
  (red dashed) complete waveform generated by a non-geodesic prograde orbit with initial 
  eccentricity $e_0=0.5$ and semilatus rectum $p_0=5.9$ around a Kerr black hole with $\ha=0.4$. 
  The vertical line marks the merger time. In the last panel of each mode, we report the 
  relative difference of the amplitude (orange) and frequency (blue) together with the 
  phase difference (light blue) in radians.}
\end{figure*}
This problem can be fixed increasing the flexibility of the amplitude template.
Referring from now on to $\bar{h}$ as the QNM-rescaled ringdown waveform 
of Ref.~\cite{Damour:2014yha} (see Eq.~(1) therein) we adopt here the following 
function to model the amplitude
\be
\label{eq:templateA}
A_{\bar{h}} (\tau)  = \left( \frac{c_1^A}{1 + e^{-c_2^A \tau + c_3^A}} + c_4^A \right)^{\frac{1}{c_5^A}} \ ,
\ee
while for the phase we just follow the prescription of ~\cite{Damour:2014yha} and use
\be
\label{eq:templatePhi}
\phi_{\bar{h}}(\tau) = -c_1^\phi \ln \left( \frac{1+c_3^\phi e^{-c_2^\phi \tau} + c_4^\phi e^{-2c_2^\phi \tau}}{1+c_3^\phi + c_4^\phi} \right) ,
\ee
where $\tau\equiv t-t_{\rm mrg}$. The merger time $t_{\rm mrg}$ is defined as the amplitude peak of the 
quadrupolar waveform. The only parameters to be fitted on numerical data are $c_2^A$, $c_3^A$, 
$c_3^\phi$ and $c_4^\phi$, while the others are determined requiring the correct late-time behavior.
Note however that now the amplitude is fitted using {\it two} parameters, contrary to Ref.~\cite{Damour:2014yha},
that could employ a single fitting parameter. The constraints given by the late-time behavior of the ringdown are
\begin{align}
c_1^A &= \frac{c_5^A \alpha_{1}}{c_2^A} (A_{\rm mrg})^{c_5^A} e^{-c_3^A} \left(1 + e^{c_3^A}\right)^2,\\
c_4^A &= (A_{\rm mrg})^{c_5^A} - \frac{c_1^A}{1 + e^{c_3^A}}, \\
c_5^A &= - \frac{\ddot{A}_{\rm mrg}}{A_{\rm mrg} \alpha_{1}^2} + \frac{c_2^A}{\alpha_{1}} \frac{e^{c_3^A} - 1}{1 + e^{c_3^A}} \ ,\\
c_2^\phi &= \alpha_{21},\\
c_1^\phi &= \dfrac{1 + c_3^\phi + c_4^\phi}{c_2^\phi(c_3^\phi + 2 c_4^\phi)} \Delta \omega_{\rm mrg}, 
\end{align}
where $\alpha_i$ are the real parts of the QNM complex frequencies $\sigma_i\equiv \alpha_i + {\rm i}\omega_i$ 
(i.e., the inverse of the damping time), $\alpha_{21}\equiv \alpha_2-\alpha_1$, $A_{\rm mrg}$ is the amplitude 
at merger, $\ddot{A}_{\rm mrg}$ its second time-derivative, 
$\Delta \omega_{\rm mrg} \equiv \omega_1 - M \omega_{\rm mrg}$ is the difference 
between the frequency at merger $\omega_{\rm mrg}$ and the imaginary part of the 
fundamental QNM frequency $\omega_1$. The red curves in Fig.~\ref{fig:ringdown_oldA} 
illustrate the accuracy of the new template amplitude.
In principle, any EOB-based model for coalescing black hole binaries, like \TEOBResumS{}~\cite{Nagar:2020pcj},
should correctly incorporate the test-mass limit. Due to this problem in the original waveform
amplitude template, this is not the case of \TEOBResumS{}, although large-mass-ratio waveforms
look qualitatively and quantitatively essentially correct because of the inclusion of test-particle-informed
fits for $(A_{\rm mrg},\omega_{\rm mrg})$. Still, to guarantee that the NR-informed description of the 
postmerger-ringdown phase currently incorporated in \TEOBResumS{} is smoothly connected to the test-particle 
limit, the (multipolar) ringdown model of Ref.~\cite{Nagar:2019wds,Nagar:2020pcj} will have to be 
updated using the new amplitude template described here.

The amplitude and frequency at merger are extracted from numerical data and fitted.
For the subdominant modes, the same procedure is followed, but all the quantities have to be 
evaluated at $t_{\rm peak}^\lm = t_{\rm mrg} + \Delta t_\lm$, where $\Delta t_\lm \geq 0 $ is 
the delay of each peak respect to the dominant mode. 
The $\Delta t_\lm$ are also extracted from numerical data and are then  
fitted over the parameter space. To construct a complete ringdown model 
we have employed 97 numerical insplunge simulations for different values 
of eccentricity and spin. The global fits of the primary parameters found with the phase 
and amplitude templates are performed over $(\ha, e) = [-0.6, 0.8] \times [0, 0.9]$, 
even if for high positive spins the global fits are reliable only at moderate or low eccentricity. 
Details about this fit will be presented in a forthcoming work.

In order to smoothly match the insplunge waveform to the postmerger-ringdown description,  
the analytical waveform is improved using a Next-to-Quasi-Circular (NQC) correction factor,
following prescriptions that are standard within the EOB model. The multipolar waveform
reads
\be
h_\lm = h_\lm^{(N, \epsilon)} \hat{h}^{(\epsilon)}_\lm \hat{h}_\lm^{\rm NQC},
\ee
where $h_\lm^{(N, \epsilon)}$ indicates the (general) Newtonian prefactor, $\hat{h}^{(\epsilon)}_\lm$ 
is the resummed
(circular) PN correction and $\hat{h}_\lm^{\rm NQC}$ is the additional correction.
To exploit at best the action of NQC corrections, we use here 3 effective parameters in the
amplitude and 3 parameters in the phase, so that we have
\begin{equation}
\hat{h}^{\rm NQC}_\lm = \left( 1 + \sum_{i=1}^{3}  a_i^\lm n_i \right) \exp{ \left( {\rm i}  \sum_{i=1}^{3} b_i^\lm n_i \right) },
\label{eq:nqc}
\end{equation}
where the functions $n_i$ are:
\begin{subequations}
\begin{align}
n_1 & = \frac{p_{r_*}^2}{(r\Omega)^2}, \\
n_2 & = \frac{\ddot{r}}{r \Omega^2}, \\
n_3 & = n_1 \, p_{r_*}^2, \\
n_4 & = \frac{p_{r_*}}{r \Omega}, \\
n_5 & = n_4 \, \Omega^{2/3}, \\
n_6 & = n_5 \, p_{r_*}^2.
\end{align}
\end{subequations}
The only exception is the $\l=m=2$ mode, where we use $n_5 = n_4 \, r^2 \Omega^2$.
The $a_i^\lm$ coefficients are determined imposing continuity conditions between the EOB insplunge waveform $h_\lm^{\rm EOB}$ and the ringdown solution $h_\lm^{\rm rng}$: 
\begin{subequations}
\begin{align}
            A_\lm^{\rm EOB}(t_{\rm NQC}) & =             A_\lm^{\rm rng}(t_{\rm NQC}),  \\
      \dot{A}_\lm^{\rm EOB}(t_{\rm NQC}) & =       \dot{A}_\lm^{\rm rng}(t_{\rm NQC}),  \\
     \ddot{A}_\lm^{\rm EOB}(t_{\rm NQC}) & =      \ddot{A}_\lm^{\rm rng}(t_{\rm NQC}),  \\
       \omega_\lm^{\rm EOB}(t_{\rm NQC}) & =        \omega_\lm^{\rm rng}(t_{\rm NQC}),  \\
 \dot{\omega}_\lm^{\rm EOB}(t_{\rm NQC}) & =  \dot{\omega}_\lm^{\rm rng}(t_{\rm NQC}),  \\
\ddot{\omega}_\lm^{\rm EOB}(t_{\rm NQC}) & = \ddot{\omega}_\lm^{\rm rng}(t_{\rm NQC}),
\end{align}
\end{subequations}
where $t_{\rm NQC} = t_{\rm peak}^\lm + 2$.
The merger time $t_{\rm mrg}$ is determined using the peak of the orbital 
frequency following the same prescription adopted for the comparable mass
case in the \TEOBResumS{} model~\cite{Damour:2014sva}, 
i.e. using
\begin{equation}
t_{\rm mrg} = t^{\rm peak}_{\Omega_{\rm orb}} - 3 ,
\end{equation}
where $\Omega_{\rm orb}$ is obtained 
from the orbital frequency $\Omega$, see Eq.~\eqref{eq:freq}, 
removing the spin-orbit contribution. It was pointed out long ago in Ref.~\cite{Harms:2014dqa} that, 
in the transition from inspiral to plunge, the peak $t_{\Omega_{\rm orb}}^{\rm peak}$, 
is very close to the peak of the $\ell=m=2$ waveform mode (for any value of the BH spin)
and as such it offers an excellent reference point to attach the ringdown part when
constructing EOB models. This observation is one of the key features behind the
robustness and simplicity of the \TEOBResumS{} waveform model~\cite{Damour:2014sva}.
All the details and further improvements of 
the whole waveform model will  be discussed elsewhere. Here we just want to emphasize that the waveform 
prescriptions analyzed in this paper are reliable also during the plunge and that 
it is possible to compute
complete EOB waveform incorporating merger and ringdown {\it also} in the case of eccentric inspirals.
As a showcase, we report three circular configurations with $\ha = 0, \pm0.2$ 
in Fig.~\ref{fig:insplunge_circ} and 
an eccentric case with $\ha=0.4$ and $e_0=0.5$ in Fig.~\ref{fig:insplunge_wave}. 
In this second case we also show modes $(2,1)$ and $(3,3)$ completed through 
merger and ringdown. However, consider that the post-merger parameters for the $(2,1)$ mode 
are fitted only over circular data.
For the eccentric dynamics we use the angular radiation reaction of Eq.~\eqref{eq:Fphi_ecc_old}.

\subsection{Hyperbolic captures}
\label{subsec:hyp}
%
\begin{figure*}[t]
	\center
	\hspace{0.5cm}
	\includegraphics[width=0.21\textwidth]{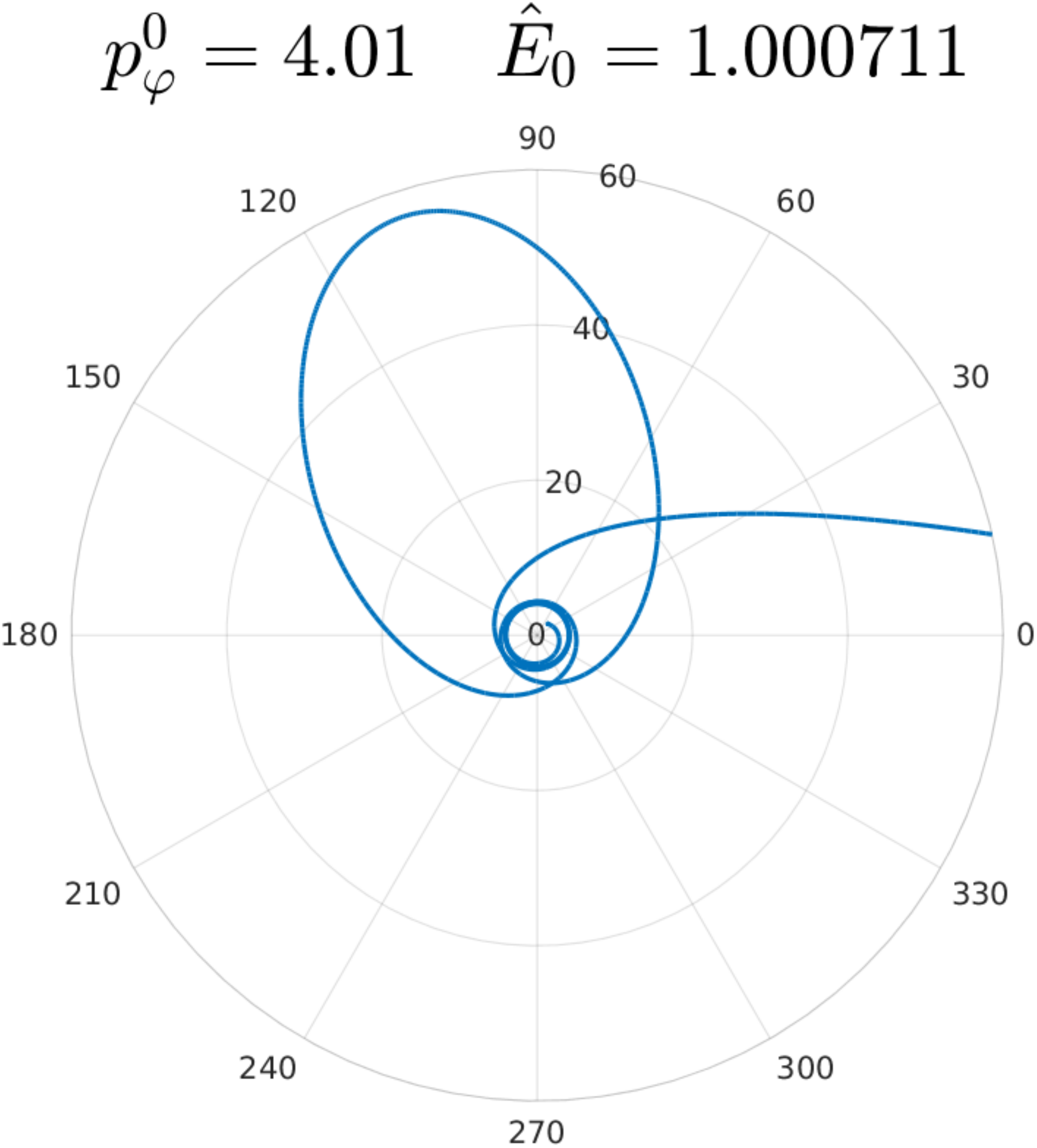}
	\hspace{2.0cm}
	\includegraphics[width=0.21\textwidth]{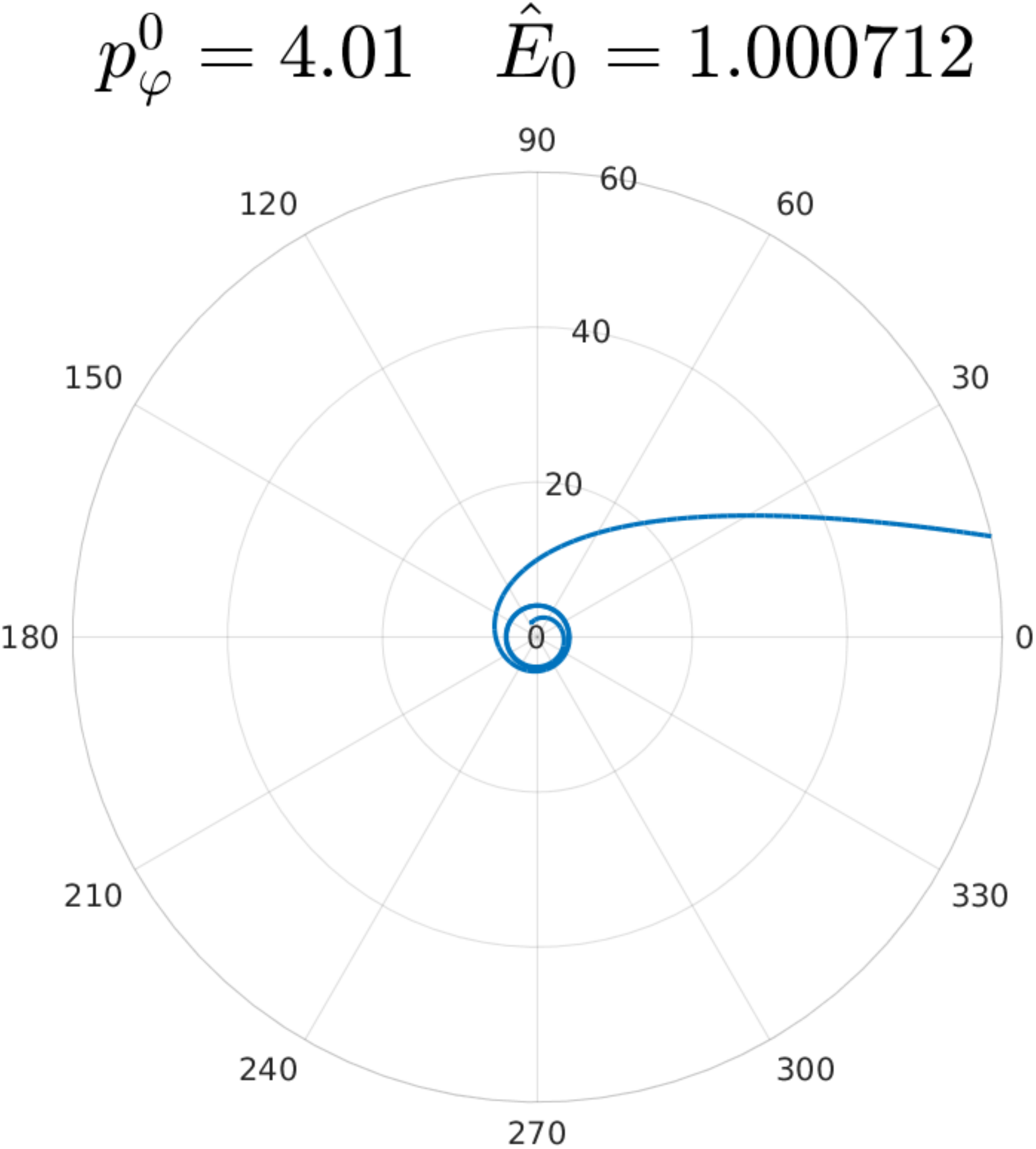} 
	\hspace{2.0cm}
	\includegraphics[width=0.21\textwidth]{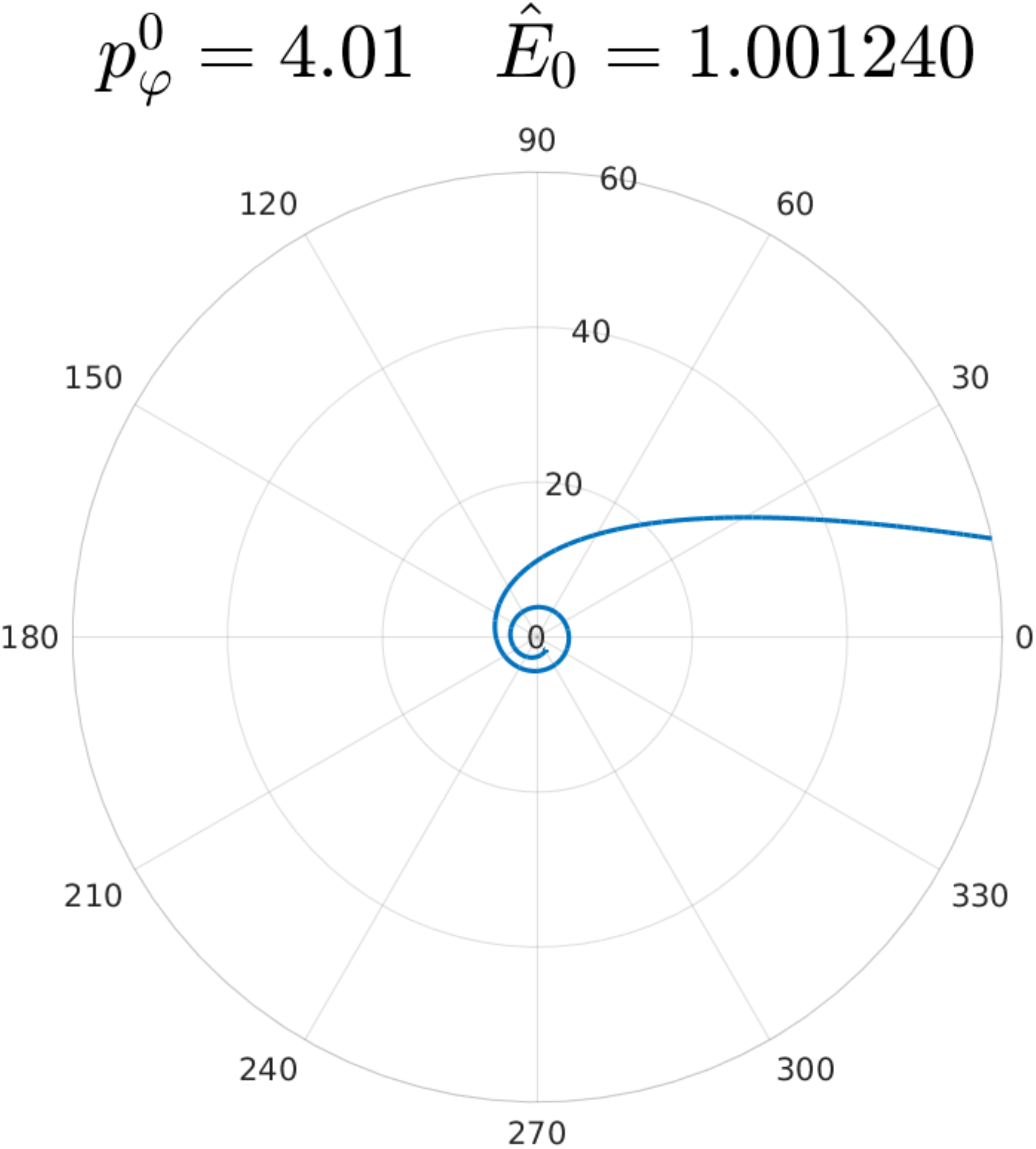} \\
	\vspace{0.5cm}
	\includegraphics[width=0.32\textwidth,height=5.6cm]{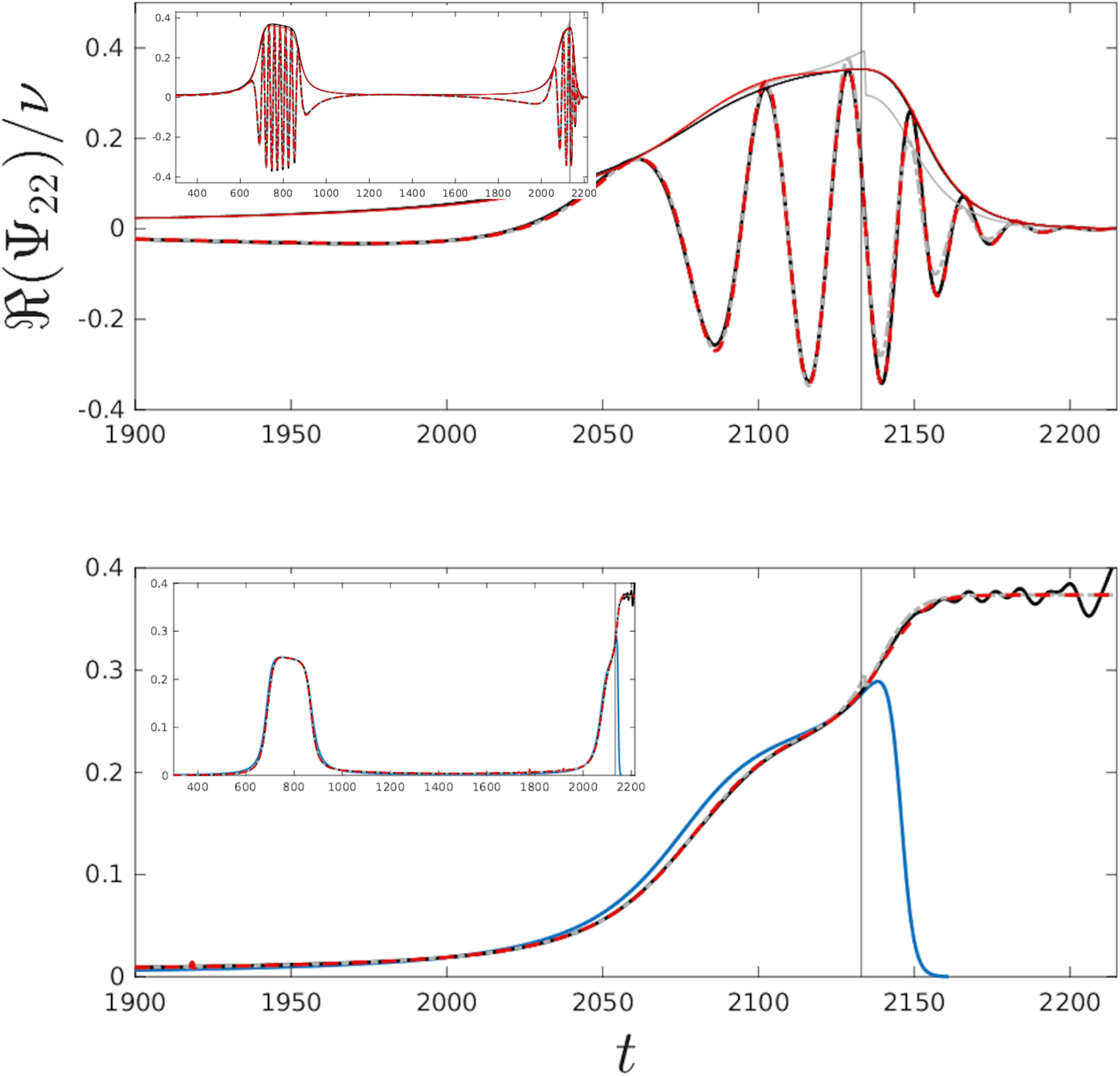}
	\includegraphics[width=0.32\textwidth,height=5.6cm]{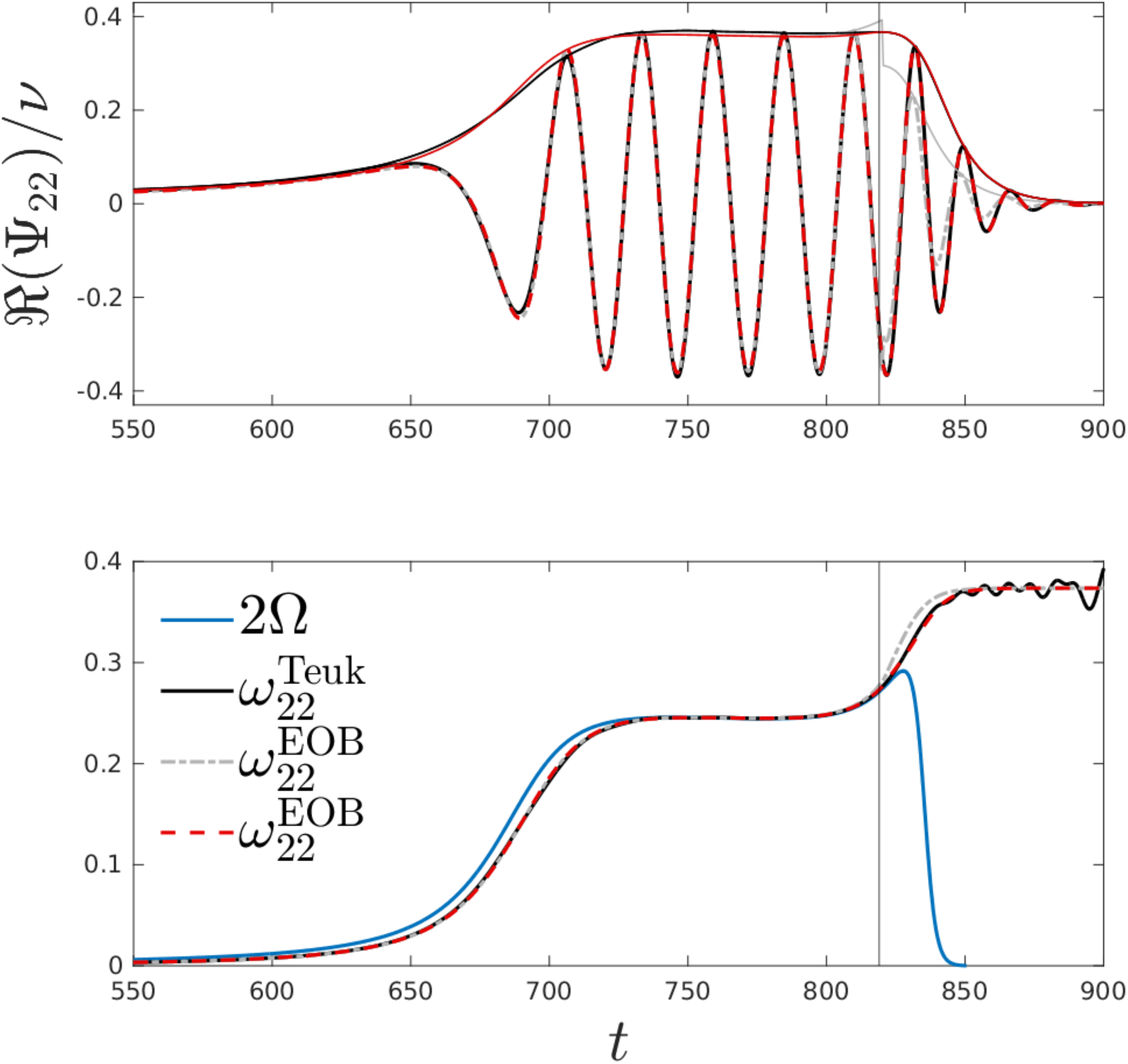} 
	\includegraphics[width=0.32\textwidth,height=5.6cm]{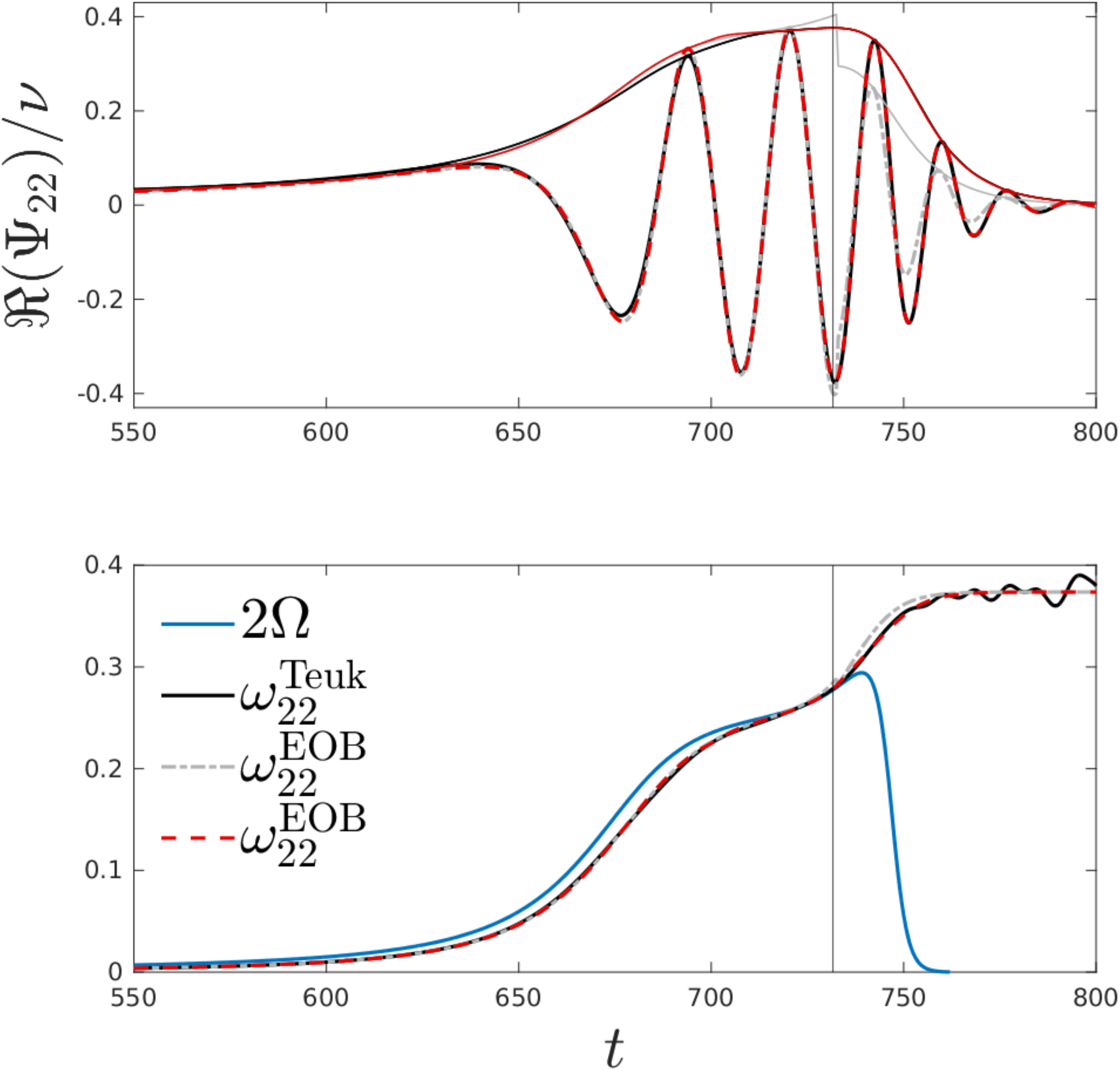} \\
	\caption{\label{fig:hyp_wave} Upper panels: trajectories
	for nonspinning dynamical captures, with $\nu=10^{-2}$, $p_\varphi^0=4.01$,
	and $\hat{E}_0 = \left(1.000711, 1.000712, 1.001240\right)$. 
	The trajectories start from $r_0=120$, but we show them from
	$r =  60$ in order to highlight the last part. 	
	Middle panels: corresponding quadrupolar waveforms.	
	The black line is the numerical result from \texttt{Teukode}, 
	while the red and grey lines are obtained with the EOB model using different prescriptions 
	for the modelization of the ringdown. See text for more details. 
	Bottom panels: frequency comparisons. The vertical lines mark the merger time 
	(i.e. the peak of the $\ell=m=2$ waveform amplitude), that in the hyperbolic case 
	is extracted directly from numerical data.}
\end{figure*}
Recently, {\TEOBResumS} has been generalized so to faithfully model also
hyperbolic encounters and dynamical capture black hole binaries~\cite{Nagar:2020xsk, Nagar:2021gss}. 
Nonetheless,the ringdown model used in these cases is the same adopted for quasi-circular waveforms,
and it might be attached to the inspiral waveform without NQC corrections~\cite{Gamba:2021gap}.
Despite these simplifications, this model allowed for a robust interpretation of GW190521~\cite{Abbott:2020tfl} 
as the outcome of the dynamical capture of two black holes~\cite{Gamba:2021gap}.

Since the test-particle limit provides a useful laboratory to test the EOB prescriptions, 
we will now focus on hyperbolic captures in the test-mass limit to discuss various ringdown
implementation and complement the information given in Ref.~\cite{Gamba:2021gap}.
The EOB dynamics and waveforms of this section are obtained with the EOB model exposed above, 
using the angular radiation reaction of Eq.~\eqref{eq:Fphi_ecc_old}, that is also the one
used for the analysis of GW190521 in Ref.~\cite{Gamba:2021gap}. 
Nonetheless, the definitions of eccentricity and semilatus rectum provided 
in Eqs.~\eqref{eq:ep_definition} are no longer valid since the two radial turning points are not 
defined for unbound motion. Therefore, as in Ref.~\cite{Nagar:2020xsk}, in this case we provide initial 
data to the dynamics directly providing the energy $\hat{E}_0$  and the angular momentum $p_{\varphi}^0$. 
Practically, we choose $p_{\varphi}^0$ and then pick $1 < \hat{E}_0 < \hat{E}^{\rm max}_0$, where  
$\hat{E}^{\rm max}_0$ is the square root of the peak of the Schwarzschild effective potential 
\be
V^{\rm Schw}_{\rm eff} = A(r)(1+p_\varphi^2 u^2).
\ee 
We choose this region of the parameter space because it
is the one that yields the most interesting phenomenologies. 
In fact, for $\hat{E}_0 > \hat{E}_0^{\rm max}$ there is always a direct plunge, 
while in the other case it is possible to have many close passages before merger. 
For example, choosing $\nu=10^{-2}$ and starting the dynamics at $r_0=120$ with 
$p_\varphi^0 = 4.08$ and $\hat{E}_0 = 1.00002$ leads to 11 close passages before merger.

It is not our aim here to carry out a systematic analysis of the parameter space as done for bound orbits, 
so we will only focus our discussion on a few, illustrative, cases. Moreover, while in Sec.~\ref{subsec:insplunge} 
we have considered $\nu=10^{-3}$, here we use $\nu=10^{-2}$. This makes the dynamics with multiple 
encounters shorter and the corresponding numerical waveforms require less computational time.
This is not a big deal at the moment, since we want to specifically focus on the ringdown part.
We consider three $\nu=10^{-2}$ configurations, with 
$r_0 = 120$, $p_\varphi^0=4.01$, and different values of energy:  $\hat{E}_0 =\left(1.000711, 
1.000712, 1.001240\right)$ ($\hat{E}^{\rm max}_0 \simeq 1.00125$). 
The first one is a double encounter, while the others are single encounters. 
The trajectories, the quadrupolar waveforms and the corresponding frequencies are 
shown in Fig.~\ref{fig:hyp_wave}. The black waveform is as usual the numerical result, while
on the analytical side we consider two different EOB waveforms colored in grey and red.
The former is obtained attaching the \emph{circular} ringdown to the analytical waveform, 
similarly to what is done for hyperbolic encounters in Refs.~\cite{Nagar:2020xsk, Nagar:2021gss}, 
but {\it without} using NQC corrections as in Ref.~\cite{Gamba:2021gap}.
The primary templates for the phase and the amplitude used in Fig.~\ref{fig:hyp_wave} are 
the ones described in Sec.~\ref{subsec:insplunge}. 
For the red waveform we consider the same templates, but we extract the fit parameters
and the numerical quantities discussed in Sec.~\ref{subsec:insplunge} directly from the numerical
waveforms. A similar use of the numerical data is also done to determine the NQC coefficients $(a_i,b_i)$
of Eq.~\eqref{eq:nqc}. Finally, also the merger time $t_{\rm mrg}$ is taken to be precisely the numerical
one, since the simple prescription $t_{\rm mrg} = t^{\rm peak}_{\Omega_{\rm orb}} - 3$ used for bound orbits 
is found to be inaccurate in the hyperbolic case.
In future work we plan a systematic campaign of simulations of hyperbolic encounters so to inform
suitable analytical representations of both $t_{\rm mrg}$ and the ringdown parameters valid for
any configurations, analogously to the case of bound orbits.
In any case, as a proof of principle, the improvement introduced in the red waveform by the use of 
more precise parameters is clearly visible either before and after the merger. 
These comparisons show that including numerical information in the model
from hyperbolic simulations could greatly enhance the analytical descriptions of waveforms emitted by 
the dynamical capture of two black holes. In particular, a systematic coverage of black hole binaries 
undergoing dynamical encounters using numerical relativity~\cite{Gold:2012tk} will be 
instrumental to improve the waveform model \TEOBResumS{} for these configurations~\cite{Nagar:2020xsk}. 
This is expected to enhance the current analysis of GW190521 under the dynamical 
capture hypothesis~\cite{Gamba:2021gap}.

\section{Conclusions}
\label{sec:conclusions}
In this paper we have analyzed in detail and improved the proposal of~\cite{Chiaramello:2020ehz}
for incorporating eccentricity effects within the EOB waveform and radiation reaction.
We focused our analysis on the large mass ratio limit. 
We  have tested the performance of the analytical prescription for the fluxes 
over a significant portion of the parameter space by comparing it with highly accurate
waveforms and fluxes obtained solving numerically the Teukolsky equation using {\Teukode}. 
We have strengthened the PN corrections to the waveform amplitude using a proper 
resummation of the $\l=7$ and $\l=8$ modes. We have also improved the residual phase of 
the tail factor introducing 7.5~PN information in the $\delta_\lm$.

The advances introduced in this work provide an analytical EOB prescription for the waveform and 
fluxes that can be used in the description of EMRIs {\it without the need} of solving numerically
the computationally-expensive Teukolsky equation. The analytical waveform is accurate over 
a large portion of the parameter space, as shown in Sec.~\ref{sec:waves}. Moreover, 
the systematic analysis of the fluxes of Sec.~\ref{sec:fluxes} has shown that the radiation 
reaction is reliable with errors of the few percents for moderate eccentricities ($e\lesssim 0.3$) 
and spin-parameters not too high. In Sec.~\ref{subsec:insplunge}, we have also shown a 
preliminary work for the complete EOB waveform, from inspiral to ringdown, that we will discuss 
in more detail in future studies. In Sec.~\ref{subsec:hyp} we have also shown some hyperbolic
encounters.

The work presented here should be seen as a first step toward modeling eccentric EMRIs 
within the EOB formalism, improving on previous work that was limited to quasi-circular 
configurations~\cite{Yunes:2010zj}.
Two immediate next steps will be considered in future work.
(i)
The inclusion of Gravitational Self-Force (GSF) results concerning the central EOB potentials, 
$(A,D,Q)$. Similar approaches have been proposed in other 
works~\cite{Barausse:2011dq, Antonelli:2019fmq}. It could be similarly implemented using
using the Hamiltonian of \TEOBResumS{}, that is a mass-ratio deformation 
of the one of a (spinning) test-particle around a Kerr black hole, where the 5PN resummed potentials are  
replaced by the corresponding one obtained from GSF knowledge~\cite{Akcay:2015pjz}. Together with 
the radiation reaction discussed here, this analytical improvement would pave 
the way to a fully GSF-informed EOB 
model for eccentric EMRIs on equatorial orbits, analogously to NR-informed EOB models 
for coalescing black hole binaries.
(ii) The fluxes and waveform will have to be improved incorporating additional 
PN information~\cite{Mishra:2015bqa} {\it beyond}
the Newtonian prefactors considered here. This will hopefully allow one to reduce 
the analytical/numerical differences that we found when eccentricities are large.

A GSF-faithful EOB model can be used to generate a wide bank of EMRI waveforms.
It is however unlikely that the efficiency of the waveform generation will be sufficient
for direct use in parameter estimation (see in particular discussion in Appendix~C of Ref.~\cite{Nagar:2021gss}).
Nonetheless, these EOB templates can be used to create a fast and accurate surrogate using
machine learning techniques, similarly to what has been done in Ref.~\cite{Schmidt:2020yuu} for comparable 
mass binaries (up to mass ratio $q=20$). Such surrogate can then be employed for parameter 
estimation of EMRIs to be detected by LISA, as a complementary tool to other frameworks~\cite{Katz:2021yft, Hughes:2021exa}.

\section{Acknowledgement}
S.~B. acknowledges support by the EU H2020 under ERC Starting Grant, no.~BinGraSp-714626.  
The numerical simulations using {\Teukode} were performed on the Virgo ``Tullio'' server 
in Torino, supported by INFN.
We are grateful to S.~Hughes for the numerical data behind Fig.~\ref{fig:rho78_iResum} 
and to D.~Chiaramello for the preliminary work on the analytical model of the ringdown
and for the 
{\RWZ}'s simulations used in Table~\ref{tab:TeukRWZcheck} and Table~\ref{tab:schw_fluxes}.

\appendix

\section{Circular EOB}
\label{appendix:circular_eob}
\subsection{Multipolar waveform factorization}
In Eq.~\eqref{eq:eobwave} we have recalled the multipolar factorization of the EOB waveform. 
Here we briefly expose the factors in more details, without assuming an high mass-ratio. 
The Newtonian factor for \textit{circularized} binaries is explicitly given by
\begin{equation}
h_{\lm}^{(N, \epsilon)}=\frac{M \nu}{D_L} n_{\lm}^{(\epsilon)} c_{\l+\epsilon}(\nu) x^{\frac{(\l+\epsilon)}{2}} Y^{\l-\epsilon,-m}\left(\frac{\pi}{2}, \varphi\right) ,
\end{equation}
where $D_L$ is the luminosity distance, $x=\Omega^{2/3}$ and the coefficients are given by
\begin{align}
n_{\lm}^{(0)}&=(i m)^{\l} \frac{8 \pi}{(2 \l+1) ! !} \sqrt{\frac{(\l+1)(\l+2)}{\l(\l-1)}} ,\nonumber \\
n_{\lm}^{(1)}&=-(i m)^{\l} \frac{16 \pi i}{(2 \l+1) ! !} \sqrt{\frac{(2 \l+1)(\l+2)\left(\l^{2}-m^{2}\right)}{(2 \l-1)(\l+1) \l(\l-1)}} ,\nonumber \\
c_{\l + \epsilon} & = \left(\frac{m_2}{M}\right)^{\l + \epsilon - 1} + (-)^{\l+\epsilon} \left(\frac{m_1}{M}\right)^{\l + \epsilon - 1} \underrel{\nu\to 0}{=} \pm 1, \nonumber
\end{align}

The factors of the PN correction $\hat{h}^{(\epsilon)}_{\lm}$ are~\cite{Damour:2014sva, Damour:2008gu}:
\begin{itemize}
\item the source term $\hat{S}_{\rm eff}^{(\epsilon)}$, whose expression depends on parity:
\begin{subequations}
\begin{align}
\hat{S}_{\rm eff}^{(0)} &= \HKerr , \\
\hat{S}_{\rm eff}^{(1)} &= \frac{p_\varphi}{r_\Omega v_\varphi} .
\end{align}
\end{subequations}
\item the leading contribution of the tail factor $T_{\lm}$, generated by the 
backscattering of the GWs with the Kerr background
\begin{equation}
T_{\lm} = \frac{\Gamma(\l+1- 2 i k )}{\Gamma\left(\l+1\right)} e^{\pi k} e^{2 i k \ln\left(2 k r_0 \right)} , \nonumber
\end{equation}
where $k=m \Omega$, $r_0 = 2 /\sqrt{e}$, and $\Gamma(z)$ 
is the Euler Gamma function.
\item the residual phase of the tail factor $e^{i \delta_{\lm}}$, that takes into account the 
fact that $T_{\lm}$ includes only the leading contribution. We have analyzed these phases 
in Sec.~\ref{sec:deltalm}.
\item the residual relativistic amplitude corrections $\rho_{\lm}$, that are resummed 
following Refs.~\cite{Nagar:2016ayt, Messina:2018ghh}. In Sec.~\ref{sec:resum_l7_l8} 
we have extended this resummation scheme to $\l=7$ and $\l=8$.
\end{itemize}

In order to improve the agreement with numerical data during the plunge, 
it is useful to replace $x=\Omega^{2/3}$ with $x=r_\Omega^2  \Omega^2 \equiv v_\varphi^2 $, where
the spin-informed radius $r_\Omega$ is given by the third Kepler law 
generalized for circular orbits in Kerr spacetime~\cite{Bardeen:1972fi} (see Eq.~\eqref{eq:rOmg}).
The new $x$-parameter is more reliable for noncircular motion and it is used both
in the Newtonian prefactors $h_{\lm}^{(N, \epsilon)}$ and in the amplitude corrections $\rho_\lm(x)$.

\subsection{Angular radiation reaction}
\label{appendixsec:Fphi_circ}
In the circular case, Eq.~\eqref{eq:enbalance} reduces to
\begin{align}
\dot{J} &= - \F_\varphi , \nonumber \\
\dot{E} &= \Omega \dot{J}. \nonumber
\end{align}
Decomposing the energy in multipoles, we have
\begin{align}
\dot{J} = \frac{1}{\Omega} \sum_{\l=2}^{\infty} \sum_{m=1}^{\l} F_{\lm} & = \frac{2}{16 \pi \Omega}
\sum_{\l=2}^{\infty} \sum_{m=1}^{\l} |\dot{h}_{\lm}|^2 \nonumber \\ 
& = \frac{1}{8 \pi} \sum_{\l=2}^{\infty} \sum_{m=1}^{\l} m^2\Omega |h_{\lm}|^2, \nonumber
\end{align}
then factorizing the Newtonian $(2,2)$ contribution, we can write the angular radiation 
reaction as
\begin{equation}
\hat{\F}_\varphi = - \frac{\dot{J}}{\nu} = - \frac{32}{5} \nu x^{7/2} \hat{f} ,
\end{equation}
where 
\begin{align}
\label{eq:hatf}
\hat{f} & = \sum_{\l=2}^{\infty} \sum_{m=1}^{\l} (F_{22}^N)^{-1} F_{\lm}  = \sum_{\l=2}^{\infty} \sum_{m=1}^{\l} \hat{F}_\lm  ,\\
F_\lm &  = \frac{1}{8\pi} m^2 \Omega^2 |h_{\lm}|^2,  \label{eq:Flm} \\
F_{22}^N & = \frac{32}{5} x^5.
\end{align} 
Finally, we can use the third Kepler law generalized to Kerr spacetime 
($r_\Omega^3 \Omega^2=1$) to write Eq.~\eqref{eq:circFphi}.

\section{Newtonian noncircular expressions}
\label{appendix:newtprefs}
In Sec.~\ref{sec:eobmodel} we have discussed the Newtonian noncircular corrections to the waveform, 
$\hat{h}_{\lm}^{(N, \epsilon)_{\rm nc}}$, and the noncircular Newtonian prefactors for the
angular radiation reaction, $\fnp{,\lm}$. 
Here we report them explicitly for the $(\l,m)=(2,2), (2,1), (3,3), (3,2), (4,4)$ multipoles:

\begin{widetext}

\begin{align}
\hat{h}_{22}^{(N, 0)_{\rm nc}} = 1&-\frac{\dot{r}^2}{2 r^2 \Omega ^2}-\frac{\ddot{r}}{2 r \Omega ^2} + i \left(\frac{2  \dot{r}}{r \Omega }+\frac{ {\dot{\Omega}}}{2 \Omega ^2} \right) \nonumber , \\
%
\hat{h}_{21}^{(N, 1)_{\rm nc}} = 1&-\frac{6 \dot{r}^2}{r^2 \Omega ^2}-\frac{3 \ddot{r}}{r \Omega ^2}-\frac{6 \dot{r} {\dot{\Omega}}}{r \Omega ^3}-\frac{{\ddot{\Omega}}}{\Omega ^3} + i \left(\frac{6  \dot{r}}{r \Omega }+\frac{3  {\dot{\Omega}}}{\Omega ^2} \right) \nonumber , \\
%
%
\hat{h}_{33}^{(N, 0)_{\rm nc}} = 1&-\frac{2 \dot{r}^2}{r^2 \Omega ^2}-\frac{\ddot{r}}{r \Omega ^2}-\frac{\dot{r} {\dot{\Omega}}}{r \Omega ^3}-\frac{{\ddot{\Omega}}}{9 \Omega ^3} + i \left(-\frac{2  \dot{r}^3}{9 r^3 \Omega ^3}-\frac{2  \ddot{r} \dot{r}}{3 r^2 \Omega ^3}-\frac{ {r^{(3)}}}{9 r \Omega ^3}+\frac{3  \dot{r}}{r \Omega }+\frac{ {\dot{\Omega}}}{\Omega ^2} \right) \nonumber , \\
%
\hat{h}_{32}^{(N, 1)_{\rm nc}} = 1&-\frac{9 \dot{r}^2}{r^2 \Omega ^2}-\frac{3 \ddot{r}}{r \Omega ^2}-\frac{9 \dot{r} {\dot{\Omega}}}{r \Omega ^3}-\frac{3 {\dot{\Omega}}^2}{4 \Omega ^4}-\frac{{\ddot{\Omega}}}{\Omega ^3} + i \left(-\frac{3  \dot{r}^3}{r^3 \Omega ^3}-\frac{9  \ddot{r} \dot{r}}{2 r^2 \Omega ^3}-\frac{9  \dot{r}^2 {\dot{\Omega}}}{2 r^2 \Omega ^4}-\frac{3  \ddot{r} {\dot{\Omega}}}{2 r \Omega ^4}\right. \nonumber \\
&\left. -\frac{ {r^{(3)}}}{2 r \Omega ^3}-\frac{3  \dot{r} {\ddot{\Omega}}}{2 r \Omega ^4}+\frac{6  \dot{r}}{r \Omega }-\frac{ {\Omega^{(3)}}}{8 \Omega ^4}+\frac{3  {\dot{\Omega}}}{\Omega ^2} \right) \nonumber , \\
%
%
\hat{h}_{44}^{(N, 0)_{\rm nc}} = 1&+\frac{3 \dot{r}^4}{32 r^4 \Omega ^4}+\frac{9 \ddot{r} \dot{r}^2}{16 r^3 \Omega ^4}+\frac{9 \ddot{r}^2}{64 r^2 \Omega ^4}+\frac{3 {r^{(3)}} \dot{r}}{16 r^2 \Omega ^4}-\frac{9 \dot{r}^2}{2 r^2 \Omega ^2}-\frac{3 \ddot{r}}{2 r \Omega ^2}+\frac{{r^{(4)}}}{64 r \Omega ^4}-\frac{3 \dot{r} {\dot{\Omega}}}{r \Omega ^3}-\frac{3 {\dot{\Omega}}^2}{16 \Omega ^4} \nonumber \\
&-\frac{{\ddot{\Omega}}}{4 \Omega ^3} + i \left(-\frac{3  \dot{r}^3}{2 r^3 \Omega ^3}-\frac{9  \ddot{r} \dot{r}}{4 r^2 \Omega ^3}-\frac{9  \dot{r}^2 {\dot{\Omega}}}{8 r^2 \Omega ^4}-\frac{3  \ddot{r} {\dot{\Omega}}}{8 r \Omega ^4}-\frac{ {r^{(3)}}}{4 r \Omega ^3}-\frac{ \dot{r} {\ddot{\Omega}}}{4 r \Omega ^4}+\frac{4  \dot{r}}{r \Omega }-\frac{ {\Omega^{(3)}}}{64 \Omega ^4}+\frac{3  {\dot{\Omega}}}{2 \Omega ^2} \right) \nonumber ,
\end{align}

\begin{align}
\fnp{,22} = 1&+\frac{3 \dot{r}^4}{4 r^4 \Omega ^4}+\frac{3 \dot{r}^3 {\dot{\Omega}}}{4 r^3 \Omega ^5}+\frac{3 \ddot{r}^2}{4 r^2 \Omega ^4}+\frac{3 \ddot{r} \dot{r} {\dot{\Omega}}}{8 r^2 \Omega ^5}-\frac{{r^{(3)}} \dot{r}}{2 r^2 \Omega ^4}+\frac{\dot{r}^2 {\ddot{\Omega}}}{8 r^2 \Omega ^5}+\frac{4 \dot{r}^2}{r^2 \Omega ^2}+\frac{\ddot{r} {\ddot{\Omega}}}{8 r \Omega ^5}-\frac{2 \ddot{r}}{r \Omega ^2} \nonumber \\
&-\frac{{r^{(3)}} {\dot{\Omega}}}{8 r \Omega ^5}+\frac{3 \dot{r} {\dot{\Omega}}}{r \Omega ^3}+\frac{3 {\dot{\Omega}}^2}{4 \Omega ^4}-\frac{{\ddot{\Omega}}}{4 \Omega ^3} \nonumber , \\
%
\fnp{,21} = 1&+\frac{72 \dot{r}^4}{r^4 \Omega ^4}+\frac{144 \dot{r}^3 {\dot{\Omega}}}{r^3 \Omega ^5}+\frac{27 \ddot{r}^2}{r^2 \Omega ^4}+\frac{27 \ddot{r} \dot{r} {\dot{\Omega}}}{r^2 \Omega ^5}-\frac{18 {r^{(3)}} \dot{r}}{r^2 \Omega ^4}+\frac{126 \dot{r}^2 {\dot{\Omega}}^2}{r^2 \Omega ^6}-\frac{12 \dot{r}^2 {\ddot{\Omega}}}{r^2 \Omega ^5}+\frac{30 \dot{r}^2}{r^2 \Omega ^2}-\frac{18 \ddot{r} {\dot{\Omega}}^2}{r \Omega ^6} \nonumber \\
&+\frac{21 \ddot{r} {\ddot{\Omega}}}{r \Omega ^5}-\frac{12 \ddot{r}}{r \Omega ^2}-\frac{9 {r^{(3)}} {\dot{\Omega}}}{r \Omega ^5}+\frac{18 \dot{r} {\dot{\Omega}}^3}{r \Omega ^7}+\frac{24 \dot{r} {\ddot{\Omega}} {\dot{\Omega}}}{r \Omega ^6}-\frac{6 \dot{r} {\Omega^{(3)}}}{r \Omega ^5}+\frac{30 \dot{r} {\dot{\Omega}}}{r \Omega ^3}+\frac{3 {\ddot{\Omega}} {\dot{\Omega}}^2}{\Omega ^7}+\frac{4 {\ddot{\Omega}}^2}{\Omega ^6} \nonumber \\
&-\frac{3 {\Omega^{(3)}} {\dot{\Omega}}}{\Omega ^6}+\frac{15 {\dot{\Omega}}^2}{\Omega ^4}-\frac{5 {\ddot{\Omega}}}{\Omega ^3} \nonumber , \\
%
%
\fnp{,33} = 1&+\frac{16 \dot{r}^6}{81 r^6 \Omega ^6}+\frac{8 \ddot{r} \dot{r}^4}{27 r^5 \Omega ^6}+\frac{8 \dot{r}^5 {\dot{\Omega}}}{27 r^5 \Omega ^7}+\frac{8 \ddot{r}^2 \dot{r}^2}{9 r^4 \Omega ^6}+\frac{16 \ddot{r} \dot{r}^3 {\dot{\Omega}}}{27 r^4 \Omega ^7}-\frac{32 {r^{(3)}} \dot{r}^3}{81 r^4 \Omega ^6}+\frac{8 \dot{r}^4 {\ddot{\Omega}}}{81 r^4 \Omega ^7}+\frac{40 \dot{r}^4}{9 r^4 \Omega ^4}-\frac{2 \ddot{r}^3}{9 r^3 \Omega ^6} \nonumber \\
&+\frac{2 \ddot{r}^2 \dot{r} {\dot{\Omega}}}{9 r^3 \Omega ^7}+\frac{8 \ddot{r} {r^{(3)}} \dot{r}}{27 r^3 \Omega ^6}+\frac{20 \ddot{r} \dot{r}^2 {\ddot{\Omega}}}{81 r^3 \Omega ^7}-\frac{20 \ddot{r} \dot{r}^2}{9 r^3 \Omega ^4}-\frac{4 {r^{(3)}} \dot{r}^2 {\dot{\Omega}}}{27 r^3 \Omega ^7}-\frac{2 {r^{(4)}} \dot{r}^2}{27 r^3 \Omega ^6}+\frac{2 \dot{r}^3 {\Omega^{(3)}}}{243 r^3 \Omega ^7}+\frac{20 \dot{r}^3 {\dot{\Omega}}}{3 r^3 \Omega ^5}-\frac{2 \ddot{r}^2 {\ddot{\Omega}}}{81 r^2 \Omega ^7} \nonumber \\
&+\frac{20 \ddot{r}^2}{9 r^2 \Omega ^4}+\frac{2 \ddot{r} {r^{(3)}} {\dot{\Omega}}}{27 r^2 \Omega ^7}-\frac{\ddot{r} {r^{(4)}}}{27 r^2 \Omega ^6}+\frac{2 \ddot{r} \dot{r} {\Omega^{(3)}}}{81 r^2 \Omega ^7}+\frac{4 {r^{(3)}}^2}{81 r^2 \Omega ^6}+\frac{4 {r^{(3)}} \dot{r} {\ddot{\Omega}}}{243 r^2 \Omega ^7}-\frac{40 {r^{(3)}} \dot{r}}{27 r^2 \Omega ^4}-\frac{{r^{(4)}} \dot{r} {\dot{\Omega}}}{27 r^2 \Omega ^7}+\frac{10 \dot{r}^2 {\dot{\Omega}}^2}{3 r^2 \Omega ^6} \nonumber \\
&+\frac{6 \dot{r}^2}{r^2 \Omega ^2}-\frac{\ddot{r} {\dot{\Omega}}^2}{3 r \Omega ^6}+\frac{2 \ddot{r} {\ddot{\Omega}}}{3 r \Omega ^5}-\frac{3 \ddot{r}}{r \Omega ^2}+\frac{{r^{(3)}} {\Omega^{(3)}}}{243 r \Omega ^7}-\frac{2 {r^{(3)}} {\dot{\Omega}}}{3 r \Omega ^5}-\frac{{r^{(4)}} {\ddot{\Omega}}}{243 r \Omega ^7}+\frac{{r^{(4)}}}{27 r \Omega ^4}+\frac{\dot{r} {\dot{\Omega}}^3}{3 r \Omega ^7} \nonumber \\
&+\frac{4 \dot{r} {\ddot{\Omega}} {\dot{\Omega}}}{9 r \Omega ^6}-\frac{\dot{r} {\Omega^{(3)}}}{9 r \Omega ^5}+\frac{5 \dot{r} {\dot{\Omega}}}{r \Omega ^3}+\frac{{\ddot{\Omega}} {\dot{\Omega}}^2}{27 \Omega ^7}+\frac{4 {\ddot{\Omega}}^2}{81 \Omega ^6}-\frac{{\Omega^{(3)}} {\dot{\Omega}}}{27 \Omega ^6}+\frac{5 {\dot{\Omega}}^2}{3 \Omega ^4}-\frac{5 {\ddot{\Omega}}}{9 \Omega ^3} \nonumber , \\
%
\fnp{,32} = 1&+\frac{45 \dot{r}^6}{2 r^6 \Omega ^6}+\frac{135 {\dot{\Omega}} \dot{r}^5}{2 r^5 \Omega ^7}+\frac{675 {\dot{\Omega}}^2 \dot{r}^4}{8 r^4 \Omega ^8}+\frac{135 \dot{r}^4}{2 r^4 \Omega ^4}+\frac{45 \ddot{r} \dot{r}^4}{2 r^5 \Omega ^6}+\frac{45 {\dot{\Omega}}^3 \dot{r}^3}{2 r^3 \Omega ^9}-\frac{45 {\Omega^{(3)}} \dot{r}^3}{8 r^3 \Omega ^7}+\frac{135 {\ddot{\Omega}} {\dot{\Omega}} \dot{r}^3}{4 r^3 \Omega ^8} \nonumber \\
&+\frac{135 {\dot{\Omega}} \dot{r}^3}{r^3 \Omega ^5}+\frac{135 \ddot{r} {\dot{\Omega}} \dot{r}^3}{2 r^4 \Omega ^7}-\frac{15 {r^{(3)}} \dot{r}^3}{r^4 \Omega ^6}+\frac{15 {\ddot{\Omega}}^2 \dot{r}^2}{2 r^2 \Omega ^8}+\frac{45 {\ddot{\Omega}} {\dot{\Omega}}^2 \dot{r}^2}{4 r^2 \Omega ^9}+\frac{495 {\dot{\Omega}}^2 \dot{r}^2}{4 r^2 \Omega ^6}+\frac{81 \ddot{r} {\dot{\Omega}}^2 \dot{r}^2}{4 r^3 \Omega ^8}-\frac{15 {\ddot{\Omega}} \dot{r}^2}{r^2 \Omega ^5} \nonumber \\
&+\frac{27 \ddot{r} {\ddot{\Omega}} \dot{r}^2}{r^3 \Omega ^7}-\frac{9 {\Omega^{(4)}} \dot{r}^2}{16 r^2 \Omega ^7}-\frac{45 {\Omega^{(3)}} {\dot{\Omega}} \dot{r}^2}{16 r^2 \Omega ^8}-\frac{27 {r^{(3)}} {\dot{\Omega}} \dot{r}^2}{2 r^3 \Omega ^7}+\frac{21 \dot{r}^2}{r^2 \Omega ^2}-\frac{27 \ddot{r} \dot{r}^2}{r^3 \Omega ^4}+\frac{135 \ddot{r}^2 \dot{r}^2}{4 r^4 \Omega ^6}-\frac{9 {r^{(4)}} \dot{r}^2}{4 r^3 \Omega ^6} \nonumber \\
&+\frac{135 {\dot{\Omega}}^3 \dot{r}}{4 r \Omega ^7}+\frac{9 \ddot{r} {\dot{\Omega}}^3 \dot{r}}{4 r^2 \Omega ^9}-\frac{33 {r^{(3)}} {\dot{\Omega}}^2 \dot{r}}{4 r^2 \Omega ^8}+\frac{4 {r^{(3)}} {\ddot{\Omega}} \dot{r}}{r^2 \Omega ^7}+\frac{15 {\ddot{\Omega}} {\Omega^{(3)}} \dot{r}}{16 r \Omega ^8}-\frac{15 {\Omega^{(3)}} \dot{r}}{4 r \Omega ^5}+\frac{33 \ddot{r} {\Omega^{(3)}} \dot{r}}{16 r^2 \Omega ^7}+\frac{15 {\ddot{\Omega}}^2 {\dot{\Omega}} \dot{r}}{8 r \Omega ^9} \nonumber \\
&+\frac{15 {\ddot{\Omega}} {\dot{\Omega}} \dot{r}}{2 r \Omega ^6}+\frac{69 \ddot{r} {\ddot{\Omega}} {\dot{\Omega}} \dot{r}}{8 r^2 \Omega ^8}-\frac{9 {\Omega^{(4)}} {\dot{\Omega}} \dot{r}}{16 r \Omega ^8}+\frac{21 {\dot{\Omega}} \dot{r}}{r \Omega ^3}-\frac{15 \ddot{r} {\dot{\Omega}} \dot{r}}{2 r^2 \Omega ^5}+\frac{81 \ddot{r}^2 {\dot{\Omega}} \dot{r}}{4 r^3 \Omega ^7}-\frac{9 {r^{(4)}} {\dot{\Omega}} \dot{r}}{4 r^2 \Omega ^7}-\frac{13 {r^{(3)}} \dot{r}}{r^2 \Omega ^4} \nonumber \\
&+\frac{9 \ddot{r} {r^{(3)}} \dot{r}}{r^3 \Omega ^6}+\frac{45 {\dot{\Omega}}^4}{16 \Omega ^8}-\frac{3 {r^{(3)}} {\dot{\Omega}}^3}{4 r \Omega ^9}+\frac{5 {\ddot{\Omega}}^2}{2 \Omega ^6}-\frac{3 \ddot{r} {\ddot{\Omega}}^2}{2 r \Omega ^8}+\frac{5 {\Omega^{(3)}}^2}{64 \Omega ^8}+\frac{15 {\ddot{\Omega}} {\dot{\Omega}}^2}{8 \Omega ^7}+\frac{3 \ddot{r} {\ddot{\Omega}} {\dot{\Omega}}^2}{4 r \Omega ^9} \nonumber \\
&-\frac{3 {\Omega^{(4)}} {\dot{\Omega}}^2}{64 \Omega ^9}+\frac{21 {\dot{\Omega}}^2}{2 \Omega ^4}-\frac{75 \ddot{r} {\dot{\Omega}}^2}{4 r \Omega ^6}+\frac{189 \ddot{r}^2 {\dot{\Omega}}^2}{16 r^2 \Omega ^8}-\frac{3 {r^{(4)}} {\dot{\Omega}}^2}{16 r \Omega ^8}-\frac{7 {\ddot{\Omega}}}{2 \Omega ^3}+\frac{15 \ddot{r} {\ddot{\Omega}}}{r \Omega ^5}-\frac{27 \ddot{r}^2 {\ddot{\Omega}}}{4 r^2 \Omega ^7} \nonumber \\
&-\frac{{r^{(4)}} {\ddot{\Omega}}}{4 r \Omega ^7}+\frac{9 {r^{(3)}} {\Omega^{(3)}}}{16 r \Omega ^7}-\frac{{\ddot{\Omega}} {\Omega^{(4)}}}{16 \Omega ^8}+\frac{{\Omega^{(4)}}}{16 \Omega ^5}-\frac{3 \ddot{r} {\Omega^{(4)}}}{16 r \Omega ^7}-\frac{3 {r^{(3)}} {\ddot{\Omega}} {\dot{\Omega}}}{8 r \Omega ^8}+\frac{5 {\ddot{\Omega}} {\Omega^{(3)}} {\dot{\Omega}}}{32 \Omega ^9}-\frac{5 {\Omega^{(3)}} {\dot{\Omega}}}{2 \Omega ^6} \nonumber \\
&+\frac{33 \ddot{r} {\Omega^{(3)}} {\dot{\Omega}}}{16 r \Omega ^8}-\frac{15 {r^{(3)}} {\dot{\Omega}}}{2 r \Omega ^5}+\frac{9 \ddot{r} {r^{(3)}} {\dot{\Omega}}}{2 r^2 \Omega ^7}-\frac{9 \ddot{r}}{r \Omega ^2}+\frac{81 \ddot{r}^2}{4 r^2 \Omega ^4}+\frac{{r^{(4)}}}{4 r \Omega ^4}-\frac{27 \ddot{r}^3}{4 r^3 \Omega ^6}+\frac{{r^{(3)}}^2}{r^2 \Omega ^6}-\frac{3 \ddot{r} {r^{(4)}}}{4 r^2 \Omega ^6} \nonumber , \\
%
\fnp{,44} = 1&+\frac{45 \dot{r}^8}{1024 r^8 \Omega ^8}+\frac{45 {\dot{\Omega}} \dot{r}^7}{512 r^7 \Omega ^9}+\frac{45 {\ddot{\Omega}} \dot{r}^6}{1024 r^6 \Omega ^9}+\frac{45 \dot{r}^6}{16 r^6 \Omega ^6}+\frac{45 \ddot{r} \dot{r}^6}{256 r^7 \Omega ^8}+\frac{15 {\Omega^{(3)}} \dot{r}^5}{2048 r^5 \Omega ^9}+\frac{405 {\dot{\Omega}} \dot{r}^5}{64 r^5 \Omega ^7} \nonumber \\
&+\frac{405 \ddot{r} {\dot{\Omega}} \dot{r}^5}{1024 r^6 \Omega ^9}-\frac{45 {r^{(3)}} \dot{r}^5}{256 r^6 \Omega ^8}+\frac{1215 {\dot{\Omega}}^2 \dot{r}^4}{256 r^4 \Omega ^8}+\frac{135 {\ddot{\Omega}} \dot{r}^4}{256 r^4 \Omega ^7}+\frac{225 \ddot{r} {\ddot{\Omega}} \dot{r}^4}{1024 r^5 \Omega ^9}+\frac{3 {\Omega^{(4)}} \dot{r}^4}{8192 r^4 \Omega ^9}-\frac{75 {r^{(3)}} {\dot{\Omega}} \dot{r}^4}{1024 r^5 \Omega ^9} \nonumber \\
&+\frac{189 \dot{r}^4}{16 r^4 \Omega ^4}+\frac{675 \ddot{r}^2 \dot{r}^4}{1024 r^6 \Omega ^8}-\frac{75 {r^{(4)}} \dot{r}^4}{1024 r^5 \Omega ^8}+\frac{225 {\dot{\Omega}}^3 \dot{r}^3}{256 r^3 \Omega ^9}+\frac{15 {r^{(3)}} {\ddot{\Omega}} \dot{r}^3}{512 r^4 \Omega ^9}-\frac{45 {\Omega^{(3)}} \dot{r}^3}{256 r^3 \Omega ^7}+\frac{165 \ddot{r} {\Omega^{(3)}} \dot{r}^3}{4096 r^4 \Omega ^9} \nonumber \\
&+\frac{45 {\ddot{\Omega}} {\dot{\Omega}} \dot{r}^3}{32 r^3 \Omega ^8}+\frac{315 {\dot{\Omega}} \dot{r}^3}{16 r^3 \Omega ^5}+\frac{405 \ddot{r} {\dot{\Omega}} \dot{r}^3}{128 r^4 \Omega ^7}+\frac{225 \ddot{r}^2 {\dot{\Omega}} \dot{r}^3}{512 r^5 \Omega ^9}-\frac{105 {r^{(4)}} {\dot{\Omega}} \dot{r}^3}{2048 r^4 \Omega ^9}-\frac{45 {r^{(3)}} \dot{r}^3}{16 r^4 \Omega ^6}+\frac{75 \ddot{r} {r^{(3)}} \dot{r}^3}{256 r^5 \Omega ^8} \nonumber \\
&-\frac{3 {r^{(5)}} \dot{r}^3}{512 r^4 \Omega ^8}+\frac{25 {\ddot{\Omega}}^2 \dot{r}^2}{128 r^2 \Omega ^8}+\frac{165 {\ddot{\Omega}} {\dot{\Omega}}^2 \dot{r}^2}{512 r^2 \Omega ^9}+\frac{105 {\dot{\Omega}}^2 \dot{r}^2}{8 r^2 \Omega ^6}+\frac{135 \ddot{r} {\dot{\Omega}}^2 \dot{r}^2}{128 r^3 \Omega ^8}-\frac{35 {\ddot{\Omega}} \dot{r}^2}{32 r^2 \Omega ^5}+\frac{225 \ddot{r} {\ddot{\Omega}} \dot{r}^2}{128 r^3 \Omega ^7} \nonumber \\
&+\frac{135 \ddot{r}^2 {\ddot{\Omega}} \dot{r}^2}{2048 r^4 \Omega ^9}-\frac{15 {r^{(4)}} {\ddot{\Omega}} \dot{r}^2}{2048 r^3 \Omega ^9}+\frac{45 {r^{(3)}} {\Omega^{(3)}} \dot{r}^2}{4096 r^3 \Omega ^9}-\frac{9 {\Omega^{(4)}} \dot{r}^2}{512 r^2 \Omega ^7}+\frac{9 \ddot{r} {\Omega^{(4)}} \dot{r}^2}{4096 r^3 \Omega ^9}-\frac{15 {\Omega^{(3)}} {\dot{\Omega}} \dot{r}^2}{256 r^2 \Omega ^8}-\frac{315 {r^{(3)}} {\dot{\Omega}} \dot{r}^2}{128 r^3 \Omega ^7} \nonumber \\
&+\frac{135 \ddot{r} {r^{(3)}} {\dot{\Omega}} \dot{r}^2}{1024 r^4 \Omega ^9}-\frac{9 {r^{(5)}} {\dot{\Omega}} \dot{r}^2}{2048 r^3 \Omega ^9}+\frac{8 \dot{r}^2}{r^2 \Omega ^2}-\frac{63 \ddot{r} \dot{r}^2}{8 r^3 \Omega ^4}+\frac{135 \ddot{r}^2 \dot{r}^2}{32 r^4 \Omega ^6}-\frac{15 {r^{(4)}} \dot{r}^2}{64 r^3 \Omega ^6}+\frac{15 {r^{(3)}}^2 \dot{r}^2}{128 r^4 \Omega ^8} \nonumber \\
&-\frac{45 \ddot{r} {r^{(4)}} \dot{r}^2}{1024 r^4 \Omega ^8}+\frac{45 {\dot{\Omega}}^3 \dot{r}}{16 r \Omega ^7}+\frac{45 \ddot{r} {\dot{\Omega}}^3 \dot{r}}{512 r^2 \Omega ^9}-\frac{75 {r^{(3)}} {\dot{\Omega}}^2 \dot{r}}{128 r^2 \Omega ^8}+\frac{15 {r^{(3)}} {\ddot{\Omega}} \dot{r}}{128 r^2 \Omega ^7}-\frac{{r^{(5)}} {\ddot{\Omega}} \dot{r}}{1024 r^2 \Omega ^9}+\frac{5 {\ddot{\Omega}} {\Omega^{(3)}} \dot{r}}{256 r \Omega ^8} \nonumber \\
&-\frac{5 {\Omega^{(3)}} \dot{r}}{16 r \Omega ^5}+\frac{75 \ddot{r} {\Omega^{(3)}} \dot{r}}{512 r^2 \Omega ^7}+\frac{45 \ddot{r}^2 {\Omega^{(3)}} \dot{r}}{8192 r^3 \Omega ^9}+\frac{5 {r^{(4)}} {\Omega^{(3)}} \dot{r}}{16384 r^2 \Omega ^9}+\frac{3 {r^{(3)}} {\Omega^{(4)}} \dot{r}}{4096 r^2 \Omega ^9}+\frac{5 {\ddot{\Omega}}^2 {\dot{\Omega}} \dot{r}}{128 r \Omega ^9}+\frac{5 {\ddot{\Omega}} {\dot{\Omega}} \dot{r}}{8 r \Omega ^6} \nonumber \\
&+\frac{15 \ddot{r} {\ddot{\Omega}} {\dot{\Omega}} \dot{r}}{32 r^2 \Omega ^8}-\frac{3 {\Omega^{(4)}} {\dot{\Omega}} \dot{r}}{256 r \Omega ^8}+\frac{7 {\dot{\Omega}} \dot{r}}{r \Omega ^3}-\frac{105 \ddot{r} {\dot{\Omega}} \dot{r}}{32 r^2 \Omega ^5}+\frac{405 \ddot{r}^2 {\dot{\Omega}} \dot{r}}{256 r^3 \Omega ^7}-\frac{135 {r^{(4)}} {\dot{\Omega}} \dot{r}}{512 r^2 \Omega ^7}+\frac{135 \ddot{r}^3 {\dot{\Omega}} \dot{r}}{2048 r^4 \Omega ^9} \nonumber \\
&+\frac{15 {r^{(3)}}^2 {\dot{\Omega}} \dot{r}}{512 r^3 \Omega ^9}-\frac{21 {r^{(3)}} \dot{r}}{8 r^2 \Omega ^4}+\frac{45 \ddot{r} {r^{(3)}} \dot{r}}{32 r^3 \Omega ^6}+\frac{{r^{(5)}} \dot{r}}{64 r^2 \Omega ^6}-\frac{45 \ddot{r}^2 {r^{(3)}} \dot{r}}{512 r^4 \Omega ^8}+\frac{15 {r^{(3)}} {r^{(4)}} \dot{r}}{1024 r^3 \Omega ^8}-\frac{9 \ddot{r} {r^{(5)}} \dot{r}}{1024 r^3 \Omega ^8} \nonumber \\
&+\frac{45 {\dot{\Omega}}^4}{256 \Omega ^8}-\frac{15 {r^{(3)}} {\dot{\Omega}}^3}{512 r \Omega ^9}+\frac{5 {\ddot{\Omega}}^2}{32 \Omega ^6}-\frac{5 \ddot{r} {\ddot{\Omega}}^2}{128 r \Omega ^8}+\frac{5 {\Omega^{(3)}}^2}{4096 \Omega ^8}+\frac{15 {\ddot{\Omega}} {\dot{\Omega}}^2}{128 \Omega ^7}+\frac{15 \ddot{r} {\ddot{\Omega}} {\dot{\Omega}}^2}{512 r \Omega ^9} \nonumber \\
&-\frac{3 {\Omega^{(4)}} {\dot{\Omega}}^2}{4096 \Omega ^9}+\frac{21 {\dot{\Omega}}^2}{8 \Omega ^4}-\frac{15 \ddot{r} {\dot{\Omega}}^2}{8 r \Omega ^6}+\frac{225 \ddot{r}^2 {\dot{\Omega}}^2}{512 r^2 \Omega ^8}-\frac{15 {r^{(4)}} {\dot{\Omega}}^2}{512 r \Omega ^8}-\frac{7 {\ddot{\Omega}}}{8 \Omega ^3}+\frac{55 \ddot{r} {\ddot{\Omega}}}{32 r \Omega ^5} \nonumber \\
&-\frac{255 \ddot{r}^2 {\ddot{\Omega}}}{512 r^2 \Omega ^7}-\frac{15 {r^{(4)}} {\ddot{\Omega}}}{512 r \Omega ^7}+\frac{45 \ddot{r}^3 {\ddot{\Omega}}}{2048 r^3 \Omega ^9}+\frac{5 \ddot{r} {r^{(4)}} {\ddot{\Omega}}}{2048 r^2 \Omega ^9}+\frac{15 {r^{(3)}} {\Omega^{(3)}}}{512 r \Omega ^7}-\frac{15 \ddot{r} {r^{(3)}} {\Omega^{(3)}}}{8192 r^2 \Omega ^9}-\frac{{r^{(5)}} {\Omega^{(3)}}}{16384 r \Omega ^9} \nonumber \\
&-\frac{{\ddot{\Omega}} {\Omega^{(4)}}}{1024 \Omega ^8}+\frac{{\Omega^{(4)}}}{256 \Omega ^5}-\frac{3 \ddot{r} {\Omega^{(4)}}}{512 r \Omega ^7}+\frac{9 \ddot{r}^2 {\Omega^{(4)}}}{16384 r^2 \Omega ^9}+\frac{{r^{(4)}} {\Omega^{(4)}}}{16384 r \Omega ^9}+\frac{5 {\ddot{\Omega}} {\Omega^{(3)}} {\dot{\Omega}}}{2048 \Omega ^9}-\frac{5 {\Omega^{(3)}} {\dot{\Omega}}}{32 \Omega ^6} \nonumber \\
&+\frac{15 \ddot{r} {\Omega^{(3)}} {\dot{\Omega}}}{256 r \Omega ^8}-\frac{45 {r^{(3)}} {\dot{\Omega}}}{32 r \Omega ^5}+\frac{165 \ddot{r} {r^{(3)}} {\dot{\Omega}}}{256 r^2 \Omega ^7}+\frac{3 {r^{(5)}} {\dot{\Omega}}}{512 r \Omega ^7}-\frac{45 \ddot{r}^2 {r^{(3)}} {\dot{\Omega}}}{2048 r^3 \Omega ^9}+\frac{5 {r^{(3)}} {r^{(4)}} {\dot{\Omega}}}{2048 r^2 \Omega ^9}-\frac{3 \ddot{r} {r^{(5)}} {\dot{\Omega}}}{2048 r^2 \Omega ^9} \nonumber \\
&-\frac{4 \ddot{r}}{r \Omega ^2}+\frac{147 \ddot{r}^2}{32 r^2 \Omega ^4}+\frac{3 {r^{(4)}}}{32 r \Omega ^4}-\frac{45 \ddot{r}^3}{32 r^3 \Omega ^6}+\frac{5 {r^{(3)}}^2}{32 r^2 \Omega ^6}-\frac{5 \ddot{r} {r^{(4)}}}{32 r^2 \Omega ^6}+\frac{405 \ddot{r}^4}{4096 r^4 \Omega ^8} \nonumber \\
&-\frac{15 \ddot{r} {r^{(3)}}^2}{512 r^3 \Omega ^8}+\frac{5 {r^{(4)}}^2}{4096 r^2 \Omega ^8}+\frac{45 \ddot{r}^2 {r^{(4)}}}{2048 r^3 \Omega ^8}-\frac{{r^{(3)}} {r^{(5)}}}{1024 r^2 \Omega ^8} \nonumber .
\end{align}

\end{widetext}

\section{Waveform multipoles}
\label{appendix:subdominant_wave}
\begin{figure}[t]
  \center  
  \includegraphics[width=0.22\textwidth]{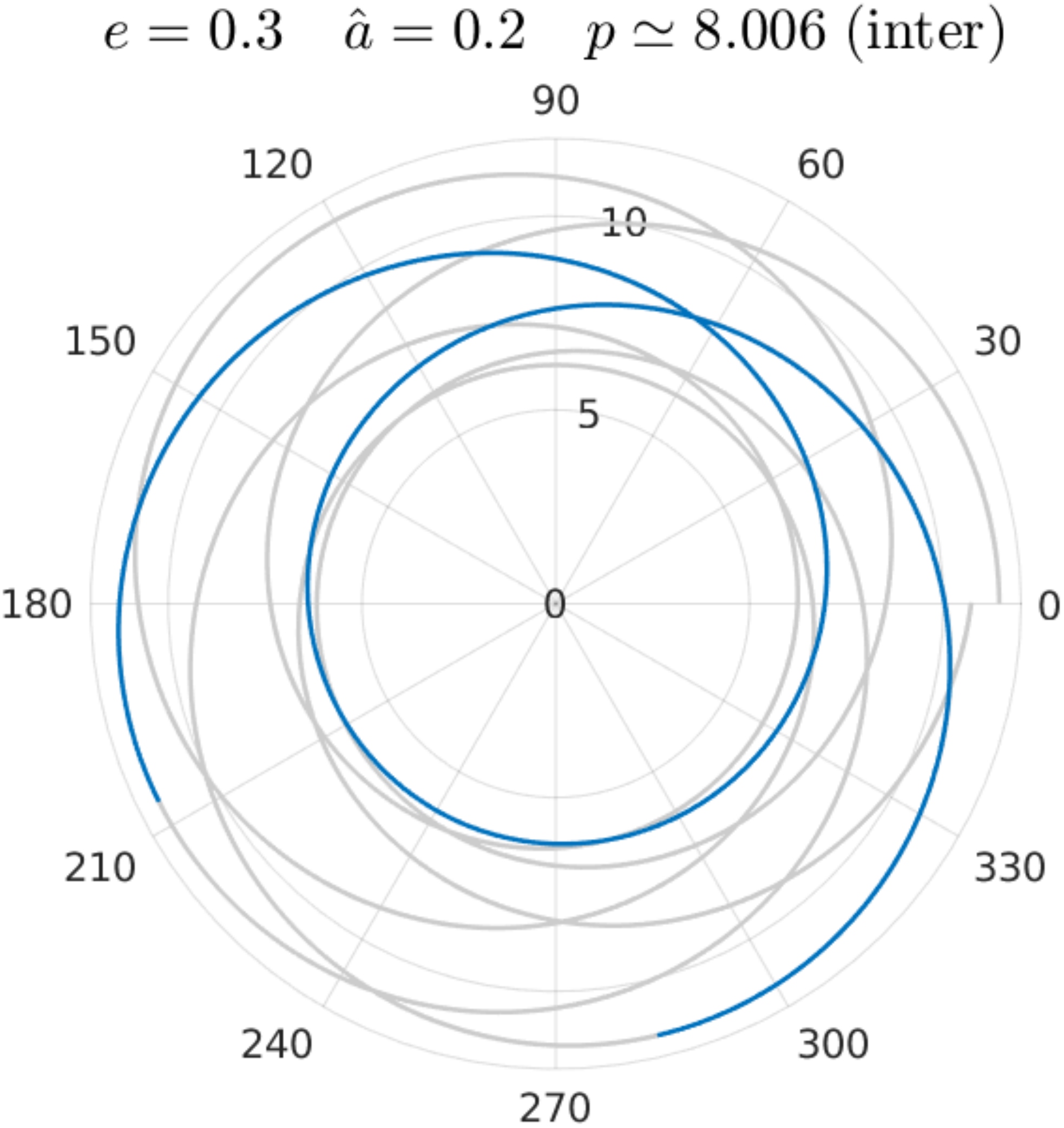}
  \quad
  \includegraphics[width=0.22\textwidth]{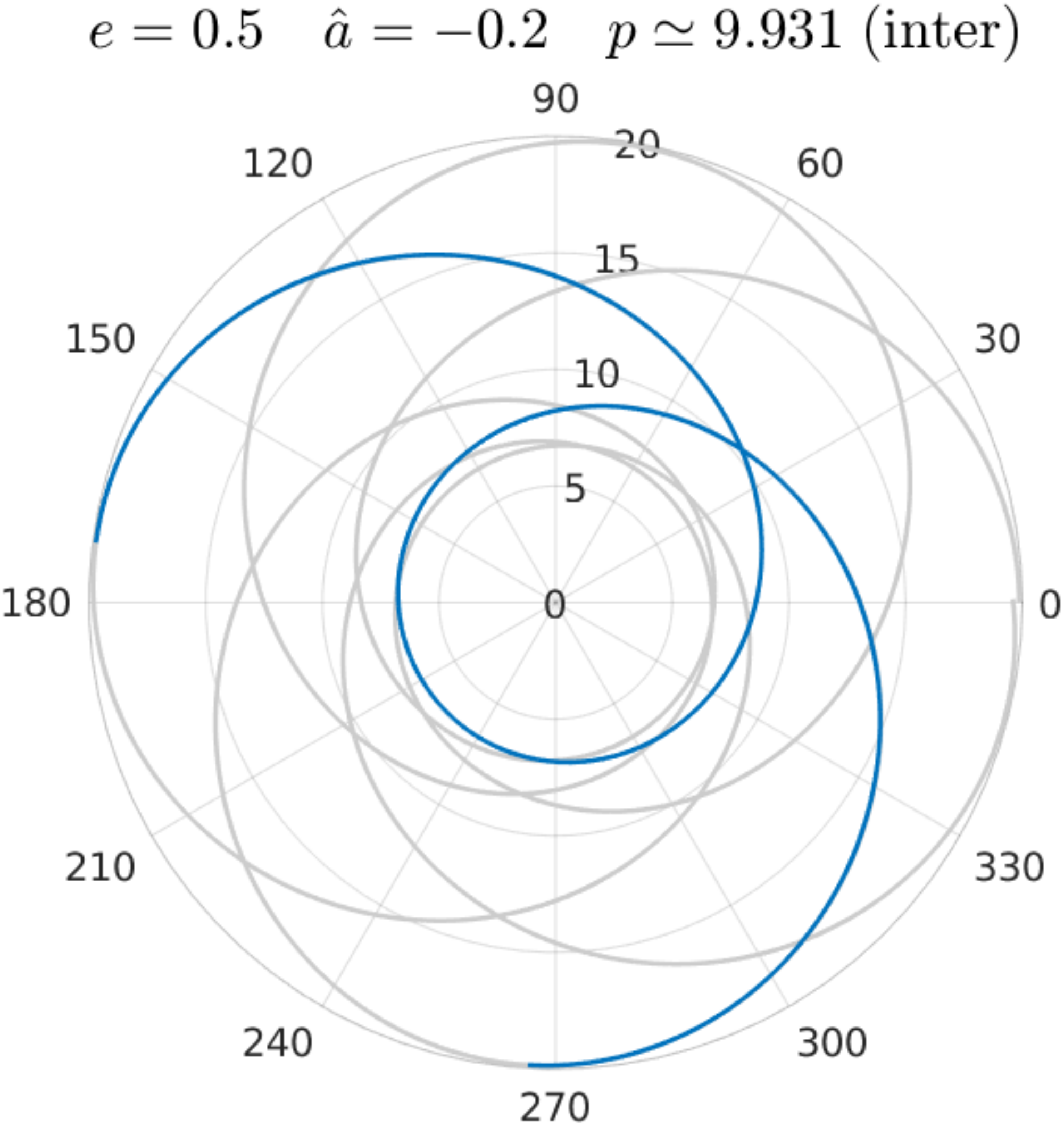} \\
  \caption{\label{fig:geo_dynamics_manymodes} Geodesic equatorial orbits for two different 
  ($e$, $\ha$, $p$) configurations. We have 
  highlighted one radial orbit for each configuration. The corresponding
  waveforms are shown in Fig.~\ref{fig:eobwave_manymodes1} and Fig.~\ref{fig:eobwave_manymodes2}.}
\end{figure}
\begin{figure*}
  \center  
  \includegraphics[width=0.24\textwidth,height=3cm]{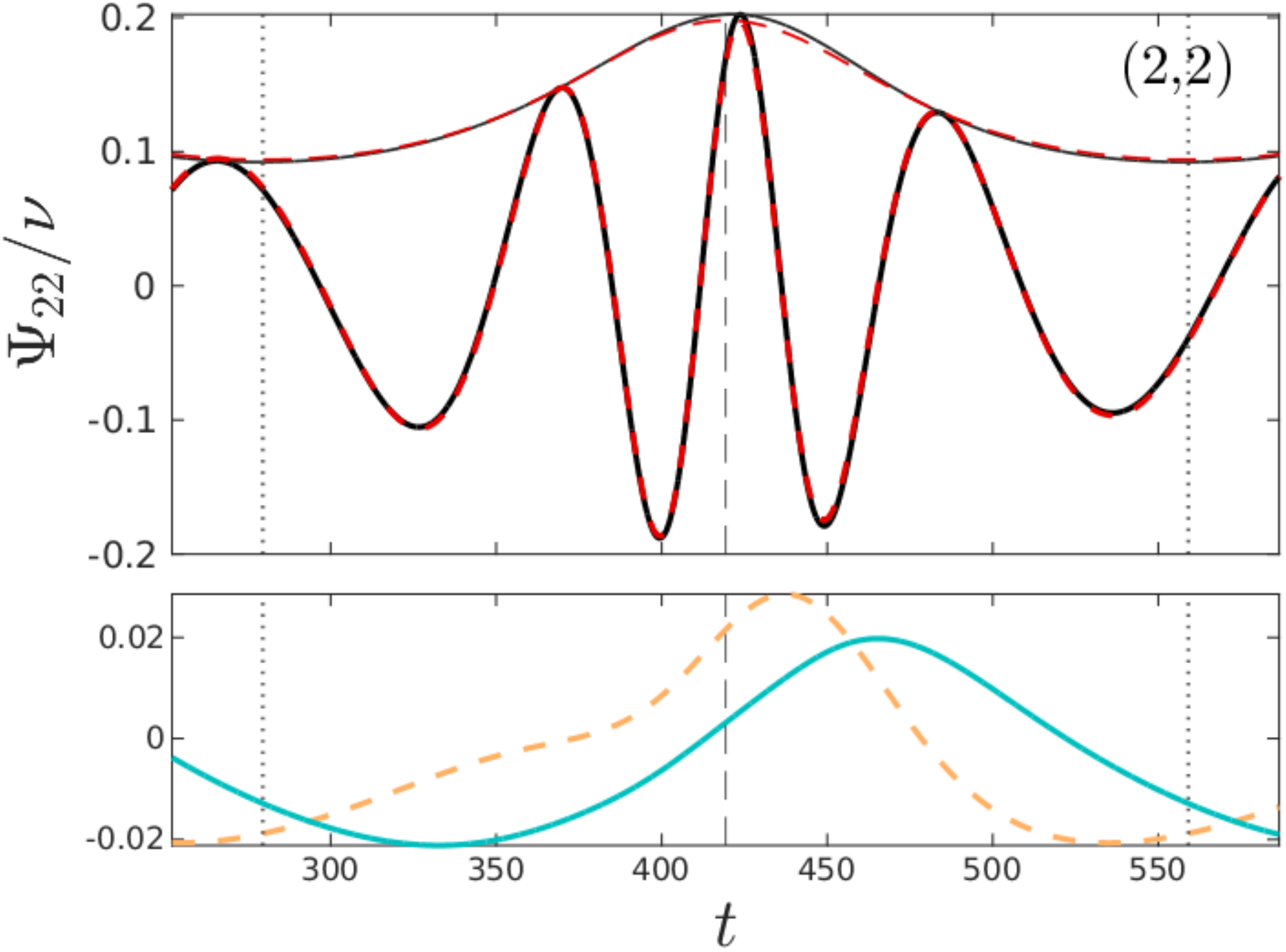}
  \includegraphics[width=0.24\textwidth,height=3cm]{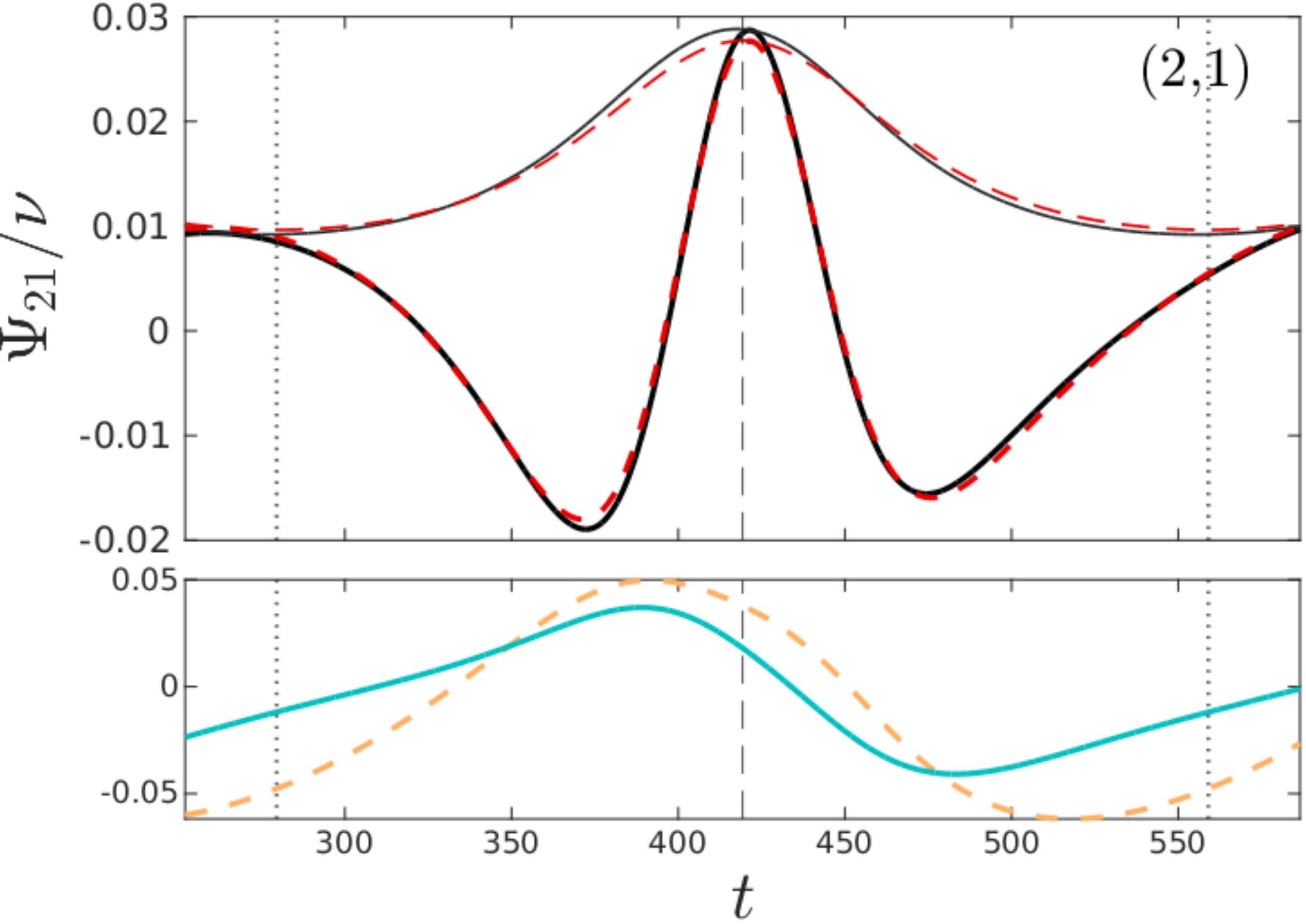}
  \includegraphics[width=0.24\textwidth,height=3cm]{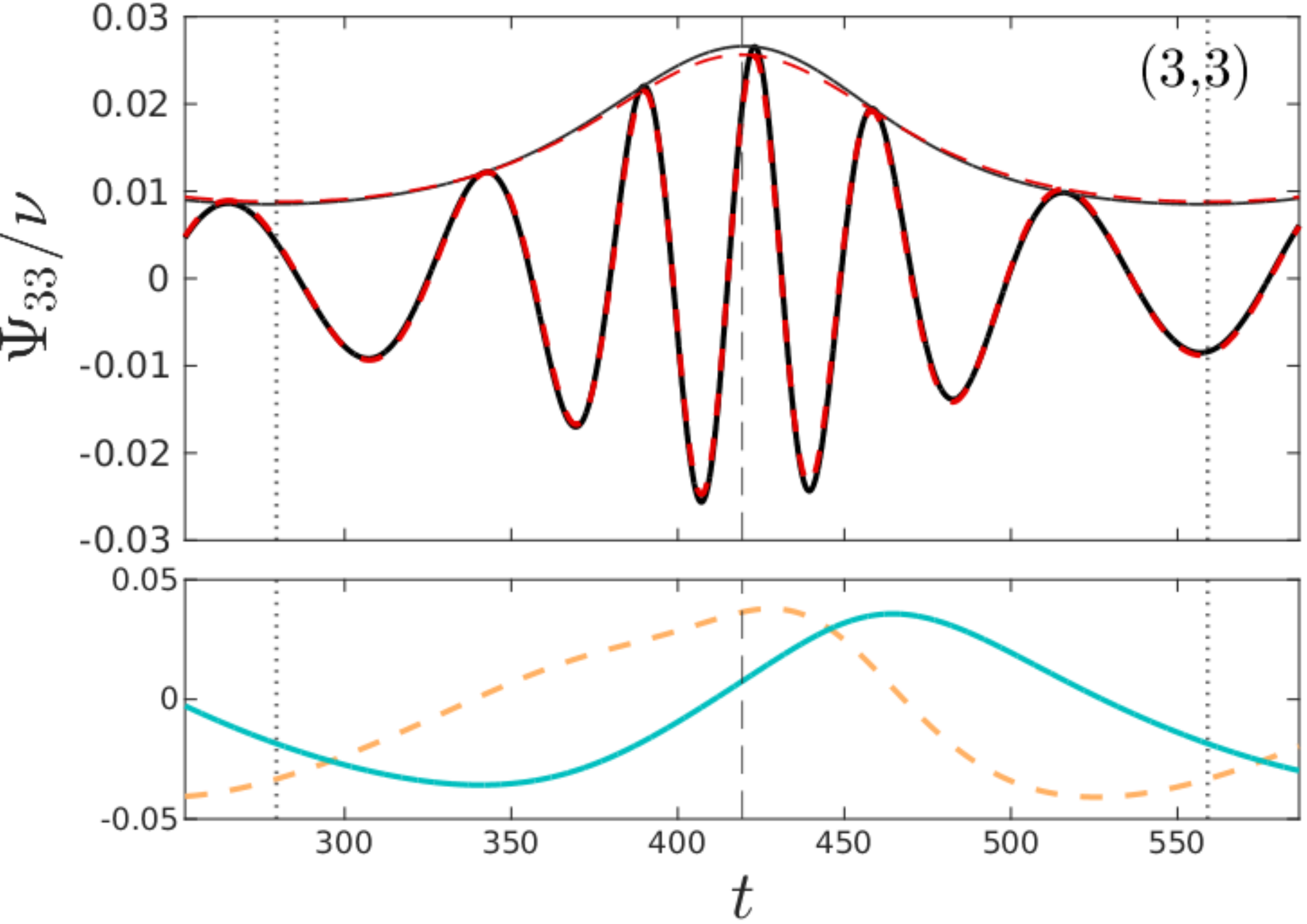}
  \includegraphics[width=0.24\textwidth,height=3.1cm]{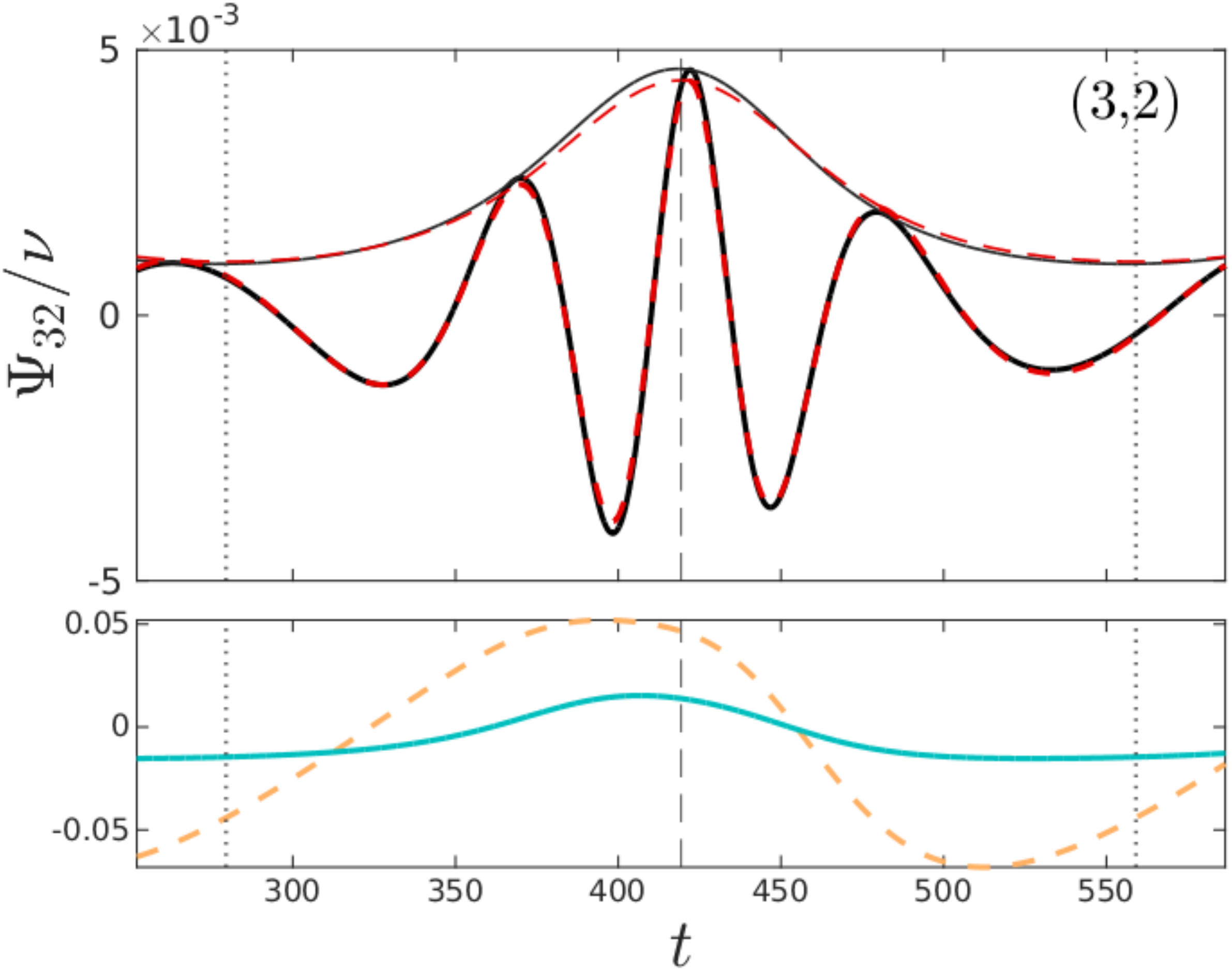} \\
  
  \includegraphics[width=0.24\textwidth,height=3cm]{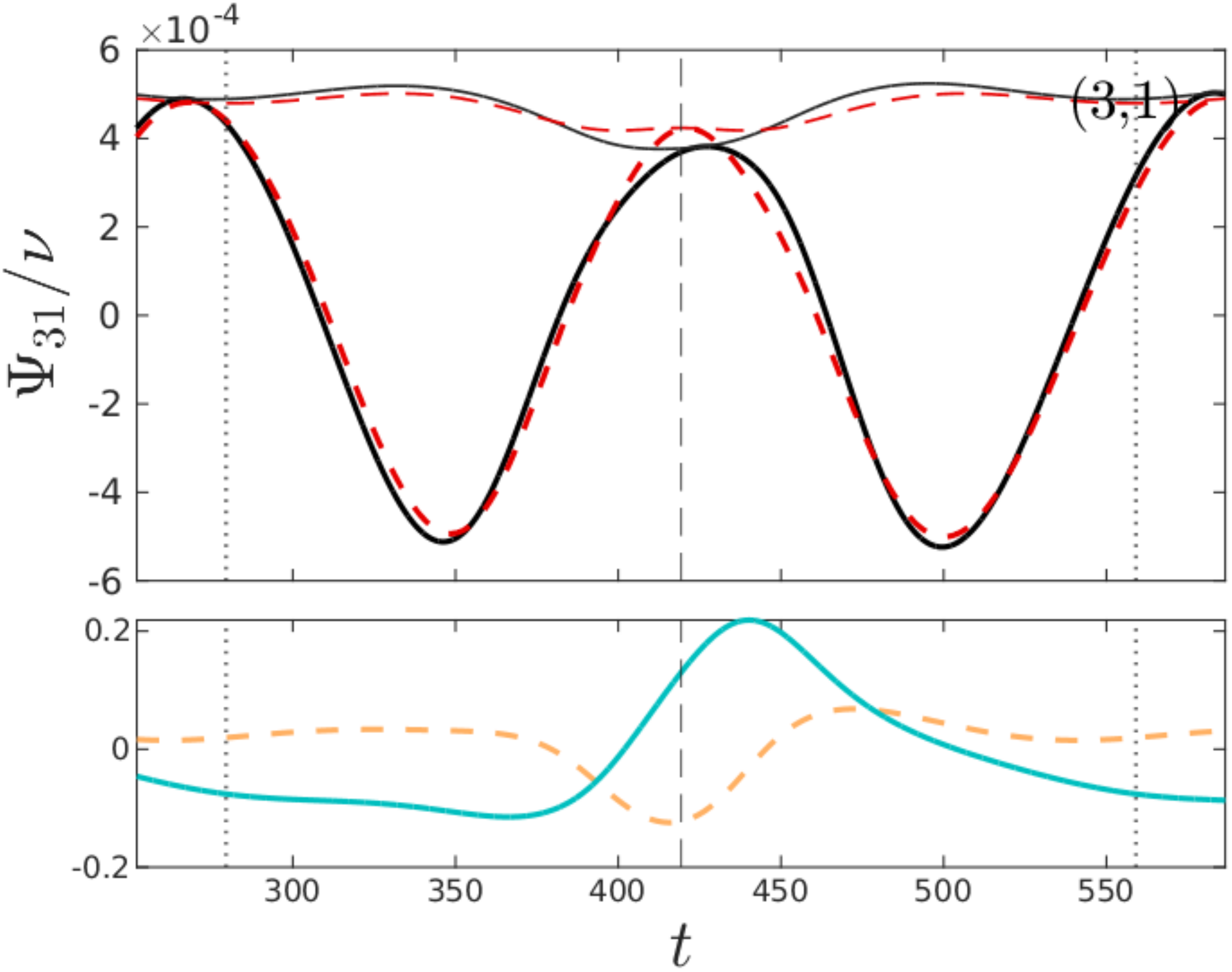} 
  \includegraphics[width=0.24\textwidth,height=3cm]{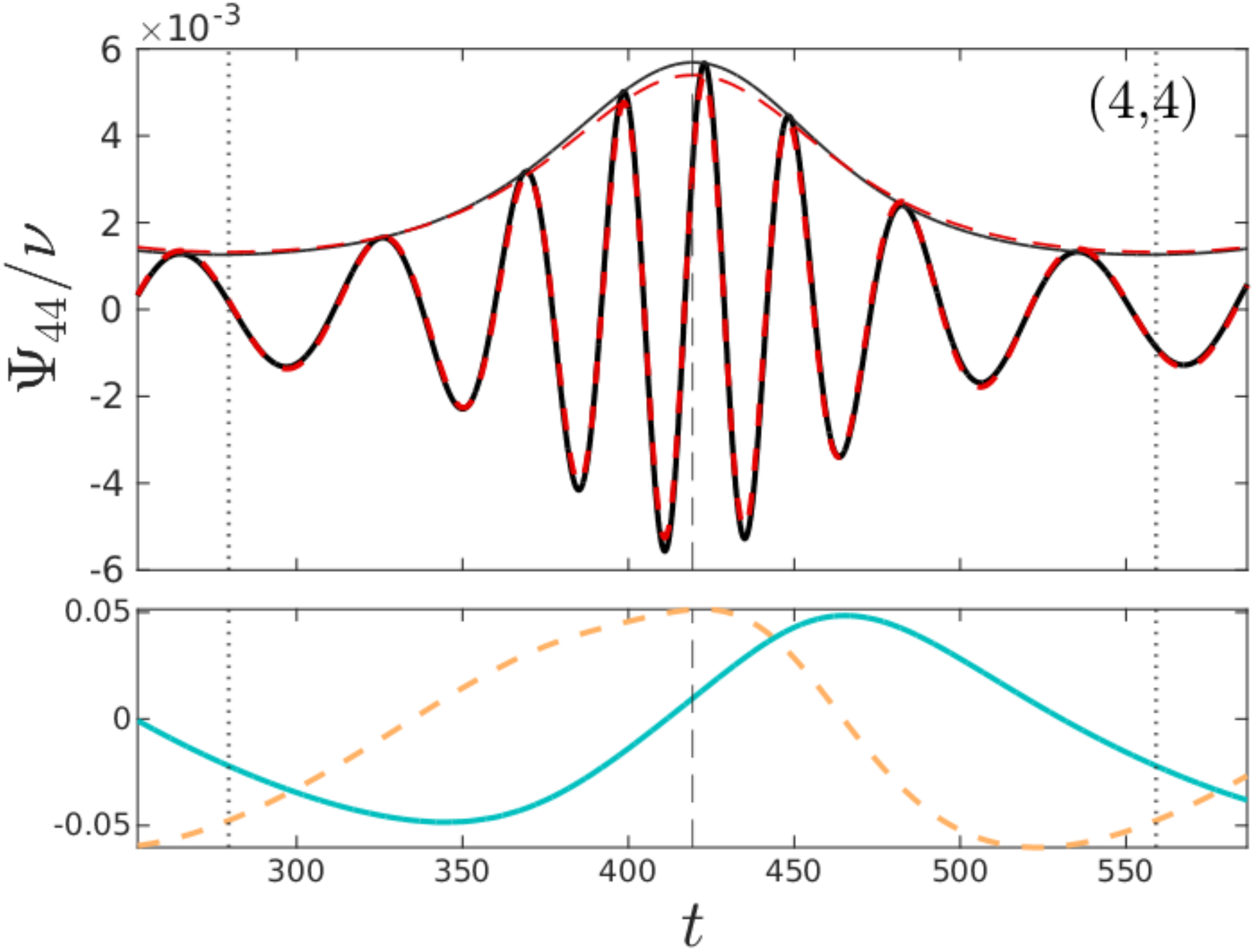} 
  \includegraphics[width=0.24\textwidth,height=3cm]{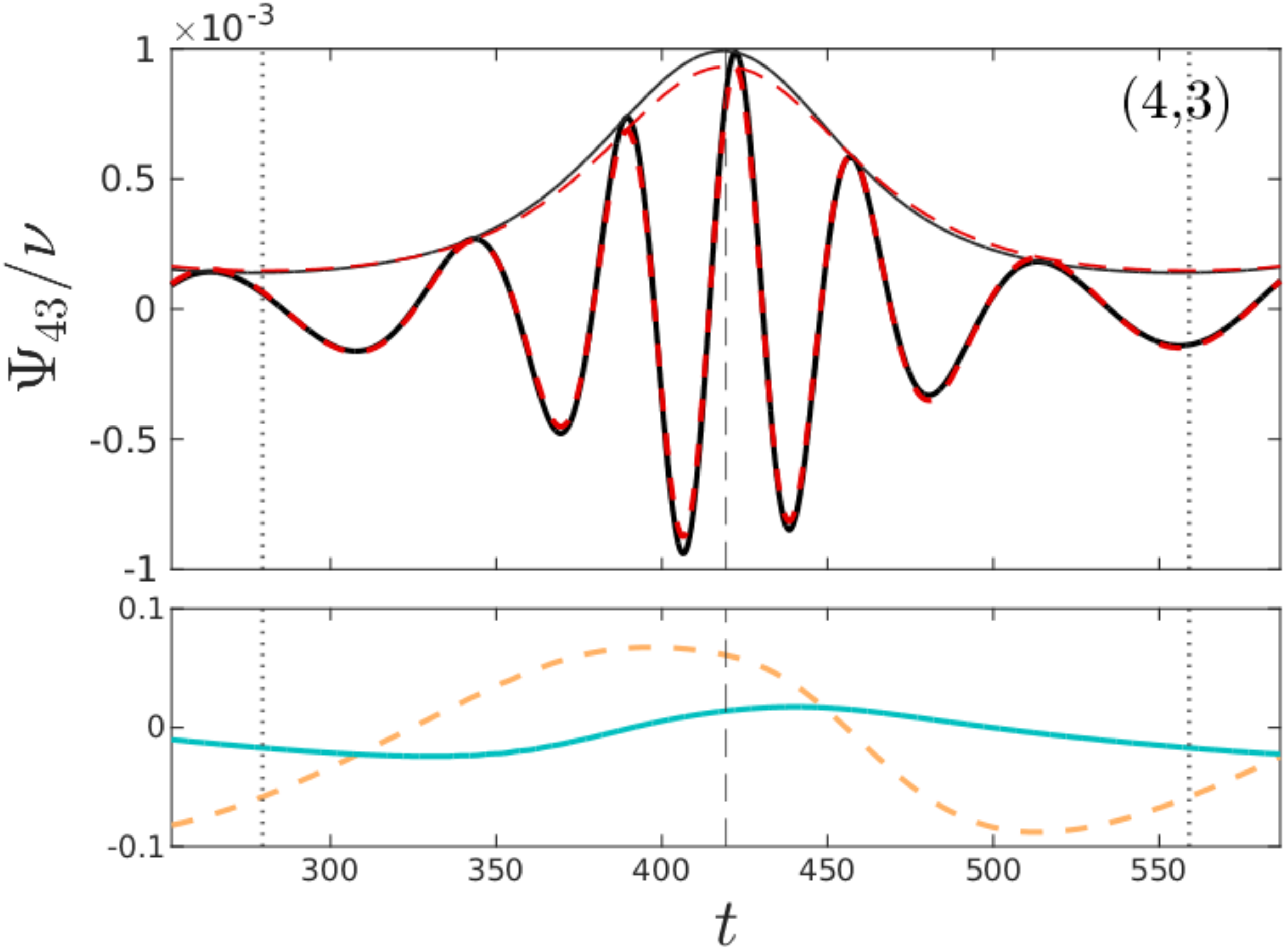} 
  \includegraphics[width=0.24\textwidth,height=3cm]{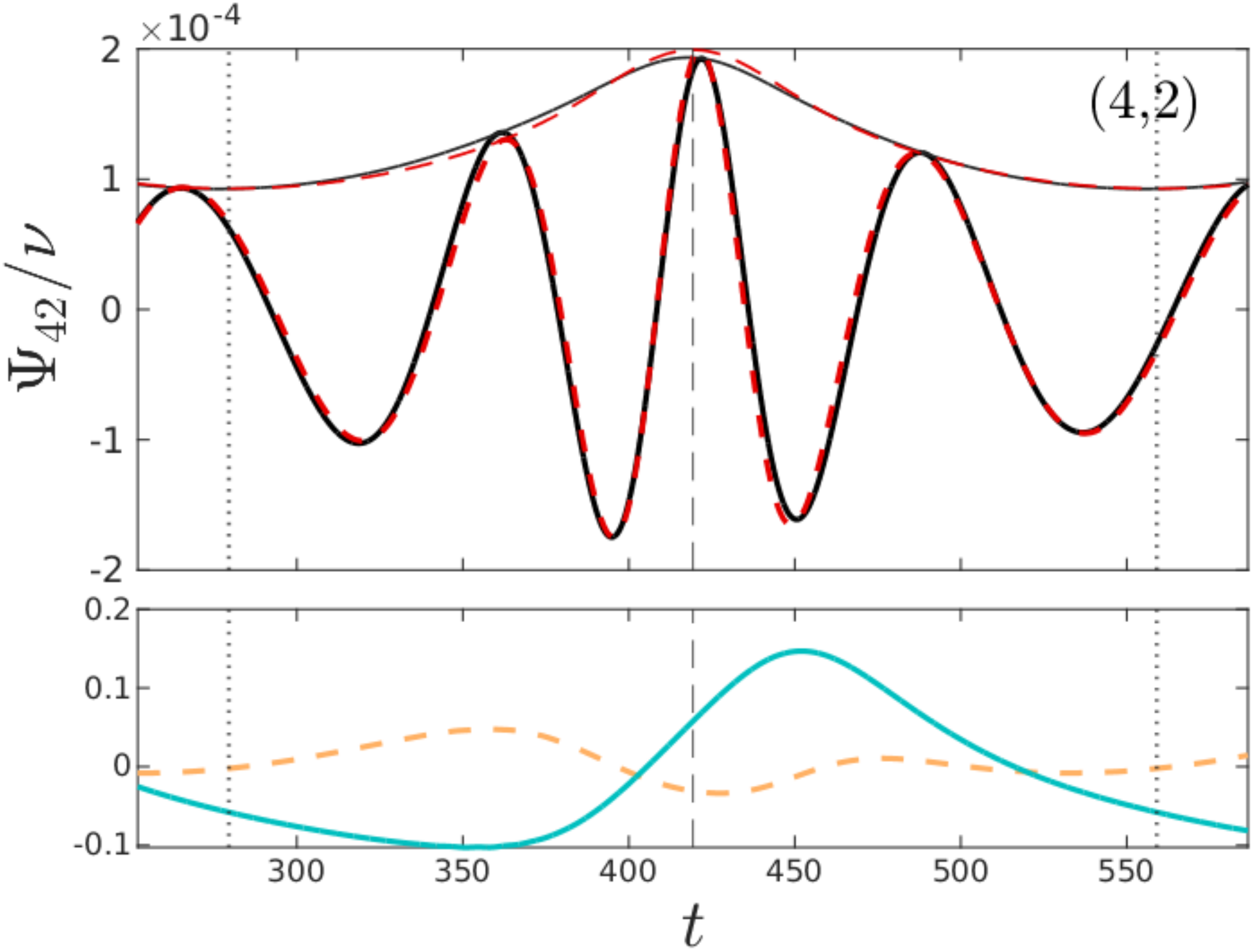} \\
  
  \includegraphics[width=0.24\textwidth,height=3cm]{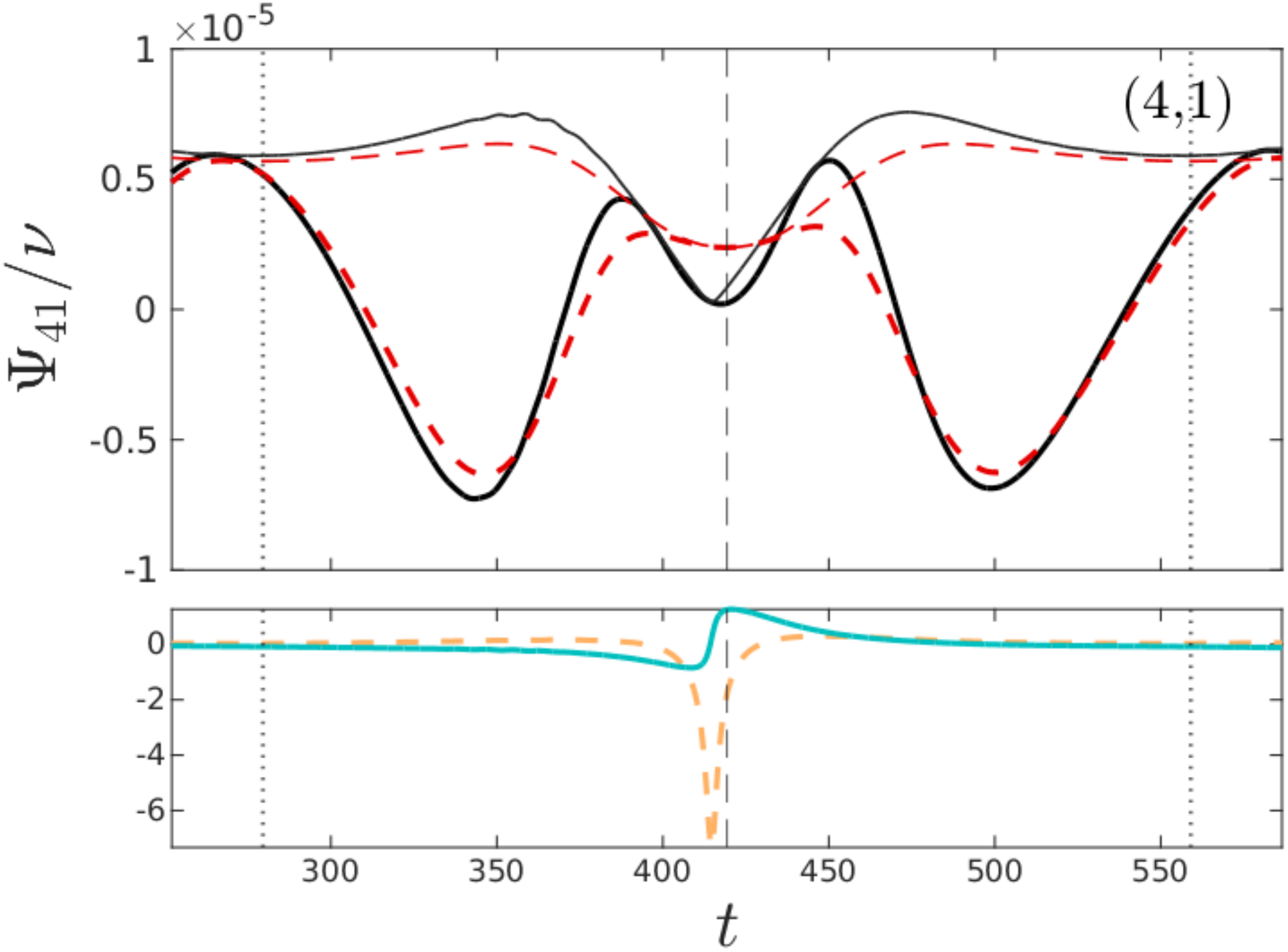}
  \includegraphics[width=0.24\textwidth,height=3cm]{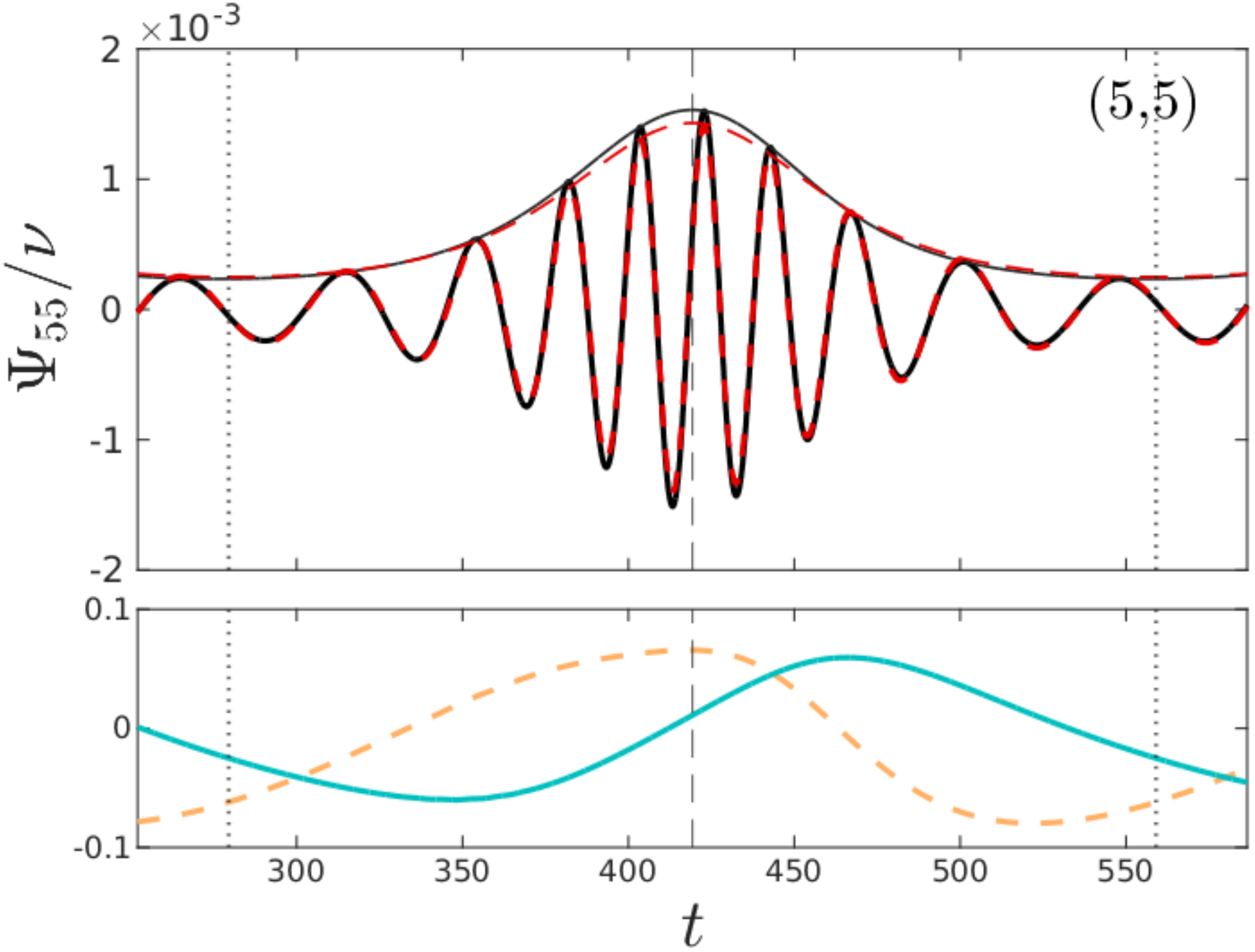}
  \includegraphics[width=0.24\textwidth,height=3cm]{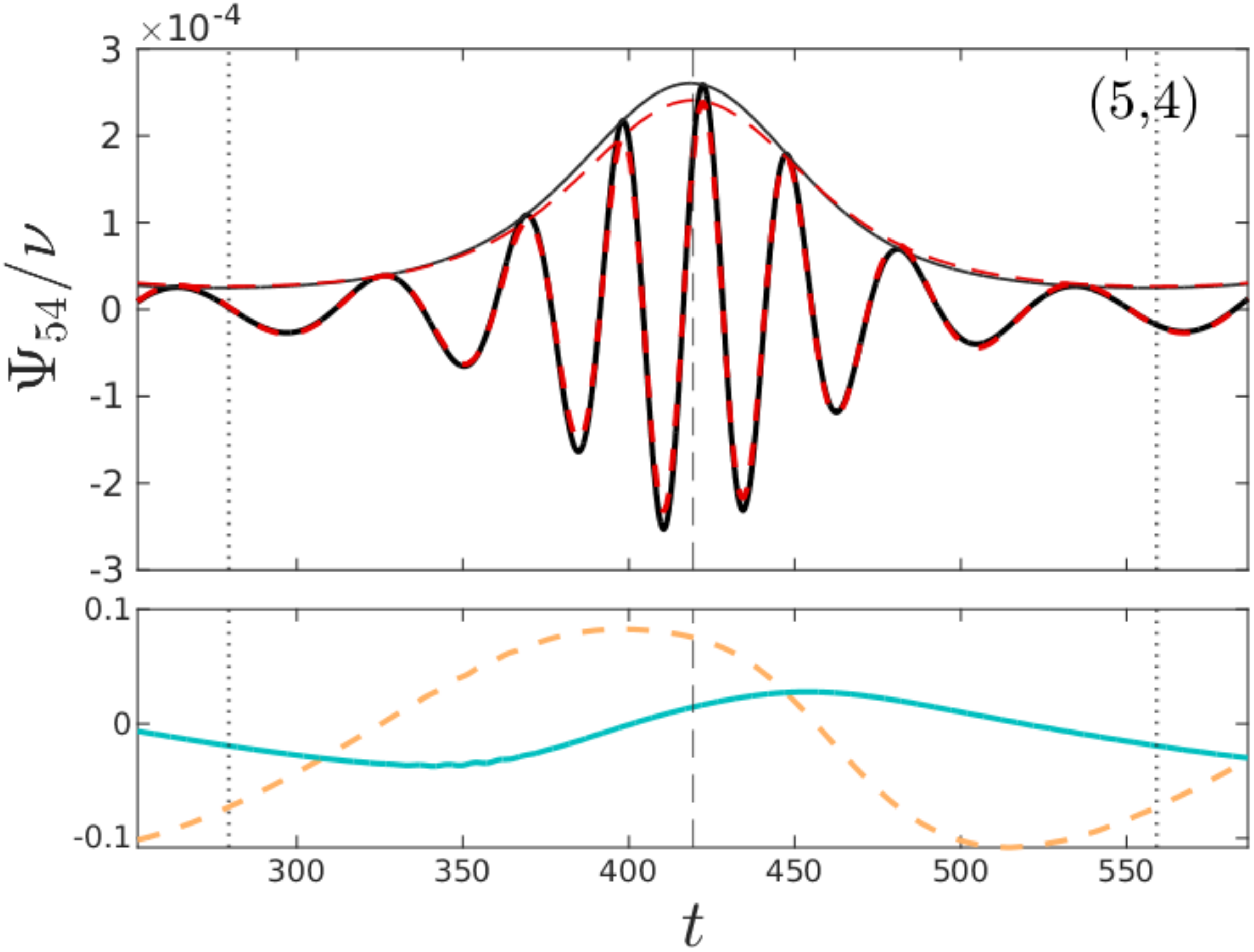}
  \includegraphics[width=0.24\textwidth,height=3cm]{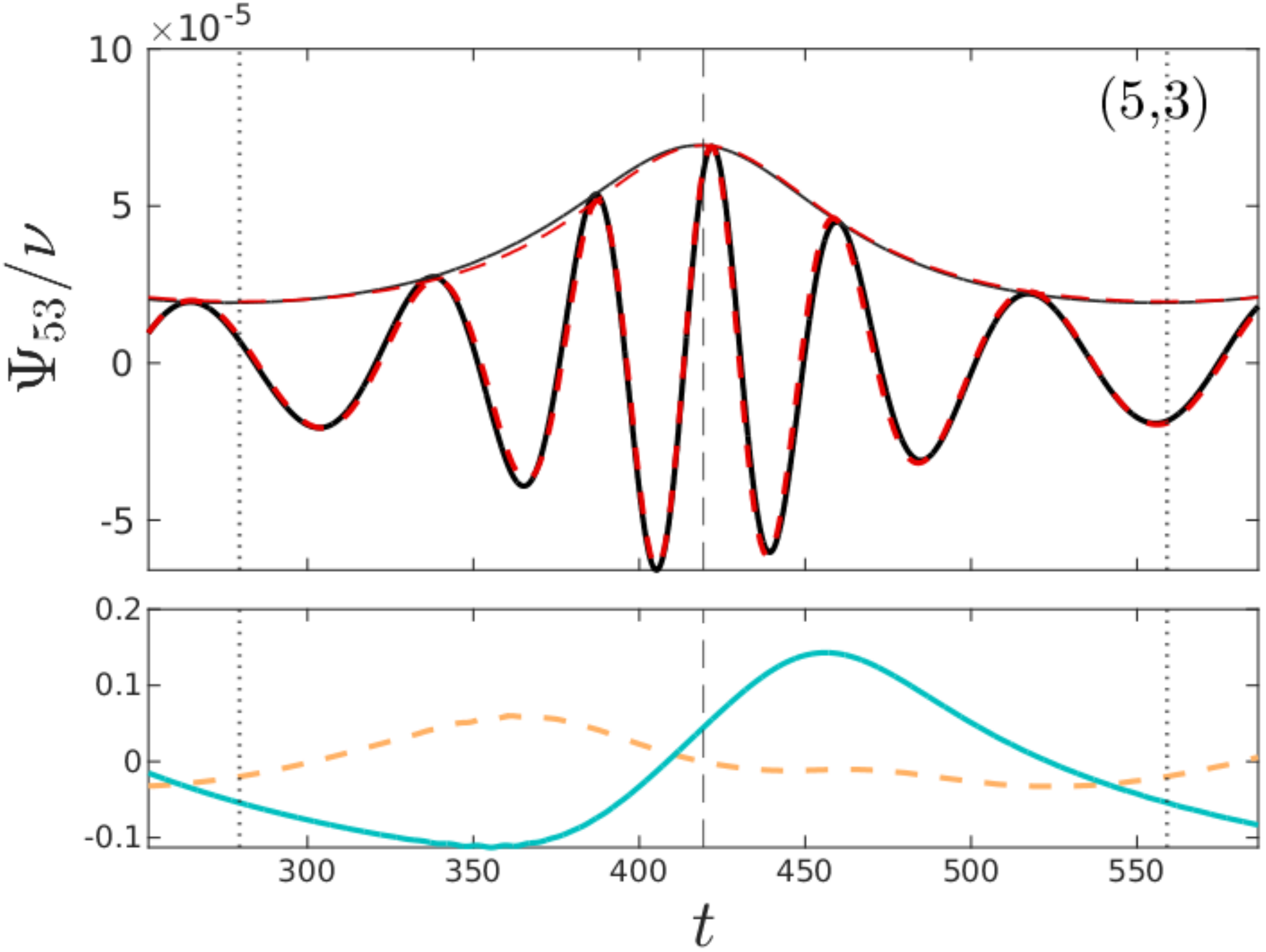} \\
  
  \includegraphics[width=0.24\textwidth,height=3cm]{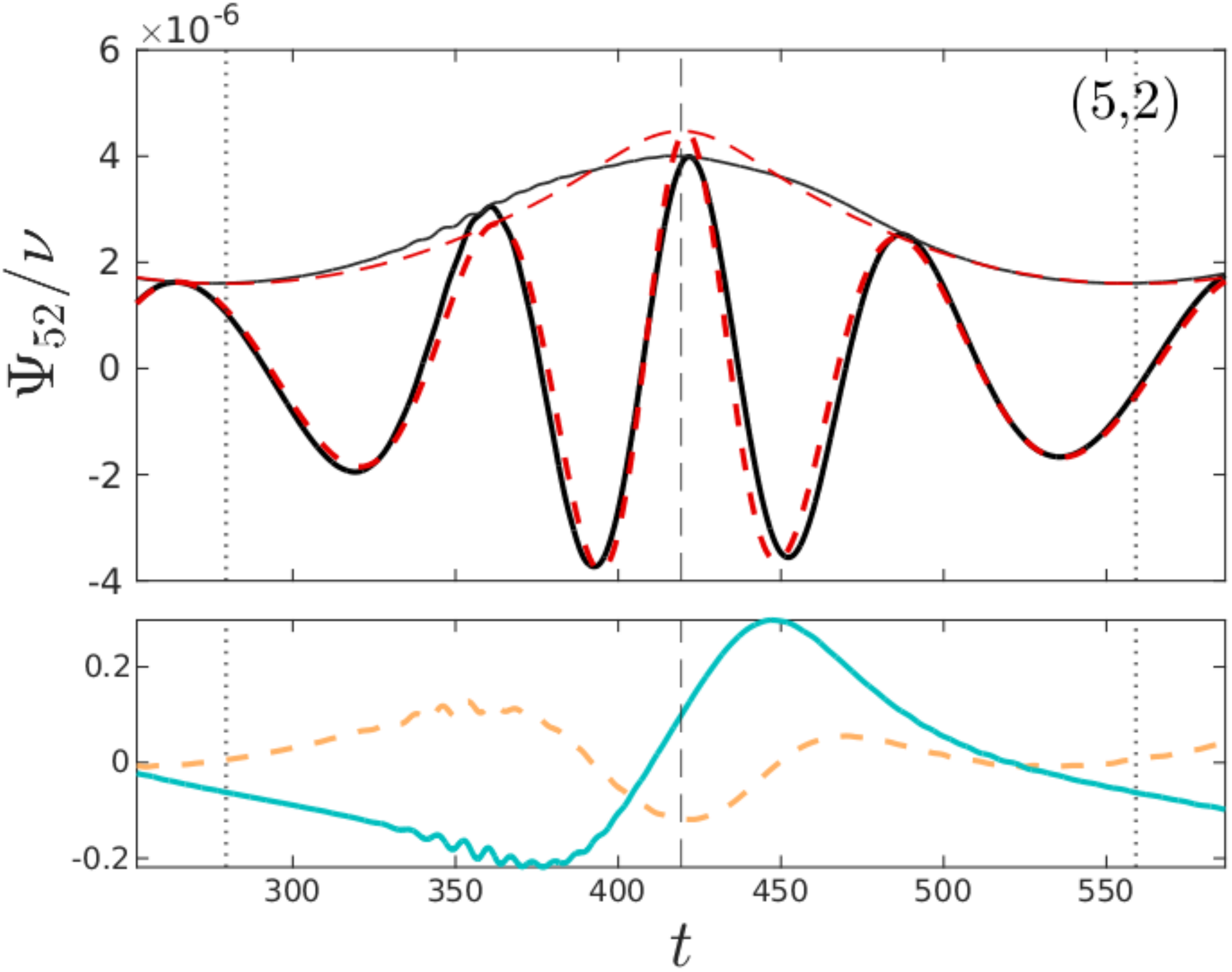}
  \includegraphics[width=0.24\textwidth,height=3cm]{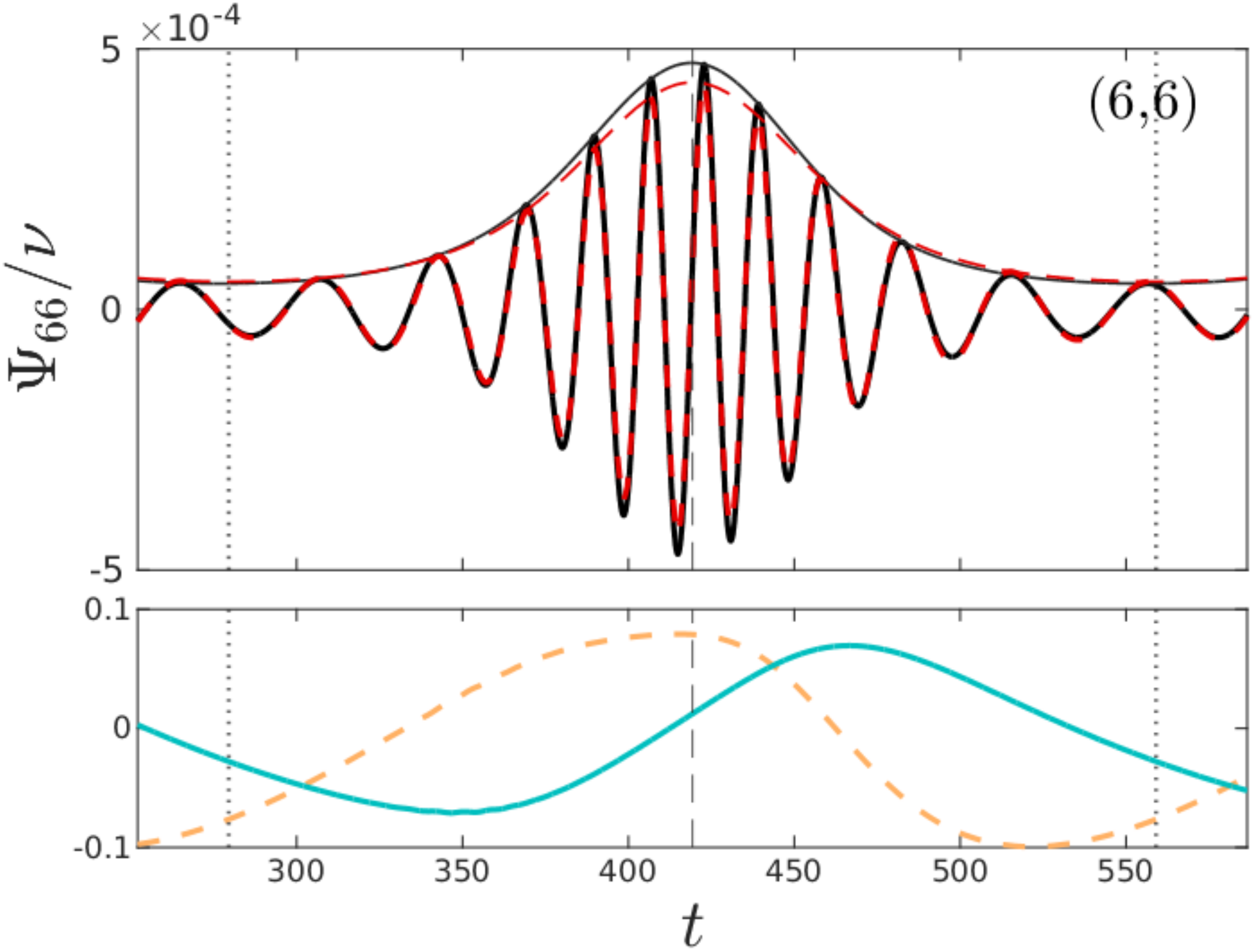}
  \includegraphics[width=0.24\textwidth,height=3cm]{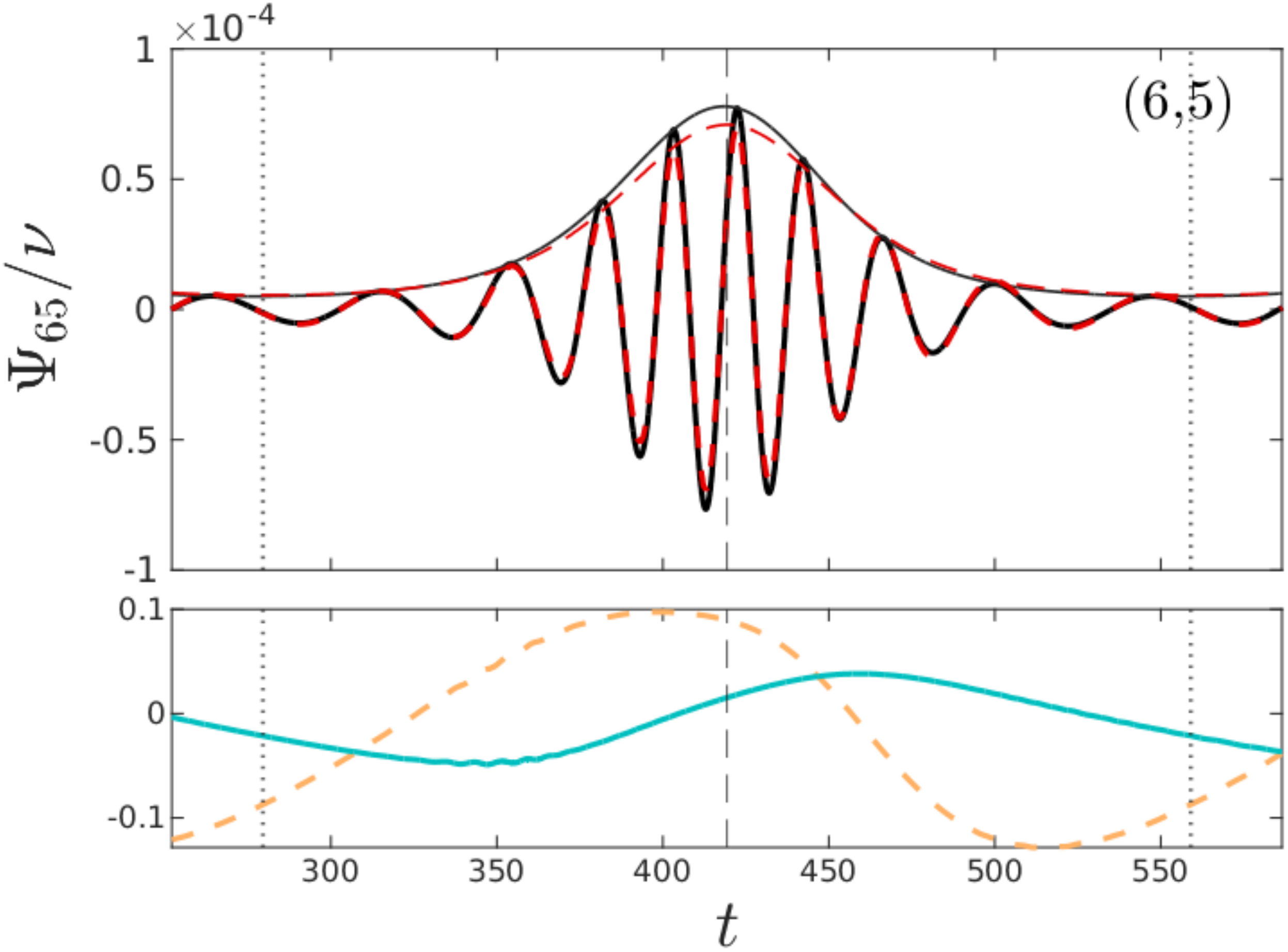}
  \includegraphics[width=0.24\textwidth,height=3cm]{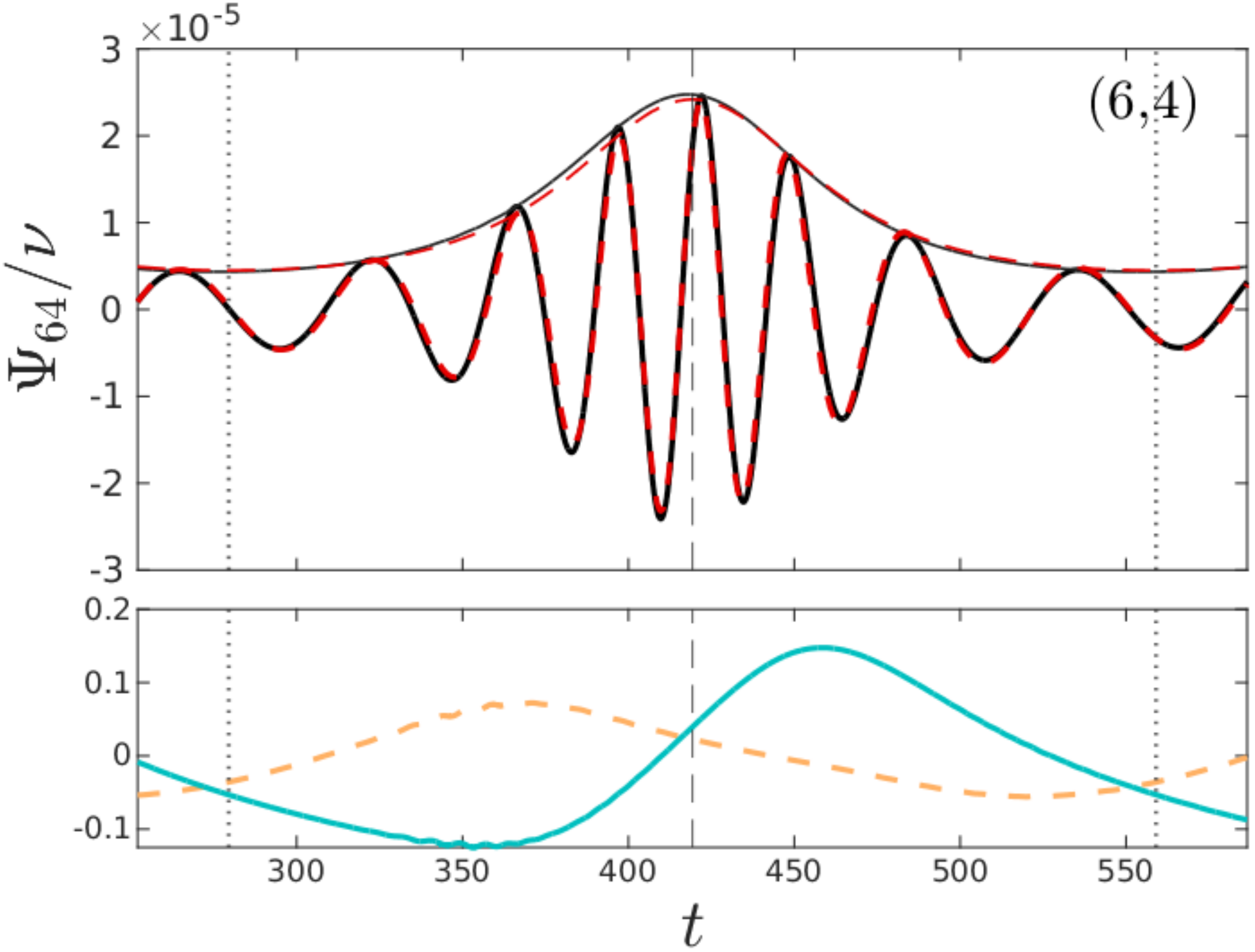} \\
  
  \includegraphics[width=0.24\textwidth,height=3cm]{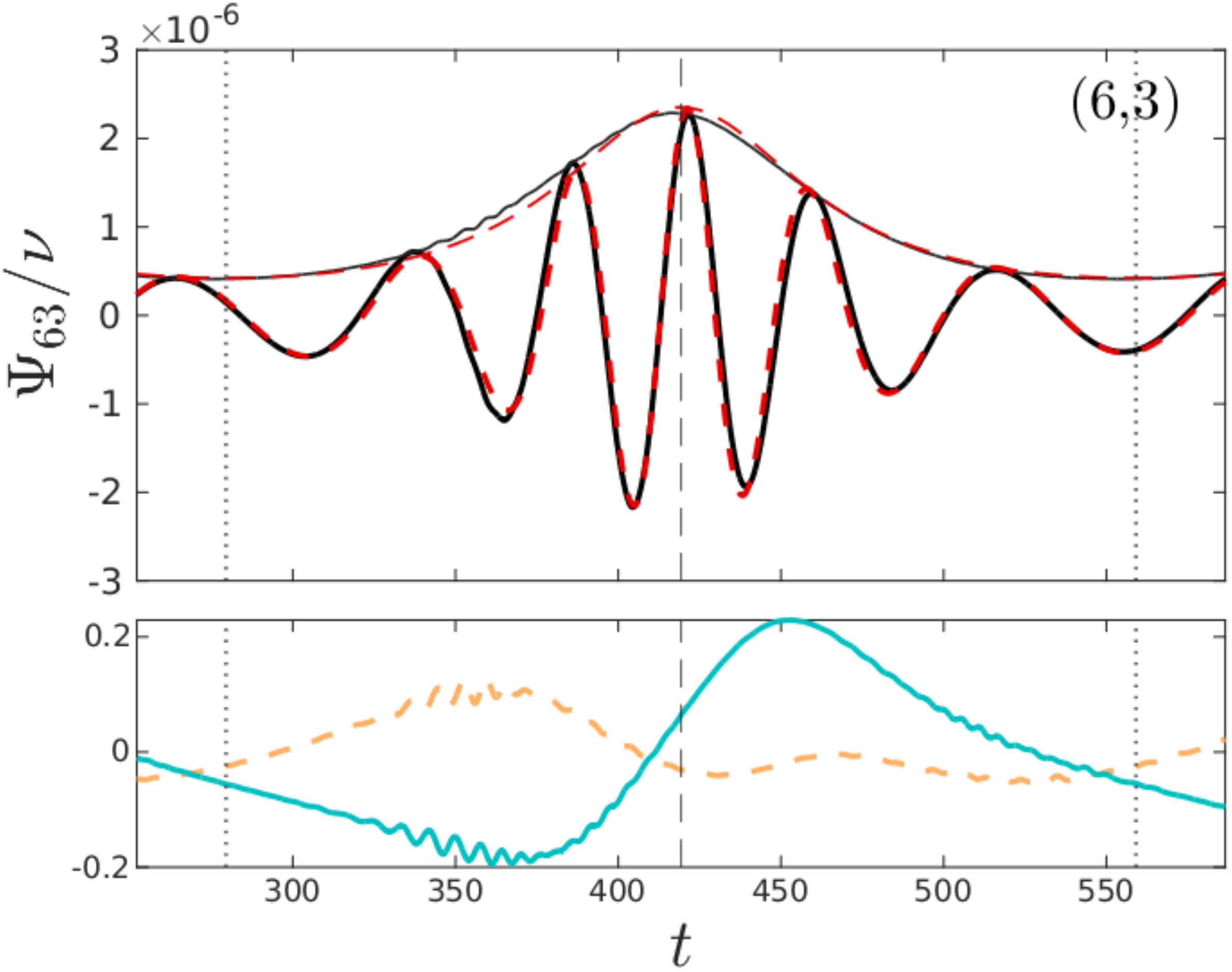}
  \includegraphics[width=0.24\textwidth,height=3cm]{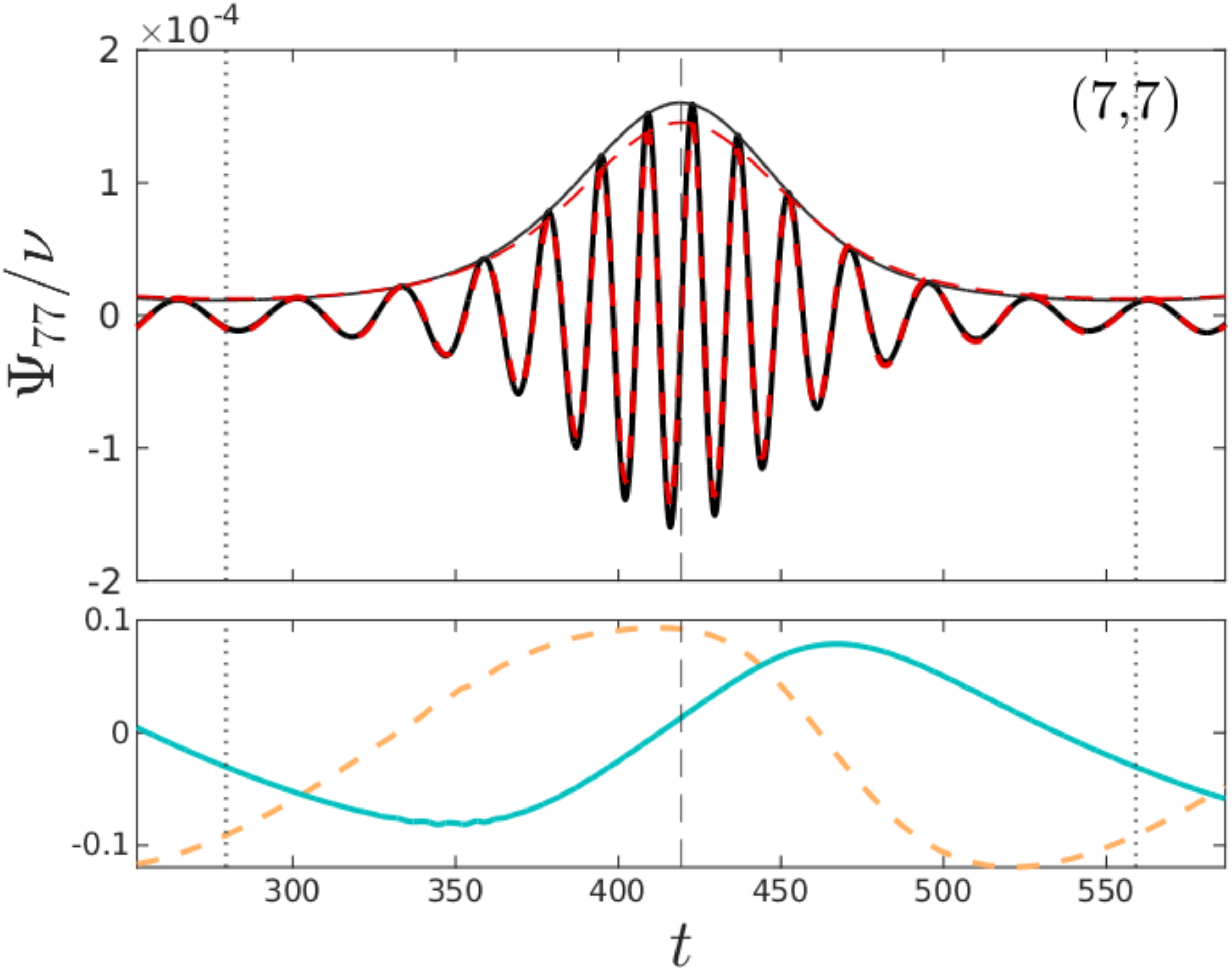}
  \includegraphics[width=0.24\textwidth,height=3cm]{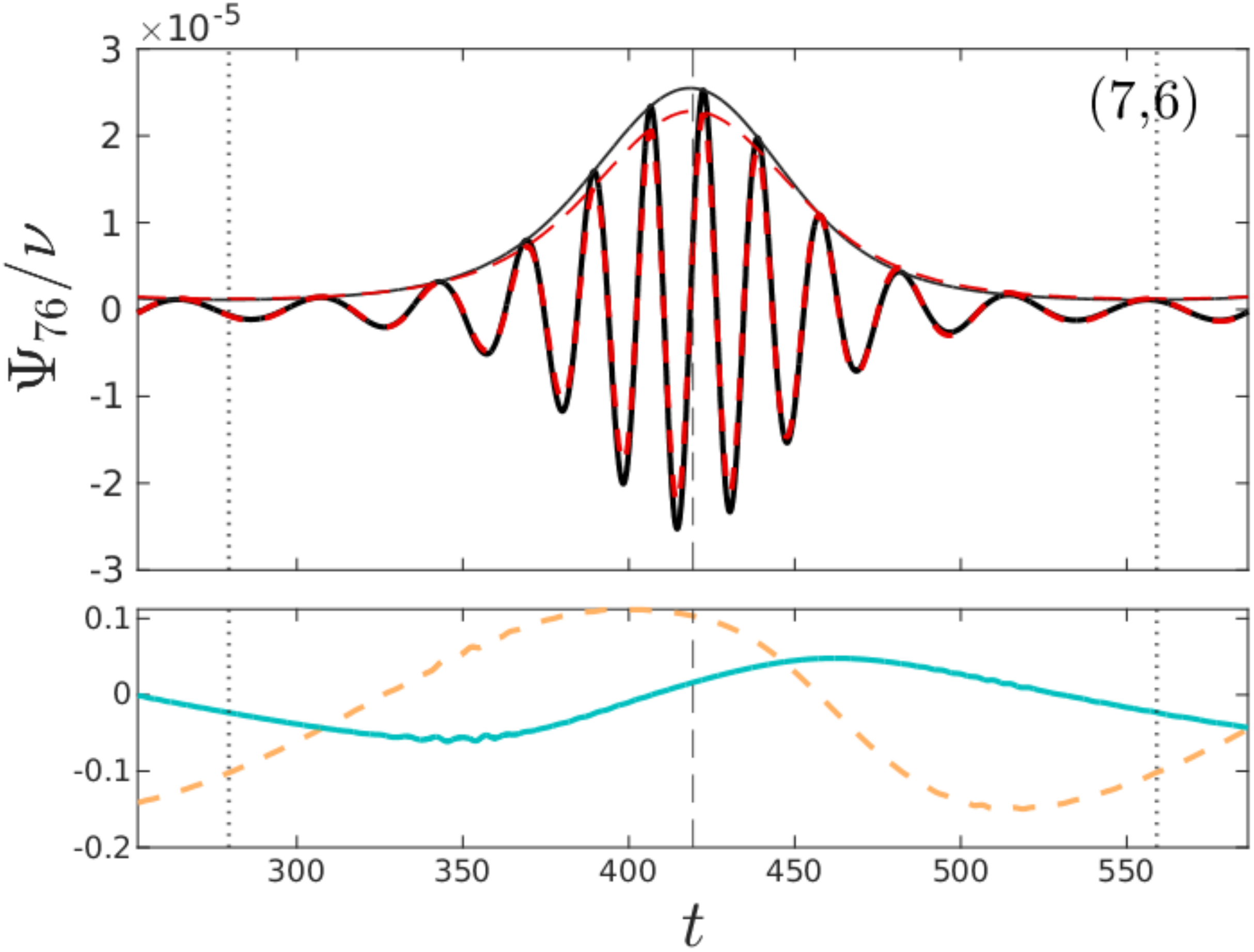}
  \includegraphics[width=0.24\textwidth,height=3cm]{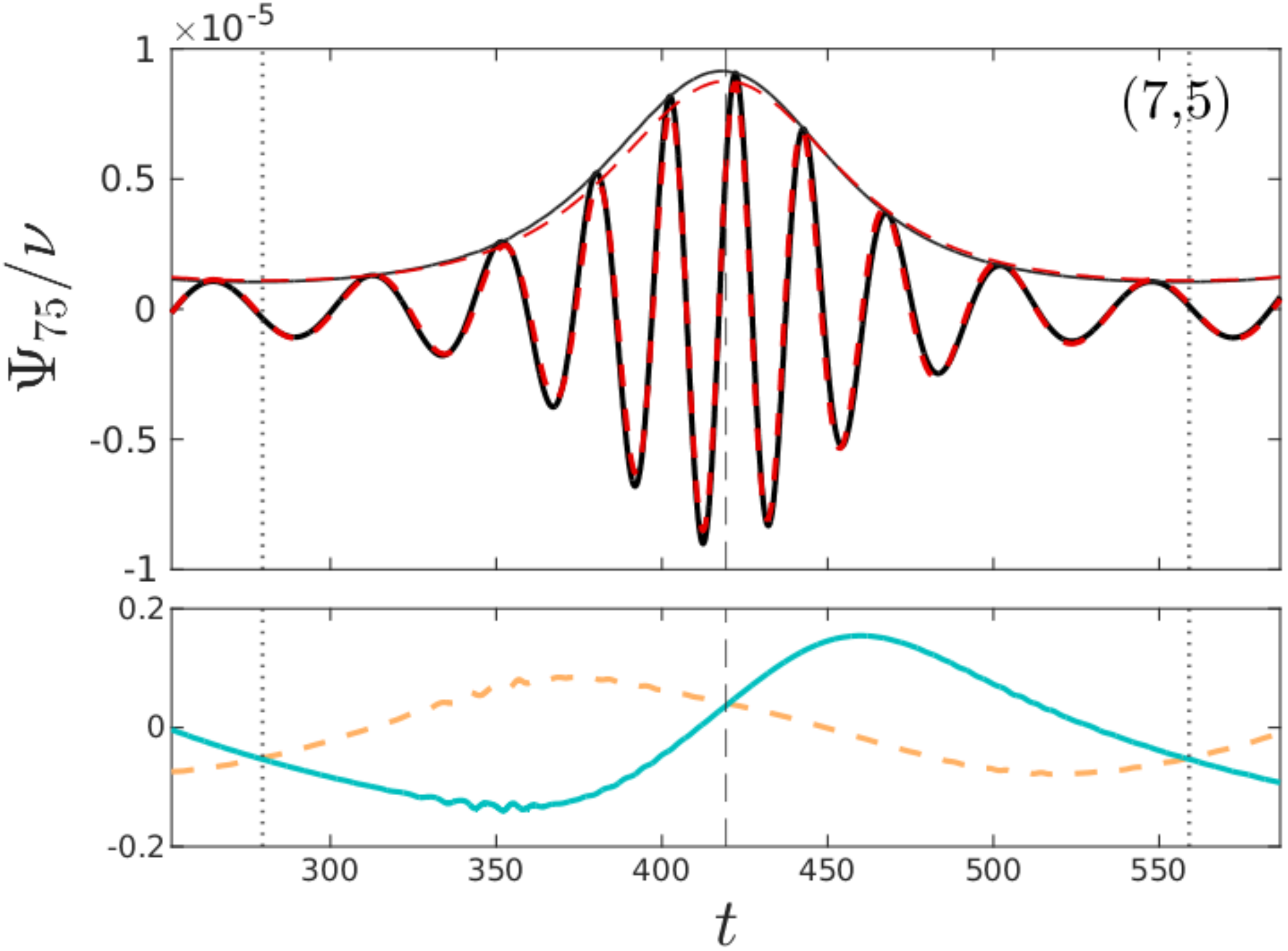} \\
  
  \includegraphics[width=0.24\textwidth,height=3cm]{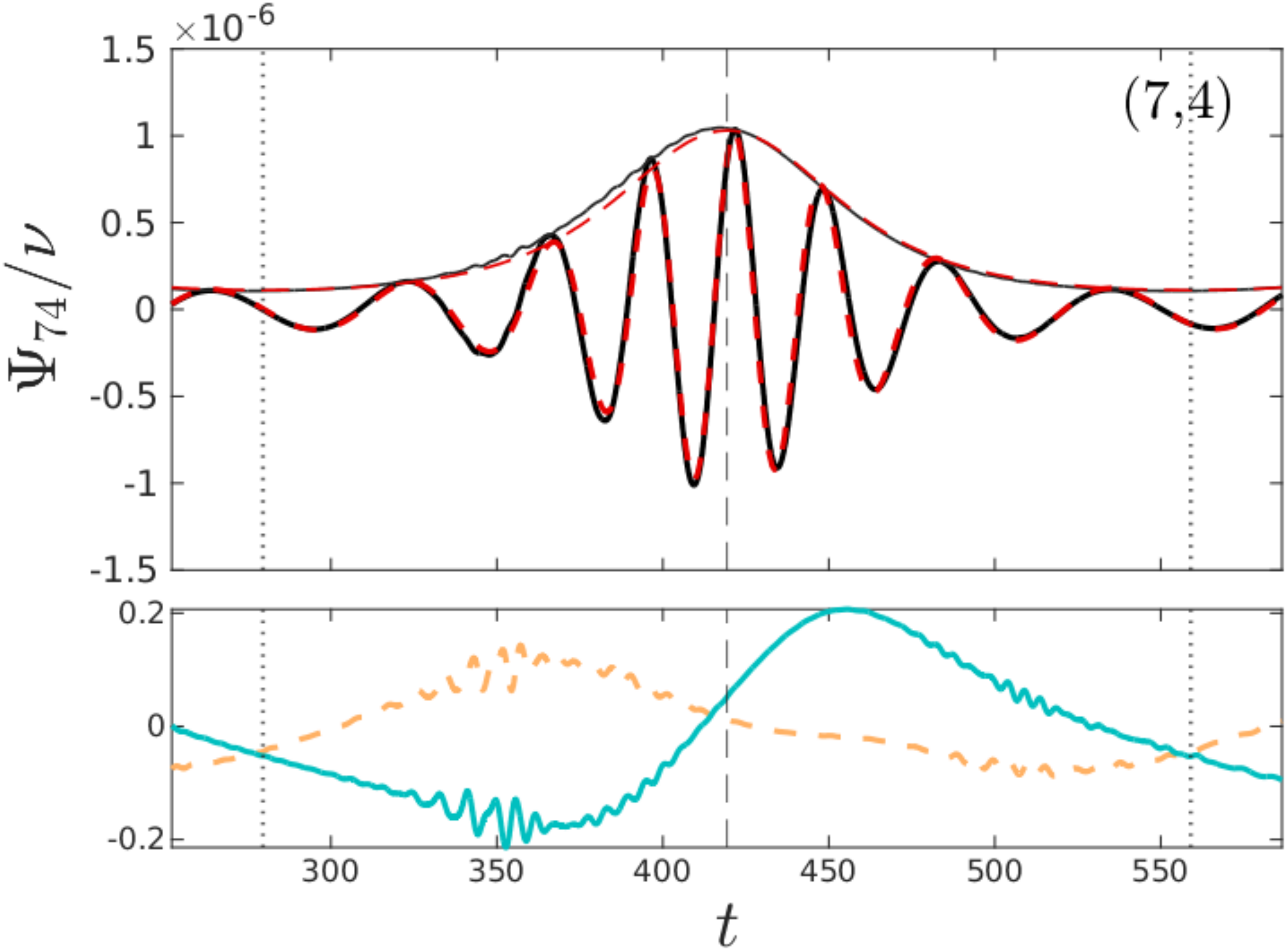}
  \includegraphics[width=0.24\textwidth,height=3cm]{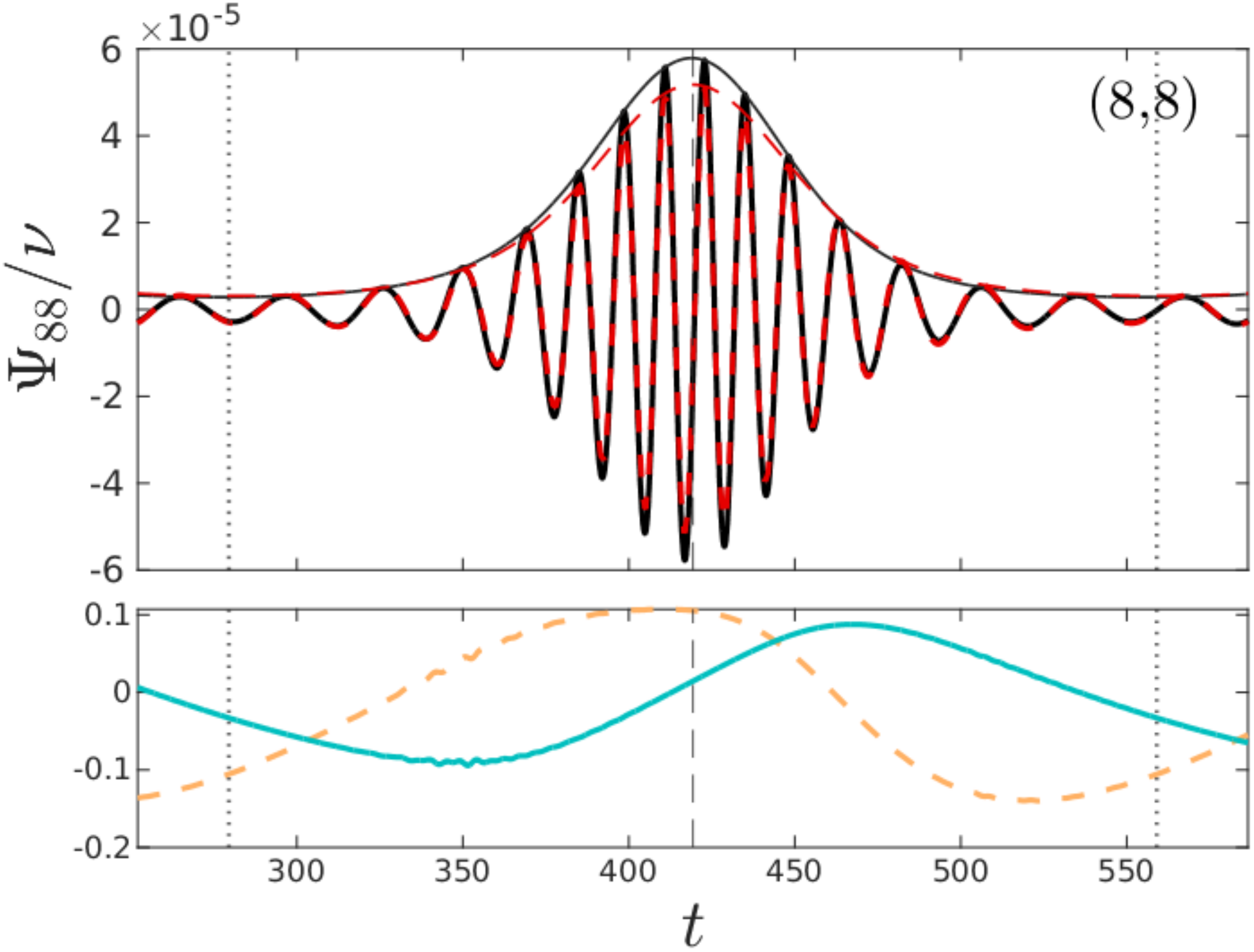}
  \includegraphics[width=0.24\textwidth,height=3cm]{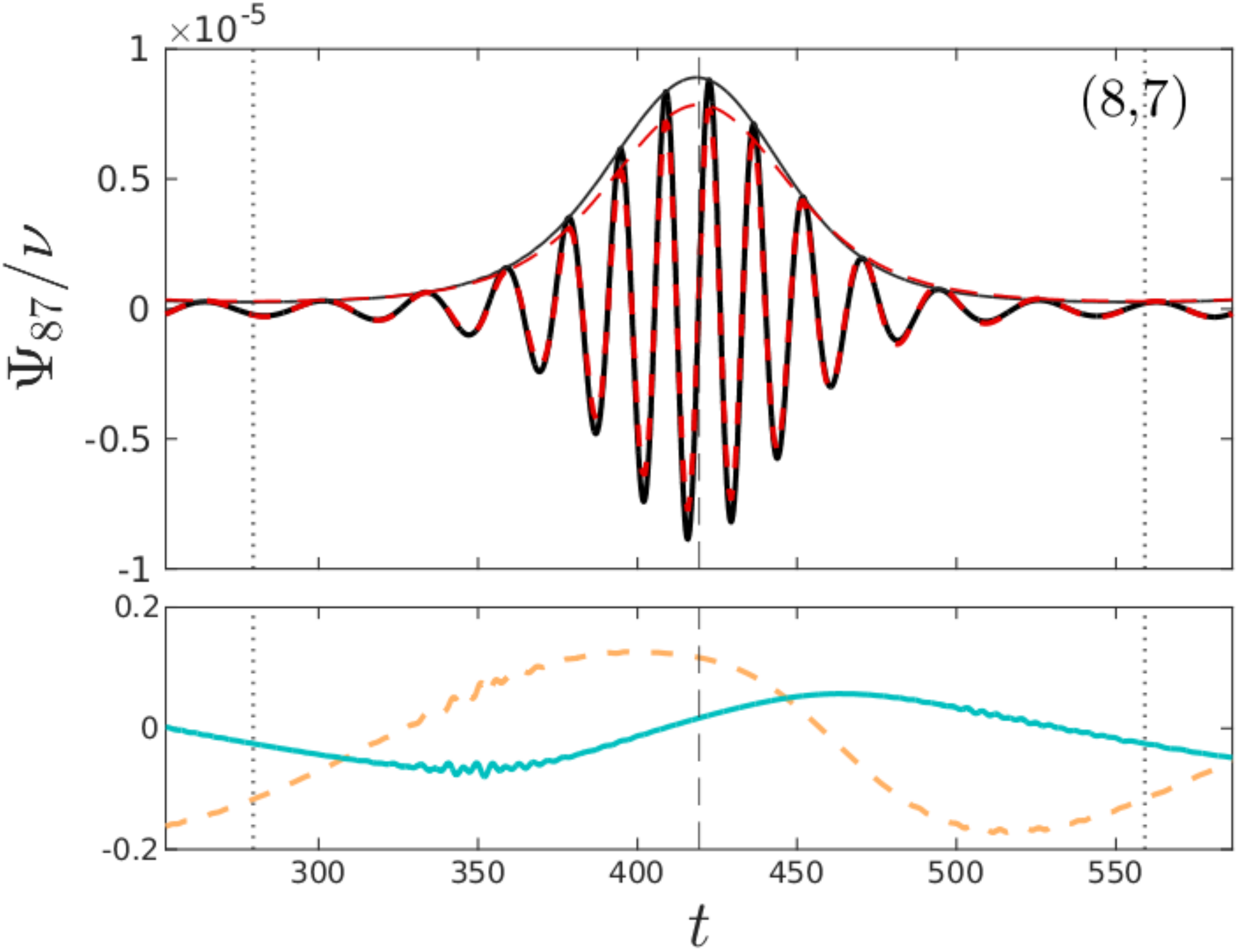}
  \includegraphics[width=0.24\textwidth,height=3cm]{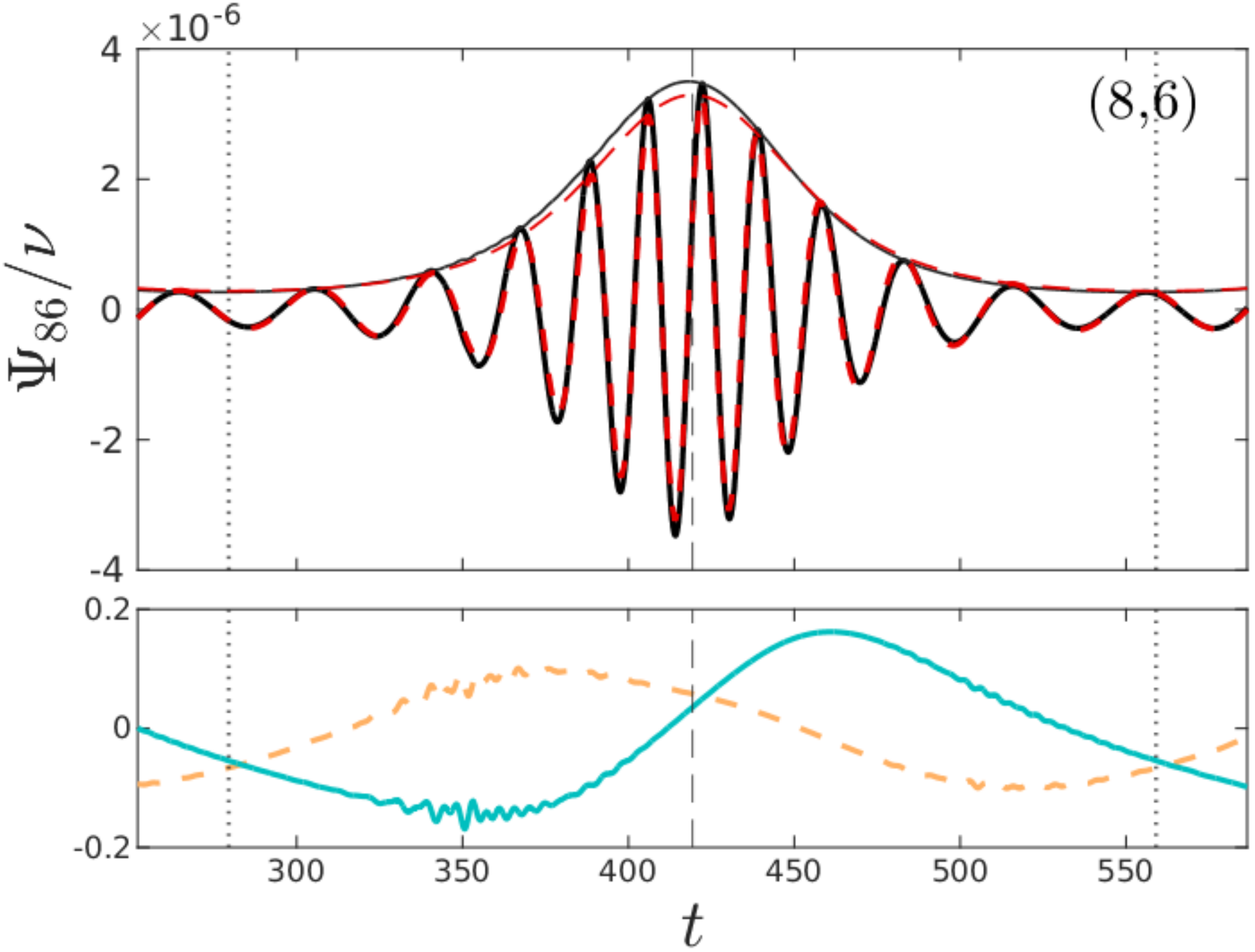}
  \caption{\label{fig:eobwave_manymodes1} Numerical (black) and EOB (dashed-red)	
  waveform multipoles for the intermediate simulation with $e=0.3$ and $\ha = 0.2$,
  whose orbits are shown in Fig.~\ref{fig:geo_dynamics_manymodes}.  
  For each multipole,  we report the relative amplitude difference (dashed orange) and the phase 
  difference (light blue). The vertical lines mark the periastron 
  and the apastron.}
\end{figure*}
\begin{figure*}
  \center  
  \includegraphics[width=0.24\textwidth,height=3cm]{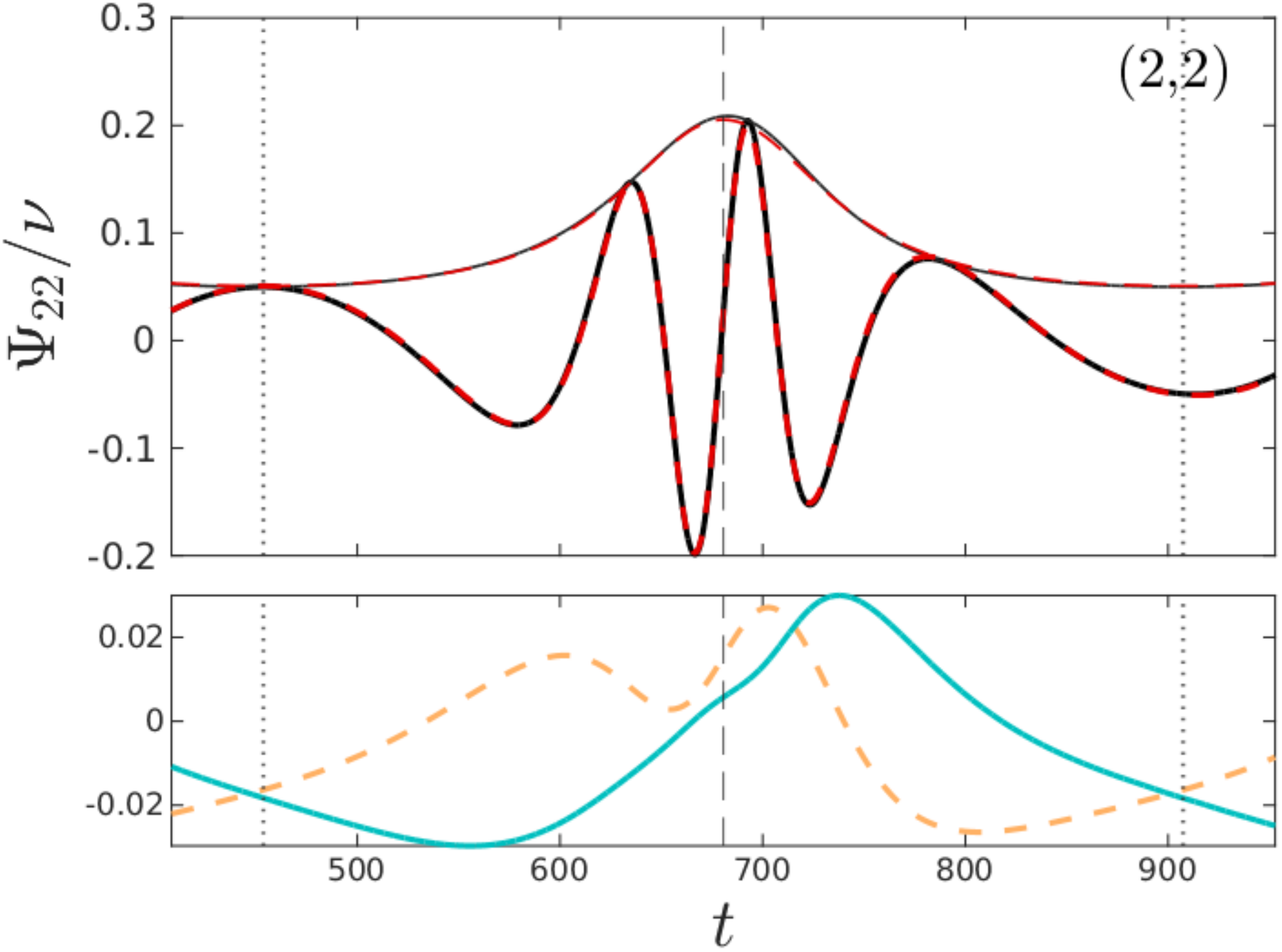}
  \includegraphics[width=0.24\textwidth,height=3cm]{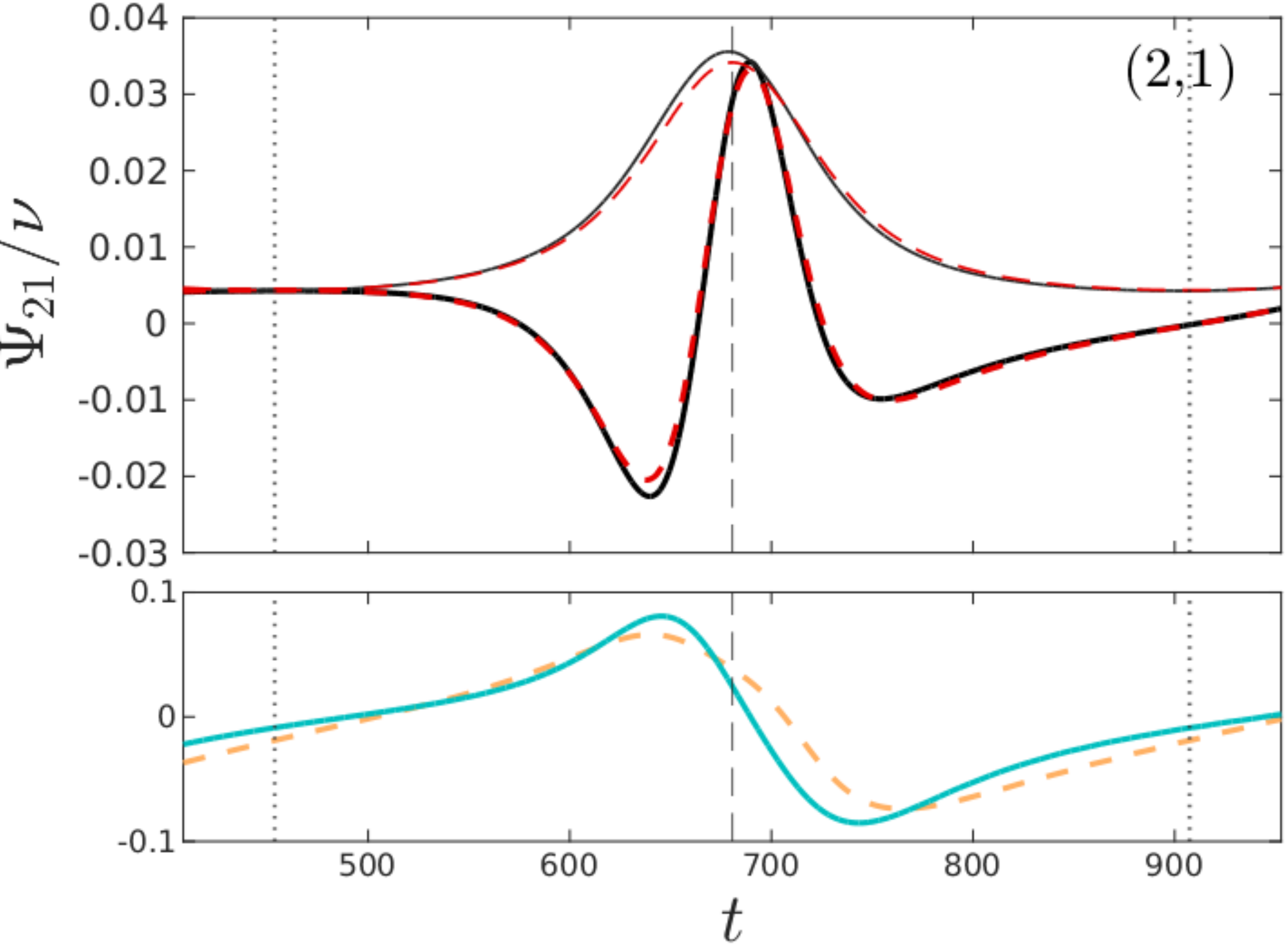}
  \includegraphics[width=0.24\textwidth,height=3cm]{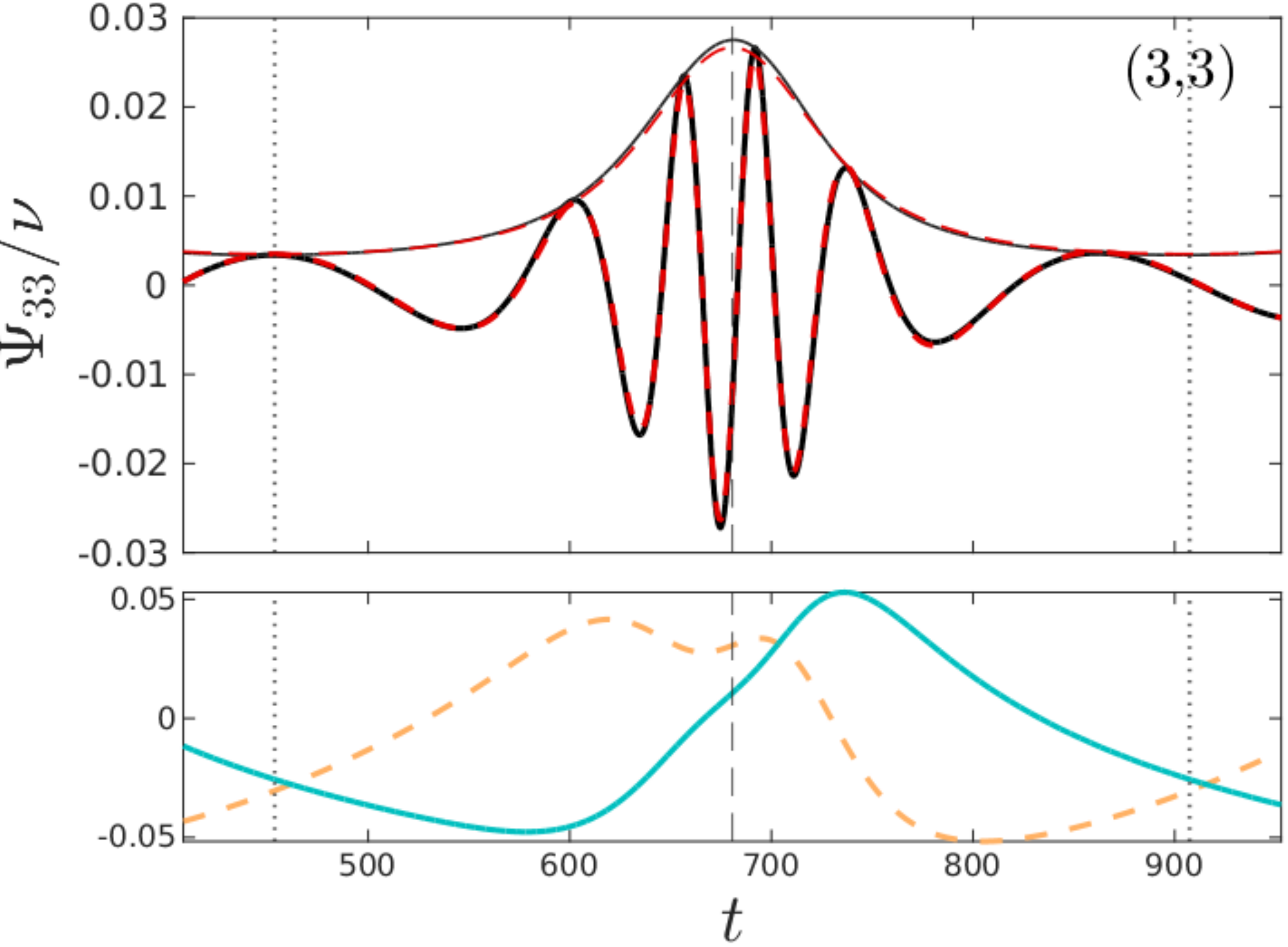}
  \includegraphics[width=0.24\textwidth,height=3.1cm]{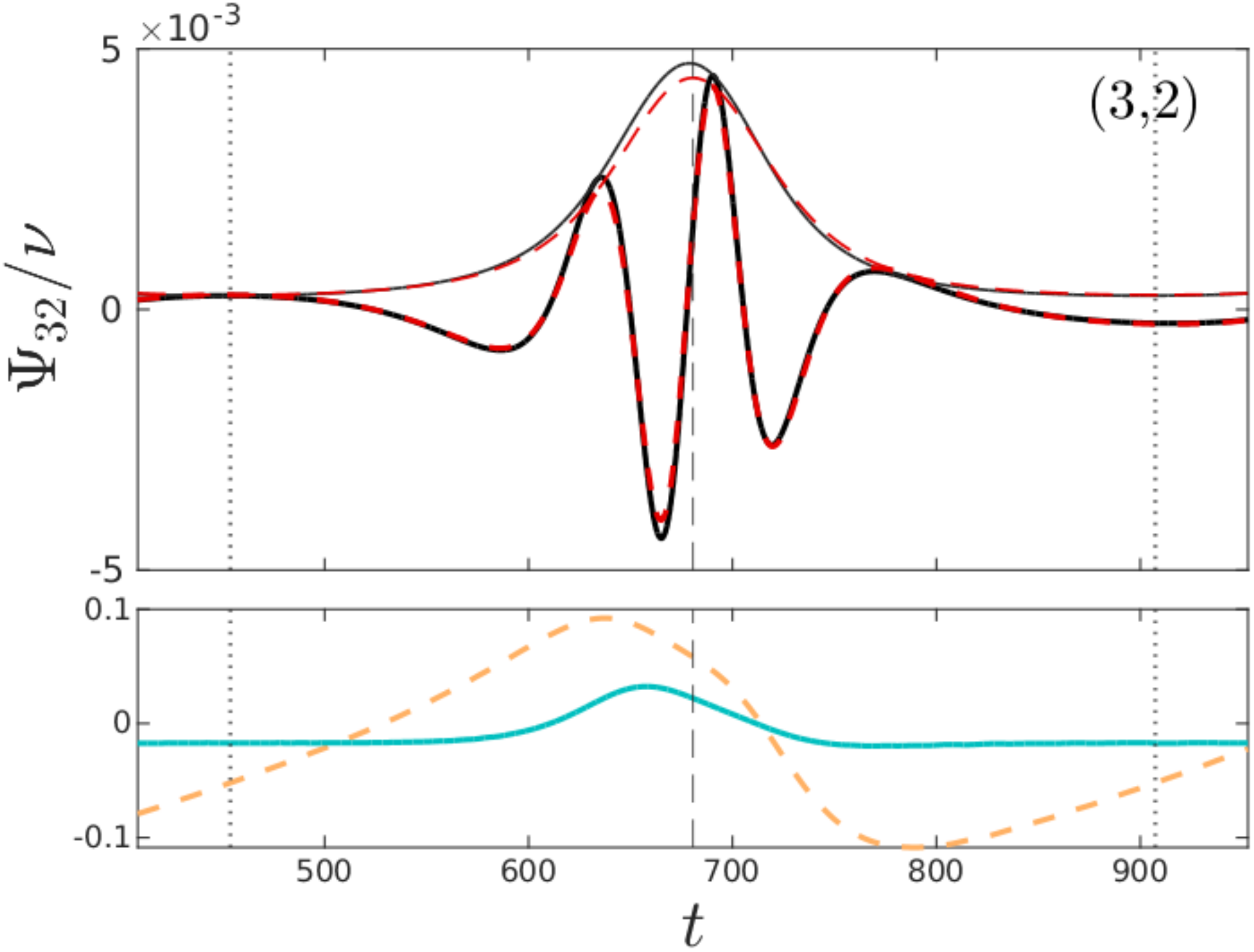} \\
  
  \includegraphics[width=0.24\textwidth,height=3cm]{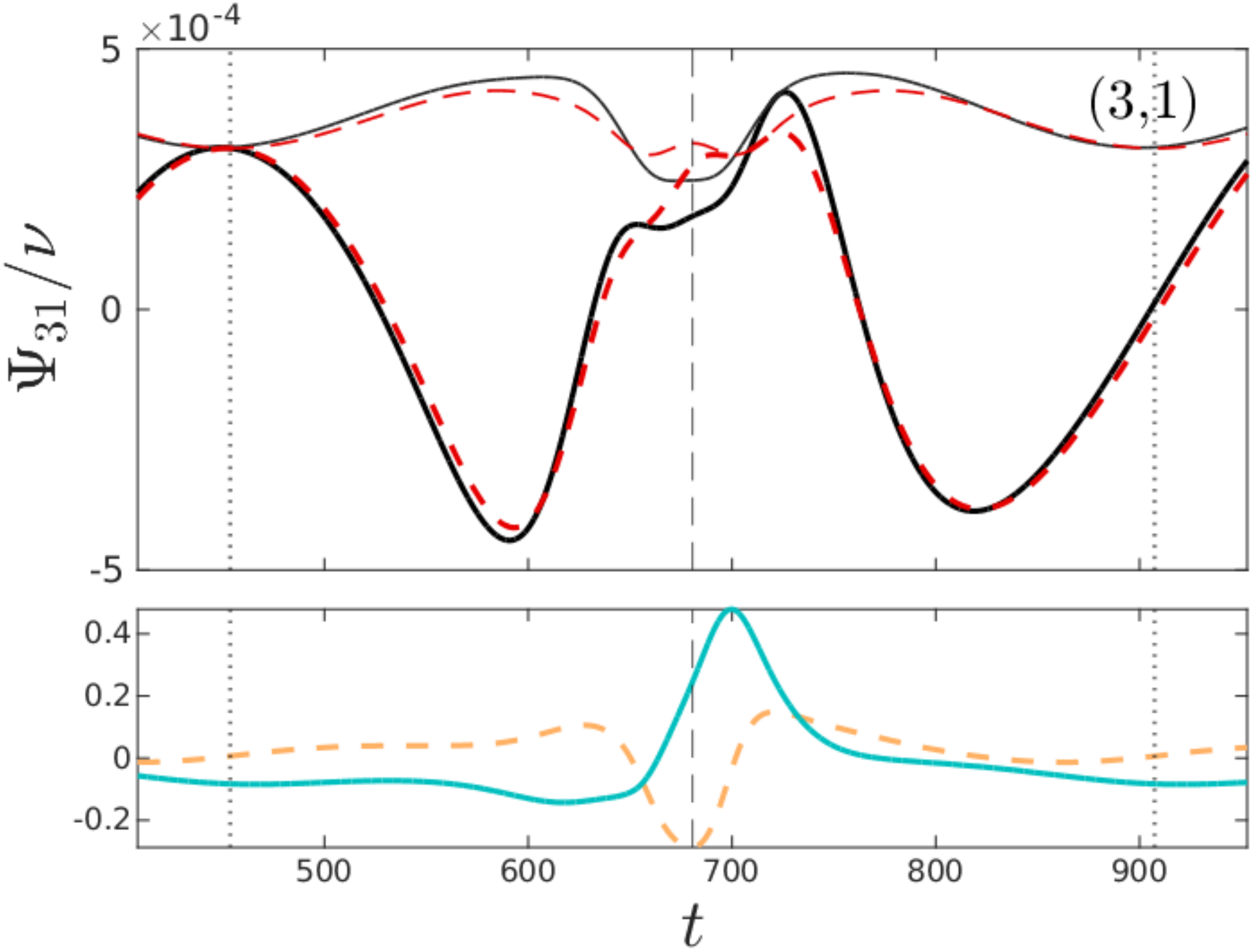} 
  \includegraphics[width=0.24\textwidth,height=3cm]{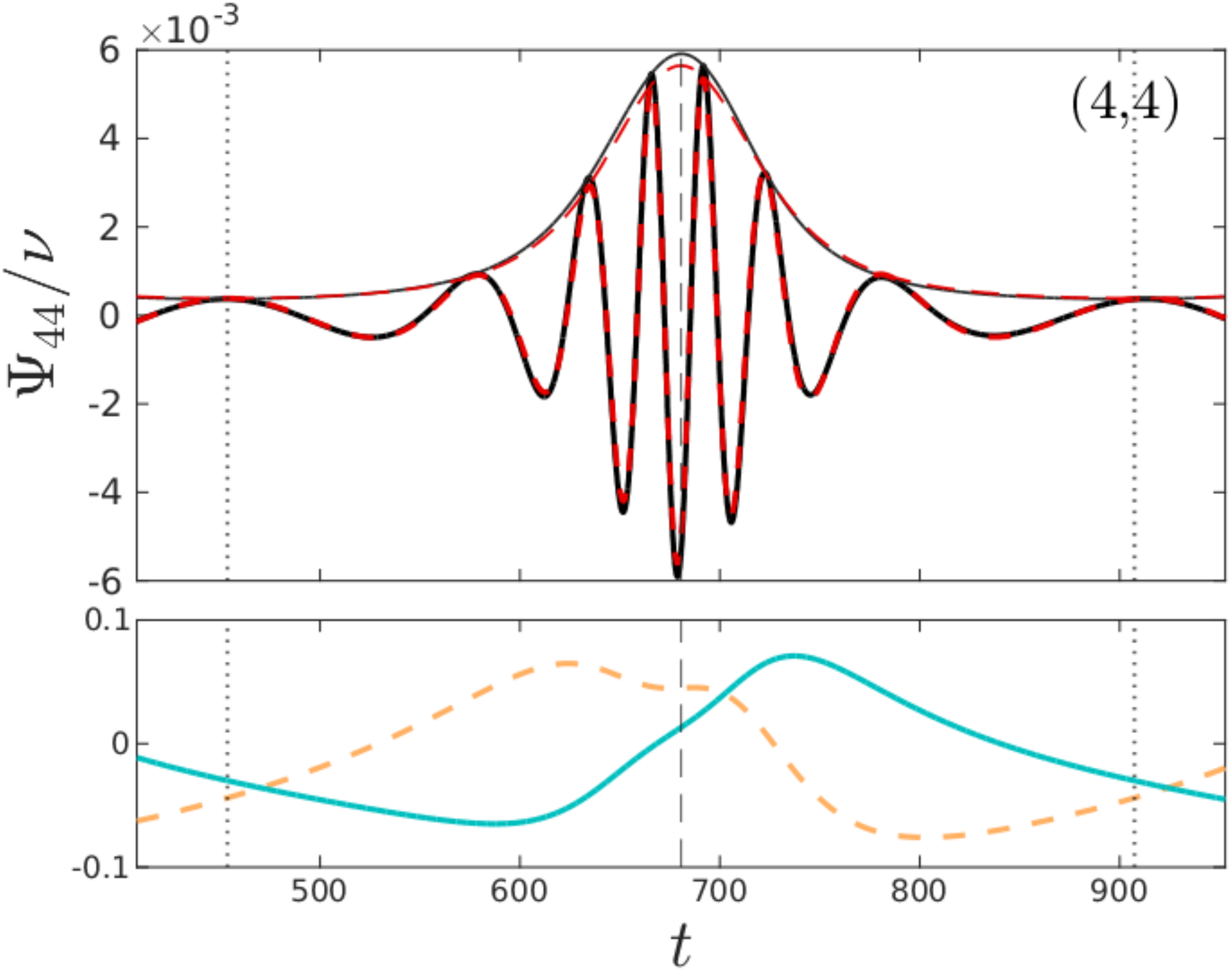} 
  \includegraphics[width=0.24\textwidth,height=3cm]{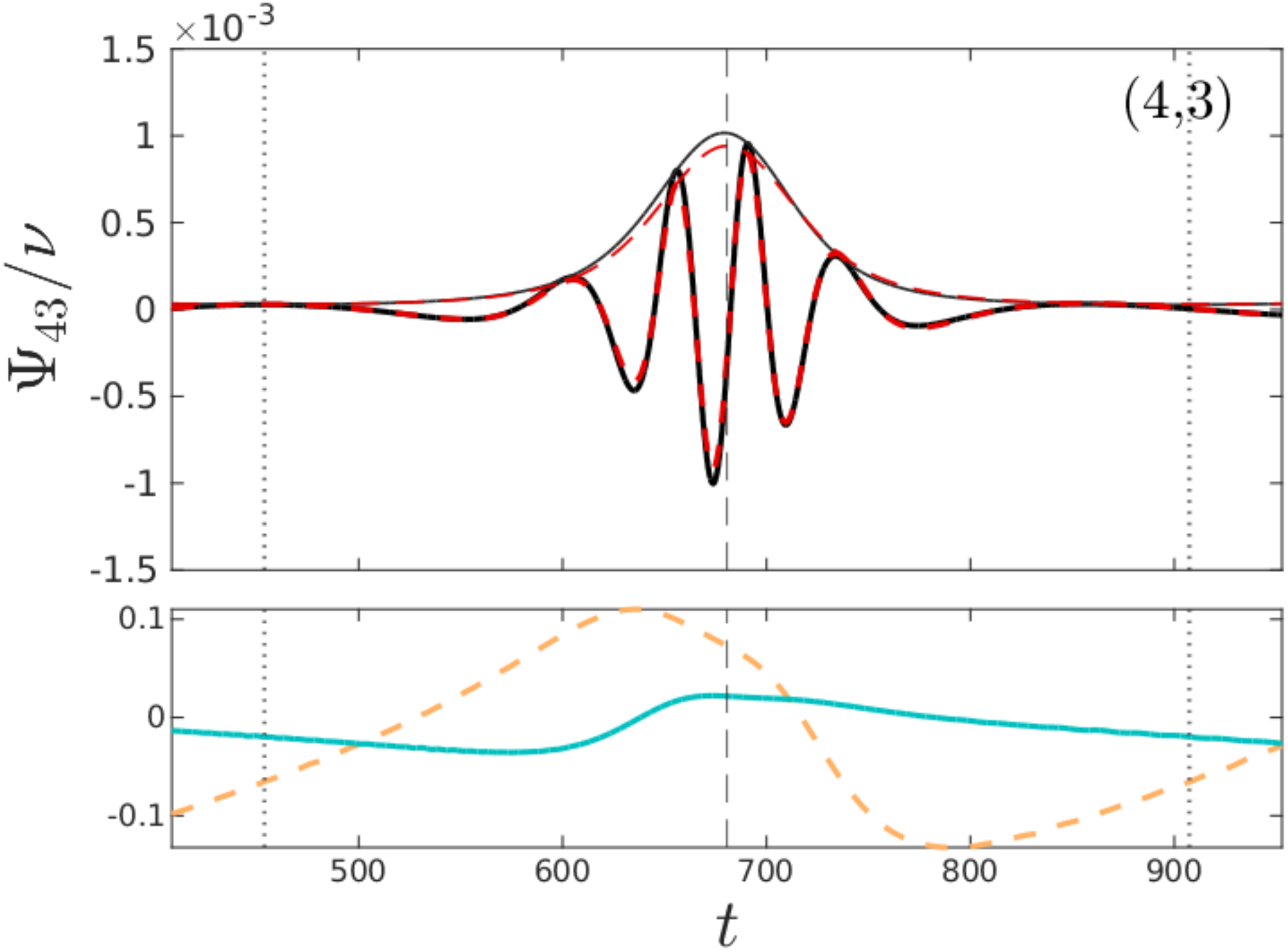} 
  \includegraphics[width=0.24\textwidth,height=3cm]{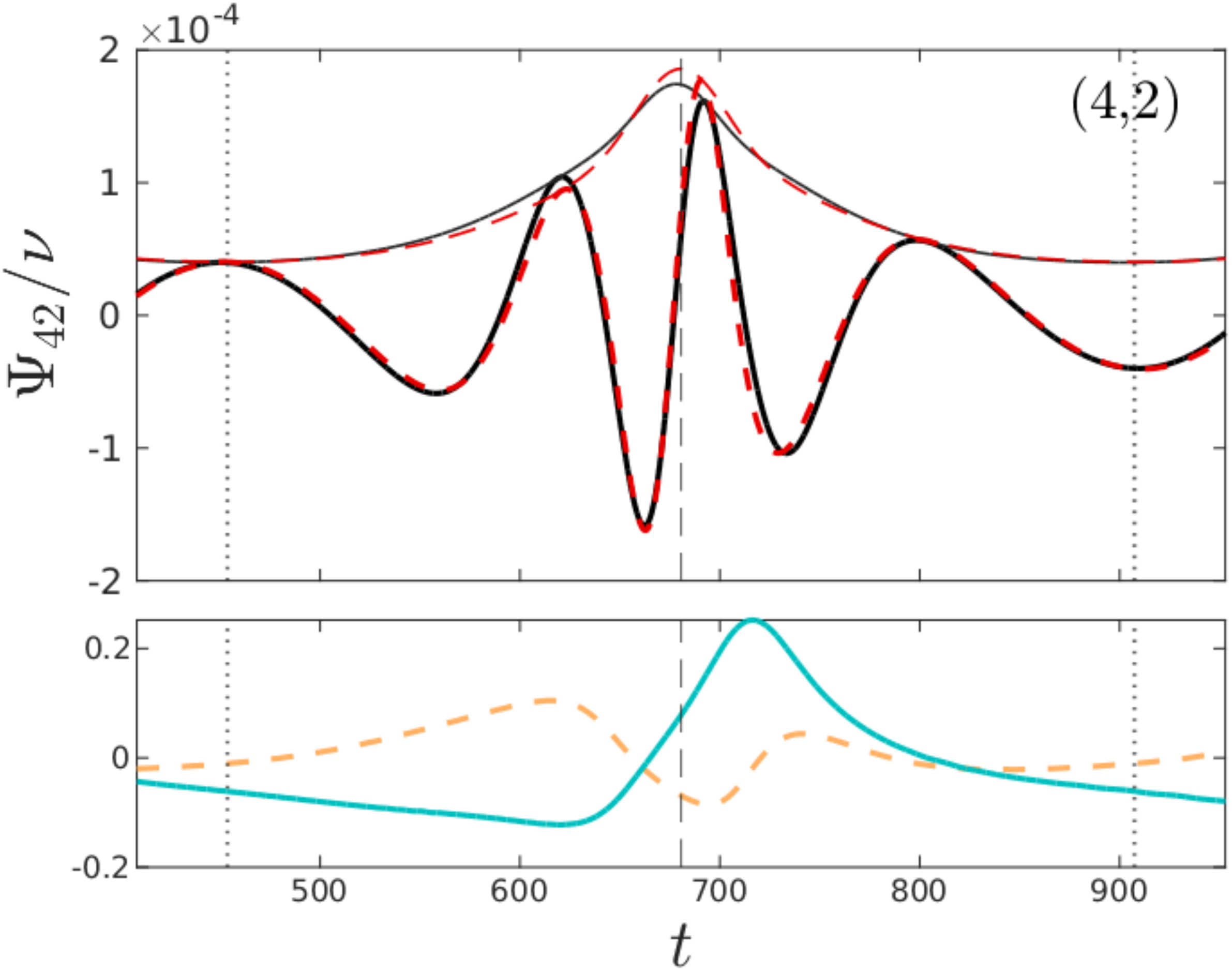} \\
  
  \includegraphics[width=0.24\textwidth,height=3cm]{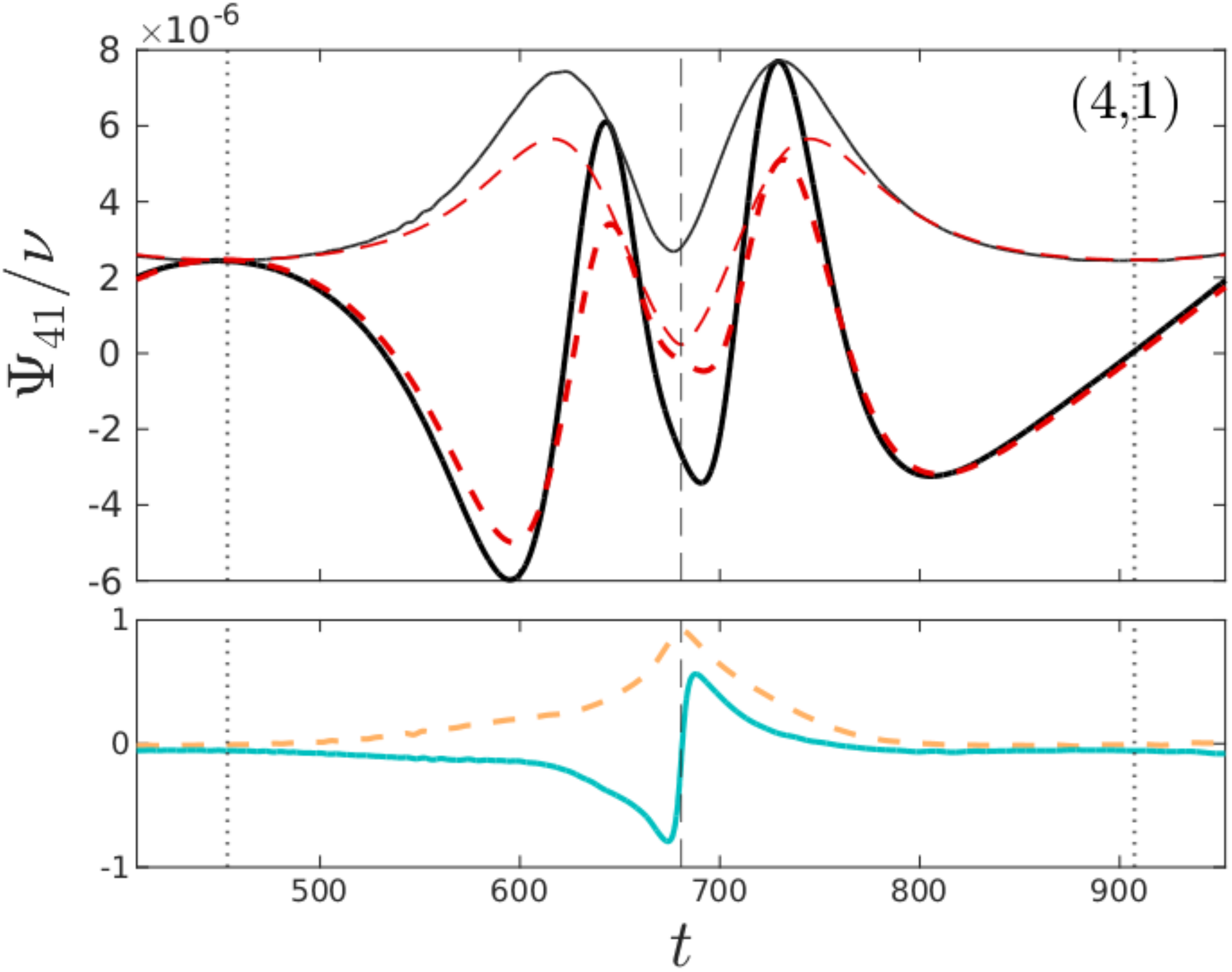}
  \includegraphics[width=0.24\textwidth,height=3cm]{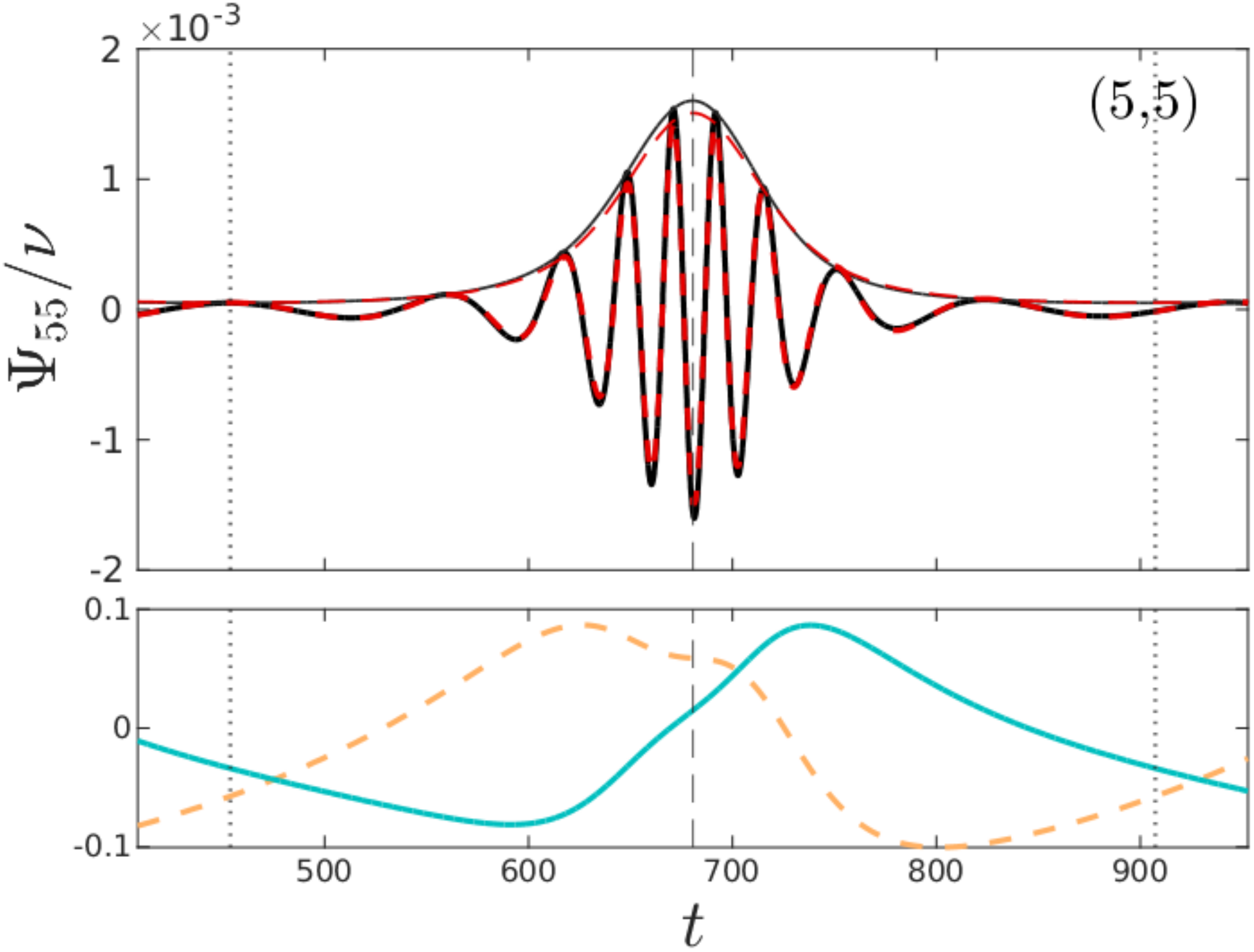}
  \includegraphics[width=0.24\textwidth,height=3cm]{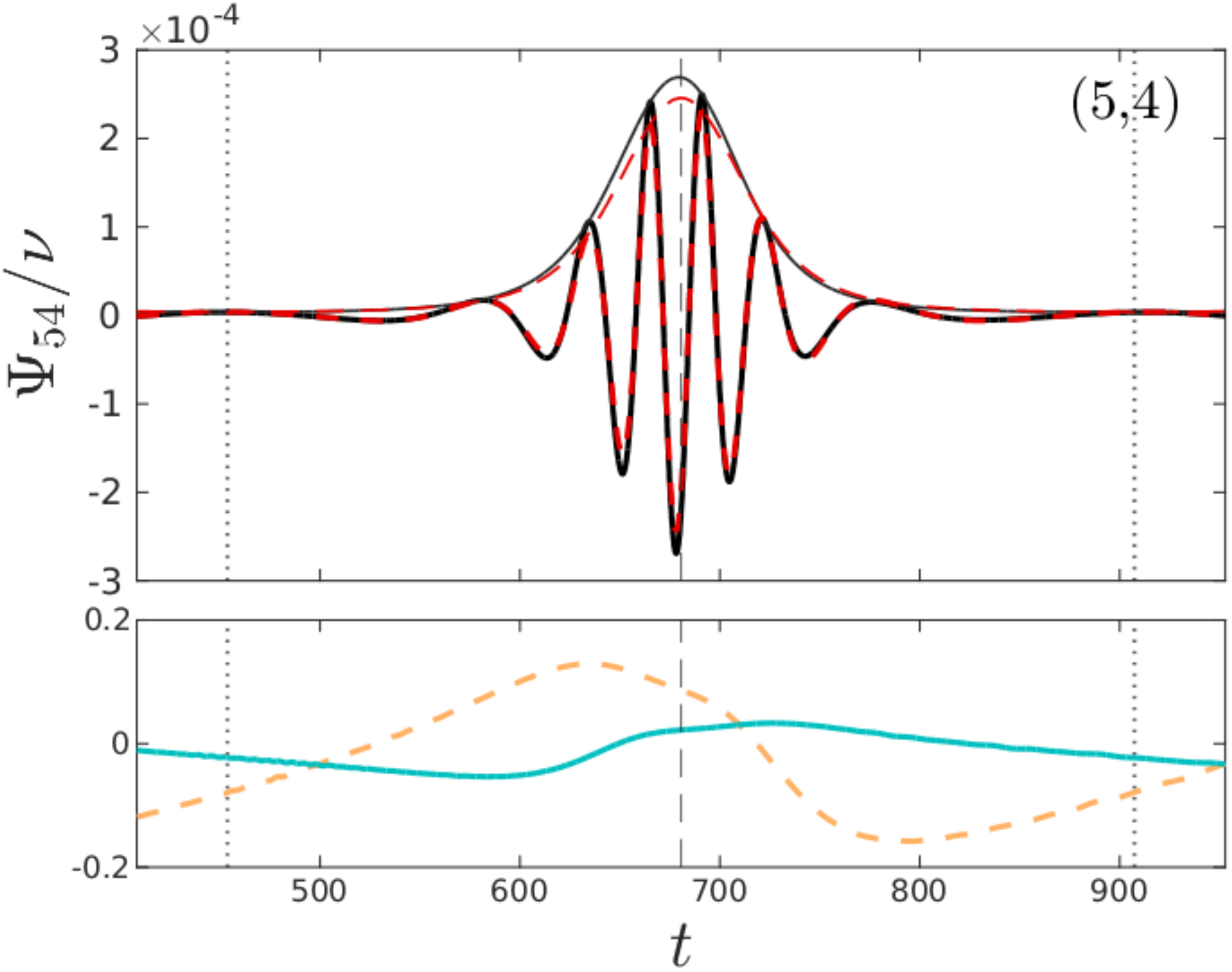}
  \includegraphics[width=0.24\textwidth,height=3cm]{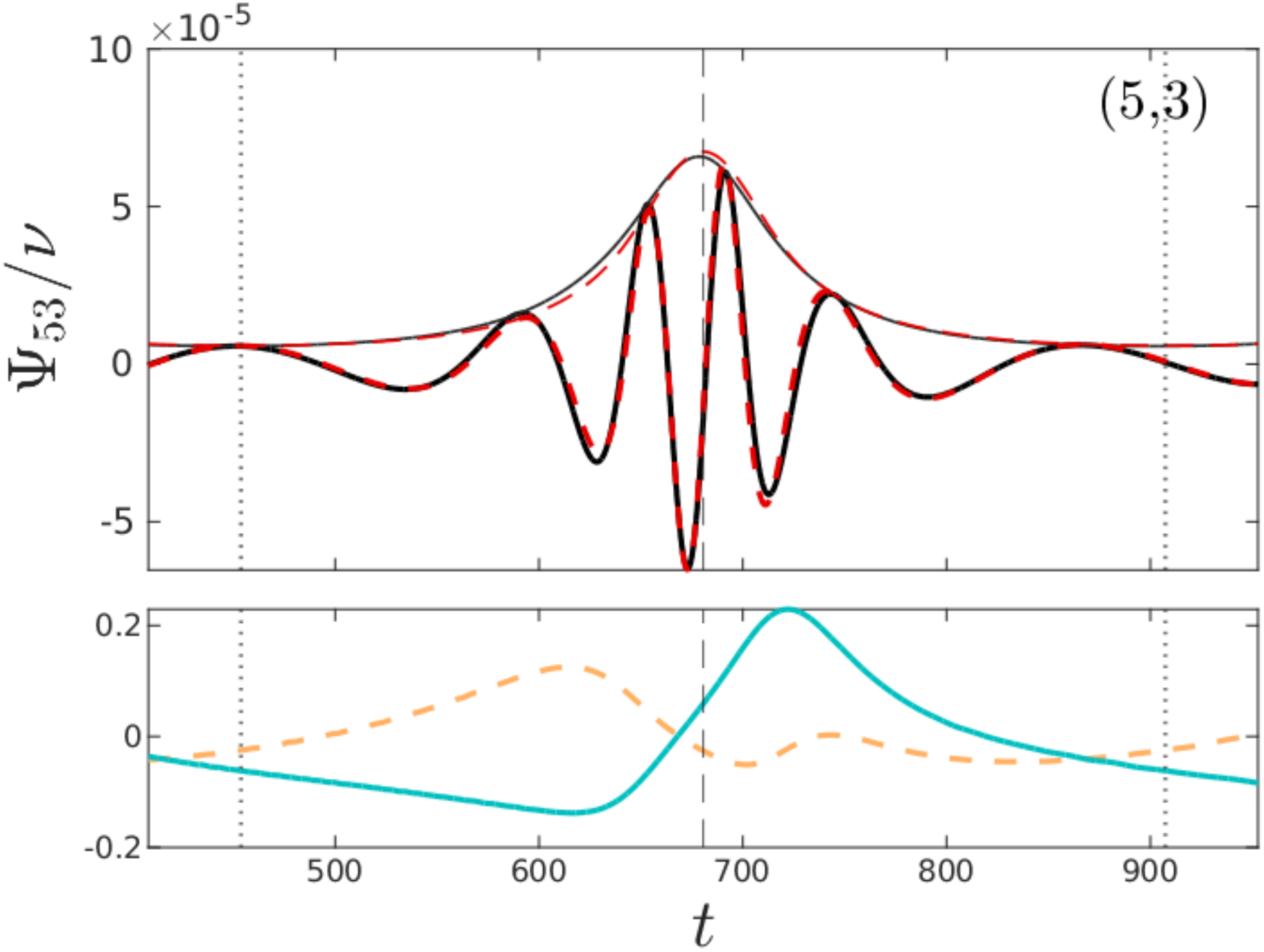} \\
 
  \includegraphics[width=0.24\textwidth,height=3cm]{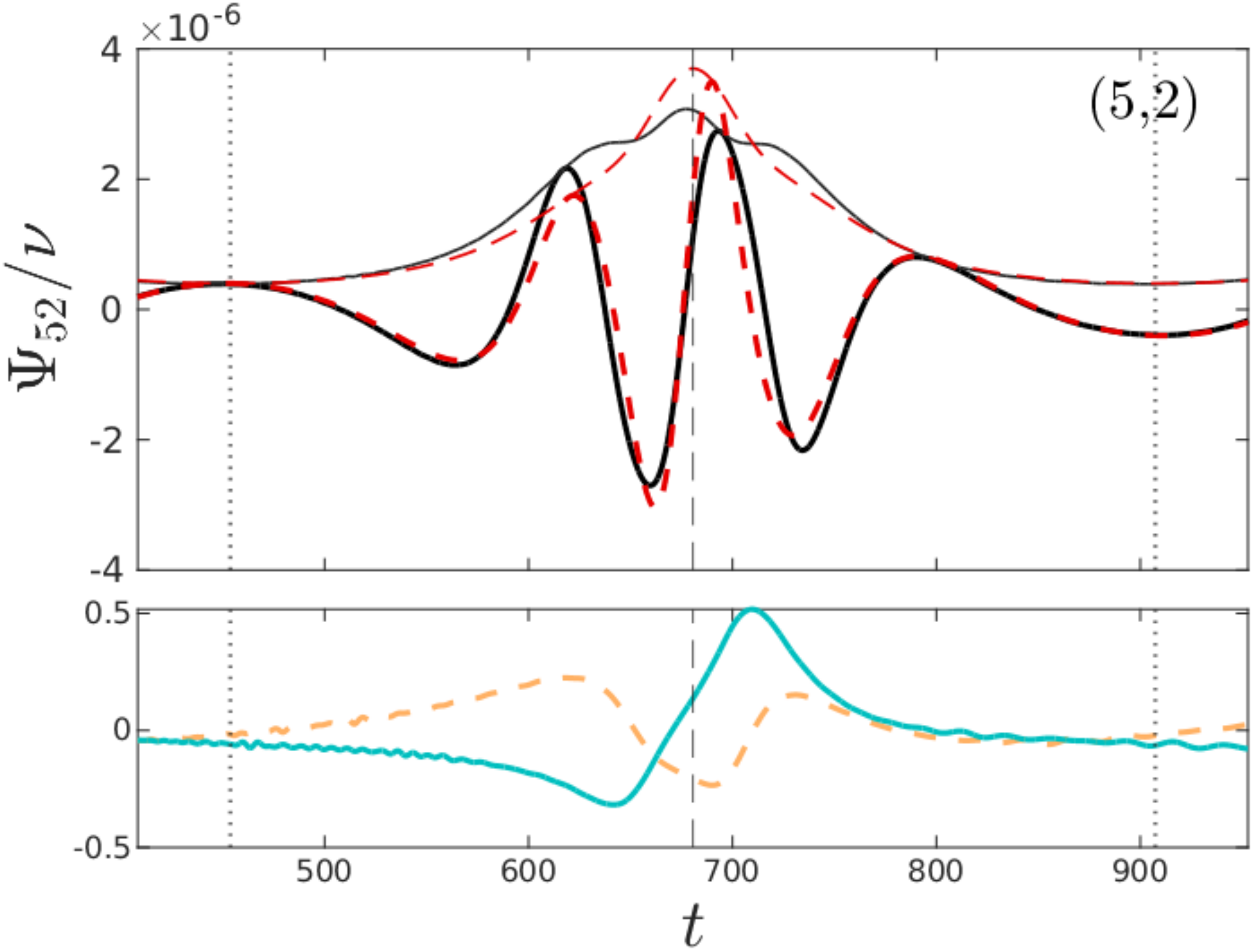}
  \includegraphics[width=0.24\textwidth,height=3cm]{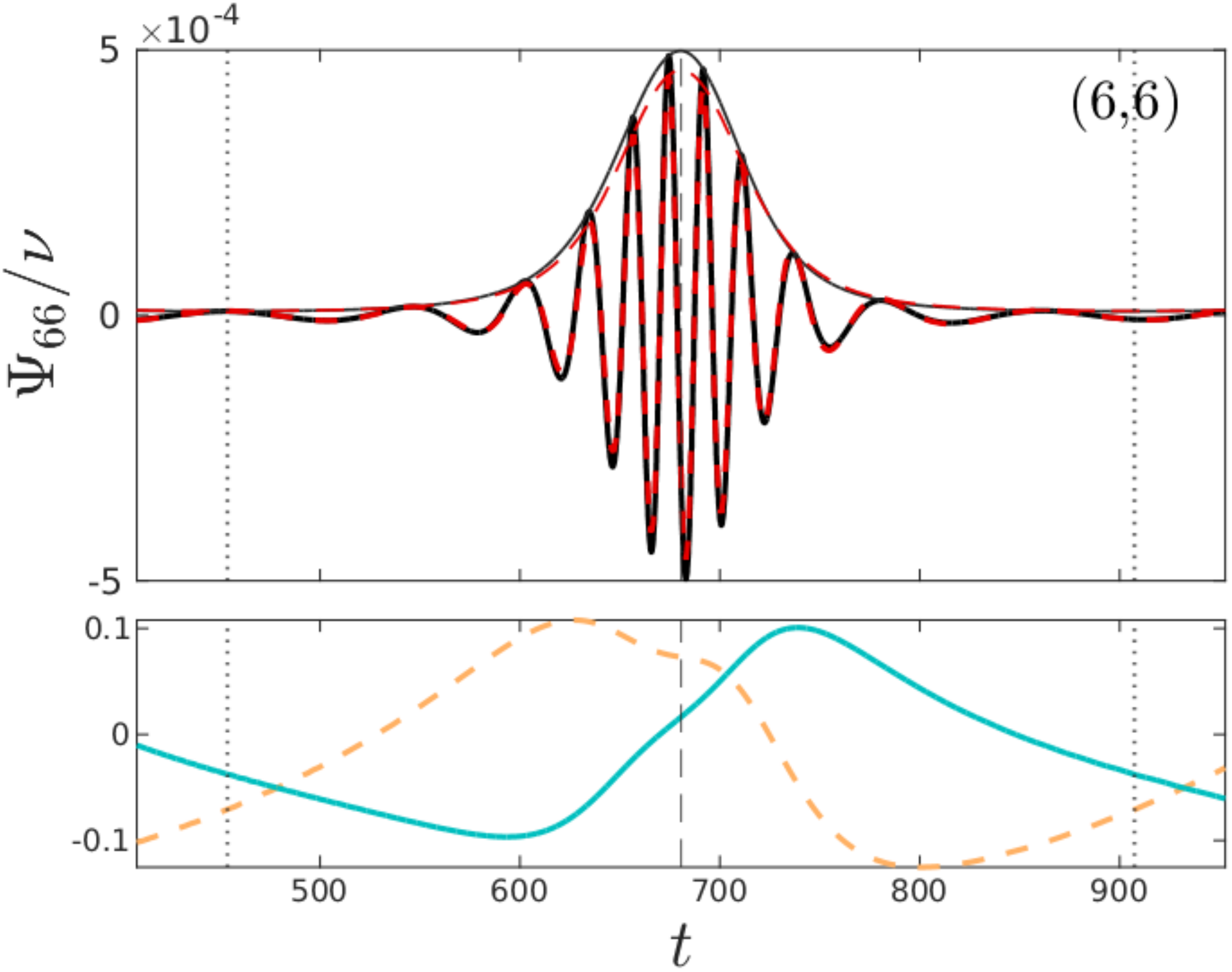}
  \includegraphics[width=0.24\textwidth,height=3cm]{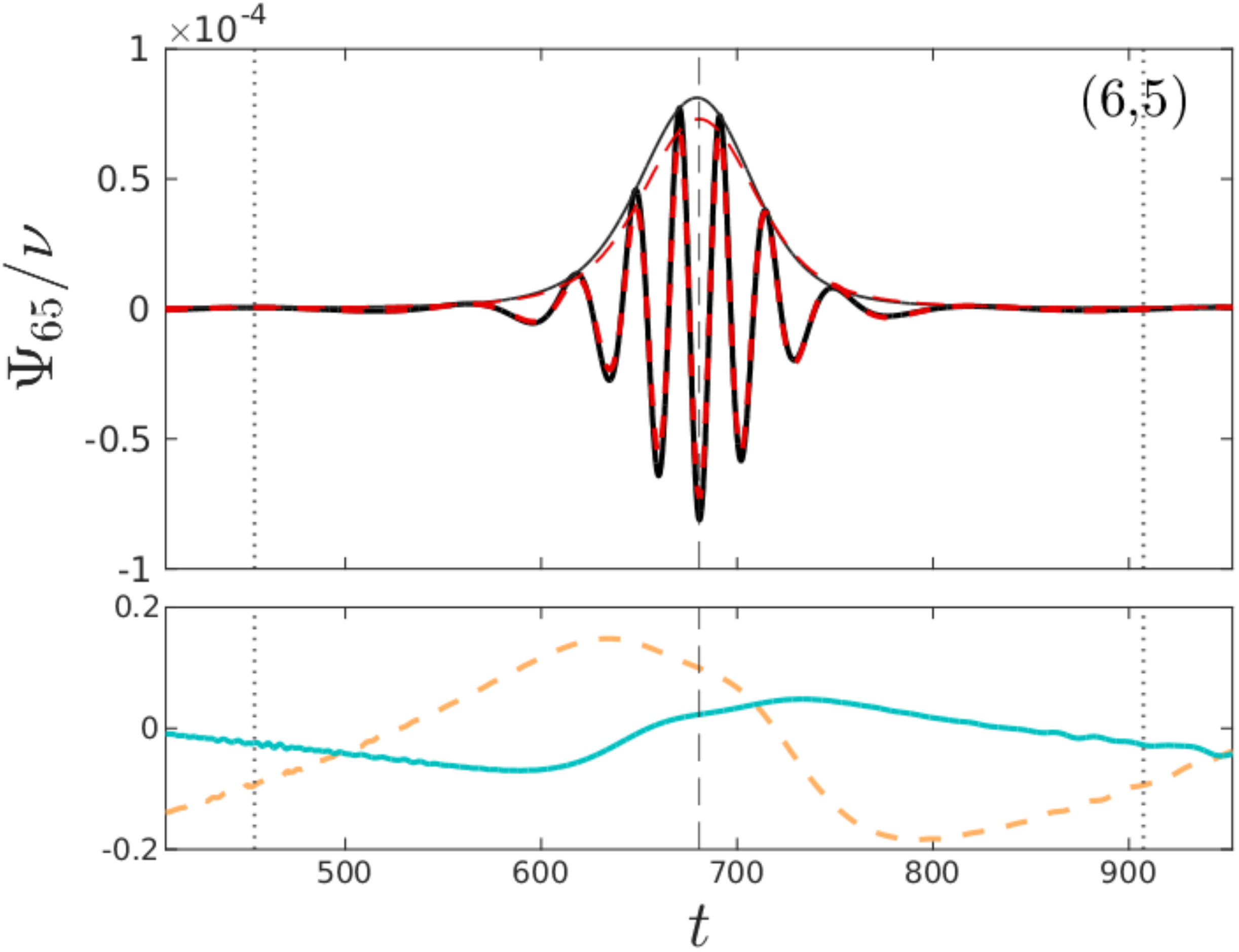}
  \includegraphics[width=0.24\textwidth,height=3cm]{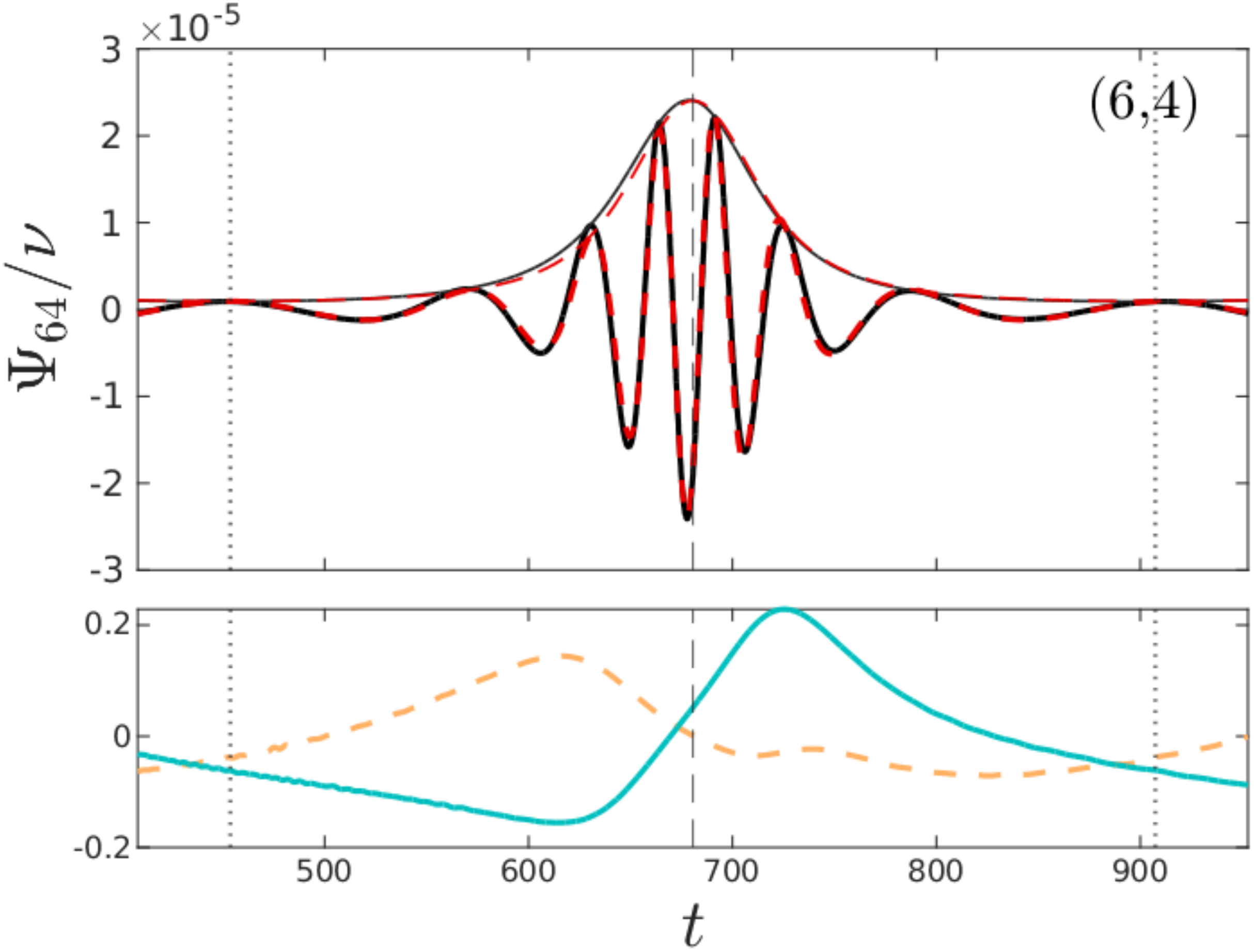} \\
  
  \includegraphics[width=0.24\textwidth,height=3cm]{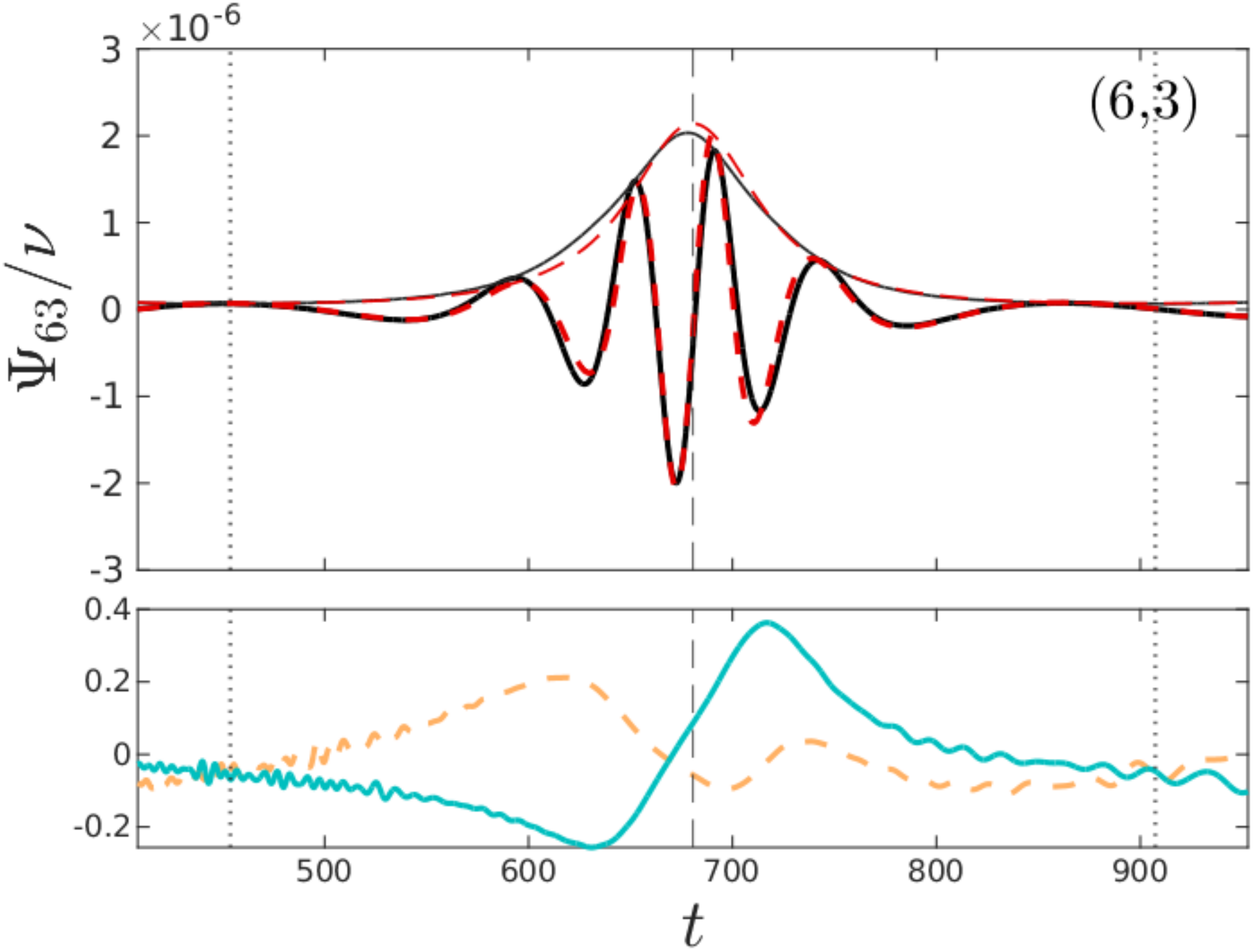}
  \includegraphics[width=0.24\textwidth,height=3cm]{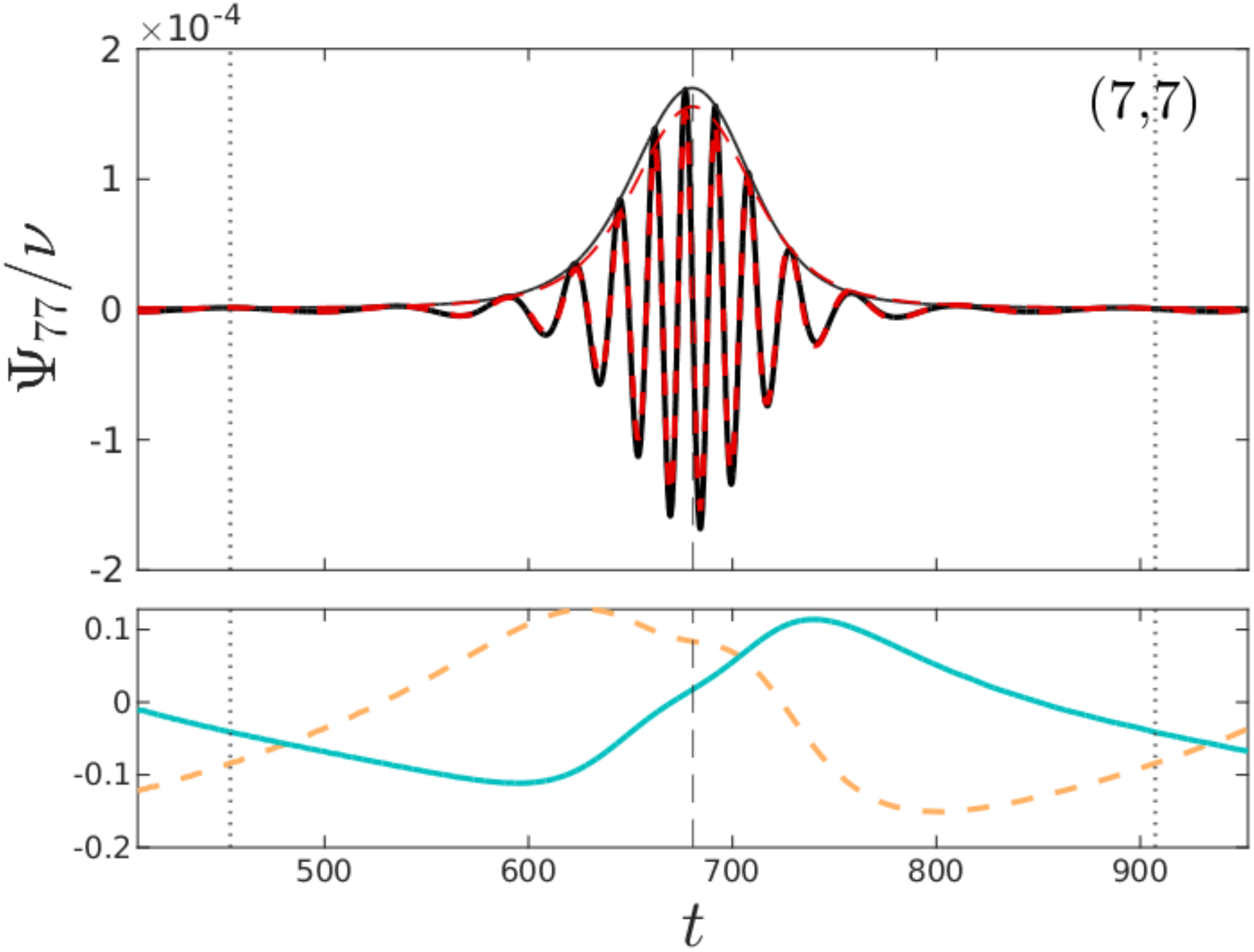}
  \includegraphics[width=0.24\textwidth,height=3cm]{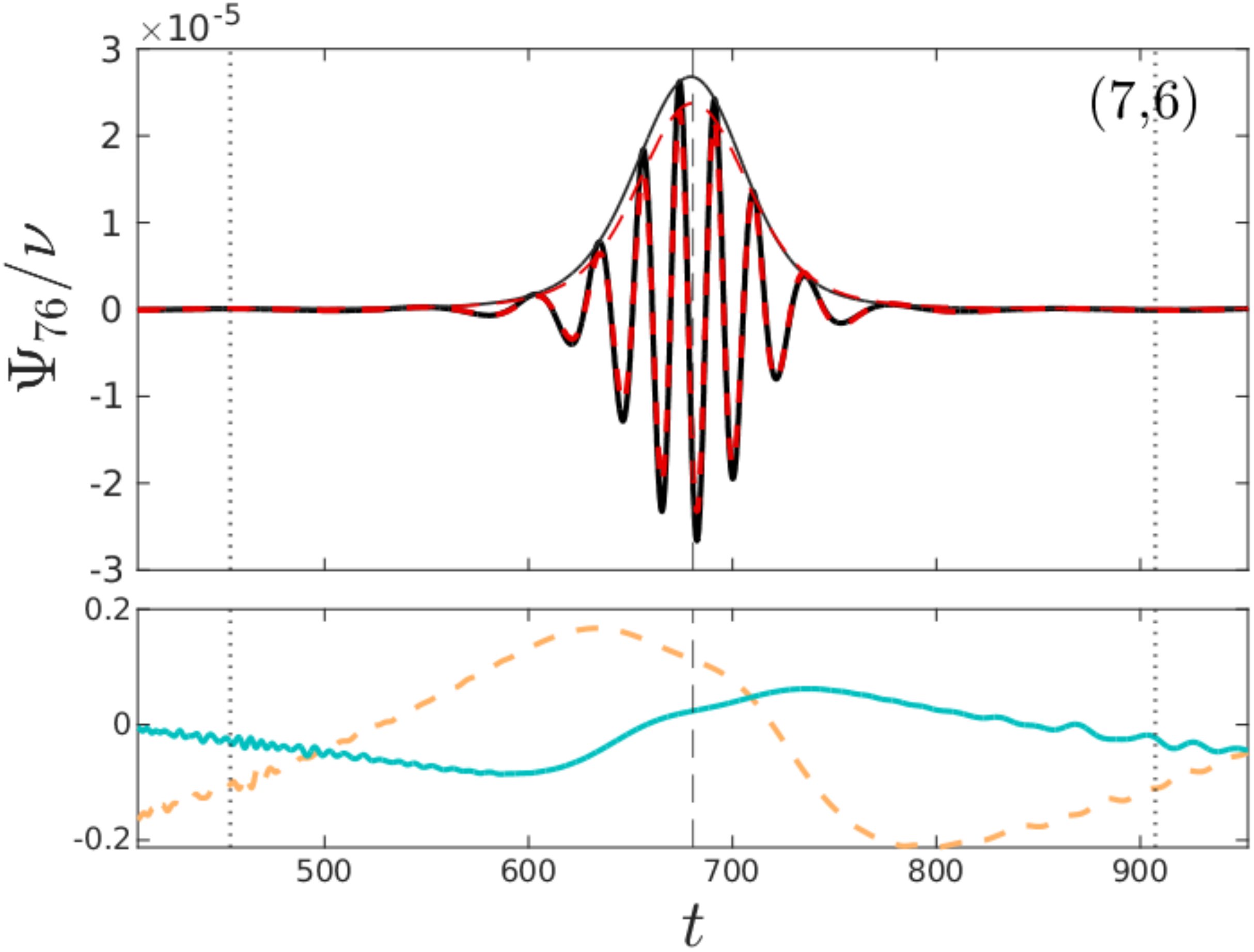}
  \includegraphics[width=0.24\textwidth,height=3cm]{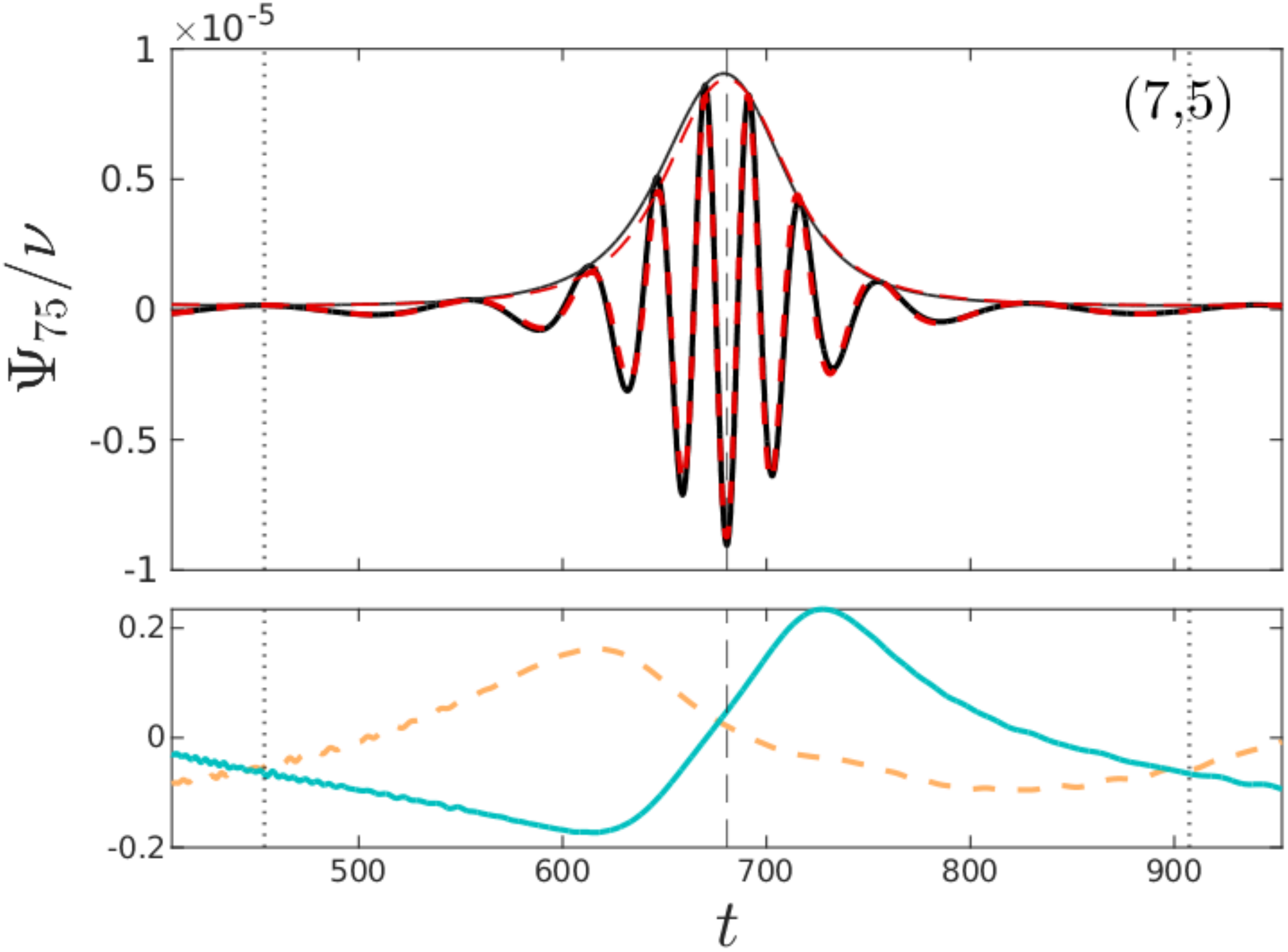} \\
  
  \includegraphics[width=0.24\textwidth,height=3cm]{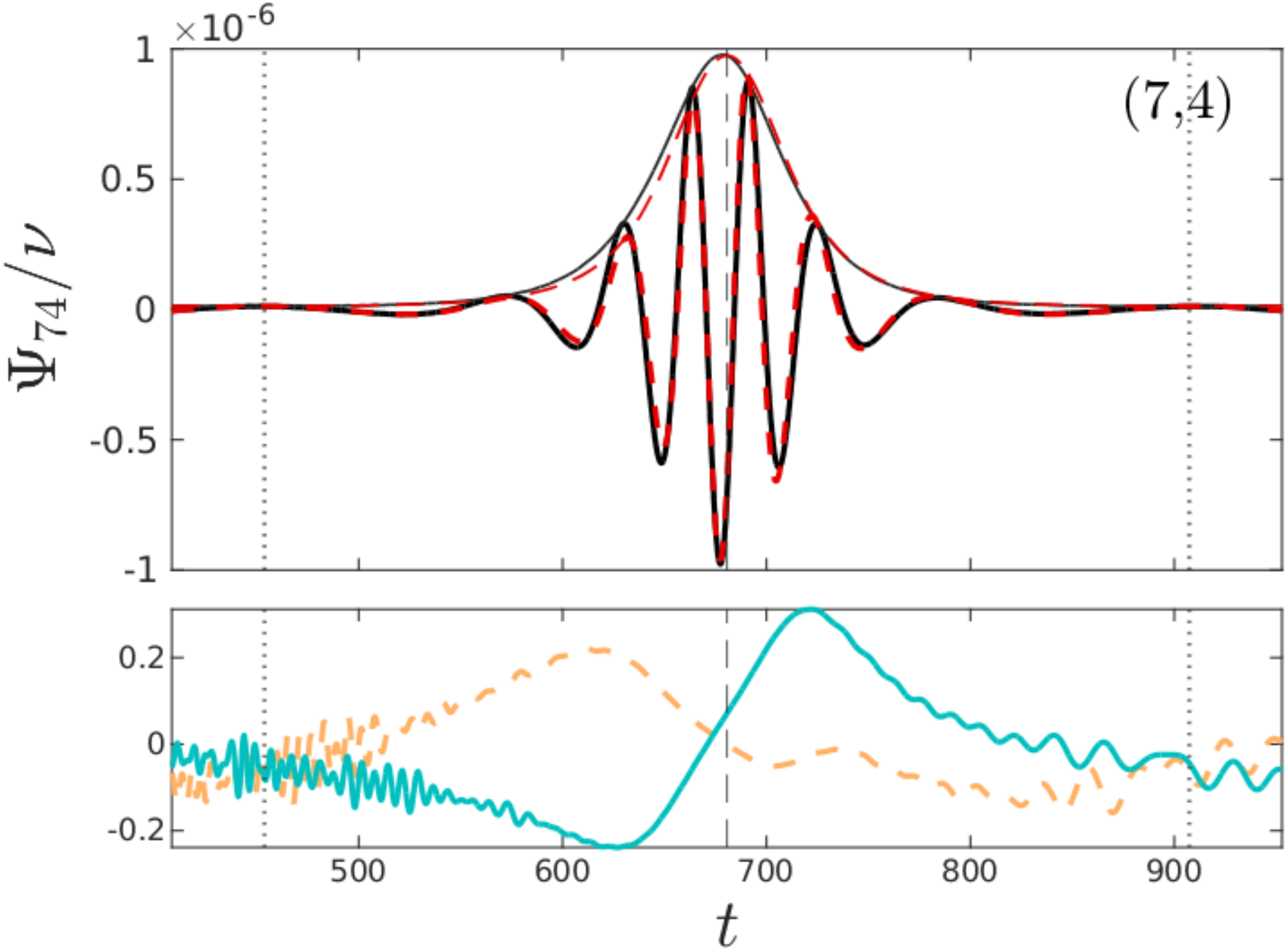}
  \includegraphics[width=0.24\textwidth,height=3cm]{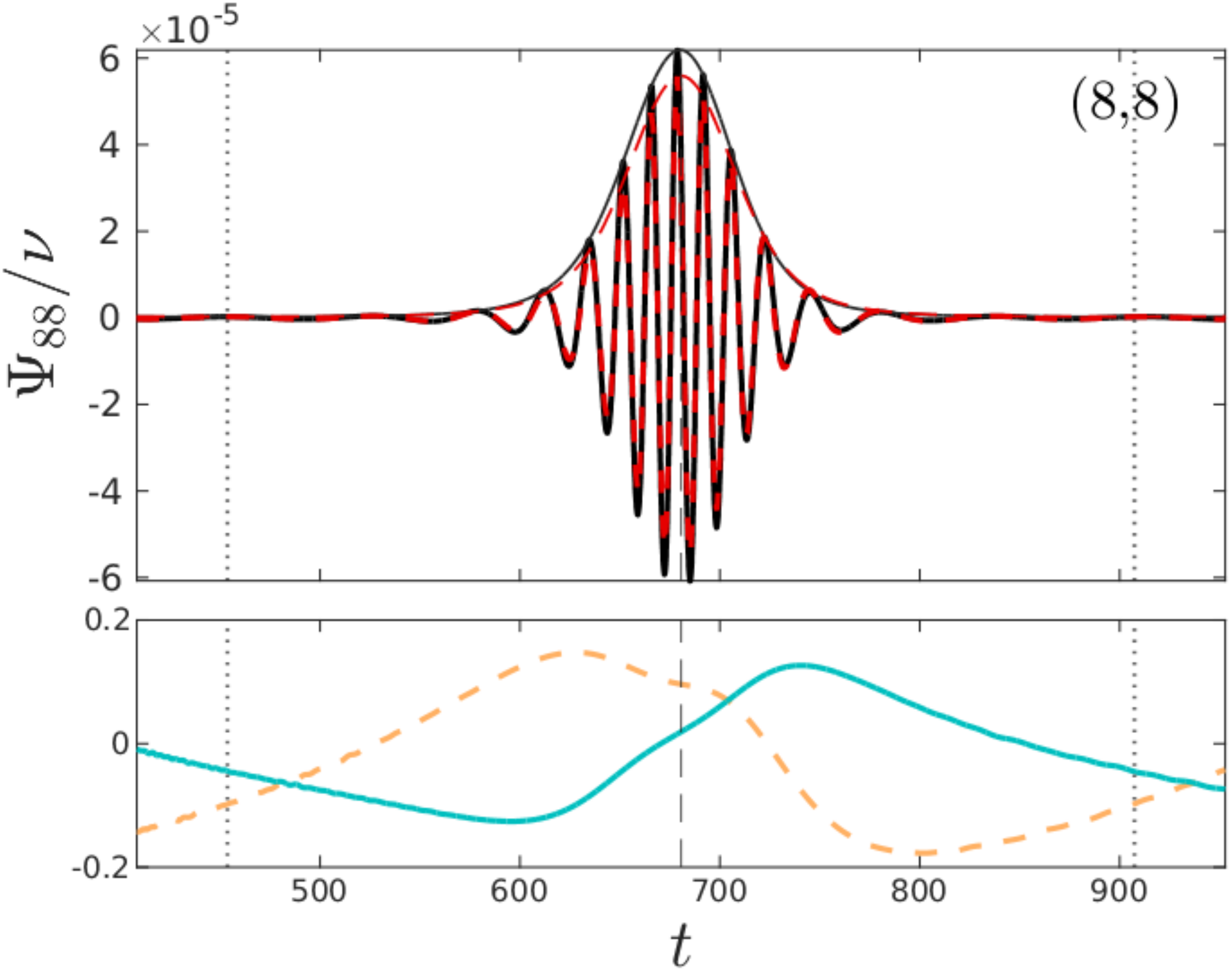}
  \includegraphics[width=0.24\textwidth,height=3cm]{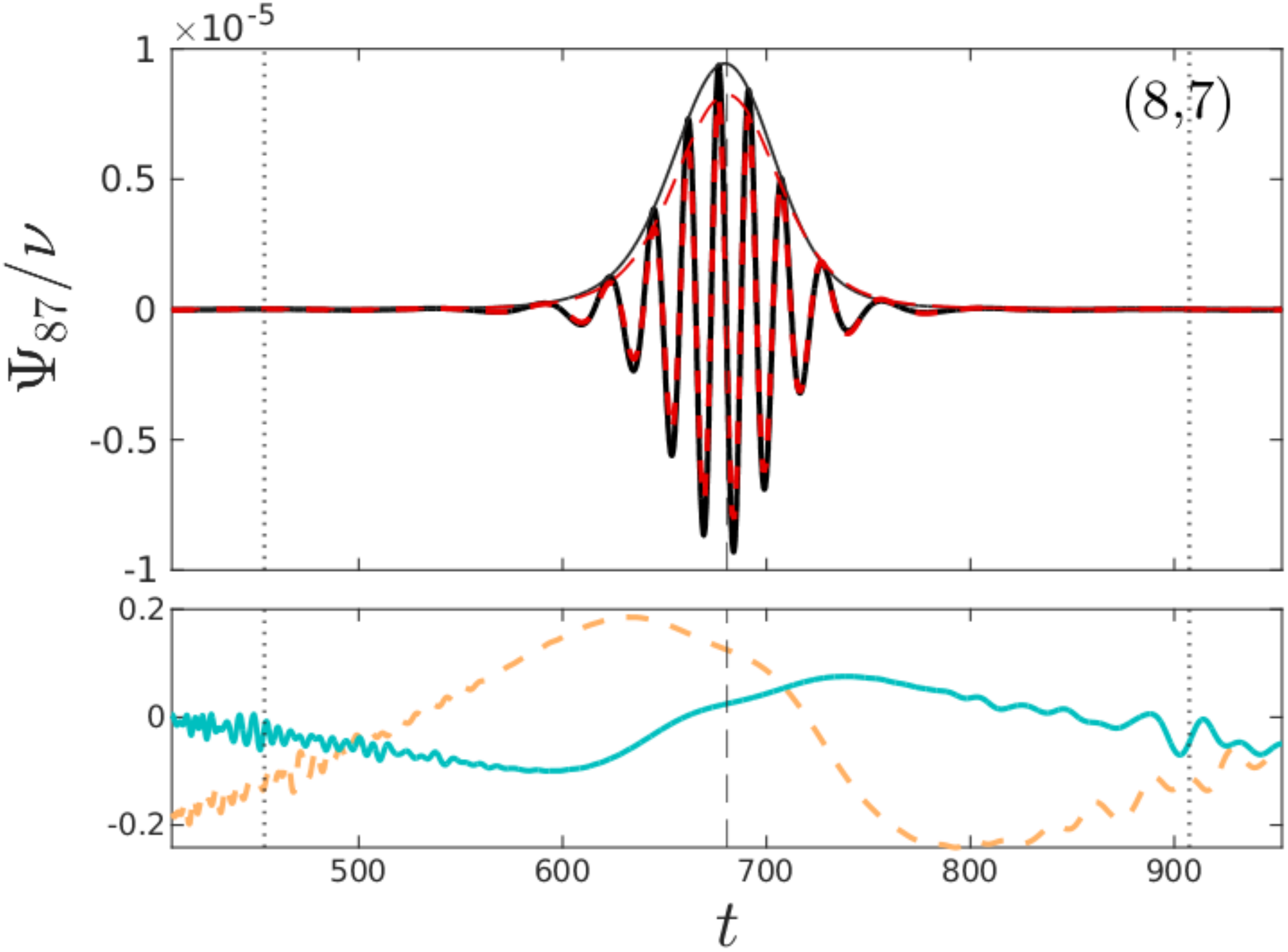}
  \includegraphics[width=0.24\textwidth,height=3cm]{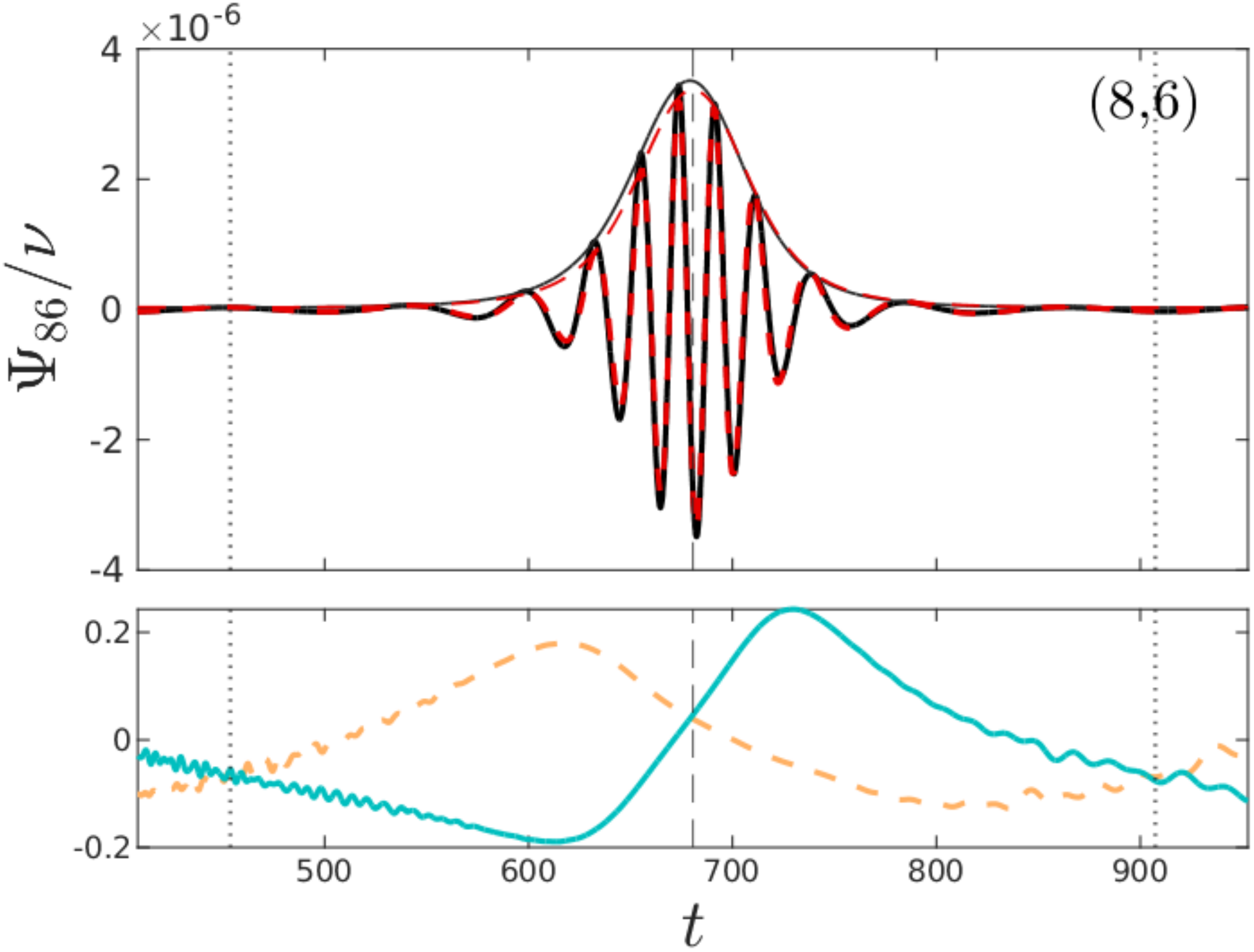}
  \caption{\label{fig:eobwave_manymodes2} Numerical (black) and EOB (dashed-red)	
  waveform multipoles for the intermediate simulation with $e=0.5$ and $\ha = -0.2$,
  whose orbits are shown in Fig.~\ref{fig:geo_dynamics_manymodes}.  
  For each multipole, we report the relative amplitude difference (dashed orange) and the phase 
  difference (light blue). The vertical lines mark the periastron 
  and the apastron.}
\end{figure*}

Since in the main discussion we have analyzed in detail only the quadrupolar waveform and in 
Fig.~\ref{fig:eobwaves} we have shown only the $\l=m=2$ and $\l=m=4$ modes, in this appendix we
report almost all the modes up to $\l=8$ for two eccentric cases. 
The two geodesic dynamics considered are shown in  
Fig.~\ref{fig:geo_dynamics_manymodes}, while the waveform multipoles are reported 
in Fig.~\ref{fig:eobwave_manymodes1} and Fig.~\ref{fig:eobwave_manymodes2}.

\section{Analytical/numerical fluxes disagreement}
\label{appendix:Fluxes}

In this Appendix we list all the numerical simulations analyzed in this work and 
we compare the corresponding averaged fluxes.
In Table~\ref{tab:simulations_ecc1}, Table~\ref{tab:simulations_ecc2} and 
Table~\ref{tab:simulations_ecc3} we report all the numerical energy and angular momentum
averaged fluxes and the corresponding analytical fluxes $\av{\FNP}$ computed using the 
standard radiation reaction $\FphiNP$ of Eq.~\eqref{eq:Fphi_ecc}. 
The numerical energy fluxes are computed without considering the modes with $m=0$
in order to be coherent with the analytical results.
We also recall that these relative differences are plotted against 
the spin $\ha$ in Fig.~\ref{fig:reldiff_rainbow_FNP} 
and against $r_--r_{\rm LR}$ in Fig.~\ref{fig:reldiff_harlequin}. 

\begin{figure*}[t]
	\center
 	\includegraphics[width=0.32\textwidth,height=3.2cm]{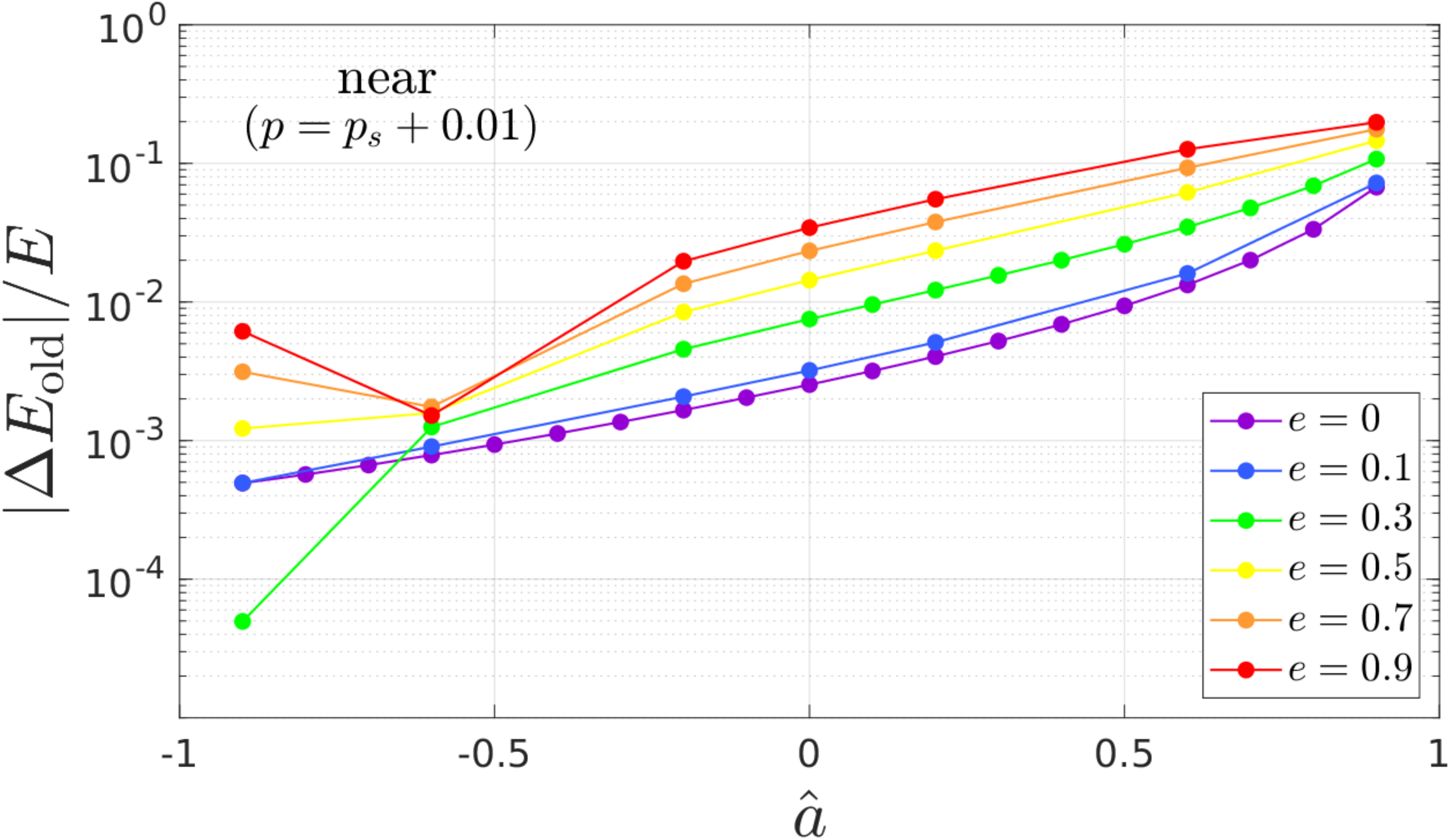}
	\includegraphics[width=0.32\textwidth,height=3.2cm]{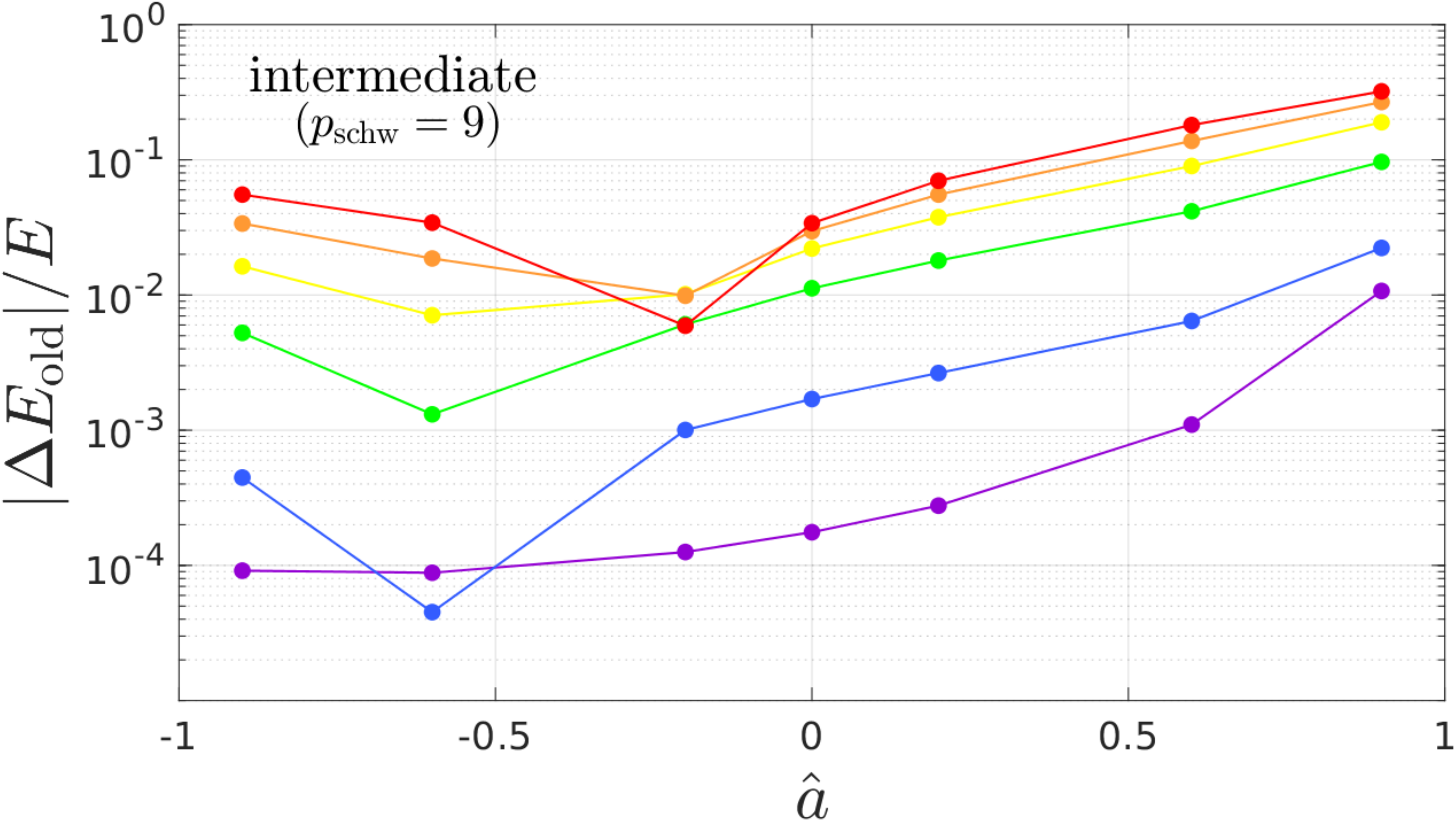}
	\includegraphics[width=0.32\textwidth,height=3.2cm]{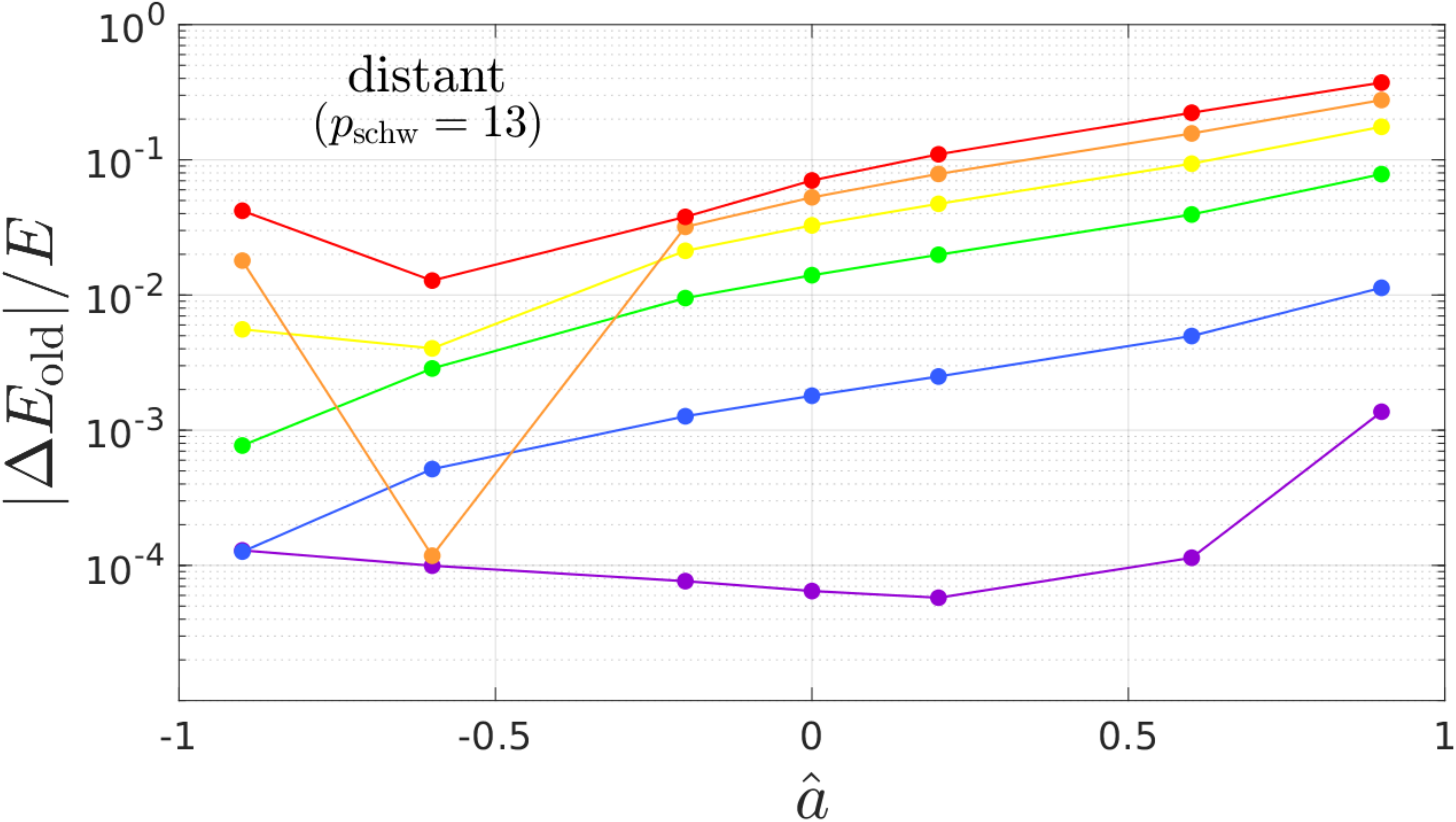} \\ 
	\includegraphics[width=0.32\textwidth,height=3.2cm]{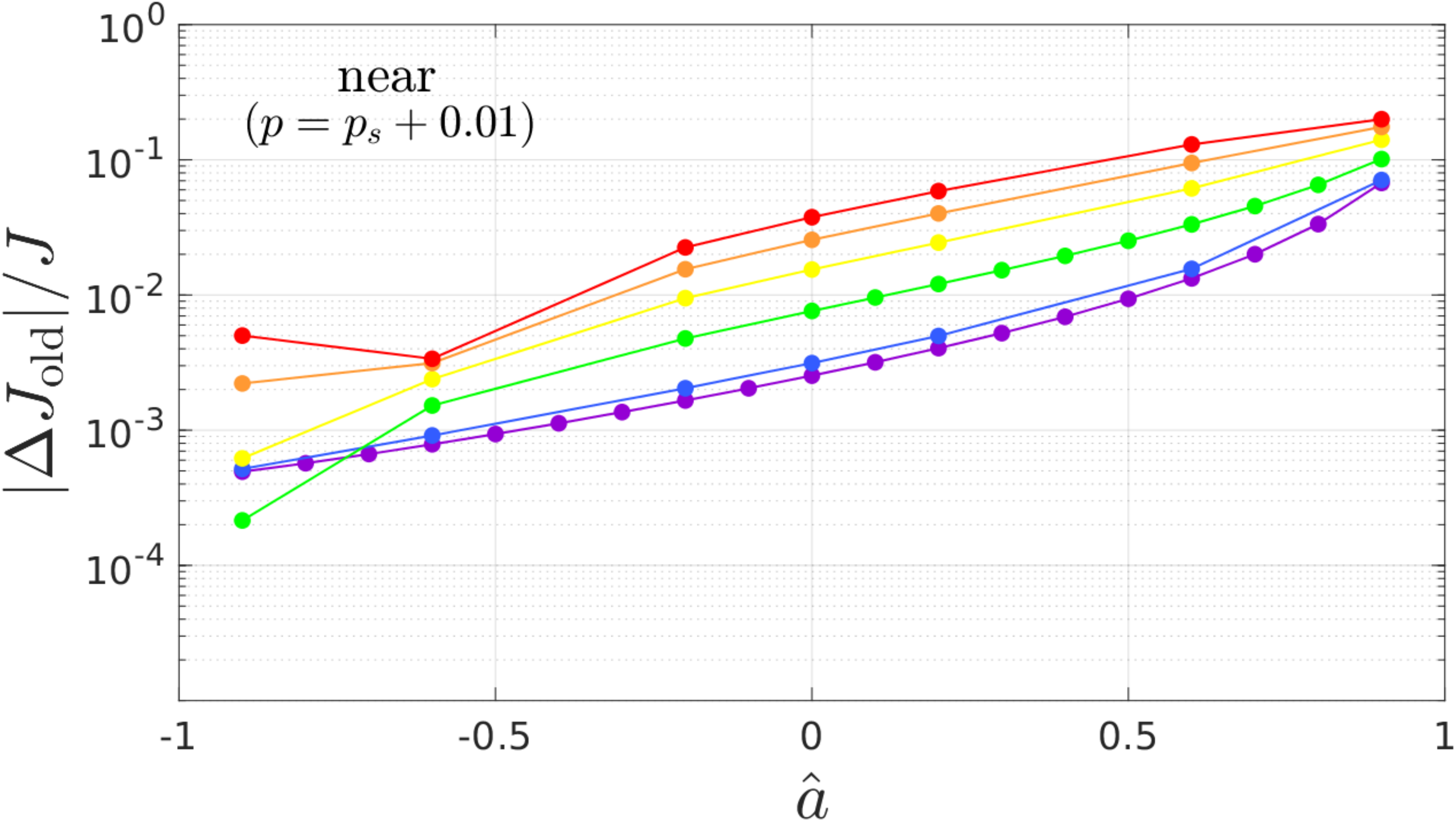}
	\includegraphics[width=0.32\textwidth,height=3.2cm]{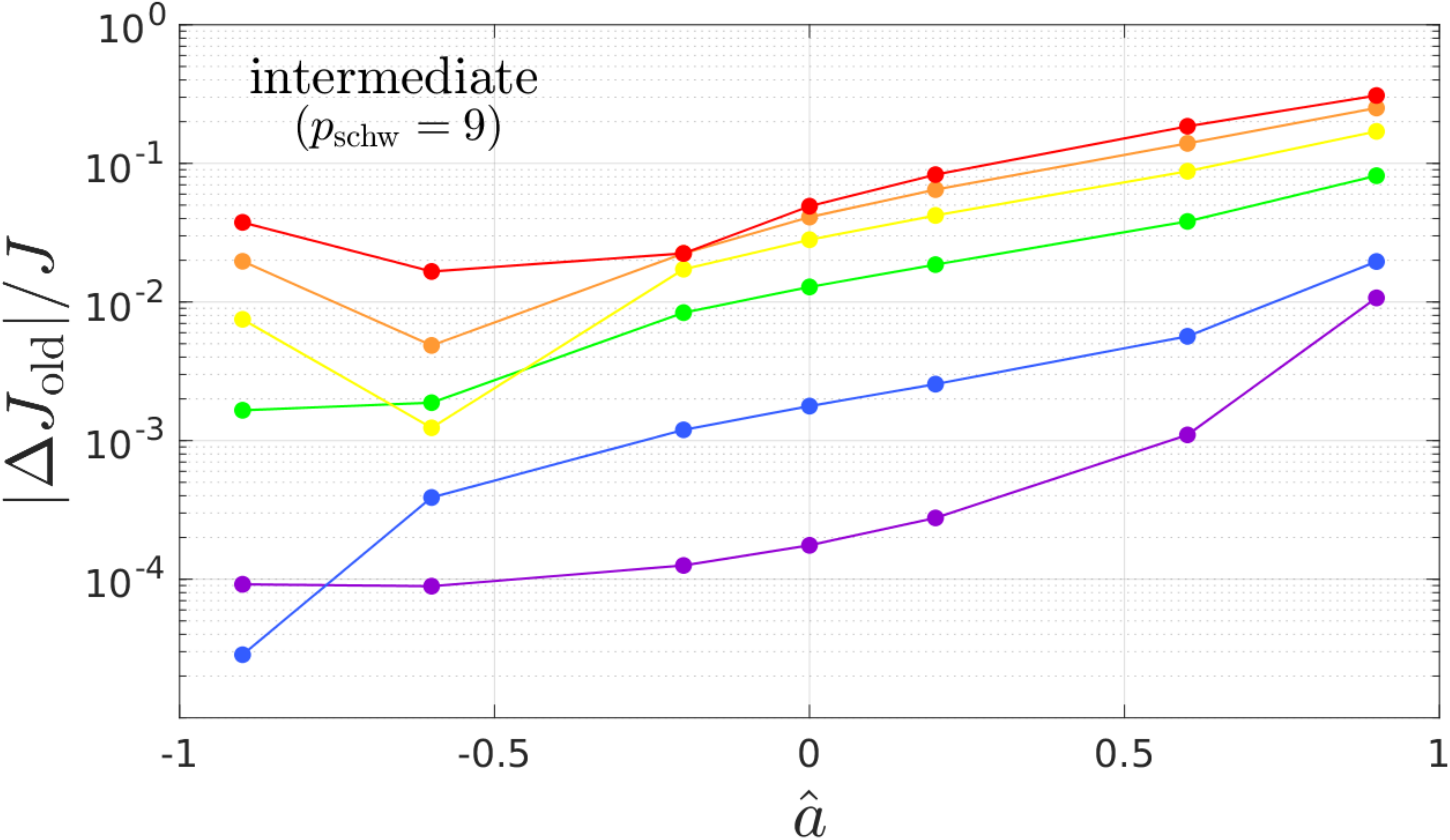}
	\includegraphics[width=0.32\textwidth,height=3.2cm]{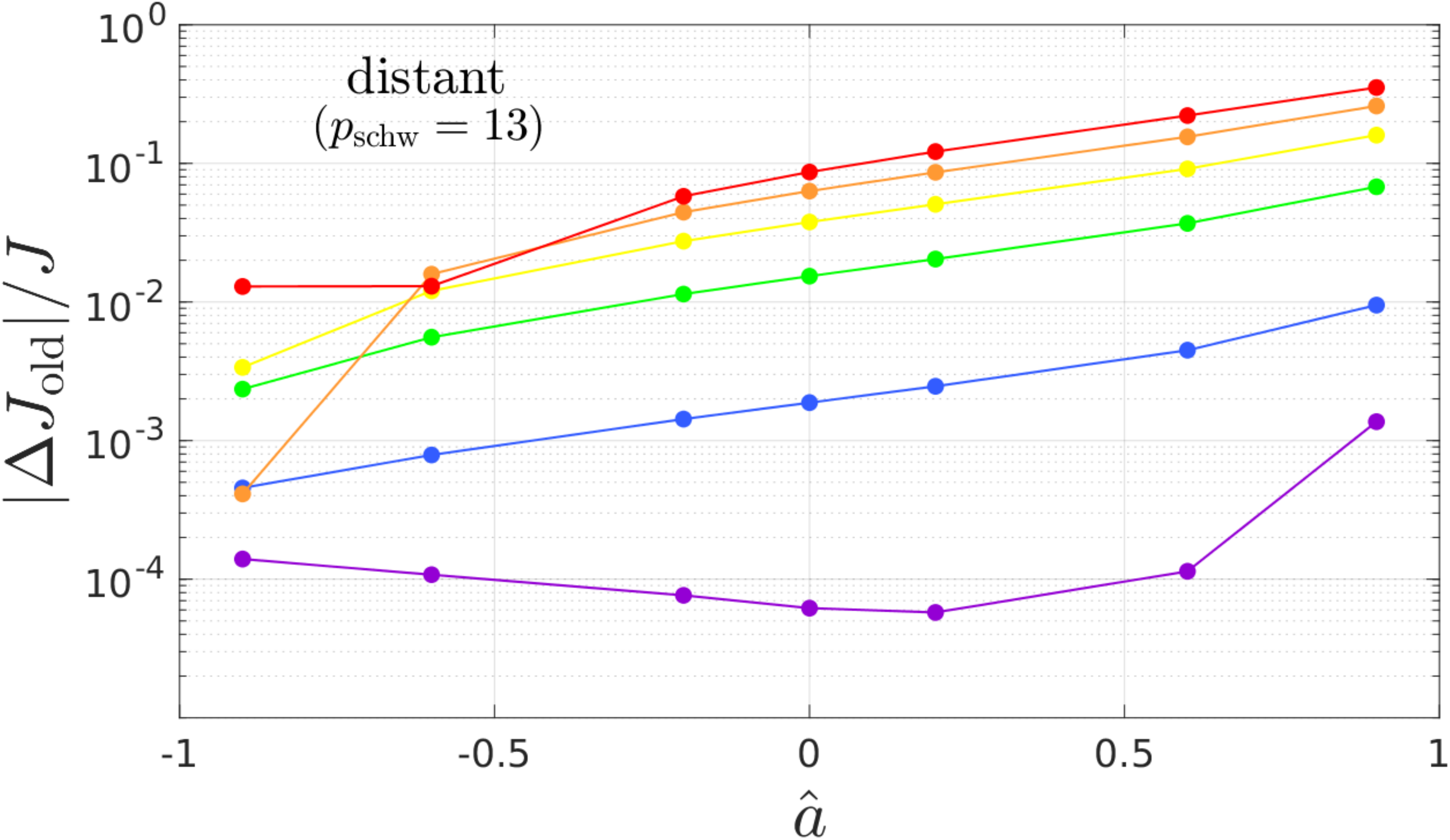} \\

	\caption{\label{fig:reldiff_rainbow_Fold} Relative differences between numerical and analytical
	   averaged fluxes plotted against the spin (absolute value, logarithmic scale). 
	   Here we consider  $\FNPold$, the fluxes computed with the angular radiation reaction 
	   of Eq.~\eqref{eq:Fphi_ecc_old}. 
	   The analogous plots for $\FNP$ are shown in Fig.~\ref{fig:reldiff_rainbow_FNP}.
	   }
\end{figure*}

\begin{figure*}[t]
	\center
	\includegraphics[width=0.32\textwidth,height=3.2cm]{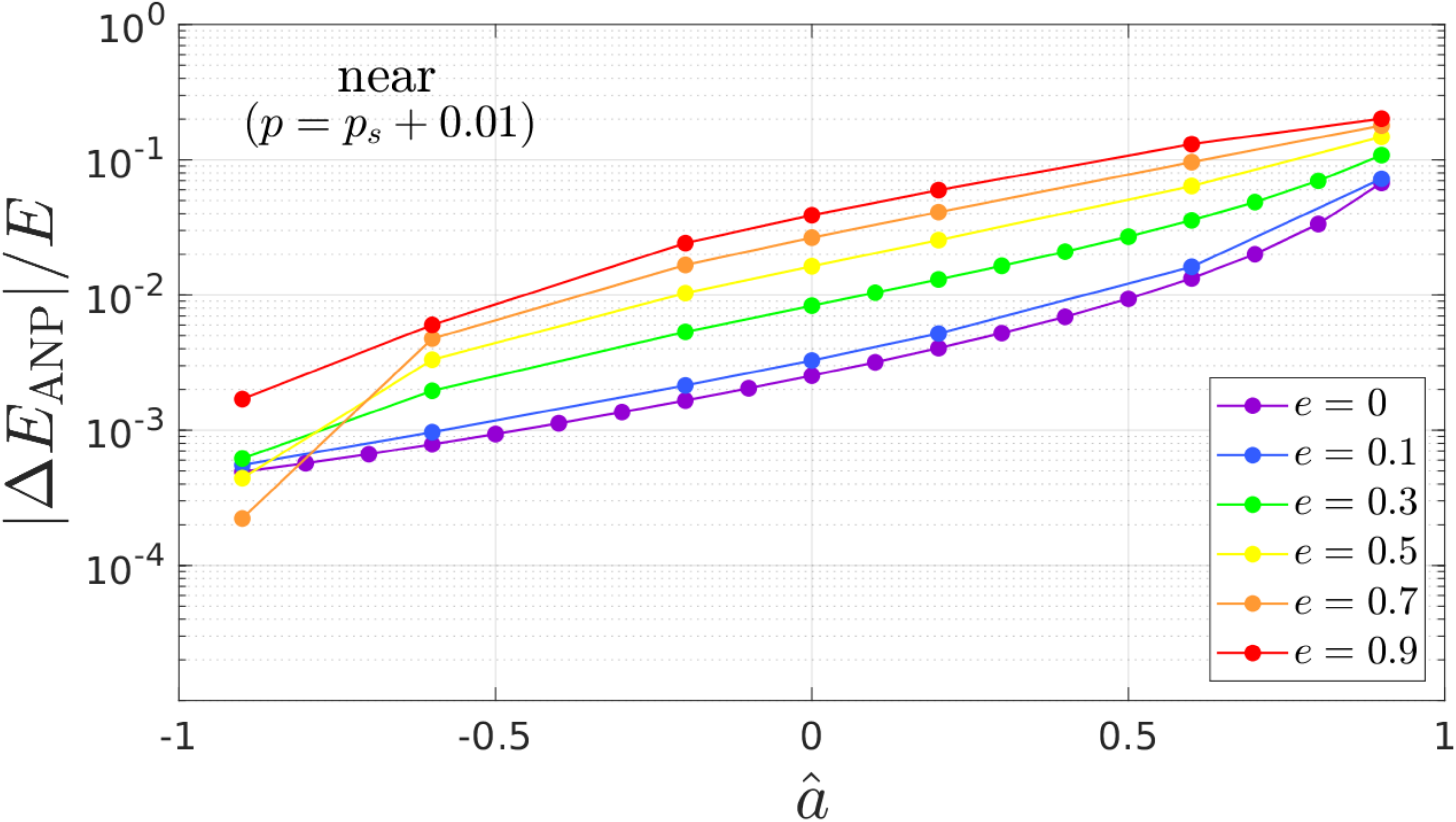}
	\includegraphics[width=0.32\textwidth,height=3.2cm]{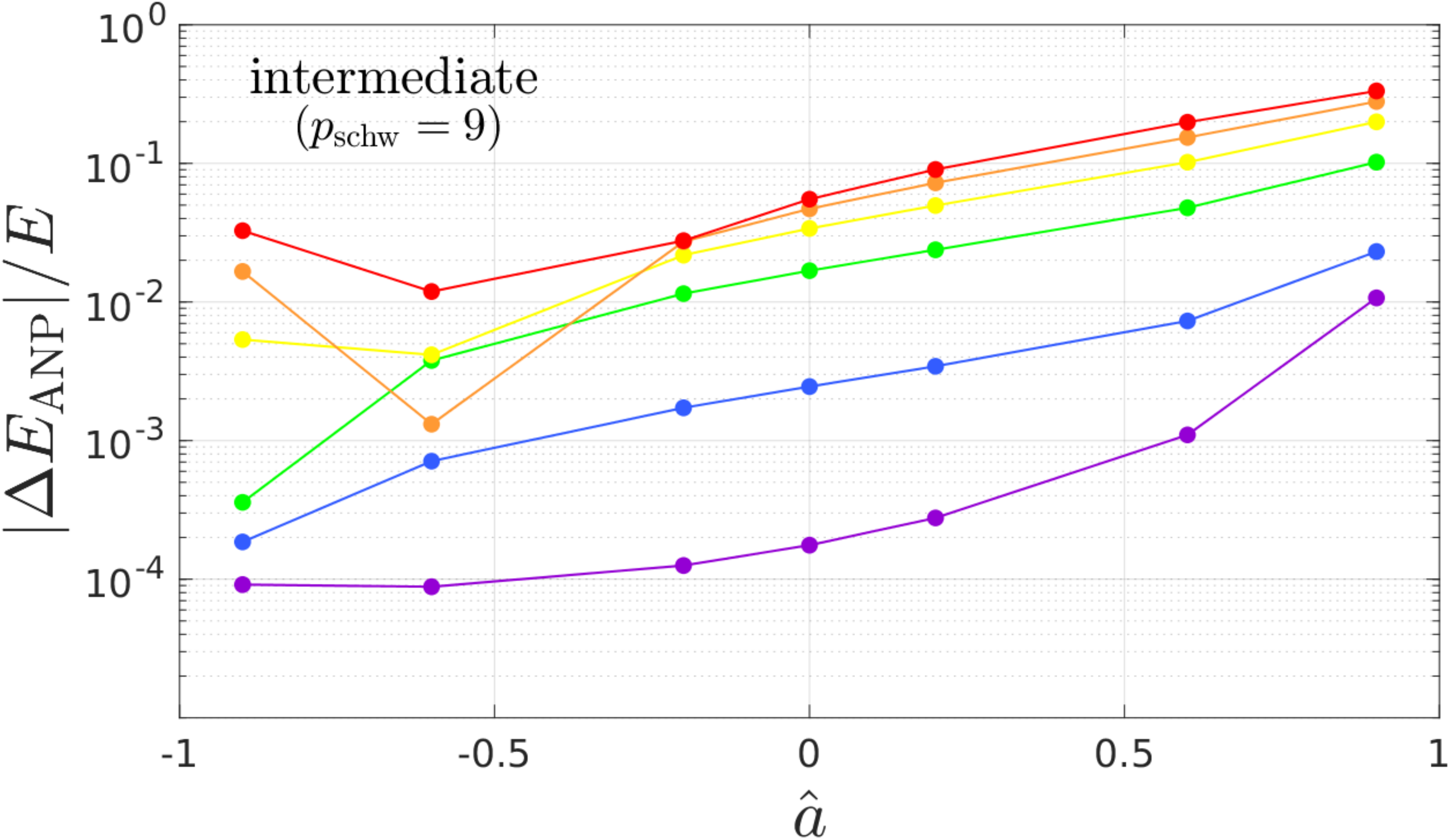}
	\includegraphics[width=0.32\textwidth,height=3.2cm]{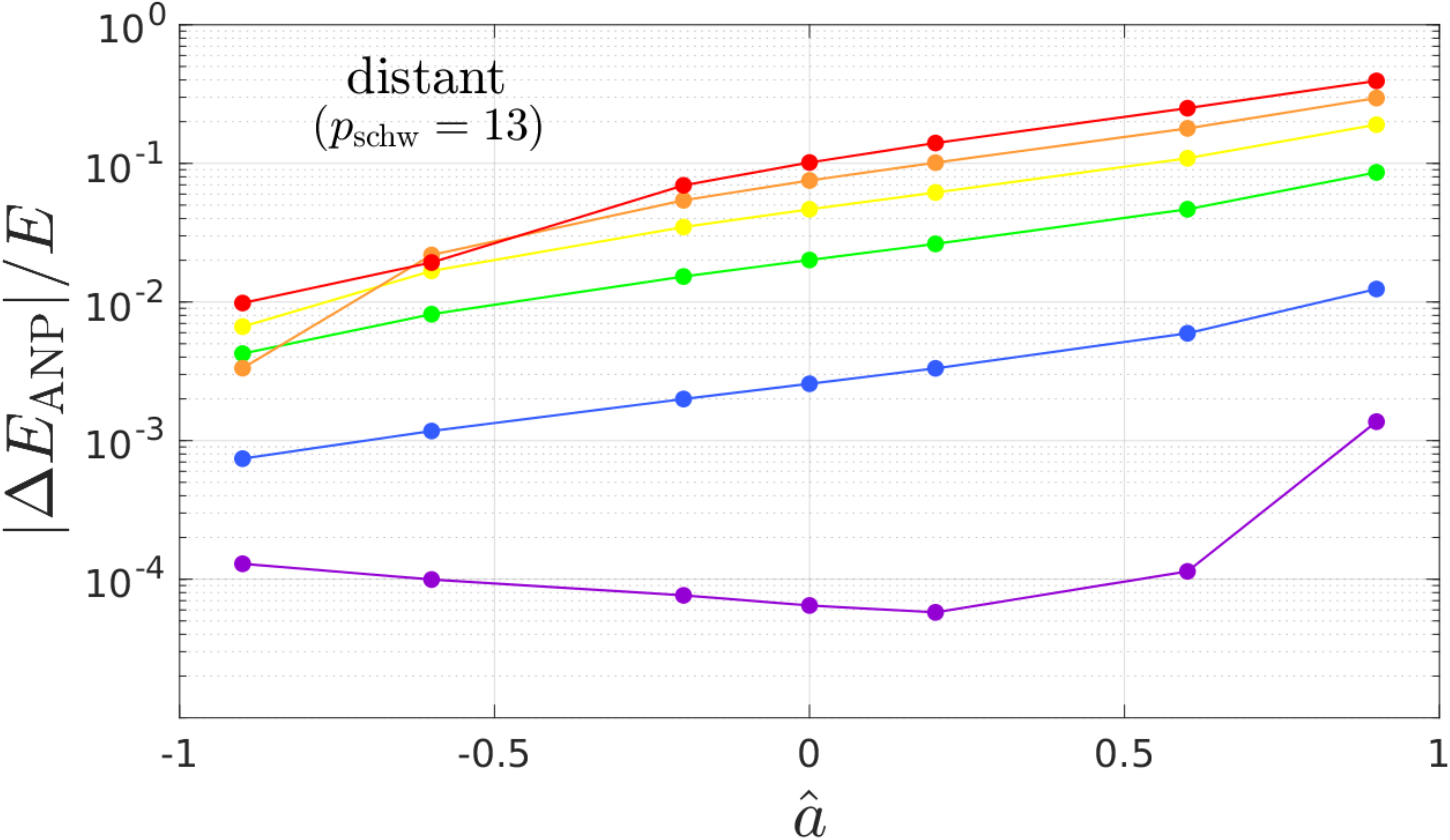} \\ 
	\includegraphics[width=0.32\textwidth,height=3.2cm]{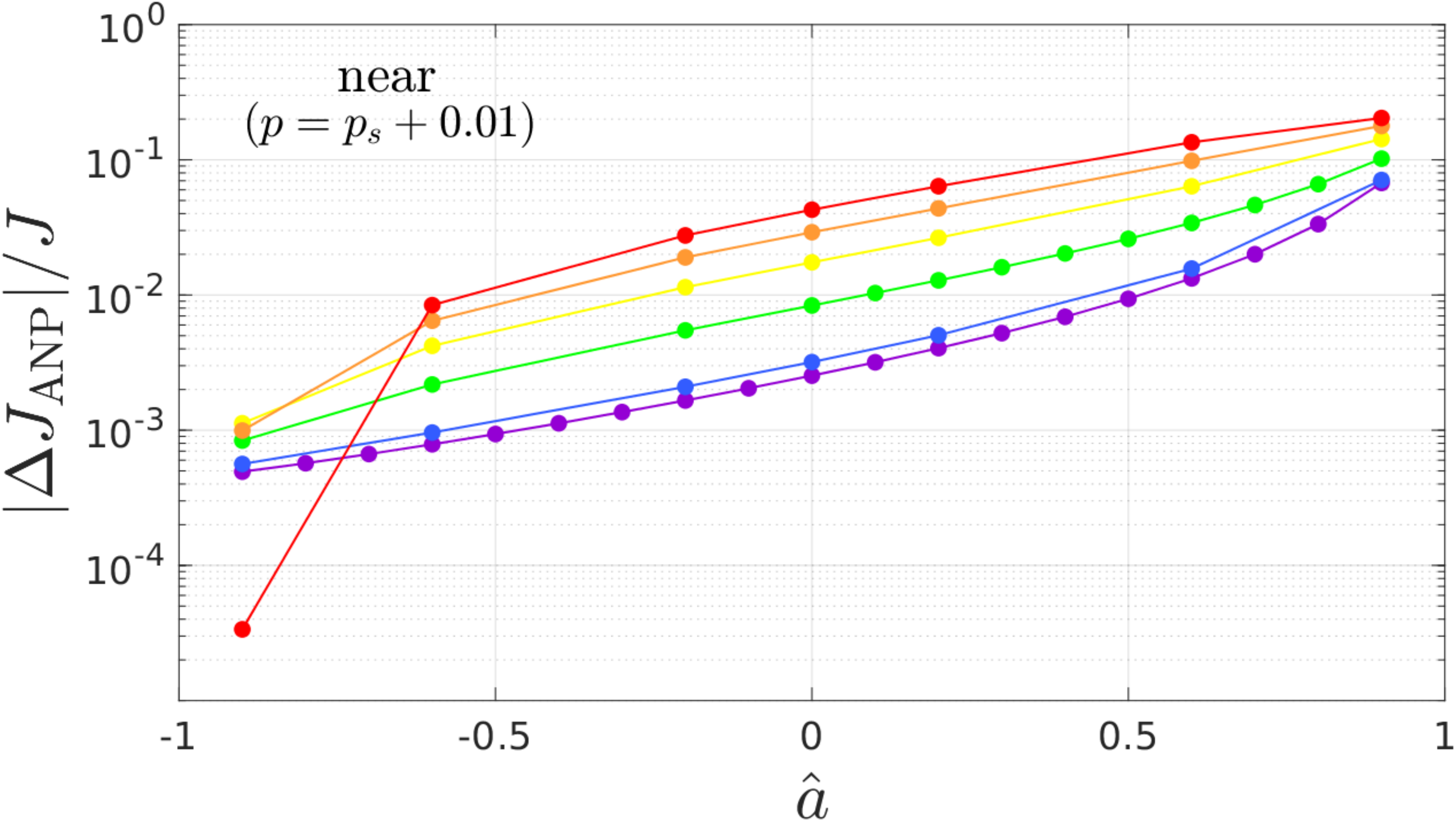}
	\includegraphics[width=0.32\textwidth,height=3.2cm]{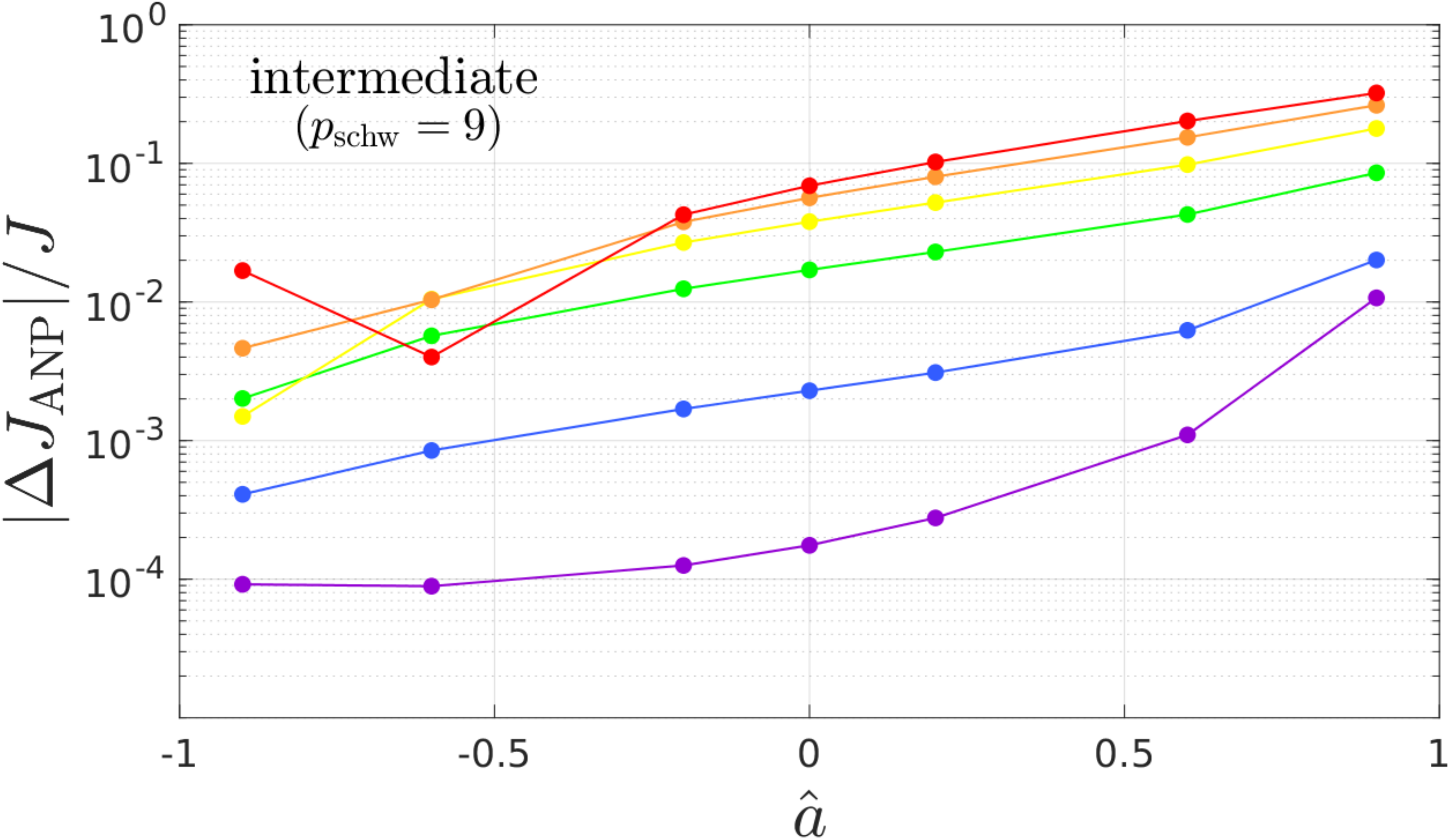}
	\includegraphics[width=0.32\textwidth,height=3.2cm]{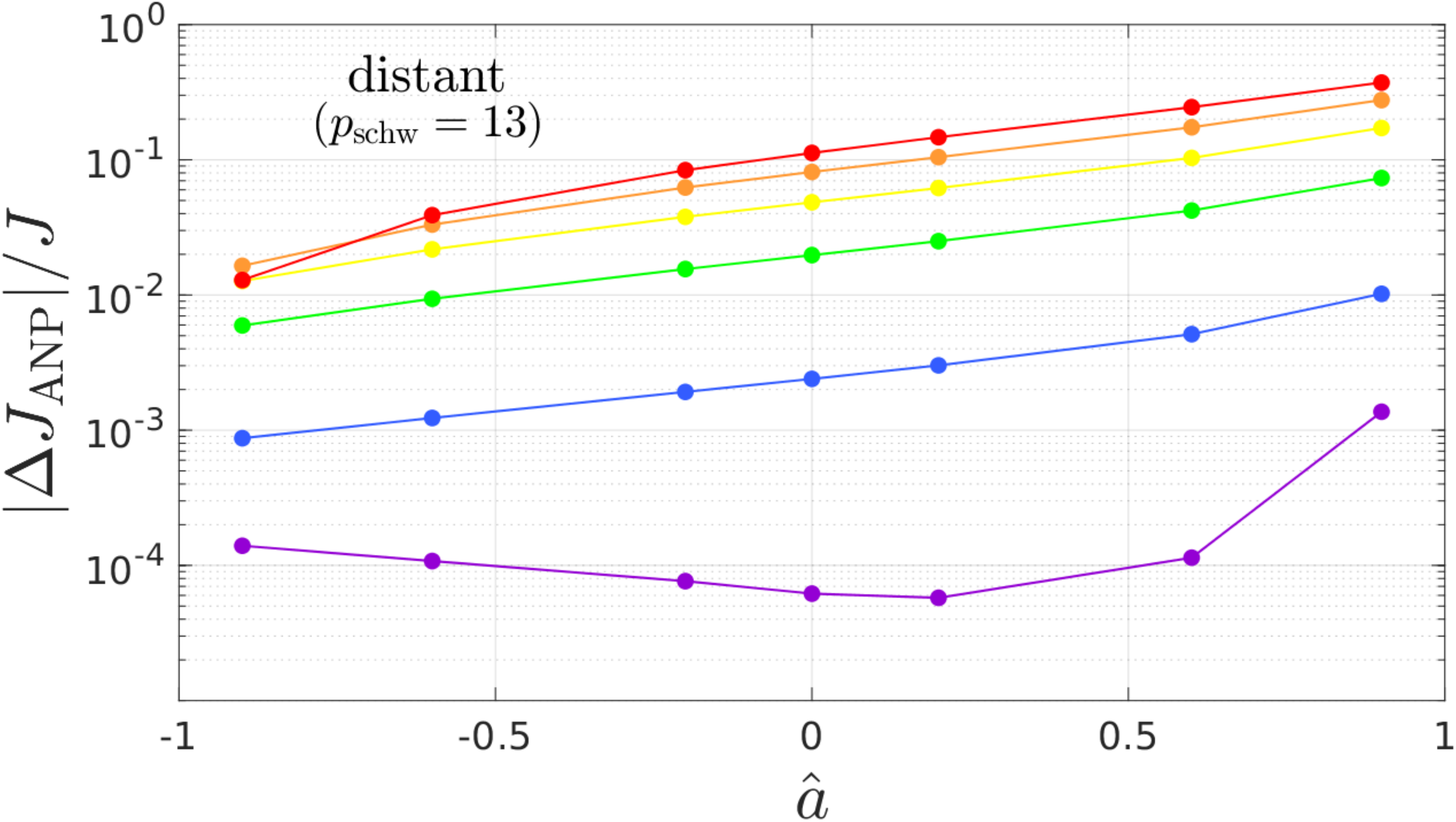} \\
	
	\caption{\label{fig:reldiff_rainbow_FANP} Analogous of Fig.~\ref{fig:reldiff_rainbow_Fold}
	for the $\FANP$, the fluxes obtained with the angular radiation 
	reaction of Eq.~\eqref{eq:Fphi_ANP}.}
\end{figure*}

\begin{figure*}[t]
	\center
	\includegraphics[width=0.32\textwidth,height=3.2cm]{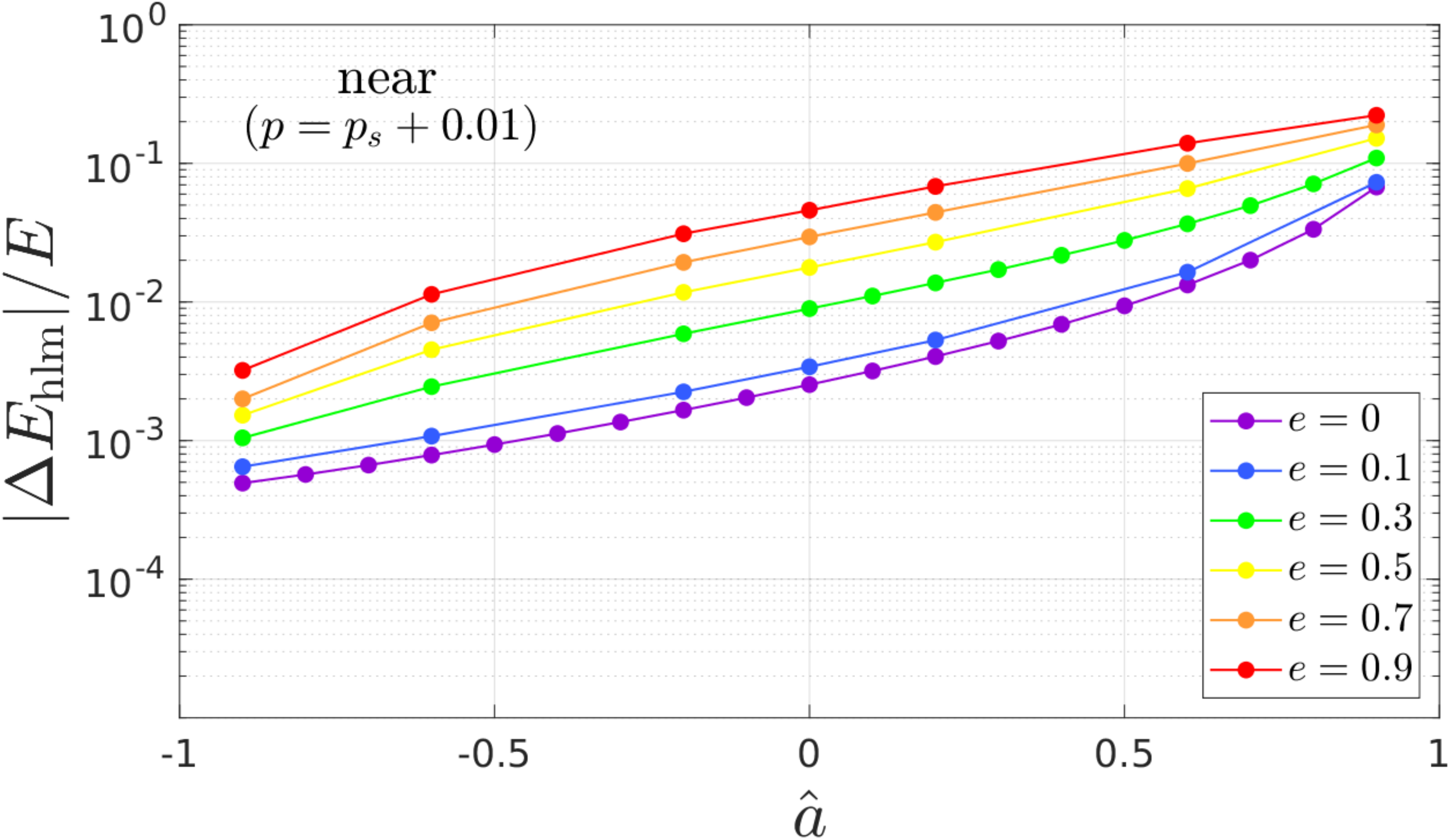}
	\includegraphics[width=0.32\textwidth,height=3.2cm]{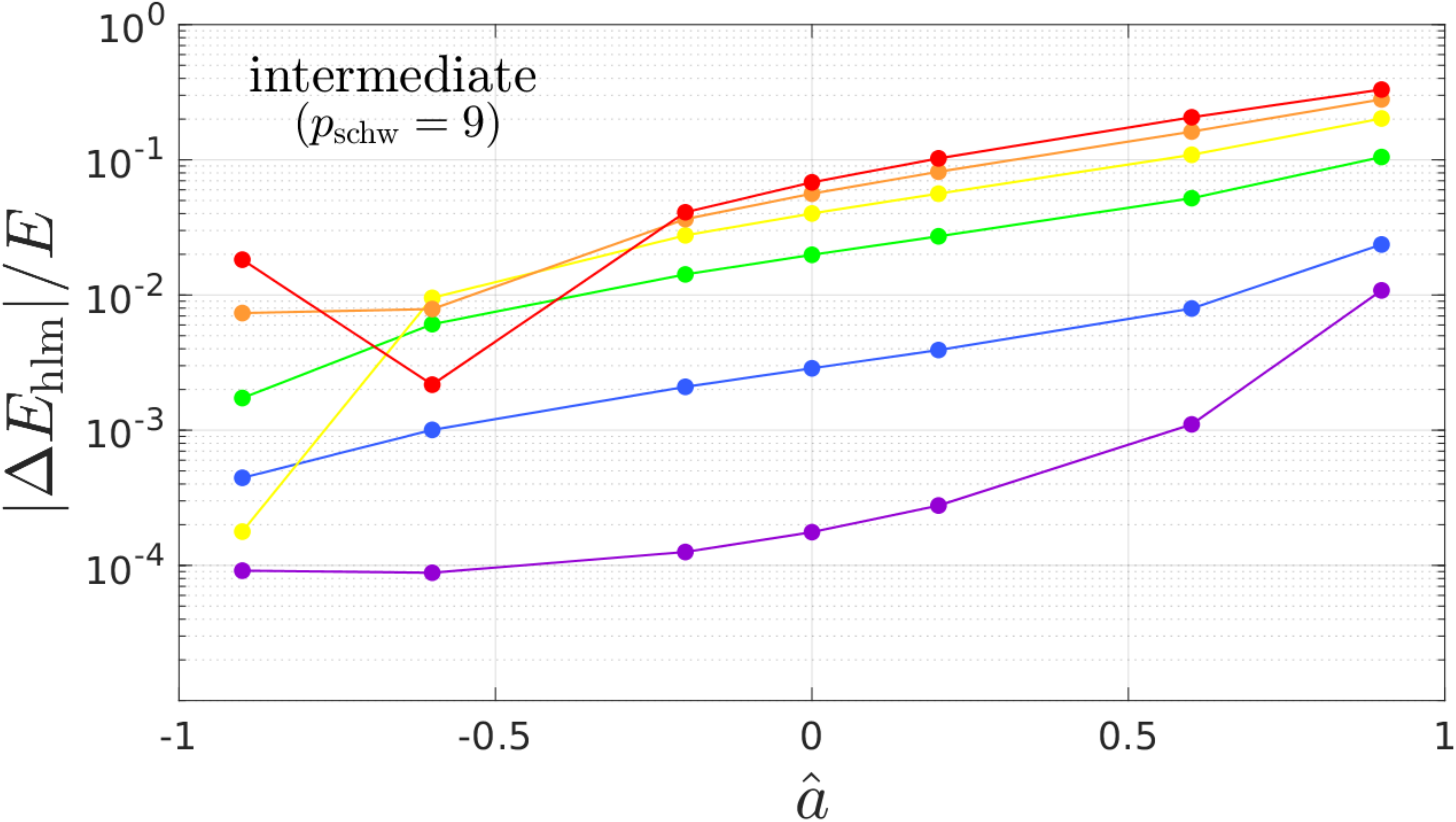}
	\includegraphics[width=0.32\textwidth,height=3.2cm]{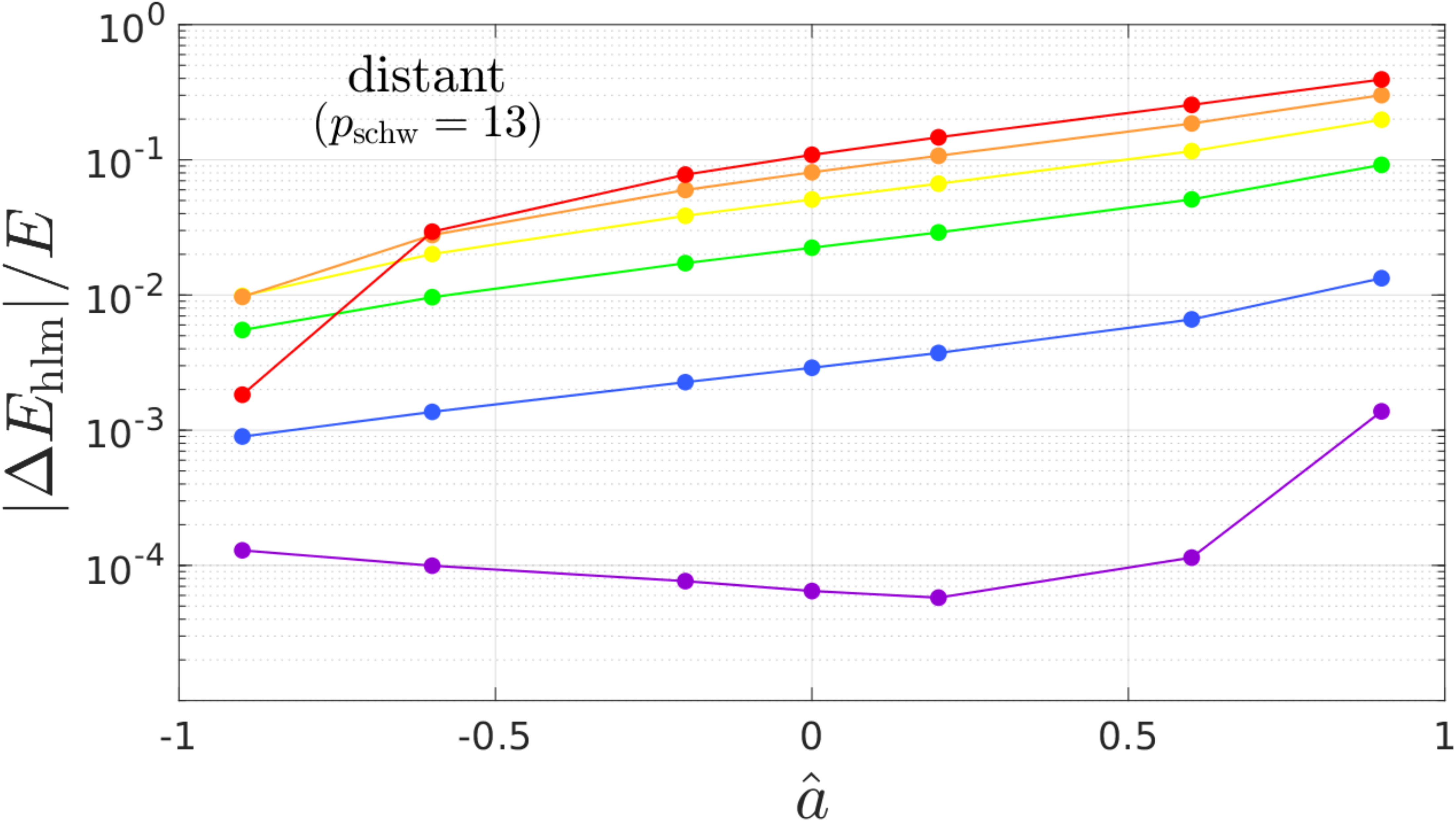} \\ 
	\includegraphics[width=0.32\textwidth,height=3.2cm]{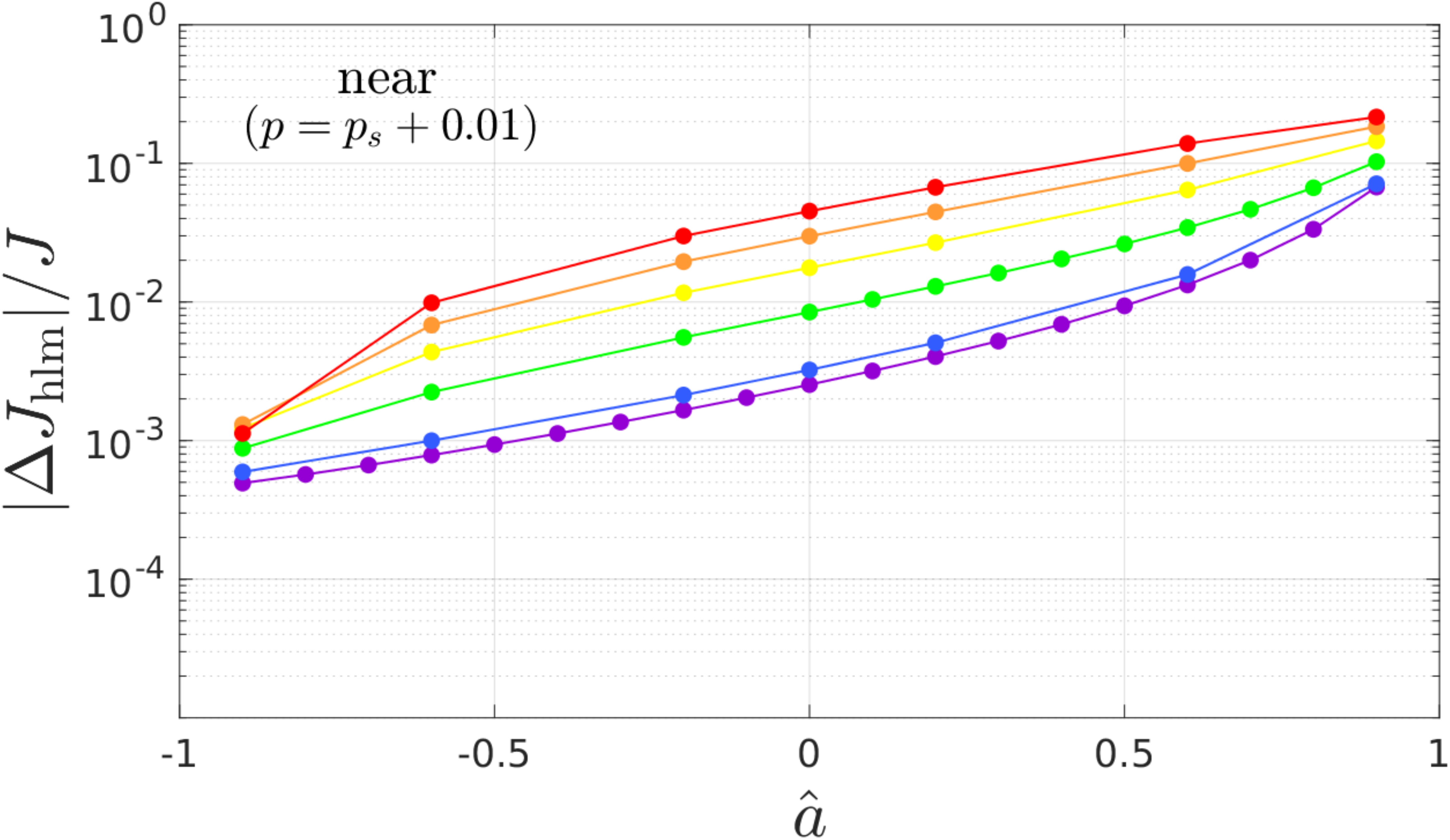}
	\includegraphics[width=0.32\textwidth,height=3.2cm]{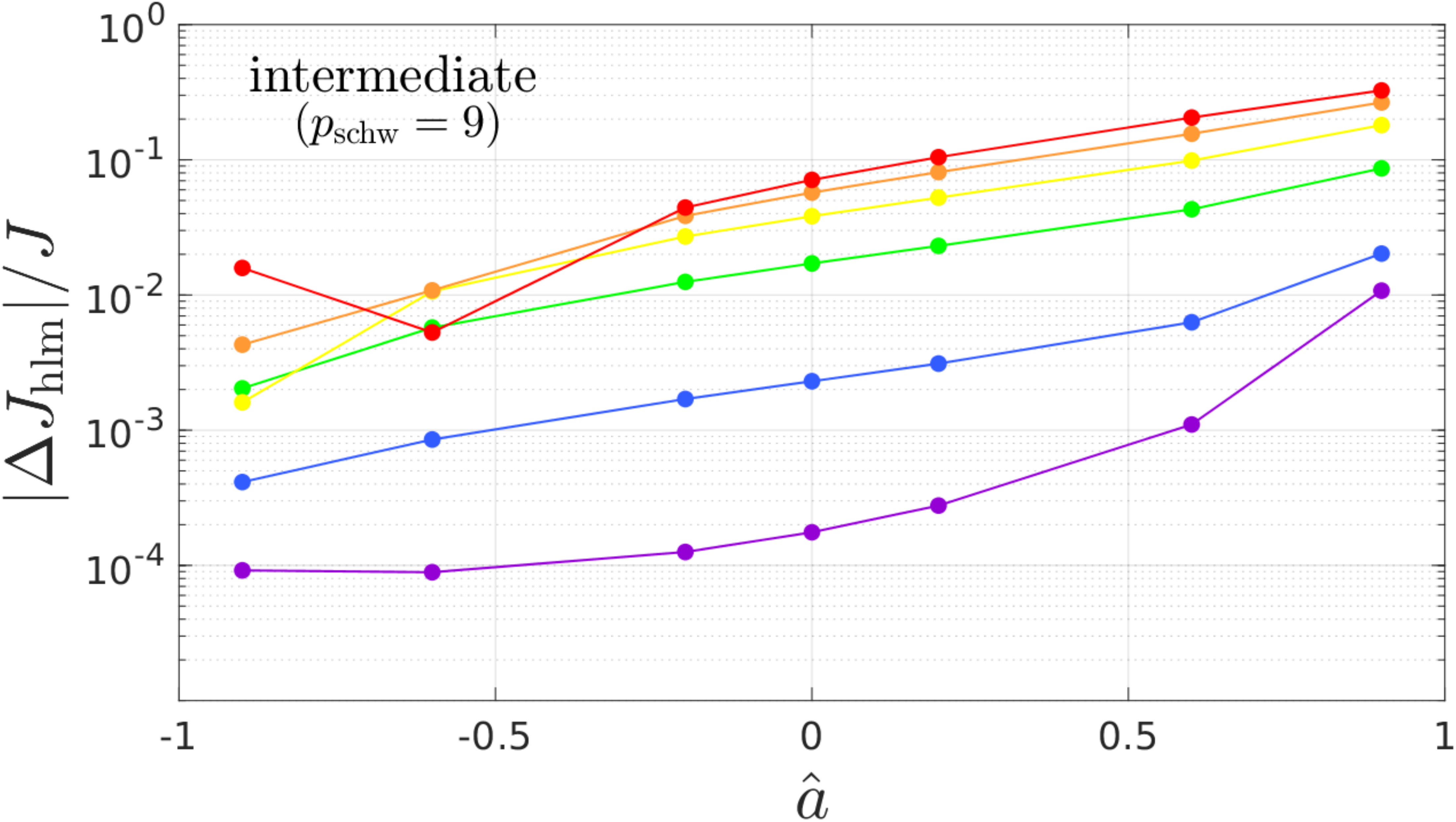}
	\includegraphics[width=0.32\textwidth,height=3.2cm]{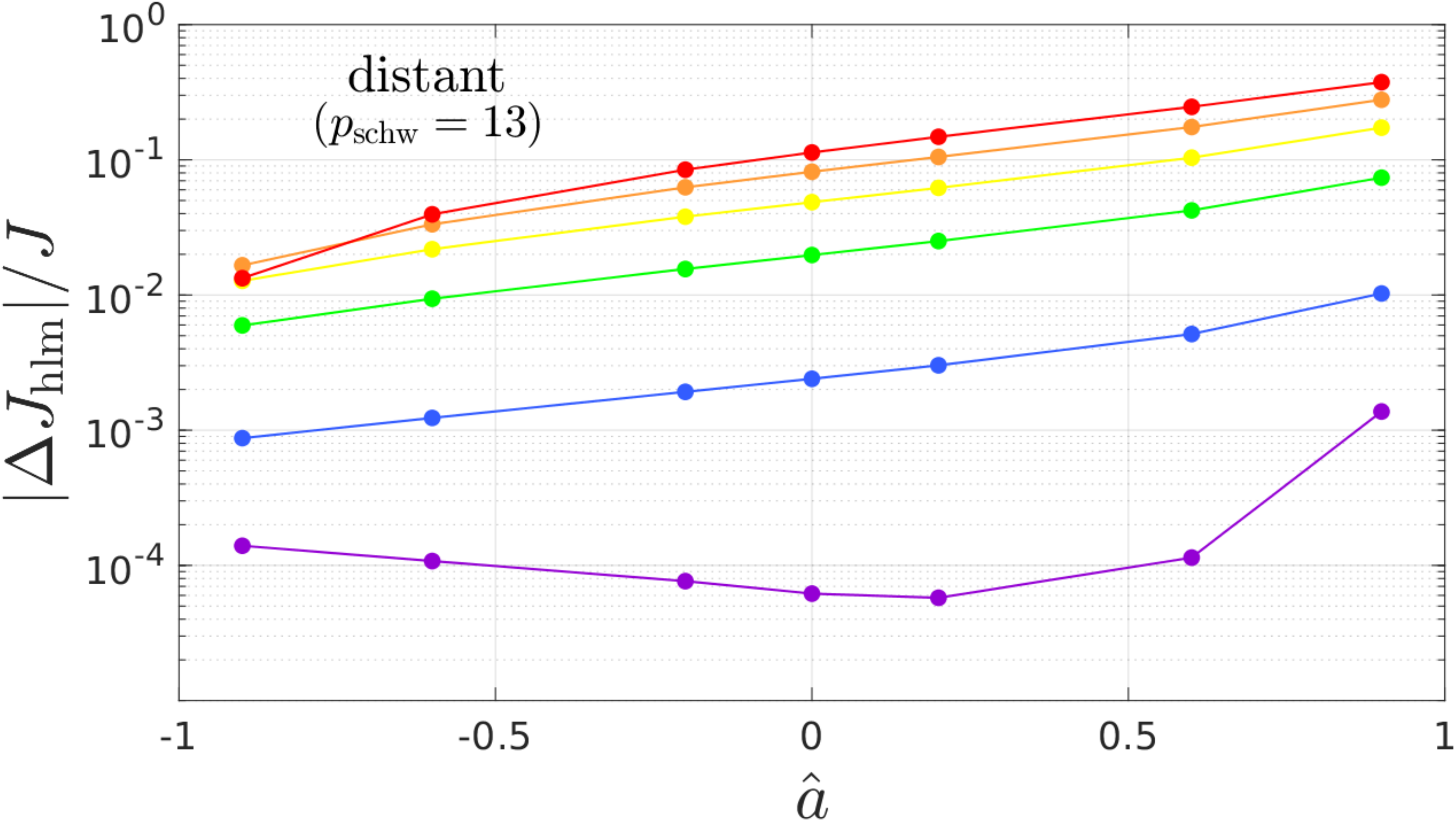} \\
	
	\caption{\label{fig:reldiff_rainbow_Fhlm} Analogous of Fig.~\ref{fig:reldiff_rainbow_Fold}
	for the fluxes computed using the EOB waveform and Eq.~\eqref{eq:fluxes_infty}.}
\end{figure*}

\begin{figure*}[]
	\center
	\includegraphics[width=0.24\textwidth,height=3.5cm]{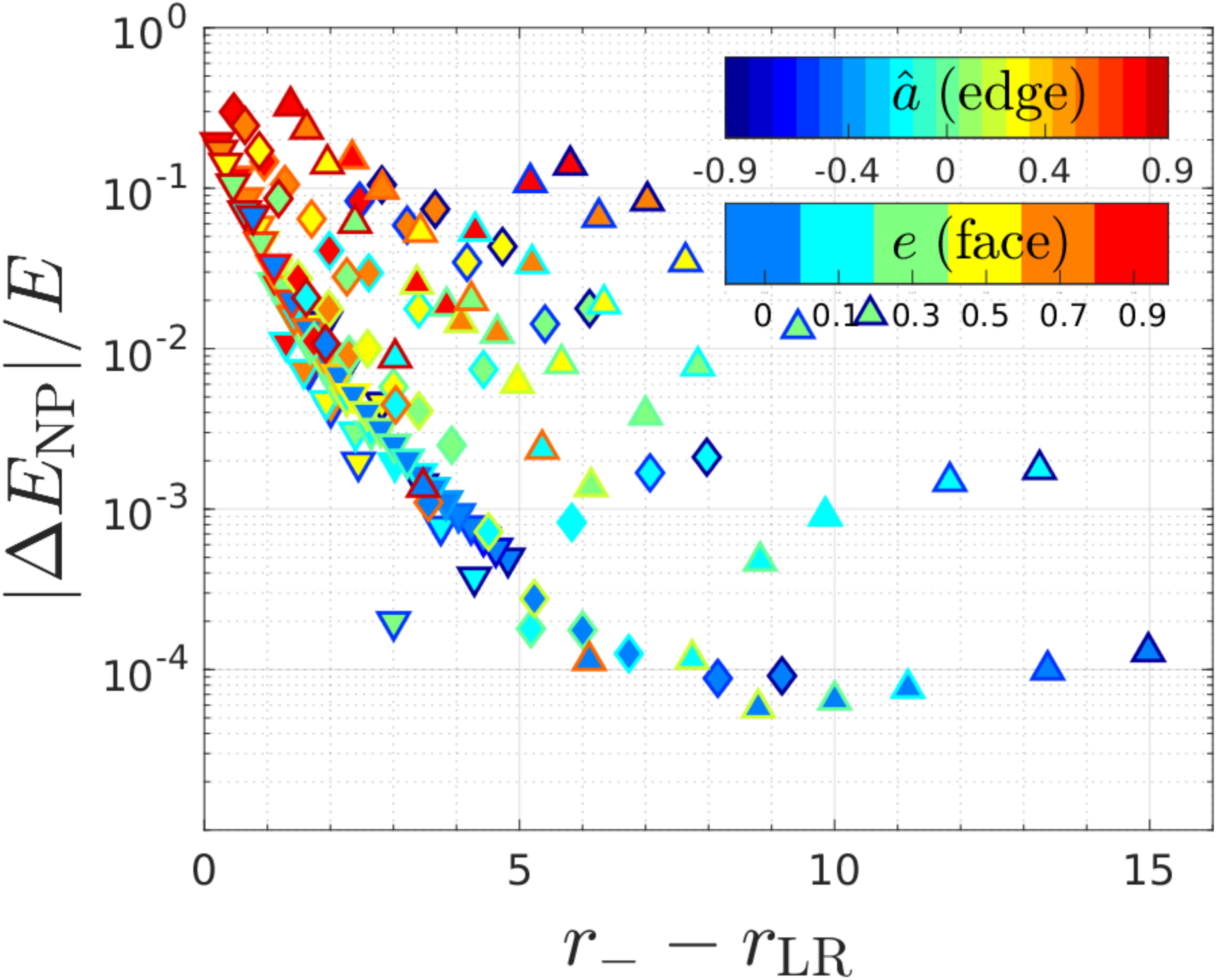}
	\includegraphics[width=0.24\textwidth,height=3.5cm]{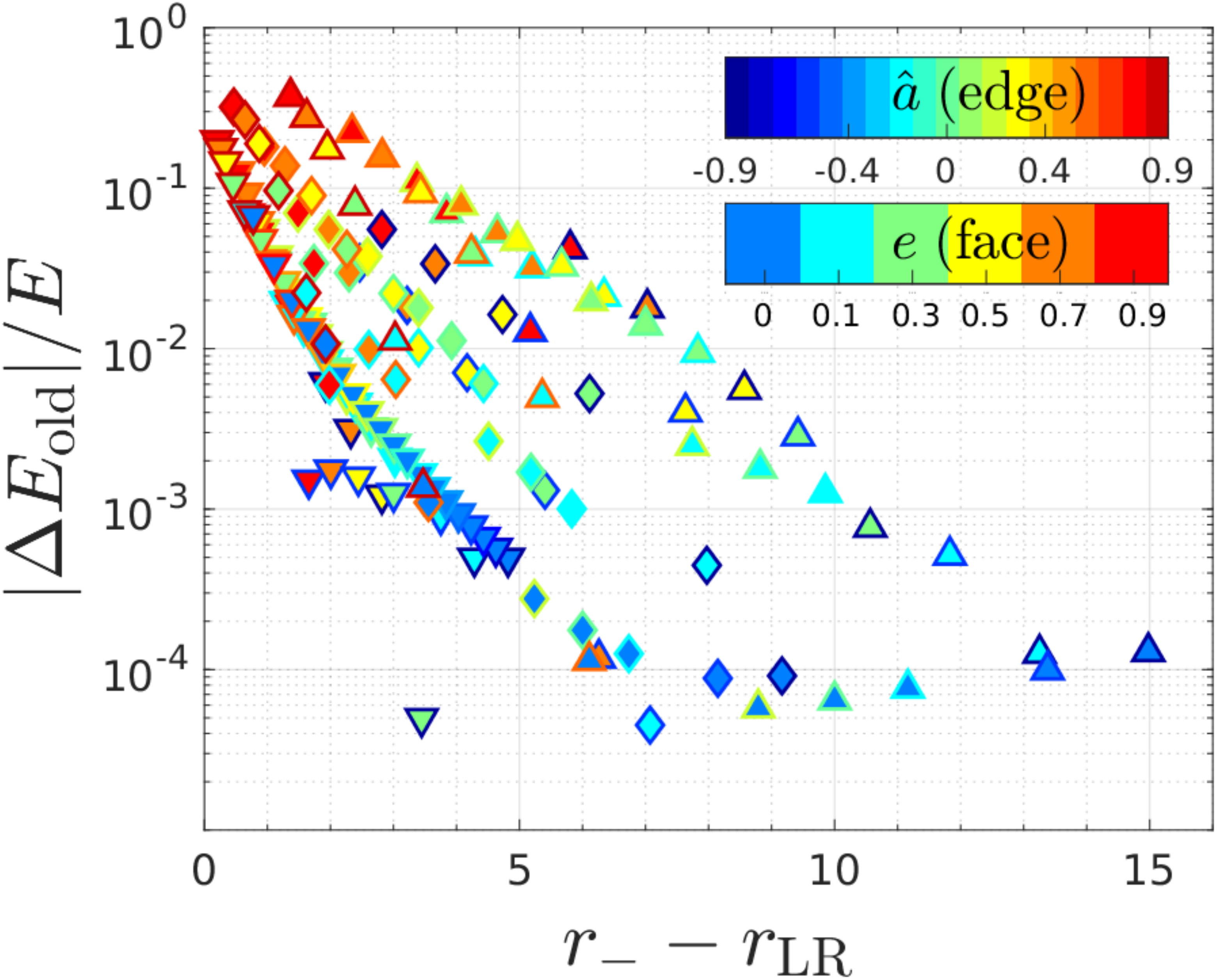}
	\includegraphics[width=0.24\textwidth,height=3.5cm]{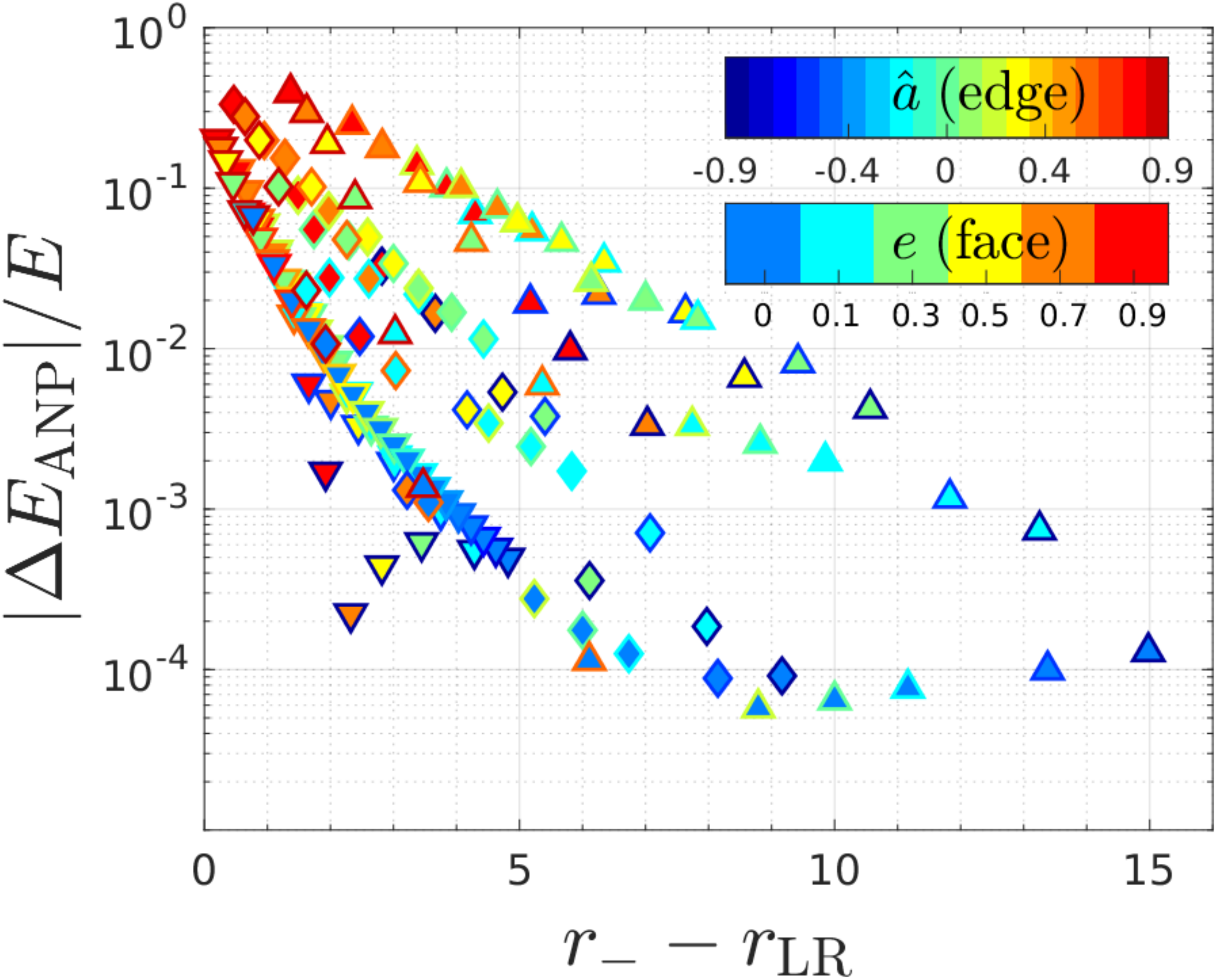}
	\includegraphics[width=0.24\textwidth,height=3.5cm]{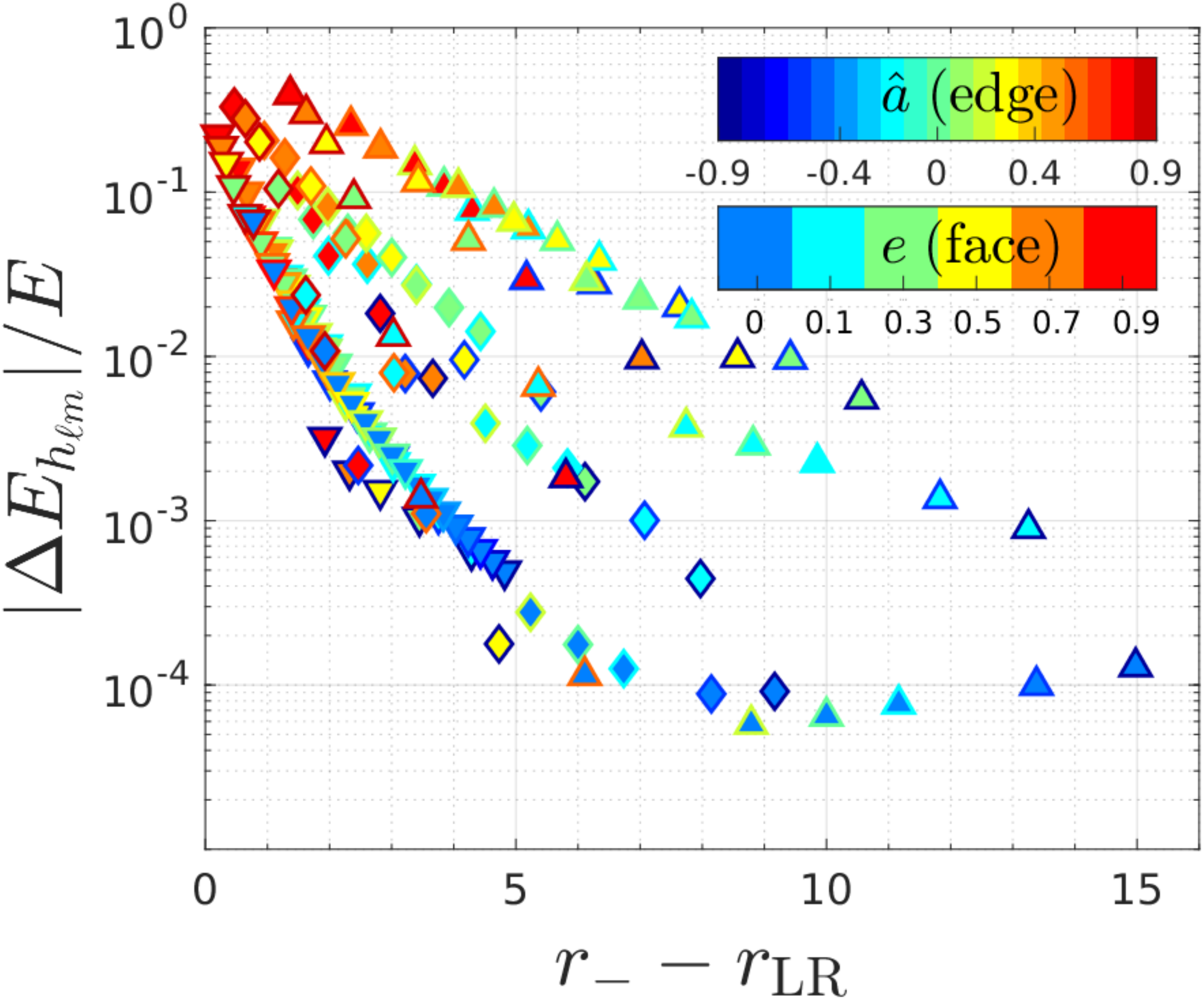}\\
	\includegraphics[width=0.24\textwidth,height=3.5cm]{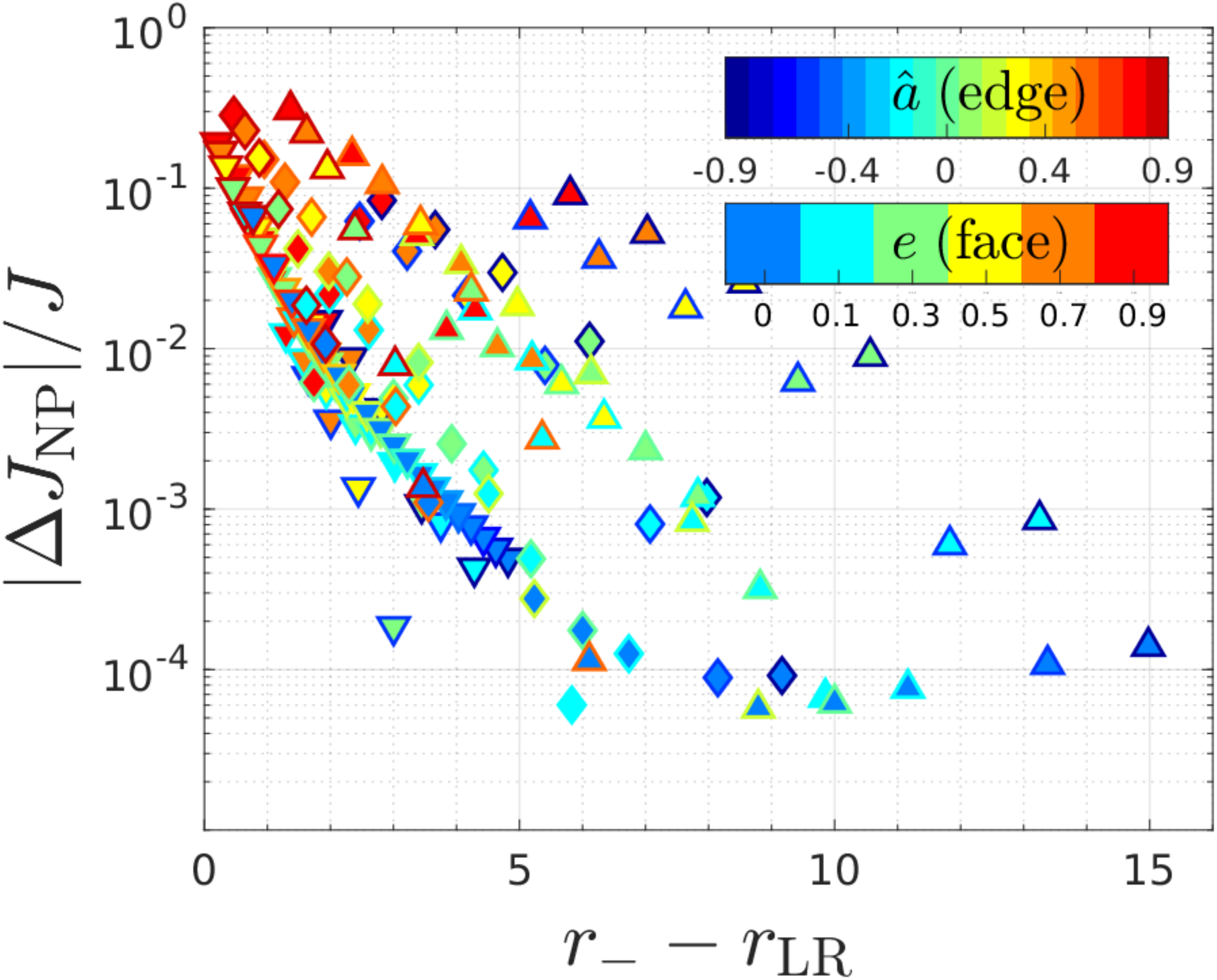}
	\includegraphics[width=0.24\textwidth,height=3.5cm]{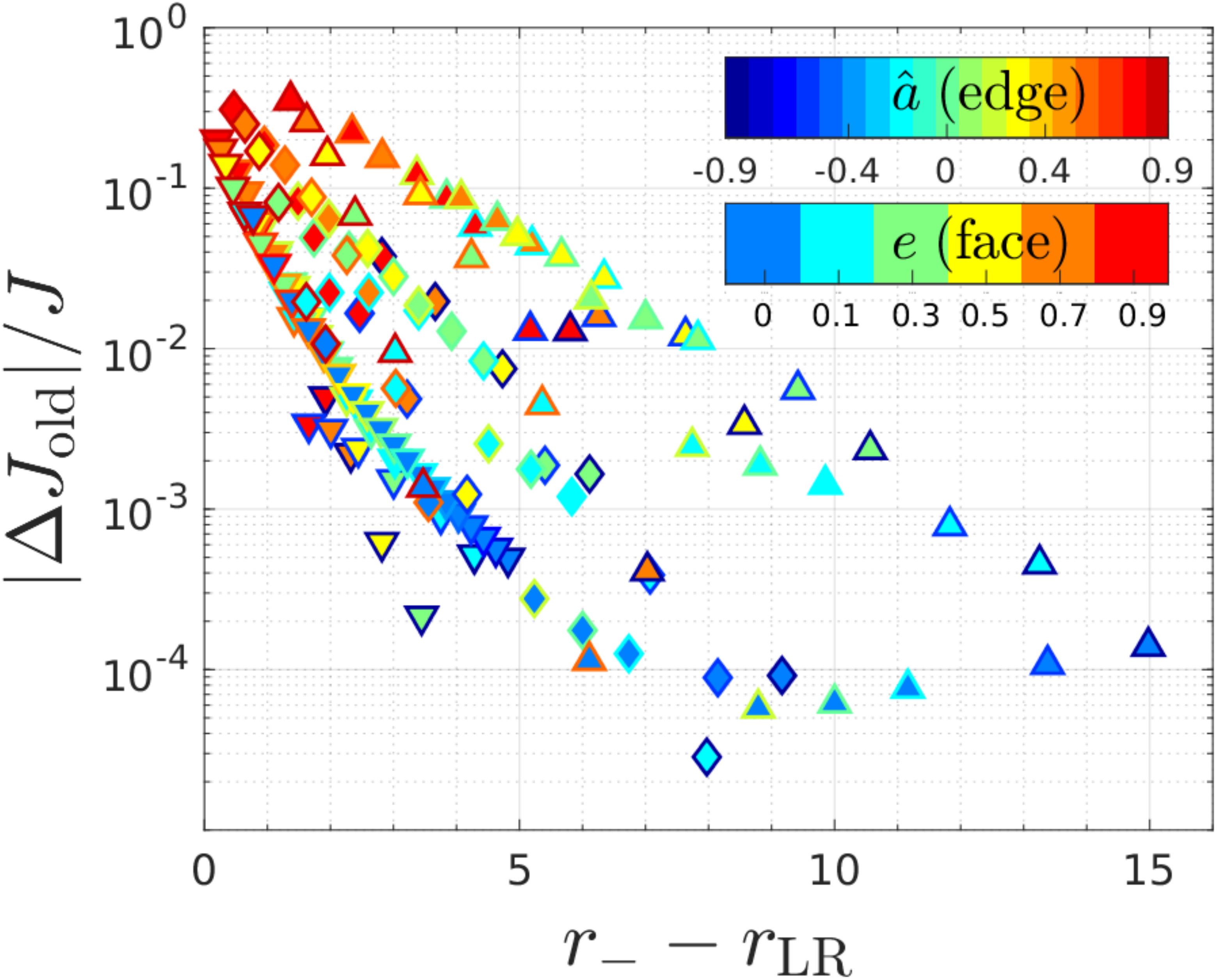}
	\includegraphics[width=0.24\textwidth,height=3.5cm]{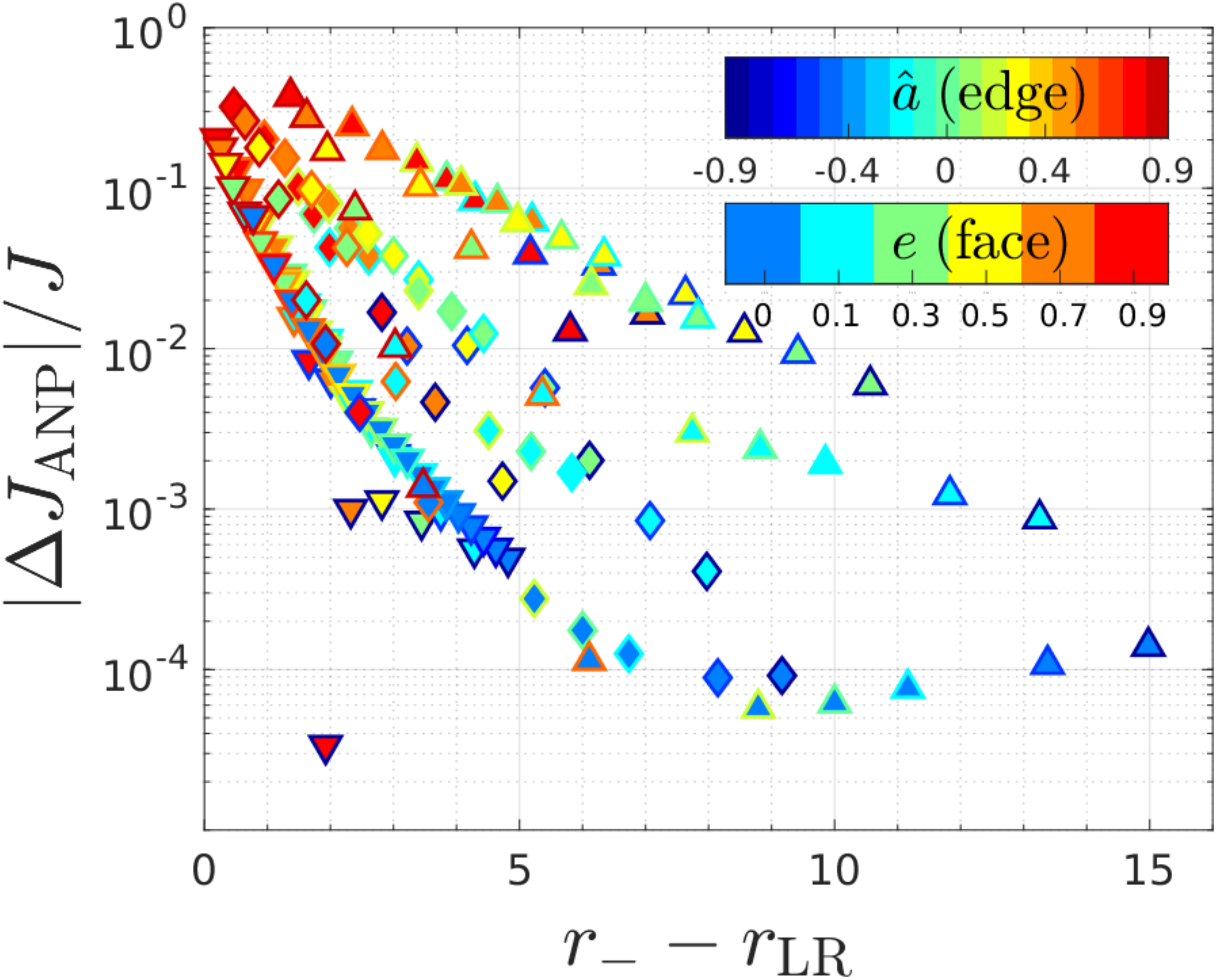}
	\includegraphics[width=0.24\textwidth,height=3.5cm]{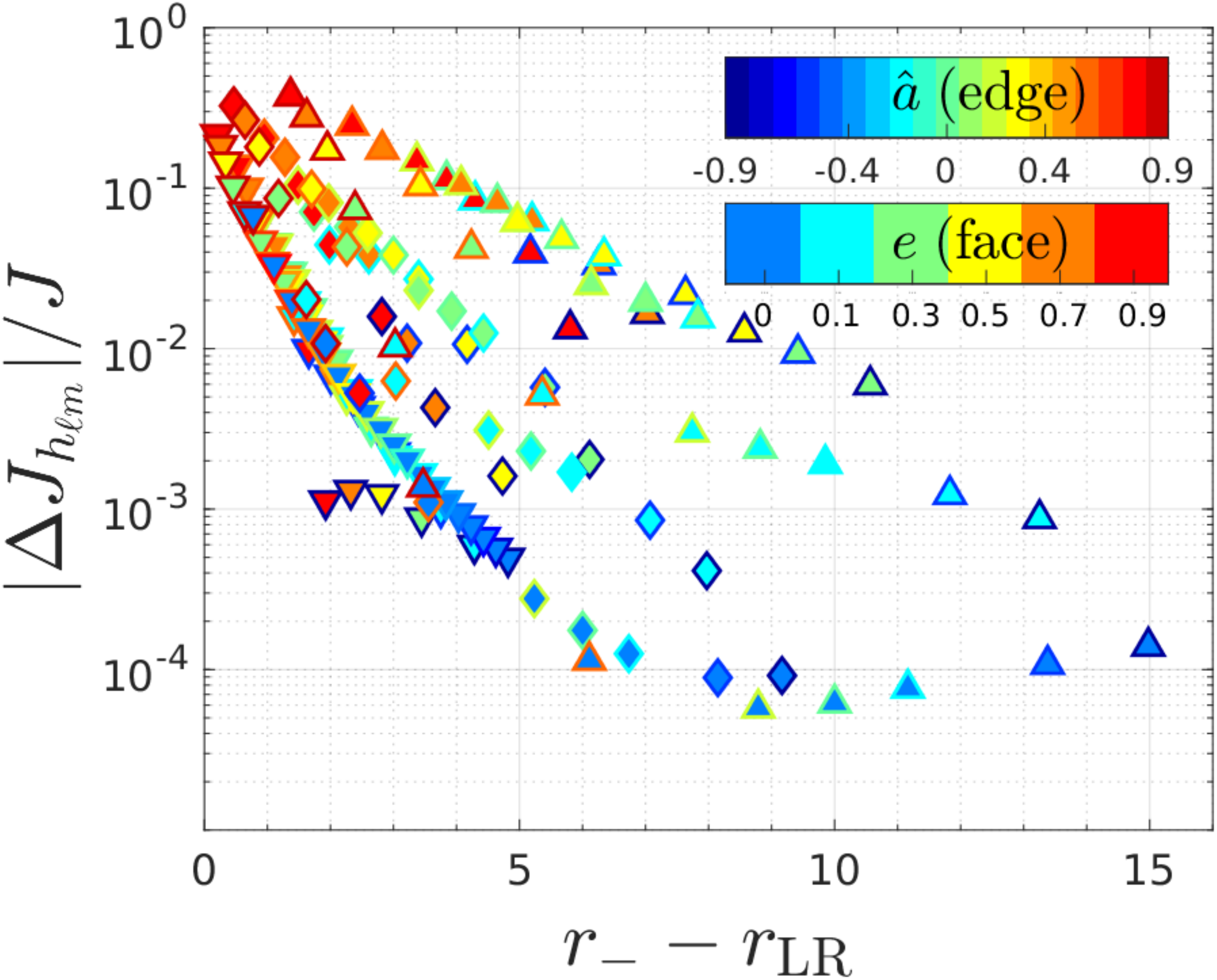}\\
	\caption{\label{fig:reldiff_harlequin_all_log} 
	Absolute values of the relative difference between numerical and analytical 
	fluxes averaged along a radial orbits for all the analytical prescriptions studied in 
	this work (see Sec.~\ref{sec:radreac} for more details). 
	The face color of the markers indicates the eccentricity, while the edge color indicates the spin.
    The shape of the markers is related to the rule used for the semilatus rectum: the reverse 
	triangle indicates the near simulations ($p = p_s + 0.01$), the diamond is for the 
	intermediate simulations ( $p=9\;p_s(e,\ha)/p_s(e,0)$ ) and the triangle pointing upward 
	indicates the distant simulations ($p=13\;p_s(e,\ha)/p_s(e,0)$).}
\end{figure*}

\begin{figure*}[]
	\center
	\includegraphics[width=0.24\textwidth,height=3.5cm]{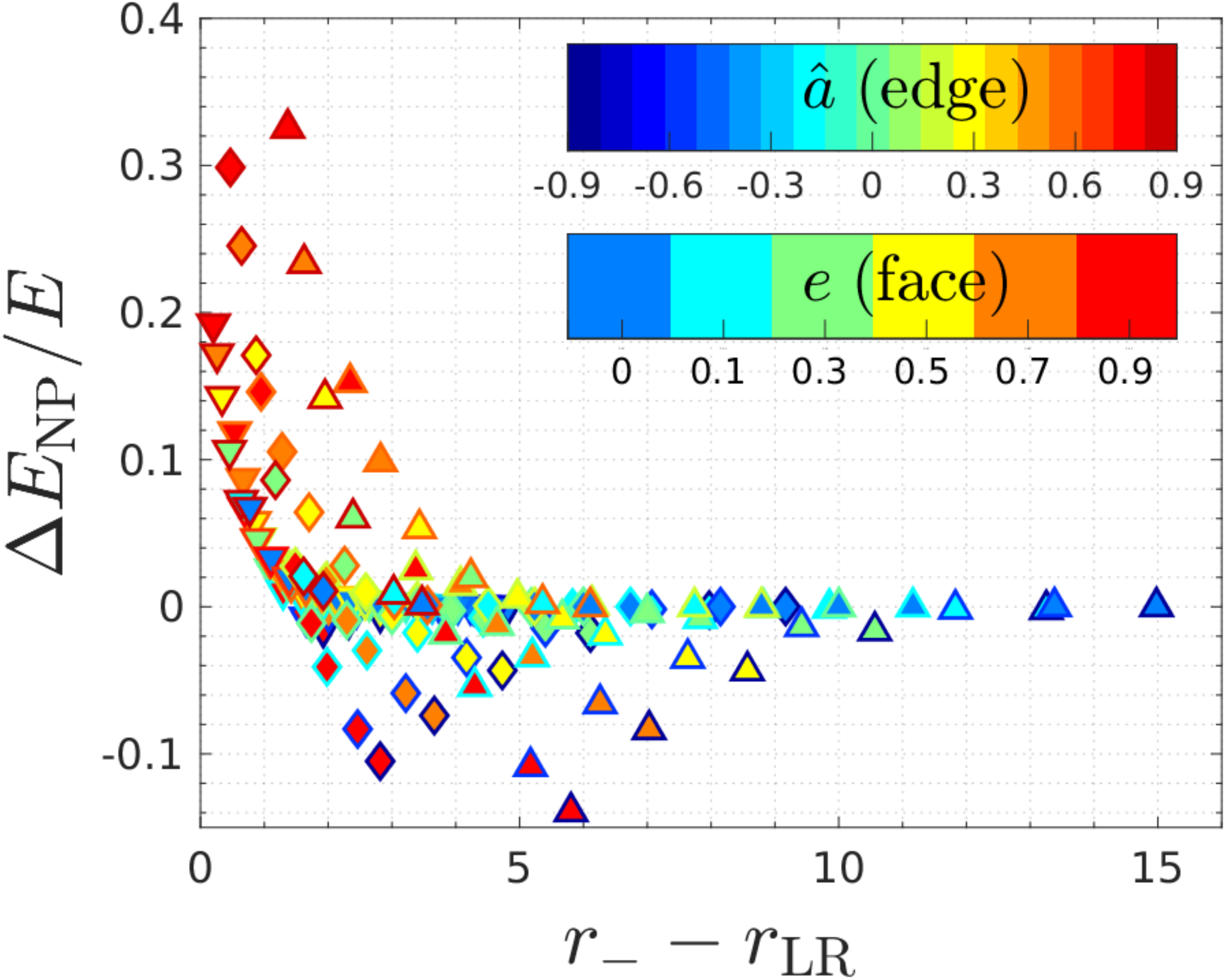}
	\includegraphics[width=0.24\textwidth,height=3.5cm]{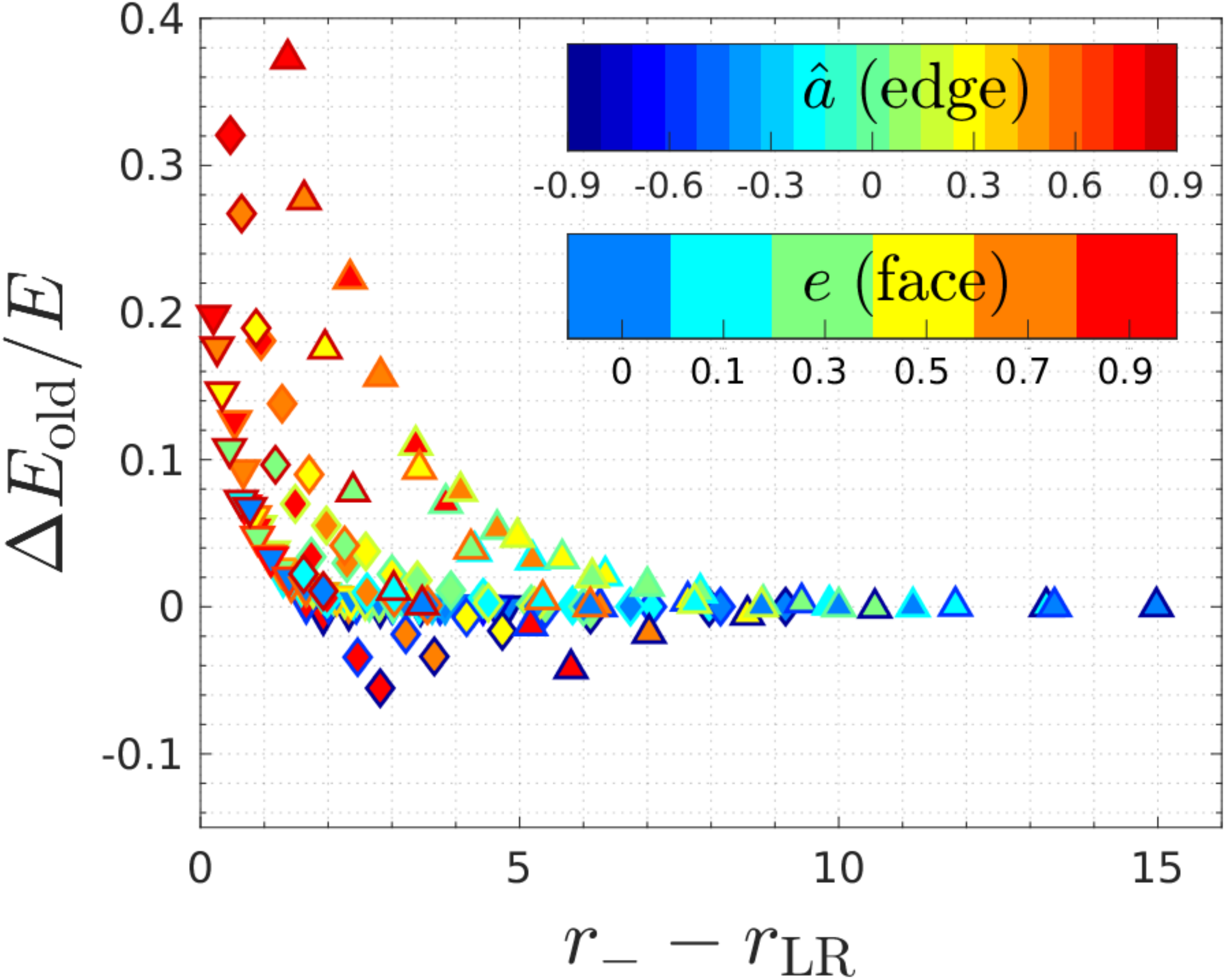}
	\includegraphics[width=0.24\textwidth,height=3.5cm]{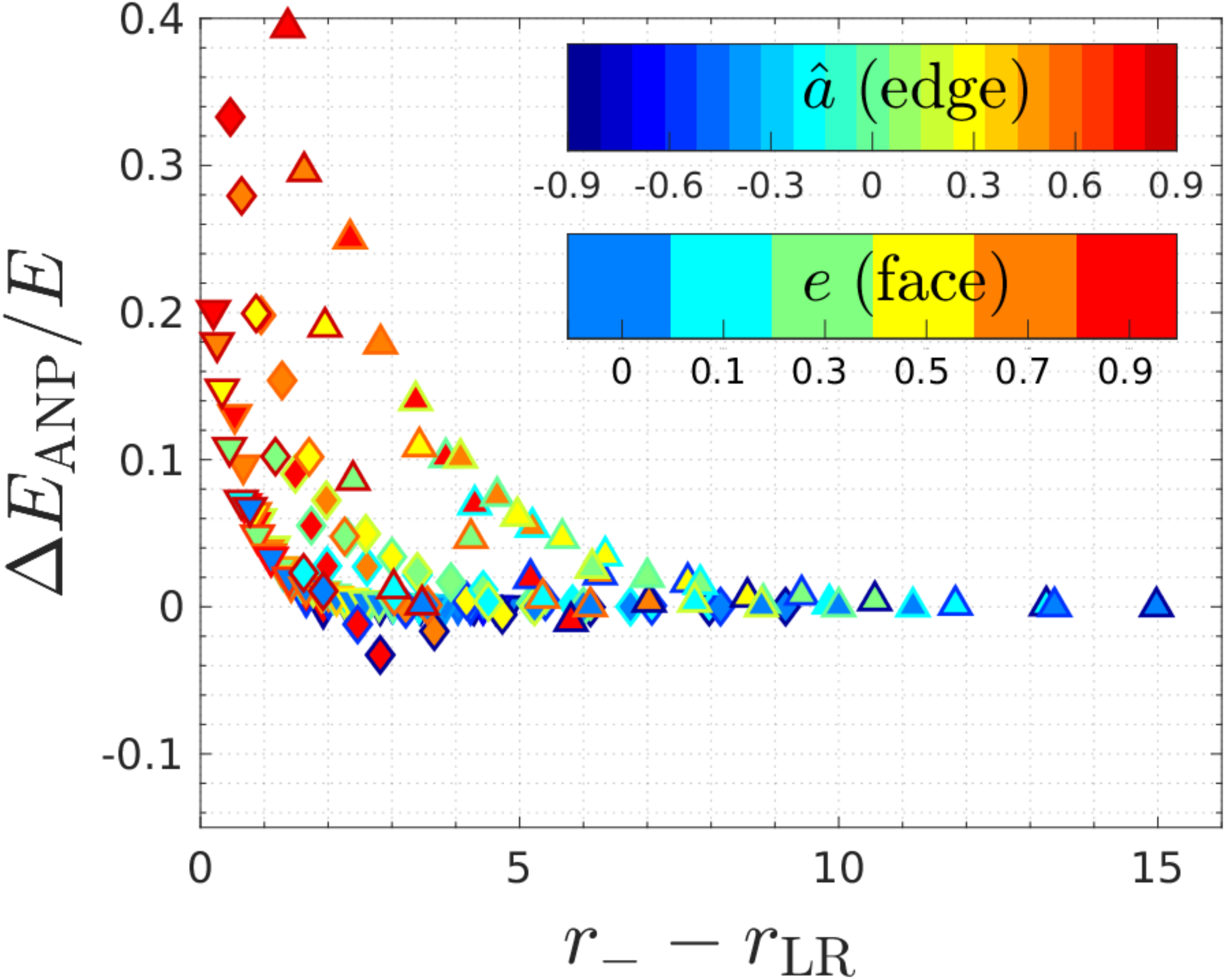}
	\includegraphics[width=0.24\textwidth,height=3.5cm]{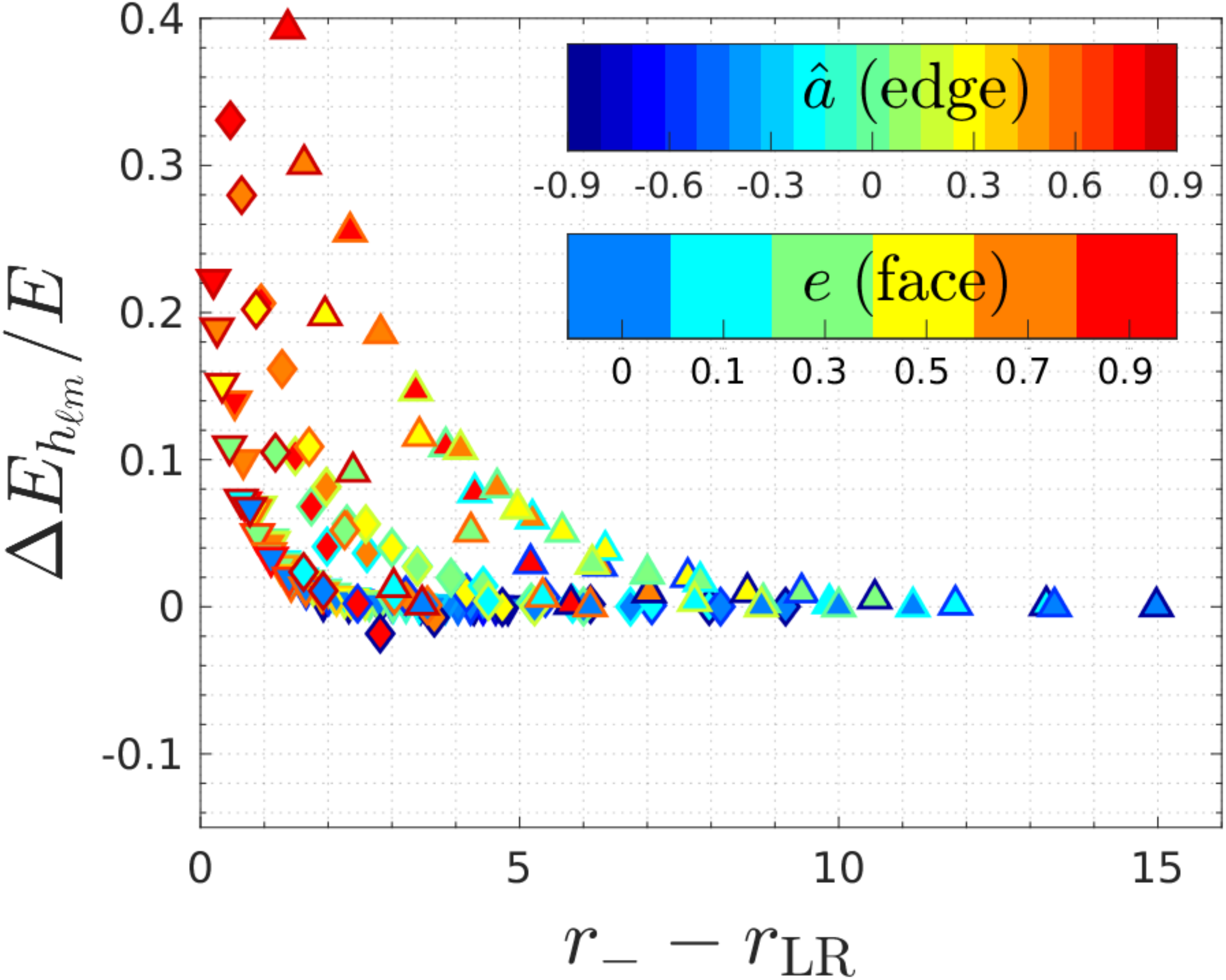}\\
	\includegraphics[width=0.24\textwidth,height=3.5cm]{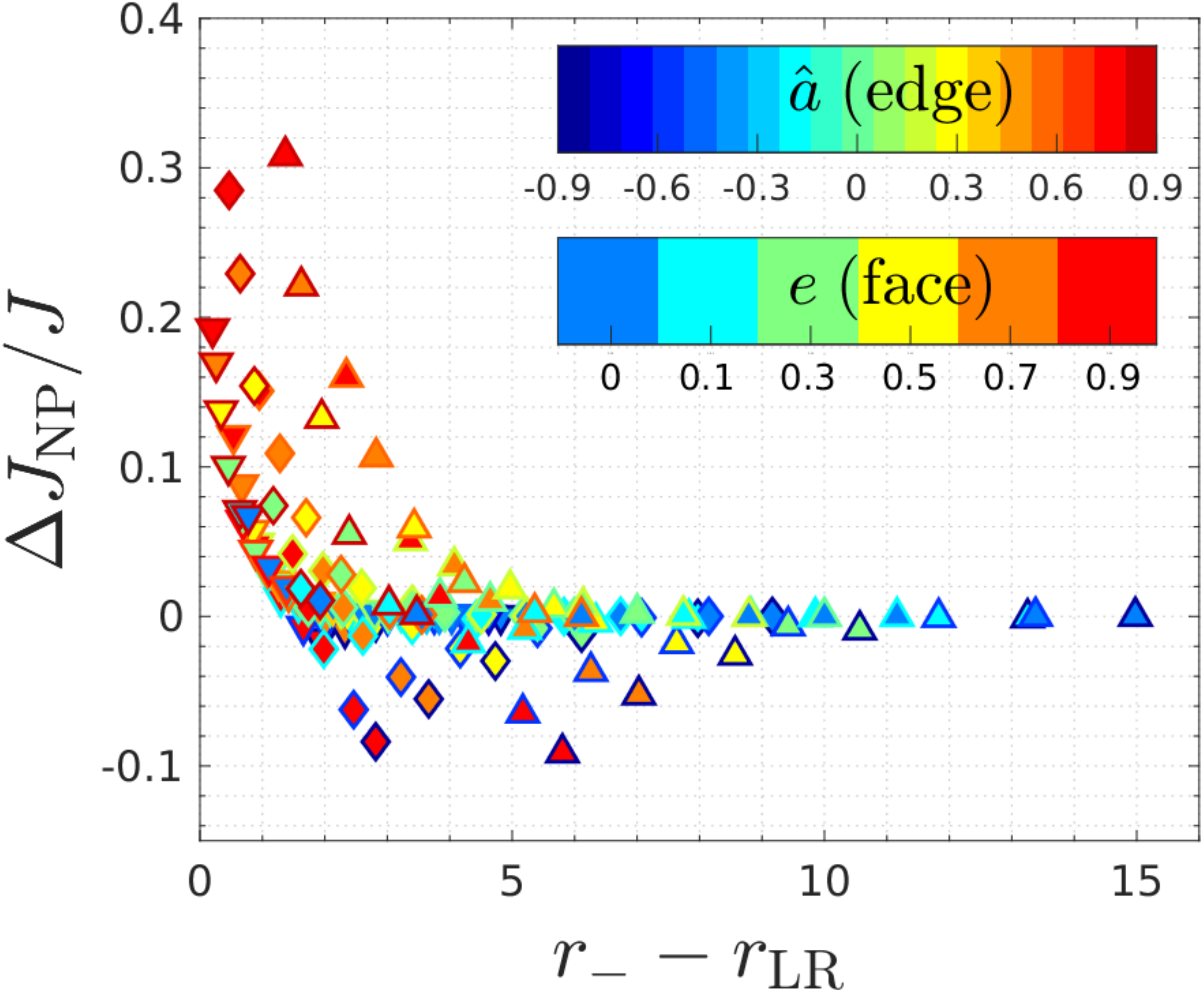}
	\includegraphics[width=0.24\textwidth,height=3.5cm]{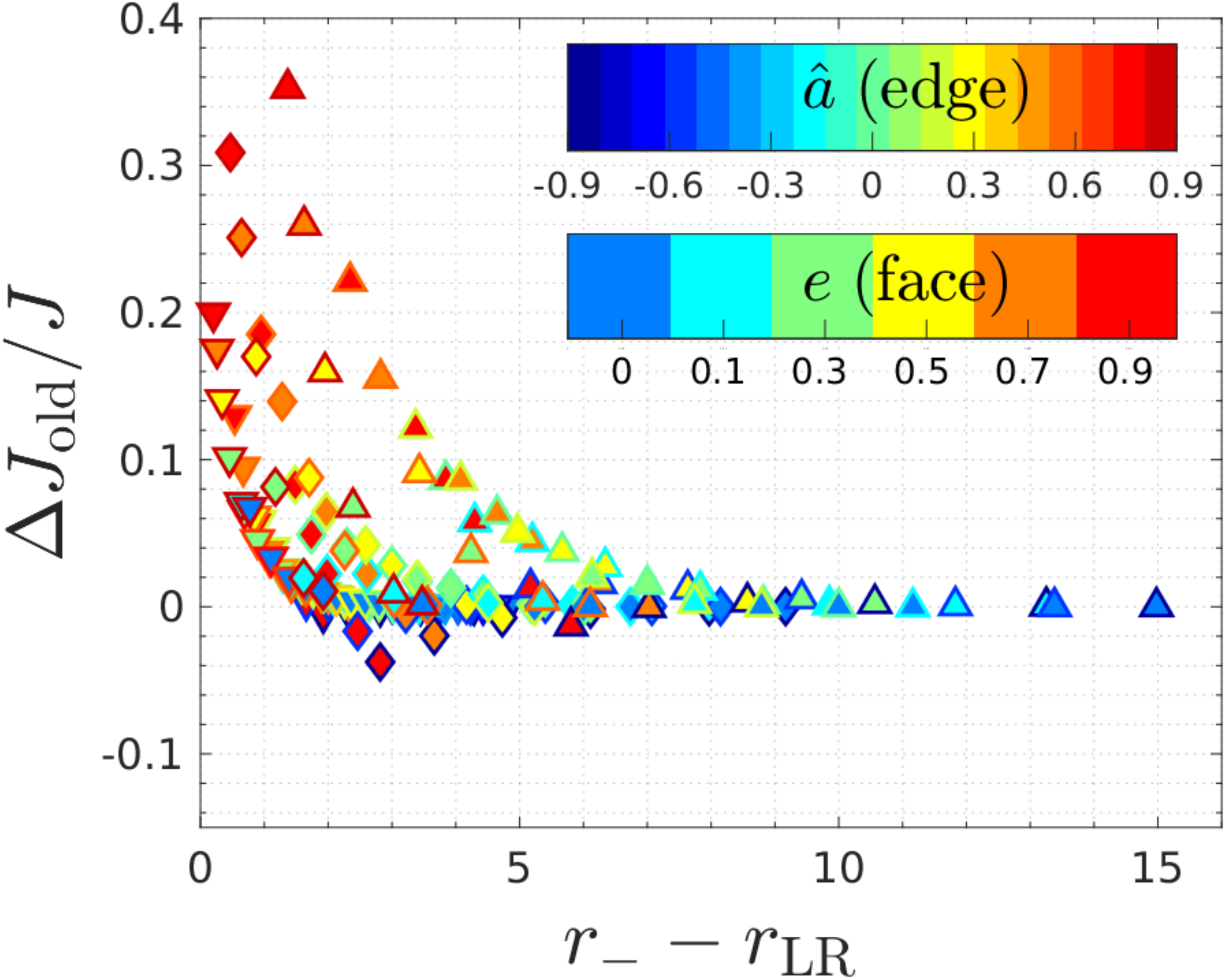}
	\includegraphics[width=0.24\textwidth,height=3.5cm]{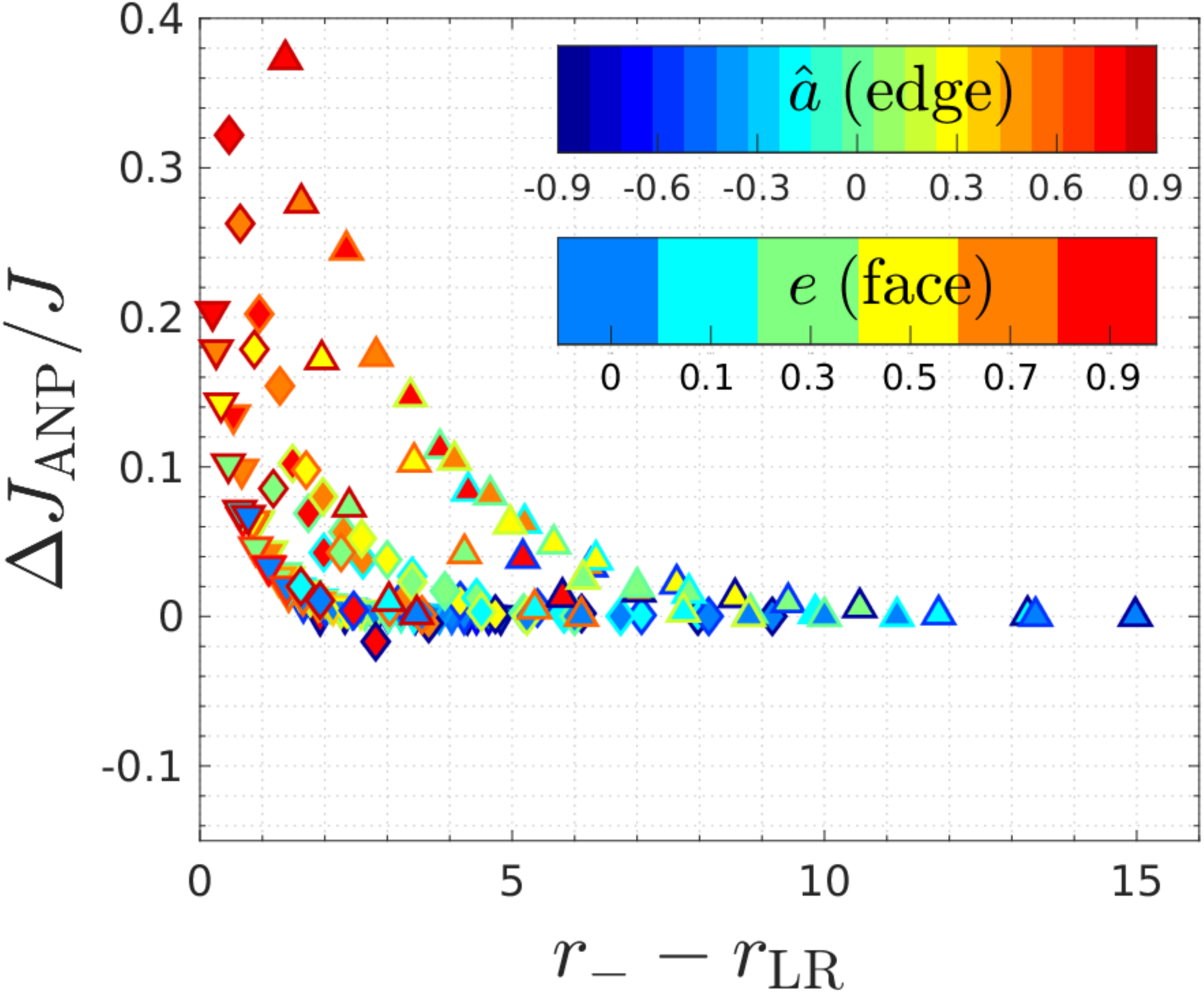}
	\includegraphics[width=0.24\textwidth,height=3.5cm]{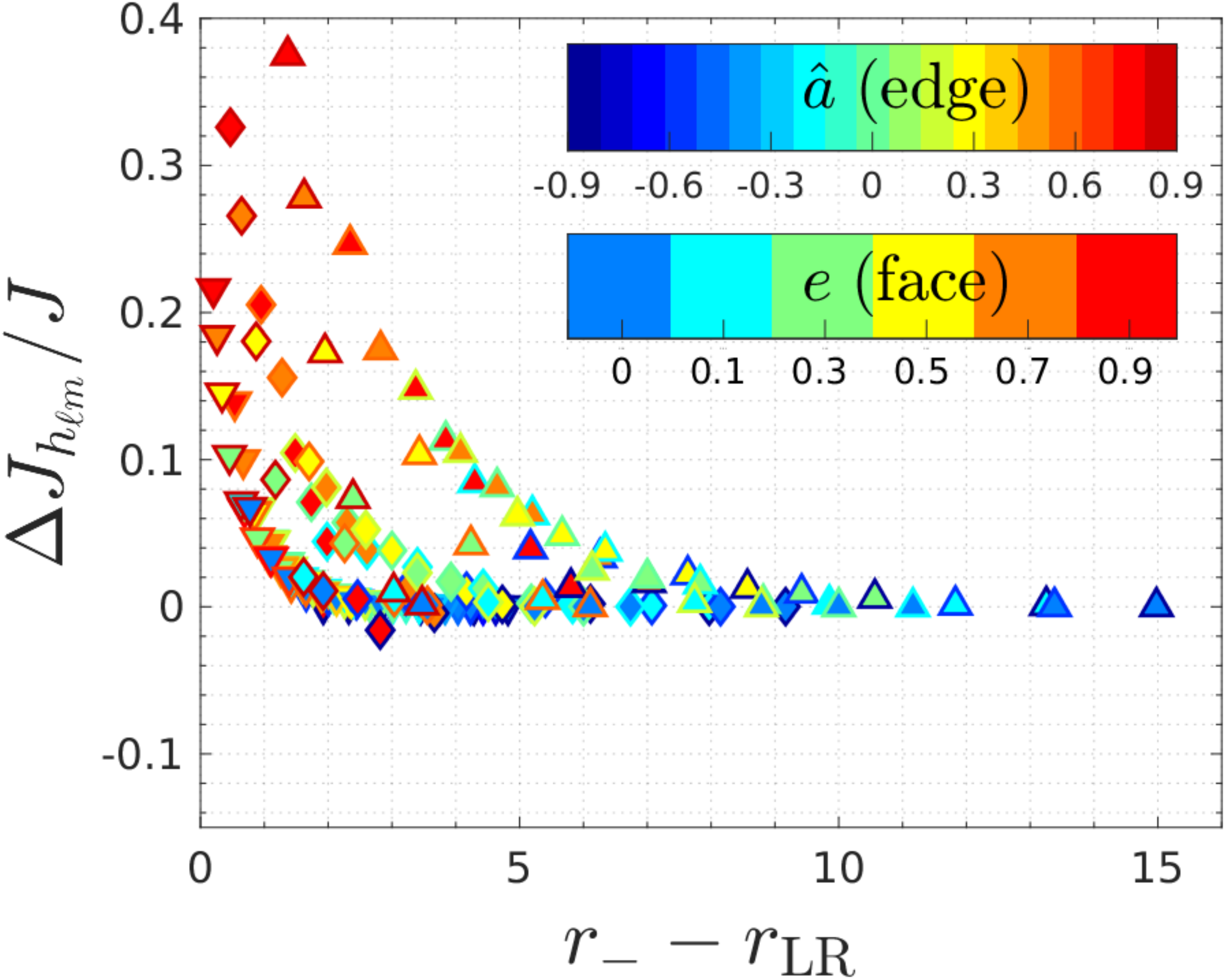}\\
	\caption{\label{fig:reldiff_harlequin_all} 
	Same relative differences 
	$\Delta F_{\rm analyt}/F = (\av{\Fteuk} - \av{\dot{F}_{\rm analyt}})/\av{\Fteuk} $ 
	of Fig.~\ref{fig:reldiff_harlequin_all_log}, but without absolute value and logscale.}
\end{figure*}

We also report the analytical/disagreement for the other analytical prescriptions:
$\av{\FNPold}$, $\av{\FANP}$, and $\av{\Fhlm}$ (see Sec.~\ref{sec:radreac} for more details). 
We plot them against the spin $\ha$ in 
Fig.~\ref{fig:reldiff_rainbow_Fold}, Fig.~\ref{fig:reldiff_rainbow_FANP} 
and Fig.~\ref{fig:reldiff_rainbow_Fhlm} and against $r_--r_{\rm LR}$ in
Fig.~\ref{fig:reldiff_harlequin_all_log} and
Fig.~\ref{fig:reldiff_harlequin_all}. The numerical values of the relative differences between numerical 
and analytical averaged fluxes obtained adopting the different analytical prescriptions
can be found in in Table~\ref{tab:AllFluxesDiff1} and Table~\ref{tab:AllFluxesDiff2}.

\begin{table*}[th]
   \caption{\label{tab:simulations_ecc1} List of simulations used in our tests with 
   corresponding numerical/analytical averaged fluxes and relative differences. 
   For each eccentricity, there 
   are three blocks of simulations, one for each class of semilatera recta: 
   near, intermediate and distant. The semilatus recta are truncated to the third 
   decimal, see Sec.~\ref{subsec:144sims}.}
   \begin{center}
     \begin{ruledtabular}
\begin{tabular}{ c r c c | c c r | c c r } 
$e$ & \multicolumn{1}{c}{$\ha$} & $p$ & $p_s$ & $\av{\Eteuk} $ & 
$\av{\ENP}$ & \multicolumn{1}{c|}{$\Delta E_{\rm NP}/E$} & 
$\av{\Jteuk} $ & $\av{\JNP}$ 
& \multicolumn{1}{c}{$\Delta J_{\rm NP}/J$} \\
\hline
\hline
$0.0$ & $-0.9$ & $8.727$ & $8.717$ & $1.732\cdot 10^{-4}$ & $1.731\cdot 10^{-4}$ & $4.9\cdot 10^{-4}$ & $4.309\cdot 10^{-3}$ & $4.307\cdot 10^{-3}$ & $4.9\cdot 10^{-4}$ \\
$0.0$ & $-0.8$ & $8.442$ & $8.432$ & $2.017\cdot 10^{-4}$ & $2.016\cdot 10^{-4}$ & $5.7\cdot 10^{-4}$ & $4.786\cdot 10^{-3}$ & $4.783\cdot 10^{-3}$ & $5.7\cdot 10^{-4}$ \\
$0.0$ & $-0.7$ & $8.153$ & $8.143$ & $2.365\cdot 10^{-4}$ & $2.364\cdot 10^{-4}$ & $6.7\cdot 10^{-4}$ & $5.341\cdot 10^{-3}$ & $5.337\cdot 10^{-3}$ & $6.7\cdot 10^{-4}$ \\
$0.0$ & $-0.6$ & $7.861$ & $7.851$ & $2.794\cdot 10^{-4}$ & $2.792\cdot 10^{-4}$ & $7.9\cdot 10^{-4}$ & $5.991\cdot 10^{-3}$ & $5.986\cdot 10^{-3}$ & $7.9\cdot 10^{-4}$ \\
$0.0$ & $-0.5$ & $7.565$ & $7.555$ & $3.326\cdot 10^{-4}$ & $3.323\cdot 10^{-4}$ & $9.4\cdot 10^{-4}$ & $6.754\cdot 10^{-3}$ & $6.748\cdot 10^{-3}$ & $9.4\cdot 10^{-4}$ \\
$0.0$ & $-0.4$ & $7.264$ & $7.254$ & $3.995\cdot 10^{-4}$ & $3.990\cdot 10^{-4}$ & $1.1\cdot 10^{-3}$ & $7.661\cdot 10^{-3}$ & $7.653\cdot 10^{-3}$ & $1.1\cdot 10^{-3}$ \\
$0.0$ & $-0.3$ & $6.959$ & $6.949$ & $4.843\cdot 10^{-4}$ & $4.836\cdot 10^{-4}$ & $1.4\cdot 10^{-3}$ & $8.745\cdot 10^{-3}$ & $8.734\cdot 10^{-3}$ & $1.4\cdot 10^{-3}$ \\
$0.0$ & $-0.2$ & $6.649$ & $6.639$ & $5.938\cdot 10^{-4}$ & $5.928\cdot 10^{-4}$ & $1.7\cdot 10^{-3}$ & $1.006\cdot 10^{-2}$ & $1.004\cdot 10^{-2}$ & $1.7\cdot 10^{-3}$ \\
$0.0$ & $-0.1$ & $6.333$ & $6.323$ & $7.374\cdot 10^{-4}$ & $7.359\cdot 10^{-4}$ & $2.0\cdot 10^{-3}$ & $1.168\cdot 10^{-2}$ & $1.165\cdot 10^{-2}$ & $2.0\cdot 10^{-3}$ \\
$0.0$ & $0.0$ & $6.010$ & $6.000$ & $9.287\cdot 10^{-4}$ & $9.264\cdot 10^{-4}$ & $2.5\cdot 10^{-3}$ & $1.368\cdot 10^{-2}$ & $1.365\cdot 10^{-2}$ & $2.5\cdot 10^{-3}$ \\
$0.0$ & $0.1$ & $5.679$ & $5.669$ & $1.189\cdot 10^{-3}$ & $1.185\cdot 10^{-3}$ & $3.2\cdot 10^{-3}$ & $1.621\cdot 10^{-2}$ & $1.616\cdot 10^{-2}$ & $3.2\cdot 10^{-3}$ \\
$0.0$ & $0.2$ & $5.339$ & $5.329$ & $1.553\cdot 10^{-3}$ & $1.547\cdot 10^{-3}$ & $4.0\cdot 10^{-3}$ & $1.947\cdot 10^{-2}$ & $1.939\cdot 10^{-2}$ & $4.0\cdot 10^{-3}$ \\
$0.0$ & $0.3$ & $4.989$ & $4.979$ & $2.075\cdot 10^{-3}$ & $2.064\cdot 10^{-3}$ & $5.2\cdot 10^{-3}$ & $2.375\cdot 10^{-2}$ & $2.362\cdot 10^{-2}$ & $5.2\cdot 10^{-3}$ \\
$0.0$ & $0.4$ & $4.624$ & $4.614$ & $2.857\cdot 10^{-3}$ & $2.837\cdot 10^{-3}$ & $6.9\cdot 10^{-3}$ & $2.955\cdot 10^{-2}$ & $2.935\cdot 10^{-2}$ & $6.9\cdot 10^{-3}$ \\
$0.0$ & $0.5$ & $4.243$ & $4.233$ & $4.072\cdot 10^{-3}$ & $4.034\cdot 10^{-3}$ & $9.4\cdot 10^{-3}$ & $3.763\cdot 10^{-2}$ & $3.728\cdot 10^{-2}$ & $9.4\cdot 10^{-3}$ \\
$0.0$ & $0.6$ & $3.839$ & $3.829$ & $6.089\cdot 10^{-3}$ & $6.008\cdot 10^{-3}$ & $1.3\cdot 10^{-2}$ & $4.945\cdot 10^{-2}$ & $4.880\cdot 10^{-2}$ & $1.3\cdot 10^{-2}$ \\
$0.0$ & $0.7$ & $3.403$ & $3.393$ & $9.715\cdot 10^{-3}$ & $9.520\cdot 10^{-3}$ & $2.0\cdot 10^{-2}$ & $6.779\cdot 10^{-2}$ & $6.643\cdot 10^{-2}$ & $2.0\cdot 10^{-2}$ \\
$0.0$ & $0.8$ & $2.917$ & $2.907$ & $1.709\cdot 10^{-2}$ & $1.652\cdot 10^{-2}$ & $3.3\cdot 10^{-2}$ & $9.882\cdot 10^{-2}$ & $9.551\cdot 10^{-2}$ & $3.3\cdot 10^{-2}$ \\
$0.0$ & $0.9$ & $2.331$ & $2.321$ & $3.552\cdot 10^{-2}$ & $3.312\cdot 10^{-2}$ & $6.7\cdot 10^{-2}$ & $1.583\cdot 10^{-1}$ & $1.477\cdot 10^{-1}$ & $6.7\cdot 10^{-2}$ \\
\hline
$0.0$ & $-0.9$ & $13.076$ & $8.717$ & $1.837\cdot 10^{-5}$ & $1.837\cdot 10^{-5}$ & $9.1\cdot 10^{-5}$ & $8.520\cdot 10^{-4}$ & $8.519\cdot 10^{-4}$ & $9.2\cdot 10^{-5}$ \\
$0.0$ & $-0.6$ & $11.776$ & $7.851$ & $3.015\cdot 10^{-5}$ & $3.015\cdot 10^{-5}$ & $8.8\cdot 10^{-5}$ & $1.200\cdot 10^{-3}$ & $1.200\cdot 10^{-3}$ & $8.9\cdot 10^{-5}$ \\
$0.0$ & $-0.2$ & $9.959$ & $6.639$ & $6.618\cdot 10^{-5}$ & $6.617\cdot 10^{-5}$ & $1.3\cdot 10^{-4}$ & $2.066\cdot 10^{-3}$ & $2.066\cdot 10^{-3}$ & $1.3\cdot 10^{-4}$ \\
$0.0$ & $0.0$ & $9.000$ & $6.000$ & $1.059\cdot 10^{-4}$ & $1.059\cdot 10^{-4}$ & $1.8\cdot 10^{-4}$ & $2.859\cdot 10^{-3}$ & $2.859\cdot 10^{-3}$ & $1.8\cdot 10^{-4}$ \\
$0.0$ & $0.2$ & $7.994$ & $5.329$ & $1.827\cdot 10^{-4}$ & $1.827\cdot 10^{-4}$ & $2.8\cdot 10^{-4}$ & $4.167\cdot 10^{-3}$ & $4.166\cdot 10^{-3}$ & $2.8\cdot 10^{-4}$ \\
$0.0$ & $0.6$ & $5.744$ & $3.829$ & $8.083\cdot 10^{-4}$ & $8.074\cdot 10^{-4}$ & $1.1\cdot 10^{-3}$ & $1.161\cdot 10^{-2}$ & $1.160\cdot 10^{-2}$ & $1.1\cdot 10^{-3}$ \\
$0.0$ & $0.9$ & $3.481$ & $2.321$ & $6.623\cdot 10^{-3}$ & $6.552\cdot 10^{-3}$ & $1.1\cdot 10^{-2}$ & $4.898\cdot 10^{-2}$ & $4.846\cdot 10^{-2}$ & $1.1\cdot 10^{-2}$ \\
\hline
$0.0$ & $-0.9$ & $18.888$ & $8.717$ & $2.702\cdot 10^{-6}$ & $2.702\cdot 10^{-6}$ & $1.3\cdot 10^{-4}$ & $2.194\cdot 10^{-4}$ & $2.194\cdot 10^{-4}$ & $1.4\cdot 10^{-4}$ \\
$0.0$ & $-0.6$ & $17.010$ & $7.851$ & $4.477\cdot 10^{-6}$ & $4.476\cdot 10^{-6}$ & $9.9\cdot 10^{-5}$ & $3.114\cdot 10^{-4}$ & $3.114\cdot 10^{-4}$ & $1.1\cdot 10^{-4}$ \\
$0.0$ & $-0.2$ & $14.385$ & $6.639$ & $1.000\cdot 10^{-5}$ & $1.000\cdot 10^{-5}$ & $7.6\cdot 10^{-5}$ & $5.438\cdot 10^{-4}$ & $5.438\cdot 10^{-4}$ & $7.6\cdot 10^{-5}$ \\
$0.0$ & $0.0$ & $13.000$ & $6.000$ & $1.621\cdot 10^{-5}$ & $1.621\cdot 10^{-5}$ & $6.5\cdot 10^{-5}$ & $7.598\cdot 10^{-4}$ & $7.597\cdot 10^{-4}$ & $6.2\cdot 10^{-5}$ \\
$0.0$ & $0.2$ & $11.547$ & $5.329$ & $2.843\cdot 10^{-5}$ & $2.843\cdot 10^{-5}$ & $5.8\cdot 10^{-5}$ & $1.121\cdot 10^{-3}$ & $1.121\cdot 10^{-3}$ & $5.8\cdot 10^{-5}$ \\
$0.0$ & $0.6$ & $8.296$ & $3.829$ & $1.336\cdot 10^{-4}$ & $1.335\cdot 10^{-4}$ & $1.1\cdot 10^{-4}$ & $3.272\cdot 10^{-3}$ & $3.271\cdot 10^{-3}$ & $1.1\cdot 10^{-4}$ \\
$0.0$ & $0.9$ & $5.029$ & $2.321$ & $1.280\cdot 10^{-3}$ & $1.278\cdot 10^{-3}$ & $1.4\cdot 10^{-3}$ & $1.558\cdot 10^{-2}$ & $1.556\cdot 10^{-2}$ & $1.4\cdot 10^{-3}$ \\

\Xhline{2.75\arrayrulewidth}

$0.1$ & $-0.9$ & $9.014$ & $9.004$ & $1.870\cdot 10^{-4}$ & $1.869\cdot 10^{-4}$ & $3.8\cdot 10^{-4}$ & $4.505\cdot 10^{-3}$ & $4.503\cdot 10^{-3}$ & $4.3\cdot 10^{-4}$ \\
$0.1$ & $-0.6$ & $8.119$ & $8.109$ & $3.002\cdot 10^{-4}$ & $3.000\cdot 10^{-4}$ & $7.8\cdot 10^{-4}$ & $6.241\cdot 10^{-3}$ & $6.236\cdot 10^{-3}$ & $8.2\cdot 10^{-4}$ \\
$0.1$ & $-0.2$ & $6.869$ & $6.859$ & $6.329\cdot 10^{-4}$ & $6.317\cdot 10^{-4}$ & $1.9\cdot 10^{-3}$ & $1.042\cdot 10^{-2}$ & $1.040\cdot 10^{-2}$ & $1.9\cdot 10^{-3}$ \\
$0.1$ & $0.0$ & $6.210$ & $6.200$ & $9.845\cdot 10^{-4}$ & $9.814\cdot 10^{-4}$ & $3.1\cdot 10^{-3}$ & $1.412\cdot 10^{-2}$ & $1.408\cdot 10^{-2}$ & $3.0\cdot 10^{-3}$ \\
$0.1$ & $0.2$ & $5.518$ & $5.508$ & $1.636\cdot 10^{-3}$ & $1.628\cdot 10^{-3}$ & $5.0\cdot 10^{-3}$ & $2.001\cdot 10^{-2}$ & $1.991\cdot 10^{-2}$ & $4.9\cdot 10^{-3}$ \\
$0.1$ & $0.6$ & $3.970$ & $3.960$ & $6.300\cdot 10^{-3}$ & $6.200\cdot 10^{-3}$ & $1.6\cdot 10^{-2}$ & $5.022\cdot 10^{-2}$ & $4.945\cdot 10^{-2}$ & $1.6\cdot 10^{-2}$ \\
$0.1$ & $0.9$ & $2.415$ & $2.405$ & $3.566\cdot 10^{-2}$ & $3.308\cdot 10^{-2}$ & $7.2\cdot 10^{-2}$ & $1.578\cdot 10^{-1}$ & $1.466\cdot 10^{-1}$ & $7.1\cdot 10^{-2}$ \\
\hline
$0.1$ & $-0.9$ & $13.070$ & $9.004$ & $1.897\cdot 10^{-5}$ & $1.901\cdot 10^{-5}$ & $-2.1\cdot 10^{-3}$ & $8.583\cdot 10^{-4}$ & $8.593\cdot 10^{-4}$ & $-1.2\cdot 10^{-3}$ \\
$0.1$ & $-0.6$ & $11.772$ & $8.109$ & $3.112\cdot 10^{-5}$ & $3.117\cdot 10^{-5}$ & $-1.7\cdot 10^{-3}$ & $1.209\cdot 10^{-3}$ & $1.210\cdot 10^{-3}$ & $-8.1\cdot 10^{-4}$ \\
$0.1$ & $-0.2$ & $9.957$ & $6.859$ & $6.825\cdot 10^{-5}$ & $6.830\cdot 10^{-5}$ & $-8.3\cdot 10^{-4}$ & $2.081\cdot 10^{-3}$ & $2.081\cdot 10^{-3}$ & $-6.0\cdot 10^{-5}$ \\
$0.1$ & $0.0$ & $9.000$ & $6.200$ & $1.091\cdot 10^{-4}$ & $1.092\cdot 10^{-4}$ & $-1.8\cdot 10^{-4}$ & $2.878\cdot 10^{-3}$ & $2.877\cdot 10^{-3}$ & $4.9\cdot 10^{-4}$ \\
$0.1$ & $0.2$ & $7.996$ & $5.508$ & $1.882\cdot 10^{-4}$ & $1.880\cdot 10^{-4}$ & $7.2\cdot 10^{-4}$ & $4.194\cdot 10^{-3}$ & $4.188\cdot 10^{-3}$ & $1.2\cdot 10^{-3}$ \\
$0.1$ & $0.6$ & $5.749$ & $3.960$ & $8.299\cdot 10^{-4}$ & $8.262\cdot 10^{-4}$ & $4.5\cdot 10^{-3}$ & $1.167\cdot 10^{-2}$ & $1.162\cdot 10^{-2}$ & $4.4\cdot 10^{-3}$ \\
$0.1$ & $0.9$ & $3.491$ & $2.405$ & $6.735\cdot 10^{-3}$ & $6.596\cdot 10^{-3}$ & $2.1\cdot 10^{-2}$ & $4.903\cdot 10^{-2}$ & $4.811\cdot 10^{-2}$ & $1.9\cdot 10^{-2}$ \\
\hline
$0.1$ & $-0.9$ & $18.879$ & $9.004$ & $2.768\cdot 10^{-6}$ & $2.773\cdot 10^{-6}$ & $-1.8\cdot 10^{-3}$ & $2.196\cdot 10^{-4}$ & $2.198\cdot 10^{-4}$ & $-8.6\cdot 10^{-4}$ \\
$0.1$ & $-0.6$ & $17.004$ & $8.109$ & $4.584\cdot 10^{-6}$ & $4.590\cdot 10^{-6}$ & $-1.5\cdot 10^{-3}$ & $3.117\cdot 10^{-4}$ & $3.118\cdot 10^{-4}$ & $-6.0\cdot 10^{-4}$ \\
$0.1$ & $-0.2$ & $14.382$ & $6.859$ & $1.023\cdot 10^{-5}$ & $1.024\cdot 10^{-5}$ & $-9.1\cdot 10^{-4}$ & $5.440\cdot 10^{-4}$ & $5.441\cdot 10^{-4}$ & $-6.7\cdot 10^{-5}$ \\
$0.1$ & $0.0$ & $13.000$ & $6.200$ & $1.657\cdot 10^{-5}$ & $1.658\cdot 10^{-5}$ & $-4.7\cdot 10^{-4}$ & $7.598\cdot 10^{-4}$ & $7.596\cdot 10^{-4}$ & $3.2\cdot 10^{-4}$ \\
$0.1$ & $0.2$ & $11.549$ & $5.508$ & $2.904\cdot 10^{-5}$ & $2.904\cdot 10^{-5}$ & $1.2\cdot 10^{-4}$ & $1.121\cdot 10^{-3}$ & $1.120\cdot 10^{-3}$ & $8.4\cdot 10^{-4}$ \\
$0.1$ & $0.6$ & $8.304$ & $3.960$ & $1.360\cdot 10^{-4}$ & $1.357\cdot 10^{-4}$ & $2.4\cdot 10^{-3}$ & $3.266\cdot 10^{-3}$ & $3.257\cdot 10^{-3}$ & $2.7\cdot 10^{-3}$ \\
$0.1$ & $0.9$ & $5.043$ & $2.405$ & $1.293\cdot 10^{-3}$ & $1.281\cdot 10^{-3}$ & $8.8\cdot 10^{-3}$ & $1.549\cdot 10^{-2}$ & $1.537\cdot 10^{-2}$ & $7.9\cdot 10^{-3}$ \\
\end{tabular}
\end{ruledtabular}
\end{center}
\end{table*}

\begin{table*}[th]
   \caption{\label{tab:simulations_ecc2}  Same scheme as Table~\ref{tab:simulations_ecc1}.}
   \begin{center}
     \begin{ruledtabular}
\begin{tabular}{ c r c c | c c r | c c r } 
$e$ & \multicolumn{1}{c}{$\ha$} & $p$ & $p_s$ & $\av{\Eteuk} $ & 
$\av{\ENP}$ & \multicolumn{1}{c|}{$\Delta E_{\rm NP}/E$} & 
$\av{\Jteuk} $ & $\av{\JNP}$ 
& \multicolumn{1}{c}{$\Delta J_{\rm NP}/J$} \\
\hline
\hline
$0.3$ & $-0.9$ & $9.564$ & $9.554$ & $2.546\cdot 10^{-4}$ & $2.550\cdot 10^{-4}$ & $-1.4\cdot 10^{-3}$ & $5.324\cdot 10^{-3}$ & $5.330\cdot 10^{-3}$ & $-1.1\cdot 10^{-3}$ \\
$0.3$ & $-0.6$ & $8.622$ & $8.612$ & $4.055\cdot 10^{-4}$ & $4.056\cdot 10^{-4}$ & $-2.0\cdot 10^{-4}$ & $7.339\cdot 10^{-3}$ & $7.337\cdot 10^{-3}$ & $1.9\cdot 10^{-4}$ \\
$0.3$ & $-0.2$ & $7.305$ & $7.295$ & $8.426\cdot 10^{-4}$ & $8.400\cdot 10^{-4}$ & $3.0\cdot 10^{-3}$ & $1.215\cdot 10^{-2}$ & $1.211\cdot 10^{-2}$ & $3.3\cdot 10^{-3}$ \\
$0.3$ & $0.0$ & $6.610$ & $6.600$ & $1.298\cdot 10^{-3}$ & $1.291\cdot 10^{-3}$ & $5.9\cdot 10^{-3}$ & $1.637\cdot 10^{-2}$ & $1.627\cdot 10^{-2}$ & $6.2\cdot 10^{-3}$ \\
$0.3$ & $0.1$ & $6.250$ & $6.240$ & $1.648\cdot 10^{-3}$ & $1.635\cdot 10^{-3}$ & $7.9\cdot 10^{-3}$ & $1.929\cdot 10^{-2}$ & $1.913\cdot 10^{-2}$ & $8.1\cdot 10^{-3}$ \\
$0.3$ & $0.2$ & $5.881$ & $5.871$ & $2.131\cdot 10^{-3}$ & $2.109\cdot 10^{-3}$ & $1.1\cdot 10^{-2}$ & $2.302\cdot 10^{-2}$ & $2.278\cdot 10^{-2}$ & $1.1\cdot 10^{-2}$ \\
$0.3$ & $0.3$ & $5.500$ & $5.490$ & $2.817\cdot 10^{-3}$ & $2.777\cdot 10^{-3}$ & $1.4\cdot 10^{-2}$ & $2.789\cdot 10^{-2}$ & $2.750\cdot 10^{-2}$ & $1.4\cdot 10^{-2}$ \\
$0.3$ & $0.4$ & $5.104$ & $5.094$ & $3.824\cdot 10^{-3}$ & $3.754\cdot 10^{-3}$ & $1.8\cdot 10^{-2}$ & $3.440\cdot 10^{-2}$ & $3.378\cdot 10^{-2}$ & $1.8\cdot 10^{-2}$ \\
$0.3$ & $0.5$ & $4.689$ & $4.679$ & $5.366\cdot 10^{-3}$ & $5.235\cdot 10^{-3}$ & $2.4\cdot 10^{-2}$ & $4.339\cdot 10^{-2}$ & $4.236\cdot 10^{-2}$ & $2.4\cdot 10^{-2}$ \\
$0.3$ & $0.6$ & $4.250$ & $4.240$ & $7.862\cdot 10^{-3}$ & $7.602\cdot 10^{-3}$ & $3.3\cdot 10^{-2}$ & $5.630\cdot 10^{-2}$ & $5.451\cdot 10^{-2}$ & $3.2\cdot 10^{-2}$ \\
$0.3$ & $0.7$ & $3.777$ & $3.767$ & $1.221\cdot 10^{-2}$ & $1.165\cdot 10^{-2}$ & $4.6\cdot 10^{-2}$ & $7.590\cdot 10^{-2}$ & $7.256\cdot 10^{-2}$ & $4.4\cdot 10^{-2}$ \\
$0.3$ & $0.8$ & $3.249$ & $3.239$ & $2.065\cdot 10^{-2}$ & $1.926\cdot 10^{-2}$ & $6.7\cdot 10^{-2}$ & $1.078\cdot 10^{-1}$ & $1.009\cdot 10^{-1}$ & $6.4\cdot 10^{-2}$ \\
$0.3$ & $0.9$ & $2.615$ & $2.605$ & $4.030\cdot 10^{-2}$ & $3.602\cdot 10^{-2}$ & $1.1\cdot 10^{-1}$ & $1.662\cdot 10^{-1}$ & $1.495\cdot 10^{-1}$ & $1.0\cdot 10^{-1}$ \\
\hline
$0.3$ & $-0.9$ & $13.028$ & $9.554$ & $2.350\cdot 10^{-5}$ & $2.392\cdot 10^{-5}$ & $-1.8\cdot 10^{-2}$ & $8.995\cdot 10^{-4}$ & $9.095\cdot 10^{-4}$ & $-1.1\cdot 10^{-2}$ \\
$0.3$ & $-0.6$ & $11.743$ & $8.612$ & $3.843\cdot 10^{-5}$ & $3.898\cdot 10^{-5}$ & $-1.4\cdot 10^{-2}$ & $1.265\cdot 10^{-3}$ & $1.275\cdot 10^{-3}$ & $-7.9\cdot 10^{-3}$ \\
$0.3$ & $-0.2$ & $9.947$ & $7.295$ & $8.381\cdot 10^{-5}$ & $8.443\cdot 10^{-5}$ & $-7.4\cdot 10^{-3}$ & $2.174\cdot 10^{-3}$ & $2.178\cdot 10^{-3}$ & $-1.7\cdot 10^{-3}$ \\
$0.3$ & $0.0$ & $9.000$ & $6.600$ & $1.335\cdot 10^{-4}$ & $1.338\cdot 10^{-4}$ & $-2.5\cdot 10^{-3}$ & $3.004\cdot 10^{-3}$ & $2.996\cdot 10^{-3}$ & $2.6\cdot 10^{-3}$ \\
$0.3$ & $0.2$ & $8.006$ & $5.871$ & $2.289\cdot 10^{-4}$ & $2.280\cdot 10^{-4}$ & $4.1\cdot 10^{-3}$ & $4.369\cdot 10^{-3}$ & $4.333\cdot 10^{-3}$ & $8.2\cdot 10^{-3}$ \\
$0.3$ & $0.6$ & $5.782$ & $4.240$ & $9.889\cdot 10^{-4}$ & $9.612\cdot 10^{-4}$ & $2.8\cdot 10^{-2}$ & $1.207\cdot 10^{-2}$ & $1.173\cdot 10^{-2}$ & $2.8\cdot 10^{-2}$ \\
$0.3$ & $0.9$ & $3.553$ & $2.605$ & $7.574\cdot 10^{-3}$ & $6.922\cdot 10^{-3}$ & $8.6\cdot 10^{-2}$ & $4.941\cdot 10^{-2}$ & $4.575\cdot 10^{-2}$ & $7.4\cdot 10^{-2}$ \\
\hline
$0.3$ & $-0.9$ & $18.818$ & $9.554$ & $3.231\cdot 10^{-6}$ & $3.284\cdot 10^{-6}$ & $-1.6\cdot 10^{-2}$ & $2.191\cdot 10^{-4}$ & $2.211\cdot 10^{-4}$ & $-9.0\cdot 10^{-3}$ \\
$0.3$ & $-0.6$ & $16.962$ & $8.612$ & $5.333\cdot 10^{-6}$ & $5.405\cdot 10^{-6}$ & $-1.3\cdot 10^{-2}$ & $3.105\cdot 10^{-4}$ & $3.124\cdot 10^{-4}$ & $-6.3\cdot 10^{-3}$ \\
$0.3$ & $-0.2$ & $14.368$ & $7.295$ & $1.184\cdot 10^{-5}$ & $1.193\cdot 10^{-5}$ & $-7.8\cdot 10^{-3}$ & $5.407\cdot 10^{-4}$ & $5.413\cdot 10^{-4}$ & $-1.2\cdot 10^{-3}$ \\
$0.3$ & $0.0$ & $13.000$ & $6.600$ & $1.910\cdot 10^{-5}$ & $1.918\cdot 10^{-5}$ & $-3.9\cdot 10^{-3}$ & $7.540\cdot 10^{-4}$ & $7.522\cdot 10^{-4}$ & $2.3\cdot 10^{-3}$ \\
$0.3$ & $0.2$ & $11.564$ & $5.871$ & $3.331\cdot 10^{-5}$ & $3.327\cdot 10^{-5}$ & $1.4\cdot 10^{-3}$ & $1.110\cdot 10^{-3}$ & $1.102\cdot 10^{-3}$ & $7.0\cdot 10^{-3}$ \\
$0.3$ & $0.6$ & $8.352$ & $4.240$ & $1.533\cdot 10^{-4}$ & $1.503\cdot 10^{-4}$ & $2.0\cdot 10^{-2}$ & $3.211\cdot 10^{-3}$ & $3.137\cdot 10^{-3}$ & $2.3\cdot 10^{-2}$ \\
$0.3$ & $0.9$ & $5.132$ & $2.605$ & $1.389\cdot 10^{-3}$ & $1.305\cdot 10^{-3}$ & $6.1\cdot 10^{-2}$ & $1.489\cdot 10^{-2}$ & $1.407\cdot 10^{-2}$ & $5.5\cdot 10^{-2}$ \\

\Xhline{2.5\arrayrulewidth}

$0.5$ & $-0.9$ & $10.089$ & $10.079$ & $3.287\cdot 10^{-4}$ & $3.302\cdot 10^{-4}$ & $-4.6\cdot 10^{-3}$ & $5.933\cdot 10^{-3}$ & $5.958\cdot 10^{-3}$ & $-4.2\cdot 10^{-3}$ \\
$0.5$ & $-0.6$ & $9.107$ & $9.097$ & $5.198\cdot 10^{-4}$ & $5.208\cdot 10^{-4}$ & $-2.0\cdot 10^{-3}$ & $8.149\cdot 10^{-3}$ & $8.160\cdot 10^{-3}$ & $-1.4\cdot 10^{-3}$ \\
$0.5$ & $-0.2$ & $7.734$ & $7.724$ & $1.067\cdot 10^{-3}$ & $1.062\cdot 10^{-3}$ & $4.7\cdot 10^{-3}$ & $1.341\cdot 10^{-2}$ & $1.334\cdot 10^{-2}$ & $5.5\cdot 10^{-3}$ \\
$0.5$ & $0.0$ & $7.010$ & $7.000$ & $1.630\cdot 10^{-3}$ & $1.613\cdot 10^{-3}$ & $1.1\cdot 10^{-2}$ & $1.799\cdot 10^{-2}$ & $1.779\cdot 10^{-2}$ & $1.1\cdot 10^{-2}$ \\
$0.5$ & $0.2$ & $6.250$ & $6.240$ & $2.647\cdot 10^{-3}$ & $2.595\cdot 10^{-3}$ & $2.0\cdot 10^{-2}$ & $2.517\cdot 10^{-2}$ & $2.466\cdot 10^{-2}$ & $2.0\cdot 10^{-2}$ \\
$0.5$ & $0.6$ & $4.548$ & $4.538$ & $9.403\cdot 10^{-3}$ & $8.859\cdot 10^{-3}$ & $5.8\cdot 10^{-2}$ & $6.039\cdot 10^{-2}$ & $5.693\cdot 10^{-2}$ & $5.7\cdot 10^{-2}$ \\
$0.5$ & $0.9$ & $2.843$ & $2.833$ & $4.424\cdot 10^{-2}$ & $3.792\cdot 10^{-2}$ & $1.4\cdot 10^{-1}$ & $1.696\cdot 10^{-1}$ & $1.463\cdot 10^{-1}$ & $1.4\cdot 10^{-1}$ \\
\hline
$0.5$ & $-0.9$ & $12.959$ & $10.079$ & $3.078\cdot 10^{-5}$ & $3.211\cdot 10^{-5}$ & $-4.3\cdot 10^{-2}$ & $9.314\cdot 10^{-4}$ & $9.592\cdot 10^{-4}$ & $-3.0\cdot 10^{-2}$ \\
$0.5$ & $-0.6$ & $11.697$ & $9.097$ & $5.009\cdot 10^{-5}$ & $5.182\cdot 10^{-5}$ & $-3.5\cdot 10^{-2}$ & $1.309\cdot 10^{-3}$ & $1.337\cdot 10^{-3}$ & $-2.1\cdot 10^{-2}$ \\
$0.5$ & $-0.2$ & $9.931$ & $7.724$ & $1.082\cdot 10^{-4}$ & $1.102\cdot 10^{-4}$ & $-1.8\cdot 10^{-2}$ & $2.243\cdot 10^{-3}$ & $2.256\cdot 10^{-3}$ & $-5.9\cdot 10^{-3}$ \\
$0.5$ & $0.0$ & $9.000$ & $7.000$ & $1.713\cdot 10^{-4}$ & $1.723\cdot 10^{-4}$ & $-5.8\cdot 10^{-3}$ & $3.093\cdot 10^{-3}$ & $3.078\cdot 10^{-3}$ & $4.9\cdot 10^{-3}$ \\
$0.5$ & $0.2$ & $8.022$ & $6.240$ & $2.914\cdot 10^{-4}$ & $2.885\cdot 10^{-4}$ & $1.0\cdot 10^{-2}$ & $4.488\cdot 10^{-3}$ & $4.403\cdot 10^{-3}$ & $1.9\cdot 10^{-2}$ \\
$0.5$ & $0.6$ & $5.834$ & $4.538$ & $1.222\cdot 10^{-3}$ & $1.143\cdot 10^{-3}$ & $6.4\cdot 10^{-2}$ & $1.228\cdot 10^{-2}$ & $1.147\cdot 10^{-2}$ & $6.6\cdot 10^{-2}$ \\
$0.5$ & $0.9$ & $3.643$ & $2.833$ & $8.674\cdot 10^{-3}$ & $7.191\cdot 10^{-3}$ & $1.7\cdot 10^{-1}$ & $4.867\cdot 10^{-2}$ & $4.117\cdot 10^{-2}$ & $1.5\cdot 10^{-1}$ \\
\hline
$0.5$ & $-0.9$ & $18.718$ & $10.079$ & $3.808\cdot 10^{-6}$ & $3.973\cdot 10^{-6}$ & $-4.3\cdot 10^{-2}$ & $2.062\cdot 10^{-4}$ & $2.115\cdot 10^{-4}$ & $-2.6\cdot 10^{-2}$ \\
$0.5$ & $-0.6$ & $16.895$ & $9.097$ & $6.255\cdot 10^{-6}$ & $6.473\cdot 10^{-6}$ & $-3.5\cdot 10^{-2}$ & $2.915\cdot 10^{-4}$ & $2.968\cdot 10^{-4}$ & $-1.8\cdot 10^{-2}$ \\
$0.5$ & $-0.2$ & $14.345$ & $7.724$ & $1.377\cdot 10^{-5}$ & $1.403\cdot 10^{-5}$ & $-1.9\cdot 10^{-2}$ & $5.058\cdot 10^{-4}$ & $5.077\cdot 10^{-4}$ & $-3.7\cdot 10^{-3}$ \\
$0.5$ & $0.0$ & $13.000$ & $7.000$ & $2.208\cdot 10^{-5}$ & $2.226\cdot 10^{-5}$ & $-8.1\cdot 10^{-3}$ & $7.038\cdot 10^{-4}$ & $6.995\cdot 10^{-4}$ & $6.0\cdot 10^{-3}$ \\
$0.5$ & $0.2$ & $11.588$ & $6.240$ & $3.822\cdot 10^{-5}$ & $3.799\cdot 10^{-5}$ & $6.1\cdot 10^{-3}$ & $1.033\cdot 10^{-3}$ & $1.014\cdot 10^{-3}$ & $1.9\cdot 10^{-2}$ \\
$0.5$ & $0.6$ & $8.427$ & $4.538$ & $1.715\cdot 10^{-4}$ & $1.624\cdot 10^{-4}$ & $5.3\cdot 10^{-2}$ & $2.960\cdot 10^{-3}$ & $2.785\cdot 10^{-3}$ & $5.9\cdot 10^{-2}$ \\
$0.5$ & $0.9$ & $5.262$ & $2.833$ & $1.467\cdot 10^{-3}$ & $1.259\cdot 10^{-3}$ & $1.4\cdot 10^{-1}$ & $1.343\cdot 10^{-2}$ & $1.165\cdot 10^{-2}$ & $1.3\cdot 10^{-1}$ \\
\end{tabular}
\end{ruledtabular}
\end{center}
\end{table*}

\begin{table*}[th]
   \caption{\label{tab:simulations_ecc3} Same scheme as Table~\ref{tab:simulations_ecc1}}
   \begin{center}
     \begin{ruledtabular}
\begin{tabular}{ c r c c | c c r | c c r } 
$e$ & \multicolumn{1}{c}{$\ha$} & $p$ & $p_s$ & $\av{\Eteuk} $ & 
$\av{\ENP}$ & \multicolumn{1}{c|}{$\Delta E_{\rm NP}/E$} & 
$\av{\Jteuk} $ & $\av{\JNP}$ 
& \multicolumn{1}{c}{$\Delta J_{\rm NP}/J$} \\
\hline
\hline
$0.7$ & $-0.9$ & $10.595$ & $10.585$ & $3.364\cdot 10^{-4}$ & $3.394\cdot 10^{-4}$ & $-8.9\cdot 10^{-3}$ & $5.335\cdot 10^{-3}$ & $5.381\cdot 10^{-3}$ & $-8.8\cdot 10^{-3}$ \\
$0.7$ & $-0.6$ & $9.580$ & $9.570$ & $5.294\cdot 10^{-4}$ & $5.316\cdot 10^{-4}$ & $-4.2\cdot 10^{-3}$ & $7.320\cdot 10^{-3}$ & $7.346\cdot 10^{-3}$ & $-3.6\cdot 10^{-3}$ \\
$0.7$ & $-0.2$ & $8.160$ & $8.150$ & $1.077\cdot 10^{-3}$ & $1.070\cdot 10^{-3}$ & $7.4\cdot 10^{-3}$ & $1.203\cdot 10^{-2}$ & $1.193\cdot 10^{-2}$ & $8.6\cdot 10^{-3}$ \\
$0.7$ & $0.0$ & $7.410$ & $7.400$ & $1.637\cdot 10^{-3}$ & $1.609\cdot 10^{-3}$ & $1.7\cdot 10^{-2}$ & $1.613\cdot 10^{-2}$ & $1.583\cdot 10^{-2}$ & $1.9\cdot 10^{-2}$ \\
$0.7$ & $0.2$ & $6.622$ & $6.612$ & $2.641\cdot 10^{-3}$ & $2.558\cdot 10^{-3}$ & $3.2\cdot 10^{-2}$ & $2.255\cdot 10^{-2}$ & $2.181\cdot 10^{-2}$ & $3.3\cdot 10^{-2}$ \\
$0.7$ & $0.6$ & $4.858$ & $4.848$ & $9.173\cdot 10^{-3}$ & $8.375\cdot 10^{-3}$ & $8.7\cdot 10^{-2}$ & $5.390\cdot 10^{-2}$ & $4.916\cdot 10^{-2}$ & $8.8\cdot 10^{-2}$ \\
$0.7$ & $0.9$ & $3.088$ & $3.078$ & $4.151\cdot 10^{-2}$ & $3.437\cdot 10^{-2}$ & $1.7\cdot 10^{-1}$ & $1.504\cdot 10^{-1}$ & $1.249\cdot 10^{-1}$ & $1.7\cdot 10^{-1}$ \\
\hline
$0.7$ & $-0.9$ & $12.873$ & $10.585$ & $3.427\cdot 10^{-5}$ & $3.680\cdot 10^{-5}$ & $-7.4\cdot 10^{-2}$ & $8.231\cdot 10^{-4}$ & $8.685\cdot 10^{-4}$ & $-5.5\cdot 10^{-2}$ \\
$0.7$ & $-0.6$ & $11.639$ & $9.570$ & $5.545\cdot 10^{-5}$ & $5.871\cdot 10^{-5}$ & $-5.9\cdot 10^{-2}$ & $1.155\cdot 10^{-3}$ & $1.201\cdot 10^{-3}$ & $-4.1\cdot 10^{-2}$ \\
$0.7$ & $-0.2$ & $9.912$ & $8.150$ & $1.186\cdot 10^{-4}$ & $1.222\cdot 10^{-4}$ & $-3.0\cdot 10^{-2}$ & $1.974\cdot 10^{-3}$ & $2.000\cdot 10^{-3}$ & $-1.3\cdot 10^{-2}$ \\
$0.7$ & $0.0$ & $9.000$ & $7.400$ & $1.865\cdot 10^{-4}$ & $1.882\cdot 10^{-4}$ & $-9.2\cdot 10^{-3}$ & $2.719\cdot 10^{-3}$ & $2.702\cdot 10^{-3}$ & $5.9\cdot 10^{-3}$ \\
$0.7$ & $0.2$ & $8.042$ & $6.612$ & $3.144\cdot 10^{-4}$ & $3.089\cdot 10^{-4}$ & $1.8\cdot 10^{-2}$ & $3.937\cdot 10^{-3}$ & $3.817\cdot 10^{-3}$ & $3.0\cdot 10^{-2}$ \\
$0.7$ & $0.6$ & $5.896$ & $4.848$ & $1.278\cdot 10^{-3}$ & $1.143\cdot 10^{-3}$ & $1.1\cdot 10^{-1}$ & $1.068\cdot 10^{-2}$ & $9.517\cdot 10^{-3}$ & $1.1\cdot 10^{-1}$ \\
$0.7$ & $0.9$ & $3.744$ & $3.078$ & $8.476\cdot 10^{-3}$ & $6.396\cdot 10^{-3}$ & $2.5\cdot 10^{-1}$ & $4.133\cdot 10^{-2}$ & $3.186\cdot 10^{-2}$ & $2.3\cdot 10^{-1}$ \\
\hline
$0.7$ & $-0.9$ & $18.595$ & $10.585$ & $3.648\cdot 10^{-6}$ & $3.952\cdot 10^{-6}$ & $-8.3\cdot 10^{-2}$ & $1.578\cdot 10^{-4}$ & $1.660\cdot 10^{-4}$ & $-5.2\cdot 10^{-2}$ \\
$0.7$ & $-0.6$ & $16.812$ & $9.570$ & $5.957\cdot 10^{-6}$ & $6.349\cdot 10^{-6}$ & $-6.6\cdot 10^{-2}$ & $2.225\cdot 10^{-4}$ & $2.307\cdot 10^{-4}$ & $-3.7\cdot 10^{-2}$ \\
$0.7$ & $-0.2$ & $14.317$ & $8.150$ & $1.298\cdot 10^{-5}$ & $1.341\cdot 10^{-5}$ & $-3.4\cdot 10^{-2}$ & $3.846\cdot 10^{-4}$ & $3.879\cdot 10^{-4}$ & $-8.5\cdot 10^{-3}$ \\
$0.7$ & $0.0$ & $13.000$ & $7.400$ & $2.067\cdot 10^{-5}$ & $2.093\cdot 10^{-5}$ & $-1.2\cdot 10^{-2}$ & $5.338\cdot 10^{-4}$ & $5.283\cdot 10^{-4}$ & $1.0\cdot 10^{-2}$ \\
$0.7$ & $0.2$ & $11.616$ & $6.612$ & $3.548\cdot 10^{-5}$ & $3.497\cdot 10^{-5}$ & $1.5\cdot 10^{-2}$ & $7.815\cdot 10^{-4}$ & $7.550\cdot 10^{-4}$ & $3.4\cdot 10^{-2}$ \\
$0.7$ & $0.6$ & $8.517$ & $4.848$ & $1.550\cdot 10^{-4}$ & $1.397\cdot 10^{-4}$ & $9.9\cdot 10^{-2}$ & $2.219\cdot 10^{-3}$ & $1.982\cdot 10^{-3}$ & $1.1\cdot 10^{-1}$ \\
$0.7$ & $0.9$ & $5.408$ & $3.078$ & $1.257\cdot 10^{-3}$ & $9.635\cdot 10^{-4}$ & $2.3\cdot 10^{-1}$ & $9.906\cdot 10^{-3}$ & $7.716\cdot 10^{-3}$ & $2.2\cdot 10^{-1}$ \\

\Xhline{2.5\arrayrulewidth}

$0.9$ & $-0.9$ & $11.084$ & $11.074$ & $1.439\cdot 10^{-4}$ & $1.460\cdot 10^{-4}$ & $-1.5\cdot 10^{-2}$ & $2.040\cdot 10^{-3}$ & $2.071\cdot 10^{-3}$ & $-1.5\cdot 10^{-2}$ \\
$0.9$ & $-0.6$ & $10.042$ & $10.032$ & $2.261\cdot 10^{-4}$ & $2.277\cdot 10^{-4}$ & $-7.2\cdot 10^{-3}$ & $2.807\cdot 10^{-3}$ & $2.826\cdot 10^{-3}$ & $-6.7\cdot 10^{-3}$ \\
$0.9$ & $-0.2$ & $8.582$ & $8.572$ & $4.594\cdot 10^{-4}$ & $4.544\cdot 10^{-4}$ & $1.1\cdot 10^{-2}$ & $4.643\cdot 10^{-3}$ & $4.585\cdot 10^{-3}$ & $1.2\cdot 10^{-2}$ \\
$0.9$ & $0.0$ & $7.810$ & $7.800$ & $6.979\cdot 10^{-4}$ & $6.799\cdot 10^{-4}$ & $2.6\cdot 10^{-2}$ & $6.253\cdot 10^{-3}$ & $6.080\cdot 10^{-3}$ & $2.8\cdot 10^{-2}$ \\
$0.9$ & $0.2$ & $6.999$ & $6.989$ & $1.126\cdot 10^{-3}$ & $1.073\cdot 10^{-3}$ & $4.7\cdot 10^{-2}$ & $8.798\cdot 10^{-3}$ & $8.368\cdot 10^{-3}$ & $4.9\cdot 10^{-2}$ \\
$0.9$ & $0.6$ & $5.177$ & $5.167$ & $3.935\cdot 10^{-3}$ & $3.467\cdot 10^{-3}$ & $1.2\cdot 10^{-1}$ & $2.155\cdot 10^{-2}$ & $1.894\cdot 10^{-2}$ & $1.2\cdot 10^{-1}$ \\
$0.9$ & $0.9$ & $3.344$ & $3.334$ & $1.877\cdot 10^{-2}$ & $1.516\cdot 10^{-2}$ & $1.9\cdot 10^{-1}$ & $6.525\cdot 10^{-2}$ & $5.270\cdot 10^{-2}$ & $1.9\cdot 10^{-1}$ \\
\hline
$0.9$ & $-0.9$ & $12.778$ & $11.074$ & $1.704\cdot 10^{-5}$ & $1.883\cdot 10^{-5}$ & $-1.0\cdot 10^{-1}$ & $3.320\cdot 10^{-4}$ & $3.598\cdot 10^{-4}$ & $-8.4\cdot 10^{-2}$ \\
$0.9$ & $-0.6$ & $11.576$ & $10.032$ & $2.741\cdot 10^{-5}$ & $2.969\cdot 10^{-5}$ & $-8.3\cdot 10^{-2}$ & $4.654\cdot 10^{-4}$ & $4.944\cdot 10^{-4}$ & $-6.2\cdot 10^{-2}$ \\
$0.9$ & $-0.2$ & $9.891$ & $8.572$ & $5.806\cdot 10^{-5}$ & $6.043\cdot 10^{-5}$ & $-4.1\cdot 10^{-2}$ & $7.946\cdot 10^{-4}$ & $8.120\cdot 10^{-4}$ & $-2.2\cdot 10^{-2}$ \\
$0.9$ & $0.0$ & $9.000$ & $7.800$ & $9.068\cdot 10^{-5}$ & $9.169\cdot 10^{-5}$ & $-1.1\cdot 10^{-2}$ & $1.093\cdot 10^{-3}$ & $1.087\cdot 10^{-3}$ & $6.1\cdot 10^{-3}$ \\
$0.9$ & $0.2$ & $8.064$ & $6.989$ & $1.516\cdot 10^{-4}$ & $1.475\cdot 10^{-4}$ & $2.7\cdot 10^{-2}$ & $1.582\cdot 10^{-3}$ & $1.515\cdot 10^{-3}$ & $4.2\cdot 10^{-2}$ \\
$0.9$ & $0.6$ & $5.962$ & $5.167$ & $5.996\cdot 10^{-4}$ & $5.120\cdot 10^{-4}$ & $1.5\cdot 10^{-1}$ & $4.278\cdot 10^{-3}$ & $3.633\cdot 10^{-3}$ & $1.5\cdot 10^{-1}$ \\
$0.9$ & $0.9$ & $3.847$ & $3.334$ & $3.818\cdot 10^{-3}$ & $2.678\cdot 10^{-3}$ & $3.0\cdot 10^{-1}$ & $1.656\cdot 10^{-2}$ & $1.184\cdot 10^{-2}$ & $2.8\cdot 10^{-1}$ \\
\hline
$0.9$ & $-0.9$ & $18.457$ & $11.074$ & $1.474\cdot 10^{-6}$ & $1.679\cdot 10^{-6}$ & $-1.4\cdot 10^{-1}$ & $5.209\cdot 10^{-5}$ & $5.684\cdot 10^{-5}$ & $-9.1\cdot 10^{-2}$ \\
$0.9$ & $-0.6$ & $16.720$ & $10.032$ & $2.391\cdot 10^{-6}$ & $2.651\cdot 10^{-6}$ & $-1.1\cdot 10^{-1}$ & $7.330\cdot 10^{-5}$ & $7.803\cdot 10^{-5}$ & $-6.5\cdot 10^{-2}$ \\
$0.9$ & $-0.2$ & $14.287$ & $8.572$ & $5.151\cdot 10^{-6}$ & $5.430\cdot 10^{-6}$ & $-5.4\cdot 10^{-2}$ & $1.262\cdot 10^{-4}$ & $1.284\cdot 10^{-4}$ & $-1.7\cdot 10^{-2}$ \\
$0.9$ & $0.0$ & $13.000$ & $7.800$ & $8.148\cdot 10^{-6}$ & $8.299\cdot 10^{-6}$ & $-1.8\cdot 10^{-2}$ & $1.747\cdot 10^{-4}$ & $1.724\cdot 10^{-4}$ & $1.3\cdot 10^{-2}$ \\
$0.9$ & $0.2$ & $11.648$ & $6.989$ & $1.387\cdot 10^{-5}$ & $1.352\cdot 10^{-5}$ & $2.5\cdot 10^{-2}$ & $2.550\cdot 10^{-4}$ & $2.421\cdot 10^{-4}$ & $5.1\cdot 10^{-2}$ \\
$0.9$ & $0.6$ & $8.612$ & $5.167$ & $5.904\cdot 10^{-5}$ & $5.003\cdot 10^{-5}$ & $1.5\cdot 10^{-1}$ & $7.194\cdot 10^{-4}$ & $6.041\cdot 10^{-4}$ & $1.6\cdot 10^{-1}$ \\
$0.9$ & $0.9$ & $5.557$ & $3.334$ & $4.602\cdot 10^{-4}$ & $3.105\cdot 10^{-4}$ & $3.3\cdot 10^{-1}$ & $3.196\cdot 10^{-3}$ & $2.211\cdot 10^{-3}$ & $3.1\cdot 10^{-1}$ \\
\end{tabular}
\end{ruledtabular}
\end{center}
\end{table*}


\begin{table*}[th]
   \caption{\label{tab:AllFluxesDiff1} Analytical/numerical relative differences for the energy and 
   angular momentum fluxes: $\Delta F_{\rm analytic}/F = (\av{\Fteuk}
   - \av{\dot{F}_{\rm analytic}})/\av{\Fteuk} $. 
   We report both the fluxes computed from the radiation reactions and from the EOB waveform for 
   all the eccentric simulations. Black: $\Delta F/F<1\%$, 
   \textcolor{colortab1}{blue}: $1\% \leq \Delta F/F<5\%$, 
   \textcolor{colortab2}{red} : $5\%\leq \Delta F/F$. }
   \begin{center}
     \begin{ruledtabular}
\begin{tabular}{ c r c c | r r r r | r r r r} 
$e$ & \multicolumn{1}{c}{$\ha$} & $p$ & $p_s$       & \multicolumn{1}{c}{$\Delta E_{\rm NP}/E $}  &
\multicolumn{1}{c}{$\Delta E_{\rm old}/E $} & \multicolumn{1}{c}{$\Delta E_{\rm ANP}/E $} &
\multicolumn{1}{c|}{$\Delta E_{h_\lm}/E $}  & \multicolumn{1}{c}{$\Delta J_{\rm NP}/J $}   &
\multicolumn{1}{c}{$\Delta J_{\rm old}/J $} & \multicolumn{1}{c}{$\Delta J_{\rm ANP}/J $}&
\multicolumn{1}{c}{$\Delta J_{h_\lm}/J  $}\\
\hline
\hline 
$0.1$ & $-0.9$ & $9.014$ & $9.004$ & $  3.8\cdot 10^{-4} $ & $  4.9\cdot 10^{-4} $ & $  5.5\cdot 10^{-4} $ & $  6.5\cdot 10^{-4} $ & $  4.3\cdot 10^{-4} $ & $  5.2\cdot 10^{-4} $ & $  5.6\cdot 10^{-4} $ & $  5.9\cdot 10^{-4} $ \\
$0.1$ & $-0.6$ & $8.119$ & $8.109$ & $  7.8\cdot 10^{-4} $ & $  9.0\cdot 10^{-4} $ & $  9.7\cdot 10^{-4} $ & $  1.1\cdot 10^{-3} $ & $  8.2\cdot 10^{-4} $ & $  9.1\cdot 10^{-4} $ & $  9.6\cdot 10^{-4} $ & $  1.0\cdot 10^{-3} $ \\
$0.1$ & $-0.2$ & $6.869$ & $6.859$ & $  1.9\cdot 10^{-3} $ & $  2.1\cdot 10^{-3} $ & $  2.1\cdot 10^{-3} $ & $  2.2\cdot 10^{-3} $ & $  1.9\cdot 10^{-3} $ & $  2.0\cdot 10^{-3} $ & $  2.1\cdot 10^{-3} $ & $  2.1\cdot 10^{-3} $ \\
$0.1$ & $0.0$ & $6.210$ & $6.200$ & $  3.1\cdot 10^{-3} $ & $  3.2\cdot 10^{-3} $ & $  3.3\cdot 10^{-3} $ & $  3.4\cdot 10^{-3} $ & $  3.0\cdot 10^{-3} $ & $  3.1\cdot 10^{-3} $ & $  3.2\cdot 10^{-3} $ & $  3.2\cdot 10^{-3} $ \\
$0.1$ & $0.2$ & $5.518$ & $5.508$ & $  5.0\cdot 10^{-3} $ & $  5.1\cdot 10^{-3} $ & $  5.2\cdot 10^{-3} $ & $  5.3\cdot 10^{-3} $ & $  4.9\cdot 10^{-3} $ & $  5.0\cdot 10^{-3} $ & $  5.0\cdot 10^{-3} $ & $  5.1\cdot 10^{-3} $ \\
$0.1$ & $0.6$ & $3.970$ & $3.960$ & $ \color{colortab1} 1.6\cdot 10^{-2} $ & $ \color{colortab1} 1.6\cdot 10^{-2} $ & $ \color{colortab1} 1.6\cdot 10^{-2} $ & $ \color{colortab1} 1.6\cdot 10^{-2} $ & $ \color{colortab1} 1.6\cdot 10^{-2} $ & $ \color{colortab1} 1.6\cdot 10^{-2} $ & $ \color{colortab1} 1.6\cdot 10^{-2} $ & $ \color{colortab1} 1.6\cdot 10^{-2} $ \\
$0.1$ & $0.9$ & $2.415$ & $2.405$ & $ \color{colortab2} 7.2\cdot 10^{-2} $ & $ \color{colortab2} 7.2\cdot 10^{-2} $ & $ \color{colortab2} 7.2\cdot 10^{-2} $ & $ \color{colortab2} 7.3\cdot 10^{-2} $ & $ \color{colortab2} 7.1\cdot 10^{-2} $ & $ \color{colortab2} 7.1\cdot 10^{-2} $ & $ \color{colortab2} 7.1\cdot 10^{-2} $ & $ \color{colortab2} 7.1\cdot 10^{-2} $ \\
\hline
$0.1$ & $-0.9$ & $13.070$ & $9.004$ & $  -2.1\cdot 10^{-3} $ & $  -4.5\cdot 10^{-4} $ & $  1.9\cdot 10^{-4} $ & $  4.4\cdot 10^{-4} $ & $  -1.2\cdot 10^{-3} $ & $  -2.9\cdot 10^{-5} $ & $  4.1\cdot 10^{-4} $ & $  4.1\cdot 10^{-4} $ \\
$0.1$ & $-0.6$ & $11.772$ & $8.109$ & $  -1.7\cdot 10^{-3} $ & $  4.5\cdot 10^{-5} $ & $  7.1\cdot 10^{-4} $ & $  1.0\cdot 10^{-3} $ & $  -8.1\cdot 10^{-4} $ & $  3.9\cdot 10^{-4} $ & $  8.5\cdot 10^{-4} $ & $  8.5\cdot 10^{-4} $ \\
$0.1$ & $-0.2$ & $9.957$ & $6.859$ & $  -8.3\cdot 10^{-4} $ & $  1.0\cdot 10^{-3} $ & $  1.7\cdot 10^{-3} $ & $  2.1\cdot 10^{-3} $ & $  -6.0\cdot 10^{-5} $ & $  1.2\cdot 10^{-3} $ & $  1.7\cdot 10^{-3} $ & $  1.7\cdot 10^{-3} $ \\
$0.1$ & $0.0$ & $9.000$ & $6.200$ & $  -1.8\cdot 10^{-4} $ & $  1.7\cdot 10^{-3} $ & $  2.5\cdot 10^{-3} $ & $  2.9\cdot 10^{-3} $ & $  4.9\cdot 10^{-4} $ & $  1.8\cdot 10^{-3} $ & $  2.3\cdot 10^{-3} $ & $  2.3\cdot 10^{-3} $ \\
$0.1$ & $0.2$ & $7.996$ & $5.508$ & $  7.2\cdot 10^{-4} $ & $  2.6\cdot 10^{-3} $ & $  3.4\cdot 10^{-3} $ & $  3.9\cdot 10^{-3} $ & $  1.2\cdot 10^{-3} $ & $  2.6\cdot 10^{-3} $ & $  3.1\cdot 10^{-3} $ & $  3.1\cdot 10^{-3} $ \\
$0.1$ & $0.6$ & $5.749$ & $3.960$ & $  4.5\cdot 10^{-3} $ & $  6.4\cdot 10^{-3} $ & $  7.3\cdot 10^{-3} $ & $  7.9\cdot 10^{-3} $ & $  4.4\cdot 10^{-3} $ & $  5.7\cdot 10^{-3} $ & $  6.2\cdot 10^{-3} $ & $  6.3\cdot 10^{-3} $ \\
$0.1$ & $0.9$ & $3.491$ & $2.405$ & $ \color{colortab1} 2.1\cdot 10^{-2} $ & $ \color{colortab1} 2.2\cdot 10^{-2} $ & $ \color{colortab1} 2.3\cdot 10^{-2} $ & $ \color{colortab1} 2.4\cdot 10^{-2} $ & $ \color{colortab1} 1.9\cdot 10^{-2} $ & $ \color{colortab1} 2.0\cdot 10^{-2} $ & $ \color{colortab1} 2.0\cdot 10^{-2} $ & $ \color{colortab1} 2.0\cdot 10^{-2} $ \\
\hline
$0.1$ & $-0.9$ & $18.879$ & $9.004$ & $  -1.8\cdot 10^{-3} $ & $  1.3\cdot 10^{-4} $ & $  7.4\cdot 10^{-4} $ & $  9.0\cdot 10^{-4} $ & $  -8.6\cdot 10^{-4} $ & $  4.6\cdot 10^{-4} $ & $  8.7\cdot 10^{-4} $ & $  8.7\cdot 10^{-4} $ \\
$0.1$ & $-0.6$ & $17.004$ & $8.109$ & $  -1.5\cdot 10^{-3} $ & $  5.2\cdot 10^{-4} $ & $  1.2\cdot 10^{-3} $ & $  1.4\cdot 10^{-3} $ & $  -6.0\cdot 10^{-4} $ & $  7.9\cdot 10^{-4} $ & $  1.2\cdot 10^{-3} $ & $  1.2\cdot 10^{-3} $ \\
$0.1$ & $-0.2$ & $14.382$ & $6.859$ & $  -9.1\cdot 10^{-4} $ & $  1.3\cdot 10^{-3} $ & $  2.0\cdot 10^{-3} $ & $  2.3\cdot 10^{-3} $ & $  -6.7\cdot 10^{-5} $ & $  1.4\cdot 10^{-3} $ & $  1.9\cdot 10^{-3} $ & $  1.9\cdot 10^{-3} $ \\
$0.1$ & $0.0$ & $13.000$ & $6.200$ & $  -4.7\cdot 10^{-4} $ & $  1.8\cdot 10^{-3} $ & $  2.6\cdot 10^{-3} $ & $  2.9\cdot 10^{-3} $ & $  3.2\cdot 10^{-4} $ & $  1.9\cdot 10^{-3} $ & $  2.4\cdot 10^{-3} $ & $  2.4\cdot 10^{-3} $ \\
$0.1$ & $0.2$ & $11.549$ & $5.508$ & $  1.2\cdot 10^{-4} $ & $  2.5\cdot 10^{-3} $ & $  3.3\cdot 10^{-3} $ & $  3.7\cdot 10^{-3} $ & $  8.4\cdot 10^{-4} $ & $  2.5\cdot 10^{-3} $ & $  3.0\cdot 10^{-3} $ & $  3.0\cdot 10^{-3} $ \\
$0.1$ & $0.6$ & $8.304$ & $3.960$ & $  2.4\cdot 10^{-3} $ & $  5.0\cdot 10^{-3} $ & $  5.9\cdot 10^{-3} $ & $  6.6\cdot 10^{-3} $ & $  2.7\cdot 10^{-3} $ & $  4.5\cdot 10^{-3} $ & $  5.1\cdot 10^{-3} $ & $  5.1\cdot 10^{-3} $ \\
$0.1$ & $0.9$ & $5.043$ & $2.405$ & $  8.8\cdot 10^{-3} $ & $ \color{colortab1} 1.1\cdot 10^{-2} $ & $ \color{colortab1} 1.2\cdot 10^{-2} $ & $ \color{colortab1} 1.3\cdot 10^{-2} $ & $  7.9\cdot 10^{-3} $ & $  9.5\cdot 10^{-3} $ & $ \color{colortab1} 1.0\cdot 10^{-2} $ & $ \color{colortab1} 1.0\cdot 10^{-2} $ \\

\Xhline{2.5\arrayrulewidth}

$0.3$ & $-0.9$ & $9.564$ & $9.554$ & $  -1.4\cdot 10^{-3} $ & $  -5.0\cdot 10^{-5} $ & $  6.2\cdot 10^{-4} $ & $  1.0\cdot 10^{-3} $ & $  -1.1\cdot 10^{-3} $ & $  2.1\cdot 10^{-4} $ & $  8.4\cdot 10^{-4} $ & $  8.8\cdot 10^{-4} $ \\
$0.3$ & $-0.6$ & $8.622$ & $8.612$ & $  -2.0\cdot 10^{-4} $ & $  1.3\cdot 10^{-3} $ & $  2.0\cdot 10^{-3} $ & $  2.5\cdot 10^{-3} $ & $  1.9\cdot 10^{-4} $ & $  1.5\cdot 10^{-3} $ & $  2.2\cdot 10^{-3} $ & $  2.2\cdot 10^{-3} $ \\
$0.3$ & $-0.2$ & $7.305$ & $7.295$ & $  3.0\cdot 10^{-3} $ & $  4.6\cdot 10^{-3} $ & $  5.3\cdot 10^{-3} $ & $  5.9\cdot 10^{-3} $ & $  3.3\cdot 10^{-3} $ & $  4.8\cdot 10^{-3} $ & $  5.5\cdot 10^{-3} $ & $  5.5\cdot 10^{-3} $ \\
$0.3$ & $0.0$ & $6.610$ & $6.600$ & $  5.9\cdot 10^{-3} $ & $  7.5\cdot 10^{-3} $ & $  8.3\cdot 10^{-3} $ & $  9.0\cdot 10^{-3} $ & $  6.2\cdot 10^{-3} $ & $  7.6\cdot 10^{-3} $ & $  8.4\cdot 10^{-3} $ & $  8.5\cdot 10^{-3} $ \\
$0.3$ & $0.1$ & $6.250$ & $6.240$ & $  7.9\cdot 10^{-3} $ & $  9.6\cdot 10^{-3} $ & $ \color{colortab1} 1.0\cdot 10^{-2} $ & $ \color{colortab1} 1.1\cdot 10^{-2} $ & $  8.1\cdot 10^{-3} $ & $  9.6\cdot 10^{-3} $ & $ \color{colortab1} 1.0\cdot 10^{-2} $ & $ \color{colortab1} 1.0\cdot 10^{-2} $ \\
$0.3$ & $0.2$ & $5.881$ & $5.871$ & $ \color{colortab1} 1.1\cdot 10^{-2} $ & $ \color{colortab1} 1.2\cdot 10^{-2} $ & $ \color{colortab1} 1.3\cdot 10^{-2} $ & $ \color{colortab1} 1.4\cdot 10^{-2} $ & $ \color{colortab1} 1.1\cdot 10^{-2} $ & $ \color{colortab1} 1.2\cdot 10^{-2} $ & $ \color{colortab1} 1.3\cdot 10^{-2} $ & $ \color{colortab1} 1.3\cdot 10^{-2} $ \\
$0.3$ & $0.3$ & $5.500$ & $5.490$ & $ \color{colortab1} 1.4\cdot 10^{-2} $ & $ \color{colortab1} 1.6\cdot 10^{-2} $ & $ \color{colortab1} 1.6\cdot 10^{-2} $ & $ \color{colortab1} 1.7\cdot 10^{-2} $ & $ \color{colortab1} 1.4\cdot 10^{-2} $ & $ \color{colortab1} 1.5\cdot 10^{-2} $ & $ \color{colortab1} 1.6\cdot 10^{-2} $ & $ \color{colortab1} 1.6\cdot 10^{-2} $ \\
$0.3$ & $0.4$ & $5.104$ & $5.094$ & $ \color{colortab1} 1.8\cdot 10^{-2} $ & $ \color{colortab1} 2.0\cdot 10^{-2} $ & $ \color{colortab1} 2.1\cdot 10^{-2} $ & $ \color{colortab1} 2.2\cdot 10^{-2} $ & $ \color{colortab1} 1.8\cdot 10^{-2} $ & $ \color{colortab1} 1.9\cdot 10^{-2} $ & $ \color{colortab1} 2.0\cdot 10^{-2} $ & $ \color{colortab1} 2.0\cdot 10^{-2} $ \\
$0.3$ & $0.5$ & $4.689$ & $4.679$ & $ \color{colortab1} 2.4\cdot 10^{-2} $ & $ \color{colortab1} 2.6\cdot 10^{-2} $ & $ \color{colortab1} 2.7\cdot 10^{-2} $ & $ \color{colortab1} 2.8\cdot 10^{-2} $ & $ \color{colortab1} 2.4\cdot 10^{-2} $ & $ \color{colortab1} 2.5\cdot 10^{-2} $ & $ \color{colortab1} 2.6\cdot 10^{-2} $ & $ \color{colortab1} 2.6\cdot 10^{-2} $ \\
$0.3$ & $0.6$ & $4.250$ & $4.240$ & $ \color{colortab1} 3.3\cdot 10^{-2} $ & $ \color{colortab1} 3.5\cdot 10^{-2} $ & $ \color{colortab1} 3.6\cdot 10^{-2} $ & $ \color{colortab1} 3.7\cdot 10^{-2} $ & $ \color{colortab1} 3.2\cdot 10^{-2} $ & $ \color{colortab1} 3.3\cdot 10^{-2} $ & $ \color{colortab1} 3.4\cdot 10^{-2} $ & $ \color{colortab1} 3.4\cdot 10^{-2} $ \\
$0.3$ & $0.7$ & $3.777$ & $3.767$ & $ \color{colortab1} 4.6\cdot 10^{-2} $ & $ \color{colortab1} 4.8\cdot 10^{-2} $ & $ \color{colortab1} 4.9\cdot 10^{-2} $ & $ \color{colortab1} 5.0\cdot 10^{-2} $ & $ \color{colortab1} 4.4\cdot 10^{-2} $ & $ \color{colortab1} 4.5\cdot 10^{-2} $ & $ \color{colortab1} 4.6\cdot 10^{-2} $ & $ \color{colortab1} 4.7\cdot 10^{-2} $ \\
$0.3$ & $0.8$ & $3.249$ & $3.239$ & $ \color{colortab2} 6.7\cdot 10^{-2} $ & $ \color{colortab2} 6.9\cdot 10^{-2} $ & $ \color{colortab2} 7.0\cdot 10^{-2} $ & $ \color{colortab2} 7.1\cdot 10^{-2} $ & $ \color{colortab2} 6.4\cdot 10^{-2} $ & $ \color{colortab2} 6.5\cdot 10^{-2} $ & $ \color{colortab2} 6.6\cdot 10^{-2} $ & $ \color{colortab2} 6.7\cdot 10^{-2} $ \\
$0.3$ & $0.9$ & $2.615$ & $2.605$ & $ \color{colortab2} 1.1\cdot 10^{-1} $ & $ \color{colortab2} 1.1\cdot 10^{-1} $ & $ \color{colortab2} 1.1\cdot 10^{-1} $ & $ \color{colortab2} 1.1\cdot 10^{-1} $ & $ \color{colortab2} 1.0\cdot 10^{-1} $ & $ \color{colortab2} 1.0\cdot 10^{-1} $ & $ \color{colortab2} 1.0\cdot 10^{-1} $ & $ \color{colortab2} 1.0\cdot 10^{-1} $ \\
\hline
$0.3$ & $-0.9$ & $13.028$ & $9.554$ & $ \color{colortab1} -1.8\cdot 10^{-2} $ & $  -5.2\cdot 10^{-3} $ & $  -3.6\cdot 10^{-4} $ & $  1.7\cdot 10^{-3} $ & $ \color{colortab1} -1.1\cdot 10^{-2} $ & $  -1.7\cdot 10^{-3} $ & $  2.0\cdot 10^{-3} $ & $  2.0\cdot 10^{-3} $ \\
$0.3$ & $-0.6$ & $11.743$ & $8.612$ & $ \color{colortab1} -1.4\cdot 10^{-2} $ & $  -1.3\cdot 10^{-3} $ & $  3.8\cdot 10^{-3} $ & $  6.1\cdot 10^{-3} $ & $  -7.9\cdot 10^{-3} $ & $  1.9\cdot 10^{-3} $ & $  5.7\cdot 10^{-3} $ & $  5.7\cdot 10^{-3} $ \\
$0.3$ & $-0.2$ & $9.947$ & $7.295$ & $  -7.4\cdot 10^{-3} $ & $  6.1\cdot 10^{-3} $ & $ \color{colortab1} 1.1\cdot 10^{-2} $ & $ \color{colortab1} 1.4\cdot 10^{-2} $ & $  -1.7\cdot 10^{-3} $ & $  8.4\cdot 10^{-3} $ & $ \color{colortab1} 1.2\cdot 10^{-2} $ & $ \color{colortab1} 1.3\cdot 10^{-2} $ \\
$0.3$ & $0.0$ & $9.000$ & $6.600$ & $  -2.5\cdot 10^{-3} $ & $ \color{colortab1} 1.1\cdot 10^{-2} $ & $ \color{colortab1} 1.7\cdot 10^{-2} $ & $ \color{colortab1} 2.0\cdot 10^{-2} $ & $  2.6\cdot 10^{-3} $ & $ \color{colortab1} 1.3\cdot 10^{-2} $ & $ \color{colortab1} 1.7\cdot 10^{-2} $ & $ \color{colortab1} 1.7\cdot 10^{-2} $ \\
$0.3$ & $0.2$ & $8.006$ & $5.871$ & $  4.1\cdot 10^{-3} $ & $ \color{colortab1} 1.8\cdot 10^{-2} $ & $ \color{colortab1} 2.4\cdot 10^{-2} $ & $ \color{colortab1} 2.7\cdot 10^{-2} $ & $  8.2\cdot 10^{-3} $ & $ \color{colortab1} 1.9\cdot 10^{-2} $ & $ \color{colortab1} 2.3\cdot 10^{-2} $ & $ \color{colortab1} 2.3\cdot 10^{-2} $ \\
$0.3$ & $0.6$ & $5.782$ & $4.240$ & $ \color{colortab1} 2.8\cdot 10^{-2} $ & $ \color{colortab1} 4.2\cdot 10^{-2} $ & $ \color{colortab1} 4.8\cdot 10^{-2} $ & $ \color{colortab2} 5.2\cdot 10^{-2} $ & $ \color{colortab1} 2.8\cdot 10^{-2} $ & $ \color{colortab1} 3.8\cdot 10^{-2} $ & $ \color{colortab1} 4.3\cdot 10^{-2} $ & $ \color{colortab1} 4.3\cdot 10^{-2} $ \\
$0.3$ & $0.9$ & $3.553$ & $2.605$ & $ \color{colortab2} 8.6\cdot 10^{-2} $ & $ \color{colortab2} 9.7\cdot 10^{-2} $ & $ \color{colortab2} 1.0\cdot 10^{-1} $ & $ \color{colortab2} 1.0\cdot 10^{-1} $ & $ \color{colortab2} 7.4\cdot 10^{-2} $ & $ \color{colortab2} 8.1\cdot 10^{-2} $ & $ \color{colortab2} 8.5\cdot 10^{-2} $ & $ \color{colortab2} 8.6\cdot 10^{-2} $ \\
\hline
$0.3$ & $-0.9$ & $18.818$ & $9.554$ & $ \color{colortab1} -1.6\cdot 10^{-2} $ & $  -7.7\cdot 10^{-4} $ & $  4.2\cdot 10^{-3} $ & $  5.5\cdot 10^{-3} $ & $  -9.0\cdot 10^{-3} $ & $  2.4\cdot 10^{-3} $ & $  5.9\cdot 10^{-3} $ & $  6.0\cdot 10^{-3} $ \\
$0.3$ & $-0.6$ & $16.962$ & $8.612$ & $ \color{colortab1} -1.3\cdot 10^{-2} $ & $  2.9\cdot 10^{-3} $ & $  8.2\cdot 10^{-3} $ & $  9.6\cdot 10^{-3} $ & $  -6.3\cdot 10^{-3} $ & $  5.6\cdot 10^{-3} $ & $  9.4\cdot 10^{-3} $ & $  9.4\cdot 10^{-3} $ \\
$0.3$ & $-0.2$ & $14.368$ & $7.295$ & $  -7.8\cdot 10^{-3} $ & $  9.5\cdot 10^{-3} $ & $ \color{colortab1} 1.5\cdot 10^{-2} $ & $ \color{colortab1} 1.7\cdot 10^{-2} $ & $  -1.2\cdot 10^{-3} $ & $ \color{colortab1} 1.1\cdot 10^{-2} $ & $ \color{colortab1} 1.6\cdot 10^{-2} $ & $ \color{colortab1} 1.6\cdot 10^{-2} $ \\
$0.3$ & $0.0$ & $13.000$ & $6.600$ & $  -3.9\cdot 10^{-3} $ & $ \color{colortab1} 1.4\cdot 10^{-2} $ & $ \color{colortab1} 2.0\cdot 10^{-2} $ & $ \color{colortab1} 2.2\cdot 10^{-2} $ & $  2.3\cdot 10^{-3} $ & $ \color{colortab1} 1.5\cdot 10^{-2} $ & $ \color{colortab1} 2.0\cdot 10^{-2} $ & $ \color{colortab1} 2.0\cdot 10^{-2} $ \\
$0.3$ & $0.2$ & $11.564$ & $5.871$ & $  1.4\cdot 10^{-3} $ & $ \color{colortab1} 2.0\cdot 10^{-2} $ & $ \color{colortab1} 2.6\cdot 10^{-2} $ & $ \color{colortab1} 2.9\cdot 10^{-2} $ & $  7.0\cdot 10^{-3} $ & $ \color{colortab1} 2.0\cdot 10^{-2} $ & $ \color{colortab1} 2.5\cdot 10^{-2} $ & $ \color{colortab1} 2.5\cdot 10^{-2} $ \\
$0.3$ & $0.6$ & $8.352$ & $4.240$ & $ \color{colortab1} 2.0\cdot 10^{-2} $ & $ \color{colortab1} 3.9\cdot 10^{-2} $ & $ \color{colortab1} 4.7\cdot 10^{-2} $ & $ \color{colortab2} 5.1\cdot 10^{-2} $ & $ \color{colortab1} 2.3\cdot 10^{-2} $ & $ \color{colortab1} 3.7\cdot 10^{-2} $ & $ \color{colortab1} 4.2\cdot 10^{-2} $ & $ \color{colortab1} 4.2\cdot 10^{-2} $ \\
$0.3$ & $0.9$ & $5.132$ & $2.605$ & $ \color{colortab2} 6.1\cdot 10^{-2} $ & $ \color{colortab2} 7.8\cdot 10^{-2} $ & $ \color{colortab2} 8.6\cdot 10^{-2} $ & $ \color{colortab2} 9.2\cdot 10^{-2} $ & $ \color{colortab2} 5.5\cdot 10^{-2} $ & $ \color{colortab2} 6.8\cdot 10^{-2} $ & $ \color{colortab2} 7.3\cdot 10^{-2} $ & $ \color{colortab2} 7.4\cdot 10^{-2} $ \\

\Xhline{2.5\arrayrulewidth}

$0.5$ & $-0.9$ & $10.089$ & $10.079$ & $  -4.6\cdot 10^{-3} $ & $  -1.2\cdot 10^{-3} $ & $  4.4\cdot 10^{-4} $ & $  1.5\cdot 10^{-3} $ & $  -4.2\cdot 10^{-3} $ & $  -6.2\cdot 10^{-4} $ & $  1.1\cdot 10^{-3} $ & $  1.2\cdot 10^{-3} $ \\
$0.5$ & $-0.6$ & $9.107$ & $9.097$ & $  -2.0\cdot 10^{-3} $ & $  1.6\cdot 10^{-3} $ & $  3.3\cdot 10^{-3} $ & $  4.5\cdot 10^{-3} $ & $  -1.4\cdot 10^{-3} $ & $  2.4\cdot 10^{-3} $ & $  4.2\cdot 10^{-3} $ & $  4.4\cdot 10^{-3} $ \\
$0.5$ & $-0.2$ & $7.734$ & $7.724$ & $  4.7\cdot 10^{-3} $ & $  8.5\cdot 10^{-3} $ & $ \color{colortab1} 1.0\cdot 10^{-2} $ & $ \color{colortab1} 1.2\cdot 10^{-2} $ & $  5.5\cdot 10^{-3} $ & $  9.5\cdot 10^{-3} $ & $ \color{colortab1} 1.1\cdot 10^{-2} $ & $ \color{colortab1} 1.2\cdot 10^{-2} $ \\
$0.5$ & $0.0$ & $7.010$ & $7.000$ & $ \color{colortab1} 1.1\cdot 10^{-2} $ & $ \color{colortab1} 1.4\cdot 10^{-2} $ & $ \color{colortab1} 1.6\cdot 10^{-2} $ & $ \color{colortab1} 1.8\cdot 10^{-2} $ & $ \color{colortab1} 1.1\cdot 10^{-2} $ & $ \color{colortab1} 1.5\cdot 10^{-2} $ & $ \color{colortab1} 1.7\cdot 10^{-2} $ & $ \color{colortab1} 1.8\cdot 10^{-2} $ \\
$0.5$ & $0.2$ & $6.250$ & $6.240$ & $ \color{colortab1} 2.0\cdot 10^{-2} $ & $ \color{colortab1} 2.3\cdot 10^{-2} $ & $ \color{colortab1} 2.5\cdot 10^{-2} $ & $ \color{colortab1} 2.7\cdot 10^{-2} $ & $ \color{colortab1} 2.0\cdot 10^{-2} $ & $ \color{colortab1} 2.4\cdot 10^{-2} $ & $ \color{colortab1} 2.6\cdot 10^{-2} $ & $ \color{colortab1} 2.7\cdot 10^{-2} $ \\
$0.5$ & $0.6$ & $4.548$ & $4.538$ & $ \color{colortab2} 5.8\cdot 10^{-2} $ & $ \color{colortab2} 6.2\cdot 10^{-2} $ & $ \color{colortab2} 6.4\cdot 10^{-2} $ & $ \color{colortab2} 6.6\cdot 10^{-2} $ & $ \color{colortab2} 5.7\cdot 10^{-2} $ & $ \color{colortab2} 6.2\cdot 10^{-2} $ & $ \color{colortab2} 6.4\cdot 10^{-2} $ & $ \color{colortab2} 6.4\cdot 10^{-2} $ \\
$0.5$ & $0.9$ & $2.843$ & $2.833$ & $ \color{colortab2} 1.4\cdot 10^{-1} $ & $ \color{colortab2} 1.5\cdot 10^{-1} $ & $ \color{colortab2} 1.5\cdot 10^{-1} $ & $ \color{colortab2} 1.5\cdot 10^{-1} $ & $ \color{colortab2} 1.4\cdot 10^{-1} $ & $ \color{colortab2} 1.4\cdot 10^{-1} $ & $ \color{colortab2} 1.4\cdot 10^{-1} $ & $ \color{colortab2} 1.5\cdot 10^{-1} $ \\
\hline
$0.5$ & $-0.9$ & $12.959$ & $10.079$ & $ \color{colortab1} -4.3\cdot 10^{-2} $ & $ \color{colortab1} -1.6\cdot 10^{-2} $ & $  -5.4\cdot 10^{-3} $ & $  -1.8\cdot 10^{-4} $ & $ \color{colortab1} -3.0\cdot 10^{-2} $ & $  -7.5\cdot 10^{-3} $ & $  1.5\cdot 10^{-3} $ & $  1.6\cdot 10^{-3} $ \\
$0.5$ & $-0.6$ & $11.697$ & $9.097$ & $ \color{colortab1} -3.5\cdot 10^{-2} $ & $  -7.1\cdot 10^{-3} $ & $  4.2\cdot 10^{-3} $ & $  9.5\cdot 10^{-3} $ & $ \color{colortab1} -2.1\cdot 10^{-2} $ & $  1.2\cdot 10^{-3} $ & $ \color{colortab1} 1.0\cdot 10^{-2} $ & $ \color{colortab1} 1.1\cdot 10^{-2} $ \\
$0.5$ & $-0.2$ & $9.931$ & $7.724$ & $ \color{colortab1} -1.8\cdot 10^{-2} $ & $ \color{colortab1} 1.0\cdot 10^{-2} $ & $ \color{colortab1} 2.2\cdot 10^{-2} $ & $ \color{colortab1} 2.8\cdot 10^{-2} $ & $  -5.9\cdot 10^{-3} $ & $ \color{colortab1} 1.7\cdot 10^{-2} $ & $ \color{colortab1} 2.7\cdot 10^{-2} $ & $ \color{colortab1} 2.7\cdot 10^{-2} $ \\
$0.5$ & $0.0$ & $9.000$ & $7.000$ & $  -5.8\cdot 10^{-3} $ & $ \color{colortab1} 2.2\cdot 10^{-2} $ & $ \color{colortab1} 3.4\cdot 10^{-2} $ & $ \color{colortab1} 4.0\cdot 10^{-2} $ & $  4.9\cdot 10^{-3} $ & $ \color{colortab1} 2.8\cdot 10^{-2} $ & $ \color{colortab1} 3.8\cdot 10^{-2} $ & $ \color{colortab1} 3.8\cdot 10^{-2} $ \\
$0.5$ & $0.2$ & $8.022$ & $6.240$ & $  1.0\cdot 10^{-2} $ & $ \color{colortab1} 3.8\cdot 10^{-2} $ & $ \color{colortab1} 5.0\cdot 10^{-2} $ & $ \color{colortab2} 5.6\cdot 10^{-2} $ & $ \color{colortab1} 1.9\cdot 10^{-2} $ & $ \color{colortab1} 4.2\cdot 10^{-2} $ & $ \color{colortab2} 5.2\cdot 10^{-2} $ & $ \color{colortab2} 5.3\cdot 10^{-2} $ \\
$0.5$ & $0.6$ & $5.834$ & $4.538$ & $ \color{colortab2} 6.4\cdot 10^{-2} $ & $ \color{colortab2} 9.0\cdot 10^{-2} $ & $ \color{colortab2} 1.0\cdot 10^{-1} $ & $ \color{colortab2} 1.1\cdot 10^{-1} $ & $ \color{colortab2} 6.6\cdot 10^{-2} $ & $ \color{colortab2} 8.8\cdot 10^{-2} $ & $ \color{colortab2} 9.8\cdot 10^{-2} $ & $ \color{colortab2} 9.9\cdot 10^{-2} $ \\
$0.5$ & $0.9$ & $3.643$ & $2.833$ & $ \color{colortab2} 1.7\cdot 10^{-1} $ & $ \color{colortab2} 1.9\cdot 10^{-1} $ & $ \color{colortab2} 2.0\cdot 10^{-1} $ & $ \color{colortab2} 2.0\cdot 10^{-1} $ & $ \color{colortab2} 1.5\cdot 10^{-1} $ & $ \color{colortab2} 1.7\cdot 10^{-1} $ & $ \color{colortab2} 1.8\cdot 10^{-1} $ & $ \color{colortab2} 1.8\cdot 10^{-1} $ \\
\end{tabular}
 \end{ruledtabular}
 \end{center}
 \end{table*} 

\begin{table*}[th]
   \caption{\label{tab:AllFluxesDiff2} Same scheme as Table~\ref{tab:AllFluxesDiff1}.}
   \begin{center}
     \begin{ruledtabular}
\begin{tabular}{ c r c c | r r r r | r r r r} 
$e$ & \multicolumn{1}{c}{$\ha$} & $p$ & $p_s$       & \multicolumn{1}{c}{$\Delta E_{\rm NP}/E $}  &
\multicolumn{1}{c}{$\Delta E_{\rm old}/E $} & \multicolumn{1}{c}{$\Delta E_{\rm ANP}/E $} &
\multicolumn{1}{c|}{$\Delta E_{h_\lm}/E $}  & \multicolumn{1}{c}{$\Delta J_{\rm NP}/J $}   &
\multicolumn{1}{c}{$\Delta J_{\rm old}/J $} & \multicolumn{1}{c}{$\Delta J_{\rm ANP}/J $}&
\multicolumn{1}{c}{$\Delta J_{h_\lm}/J  $}\\
\hline
\hline 
$0.5$ & $-0.9$ & $18.718$ & $10.079$ & $ \color{colortab1} -4.3\cdot 10^{-2} $ & $  -5.6\cdot 10^{-3} $ & $  6.6\cdot 10^{-3} $ & $  9.9\cdot 10^{-3} $ & $ \color{colortab1} -2.6\cdot 10^{-2} $ & $  3.4\cdot 10^{-3} $ & $ \color{colortab1} 1.3\cdot 10^{-2} $ & $ \color{colortab1} 1.3\cdot 10^{-2} $ \\
$0.5$ & $-0.6$ & $16.895$ & $9.097$ & $ \color{colortab1} -3.5\cdot 10^{-2} $ & $  4.0\cdot 10^{-3} $ & $ \color{colortab1} 1.7\cdot 10^{-2} $ & $ \color{colortab1} 2.0\cdot 10^{-2} $ & $ \color{colortab1} -1.8\cdot 10^{-2} $ & $ \color{colortab1} 1.2\cdot 10^{-2} $ & $ \color{colortab1} 2.2\cdot 10^{-2} $ & $ \color{colortab1} 2.2\cdot 10^{-2} $ \\
$0.5$ & $-0.2$ & $14.345$ & $7.724$ & $ \color{colortab1} -1.9\cdot 10^{-2} $ & $ \color{colortab1} 2.1\cdot 10^{-2} $ & $ \color{colortab1} 3.5\cdot 10^{-2} $ & $ \color{colortab1} 3.9\cdot 10^{-2} $ & $  -3.7\cdot 10^{-3} $ & $ \color{colortab1} 2.7\cdot 10^{-2} $ & $ \color{colortab1} 3.8\cdot 10^{-2} $ & $ \color{colortab1} 3.8\cdot 10^{-2} $ \\
$0.5$ & $0.0$ & $13.000$ & $7.000$ & $  -8.1\cdot 10^{-3} $ & $ \color{colortab1} 3.3\cdot 10^{-2} $ & $ \color{colortab1} 4.7\cdot 10^{-2} $ & $ \color{colortab2} 5.1\cdot 10^{-2} $ & $  6.0\cdot 10^{-3} $ & $ \color{colortab1} 3.8\cdot 10^{-2} $ & $ \color{colortab1} 4.9\cdot 10^{-2} $ & $ \color{colortab1} 4.9\cdot 10^{-2} $ \\
$0.5$ & $0.2$ & $11.588$ & $6.240$ & $  6.1\cdot 10^{-3} $ & $ \color{colortab1} 4.7\cdot 10^{-2} $ & $ \color{colortab2} 6.2\cdot 10^{-2} $ & $ \color{colortab2} 6.7\cdot 10^{-2} $ & $ \color{colortab1} 1.9\cdot 10^{-2} $ & $ \color{colortab2} 5.1\cdot 10^{-2} $ & $ \color{colortab2} 6.2\cdot 10^{-2} $ & $ \color{colortab2} 6.2\cdot 10^{-2} $ \\
$0.5$ & $0.6$ & $8.427$ & $4.538$ & $ \color{colortab2} 5.3\cdot 10^{-2} $ & $ \color{colortab2} 9.4\cdot 10^{-2} $ & $ \color{colortab2} 1.1\cdot 10^{-1} $ & $ \color{colortab2} 1.2\cdot 10^{-1} $ & $ \color{colortab2} 5.9\cdot 10^{-2} $ & $ \color{colortab2} 9.1\cdot 10^{-2} $ & $ \color{colortab2} 1.0\cdot 10^{-1} $ & $ \color{colortab2} 1.0\cdot 10^{-1} $ \\
$0.5$ & $0.9$ & $5.262$ & $2.833$ & $ \color{colortab2} 1.4\cdot 10^{-1} $ & $ \color{colortab2} 1.8\cdot 10^{-1} $ & $ \color{colortab2} 1.9\cdot 10^{-1} $ & $ \color{colortab2} 2.0\cdot 10^{-1} $ & $ \color{colortab2} 1.3\cdot 10^{-1} $ & $ \color{colortab2} 1.6\cdot 10^{-1} $ & $ \color{colortab2} 1.7\cdot 10^{-1} $ & $ \color{colortab2} 1.7\cdot 10^{-1} $ \\

\Xhline{2.5\arrayrulewidth}

$0.7$ & $-0.9$ & $10.595$ & $10.585$ & $  -8.9\cdot 10^{-3} $ & $  -3.1\cdot 10^{-3} $ & $  -2.2\cdot 10^{-4} $ & $  2.0\cdot 10^{-3} $ & $  -8.8\cdot 10^{-3} $ & $  -2.2\cdot 10^{-3} $ & $  9.9\cdot 10^{-4} $ & $  1.3\cdot 10^{-3} $ \\
$0.7$ & $-0.6$ & $9.580$ & $9.570$ & $  -4.2\cdot 10^{-3} $ & $  1.8\cdot 10^{-3} $ & $  4.8\cdot 10^{-3} $ & $  7.1\cdot 10^{-3} $ & $  -3.6\cdot 10^{-3} $ & $  3.1\cdot 10^{-3} $ & $  6.4\cdot 10^{-3} $ & $  6.8\cdot 10^{-3} $ \\
$0.7$ & $-0.2$ & $8.160$ & $8.150$ & $  7.4\cdot 10^{-3} $ & $ \color{colortab1} 1.4\cdot 10^{-2} $ & $ \color{colortab1} 1.7\cdot 10^{-2} $ & $ \color{colortab1} 1.9\cdot 10^{-2} $ & $  8.6\cdot 10^{-3} $ & $ \color{colortab1} 1.6\cdot 10^{-2} $ & $ \color{colortab1} 1.9\cdot 10^{-2} $ & $ \color{colortab1} 2.0\cdot 10^{-2} $ \\
$0.7$ & $0.0$ & $7.410$ & $7.400$ & $ \color{colortab1} 1.7\cdot 10^{-2} $ & $ \color{colortab1} 2.3\cdot 10^{-2} $ & $ \color{colortab1} 2.7\cdot 10^{-2} $ & $ \color{colortab1} 2.9\cdot 10^{-2} $ & $ \color{colortab1} 1.9\cdot 10^{-2} $ & $ \color{colortab1} 2.6\cdot 10^{-2} $ & $ \color{colortab1} 2.9\cdot 10^{-2} $ & $ \color{colortab1} 3.0\cdot 10^{-2} $ \\
$0.7$ & $0.2$ & $6.622$ & $6.612$ & $ \color{colortab1} 3.2\cdot 10^{-2} $ & $ \color{colortab1} 3.8\cdot 10^{-2} $ & $ \color{colortab1} 4.1\cdot 10^{-2} $ & $ \color{colortab1} 4.4\cdot 10^{-2} $ & $ \color{colortab1} 3.3\cdot 10^{-2} $ & $ \color{colortab1} 4.0\cdot 10^{-2} $ & $ \color{colortab1} 4.4\cdot 10^{-2} $ & $ \color{colortab1} 4.5\cdot 10^{-2} $ \\
$0.7$ & $0.6$ & $4.858$ & $4.848$ & $ \color{colortab2} 8.7\cdot 10^{-2} $ & $ \color{colortab2} 9.3\cdot 10^{-2} $ & $ \color{colortab2} 9.6\cdot 10^{-2} $ & $ \color{colortab2} 1.0\cdot 10^{-1} $ & $ \color{colortab2} 8.8\cdot 10^{-2} $ & $ \color{colortab2} 9.5\cdot 10^{-2} $ & $ \color{colortab2} 9.8\cdot 10^{-2} $ & $ \color{colortab2} 1.0\cdot 10^{-1} $ \\
$0.7$ & $0.9$ & $3.088$ & $3.078$ & $ \color{colortab2} 1.7\cdot 10^{-1} $ & $ \color{colortab2} 1.8\cdot 10^{-1} $ & $ \color{colortab2} 1.8\cdot 10^{-1} $ & $ \color{colortab2} 1.9\cdot 10^{-1} $ & $ \color{colortab2} 1.7\cdot 10^{-1} $ & $ \color{colortab2} 1.7\cdot 10^{-1} $ & $ \color{colortab2} 1.8\cdot 10^{-1} $ & $ \color{colortab2} 1.8\cdot 10^{-1} $ \\
\hline
$0.7$ & $-0.9$ & $12.873$ & $10.585$ & $ \color{colortab2} -7.4\cdot 10^{-2} $ & $ \color{colortab1} -3.4\cdot 10^{-2} $ & $ \color{colortab1} -1.7\cdot 10^{-2} $ & $  -7.4\cdot 10^{-3} $ & $ \color{colortab2} -5.5\cdot 10^{-2} $ & $ \color{colortab1} -2.0\cdot 10^{-2} $ & $  -4.6\cdot 10^{-3} $ & $  -4.3\cdot 10^{-3} $ \\
$0.7$ & $-0.6$ & $11.639$ & $9.570$ & $ \color{colortab2} -5.9\cdot 10^{-2} $ & $ \color{colortab1} -1.9\cdot 10^{-2} $ & $  -1.3\cdot 10^{-3} $ & $  7.9\cdot 10^{-3} $ & $ \color{colortab1} -4.1\cdot 10^{-2} $ & $  -4.9\cdot 10^{-3} $ & $ \color{colortab1} 1.0\cdot 10^{-2} $ & $ \color{colortab1} 1.1\cdot 10^{-2} $ \\
$0.7$ & $-0.2$ & $9.912$ & $8.150$ & $ \color{colortab1} -3.0\cdot 10^{-2} $ & $  9.9\cdot 10^{-3} $ & $ \color{colortab1} 2.7\cdot 10^{-2} $ & $ \color{colortab1} 3.6\cdot 10^{-2} $ & $ \color{colortab1} -1.3\cdot 10^{-2} $ & $ \color{colortab1} 2.2\cdot 10^{-2} $ & $ \color{colortab1} 3.8\cdot 10^{-2} $ & $ \color{colortab1} 3.8\cdot 10^{-2} $ \\
$0.7$ & $0.0$ & $9.000$ & $7.400$ & $  -9.2\cdot 10^{-3} $ & $ \color{colortab1} 3.0\cdot 10^{-2} $ & $ \color{colortab1} 4.7\cdot 10^{-2} $ & $ \color{colortab2} 5.6\cdot 10^{-2} $ & $  5.9\cdot 10^{-3} $ & $ \color{colortab1} 4.1\cdot 10^{-2} $ & $ \color{colortab2} 5.6\cdot 10^{-2} $ & $ \color{colortab2} 5.7\cdot 10^{-2} $ \\
$0.7$ & $0.2$ & $8.042$ & $6.612$ & $ \color{colortab1} 1.8\cdot 10^{-2} $ & $ \color{colortab2} 5.5\cdot 10^{-2} $ & $ \color{colortab2} 7.2\cdot 10^{-2} $ & $ \color{colortab2} 8.2\cdot 10^{-2} $ & $ \color{colortab1} 3.0\cdot 10^{-2} $ & $ \color{colortab2} 6.5\cdot 10^{-2} $ & $ \color{colortab2} 8.0\cdot 10^{-2} $ & $ \color{colortab2} 8.1\cdot 10^{-2} $ \\
$0.7$ & $0.6$ & $5.896$ & $4.848$ & $ \color{colortab2} 1.1\cdot 10^{-1} $ & $ \color{colortab2} 1.4\cdot 10^{-1} $ & $ \color{colortab2} 1.5\cdot 10^{-1} $ & $ \color{colortab2} 1.6\cdot 10^{-1} $ & $ \color{colortab2} 1.1\cdot 10^{-1} $ & $ \color{colortab2} 1.4\cdot 10^{-1} $ & $ \color{colortab2} 1.5\cdot 10^{-1} $ & $ \color{colortab2} 1.6\cdot 10^{-1} $ \\
$0.7$ & $0.9$ & $3.744$ & $3.078$ & $ \color{colortab2} 2.5\cdot 10^{-1} $ & $ \color{colortab2} 2.7\cdot 10^{-1} $ & $ \color{colortab2} 2.8\cdot 10^{-1} $ & $ \color{colortab2} 2.8\cdot 10^{-1} $ & $ \color{colortab2} 2.3\cdot 10^{-1} $ & $ \color{colortab2} 2.5\cdot 10^{-1} $ & $ \color{colortab2} 2.6\cdot 10^{-1} $ & $ \color{colortab2} 2.7\cdot 10^{-1} $ \\
\hline
$0.7$ & $-0.9$ & $18.595$ & $10.585$ & $ \color{colortab2} -8.3\cdot 10^{-2} $ & $ \color{colortab1} -1.8\cdot 10^{-2} $ & $  3.3\cdot 10^{-3} $ & $  9.7\cdot 10^{-3} $ & $ \color{colortab2} -5.2\cdot 10^{-2} $ & $  -4.1\cdot 10^{-4} $ & $ \color{colortab1} 1.6\cdot 10^{-2} $ & $ \color{colortab1} 1.7\cdot 10^{-2} $ \\
$0.7$ & $-0.6$ & $16.812$ & $9.570$ & $ \color{colortab2} -6.6\cdot 10^{-2} $ & $  1.2\cdot 10^{-4} $ & $ \color{colortab1} 2.2\cdot 10^{-2} $ & $ \color{colortab1} 2.8\cdot 10^{-2} $ & $ \color{colortab1} -3.7\cdot 10^{-2} $ & $ \color{colortab1} 1.6\cdot 10^{-2} $ & $ \color{colortab1} 3.3\cdot 10^{-2} $ & $ \color{colortab1} 3.3\cdot 10^{-2} $ \\
$0.7$ & $-0.2$ & $14.317$ & $8.150$ & $ \color{colortab1} -3.4\cdot 10^{-2} $ & $ \color{colortab1} 3.2\cdot 10^{-2} $ & $ \color{colortab2} 5.4\cdot 10^{-2} $ & $ \color{colortab2} 6.0\cdot 10^{-2} $ & $  -8.5\cdot 10^{-3} $ & $ \color{colortab1} 4.4\cdot 10^{-2} $ & $ \color{colortab2} 6.2\cdot 10^{-2} $ & $ \color{colortab2} 6.3\cdot 10^{-2} $ \\
$0.7$ & $0.0$ & $13.000$ & $7.400$ & $ \color{colortab1} -1.2\cdot 10^{-2} $ & $ \color{colortab2} 5.3\cdot 10^{-2} $ & $ \color{colortab2} 7.5\cdot 10^{-2} $ & $ \color{colortab2} 8.1\cdot 10^{-2} $ & $ \color{colortab1} 1.0\cdot 10^{-2} $ & $ \color{colortab2} 6.3\cdot 10^{-2} $ & $ \color{colortab2} 8.1\cdot 10^{-2} $ & $ \color{colortab2} 8.2\cdot 10^{-2} $ \\
$0.7$ & $0.2$ & $11.616$ & $6.612$ & $ \color{colortab1} 1.5\cdot 10^{-2} $ & $ \color{colortab2} 7.9\cdot 10^{-2} $ & $ \color{colortab2} 1.0\cdot 10^{-1} $ & $ \color{colortab2} 1.1\cdot 10^{-1} $ & $ \color{colortab1} 3.4\cdot 10^{-2} $ & $ \color{colortab2} 8.6\cdot 10^{-2} $ & $ \color{colortab2} 1.0\cdot 10^{-1} $ & $ \color{colortab2} 1.1\cdot 10^{-1} $ \\
$0.7$ & $0.6$ & $8.517$ & $4.848$ & $ \color{colortab2} 9.9\cdot 10^{-2} $ & $ \color{colortab2} 1.6\cdot 10^{-1} $ & $ \color{colortab2} 1.8\cdot 10^{-1} $ & $ \color{colortab2} 1.9\cdot 10^{-1} $ & $ \color{colortab2} 1.1\cdot 10^{-1} $ & $ \color{colortab2} 1.6\cdot 10^{-1} $ & $ \color{colortab2} 1.7\cdot 10^{-1} $ & $ \color{colortab2} 1.7\cdot 10^{-1} $ \\
$0.7$ & $0.9$ & $5.408$ & $3.078$ & $ \color{colortab2} 2.3\cdot 10^{-1} $ & $ \color{colortab2} 2.8\cdot 10^{-1} $ & $ \color{colortab2} 3.0\cdot 10^{-1} $ & $ \color{colortab2} 3.0\cdot 10^{-1} $ & $ \color{colortab2} 2.2\cdot 10^{-1} $ & $ \color{colortab2} 2.6\cdot 10^{-1} $ & $ \color{colortab2} 2.8\cdot 10^{-1} $ & $ \color{colortab2} 2.8\cdot 10^{-1} $ \\

\Xhline{2.5\arrayrulewidth}

$0.9$ & $-0.9$ & $11.084$ & $11.074$ & $ \color{colortab1} -1.5\cdot 10^{-2} $ & $  -6.1\cdot 10^{-3} $ & $  -1.7\cdot 10^{-3} $ & $  3.2\cdot 10^{-3} $ & $ \color{colortab1} -1.5\cdot 10^{-2} $ & $  -5.0\cdot 10^{-3} $ & $  -3.4\cdot 10^{-5} $ & $  1.1\cdot 10^{-3} $ \\
$0.9$ & $-0.6$ & $10.042$ & $10.032$ & $  -7.2\cdot 10^{-3} $ & $  1.5\cdot 10^{-3} $ & $  6.0\cdot 10^{-3} $ & $ \color{colortab1} 1.1\cdot 10^{-2} $ & $  -6.7\cdot 10^{-3} $ & $  3.4\cdot 10^{-3} $ & $  8.4\cdot 10^{-3} $ & $  9.9\cdot 10^{-3} $ \\
$0.9$ & $-0.2$ & $8.582$ & $8.572$ & $ \color{colortab1} 1.1\cdot 10^{-2} $ & $ \color{colortab1} 2.0\cdot 10^{-2} $ & $ \color{colortab1} 2.4\cdot 10^{-2} $ & $ \color{colortab1} 3.1\cdot 10^{-2} $ & $ \color{colortab1} 1.2\cdot 10^{-2} $ & $ \color{colortab1} 2.2\cdot 10^{-2} $ & $ \color{colortab1} 2.8\cdot 10^{-2} $ & $ \color{colortab1} 3.0\cdot 10^{-2} $ \\
$0.9$ & $0.0$ & $7.810$ & $7.800$ & $ \color{colortab1} 2.6\cdot 10^{-2} $ & $ \color{colortab1} 3.4\cdot 10^{-2} $ & $ \color{colortab1} 3.9\cdot 10^{-2} $ & $ \color{colortab1} 4.6\cdot 10^{-2} $ & $ \color{colortab1} 2.8\cdot 10^{-2} $ & $ \color{colortab1} 3.8\cdot 10^{-2} $ & $ \color{colortab1} 4.3\cdot 10^{-2} $ & $ \color{colortab1} 4.5\cdot 10^{-2} $ \\
$0.9$ & $0.2$ & $6.999$ & $6.989$ & $ \color{colortab1} 4.7\cdot 10^{-2} $ & $ \color{colortab2} 5.5\cdot 10^{-2} $ & $ \color{colortab2} 6.0\cdot 10^{-2} $ & $ \color{colortab2} 6.8\cdot 10^{-2} $ & $ \color{colortab1} 4.9\cdot 10^{-2} $ & $ \color{colortab2} 5.9\cdot 10^{-2} $ & $ \color{colortab2} 6.4\cdot 10^{-2} $ & $ \color{colortab2} 6.7\cdot 10^{-2} $ \\
$0.9$ & $0.6$ & $5.177$ & $5.167$ & $ \color{colortab2} 1.2\cdot 10^{-1} $ & $ \color{colortab2} 1.3\cdot 10^{-1} $ & $ \color{colortab2} 1.3\cdot 10^{-1} $ & $ \color{colortab2} 1.4\cdot 10^{-1} $ & $ \color{colortab2} 1.2\cdot 10^{-1} $ & $ \color{colortab2} 1.3\cdot 10^{-1} $ & $ \color{colortab2} 1.3\cdot 10^{-1} $ & $ \color{colortab2} 1.4\cdot 10^{-1} $ \\
$0.9$ & $0.9$ & $3.344$ & $3.334$ & $ \color{colortab2} 1.9\cdot 10^{-1} $ & $ \color{colortab2} 2.0\cdot 10^{-1} $ & $ \color{colortab2} 2.0\cdot 10^{-1} $ & $ \color{colortab2} 2.2\cdot 10^{-1} $ & $ \color{colortab2} 1.9\cdot 10^{-1} $ & $ \color{colortab2} 2.0\cdot 10^{-1} $ & $ \color{colortab2} 2.0\cdot 10^{-1} $ & $ \color{colortab2} 2.2\cdot 10^{-1} $ \\
\hline
$0.9$ & $-0.9$ & $12.778$ & $11.074$ & $ \color{colortab2} -1.0\cdot 10^{-1} $ & $ \color{colortab2} -5.5\cdot 10^{-2} $ & $ \color{colortab1} -3.3\cdot 10^{-2} $ & $ \color{colortab1} -1.8\cdot 10^{-2} $ & $ \color{colortab2} -8.4\cdot 10^{-2} $ & $ \color{colortab1} -3.8\cdot 10^{-2} $ & $ \color{colortab1} -1.7\cdot 10^{-2} $ & $ \color{colortab1} -1.6\cdot 10^{-2} $ \\
$0.9$ & $-0.6$ & $11.576$ & $10.032$ & $ \color{colortab2} -8.3\cdot 10^{-2} $ & $ \color{colortab1} -3.4\cdot 10^{-2} $ & $ \color{colortab1} -1.2\cdot 10^{-2} $ & $  2.2\cdot 10^{-3} $ & $ \color{colortab2} -6.2\cdot 10^{-2} $ & $ \color{colortab1} -1.7\cdot 10^{-2} $ & $  4.0\cdot 10^{-3} $ & $  5.3\cdot 10^{-3} $ \\
$0.9$ & $-0.2$ & $9.891$ & $8.572$ & $ \color{colortab1} -4.1\cdot 10^{-2} $ & $  5.9\cdot 10^{-3} $ & $ \color{colortab1} 2.8\cdot 10^{-2} $ & $ \color{colortab1} 4.1\cdot 10^{-2} $ & $ \color{colortab1} -2.2\cdot 10^{-2} $ & $ \color{colortab1} 2.2\cdot 10^{-2} $ & $ \color{colortab1} 4.3\cdot 10^{-2} $ & $ \color{colortab1} 4.4\cdot 10^{-2} $ \\
$0.9$ & $0.0$ & $9.000$ & $7.800$ & $ \color{colortab1} -1.1\cdot 10^{-2} $ & $ \color{colortab1} 3.4\cdot 10^{-2} $ & $ \color{colortab2} 5.5\cdot 10^{-2} $ & $ \color{colortab2} 6.8\cdot 10^{-2} $ & $  6.1\cdot 10^{-3} $ & $ \color{colortab1} 4.9\cdot 10^{-2} $ & $ \color{colortab2} 6.9\cdot 10^{-2} $ & $ \color{colortab2} 7.1\cdot 10^{-2} $ \\
$0.9$ & $0.2$ & $8.064$ & $6.989$ & $ \color{colortab1} 2.7\cdot 10^{-2} $ & $ \color{colortab2} 7.0\cdot 10^{-2} $ & $ \color{colortab2} 9.0\cdot 10^{-2} $ & $ \color{colortab2} 1.0\cdot 10^{-1} $ & $ \color{colortab1} 4.2\cdot 10^{-2} $ & $ \color{colortab2} 8.3\cdot 10^{-2} $ & $ \color{colortab2} 1.0\cdot 10^{-1} $ & $ \color{colortab2} 1.0\cdot 10^{-1} $ \\
$0.9$ & $0.6$ & $5.962$ & $5.167$ & $ \color{colortab2} 1.5\cdot 10^{-1} $ & $ \color{colortab2} 1.8\cdot 10^{-1} $ & $ \color{colortab2} 2.0\cdot 10^{-1} $ & $ \color{colortab2} 2.1\cdot 10^{-1} $ & $ \color{colortab2} 1.5\cdot 10^{-1} $ & $ \color{colortab2} 1.9\cdot 10^{-1} $ & $ \color{colortab2} 2.0\cdot 10^{-1} $ & $ \color{colortab2} 2.1\cdot 10^{-1} $ \\
$0.9$ & $0.9$ & $3.847$ & $3.334$ & $ \color{colortab2} 3.0\cdot 10^{-1} $ & $ \color{colortab2} 3.2\cdot 10^{-1} $ & $ \color{colortab2} 3.3\cdot 10^{-1} $ & $ \color{colortab2} 3.3\cdot 10^{-1} $ & $ \color{colortab2} 2.8\cdot 10^{-1} $ & $ \color{colortab2} 3.1\cdot 10^{-1} $ & $ \color{colortab2} 3.2\cdot 10^{-1} $ & $ \color{colortab2} 3.3\cdot 10^{-1} $ \\
\hline
$0.9$ & $-0.9$ & $18.457$ & $11.074$ & $ \color{colortab2} -1.4\cdot 10^{-1} $ & $ \color{colortab1} -4.2\cdot 10^{-2} $ & $  -9.8\cdot 10^{-3} $ & $  1.8\cdot 10^{-3} $ & $ \color{colortab2} -9.1\cdot 10^{-2} $ & $ \color{colortab1} -1.3\cdot 10^{-2} $ & $ \color{colortab1} 1.3\cdot 10^{-2} $ & $ \color{colortab1} 1.3\cdot 10^{-2} $ \\
$0.9$ & $-0.6$ & $16.720$ & $10.032$ & $ \color{colortab2} -1.1\cdot 10^{-1} $ & $ \color{colortab1} -1.3\cdot 10^{-2} $ & $ \color{colortab1} 1.9\cdot 10^{-2} $ & $ \color{colortab1} 2.9\cdot 10^{-2} $ & $ \color{colortab2} -6.5\cdot 10^{-2} $ & $ \color{colortab1} 1.3\cdot 10^{-2} $ & $ \color{colortab1} 3.9\cdot 10^{-2} $ & $ \color{colortab1} 3.9\cdot 10^{-2} $ \\
$0.9$ & $-0.2$ & $14.287$ & $8.572$ & $ \color{colortab2} -5.4\cdot 10^{-2} $ & $ \color{colortab1} 3.8\cdot 10^{-2} $ & $ \color{colortab2} 6.9\cdot 10^{-2} $ & $ \color{colortab2} 7.8\cdot 10^{-2} $ & $ \color{colortab1} -1.7\cdot 10^{-2} $ & $ \color{colortab2} 5.8\cdot 10^{-2} $ & $ \color{colortab2} 8.4\cdot 10^{-2} $ & $ \color{colortab2} 8.5\cdot 10^{-2} $ \\
$0.9$ & $0.0$ & $13.000$ & $7.800$ & $ \color{colortab1} -1.8\cdot 10^{-2} $ & $ \color{colortab2} 7.0\cdot 10^{-2} $ & $ \color{colortab2} 1.0\cdot 10^{-1} $ & $ \color{colortab2} 1.1\cdot 10^{-1} $ & $ \color{colortab1} 1.3\cdot 10^{-2} $ & $ \color{colortab2} 8.7\cdot 10^{-2} $ & $ \color{colortab2} 1.1\cdot 10^{-1} $ & $ \color{colortab2} 1.1\cdot 10^{-1} $ \\
$0.9$ & $0.2$ & $11.648$ & $6.989$ & $ \color{colortab1} 2.5\cdot 10^{-2} $ & $ \color{colortab2} 1.1\cdot 10^{-1} $ & $ \color{colortab2} 1.4\cdot 10^{-1} $ & $ \color{colortab2} 1.5\cdot 10^{-1} $ & $ \color{colortab2} 5.1\cdot 10^{-2} $ & $ \color{colortab2} 1.2\cdot 10^{-1} $ & $ \color{colortab2} 1.5\cdot 10^{-1} $ & $ \color{colortab2} 1.5\cdot 10^{-1} $ \\
$0.9$ & $0.6$ & $8.612$ & $5.167$ & $ \color{colortab2} 1.5\cdot 10^{-1} $ & $ \color{colortab2} 2.2\cdot 10^{-1} $ & $ \color{colortab2} 2.5\cdot 10^{-1} $ & $ \color{colortab2} 2.5\cdot 10^{-1} $ & $ \color{colortab2} 1.6\cdot 10^{-1} $ & $ \color{colortab2} 2.2\cdot 10^{-1} $ & $ \color{colortab2} 2.5\cdot 10^{-1} $ & $ \color{colortab2} 2.5\cdot 10^{-1} $ \\
$0.9$ & $0.9$ & $5.557$ & $3.334$ & $ \color{colortab2} 3.3\cdot 10^{-1} $ & $ \color{colortab2} 3.7\cdot 10^{-1} $ & $ \color{colortab2} 3.9\cdot 10^{-1} $ & $ \color{colortab2} 3.9\cdot 10^{-1} $ & $ \color{colortab2} 3.1\cdot 10^{-1} $ & $ \color{colortab2} 3.5\cdot 10^{-1} $ & $ \color{colortab2} 3.7\cdot 10^{-1} $ & $ \color{colortab2} 3.7\cdot 10^{-1} $ \\
\end{tabular}
 \end{ruledtabular}
 \end{center}
 \end{table*}

\section{Contribution of the higher modes to the fluxes}
\label{appendix:subdominantJ}
In this appendix we report the contribution of the $\l$-modes for all the simulations that we 
have performed. As already discussed, increasing the eccentricity and/or the spin, i.e. going 
in stronger fields, leads to an greater relative contribution of the subdominant modes. We report 
explicitly the relative $\l$-contributions to the numerical angular momentum flux
\begin{equation}
\delta \av{\dot{J}}_\l = \sum_{m=1}^{\l} \av{\dot{J}^{\rm teuk}_\lm} / \av{\dot{J}^{\rm teuk}}
\label{eq:Jl_contrib}
\end{equation} 
in Table~\ref{tab:subdominant1}, Table~\ref{tab:subdominant2} 
and Table~\ref{tab:subdominant3}. The contributions to the energy fluxes are analogous.

\begin{table*}[th]
   \caption{\label{tab:subdominant1} Relative contributions of the $\l$-modes for the angular 
   momentum computed from the {\Teukode}'s results. With $\delta \av{\dot{J}}_\l$ we indicate the 
   relative contribution to the total angular momentum flux of the $\l$-modes 
   summed together, see Eq.~\eqref{eq:Jl_contrib}. 
   Continue in Table~\ref{tab:subdominant2} and Table~\ref{tab:subdominant3}.}
   \begin{center}
     \begin{ruledtabular}
\begin{tabular}{ c r c c | c | c c c c c c c } 
$e$ & \multicolumn{1}{c}{$\ha$} & $p$ & $p_s$ & $\av{\dot{J}}$ & $\delta\av{\dot{J}}_2$ 
& $ \delta\av{\dot{J}}_3$ & $ \delta\av{\dot{J}}_4$ & $ \delta\av{\dot{J}}_5$ 
& $\delta\av{\dot{J}}_6$ & $ \delta\av{\dot{J}}_7$ & $ \delta\av{\dot{J}}_8$  \\
\hline
\hline
$    0.0 $ & $   -0.9 $ & $    8.727 $ & $    8.717 $ & $    4.309\cdot 10^{-3} $  & $   8.44\cdot 10^{-1} $ & $   1.27\cdot 10^{-1} $ & $   2.37\cdot 10^{-2} $ & $   4.80\cdot 10^{-3} $ & $   1.01\cdot 10^{-3} $ & $   2.14\cdot 10^{-4} $ & $   4.63\cdot 10^{-5} $ \\ 
$    0.0 $ & $   -0.8 $ & $    8.442 $ & $    8.432 $ & $    4.786\cdot 10^{-3} $  & $   8.39\cdot 10^{-1} $ & $   1.29\cdot 10^{-1} $ & $   2.49\cdot 10^{-2} $ & $   5.16\cdot 10^{-3} $ & $   1.11\cdot 10^{-3} $ & $   2.43\cdot 10^{-4} $ & $   5.37\cdot 10^{-5} $ \\ 
$    0.0 $ & $   -0.7 $ & $    8.153 $ & $    8.143 $ & $    5.341\cdot 10^{-3} $  & $   8.35\cdot 10^{-1} $ & $   1.32\cdot 10^{-1} $ & $   2.61\cdot 10^{-2} $ & $   5.56\cdot 10^{-3} $ & $   1.23\cdot 10^{-3} $ & $   2.76\cdot 10^{-4} $ & $   6.28\cdot 10^{-5} $ \\ 
$    0.0 $ & $   -0.6 $ & $    7.861 $ & $    7.851 $ & $    5.991\cdot 10^{-3} $  & $   8.30\cdot 10^{-1} $ & $   1.35\cdot 10^{-1} $ & $   2.74\cdot 10^{-2} $ & $   6.00\cdot 10^{-3} $ & $   1.36\cdot 10^{-3} $ & $   3.15\cdot 10^{-4} $ & $   7.38\cdot 10^{-5} $ \\ 
$    0.0 $ & $   -0.5 $ & $    7.565 $ & $    7.555 $ & $    6.754\cdot 10^{-3} $  & $   8.25\cdot 10^{-1} $ & $   1.38\cdot 10^{-1} $ & $   2.89\cdot 10^{-2} $ & $   6.51\cdot 10^{-3} $ & $   1.52\cdot 10^{-3} $ & $   3.63\cdot 10^{-4} $ & $   8.74\cdot 10^{-5} $ \\ 
$    0.0 $ & $   -0.4 $ & $    7.264 $ & $    7.254 $ & $    7.661\cdot 10^{-3} $  & $   8.19\cdot 10^{-1} $ & $   1.41\cdot 10^{-1} $ & $   3.04\cdot 10^{-2} $ & $   7.08\cdot 10^{-3} $ & $   1.71\cdot 10^{-3} $ & $   4.19\cdot 10^{-4} $ & $   1.04\cdot 10^{-4} $ \\ 
$    0.0 $ & $   -0.3 $ & $    6.959 $ & $    6.949 $ & $    8.745\cdot 10^{-3} $  & $   8.13\cdot 10^{-1} $ & $   1.45\cdot 10^{-1} $ & $   3.22\cdot 10^{-2} $ & $   7.73\cdot 10^{-3} $ & $   1.92\cdot 10^{-3} $ & $   4.89\cdot 10^{-4} $ & $   1.26\cdot 10^{-4} $ \\ 
$    0.0 $ & $   -0.2 $ & $    6.649 $ & $    6.639 $ & $    1.006\cdot 10^{-2} $  & $   8.06\cdot 10^{-1} $ & $   1.49\cdot 10^{-1} $ & $   3.42\cdot 10^{-2} $ & $   8.48\cdot 10^{-3} $ & $   2.18\cdot 10^{-3} $ & $   5.74\cdot 10^{-4} $ & $   1.53\cdot 10^{-4} $ \\ 
$    0.0 $ & $   -0.1 $ & $    6.333 $ & $    6.323 $ & $    1.168\cdot 10^{-2} $  & $   7.98\cdot 10^{-1} $ & $   1.53\cdot 10^{-1} $ & $   3.64\cdot 10^{-2} $ & $   9.35\cdot 10^{-3} $ & $   2.50\cdot 10^{-3} $ & $   6.80\cdot 10^{-4} $ & $   1.87\cdot 10^{-4} $ \\ 
$    0.0 $ & $    0.0 $ & $    6.010 $ & $    6.000 $ & $    1.368\cdot 10^{-2} $  & $   7.90\cdot 10^{-1} $ & $   1.57\cdot 10^{-1} $ & $   3.88\cdot 10^{-2} $ & $   1.04\cdot 10^{-2} $ & $   2.88\cdot 10^{-3} $ & $   8.13\cdot 10^{-4} $ & $   2.33\cdot 10^{-4} $ \\ 
$    0.0 $ & $    0.1 $ & $    5.679 $ & $    5.669 $ & $    1.621\cdot 10^{-2} $  & $   7.80\cdot 10^{-1} $ & $   1.62\cdot 10^{-1} $ & $   4.16\cdot 10^{-2} $ & $   1.16\cdot 10^{-2} $ & $   3.34\cdot 10^{-3} $ & $   9.85\cdot 10^{-4} $ & $   2.94\cdot 10^{-4} $ \\ 
$    0.0 $ & $    0.2 $ & $    5.339 $ & $    5.329 $ & $    1.947\cdot 10^{-2} $  & $   7.70\cdot 10^{-1} $ & $   1.67\cdot 10^{-1} $ & $   4.48\cdot 10^{-2} $ & $   1.30\cdot 10^{-2} $ & $   3.93\cdot 10^{-3} $ & $   1.21\cdot 10^{-3} $ & $   3.77\cdot 10^{-4} $ \\ 
$    0.0 $ & $    0.3 $ & $    4.989 $ & $    4.979 $ & $    2.375\cdot 10^{-2} $  & $   7.57\cdot 10^{-1} $ & $   1.73\cdot 10^{-1} $ & $   4.86\cdot 10^{-2} $ & $   1.48\cdot 10^{-2} $ & $   4.67\cdot 10^{-3} $ & $   1.51\cdot 10^{-3} $ & $   4.94\cdot 10^{-4} $ \\ 
$    0.0 $ & $    0.4 $ & $    4.624 $ & $    4.614 $ & $    2.955\cdot 10^{-2} $  & $   7.43\cdot 10^{-1} $ & $   1.79\cdot 10^{-1} $ & $   5.30\cdot 10^{-2} $ & $   1.70\cdot 10^{-2} $ & $   5.65\cdot 10^{-3} $ & $   1.92\cdot 10^{-3} $ & $   6.63\cdot 10^{-4} $ \\ 
$    0.0 $ & $    0.5 $ & $    4.243 $ & $    4.233 $ & $    3.763\cdot 10^{-2} $  & $   7.26\cdot 10^{-1} $ & $   1.86\cdot 10^{-1} $ & $   5.83\cdot 10^{-2} $ & $   1.98\cdot 10^{-2} $ & $   6.97\cdot 10^{-3} $ & $   2.51\cdot 10^{-3} $ & $   9.18\cdot 10^{-4} $ \\ 
$    0.0 $ & $    0.6 $ & $    3.839 $ & $    3.829 $ & $    4.945\cdot 10^{-2} $  & $   7.04\cdot 10^{-1} $ & $   1.94\cdot 10^{-1} $ & $   6.49\cdot 10^{-2} $ & $   2.35\cdot 10^{-2} $ & $   8.83\cdot 10^{-3} $ & $   3.40\cdot 10^{-3} $ & $   1.32\cdot 10^{-3} $ \\ 
$    0.0 $ & $    0.7 $ & $    3.403 $ & $    3.393 $ & $    6.779\cdot 10^{-2} $  & $   6.76\cdot 10^{-1} $ & $   2.04\cdot 10^{-1} $ & $   7.35\cdot 10^{-2} $ & $   2.87\cdot 10^{-2} $ & $   1.16\cdot 10^{-2} $ & $   4.82\cdot 10^{-3} $ & $   2.03\cdot 10^{-3} $ \\ 
$    0.0 $ & $    0.8 $ & $    2.917 $ & $    2.907 $ & $    9.882\cdot 10^{-2} $  & $   6.35\cdot 10^{-1} $ & $   2.15\cdot 10^{-1} $ & $   8.56\cdot 10^{-2} $ & $   3.66\cdot 10^{-2} $ & $   1.62\cdot 10^{-2} $ & $   7.39\cdot 10^{-3} $ & $   3.41\cdot 10^{-3} $ \\ 
$    0.0 $ & $    0.9 $ & $    2.331 $ & $    2.321 $ & $    1.583\cdot 10^{-1} $  & $   5.66\cdot 10^{-1} $ & $   2.32\cdot 10^{-1} $ & $   1.05\cdot 10^{-1} $ & $   5.11\cdot 10^{-2} $ & $   2.57\cdot 10^{-2} $ & $   1.32\cdot 10^{-2} $ & $   6.92\cdot 10^{-3} $ \\ 
\hline
$    0.0 $ & $   -0.9 $ & $   13.076 $ & $    8.717 $ & $    8.520\cdot 10^{-4} $  & $   9.00\cdot 10^{-1} $ & $   8.78\cdot 10^{-2} $ & $   1.06\cdot 10^{-2} $ & $   1.38\cdot 10^{-3} $ & $   1.85\cdot 10^{-4} $ & $   2.53\cdot 10^{-5} $ & $   3.49\cdot 10^{-6} $ \\ 
$    0.0 $ & $   -0.6 $ & $   11.776 $ & $    7.851 $ & $    1.200\cdot 10^{-3} $  & $   8.91\cdot 10^{-1} $ & $   9.49\cdot 10^{-2} $ & $   1.25\cdot 10^{-2} $ & $   1.78\cdot 10^{-3} $ & $   2.61\cdot 10^{-4} $ & $   3.90\cdot 10^{-5} $ & $   5.87\cdot 10^{-6} $ \\ 
$    0.0 $ & $   -0.2 $ & $    9.959 $ & $    6.639 $ & $    2.066\cdot 10^{-3} $  & $   8.74\cdot 10^{-1} $ & $   1.07\cdot 10^{-1} $ & $   1.62\cdot 10^{-2} $ & $   2.65\cdot 10^{-3} $ & $   4.47\cdot 10^{-4} $ & $   7.67\cdot 10^{-5} $ & $   1.33\cdot 10^{-5} $ \\ 
$    0.0 $ & $    0.0 $ & $    9.000 $ & $    6.000 $ & $    2.859\cdot 10^{-3} $  & $   8.62\cdot 10^{-1} $ & $   1.15\cdot 10^{-1} $ & $   1.89\cdot 10^{-2} $ & $   3.35\cdot 10^{-3} $ & $   6.13\cdot 10^{-4} $ & $   1.14\cdot 10^{-4} $ & $   2.16\cdot 10^{-5} $ \\ 
$    0.0 $ & $    0.2 $ & $    7.994 $ & $    5.329 $ & $    4.167\cdot 10^{-3} $  & $   8.48\cdot 10^{-1} $ & $   1.24\cdot 10^{-1} $ & $   2.24\cdot 10^{-2} $ & $   4.37\cdot 10^{-3} $ & $   8.82\cdot 10^{-4} $ & $   1.81\cdot 10^{-4} $ & $   3.76\cdot 10^{-5} $ \\ 
$    0.0 $ & $    0.6 $ & $    5.744 $ & $    3.829 $ & $    1.161\cdot 10^{-2} $  & $   8.01\cdot 10^{-1} $ & $   1.52\cdot 10^{-1} $ & $   3.54\cdot 10^{-2} $ & $   8.89\cdot 10^{-3} $ & $   2.31\cdot 10^{-3} $ & $   6.15\cdot 10^{-4} $ & $   1.65\cdot 10^{-4} $ \\ 
$    0.0 $ & $    0.9 $ & $    3.481 $ & $    2.321 $ & $    4.898\cdot 10^{-2} $  & $   7.03\cdot 10^{-1} $ & $   1.95\cdot 10^{-1} $ & $   6.49\cdot 10^{-2} $ & $   2.32\cdot 10^{-2} $ & $   8.62\cdot 10^{-3} $ & $   3.27\cdot 10^{-3} $ & $   1.26\cdot 10^{-3} $ \\ 
\hline
$    0.0 $ & $   -0.9 $ & $   18.888 $ & $    8.717 $ & $    2.194\cdot 10^{-4} $  & $   9.31\cdot 10^{-1} $ & $   6.29\cdot 10^{-2} $ & $   5.25\cdot 10^{-3} $ & $   4.70\cdot 10^{-4} $ & $   4.34\cdot 10^{-5} $ & $   4.07\cdot 10^{-6} $ & $   3.85\cdot 10^{-7} $ \\ 
$    0.0 $ & $   -0.6 $ & $   17.010 $ & $    7.851 $ & $    3.114\cdot 10^{-4} $  & $   9.24\cdot 10^{-1} $ & $   6.85\cdot 10^{-2} $ & $   6.28\cdot 10^{-3} $ & $   6.17\cdot 10^{-4} $ & $   6.26\cdot 10^{-5} $ & $   6.45\cdot 10^{-6} $ & $   6.70\cdot 10^{-7} $ \\ 
$    0.0 $ & $   -0.2 $ & $   14.385 $ & $    6.639 $ & $    5.438\cdot 10^{-4} $  & $   9.12\cdot 10^{-1} $ & $   7.83\cdot 10^{-2} $ & $   8.31\cdot 10^{-3} $ & $   9.47\cdot 10^{-4} $ & $   1.11\cdot 10^{-4} $ & $   1.33\cdot 10^{-5} $ & $   1.60\cdot 10^{-6} $ \\ 
$    0.0 $ & $    0.0 $ & $   13.000 $ & $    6.000 $ & $    7.598\cdot 10^{-4} $  & $   9.04\cdot 10^{-1} $ & $   8.47\cdot 10^{-2} $ & $   9.81\cdot 10^{-3} $ & $   1.22\cdot 10^{-3} $ & $   1.57\cdot 10^{-4} $ & $   2.05\cdot 10^{-5} $ & $   2.70\cdot 10^{-6} $ \\ 
$    0.0 $ & $    0.2 $ & $   11.547 $ & $    5.329 $ & $    1.121\cdot 10^{-3} $  & $   8.94\cdot 10^{-1} $ & $   9.27\cdot 10^{-2} $ & $   1.19\cdot 10^{-2} $ & $   1.64\cdot 10^{-3} $ & $   2.33\cdot 10^{-4} $ & $   3.36\cdot 10^{-5} $ & $   4.91\cdot 10^{-6} $ \\ 
$    0.0 $ & $    0.6 $ & $    8.296 $ & $    3.829 $ & $    3.272\cdot 10^{-3} $  & $   8.59\cdot 10^{-1} $ & $   1.17\cdot 10^{-1} $ & $   1.98\cdot 10^{-2} $ & $   3.59\cdot 10^{-3} $ & $   6.73\cdot 10^{-4} $ & $   1.28\cdot 10^{-4} $ & $   2.48\cdot 10^{-5} $ \\ 
$    0.0 $ & $    0.9 $ & $    5.029 $ & $    2.321 $ & $    1.558\cdot 10^{-2} $  & $   7.85\cdot 10^{-1} $ & $   1.60\cdot 10^{-1} $ & $   3.98\cdot 10^{-2} $ & $   1.07\cdot 10^{-2} $ & $   2.97\cdot 10^{-3} $ & $   8.43\cdot 10^{-4} $ & $   2.42\cdot 10^{-4} $ \\ 

\Xhline{2.5\arrayrulewidth}

$    0.1 $ & $   -0.9 $ & $    9.014 $ & $    9.004 $ & $    4.505\cdot 10^{-3} $  & $   8.40\cdot 10^{-1} $ & $   1.29\cdot 10^{-1} $ & $   2.48\cdot 10^{-2} $ & $   5.19\cdot 10^{-3} $ & $   1.13\cdot 10^{-3} $ & $   2.50\cdot 10^{-4} $ & $   5.64\cdot 10^{-5} $ \\ 
$    0.1 $ & $   -0.6 $ & $    8.119 $ & $    8.109 $ & $    6.241\cdot 10^{-3} $  & $   8.26\cdot 10^{-1} $ & $   1.37\cdot 10^{-1} $ & $   2.86\cdot 10^{-2} $ & $   6.46\cdot 10^{-3} $ & $   1.52\cdot 10^{-3} $ & $   3.65\cdot 10^{-4} $ & $   8.90\cdot 10^{-5} $ \\ 
$    0.1 $ & $   -0.2 $ & $    6.869 $ & $    6.859 $ & $    1.042\cdot 10^{-2} $  & $   8.02\cdot 10^{-1} $ & $   1.51\cdot 10^{-1} $ & $   3.55\cdot 10^{-2} $ & $   9.05\cdot 10^{-3} $ & $   2.41\cdot 10^{-3} $ & $   6.54\cdot 10^{-4} $ & $   1.81\cdot 10^{-4} $ \\ 
$    0.1 $ & $    0.0 $ & $    6.210 $ & $    6.200 $ & $    1.412\cdot 10^{-2} $  & $   7.85\cdot 10^{-1} $ & $   1.59\cdot 10^{-1} $ & $   4.02\cdot 10^{-2} $ & $   1.10\cdot 10^{-2} $ & $   3.14\cdot 10^{-3} $ & $   9.20\cdot 10^{-4} $ & $   2.73\cdot 10^{-4} $ \\ 
$    0.1 $ & $    0.2 $ & $    5.518 $ & $    5.508 $ & $    2.001\cdot 10^{-2} $  & $   7.65\cdot 10^{-1} $ & $   1.69\cdot 10^{-1} $ & $   4.62\cdot 10^{-2} $ & $   1.37\cdot 10^{-2} $ & $   4.26\cdot 10^{-3} $ & $   1.35\cdot 10^{-3} $ & $   4.36\cdot 10^{-4} $ \\ 
$    0.1 $ & $    0.6 $ & $    3.970 $ & $    3.960 $ & $    5.022\cdot 10^{-2} $  & $   7.00\cdot 10^{-1} $ & $   1.95\cdot 10^{-1} $ & $   6.63\cdot 10^{-2} $ & $   2.44\cdot 10^{-2} $ & $   9.37\cdot 10^{-3} $ & $   3.69\cdot 10^{-3} $ & $   1.48\cdot 10^{-3} $ \\ 
$    0.1 $ & $    0.9 $ & $    2.415 $ & $    2.405 $ & $    1.578\cdot 10^{-1} $  & $   5.63\cdot 10^{-1} $ & $   2.32\cdot 10^{-1} $ & $   1.06\cdot 10^{-1} $ & $   5.19\cdot 10^{-2} $ & $   2.64\cdot 10^{-2} $ & $   1.38\cdot 10^{-2} $ & $   7.32\cdot 10^{-3} $ \\ 
\hline
$    0.1 $ & $   -0.9 $ & $   13.070 $ & $    9.004 $ & $    8.583\cdot 10^{-4} $  & $   8.98\cdot 10^{-1} $ & $   8.92\cdot 10^{-2} $ & $   1.11\cdot 10^{-2} $ & $   1.49\cdot 10^{-3} $ & $   2.09\cdot 10^{-4} $ & $   2.99\cdot 10^{-5} $ & $   4.35\cdot 10^{-6} $ \\ 
$    0.1 $ & $   -0.6 $ & $   11.772 $ & $    8.109 $ & $    1.209\cdot 10^{-3} $  & $   8.88\cdot 10^{-1} $ & $   9.63\cdot 10^{-2} $ & $   1.31\cdot 10^{-2} $ & $   1.92\cdot 10^{-3} $ & $   2.93\cdot 10^{-4} $ & $   4.58\cdot 10^{-5} $ & $   7.29\cdot 10^{-6} $ \\ 
$    0.1 $ & $   -0.2 $ & $    9.957 $ & $    6.859 $ & $    2.081\cdot 10^{-3} $  & $   8.71\cdot 10^{-1} $ & $   1.08\cdot 10^{-1} $ & $   1.69\cdot 10^{-2} $ & $   2.84\cdot 10^{-3} $ & $   4.98\cdot 10^{-4} $ & $   8.95\cdot 10^{-5} $ & $   1.63\cdot 10^{-5} $ \\ 
$    0.1 $ & $    0.0 $ & $    9.000 $ & $    6.200 $ & $    2.878\cdot 10^{-3} $  & $   8.60\cdot 10^{-1} $ & $   1.16\cdot 10^{-1} $ & $   1.96\cdot 10^{-2} $ & $   3.58\cdot 10^{-3} $ & $   6.81\cdot 10^{-4} $ & $   1.33\cdot 10^{-4} $ & $   2.63\cdot 10^{-5} $ \\ 
$    0.1 $ & $    0.2 $ & $    7.996 $ & $    5.508 $ & $    4.194\cdot 10^{-3} $  & $   8.45\cdot 10^{-1} $ & $   1.26\cdot 10^{-1} $ & $   2.32\cdot 10^{-2} $ & $   4.66\cdot 10^{-3} $ & $   9.75\cdot 10^{-4} $ & $   2.09\cdot 10^{-4} $ & $   4.54\cdot 10^{-5} $ \\ 
$    0.1 $ & $    0.6 $ & $    5.749 $ & $    3.960 $ & $    1.167\cdot 10^{-2} $  & $   7.98\cdot 10^{-1} $ & $   1.53\cdot 10^{-1} $ & $   3.64\cdot 10^{-2} $ & $   9.38\cdot 10^{-3} $ & $   2.52\cdot 10^{-3} $ & $   6.95\cdot 10^{-4} $ & $   1.95\cdot 10^{-4} $ \\ 
$    0.1 $ & $    0.9 $ & $    3.491 $ & $    2.405 $ & $    4.903\cdot 10^{-2} $  & $   6.99\cdot 10^{-1} $ & $   1.96\cdot 10^{-1} $ & $   6.61\cdot 10^{-2} $ & $   2.41\cdot 10^{-2} $ & $   9.15\cdot 10^{-3} $ & $   3.57\cdot 10^{-3} $ & $   1.41\cdot 10^{-3} $ \\ 
\hline
$    0.1 $ & $   -0.9 $ & $   18.879 $ & $    9.004 $ & $    2.196\cdot 10^{-4} $  & $   9.30\cdot 10^{-1} $ & $   6.40\cdot 10^{-2} $ & $   5.49\cdot 10^{-3} $ & $   5.10\cdot 10^{-4} $ & $   4.92\cdot 10^{-5} $ & $   4.87\cdot 10^{-6} $ & $   4.89\cdot 10^{-7} $ \\ 
$    0.1 $ & $   -0.6 $ & $   17.004 $ & $    8.109 $ & $    3.117\cdot 10^{-4} $  & $   9.23\cdot 10^{-1} $ & $   6.97\cdot 10^{-2} $ & $   6.56\cdot 10^{-3} $ & $   6.68\cdot 10^{-4} $ & $   7.08\cdot 10^{-5} $ & $   7.67\cdot 10^{-6} $ & $   8.46\cdot 10^{-7} $ \\ 
$    0.1 $ & $   -0.2 $ & $   14.382 $ & $    6.859 $ & $    5.440\cdot 10^{-4} $  & $   9.11\cdot 10^{-1} $ & $   7.95\cdot 10^{-2} $ & $   8.66\cdot 10^{-3} $ & $   1.02\cdot 10^{-3} $ & $   1.25\cdot 10^{-4} $ & $   1.57\cdot 10^{-5} $ & $   2.01\cdot 10^{-6} $ \\ 
$    0.1 $ & $    0.0 $ & $   13.000 $ & $    6.200 $ & $    7.598\cdot 10^{-4} $  & $   9.02\cdot 10^{-1} $ & $   8.60\cdot 10^{-2} $ & $   1.02\cdot 10^{-2} $ & $   1.31\cdot 10^{-3} $ & $   1.76\cdot 10^{-4} $ & $   2.41\cdot 10^{-5} $ & $   3.35\cdot 10^{-6} $ \\ 
$    0.1 $ & $    0.2 $ & $   11.549 $ & $    5.508 $ & $    1.121\cdot 10^{-3} $  & $   8.92\cdot 10^{-1} $ & $   9.39\cdot 10^{-2} $ & $   1.23\cdot 10^{-2} $ & $   1.75\cdot 10^{-3} $ & $   2.60\cdot 10^{-4} $ & $   3.93\cdot 10^{-5} $ & $   6.06\cdot 10^{-6} $ \\ 
$    0.1 $ & $    0.6 $ & $    8.304 $ & $    3.960 $ & $    3.266\cdot 10^{-3} $  & $   8.56\cdot 10^{-1} $ & $   1.19\cdot 10^{-1} $ & $   2.04\cdot 10^{-2} $ & $   3.81\cdot 10^{-3} $ & $   7.41\cdot 10^{-4} $ & $   1.48\cdot 10^{-4} $ & $   2.99\cdot 10^{-5} $ \\ 
$    0.1 $ & $    0.9 $ & $    5.043 $ & $    2.405 $ & $    1.549\cdot 10^{-2} $  & $   7.83\cdot 10^{-1} $ & $   1.61\cdot 10^{-1} $ & $   4.08\cdot 10^{-2} $ & $   1.12\cdot 10^{-2} $ & $   3.20\cdot 10^{-3} $ & $   9.38\cdot 10^{-4} $ & $   2.80\cdot 10^{-4} $ \\ 

\end{tabular}
\end{ruledtabular}
\end{center}
\end{table*}

\begin{table*}[th]
   \caption{\label{tab:subdominant2} Same scheme as Table~\ref{tab:subdominant1}.}
   \begin{center}
     \begin{ruledtabular}
\begin{tabular}{ c r c c | c | c c c c c c c } 
$e$ & \multicolumn{1}{c}{$\ha$} & $p$ & $p_s$ & $\av{\dot{J}}$ & $\delta\av{\dot{J}}_2$ 
& $ \delta\av{\dot{J}}_3$ & $ \delta\av{\dot{J}}_4$ & $ \delta\av{\dot{J}}_5$ 
& $\delta\av{\dot{J}}_6$ & $ \delta\av{\dot{J}}_7$ & $ \delta\av{\dot{J}}_8$  \\
\hline
\hline
$    0.3 $ & $   -0.9 $ & $    9.564 $ & $    9.554 $ & $    5.324\cdot 10^{-3} $  & $   8.20\cdot 10^{-1} $ & $   1.40\cdot 10^{-1} $ & $   3.04\cdot 10^{-2} $ & $   7.22\cdot 10^{-3} $ & $   1.79\cdot 10^{-3} $ & $   4.58\cdot 10^{-4} $ & $   1.19\cdot 10^{-4} $ \\ 
$    0.3 $ & $   -0.6 $ & $    8.622 $ & $    8.612 $ & $    7.339\cdot 10^{-3} $  & $   8.05\cdot 10^{-1} $ & $   1.48\cdot 10^{-1} $ & $   3.46\cdot 10^{-2} $ & $   8.86\cdot 10^{-3} $ & $   2.37\cdot 10^{-3} $ & $   6.53\cdot 10^{-4} $ & $   1.83\cdot 10^{-4} $ \\ 
$    0.3 $ & $   -0.2 $ & $    7.305 $ & $    7.295 $ & $    1.215\cdot 10^{-2} $  & $   7.79\cdot 10^{-1} $ & $   1.61\cdot 10^{-1} $ & $   4.22\cdot 10^{-2} $ & $   1.21\cdot 10^{-2} $ & $   3.65\cdot 10^{-3} $ & $   1.13\cdot 10^{-3} $ & $   3.55\cdot 10^{-4} $ \\ 
$    0.3 $ & $    0.0 $ & $    6.610 $ & $    6.600 $ & $    1.637\cdot 10^{-2} $  & $   7.62\cdot 10^{-1} $ & $   1.69\cdot 10^{-1} $ & $   4.74\cdot 10^{-2} $ & $   1.45\cdot 10^{-2} $ & $   4.68\cdot 10^{-3} $ & $   1.55\cdot 10^{-3} $ & $   5.21\cdot 10^{-4} $ \\ 
$    0.3 $ & $    0.1 $ & $    6.250 $ & $    6.240 $ & $    1.929\cdot 10^{-2} $  & $   7.52\cdot 10^{-1} $ & $   1.73\cdot 10^{-1} $ & $   5.04\cdot 10^{-2} $ & $   1.60\cdot 10^{-2} $ & $   5.36\cdot 10^{-3} $ & $   1.84\cdot 10^{-3} $ & $   6.43\cdot 10^{-4} $ \\ 
$    0.3 $ & $    0.2 $ & $    5.881 $ & $    5.871 $ & $    2.302\cdot 10^{-2} $  & $   7.41\cdot 10^{-1} $ & $   1.78\cdot 10^{-1} $ & $   5.38\cdot 10^{-2} $ & $   1.78\cdot 10^{-2} $ & $   6.19\cdot 10^{-3} $ & $   2.21\cdot 10^{-3} $ & $   8.06\cdot 10^{-4} $ \\ 
$    0.3 $ & $    0.3 $ & $    5.500 $ & $    5.490 $ & $    2.789\cdot 10^{-2} $  & $   7.28\cdot 10^{-1} $ & $   1.83\cdot 10^{-1} $ & $   5.78\cdot 10^{-2} $ & $   2.00\cdot 10^{-2} $ & $   7.24\cdot 10^{-3} $ & $   2.70\cdot 10^{-3} $ & $   1.03\cdot 10^{-3} $ \\ 
$    0.3 $ & $    0.4 $ & $    5.104 $ & $    5.094 $ & $    3.440\cdot 10^{-2} $  & $   7.13\cdot 10^{-1} $ & $   1.89\cdot 10^{-1} $ & $   6.24\cdot 10^{-2} $ & $   2.26\cdot 10^{-2} $ & $   8.57\cdot 10^{-3} $ & $   3.35\cdot 10^{-3} $ & $   1.33\cdot 10^{-3} $ \\ 
$    0.3 $ & $    0.5 $ & $    4.689 $ & $    4.679 $ & $    4.339\cdot 10^{-2} $  & $   6.94\cdot 10^{-1} $ & $   1.95\cdot 10^{-1} $ & $   6.79\cdot 10^{-2} $ & $   2.59\cdot 10^{-2} $ & $   1.03\cdot 10^{-2} $ & $   4.25\cdot 10^{-3} $ & $   1.78\cdot 10^{-3} $ \\ 
$    0.3 $ & $    0.6 $ & $    4.250 $ & $    4.240 $ & $    5.630\cdot 10^{-2} $  & $   6.72\cdot 10^{-1} $ & $   2.03\cdot 10^{-1} $ & $   7.47\cdot 10^{-2} $ & $   3.01\cdot 10^{-2} $ & $   1.27\cdot 10^{-2} $ & $   5.55\cdot 10^{-3} $ & $   2.47\cdot 10^{-3} $ \\ 
$    0.3 $ & $    0.7 $ & $    3.777 $ & $    3.767 $ & $    7.590\cdot 10^{-2} $  & $   6.42\cdot 10^{-1} $ & $   2.11\cdot 10^{-1} $ & $   8.33\cdot 10^{-2} $ & $   3.59\cdot 10^{-2} $ & $   1.62\cdot 10^{-2} $ & $   7.55\cdot 10^{-3} $ & $   3.59\cdot 10^{-3} $ \\ 
$    0.3 $ & $    0.8 $ & $    3.249 $ & $    3.239 $ & $    1.078\cdot 10^{-1} $  & $   6.01\cdot 10^{-1} $ & $   2.21\cdot 10^{-1} $ & $   9.52\cdot 10^{-2} $ & $   4.44\cdot 10^{-2} $ & $   2.17\cdot 10^{-2} $ & $   1.10\cdot 10^{-2} $ & $   5.64\cdot 10^{-3} $ \\ 
$    0.3 $ & $    0.9 $ & $    2.615 $ & $    2.605 $ & $    1.662\cdot 10^{-1} $  & $   5.31\cdot 10^{-1} $ & $   2.34\cdot 10^{-1} $ & $   1.14\cdot 10^{-1} $ & $   5.94\cdot 10^{-2} $ & $   3.23\cdot 10^{-2} $ & $   1.81\cdot 10^{-2} $ & $   1.04\cdot 10^{-2} $ \\ 
\hline
$    0.3 $ & $   -0.9 $ & $   13.028 $ & $    9.554 $ & $    8.995\cdot 10^{-4} $  & $   8.83\cdot 10^{-1} $ & $   9.95\cdot 10^{-2} $ & $   1.46\cdot 10^{-2} $ & $   2.37\cdot 10^{-3} $ & $   4.08\cdot 10^{-4} $ & $   7.27\cdot 10^{-5} $ & $   1.32\cdot 10^{-5} $ \\ 
$    0.3 $ & $   -0.6 $ & $   11.743 $ & $    8.612 $ & $    1.265\cdot 10^{-3} $  & $   8.72\cdot 10^{-1} $ & $   1.07\cdot 10^{-1} $ & $   1.70\cdot 10^{-2} $ & $   3.01\cdot 10^{-3} $ & $   5.63\cdot 10^{-4} $ & $   1.09\cdot 10^{-4} $ & $   2.16\cdot 10^{-5} $ \\ 
$    0.3 $ & $   -0.2 $ & $    9.947 $ & $    7.295 $ & $    2.174\cdot 10^{-3} $  & $   8.53\cdot 10^{-1} $ & $   1.19\cdot 10^{-1} $ & $   2.16\cdot 10^{-2} $ & $   4.35\cdot 10^{-3} $ & $   9.26\cdot 10^{-4} $ & $   2.04\cdot 10^{-4} $ & $   4.60\cdot 10^{-5} $ \\ 
$    0.3 $ & $    0.0 $ & $    9.000 $ & $    6.600 $ & $    3.004\cdot 10^{-3} $  & $   8.41\cdot 10^{-1} $ & $   1.27\cdot 10^{-1} $ & $   2.49\cdot 10^{-2} $ & $   5.40\cdot 10^{-3} $ & $   1.24\cdot 10^{-3} $ & $   2.95\cdot 10^{-4} $ & $   7.18\cdot 10^{-5} $ \\ 
$    0.3 $ & $    0.2 $ & $    8.006 $ & $    5.871 $ & $    4.369\cdot 10^{-3} $  & $   8.25\cdot 10^{-1} $ & $   1.37\cdot 10^{-1} $ & $   2.91\cdot 10^{-2} $ & $   6.90\cdot 10^{-3} $ & $   1.73\cdot 10^{-3} $ & $   4.49\cdot 10^{-4} $ & $   1.19\cdot 10^{-4} $ \\ 
$    0.3 $ & $    0.6 $ & $    5.782 $ & $    4.240 $ & $    1.207\cdot 10^{-2} $  & $   7.73\cdot 10^{-1} $ & $   1.64\cdot 10^{-1} $ & $   4.40\cdot 10^{-2} $ & $   1.31\cdot 10^{-2} $ & $   4.13\cdot 10^{-3} $ & $   1.35\cdot 10^{-3} $ & $   4.53\cdot 10^{-4} $ \\ 
$    0.3 $ & $    0.9 $ & $    3.553 $ & $    2.605 $ & $    4.941\cdot 10^{-2} $  & $   6.69\cdot 10^{-1} $ & $   2.03\cdot 10^{-1} $ & $   7.51\cdot 10^{-2} $ & $   3.05\cdot 10^{-2} $ & $   1.31\cdot 10^{-2} $ & $   5.81\cdot 10^{-3} $ & $   2.65\cdot 10^{-3} $ \\ 
\hline
$    0.3 $ & $   -0.9 $ & $   18.818 $ & $    9.554 $ & $    2.191\cdot 10^{-4} $  & $   9.20\cdot 10^{-1} $ & $   7.19\cdot 10^{-2} $ & $   7.33\cdot 10^{-3} $ & $   8.32\cdot 10^{-4} $ & $   1.00\cdot 10^{-4} $ & $   1.25\cdot 10^{-5} $ & $   1.59\cdot 10^{-6} $ \\ 
$    0.3 $ & $   -0.6 $ & $   16.962 $ & $    8.612 $ & $    3.105\cdot 10^{-4} $  & $   9.12\cdot 10^{-1} $ & $   7.80\cdot 10^{-2} $ & $   8.68\cdot 10^{-3} $ & $   1.08\cdot 10^{-3} $ & $   1.42\cdot 10^{-4} $ & $   1.93\cdot 10^{-5} $ & $   2.68\cdot 10^{-6} $ \\ 
$    0.3 $ & $   -0.2 $ & $   14.368 $ & $    7.295 $ & $    5.407\cdot 10^{-4} $  & $   8.98\cdot 10^{-1} $ & $   8.84\cdot 10^{-2} $ & $   1.13\cdot 10^{-2} $ & $   1.61\cdot 10^{-3} $ & $   2.44\cdot 10^{-4} $ & $   3.82\cdot 10^{-5} $ & $   6.11\cdot 10^{-6} $ \\ 
$    0.3 $ & $    0.0 $ & $   13.000 $ & $    6.600 $ & $    7.540\cdot 10^{-4} $  & $   8.89\cdot 10^{-1} $ & $   9.52\cdot 10^{-2} $ & $   1.32\cdot 10^{-2} $ & $   2.05\cdot 10^{-3} $ & $   3.36\cdot 10^{-4} $ & $   5.71\cdot 10^{-5} $ & $   9.94\cdot 10^{-6} $ \\ 
$    0.3 $ & $    0.2 $ & $   11.564 $ & $    5.871 $ & $    1.110\cdot 10^{-3} $  & $   8.77\cdot 10^{-1} $ & $   1.03\cdot 10^{-1} $ & $   1.58\cdot 10^{-2} $ & $   2.69\cdot 10^{-3} $ & $   4.86\cdot 10^{-4} $ & $   9.08\cdot 10^{-5} $ & $   1.74\cdot 10^{-5} $ \\ 
$    0.3 $ & $    0.6 $ & $    8.352 $ & $    4.240 $ & $    3.211\cdot 10^{-3} $  & $   8.39\cdot 10^{-1} $ & $   1.28\cdot 10^{-1} $ & $   2.54\cdot 10^{-2} $ & $   5.57\cdot 10^{-3} $ & $   1.30\cdot 10^{-3} $ & $   3.12\cdot 10^{-4} $ & $   7.72\cdot 10^{-5} $ \\ 
$    0.3 $ & $    0.9 $ & $    5.132 $ & $    2.605 $ & $    1.489\cdot 10^{-2} $  & $   7.61\cdot 10^{-1} $ & $   1.69\cdot 10^{-1} $ & $   4.77\cdot 10^{-2} $ & $   1.49\cdot 10^{-2} $ & $   4.91\cdot 10^{-3} $ & $   1.69\cdot 10^{-3} $ & $   5.92\cdot 10^{-4} $ \\ 

\Xhline{2.5\arrayrulewidth}

$    0.5 $ & $   -0.9 $ & $   10.089 $ & $   10.079 $ & $    5.933\cdot 10^{-3} $  & $   7.96\cdot 10^{-1} $ & $   1.52\cdot 10^{-1} $ & $   3.74\cdot 10^{-2} $ & $   1.01\cdot 10^{-2} $ & $   2.84\cdot 10^{-3} $ & $   8.19\cdot 10^{-4} $ & $   2.40\cdot 10^{-4} $ \\ 
$    0.5 $ & $   -0.6 $ & $    9.107 $ & $    9.097 $ & $    8.149\cdot 10^{-3} $  & $   7.80\cdot 10^{-1} $ & $   1.60\cdot 10^{-1} $ & $   4.22\cdot 10^{-2} $ & $   1.22\cdot 10^{-2} $ & $   3.68\cdot 10^{-3} $ & $   1.14\cdot 10^{-3} $ & $   3.60\cdot 10^{-4} $ \\ 
$    0.5 $ & $   -0.2 $ & $    7.734 $ & $    7.724 $ & $    1.341\cdot 10^{-2} $  & $   7.52\cdot 10^{-1} $ & $   1.73\cdot 10^{-1} $ & $   5.06\cdot 10^{-2} $ & $   1.63\cdot 10^{-2} $ & $   5.48\cdot 10^{-3} $ & $   1.90\cdot 10^{-3} $ & $   6.68\cdot 10^{-4} $ \\ 
$    0.5 $ & $    0.0 $ & $    7.010 $ & $    7.000 $ & $    1.799\cdot 10^{-2} $  & $   7.34\cdot 10^{-1} $ & $   1.80\cdot 10^{-1} $ & $   5.61\cdot 10^{-2} $ & $   1.92\cdot 10^{-2} $ & $   6.89\cdot 10^{-3} $ & $   2.54\cdot 10^{-3} $ & $   9.54\cdot 10^{-4} $ \\ 
$    0.5 $ & $    0.2 $ & $    6.250 $ & $    6.240 $ & $    2.517\cdot 10^{-2} $  & $   7.12\cdot 10^{-1} $ & $   1.88\cdot 10^{-1} $ & $   6.30\cdot 10^{-2} $ & $   2.31\cdot 10^{-2} $ & $   8.90\cdot 10^{-3} $ & $   3.53\cdot 10^{-3} $ & $   1.42\cdot 10^{-3} $ \\ 
$    0.5 $ & $    0.6 $ & $    4.548 $ & $    4.538 $ & $    6.039\cdot 10^{-2} $  & $   6.39\cdot 10^{-1} $ & $   2.10\cdot 10^{-1} $ & $   8.45\cdot 10^{-2} $ & $   3.71\cdot 10^{-2} $ & $   1.71\cdot 10^{-2} $ & $   8.13\cdot 10^{-3} $ & $   3.94\cdot 10^{-3} $ \\ 
$    0.5 $ & $    0.9 $ & $    2.843 $ & $    2.833 $ & $    1.696\cdot 10^{-1} $  & $   4.94\cdot 10^{-1} $ & $   2.37\cdot 10^{-1} $ & $   1.23\cdot 10^{-1} $ & $   6.82\cdot 10^{-2} $ & $   3.94\cdot 10^{-2} $ & $   2.34\cdot 10^{-2} $ & $   1.42\cdot 10^{-2} $ \\ 
\hline
$    0.5 $ & $   -0.9 $ & $   12.959 $ & $   10.079 $ & $    9.314\cdot 10^{-4} $  & $   8.58\cdot 10^{-1} $ & $   1.16\cdot 10^{-1} $ & $   2.08\cdot 10^{-2} $ & $   4.15\cdot 10^{-3} $ & $   8.74\cdot 10^{-4} $ & $   1.90\cdot 10^{-4} $ & $   4.23\cdot 10^{-5} $ \\ 
$    0.5 $ & $   -0.6 $ & $   11.697 $ & $    9.097 $ & $    1.309\cdot 10^{-3} $  & $   8.45\cdot 10^{-1} $ & $   1.24\cdot 10^{-1} $ & $   2.40\cdot 10^{-2} $ & $   5.17\cdot 10^{-3} $ & $   1.18\cdot 10^{-3} $ & $   2.77\cdot 10^{-4} $ & $   6.67\cdot 10^{-5} $ \\ 
$    0.5 $ & $   -0.2 $ & $    9.931 $ & $    7.724 $ & $    2.243\cdot 10^{-3} $  & $   8.24\cdot 10^{-1} $ & $   1.37\cdot 10^{-1} $ & $   2.98\cdot 10^{-2} $ & $   7.25\cdot 10^{-3} $ & $   1.86\cdot 10^{-3} $ & $   4.95\cdot 10^{-4} $ & $   1.35\cdot 10^{-4} $ \\ 
$    0.5 $ & $    0.0 $ & $    9.000 $ & $    7.000 $ & $    3.093\cdot 10^{-3} $  & $   8.10\cdot 10^{-1} $ & $   1.45\cdot 10^{-1} $ & $   3.38\cdot 10^{-2} $ & $   8.82\cdot 10^{-3} $ & $   2.43\cdot 10^{-3} $ & $   6.94\cdot 10^{-4} $ & $   2.03\cdot 10^{-4} $ \\ 
$    0.5 $ & $    0.2 $ & $    8.022 $ & $    6.240 $ & $    4.488\cdot 10^{-3} $  & $   7.92\cdot 10^{-1} $ & $   1.54\cdot 10^{-1} $ & $   3.90\cdot 10^{-2} $ & $   1.10\cdot 10^{-2} $ & $   3.29\cdot 10^{-3} $ & $   1.02\cdot 10^{-3} $ & $   3.22\cdot 10^{-4} $ \\ 
$    0.5 $ & $    0.6 $ & $    5.834 $ & $    4.538 $ & $    1.228\cdot 10^{-2} $  & $   7.34\cdot 10^{-1} $ & $   1.79\cdot 10^{-1} $ & $   5.60\cdot 10^{-2} $ & $   1.95\cdot 10^{-2} $ & $   7.18\cdot 10^{-3} $ & $   2.74\cdot 10^{-3} $ & $   1.07\cdot 10^{-3} $ \\ 
$    0.5 $ & $    0.9 $ & $    3.643 $ & $    2.833 $ & $    4.867\cdot 10^{-2} $  & $   6.22\cdot 10^{-1} $ & $   2.14\cdot 10^{-1} $ & $   8.89\cdot 10^{-2} $ & $   4.06\cdot 10^{-2} $ & $   1.96\cdot 10^{-2} $ & $   9.78\cdot 10^{-3} $ & $   5.01\cdot 10^{-3} $ \\ 
\hline
$    0.5 $ & $   -0.9 $ & $   18.718 $ & $   10.079 $ & $    2.062\cdot 10^{-4} $  & $   9.02\cdot 10^{-1} $ & $   8.51\cdot 10^{-2} $ & $   1.07\cdot 10^{-2} $ & $   1.51\cdot 10^{-3} $ & $   2.24\cdot 10^{-4} $ & $   3.45\cdot 10^{-5} $ & $   5.42\cdot 10^{-6} $ \\ 
$    0.5 $ & $   -0.6 $ & $   16.895 $ & $    9.097 $ & $    2.915\cdot 10^{-4} $  & $   8.93\cdot 10^{-1} $ & $   9.18\cdot 10^{-2} $ & $   1.26\cdot 10^{-2} $ & $   1.92\cdot 10^{-3} $ & $   3.11\cdot 10^{-4} $ & $   5.21\cdot 10^{-5} $ & $   8.91\cdot 10^{-6} $ \\ 
$    0.5 $ & $   -0.2 $ & $   14.345 $ & $    7.724 $ & $    5.058\cdot 10^{-4} $  & $   8.77\cdot 10^{-1} $ & $   1.03\cdot 10^{-1} $ & $   1.61\cdot 10^{-2} $ & $   2.81\cdot 10^{-3} $ & $   5.19\cdot 10^{-4} $ & $   9.91\cdot 10^{-5} $ & $   1.94\cdot 10^{-5} $ \\ 
$    0.5 $ & $    0.0 $ & $   13.000 $ & $    7.000 $ & $    7.038\cdot 10^{-4} $  & $   8.67\cdot 10^{-1} $ & $   1.10\cdot 10^{-1} $ & $   1.86\cdot 10^{-2} $ & $   3.50\cdot 10^{-3} $ & $   7.00\cdot 10^{-4} $ & $   1.45\cdot 10^{-4} $ & $   3.06\cdot 10^{-5} $ \\ 
$    0.5 $ & $    0.2 $ & $   11.588 $ & $    6.240 $ & $    1.033\cdot 10^{-3} $  & $   8.54\cdot 10^{-1} $ & $   1.19\cdot 10^{-1} $ & $   2.19\cdot 10^{-2} $ & $   4.51\cdot 10^{-3} $ & $   9.86\cdot 10^{-4} $ & $   2.23\cdot 10^{-4} $ & $   5.15\cdot 10^{-5} $ \\ 
$    0.5 $ & $    0.6 $ & $    8.427 $ & $    4.538 $ & $    2.960\cdot 10^{-3} $  & $   8.10\cdot 10^{-1} $ & $   1.44\cdot 10^{-1} $ & $   3.36\cdot 10^{-2} $ & $   8.77\cdot 10^{-3} $ & $   2.43\cdot 10^{-3} $ & $   6.95\cdot 10^{-4} $ & $   2.04\cdot 10^{-4} $ \\ 
$    0.5 $ & $    0.9 $ & $    5.262 $ & $    2.833 $ & $    1.343\cdot 10^{-2} $  & $   7.25\cdot 10^{-1} $ & $   1.82\cdot 10^{-1} $ & $   5.87\cdot 10^{-2} $ & $   2.10\cdot 10^{-2} $ & $   7.98\cdot 10^{-3} $ & $   3.14\cdot 10^{-3} $ & $   1.27\cdot 10^{-3} $ \\ 

\end{tabular}
\end{ruledtabular}
\end{center}
\end{table*}

\begin{table*}[th]
   \caption{\label{tab:subdominant3} Same scheme as Table~\ref{tab:subdominant1}.}
   \begin{center}
     \begin{ruledtabular}
\begin{tabular}{ c r c c | c | c c c c c c c } 
$e$ & \multicolumn{1}{c}{$\ha$} & $p$ & $p_s$ & $\av{\dot{J}}$ & $\delta\av{\dot{J}}_2$ 
& $ \delta\av{\dot{J}}_3$ & $ \delta\av{\dot{J}}_4$ & $ \delta\av{\dot{J}}_5$ 
& $\delta\av{\dot{J}}_6$ & $ \delta\av{\dot{J}}_7$ & $ \delta\av{\dot{J}}_8$  \\
\hline
\hline
$    0.7 $ & $   -0.9 $ & $   10.595 $ & $   10.585 $ & $    5.335\cdot 10^{-3} $  & $   7.71\cdot 10^{-1} $ & $   1.64\cdot 10^{-1} $ & $   4.50\cdot 10^{-2} $ & $   1.35\cdot 10^{-2} $ & $   4.26\cdot 10^{-3} $ & $   1.38\cdot 10^{-3} $ & $   4.52\cdot 10^{-4} $ \\ 
$    0.7 $ & $   -0.6 $ & $    9.580 $ & $    9.570 $ & $    7.320\cdot 10^{-3} $  & $   7.54\cdot 10^{-1} $ & $   1.72\cdot 10^{-1} $ & $   5.02\cdot 10^{-2} $ & $   1.61\cdot 10^{-2} $ & $   5.42\cdot 10^{-3} $ & $   1.87\cdot 10^{-3} $ & $   6.57\cdot 10^{-4} $ \\ 
$    0.7 $ & $   -0.2 $ & $    8.160 $ & $    8.150 $ & $    1.203\cdot 10^{-2} $  & $   7.25\cdot 10^{-1} $ & $   1.83\cdot 10^{-1} $ & $   5.92\cdot 10^{-2} $ & $   2.10\cdot 10^{-2} $ & $   7.80\cdot 10^{-3} $ & $   2.98\cdot 10^{-3} $ & $   1.16\cdot 10^{-3} $ \\ 
$    0.7 $ & $    0.0 $ & $    7.410 $ & $    7.400 $ & $    1.613\cdot 10^{-2} $  & $   7.06\cdot 10^{-1} $ & $   1.90\cdot 10^{-1} $ & $   6.49\cdot 10^{-2} $ & $   2.44\cdot 10^{-2} $ & $   9.60\cdot 10^{-3} $ & $   3.89\cdot 10^{-3} $ & $   1.61\cdot 10^{-3} $ \\ 
$    0.7 $ & $    0.2 $ & $    6.622 $ & $    6.612 $ & $    2.255\cdot 10^{-2} $  & $   6.82\cdot 10^{-1} $ & $   1.97\cdot 10^{-1} $ & $   7.20\cdot 10^{-2} $ & $   2.88\cdot 10^{-2} $ & $   1.21\cdot 10^{-2} $ & $   5.24\cdot 10^{-3} $ & $   2.31\cdot 10^{-3} $ \\ 
$    0.7 $ & $    0.6 $ & $    4.858 $ & $    4.848 $ & $    5.390\cdot 10^{-2} $  & $   6.07\cdot 10^{-1} $ & $   2.17\cdot 10^{-1} $ & $   9.34\cdot 10^{-2} $ & $   4.40\cdot 10^{-2} $ & $   2.18\cdot 10^{-2} $ & $   1.11\cdot 10^{-2} $ & $   5.79\cdot 10^{-3} $ \\ 
$    0.7 $ & $    0.9 $ & $    3.088 $ & $    3.078 $ & $    1.504\cdot 10^{-1} $  & $   4.60\cdot 10^{-1} $ & $   2.40\cdot 10^{-1} $ & $   1.31\cdot 10^{-1} $ & $   7.63\cdot 10^{-2} $ & $   4.61\cdot 10^{-2} $ & $   2.87\cdot 10^{-2} $ & $   1.82\cdot 10^{-2} $ \\ 
\hline
$    0.7 $ & $   -0.9 $ & $   12.873 $ & $   10.585 $ & $    8.231\cdot 10^{-4} $  & $   8.27\cdot 10^{-1} $ & $   1.35\cdot 10^{-1} $ & $   2.89\cdot 10^{-2} $ & $   6.89\cdot 10^{-3} $ & $   1.73\cdot 10^{-3} $ & $   4.49\cdot 10^{-4} $ & $   1.19\cdot 10^{-4} $ \\ 
$    0.7 $ & $   -0.6 $ & $   11.639 $ & $    9.570 $ & $    1.155\cdot 10^{-3} $  & $   8.13\cdot 10^{-1} $ & $   1.42\cdot 10^{-1} $ & $   3.29\cdot 10^{-2} $ & $   8.41\cdot 10^{-3} $ & $   2.27\cdot 10^{-3} $ & $   6.34\cdot 10^{-4} $ & $   1.81\cdot 10^{-4} $ \\ 
$    0.7 $ & $   -0.2 $ & $    9.912 $ & $    8.150 $ & $    1.974\cdot 10^{-3} $  & $   7.89\cdot 10^{-1} $ & $   1.55\cdot 10^{-1} $ & $   3.99\cdot 10^{-2} $ & $   1.14\cdot 10^{-2} $ & $   3.45\cdot 10^{-3} $ & $   1.08\cdot 10^{-3} $ & $   3.43\cdot 10^{-4} $ \\ 
$    0.7 $ & $    0.0 $ & $    9.000 $ & $    7.400 $ & $    2.719\cdot 10^{-3} $  & $   7.73\cdot 10^{-1} $ & $   1.63\cdot 10^{-1} $ & $   4.47\cdot 10^{-2} $ & $   1.36\cdot 10^{-2} $ & $   4.39\cdot 10^{-3} $ & $   1.46\cdot 10^{-3} $ & $   4.98\cdot 10^{-4} $ \\ 
$    0.7 $ & $    0.2 $ & $    8.042 $ & $    6.612 $ & $    3.937\cdot 10^{-3} $  & $   7.53\cdot 10^{-1} $ & $   1.71\cdot 10^{-1} $ & $   5.06\cdot 10^{-2} $ & $   1.66\cdot 10^{-2} $ & $   5.76\cdot 10^{-3} $ & $   2.06\cdot 10^{-3} $ & $   7.56\cdot 10^{-4} $ \\ 
$    0.7 $ & $    0.6 $ & $    5.896 $ & $    4.848 $ & $    1.068\cdot 10^{-2} $  & $   6.90\cdot 10^{-1} $ & $   1.94\cdot 10^{-1} $ & $   6.94\cdot 10^{-2} $ & $   2.75\cdot 10^{-2} $ & $   1.15\cdot 10^{-2} $ & $   4.97\cdot 10^{-3} $ & $   2.20\cdot 10^{-3} $ \\ 
$    0.7 $ & $    0.9 $ & $    3.744 $ & $    3.078 $ & $    4.133\cdot 10^{-2} $  & $   5.70\cdot 10^{-1} $ & $   2.24\cdot 10^{-1} $ & $   1.03\cdot 10^{-1} $ & $   5.18\cdot 10^{-2} $ & $   2.74\cdot 10^{-2} $ & $   1.50\cdot 10^{-2} $ & $   8.40\cdot 10^{-3} $ \\ 
\hline
$    0.7 $ & $   -0.9 $ & $   18.595 $ & $   10.585 $ & $    1.578\cdot 10^{-4} $  & $   8.81\cdot 10^{-1} $ & $   1.00\cdot 10^{-1} $ & $   1.53\cdot 10^{-2} $ & $   2.59\cdot 10^{-3} $ & $   4.63\cdot 10^{-4} $ & $   8.55\cdot 10^{-5} $ & $   1.61\cdot 10^{-5} $ \\ 
$    0.7 $ & $   -0.6 $ & $   16.812 $ & $    9.570 $ & $    2.225\cdot 10^{-4} $  & $   8.71\cdot 10^{-1} $ & $   1.08\cdot 10^{-1} $ & $   1.77\cdot 10^{-2} $ & $   3.25\cdot 10^{-3} $ & $   6.29\cdot 10^{-4} $ & $   1.26\cdot 10^{-4} $ & $   2.58\cdot 10^{-5} $ \\ 
$    0.7 $ & $   -0.2 $ & $   14.317 $ & $    8.150 $ & $    3.846\cdot 10^{-4} $  & $   8.52\cdot 10^{-1} $ & $   1.20\cdot 10^{-1} $ & $   2.23\cdot 10^{-2} $ & $   4.62\cdot 10^{-3} $ & $   1.01\cdot 10^{-3} $ & $   2.30\cdot 10^{-4} $ & $   5.32\cdot 10^{-5} $ \\ 
$    0.7 $ & $    0.0 $ & $   13.000 $ & $    7.400 $ & $    5.338\cdot 10^{-4} $  & $   8.40\cdot 10^{-1} $ & $   1.27\cdot 10^{-1} $ & $   2.54\cdot 10^{-2} $ & $   5.67\cdot 10^{-3} $ & $   1.34\cdot 10^{-3} $ & $   3.26\cdot 10^{-4} $ & $   8.14\cdot 10^{-5} $ \\ 
$    0.7 $ & $    0.2 $ & $   11.616 $ & $    6.612 $ & $    7.815\cdot 10^{-4} $  & $   8.25\cdot 10^{-1} $ & $   1.36\cdot 10^{-1} $ & $   2.95\cdot 10^{-2} $ & $   7.15\cdot 10^{-3} $ & $   1.83\cdot 10^{-3} $ & $   4.86\cdot 10^{-4} $ & $   1.32\cdot 10^{-4} $ \\ 
$    0.7 $ & $    0.6 $ & $    8.517 $ & $    4.848 $ & $    2.219\cdot 10^{-3} $  & $   7.77\cdot 10^{-1} $ & $   1.61\cdot 10^{-1} $ & $   4.34\cdot 10^{-2} $ & $   1.31\cdot 10^{-2} $ & $   4.16\cdot 10^{-3} $ & $   1.37\cdot 10^{-3} $ & $   4.63\cdot 10^{-4} $ \\ 
$    0.7 $ & $    0.9 $ & $    5.408 $ & $    3.078 $ & $    9.906\cdot 10^{-3} $  & $   6.86\cdot 10^{-1} $ & $   1.96\cdot 10^{-1} $ & $   7.07\cdot 10^{-2} $ & $   2.83\cdot 10^{-2} $ & $   1.20\cdot 10^{-2} $ & $   5.26\cdot 10^{-3} $ & $   2.36\cdot 10^{-3} $ \\ 

\Xhline{2.5\arrayrulewidth}

$    0.9 $ & $   -0.9 $ & $   11.084 $ & $   11.074 $ & $    2.040\cdot 10^{-3} $  & $   7.45\cdot 10^{-1} $ & $   1.75\cdot 10^{-1} $ & $   5.30\cdot 10^{-2} $ & $   1.76\cdot 10^{-2} $ & $   6.14\cdot 10^{-3} $ & $   2.20\cdot 10^{-3} $ & $   8.00\cdot 10^{-4} $ \\ 
$    0.9 $ & $   -0.6 $ & $   10.042 $ & $   10.032 $ & $    2.807\cdot 10^{-3} $  & $   7.28\cdot 10^{-1} $ & $   1.82\cdot 10^{-1} $ & $   5.84\cdot 10^{-2} $ & $   2.06\cdot 10^{-2} $ & $   7.64\cdot 10^{-3} $ & $   2.91\cdot 10^{-3} $ & $   1.13\cdot 10^{-3} $ \\ 
$    0.9 $ & $   -0.2 $ & $    8.582 $ & $    8.572 $ & $    4.643\cdot 10^{-3} $  & $   6.97\cdot 10^{-1} $ & $   1.92\cdot 10^{-1} $ & $   6.75\cdot 10^{-2} $ & $   2.61\cdot 10^{-2} $ & $   1.06\cdot 10^{-2} $ & $   4.44\cdot 10^{-3} $ & $   1.89\cdot 10^{-3} $ \\ 
$    0.9 $ & $    0.0 $ & $    7.810 $ & $    7.800 $ & $    6.253\cdot 10^{-3} $  & $   6.78\cdot 10^{-1} $ & $   1.98\cdot 10^{-1} $ & $   7.33\cdot 10^{-2} $ & $   2.99\cdot 10^{-2} $ & $   1.28\cdot 10^{-2} $ & $   5.65\cdot 10^{-3} $ & $   2.54\cdot 10^{-3} $ \\ 
$    0.9 $ & $    0.2 $ & $    6.999 $ & $    6.989 $ & $    8.798\cdot 10^{-3} $  & $   6.54\cdot 10^{-1} $ & $   2.04\cdot 10^{-1} $ & $   8.03\cdot 10^{-2} $ & $   3.47\cdot 10^{-2} $ & $   1.58\cdot 10^{-2} $ & $   7.38\cdot 10^{-3} $ & $   3.52\cdot 10^{-3} $ \\ 
$    0.9 $ & $    0.6 $ & $    5.177 $ & $    5.167 $ & $    2.155\cdot 10^{-2} $  & $   5.77\cdot 10^{-1} $ & $   2.21\cdot 10^{-1} $ & $   1.01\cdot 10^{-1} $ & $   5.07\cdot 10^{-2} $ & $   2.67\cdot 10^{-2} $ & $   1.45\cdot 10^{-2} $ & $   8.02\cdot 10^{-3} $ \\ 
$    0.9 $ & $    0.9 $ & $    3.344 $ & $    3.334 $ & $    6.525\cdot 10^{-2} $  & $   4.27\cdot 10^{-1} $ & $   2.42\cdot 10^{-1} $ & $   1.39\cdot 10^{-1} $ & $   8.36\cdot 10^{-2} $ & $   5.25\cdot 10^{-2} $ & $   3.40\cdot 10^{-2} $ & $   2.24\cdot 10^{-2} $ \\ 
\hline
$    0.9 $ & $   -0.9 $ & $   12.778 $ & $   11.074 $ & $    3.320\cdot 10^{-4} $  & $   7.94\cdot 10^{-1} $ & $   1.52\cdot 10^{-1} $ & $   3.84\cdot 10^{-2} $ & $   1.07\cdot 10^{-2} $ & $   3.17\cdot 10^{-3} $ & $   9.62\cdot 10^{-4} $ & $   2.99\cdot 10^{-4} $ \\ 
$    0.9 $ & $   -0.6 $ & $   11.576 $ & $   10.032 $ & $    4.654\cdot 10^{-4} $  & $   7.78\cdot 10^{-1} $ & $   1.60\cdot 10^{-1} $ & $   4.31\cdot 10^{-2} $ & $   1.29\cdot 10^{-2} $ & $   4.05\cdot 10^{-3} $ & $   1.32\cdot 10^{-3} $ & $   4.37\cdot 10^{-4} $ \\ 
$    0.9 $ & $   -0.2 $ & $    9.891 $ & $    8.572 $ & $    7.946\cdot 10^{-4} $  & $   7.51\cdot 10^{-1} $ & $   1.72\cdot 10^{-1} $ & $   5.11\cdot 10^{-2} $ & $   1.69\cdot 10^{-2} $ & $   5.88\cdot 10^{-3} $ & $   2.12\cdot 10^{-3} $ & $   7.79\cdot 10^{-4} $ \\ 
$    0.9 $ & $    0.0 $ & $    9.000 $ & $    7.800 $ & $    1.093\cdot 10^{-3} $  & $   7.34\cdot 10^{-1} $ & $   1.79\cdot 10^{-1} $ & $   5.64\cdot 10^{-2} $ & $   1.97\cdot 10^{-2} $ & $   7.28\cdot 10^{-3} $ & $   2.78\cdot 10^{-3} $ & $   1.08\cdot 10^{-3} $ \\ 
$    0.9 $ & $    0.2 $ & $    8.064 $ & $    6.989 $ & $    1.582\cdot 10^{-3} $  & $   7.13\cdot 10^{-1} $ & $   1.86\cdot 10^{-1} $ & $   6.28\cdot 10^{-2} $ & $   2.35\cdot 10^{-2} $ & $   9.25\cdot 10^{-3} $ & $   3.77\cdot 10^{-3} $ & $   1.57\cdot 10^{-3} $ \\ 
$    0.9 $ & $    0.6 $ & $    5.962 $ & $    5.167 $ & $    4.278\cdot 10^{-3} $  & $   6.46\cdot 10^{-1} $ & $   2.07\cdot 10^{-1} $ & $   8.24\cdot 10^{-2} $ & $   3.63\cdot 10^{-2} $ & $   1.69\cdot 10^{-2} $ & $   8.15\cdot 10^{-3} $ & $   4.01\cdot 10^{-3} $ \\ 
$    0.9 $ & $    0.9 $ & $    3.847 $ & $    3.334 $ & $    1.656\cdot 10^{-2} $  & $   5.19\cdot 10^{-1} $ & $   2.31\cdot 10^{-1} $ & $   1.16\cdot 10^{-1} $ & $   6.31\cdot 10^{-2} $ & $   3.60\cdot 10^{-2} $ & $   2.12\cdot 10^{-2} $ & $   1.28\cdot 10^{-2} $ \\ 
\hline
$    0.9 $ & $   -0.9 $ & $   18.457 $ & $   11.074 $ & $    5.209\cdot 10^{-5} $  & $   8.58\cdot 10^{-1} $ & $   1.16\cdot 10^{-1} $ & $   2.08\cdot 10^{-2} $ & $   4.15\cdot 10^{-3} $ & $   8.75\cdot 10^{-4} $ & $   1.91\cdot 10^{-4} $ & $   4.24\cdot 10^{-5} $ \\ 
$    0.9 $ & $   -0.6 $ & $   16.720 $ & $   10.032 $ & $    7.330\cdot 10^{-5} $  & $   8.46\cdot 10^{-1} $ & $   1.23\cdot 10^{-1} $ & $   2.38\cdot 10^{-2} $ & $   5.13\cdot 10^{-3} $ & $   1.17\cdot 10^{-3} $ & $   2.74\cdot 10^{-4} $ & $   6.56\cdot 10^{-5} $ \\ 
$    0.9 $ & $   -0.2 $ & $   14.287 $ & $    8.572 $ & $    1.262\cdot 10^{-4} $  & $   8.26\cdot 10^{-1} $ & $   1.36\cdot 10^{-1} $ & $   2.94\cdot 10^{-2} $ & $   7.09\cdot 10^{-3} $ & $   1.81\cdot 10^{-3} $ & $   4.77\cdot 10^{-4} $ & $   1.29\cdot 10^{-4} $ \\ 
$    0.9 $ & $    0.0 $ & $   13.000 $ & $    7.800 $ & $    1.747\cdot 10^{-4} $  & $   8.12\cdot 10^{-1} $ & $   1.43\cdot 10^{-1} $ & $   3.31\cdot 10^{-2} $ & $   8.56\cdot 10^{-3} $ & $   2.34\cdot 10^{-3} $ & $   6.59\cdot 10^{-4} $ & $   1.90\cdot 10^{-4} $ \\ 
$    0.9 $ & $    0.2 $ & $   11.648 $ & $    6.989 $ & $    2.550\cdot 10^{-4} $  & $   7.96\cdot 10^{-1} $ & $   1.52\cdot 10^{-1} $ & $   3.79\cdot 10^{-2} $ & $   1.06\cdot 10^{-2} $ & $   3.12\cdot 10^{-3} $ & $   9.50\cdot 10^{-4} $ & $   2.96\cdot 10^{-4} $ \\ 
$    0.9 $ & $    0.6 $ & $    8.612 $ & $    5.167 $ & $    7.194\cdot 10^{-4} $  & $   7.43\cdot 10^{-1} $ & $   1.75\cdot 10^{-1} $ & $   5.35\cdot 10^{-2} $ & $   1.82\cdot 10^{-2} $ & $   6.54\cdot 10^{-3} $ & $   2.43\cdot 10^{-3} $ & $   9.26\cdot 10^{-4} $ \\ 
$    0.9 $ & $    0.9 $ & $    5.557 $ & $    3.334 $ & $    3.196\cdot 10^{-3} $  & $   6.47\cdot 10^{-1} $ & $   2.06\cdot 10^{-1} $ & $   8.20\cdot 10^{-2} $ & $   3.60\cdot 10^{-2} $ & $   1.67\cdot 10^{-2} $ & $   8.01\cdot 10^{-3} $ & $   3.94\cdot 10^{-3} $ \\ 

\end{tabular}
\end{ruledtabular}
\end{center}
\end{table*}

\clearpage

\bibliography{refs20220330.bib,refs_loc20220330.bib}

\begin{thebibliography}{82}%
\makeatletter
\providecommand \@ifxundefined [1]{%
 \@ifx{#1\undefined}
}%
\providecommand \@ifnum [1]{%
 \ifnum #1\expandafter \@firstoftwo
 \else \expandafter \@secondoftwo
 \fi
}%
\providecommand \@ifx [1]{%
 \ifx #1\expandafter \@firstoftwo
 \else \expandafter \@secondoftwo
 \fi
}%
\providecommand \natexlab [1]{#1}%
\providecommand \enquote  [1]{``#1''}%
\providecommand \bibnamefont  [1]{#1}%
\providecommand \bibfnamefont [1]{#1}%
\providecommand \citenamefont [1]{#1}%
\providecommand \href@noop [0]{\@secondoftwo}%
\providecommand \href [0]{\begingroup \@sanitize@url \@href}%
\providecommand \@href[1]{\@@startlink{#1}\@@href}%
\providecommand \@@href[1]{\endgroup#1\@@endlink}%
\providecommand \@sanitize@url [0]{\catcode `\\12\catcode `\$12\catcode
  `\&12\catcode `\#12\catcode `\^12\catcode `\_12\catcode `\%12\relax}%
\providecommand \@@startlink[1]{}%
\providecommand \@@endlink[0]{}%
\providecommand \url  [0]{\begingroup\@sanitize@url \@url }%
\providecommand \@url [1]{\endgroup\@href {#1}{\urlprefix }}%
\providecommand \urlprefix  [0]{URL }%
\providecommand \Eprint [0]{\href }%
\providecommand \doibase [0]{http://dx.doi.org/}%
\providecommand \selectlanguage [0]{\@gobble}%
\providecommand \bibinfo  [0]{\@secondoftwo}%
\providecommand \bibfield  [0]{\@secondoftwo}%
\providecommand \translation [1]{[#1]}%
\providecommand \BibitemOpen [0]{}%
\providecommand \bibitemStop [0]{}%
\providecommand \bibitemNoStop [0]{.\EOS\space}%
\providecommand \EOS [0]{\spacefactor3000\relax}%
\providecommand \BibitemShut  [1]{\csname bibitem#1\endcsname}%
\let\auto@bib@innerbib\@empty
\bibitem [{\citenamefont {Acernese}\ \emph {et~al.}(2015)\citenamefont
  {Acernese} \emph {et~al.}}]{TheVirgo:2014hva}%
  \BibitemOpen
  \bibfield  {author} {\bibinfo {author} {\bibfnamefont {F.}~\bibnamefont
  {Acernese}} \emph {et~al.} (\bibinfo {collaboration} {VIRGO}),\ }\href
  {\doibase 10.1088/0264-9381/32/2/024001} {\bibfield  {journal} {\bibinfo
  {journal} {Class. Quant. Grav.}\ }\textbf {\bibinfo {volume} {32}},\ \bibinfo
  {pages} {024001} (\bibinfo {year} {2015})},\ \Eprint
  {http://arxiv.org/abs/1408.3978} {arXiv:1408.3978 [gr-qc]} \BibitemShut
  {NoStop}%
\bibitem [{\citenamefont {Aasi}\ \emph {et~al.}(2015)\citenamefont {Aasi} \emph
  {et~al.}}]{TheLIGOScientific:2014jea}%
  \BibitemOpen
  \bibfield  {author} {\bibinfo {author} {\bibfnamefont {J.}~\bibnamefont
  {Aasi}} \emph {et~al.} (\bibinfo {collaboration} {LIGO Scientific}),\ }\href
  {\doibase 10.1088/0264-9381/32/7/074001} {\bibfield  {journal} {\bibinfo
  {journal} {Class. Quant. Grav.}\ }\textbf {\bibinfo {volume} {32}},\ \bibinfo
  {pages} {074001} (\bibinfo {year} {2015})},\ \Eprint
  {http://arxiv.org/abs/1411.4547} {arXiv:1411.4547 [gr-qc]} \BibitemShut
  {NoStop}%
\bibitem [{\citenamefont {Abbott}\ \emph {et~al.}(2021)\citenamefont {Abbott}
  \emph {et~al.}}]{LIGOScientific:2020ibl}%
  \BibitemOpen
  \bibfield  {author} {\bibinfo {author} {\bibfnamefont {R.}~\bibnamefont
  {Abbott}} \emph {et~al.} (\bibinfo {collaboration} {LIGO Scientific,
  Virgo}),\ }\href {\doibase 10.1103/PhysRevX.11.021053} {\bibfield  {journal}
  {\bibinfo  {journal} {Phys. Rev. X}\ }\textbf {\bibinfo {volume} {11}},\
  \bibinfo {pages} {021053} (\bibinfo {year} {2021})},\ \Eprint
  {http://arxiv.org/abs/2010.14527} {arXiv:2010.14527 [gr-qc]} \BibitemShut
  {NoStop}%
\bibitem [{\citenamefont {Mukherjee}\ \emph {et~al.}(2020)\citenamefont
  {Mukherjee}, \citenamefont {Mitra},\ and\ \citenamefont
  {Chatterjee}}]{Mukherjee:2020hnm}%
  \BibitemOpen
  \bibfield  {author} {\bibinfo {author} {\bibfnamefont {S.}~\bibnamefont
  {Mukherjee}}, \bibinfo {author} {\bibfnamefont {S.}~\bibnamefont {Mitra}}, \
  and\ \bibinfo {author} {\bibfnamefont {S.}~\bibnamefont {Chatterjee}},\
  }\href@noop {} {\  (\bibinfo {year} {2020})},\ \Eprint
  {http://arxiv.org/abs/2010.00916} {arXiv:2010.00916 [gr-qc]} \BibitemShut
  {NoStop}%
\bibitem [{\citenamefont {Romero-Shaw}\ \emph {et~al.}(2019)\citenamefont
  {Romero-Shaw}, \citenamefont {Lasky},\ and\ \citenamefont
  {Thrane}}]{Romero-Shaw:2019itr}%
  \BibitemOpen
  \bibfield  {author} {\bibinfo {author} {\bibfnamefont {I.~M.}\ \bibnamefont
  {Romero-Shaw}}, \bibinfo {author} {\bibfnamefont {P.~D.}\ \bibnamefont
  {Lasky}}, \ and\ \bibinfo {author} {\bibfnamefont {E.}~\bibnamefont
  {Thrane}},\ }\href {\doibase 10.1093/mnras/stz2996} {\bibfield  {journal}
  {\bibinfo  {journal} {Mon. Not. Roy. Astron. Soc.}\ }\textbf {\bibinfo
  {volume} {490}},\ \bibinfo {pages} {5210} (\bibinfo {year} {2019})},\ \Eprint
  {http://arxiv.org/abs/1909.05466} {arXiv:1909.05466 [astro-ph.HE]}
  \BibitemShut {NoStop}%
\bibitem [{\citenamefont {Abbott}\ \emph {et~al.}(2020)\citenamefont {Abbott}
  \emph {et~al.}}]{Abbott:2020tfl}%
  \BibitemOpen
  \bibfield  {author} {\bibinfo {author} {\bibfnamefont {R.}~\bibnamefont
  {Abbott}} \emph {et~al.} (\bibinfo {collaboration} {LIGO Scientific,
  Virgo}),\ }\href {\doibase 10.1103/PhysRevLett.125.101102} {\bibfield
  {journal} {\bibinfo  {journal} {Phys. Rev. Lett.}\ }\textbf {\bibinfo
  {volume} {125}},\ \bibinfo {pages} {101102} (\bibinfo {year} {2020})},\
  \Eprint {http://arxiv.org/abs/2009.01075} {arXiv:2009.01075 [gr-qc]}
  \BibitemShut {NoStop}%
\bibitem [{\citenamefont {Romero-Shaw}\ \emph {et~al.}(2020)\citenamefont
  {Romero-Shaw}, \citenamefont {Lasky}, \citenamefont {Thrane},\ and\
  \citenamefont {Bustillo}}]{Romero-Shaw:2020thy}%
  \BibitemOpen
  \bibfield  {author} {\bibinfo {author} {\bibfnamefont {I.~M.}\ \bibnamefont
  {Romero-Shaw}}, \bibinfo {author} {\bibfnamefont {P.~D.}\ \bibnamefont
  {Lasky}}, \bibinfo {author} {\bibfnamefont {E.}~\bibnamefont {Thrane}}, \
  and\ \bibinfo {author} {\bibfnamefont {J.~C.}\ \bibnamefont {Bustillo}},\
  }\href {\doibase 10.3847/2041-8213/abbe26} {\bibfield  {journal} {\bibinfo
  {journal} {Astrophys. J. Lett.}\ }\textbf {\bibinfo {volume} {903}},\
  \bibinfo {pages} {L5} (\bibinfo {year} {2020})},\ \Eprint
  {http://arxiv.org/abs/2009.04771} {arXiv:2009.04771 [astro-ph.HE]}
  \BibitemShut {NoStop}%
\bibitem [{\citenamefont {Amaro-Seoane}(2018)}]{Amaro-Seoane:2018gbb}%
  \BibitemOpen
  \bibfield  {author} {\bibinfo {author} {\bibfnamefont {P.}~\bibnamefont
  {Amaro-Seoane}},\ }\href {\doibase 10.1103/PhysRevD.98.063018} {\bibfield
  {journal} {\bibinfo  {journal} {Phys. Rev. D}\ }\textbf {\bibinfo {volume}
  {98}},\ \bibinfo {pages} {063018} (\bibinfo {year} {2018})},\ \Eprint
  {http://arxiv.org/abs/1807.03824} {arXiv:1807.03824 [astro-ph.HE]}
  \BibitemShut {NoStop}%
\bibitem [{\citenamefont {Huerta}\ \emph {et~al.}(2018)\citenamefont {Huerta}
  \emph {et~al.}}]{Huerta:2017kez}%
  \BibitemOpen
  \bibfield  {author} {\bibinfo {author} {\bibfnamefont {E.~A.}\ \bibnamefont
  {Huerta}} \emph {et~al.},\ }\href {\doibase 10.1103/PhysRevD.97.024031}
  {\bibfield  {journal} {\bibinfo  {journal} {Phys. Rev.}\ }\textbf {\bibinfo
  {volume} {D97}},\ \bibinfo {pages} {024031} (\bibinfo {year} {2018})},\
  \Eprint {http://arxiv.org/abs/1711.06276} {arXiv:1711.06276 [gr-qc]}
  \BibitemShut {NoStop}%
\bibitem [{\citenamefont {Cao}\ and\ \citenamefont {Han}(2017)}]{Cao:2017ndf}%
  \BibitemOpen
  \bibfield  {author} {\bibinfo {author} {\bibfnamefont {Z.}~\bibnamefont
  {Cao}}\ and\ \bibinfo {author} {\bibfnamefont {W.-B.}\ \bibnamefont {Han}},\
  }\href {\doibase 10.1103/PhysRevD.96.044028} {\bibfield  {journal} {\bibinfo
  {journal} {Phys. Rev.}\ }\textbf {\bibinfo {volume} {D96}},\ \bibinfo {pages}
  {044028} (\bibinfo {year} {2017})},\ \Eprint
  {http://arxiv.org/abs/1708.00166} {arXiv:1708.00166 [gr-qc]} \BibitemShut
  {NoStop}%
\bibitem [{\citenamefont {Liu}\ \emph {et~al.}(2019)\citenamefont {Liu},
  \citenamefont {Cao},\ and\ \citenamefont {Shao}}]{Liu:2019jpg}%
  \BibitemOpen
  \bibfield  {author} {\bibinfo {author} {\bibfnamefont {X.}~\bibnamefont
  {Liu}}, \bibinfo {author} {\bibfnamefont {Z.}~\bibnamefont {Cao}}, \ and\
  \bibinfo {author} {\bibfnamefont {L.}~\bibnamefont {Shao}},\ }\href@noop {}
  {\  (\bibinfo {year} {2019})},\ \Eprint {http://arxiv.org/abs/1910.00784}
  {arXiv:1910.00784 [gr-qc]} \BibitemShut {NoStop}%
\bibitem [{\citenamefont {Liu}\ \emph {et~al.}(2021)\citenamefont {Liu},
  \citenamefont {Cao},\ and\ \citenamefont {Zhu}}]{Liu:2021pkr}%
  \BibitemOpen
  \bibfield  {author} {\bibinfo {author} {\bibfnamefont {X.}~\bibnamefont
  {Liu}}, \bibinfo {author} {\bibfnamefont {Z.}~\bibnamefont {Cao}}, \ and\
  \bibinfo {author} {\bibfnamefont {Z.-H.}\ \bibnamefont {Zhu}},\ }\href@noop
  {} {\  (\bibinfo {year} {2021})},\ \Eprint {http://arxiv.org/abs/2102.08614}
  {arXiv:2102.08614 [gr-qc]} \BibitemShut {NoStop}%
\bibitem [{\citenamefont {Islam}\ \emph {et~al.}(2021)\citenamefont {Islam},
  \citenamefont {Varma}, \citenamefont {Lodman}, \citenamefont {Field},
  \citenamefont {Khanna}, \citenamefont {Scheel}, \citenamefont {Pfeiffer},
  \citenamefont {Gerosa},\ and\ \citenamefont {Kidder}}]{Islam:2021mha}%
  \BibitemOpen
  \bibfield  {author} {\bibinfo {author} {\bibfnamefont {T.}~\bibnamefont
  {Islam}}, \bibinfo {author} {\bibfnamefont {V.}~\bibnamefont {Varma}},
  \bibinfo {author} {\bibfnamefont {J.}~\bibnamefont {Lodman}}, \bibinfo
  {author} {\bibfnamefont {S.~E.}\ \bibnamefont {Field}}, \bibinfo {author}
  {\bibfnamefont {G.}~\bibnamefont {Khanna}}, \bibinfo {author} {\bibfnamefont
  {M.~A.}\ \bibnamefont {Scheel}}, \bibinfo {author} {\bibfnamefont {H.~P.}\
  \bibnamefont {Pfeiffer}}, \bibinfo {author} {\bibfnamefont {D.}~\bibnamefont
  {Gerosa}}, \ and\ \bibinfo {author} {\bibfnamefont {L.~E.}\ \bibnamefont
  {Kidder}},\ }\href@noop {} {\  (\bibinfo {year} {2021})},\ \Eprint
  {http://arxiv.org/abs/2101.11798} {arXiv:2101.11798 [gr-qc]} \BibitemShut
  {NoStop}%
\bibitem [{\citenamefont {Chiaramello}\ and\ \citenamefont
  {Nagar}(2020)}]{Chiaramello:2020ehz}%
  \BibitemOpen
  \bibfield  {author} {\bibinfo {author} {\bibfnamefont {D.}~\bibnamefont
  {Chiaramello}}\ and\ \bibinfo {author} {\bibfnamefont {A.}~\bibnamefont
  {Nagar}},\ }\href {\doibase 10.1103/PhysRevD.101.101501} {\bibfield
  {journal} {\bibinfo  {journal} {Phys. Rev. D}\ }\textbf {\bibinfo {volume}
  {101}},\ \bibinfo {pages} {101501} (\bibinfo {year} {2020})},\ \Eprint
  {http://arxiv.org/abs/2001.11736} {arXiv:2001.11736 [gr-qc]} \BibitemShut
  {NoStop}%
\bibitem [{\citenamefont {Nagar}\ \emph
  {et~al.}(2021{\natexlab{a}})\citenamefont {Nagar}, \citenamefont {Rettegno},
  \citenamefont {Gamba},\ and\ \citenamefont {Bernuzzi}}]{Nagar:2020xsk}%
  \BibitemOpen
  \bibfield  {author} {\bibinfo {author} {\bibfnamefont {A.}~\bibnamefont
  {Nagar}}, \bibinfo {author} {\bibfnamefont {P.}~\bibnamefont {Rettegno}},
  \bibinfo {author} {\bibfnamefont {R.}~\bibnamefont {Gamba}}, \ and\ \bibinfo
  {author} {\bibfnamefont {S.}~\bibnamefont {Bernuzzi}},\ }\href {\doibase
  10.1103/PhysRevD.103.064013} {\bibfield  {journal} {\bibinfo  {journal}
  {Phys. Rev. D}\ }\textbf {\bibinfo {volume} {103}},\ \bibinfo {pages}
  {064013} (\bibinfo {year} {2021}{\natexlab{a}})},\ \Eprint
  {http://arxiv.org/abs/2009.12857} {arXiv:2009.12857 [gr-qc]} \BibitemShut
  {NoStop}%
\bibitem [{\citenamefont {Nagar}\ \emph
  {et~al.}(2021{\natexlab{b}})\citenamefont {Nagar}, \citenamefont {Bonino},\
  and\ \citenamefont {Rettegno}}]{Nagar:2021gss}%
  \BibitemOpen
  \bibfield  {author} {\bibinfo {author} {\bibfnamefont {A.}~\bibnamefont
  {Nagar}}, \bibinfo {author} {\bibfnamefont {A.}~\bibnamefont {Bonino}}, \
  and\ \bibinfo {author} {\bibfnamefont {P.}~\bibnamefont {Rettegno}},\ }\href
  {\doibase 10.1103/PhysRevD.103.104021} {\bibfield  {journal} {\bibinfo
  {journal} {Phys. Rev. D}\ }\textbf {\bibinfo {volume} {103}},\ \bibinfo
  {pages} {104021} (\bibinfo {year} {2021}{\natexlab{b}})},\ \Eprint
  {http://arxiv.org/abs/2101.08624} {arXiv:2101.08624 [gr-qc]} \BibitemShut
  {NoStop}%
\bibitem [{\citenamefont {Tanay}\ \emph {et~al.}(2016)\citenamefont {Tanay},
  \citenamefont {Haney},\ and\ \citenamefont {Gopakumar}}]{Tanay:2016zog}%
  \BibitemOpen
  \bibfield  {author} {\bibinfo {author} {\bibfnamefont {S.}~\bibnamefont
  {Tanay}}, \bibinfo {author} {\bibfnamefont {M.}~\bibnamefont {Haney}}, \ and\
  \bibinfo {author} {\bibfnamefont {A.}~\bibnamefont {Gopakumar}},\ }\href
  {\doibase 10.1103/PhysRevD.93.064031} {\bibfield  {journal} {\bibinfo
  {journal} {Phys. Rev. D}\ }\textbf {\bibinfo {volume} {93}},\ \bibinfo
  {pages} {064031} (\bibinfo {year} {2016})},\ \Eprint
  {http://arxiv.org/abs/1602.03081} {arXiv:1602.03081 [gr-qc]} \BibitemShut
  {NoStop}%
\bibitem [{\citenamefont {Moore}\ and\ \citenamefont
  {Yunes}(2019)}]{Moore:2019xkm}%
  \BibitemOpen
  \bibfield  {author} {\bibinfo {author} {\bibfnamefont {B.}~\bibnamefont
  {Moore}}\ and\ \bibinfo {author} {\bibfnamefont {N.}~\bibnamefont {Yunes}},\
  }\href {\doibase 10.1088/1361-6382/ab3778} {\bibfield  {journal} {\bibinfo
  {journal} {Class. Quant. Grav.}\ }\textbf {\bibinfo {volume} {36}},\ \bibinfo
  {pages} {185003} (\bibinfo {year} {2019})},\ \Eprint
  {http://arxiv.org/abs/1903.05203} {arXiv:1903.05203 [gr-qc]} \BibitemShut
  {NoStop}%
\bibitem [{\citenamefont {Boetzel}\ \emph {et~al.}(2019)\citenamefont
  {Boetzel}, \citenamefont {Mishra}, \citenamefont {Faye}, \citenamefont
  {Gopakumar},\ and\ \citenamefont {Iyer}}]{Boetzel:2019nfw}%
  \BibitemOpen
  \bibfield  {author} {\bibinfo {author} {\bibfnamefont {Y.}~\bibnamefont
  {Boetzel}}, \bibinfo {author} {\bibfnamefont {C.~K.}\ \bibnamefont {Mishra}},
  \bibinfo {author} {\bibfnamefont {G.}~\bibnamefont {Faye}}, \bibinfo {author}
  {\bibfnamefont {A.}~\bibnamefont {Gopakumar}}, \ and\ \bibinfo {author}
  {\bibfnamefont {B.~R.}\ \bibnamefont {Iyer}},\ }\href {\doibase
  10.1103/PhysRevD.100.044018} {\bibfield  {journal} {\bibinfo  {journal}
  {Phys. Rev. D}\ }\textbf {\bibinfo {volume} {100}},\ \bibinfo {pages}
  {044018} (\bibinfo {year} {2019})},\ \Eprint
  {http://arxiv.org/abs/1904.11814} {arXiv:1904.11814 [gr-qc]} \BibitemShut
  {NoStop}%
\bibitem [{\citenamefont {Tiwari}\ \emph {et~al.}(2019)\citenamefont {Tiwari},
  \citenamefont {Achamveedu}, \citenamefont {Haney},\ and\ \citenamefont
  {Hemantakumar}}]{Tiwari:2019jtz}%
  \BibitemOpen
  \bibfield  {author} {\bibinfo {author} {\bibfnamefont {S.}~\bibnamefont
  {Tiwari}}, \bibinfo {author} {\bibfnamefont {G.}~\bibnamefont {Achamveedu}},
  \bibinfo {author} {\bibfnamefont {M.}~\bibnamefont {Haney}}, \ and\ \bibinfo
  {author} {\bibfnamefont {P.}~\bibnamefont {Hemantakumar}},\ }\href {\doibase
  10.1103/PhysRevD.99.124008} {\bibfield  {journal} {\bibinfo  {journal} {Phys.
  Rev.}\ }\textbf {\bibinfo {volume} {D99}},\ \bibinfo {pages} {124008}
  (\bibinfo {year} {2019})},\ \Eprint {http://arxiv.org/abs/1905.07956}
  {arXiv:1905.07956 [gr-qc]} \BibitemShut {NoStop}%
\bibitem [{\citenamefont {Tiwari}\ and\ \citenamefont
  {Gopakumar}(2020)}]{Tiwari:2020hsu}%
  \BibitemOpen
  \bibfield  {author} {\bibinfo {author} {\bibfnamefont {S.}~\bibnamefont
  {Tiwari}}\ and\ \bibinfo {author} {\bibfnamefont {A.}~\bibnamefont
  {Gopakumar}},\ }\href {\doibase 10.1103/PhysRevD.102.084042} {\bibfield
  {journal} {\bibinfo  {journal} {Phys. Rev. D}\ }\textbf {\bibinfo {volume}
  {102}},\ \bibinfo {pages} {084042} (\bibinfo {year} {2020})},\ \Eprint
  {http://arxiv.org/abs/2009.11333} {arXiv:2009.11333 [gr-qc]} \BibitemShut
  {NoStop}%
\bibitem [{\citenamefont {Nagar}\ \emph {et~al.}(2007)\citenamefont {Nagar},
  \citenamefont {Damour},\ and\ \citenamefont {Tartaglia}}]{Nagar:2006xv}%
  \BibitemOpen
  \bibfield  {author} {\bibinfo {author} {\bibfnamefont {A.}~\bibnamefont
  {Nagar}}, \bibinfo {author} {\bibfnamefont {T.}~\bibnamefont {Damour}}, \
  and\ \bibinfo {author} {\bibfnamefont {A.}~\bibnamefont {Tartaglia}},\ }\href
  {\doibase 10.1088/0264-9381/24/12/S08} {\bibfield  {journal} {\bibinfo
  {journal} {Class. Quant. Grav.}\ }\textbf {\bibinfo {volume} {24}},\ \bibinfo
  {pages} {S109} (\bibinfo {year} {2007})},\ \Eprint
  {http://arxiv.org/abs/gr-qc/0612096} {arXiv:gr-qc/0612096} \BibitemShut
  {NoStop}%
\bibitem [{\citenamefont {Damour}\ and\ \citenamefont
  {Nagar}(2007)}]{Damour:2007xr}%
  \BibitemOpen
  \bibfield  {author} {\bibinfo {author} {\bibfnamefont {T.}~\bibnamefont
  {Damour}}\ and\ \bibinfo {author} {\bibfnamefont {A.}~\bibnamefont {Nagar}},\
  }\href {\doibase 10.1103/PhysRevD.76.064028} {\bibfield  {journal} {\bibinfo
  {journal} {Phys. Rev.}\ }\textbf {\bibinfo {volume} {D76}},\ \bibinfo {pages}
  {064028} (\bibinfo {year} {2007})},\ \Eprint {http://arxiv.org/abs/0705.2519}
  {arXiv:0705.2519 [gr-qc]} \BibitemShut {NoStop}%
\bibitem [{\citenamefont {Bernuzzi}\ and\ \citenamefont
  {Nagar}(2010)}]{Bernuzzi:2010ty}%
  \BibitemOpen
  \bibfield  {author} {\bibinfo {author} {\bibfnamefont {S.}~\bibnamefont
  {Bernuzzi}}\ and\ \bibinfo {author} {\bibfnamefont {A.}~\bibnamefont
  {Nagar}},\ }\href {\doibase 10.1103/PhysRevD.81.084056} {\bibfield  {journal}
  {\bibinfo  {journal} {Phys. Rev.}\ }\textbf {\bibinfo {volume} {D81}},\
  \bibinfo {pages} {084056} (\bibinfo {year} {2010})},\ \Eprint
  {http://arxiv.org/abs/1003.0597} {arXiv:1003.0597 [gr-qc]} \BibitemShut
  {NoStop}%
\bibitem [{\citenamefont {Bernuzzi}\ \emph
  {et~al.}(2011{\natexlab{a}})\citenamefont {Bernuzzi}, \citenamefont {Nagar},\
  and\ \citenamefont {Zenginoglu}}]{Bernuzzi:2010xj}%
  \BibitemOpen
  \bibfield  {author} {\bibinfo {author} {\bibfnamefont {S.}~\bibnamefont
  {Bernuzzi}}, \bibinfo {author} {\bibfnamefont {A.}~\bibnamefont {Nagar}}, \
  and\ \bibinfo {author} {\bibfnamefont {A.}~\bibnamefont {Zenginoglu}},\
  }\href {\doibase 10.1103/PhysRevD.83.064010} {\bibfield  {journal} {\bibinfo
  {journal} {Phys.Rev.}\ }\textbf {\bibinfo {volume} {D83}},\ \bibinfo {pages}
  {064010} (\bibinfo {year} {2011}{\natexlab{a}})},\ \Eprint
  {http://arxiv.org/abs/1012.2456} {arXiv:1012.2456 [gr-qc]} \BibitemShut
  {NoStop}%
\bibitem [{\citenamefont {Bernuzzi}\ \emph
  {et~al.}(2011{\natexlab{b}})\citenamefont {Bernuzzi}, \citenamefont {Nagar},\
  and\ \citenamefont {Zenginoglu}}]{Bernuzzi:2011aj}%
  \BibitemOpen
  \bibfield  {author} {\bibinfo {author} {\bibfnamefont {S.}~\bibnamefont
  {Bernuzzi}}, \bibinfo {author} {\bibfnamefont {A.}~\bibnamefont {Nagar}}, \
  and\ \bibinfo {author} {\bibfnamefont {A.}~\bibnamefont {Zenginoglu}},\
  }\href {\doibase 10.1103/PhysRevD.84.084026} {\bibfield  {journal} {\bibinfo
  {journal} {Phys.Rev.}\ }\textbf {\bibinfo {volume} {D84}},\ \bibinfo {pages}
  {084026} (\bibinfo {year} {2011}{\natexlab{b}})},\ \Eprint
  {http://arxiv.org/abs/1107.5402} {arXiv:1107.5402 [gr-qc]} \BibitemShut
  {NoStop}%
\bibitem [{\citenamefont {Bernuzzi}\ \emph {et~al.}(2012)\citenamefont
  {Bernuzzi}, \citenamefont {Nagar},\ and\ \citenamefont
  {Zenginoglu}}]{Bernuzzi:2012ku}%
  \BibitemOpen
  \bibfield  {author} {\bibinfo {author} {\bibfnamefont {S.}~\bibnamefont
  {Bernuzzi}}, \bibinfo {author} {\bibfnamefont {A.}~\bibnamefont {Nagar}}, \
  and\ \bibinfo {author} {\bibfnamefont {A.}~\bibnamefont {Zenginoglu}},\
  }\href {\doibase 10.1103/PhysRevD.86.104038} {\bibfield  {journal} {\bibinfo
  {journal} {Phys.Rev.}\ }\textbf {\bibinfo {volume} {D86}},\ \bibinfo {pages}
  {104038} (\bibinfo {year} {2012})},\ \Eprint {http://arxiv.org/abs/1207.0769}
  {arXiv:1207.0769 [gr-qc]} \BibitemShut {NoStop}%
\bibitem [{\citenamefont {Harms}\ \emph {et~al.}(2014)\citenamefont {Harms},
  \citenamefont {Bernuzzi}, \citenamefont {Nagar},\ and\ \citenamefont
  {Zenginoglu}}]{Harms:2014dqa}%
  \BibitemOpen
  \bibfield  {author} {\bibinfo {author} {\bibfnamefont {E.}~\bibnamefont
  {Harms}}, \bibinfo {author} {\bibfnamefont {S.}~\bibnamefont {Bernuzzi}},
  \bibinfo {author} {\bibfnamefont {A.}~\bibnamefont {Nagar}}, \ and\ \bibinfo
  {author} {\bibfnamefont {A.}~\bibnamefont {Zenginoglu}},\ }\href {\doibase
  10.1088/0264-9381/31/24/245004} {\bibfield  {journal} {\bibinfo  {journal}
  {Class.Quant.Grav.}\ }\textbf {\bibinfo {volume} {31}},\ \bibinfo {pages}
  {245004} (\bibinfo {year} {2014})},\ \Eprint {http://arxiv.org/abs/1406.5983}
  {arXiv:1406.5983 [gr-qc]} \BibitemShut {NoStop}%
\bibitem [{\citenamefont {Nagar}\ \emph {et~al.}(2014)\citenamefont {Nagar},
  \citenamefont {Harms}, \citenamefont {Bernuzzi},\ and\ \citenamefont
  {Zenginoğlu}}]{Nagar:2014kha}%
  \BibitemOpen
  \bibfield  {author} {\bibinfo {author} {\bibfnamefont {A.}~\bibnamefont
  {Nagar}}, \bibinfo {author} {\bibfnamefont {E.}~\bibnamefont {Harms}},
  \bibinfo {author} {\bibfnamefont {S.}~\bibnamefont {Bernuzzi}}, \ and\
  \bibinfo {author} {\bibfnamefont {A.}~\bibnamefont {Zenginoğlu}},\ }\href
  {\doibase 10.1103/PhysRevD.90.124086} {\bibfield  {journal} {\bibinfo
  {journal} {Phys. Rev.}\ }\textbf {\bibinfo {volume} {D90}},\ \bibinfo {pages}
  {124086} (\bibinfo {year} {2014})},\ \Eprint {http://arxiv.org/abs/1407.5033}
  {arXiv:1407.5033 [gr-qc]} \BibitemShut {NoStop}%
\bibitem [{\citenamefont {Harms}\ \emph
  {et~al.}(2016{\natexlab{a}})\citenamefont {Harms}, \citenamefont
  {Lukes-Gerakopoulos}, \citenamefont {Bernuzzi},\ and\ \citenamefont
  {Nagar}}]{Harms:2015ixa}%
  \BibitemOpen
  \bibfield  {author} {\bibinfo {author} {\bibfnamefont {E.}~\bibnamefont
  {Harms}}, \bibinfo {author} {\bibfnamefont {G.}~\bibnamefont
  {Lukes-Gerakopoulos}}, \bibinfo {author} {\bibfnamefont {S.}~\bibnamefont
  {Bernuzzi}}, \ and\ \bibinfo {author} {\bibfnamefont {A.}~\bibnamefont
  {Nagar}},\ }\href {\doibase 10.1103/PhysRevD.93.044015} {\bibfield  {journal}
  {\bibinfo  {journal} {Phys. Rev.}\ }\textbf {\bibinfo {volume} {D93}},\
  \bibinfo {pages} {044015} (\bibinfo {year} {2016}{\natexlab{a}})},\ \Eprint
  {http://arxiv.org/abs/1510.05548} {arXiv:1510.05548 [gr-qc]} \BibitemShut
  {NoStop}%
\bibitem [{\citenamefont {Harms}\ \emph
  {et~al.}(2016{\natexlab{b}})\citenamefont {Harms}, \citenamefont
  {Lukes-Gerakopoulos}, \citenamefont {Bernuzzi},\ and\ \citenamefont
  {Nagar}}]{Harms:2016ctx}%
  \BibitemOpen
  \bibfield  {author} {\bibinfo {author} {\bibfnamefont {E.}~\bibnamefont
  {Harms}}, \bibinfo {author} {\bibfnamefont {G.}~\bibnamefont
  {Lukes-Gerakopoulos}}, \bibinfo {author} {\bibfnamefont {S.}~\bibnamefont
  {Bernuzzi}}, \ and\ \bibinfo {author} {\bibfnamefont {A.}~\bibnamefont
  {Nagar}},\ }\href {\doibase 10.1103/PhysRevD.94.104010} {\bibfield  {journal}
  {\bibinfo  {journal} {Phys. Rev.}\ }\textbf {\bibinfo {volume} {D94}},\
  \bibinfo {pages} {104010} (\bibinfo {year} {2016}{\natexlab{b}})},\ \Eprint
  {http://arxiv.org/abs/1609.00356} {arXiv:1609.00356 [gr-qc]} \BibitemShut
  {NoStop}%
\bibitem [{\citenamefont {Lukes-Gerakopoulos}\ \emph
  {et~al.}(2017)\citenamefont {Lukes-Gerakopoulos}, \citenamefont {Harms},
  \citenamefont {Bernuzzi},\ and\ \citenamefont
  {Nagar}}]{Lukes-Gerakopoulos:2017vkj}%
  \BibitemOpen
  \bibfield  {author} {\bibinfo {author} {\bibfnamefont {G.}~\bibnamefont
  {Lukes-Gerakopoulos}}, \bibinfo {author} {\bibfnamefont {E.}~\bibnamefont
  {Harms}}, \bibinfo {author} {\bibfnamefont {S.}~\bibnamefont {Bernuzzi}}, \
  and\ \bibinfo {author} {\bibfnamefont {A.}~\bibnamefont {Nagar}},\ }\href
  {\doibase 10.1103/PhysRevD.96.064051} {\bibfield  {journal} {\bibinfo
  {journal} {Phys. Rev.}\ }\textbf {\bibinfo {volume} {D96}},\ \bibinfo {pages}
  {064051} (\bibinfo {year} {2017})},\ \Eprint
  {http://arxiv.org/abs/1707.07537} {arXiv:1707.07537 [gr-qc]} \BibitemShut
  {NoStop}%
\bibitem [{\citenamefont {Nagar}\ \emph
  {et~al.}(2019{\natexlab{a}})\citenamefont {Nagar}, \citenamefont {Messina},
  \citenamefont {Kavanagh}, \citenamefont {Lukes-Gerakopoulos}, \citenamefont
  {Warburton}, \citenamefont {Bernuzzi},\ and\ \citenamefont
  {Harms}}]{Nagar:2019wrt}%
  \BibitemOpen
  \bibfield  {author} {\bibinfo {author} {\bibfnamefont {A.}~\bibnamefont
  {Nagar}}, \bibinfo {author} {\bibfnamefont {F.}~\bibnamefont {Messina}},
  \bibinfo {author} {\bibfnamefont {C.}~\bibnamefont {Kavanagh}}, \bibinfo
  {author} {\bibfnamefont {G.}~\bibnamefont {Lukes-Gerakopoulos}}, \bibinfo
  {author} {\bibfnamefont {N.}~\bibnamefont {Warburton}}, \bibinfo {author}
  {\bibfnamefont {S.}~\bibnamefont {Bernuzzi}}, \ and\ \bibinfo {author}
  {\bibfnamefont {E.}~\bibnamefont {Harms}},\ }\href {\doibase
  10.1103/PhysRevD.100.104056} {\bibfield  {journal} {\bibinfo  {journal}
  {Phys. Rev.}\ }\textbf {\bibinfo {volume} {D100}},\ \bibinfo {pages} {104056}
  (\bibinfo {year} {2019}{\natexlab{a}})},\ \Eprint
  {http://arxiv.org/abs/1907.12233} {arXiv:1907.12233 [gr-qc]} \BibitemShut
  {NoStop}%
\bibitem [{\citenamefont {Damour}(2010)}]{Damour:2009sm}%
  \BibitemOpen
  \bibfield  {author} {\bibinfo {author} {\bibfnamefont {T.}~\bibnamefont
  {Damour}},\ }\href {\doibase 10.1103/PhysRevD.81.024017} {\bibfield
  {journal} {\bibinfo  {journal} {Phys. Rev.}\ }\textbf {\bibinfo {volume}
  {D81}},\ \bibinfo {pages} {024017} (\bibinfo {year} {2010})},\ \Eprint
  {http://arxiv.org/abs/0910.5533} {arXiv:0910.5533 [gr-qc]} \BibitemShut
  {NoStop}%
\bibitem [{\citenamefont {Barack}\ \emph {et~al.}(2010)\citenamefont {Barack},
  \citenamefont {Damour},\ and\ \citenamefont {Sago}}]{Barack:2010ny}%
  \BibitemOpen
  \bibfield  {author} {\bibinfo {author} {\bibfnamefont {L.}~\bibnamefont
  {Barack}}, \bibinfo {author} {\bibfnamefont {T.}~\bibnamefont {Damour}}, \
  and\ \bibinfo {author} {\bibfnamefont {N.}~\bibnamefont {Sago}},\ }\href
  {\doibase 10.1103/PhysRevD.82.084036} {\bibfield  {journal} {\bibinfo
  {journal} {Phys.Rev.}\ }\textbf {\bibinfo {volume} {D82}},\ \bibinfo {pages}
  {084036} (\bibinfo {year} {2010})},\ \Eprint {http://arxiv.org/abs/1008.0935}
  {arXiv:1008.0935 [gr-qc]} \BibitemShut {NoStop}%
\bibitem [{\citenamefont {Akcay}\ \emph {et~al.}(2012)\citenamefont {Akcay},
  \citenamefont {Barack}, \citenamefont {Damour},\ and\ \citenamefont
  {Sago}}]{Akcay:2012ea}%
  \BibitemOpen
  \bibfield  {author} {\bibinfo {author} {\bibfnamefont {S.}~\bibnamefont
  {Akcay}}, \bibinfo {author} {\bibfnamefont {L.}~\bibnamefont {Barack}},
  \bibinfo {author} {\bibfnamefont {T.}~\bibnamefont {Damour}}, \ and\ \bibinfo
  {author} {\bibfnamefont {N.}~\bibnamefont {Sago}},\ }\href {\doibase
  10.1103/PhysRevD.86.104041} {\bibfield  {journal} {\bibinfo  {journal} {Phys.
  Rev.}\ }\textbf {\bibinfo {volume} {D86}},\ \bibinfo {pages} {104041}
  (\bibinfo {year} {2012})},\ \Eprint {http://arxiv.org/abs/1209.0964}
  {arXiv:1209.0964 [gr-qc]} \BibitemShut {NoStop}%
\bibitem [{\citenamefont {Bini}\ and\ \citenamefont
  {Damour}(2014)}]{Bini:2014ica}%
  \BibitemOpen
  \bibfield  {author} {\bibinfo {author} {\bibfnamefont {D.}~\bibnamefont
  {Bini}}\ and\ \bibinfo {author} {\bibfnamefont {T.}~\bibnamefont {Damour}},\
  }\href {\doibase 10.1103/PhysRevD.90.024039} {\bibfield  {journal} {\bibinfo
  {journal} {Phys.Rev.}\ }\textbf {\bibinfo {volume} {D90}},\ \bibinfo {pages}
  {024039} (\bibinfo {year} {2014})},\ \Eprint {http://arxiv.org/abs/1404.2747}
  {arXiv:1404.2747 [gr-qc]} \BibitemShut {NoStop}%
\bibitem [{\citenamefont {Bini}\ and\ \citenamefont
  {Damour}(2015)}]{Bini:2015mza}%
  \BibitemOpen
  \bibfield  {author} {\bibinfo {author} {\bibfnamefont {D.}~\bibnamefont
  {Bini}}\ and\ \bibinfo {author} {\bibfnamefont {T.}~\bibnamefont {Damour}},\
  }\href {\doibase 10.1103/PhysRevD.91.064064} {\bibfield  {journal} {\bibinfo
  {journal} {Phys. Rev. D}\ }\textbf {\bibinfo {volume} {91}},\ \bibinfo
  {pages} {064064} (\bibinfo {year} {2015})},\ \Eprint
  {http://arxiv.org/abs/1503.01272} {arXiv:1503.01272 [gr-qc]} \BibitemShut
  {NoStop}%
\bibitem [{\citenamefont {Akcay}\ and\ \citenamefont {van~de
  Meent}(2016)}]{Akcay:2015pjz}%
  \BibitemOpen
  \bibfield  {author} {\bibinfo {author} {\bibfnamefont {S.}~\bibnamefont
  {Akcay}}\ and\ \bibinfo {author} {\bibfnamefont {M.}~\bibnamefont {van~de
  Meent}},\ }\href {\doibase 10.1103/PhysRevD.93.064063} {\bibfield  {journal}
  {\bibinfo  {journal} {Phys. Rev.}\ }\textbf {\bibinfo {volume} {D93}},\
  \bibinfo {pages} {064063} (\bibinfo {year} {2016})},\ \Eprint
  {http://arxiv.org/abs/1512.03392} {arXiv:1512.03392 [gr-qc]} \BibitemShut
  {NoStop}%
\bibitem [{\citenamefont {Barack}\ \emph {et~al.}(2019)\citenamefont {Barack},
  \citenamefont {Colleoni}, \citenamefont {Damour}, \citenamefont {Isoyama},\
  and\ \citenamefont {Sago}}]{Barack:2019agd}%
  \BibitemOpen
  \bibfield  {author} {\bibinfo {author} {\bibfnamefont {L.}~\bibnamefont
  {Barack}}, \bibinfo {author} {\bibfnamefont {M.}~\bibnamefont {Colleoni}},
  \bibinfo {author} {\bibfnamefont {T.}~\bibnamefont {Damour}}, \bibinfo
  {author} {\bibfnamefont {S.}~\bibnamefont {Isoyama}}, \ and\ \bibinfo
  {author} {\bibfnamefont {N.}~\bibnamefont {Sago}},\ }\href {\doibase
  10.1103/PhysRevD.100.124015} {\bibfield  {journal} {\bibinfo  {journal}
  {Phys. Rev.}\ }\textbf {\bibinfo {volume} {D100}},\ \bibinfo {pages} {124015}
  (\bibinfo {year} {2019})},\ \Eprint {http://arxiv.org/abs/1909.06103}
  {arXiv:1909.06103 [gr-qc]} \BibitemShut {NoStop}%
\bibitem [{\citenamefont {Damour}\ \emph {et~al.}(2009)\citenamefont {Damour},
  \citenamefont {Iyer},\ and\ \citenamefont {Nagar}}]{Damour:2008gu}%
  \BibitemOpen
  \bibfield  {author} {\bibinfo {author} {\bibfnamefont {T.}~\bibnamefont
  {Damour}}, \bibinfo {author} {\bibfnamefont {B.~R.}\ \bibnamefont {Iyer}}, \
  and\ \bibinfo {author} {\bibfnamefont {A.}~\bibnamefont {Nagar}},\ }\href
  {\doibase 10.1103/PhysRevD.79.064004} {\bibfield  {journal} {\bibinfo
  {journal} {Phys. Rev.}\ }\textbf {\bibinfo {volume} {D79}},\ \bibinfo {pages}
  {064004} (\bibinfo {year} {2009})},\ \Eprint {http://arxiv.org/abs/0811.2069}
  {arXiv:0811.2069 [gr-qc]} \BibitemShut {NoStop}%
\bibitem [{\citenamefont {Barack}\ and\ \citenamefont
  {Pound}(2019)}]{Barack:2018yvs}%
  \BibitemOpen
  \bibfield  {author} {\bibinfo {author} {\bibfnamefont {L.}~\bibnamefont
  {Barack}}\ and\ \bibinfo {author} {\bibfnamefont {A.}~\bibnamefont {Pound}},\
  }\href {\doibase 10.1088/1361-6633/aae552} {\bibfield  {journal} {\bibinfo
  {journal} {Rept. Prog. Phys.}\ }\textbf {\bibinfo {volume} {82}},\ \bibinfo
  {pages} {016904} (\bibinfo {year} {2019})},\ \Eprint
  {http://arxiv.org/abs/1805.10385} {arXiv:1805.10385 [gr-qc]} \BibitemShut
  {NoStop}%
\bibitem [{\citenamefont {Damour}\ and\ \citenamefont
  {Nagar}(2014{\natexlab{a}})}]{Damour:2014sva}%
  \BibitemOpen
  \bibfield  {author} {\bibinfo {author} {\bibfnamefont {T.}~\bibnamefont
  {Damour}}\ and\ \bibinfo {author} {\bibfnamefont {A.}~\bibnamefont {Nagar}},\
  }\href {\doibase 10.1103/PhysRevD.90.044018} {\bibfield  {journal} {\bibinfo
  {journal} {Phys.Rev.}\ }\textbf {\bibinfo {volume} {D90}},\ \bibinfo {pages}
  {044018} (\bibinfo {year} {2014}{\natexlab{a}})},\ \Eprint
  {http://arxiv.org/abs/1406.6913} {arXiv:1406.6913 [gr-qc]} \BibitemShut
  {NoStop}%
\bibitem [{\citenamefont {O'Shaughnessy}(2003)}]{OShaughnessy:2002tbu}%
  \BibitemOpen
  \bibfield  {author} {\bibinfo {author} {\bibfnamefont {R.~W.}\ \bibnamefont
  {O'Shaughnessy}},\ }\href {\doibase 10.1103/PhysRevD.67.044004} {\bibfield
  {journal} {\bibinfo  {journal} {Phys. Rev. D}\ }\textbf {\bibinfo {volume}
  {67}},\ \bibinfo {pages} {044004} (\bibinfo {year} {2003})},\ \Eprint
  {http://arxiv.org/abs/gr-qc/0211023} {arXiv:gr-qc/0211023} \BibitemShut
  {NoStop}%
\bibitem [{\citenamefont {Stein}\ and\ \citenamefont
  {Warburton}(2020)}]{Stein:2019buj}%
  \BibitemOpen
  \bibfield  {author} {\bibinfo {author} {\bibfnamefont {L.~C.}\ \bibnamefont
  {Stein}}\ and\ \bibinfo {author} {\bibfnamefont {N.}~\bibnamefont
  {Warburton}},\ }\href {\doibase 10.1103/PhysRevD.101.064007} {\bibfield
  {journal} {\bibinfo  {journal} {Phys. Rev. D}\ }\textbf {\bibinfo {volume}
  {101}},\ \bibinfo {pages} {064007} (\bibinfo {year} {2020})},\ \Eprint
  {http://arxiv.org/abs/1912.07609} {arXiv:1912.07609 [gr-qc]} \BibitemShut
  {NoStop}%
\bibitem [{\citenamefont {Regge}\ and\ \citenamefont
  {Wheeler}(1957)}]{Regge:1957td}%
  \BibitemOpen
  \bibfield  {author} {\bibinfo {author} {\bibfnamefont {T.}~\bibnamefont
  {Regge}}\ and\ \bibinfo {author} {\bibfnamefont {J.~A.}\ \bibnamefont
  {Wheeler}},\ }\href@noop {} {\bibfield  {journal} {\bibinfo  {journal} {Phys.
  Rev.}\ }\textbf {\bibinfo {volume} {108}},\ \bibinfo {pages} {1063} (\bibinfo
  {year} {1957})}\BibitemShut {NoStop}%
\bibitem [{\citenamefont {Zerilli}(1970)}]{Zerilli:1970se}%
  \BibitemOpen
  \bibfield  {author} {\bibinfo {author} {\bibfnamefont {F.~J.}\ \bibnamefont
  {Zerilli}},\ }\href {\doibase 10.1103/PhysRevLett.24.737} {\bibfield
  {journal} {\bibinfo  {journal} {Phys. Rev. Lett.}\ }\textbf {\bibinfo
  {volume} {24}},\ \bibinfo {pages} {737} (\bibinfo {year} {1970})}\BibitemShut
  {NoStop}%
\bibitem [{\citenamefont {Nagar}\ and\ \citenamefont
  {Rezzolla}(2005)}]{Nagar:2005ea}%
  \BibitemOpen
  \bibfield  {author} {\bibinfo {author} {\bibfnamefont {A.}~\bibnamefont
  {Nagar}}\ and\ \bibinfo {author} {\bibfnamefont {L.}~\bibnamefont
  {Rezzolla}},\ }\href {\doibase 10.1088/0264-9381/22/16/R01} {\bibfield
  {journal} {\bibinfo  {journal} {Class. Quant. Grav.}\ }\textbf {\bibinfo
  {volume} {22}},\ \bibinfo {pages} {R167} (\bibinfo {year} {2005})},\ \Eprint
  {http://arxiv.org/abs/gr-qc/0502064} {arXiv:gr-qc/0502064} \BibitemShut
  {NoStop}%
\bibitem [{\citenamefont {Nagar}\ \emph {et~al.}(2004)\citenamefont {Nagar},
  \citenamefont {Diaz}, \citenamefont {Pons},\ and\ \citenamefont
  {Font}}]{Nagar:2004ns}%
  \BibitemOpen
  \bibfield  {author} {\bibinfo {author} {\bibfnamefont {A.}~\bibnamefont
  {Nagar}}, \bibinfo {author} {\bibfnamefont {G.}~\bibnamefont {Diaz}},
  \bibinfo {author} {\bibfnamefont {J.~A.}\ \bibnamefont {Pons}}, \ and\
  \bibinfo {author} {\bibfnamefont {J.~A.}\ \bibnamefont {Font}},\ }\href@noop
  {} {\bibfield  {journal} {\bibinfo  {journal} {Phys. Rev.}\ }\textbf
  {\bibinfo {volume} {D69}},\ \bibinfo {pages} {124028} (\bibinfo {year}
  {2004})},\ \Eprint {http://arxiv.org/abs/gr-qc/0403077} {gr-qc/0403077}
  \BibitemShut {NoStop}%
\bibitem [{\citenamefont {Zengino{\u g}lu}(2008)}]{Zenginoglu:2007jw}%
  \BibitemOpen
  \bibfield  {author} {\bibinfo {author} {\bibfnamefont {A.}~\bibnamefont
  {Zengino{\u g}lu}},\ }\href {\doibase 10.1088/0264-9381/25/14/145002}
  {\bibfield  {journal} {\bibinfo  {journal} {Class. Quant. Grav.}\ }\textbf
  {\bibinfo {volume} {25}},\ \bibinfo {pages} {145002} (\bibinfo {year}
  {2008})},\ \Eprint {http://arxiv.org/abs/0712.4333} {arXiv:0712.4333 [gr-qc]}
  \BibitemShut {NoStop}%
\bibitem [{\citenamefont {Zengino{\u g}lu}\ and\ \citenamefont
  {Tiglio}(2009)}]{Zenginoglu:2009hd}%
  \BibitemOpen
  \bibfield  {author} {\bibinfo {author} {\bibfnamefont {A.}~\bibnamefont
  {Zengino{\u g}lu}}\ and\ \bibinfo {author} {\bibfnamefont {M.}~\bibnamefont
  {Tiglio}},\ }\href {\doibase 10.1103/PhysRevD.80.024044} {\bibfield
  {journal} {\bibinfo  {journal} {Phys. Rev.}\ }\textbf {\bibinfo {volume}
  {D80}},\ \bibinfo {pages} {024044} (\bibinfo {year} {2009})},\ \Eprint
  {http://arxiv.org/abs/0906.3342} {arXiv:0906.3342 [gr-qc]} \BibitemShut
  {NoStop}%
\bibitem [{\citenamefont {Zengino{\u g}lu}(2011)}]{Zenginoglu:2010cq}%
  \BibitemOpen
  \bibfield  {author} {\bibinfo {author} {\bibfnamefont {A.}~\bibnamefont
  {Zengino{\u g}lu}},\ }\href {\doibase 10.1016/j.jcp.2010.12.016} {\bibfield
  {journal} {\bibinfo  {journal} {J.Comput.Phys.}\ }\textbf {\bibinfo {volume}
  {230}},\ \bibinfo {pages} {2286} (\bibinfo {year} {2011})},\ \Eprint
  {http://arxiv.org/abs/1008.3809} {arXiv:1008.3809 [math.NA]} \BibitemShut
  {NoStop}%
\bibitem [{\citenamefont {Martel}(2004)}]{Martel:2003jj}%
  \BibitemOpen
  \bibfield  {author} {\bibinfo {author} {\bibfnamefont {K.}~\bibnamefont
  {Martel}},\ }\href {\doibase 10.1103/PhysRevD.69.044025} {\bibfield
  {journal} {\bibinfo  {journal} {Phys. Rev.}\ }\textbf {\bibinfo {volume}
  {D69}},\ \bibinfo {pages} {044025} (\bibinfo {year} {2004})},\ \Eprint
  {http://arxiv.org/abs/gr-qc/0311017} {arXiv:gr-qc/0311017} \BibitemShut
  {NoStop}%
\bibitem [{\citenamefont {Barack}\ and\ \citenamefont
  {Sago}(2010)}]{Barack:2010tm}%
  \BibitemOpen
  \bibfield  {author} {\bibinfo {author} {\bibfnamefont {L.}~\bibnamefont
  {Barack}}\ and\ \bibinfo {author} {\bibfnamefont {N.}~\bibnamefont {Sago}},\
  }\href {\doibase 10.1103/PhysRevD.81.084021} {\bibfield  {journal} {\bibinfo
  {journal} {Phys. Rev.}\ }\textbf {\bibinfo {volume} {D81}},\ \bibinfo {pages}
  {084021} (\bibinfo {year} {2010})},\ \Eprint {http://arxiv.org/abs/1002.2386}
  {arXiv:1002.2386 [gr-qc]} \BibitemShut {NoStop}%
\bibitem [{\citenamefont {Fujita}\ \emph {et~al.}(2009)\citenamefont {Fujita},
  \citenamefont {Hikida},\ and\ \citenamefont {Tagoshi}}]{Fujita:2009us}%
  \BibitemOpen
  \bibfield  {author} {\bibinfo {author} {\bibfnamefont {R.}~\bibnamefont
  {Fujita}}, \bibinfo {author} {\bibfnamefont {W.}~\bibnamefont {Hikida}}, \
  and\ \bibinfo {author} {\bibfnamefont {H.}~\bibnamefont {Tagoshi}},\ }\href
  {\doibase 10.1143/PTP.121.843} {\bibfield  {journal} {\bibinfo  {journal}
  {Prog.Theor.Phys.}\ }\textbf {\bibinfo {volume} {121}},\ \bibinfo {pages}
  {843} (\bibinfo {year} {2009})},\ \Eprint {http://arxiv.org/abs/0904.3810}
  {arXiv:0904.3810 [gr-qc]} \BibitemShut {NoStop}%
\bibitem [{\citenamefont {Glampedakis}\ and\ \citenamefont
  {Kennefick}(2002)}]{Glampedakis:2002ya}%
  \BibitemOpen
  \bibfield  {author} {\bibinfo {author} {\bibfnamefont {K.}~\bibnamefont
  {Glampedakis}}\ and\ \bibinfo {author} {\bibfnamefont {D.}~\bibnamefont
  {Kennefick}},\ }\href {\doibase 10.1103/PhysRevD.66.044002} {\bibfield
  {journal} {\bibinfo  {journal} {Phys.Rev.}\ }\textbf {\bibinfo {volume}
  {D66}},\ \bibinfo {pages} {044002} (\bibinfo {year} {2002})},\ \Eprint
  {http://arxiv.org/abs/gr-qc/0203086} {arXiv:gr-qc/0203086 [gr-qc]}
  \BibitemShut {NoStop}%
\bibitem [{\citenamefont {Shibata}(1994)}]{Shibata:1994xk}%
  \BibitemOpen
  \bibfield  {author} {\bibinfo {author} {\bibfnamefont {M.}~\bibnamefont
  {Shibata}},\ }\href {\doibase 10.1103/PhysRevD.50.6297} {\bibfield  {journal}
  {\bibinfo  {journal} {Phys. Rev. D}\ }\textbf {\bibinfo {volume} {50}},\
  \bibinfo {pages} {6297} (\bibinfo {year} {1994})}\BibitemShut {NoStop}%
\bibitem [{\citenamefont {Bini}\ and\ \citenamefont
  {Damour}(2012)}]{Bini:2012ji}%
  \BibitemOpen
  \bibfield  {author} {\bibinfo {author} {\bibfnamefont {D.}~\bibnamefont
  {Bini}}\ and\ \bibinfo {author} {\bibfnamefont {T.}~\bibnamefont {Damour}},\
  }\href {\doibase 10.1103/PhysRevD.86.124012} {\bibfield  {journal} {\bibinfo
  {journal} {Phys.Rev.}\ }\textbf {\bibinfo {volume} {D86}},\ \bibinfo {pages}
  {124012} (\bibinfo {year} {2012})},\ \Eprint {http://arxiv.org/abs/1210.2834}
  {arXiv:1210.2834 [gr-qc]} \BibitemShut {NoStop}%
\bibitem [{\citenamefont {Nagar}\ and\ \citenamefont
  {Shah}(2016)}]{Nagar:2016ayt}%
  \BibitemOpen
  \bibfield  {author} {\bibinfo {author} {\bibfnamefont {A.}~\bibnamefont
  {Nagar}}\ and\ \bibinfo {author} {\bibfnamefont {A.}~\bibnamefont {Shah}},\
  }\href {\doibase 10.1103/PhysRevD.94.104017} {\bibfield  {journal} {\bibinfo
  {journal} {Phys. Rev.}\ }\textbf {\bibinfo {volume} {D94}},\ \bibinfo {pages}
  {104017} (\bibinfo {year} {2016})},\ \Eprint
  {http://arxiv.org/abs/1606.00207} {arXiv:1606.00207 [gr-qc]} \BibitemShut
  {NoStop}%
\bibitem [{\citenamefont {Messina}\ \emph {et~al.}(2018)\citenamefont
  {Messina}, \citenamefont {Maldarella},\ and\ \citenamefont
  {Nagar}}]{Messina:2018ghh}%
  \BibitemOpen
  \bibfield  {author} {\bibinfo {author} {\bibfnamefont {F.}~\bibnamefont
  {Messina}}, \bibinfo {author} {\bibfnamefont {A.}~\bibnamefont {Maldarella}},
  \ and\ \bibinfo {author} {\bibfnamefont {A.}~\bibnamefont {Nagar}},\ }\href
  {\doibase 10.1103/PhysRevD.97.084016} {\bibfield  {journal} {\bibinfo
  {journal} {Phys. Rev.}\ }\textbf {\bibinfo {volume} {D97}},\ \bibinfo {pages}
  {084016} (\bibinfo {year} {2018})},\ \Eprint
  {http://arxiv.org/abs/1801.02366} {arXiv:1801.02366 [gr-qc]} \BibitemShut
  {NoStop}%
\bibitem [{\citenamefont {Fujita}(2015)}]{Fujita:2014eta}%
  \BibitemOpen
  \bibfield  {author} {\bibinfo {author} {\bibfnamefont {R.}~\bibnamefont
  {Fujita}},\ }\href {\doibase 10.1093/ptep/ptv012} {\bibfield  {journal}
  {\bibinfo  {journal} {PTEP}\ }\textbf {\bibinfo {volume} {2015}},\ \bibinfo
  {pages} {033E01} (\bibinfo {year} {2015})},\ \Eprint
  {http://arxiv.org/abs/1412.5689} {arXiv:1412.5689 [gr-qc]} \BibitemShut
  {NoStop}%
\bibitem [{\citenamefont {Taracchini}\ \emph {et~al.}(2013)\citenamefont
  {Taracchini}, \citenamefont {Buonanno}, \citenamefont {Hughes},\ and\
  \citenamefont {Khanna}}]{Taracchini:2013wfa}%
  \BibitemOpen
  \bibfield  {author} {\bibinfo {author} {\bibfnamefont {A.}~\bibnamefont
  {Taracchini}}, \bibinfo {author} {\bibfnamefont {A.}~\bibnamefont
  {Buonanno}}, \bibinfo {author} {\bibfnamefont {S.~A.}\ \bibnamefont
  {Hughes}}, \ and\ \bibinfo {author} {\bibfnamefont {G.}~\bibnamefont
  {Khanna}},\ }\href {\doibase 10.1103/PhysRevD.88.044001} {\bibfield
  {journal} {\bibinfo  {journal} {Phys.Rev.}\ }\textbf {\bibinfo {volume}
  {D88}},\ \bibinfo {pages} {044001} (\bibinfo {year} {2013})},\ \Eprint
  {http://arxiv.org/abs/1305.2184} {arXiv:1305.2184 [gr-qc]} \BibitemShut
  {NoStop}%
\bibitem [{\citenamefont {Damour}\ \emph {et~al.}(2013)\citenamefont {Damour},
  \citenamefont {Nagar},\ and\ \citenamefont {Bernuzzi}}]{Damour:2012ky}%
  \BibitemOpen
  \bibfield  {author} {\bibinfo {author} {\bibfnamefont {T.}~\bibnamefont
  {Damour}}, \bibinfo {author} {\bibfnamefont {A.}~\bibnamefont {Nagar}}, \
  and\ \bibinfo {author} {\bibfnamefont {S.}~\bibnamefont {Bernuzzi}},\ }\href
  {\doibase 10.1103/PhysRevD.87.084035} {\bibfield  {journal} {\bibinfo
  {journal} {Phys.Rev.}\ }\textbf {\bibinfo {volume} {D87}},\ \bibinfo {pages}
  {084035} (\bibinfo {year} {2013})},\ \Eprint {http://arxiv.org/abs/1212.4357}
  {arXiv:1212.4357 [gr-qc]} \BibitemShut {NoStop}%
\bibitem [{\citenamefont {Kojima}\ and\ \citenamefont
  {Nakamura}(1984)}]{10.1143/PTP.72.494}%
  \BibitemOpen
  \bibfield  {author} {\bibinfo {author} {\bibfnamefont {Y.}~\bibnamefont
  {Kojima}}\ and\ \bibinfo {author} {\bibfnamefont {T.}~\bibnamefont
  {Nakamura}},\ }\href {\doibase 10.1143/PTP.72.494} {\bibfield  {journal}
  {\bibinfo  {journal} {Progress of Theoretical Physics}\ }\textbf {\bibinfo
  {volume} {72}},\ \bibinfo {pages} {494} (\bibinfo {year} {1984})}\BibitemShut
  {NoStop}%
\bibitem [{\citenamefont {Rifat}\ \emph {et~al.}(2019)\citenamefont {Rifat},
  \citenamefont {Khanna},\ and\ \citenamefont {Burko}}]{Rifat:2019fkt}%
  \BibitemOpen
  \bibfield  {author} {\bibinfo {author} {\bibfnamefont {N.~E.}\ \bibnamefont
  {Rifat}}, \bibinfo {author} {\bibfnamefont {G.}~\bibnamefont {Khanna}}, \
  and\ \bibinfo {author} {\bibfnamefont {L.~M.}\ \bibnamefont {Burko}},\ }\href
  {\doibase 10.1103/PhysRevResearch.1.033150} {\bibfield  {journal} {\bibinfo
  {journal} {Phys. Rev. Research.}\ }\textbf {\bibinfo {volume} {1}},\ \bibinfo
  {pages} {033150} (\bibinfo {year} {2019})},\ \Eprint
  {http://arxiv.org/abs/1910.03462} {arXiv:1910.03462 [gr-qc]} \BibitemShut
  {NoStop}%
\bibitem [{\citenamefont {Thornburg}\ \emph {et~al.}(2020)\citenamefont
  {Thornburg}, \citenamefont {Wardell},\ and\ \citenamefont {van~de
  Meent}}]{Thornburg:2019ukt}%
  \BibitemOpen
  \bibfield  {author} {\bibinfo {author} {\bibfnamefont {J.}~\bibnamefont
  {Thornburg}}, \bibinfo {author} {\bibfnamefont {B.}~\bibnamefont {Wardell}},
  \ and\ \bibinfo {author} {\bibfnamefont {M.}~\bibnamefont {van~de Meent}},\
  }\href {\doibase 10.1103/PhysRevResearch.2.013365} {\bibfield  {journal}
  {\bibinfo  {journal} {Phys. Rev. Res.}\ }\textbf {\bibinfo {volume} {2}},\
  \bibinfo {pages} {013365} (\bibinfo {year} {2020})},\ \Eprint
  {http://arxiv.org/abs/1906.06791} {arXiv:1906.06791 [gr-qc]} \BibitemShut
  {NoStop}%
\bibitem [{\citenamefont {Munna}\ \emph {et~al.}(2020)\citenamefont {Munna},
  \citenamefont {Evans}, \citenamefont {Hopper},\ and\ \citenamefont
  {Forseth}}]{Munna:2020juq}%
  \BibitemOpen
  \bibfield  {author} {\bibinfo {author} {\bibfnamefont {C.}~\bibnamefont
  {Munna}}, \bibinfo {author} {\bibfnamefont {C.~R.}\ \bibnamefont {Evans}},
  \bibinfo {author} {\bibfnamefont {S.}~\bibnamefont {Hopper}}, \ and\ \bibinfo
  {author} {\bibfnamefont {E.}~\bibnamefont {Forseth}},\ }\href {\doibase
  10.1103/PhysRevD.102.024047} {\bibfield  {journal} {\bibinfo  {journal}
  {Phys. Rev. D}\ }\textbf {\bibinfo {volume} {102}},\ \bibinfo {pages}
  {024047} (\bibinfo {year} {2020})},\ \Eprint
  {http://arxiv.org/abs/2005.03044} {arXiv:2005.03044 [gr-qc]} \BibitemShut
  {NoStop}%
\bibitem [{\citenamefont {Munna}(2020)}]{Munna:2020iju}%
  \BibitemOpen
  \bibfield  {author} {\bibinfo {author} {\bibfnamefont {C.}~\bibnamefont
  {Munna}},\ }\href {\doibase 10.1103/PhysRevD.102.124001} {\bibfield
  {journal} {\bibinfo  {journal} {Phys. Rev. D}\ }\textbf {\bibinfo {volume}
  {102}},\ \bibinfo {pages} {124001} (\bibinfo {year} {2020})},\ \Eprint
  {http://arxiv.org/abs/2008.10622} {arXiv:2008.10622 [gr-qc]} \BibitemShut
  {NoStop}%
\bibitem [{\citenamefont {Damour}\ and\ \citenamefont
  {Nagar}(2014{\natexlab{b}})}]{Damour:2014yha}%
  \BibitemOpen
  \bibfield  {author} {\bibinfo {author} {\bibfnamefont {T.}~\bibnamefont
  {Damour}}\ and\ \bibinfo {author} {\bibfnamefont {A.}~\bibnamefont {Nagar}},\
  }\href {\doibase 10.1103/PhysRevD.90.024054} {\bibfield  {journal} {\bibinfo
  {journal} {Phys.Rev.}\ }\textbf {\bibinfo {volume} {D90}},\ \bibinfo {pages}
  {024054} (\bibinfo {year} {2014}{\natexlab{b}})},\ \Eprint
  {http://arxiv.org/abs/1406.0401} {arXiv:1406.0401 [gr-qc]} \BibitemShut
  {NoStop}%
\bibitem [{\citenamefont {Yunes}\ \emph {et~al.}(2011)\citenamefont {Yunes},
  \citenamefont {Buonanno}, \citenamefont {Hughes}, \citenamefont {Pan},
  \citenamefont {Barausse} \emph {et~al.}}]{Yunes:2010zj}%
  \BibitemOpen
  \bibfield  {author} {\bibinfo {author} {\bibfnamefont {N.}~\bibnamefont
  {Yunes}}, \bibinfo {author} {\bibfnamefont {A.}~\bibnamefont {Buonanno}},
  \bibinfo {author} {\bibfnamefont {S.~A.}\ \bibnamefont {Hughes}}, \bibinfo
  {author} {\bibfnamefont {Y.}~\bibnamefont {Pan}}, \bibinfo {author}
  {\bibfnamefont {E.}~\bibnamefont {Barausse}},  \emph {et~al.},\ }\href
  {\doibase 10.1103/PhysRevD.83.044044} {\bibfield  {journal} {\bibinfo
  {journal} {Phys.Rev.}\ }\textbf {\bibinfo {volume} {D83}},\ \bibinfo {pages}
  {044044} (\bibinfo {year} {2011})},\ \Eprint {http://arxiv.org/abs/1009.6013}
  {arXiv:1009.6013 [gr-qc]} \BibitemShut {NoStop}%
\bibitem [{\citenamefont {Barausse}\ \emph
  {et~al.}(2012{\natexlab{a}})\citenamefont {Barausse}, \citenamefont
  {Buonanno}, \citenamefont {Hughes}, \citenamefont {Khanna}, \citenamefont
  {O'Sullivan} \emph {et~al.}}]{Barausse:2011kb}%
  \BibitemOpen
  \bibfield  {author} {\bibinfo {author} {\bibfnamefont {E.}~\bibnamefont
  {Barausse}}, \bibinfo {author} {\bibfnamefont {A.}~\bibnamefont {Buonanno}},
  \bibinfo {author} {\bibfnamefont {S.~A.}\ \bibnamefont {Hughes}}, \bibinfo
  {author} {\bibfnamefont {G.}~\bibnamefont {Khanna}}, \bibinfo {author}
  {\bibfnamefont {S.}~\bibnamefont {O'Sullivan}},  \emph {et~al.},\ }\href
  {\doibase 10.1103/PhysRevD.85.024046} {\bibfield  {journal} {\bibinfo
  {journal} {Phys.Rev.}\ }\textbf {\bibinfo {volume} {D85}},\ \bibinfo {pages}
  {024046} (\bibinfo {year} {2012}{\natexlab{a}})},\ \Eprint
  {http://arxiv.org/abs/1110.3081} {arXiv:1110.3081 [gr-qc]} \BibitemShut
  {NoStop}%
\bibitem [{\citenamefont {Nagar}\ \emph
  {et~al.}(2019{\natexlab{b}})\citenamefont {Nagar}, \citenamefont {Pratten},
  \citenamefont {Riemenschneider},\ and\ \citenamefont
  {Gamba}}]{Nagar:2019wds}%
  \BibitemOpen
  \bibfield  {author} {\bibinfo {author} {\bibfnamefont {A.}~\bibnamefont
  {Nagar}}, \bibinfo {author} {\bibfnamefont {G.}~\bibnamefont {Pratten}},
  \bibinfo {author} {\bibfnamefont {G.}~\bibnamefont {Riemenschneider}}, \ and\
  \bibinfo {author} {\bibfnamefont {R.}~\bibnamefont {Gamba}},\ }\href@noop {}
  {\  (\bibinfo {year} {2019}{\natexlab{b}})},\ \Eprint
  {http://arxiv.org/abs/1904.09550} {arXiv:1904.09550 [gr-qc]} \BibitemShut
  {NoStop}%
\bibitem [{\citenamefont {Nagar}\ \emph {et~al.}(2020)\citenamefont {Nagar},
  \citenamefont {Riemenschneider}, \citenamefont {Pratten}, \citenamefont
  {Rettegno},\ and\ \citenamefont {Messina}}]{Nagar:2020pcj}%
  \BibitemOpen
  \bibfield  {author} {\bibinfo {author} {\bibfnamefont {A.}~\bibnamefont
  {Nagar}}, \bibinfo {author} {\bibfnamefont {G.}~\bibnamefont
  {Riemenschneider}}, \bibinfo {author} {\bibfnamefont {G.}~\bibnamefont
  {Pratten}}, \bibinfo {author} {\bibfnamefont {P.}~\bibnamefont {Rettegno}}, \
  and\ \bibinfo {author} {\bibfnamefont {F.}~\bibnamefont {Messina}},\ }\href
  {\doibase 10.1103/PhysRevD.102.024077} {\bibfield  {journal} {\bibinfo
  {journal} {Phys. Rev. D}\ }\textbf {\bibinfo {volume} {102}},\ \bibinfo
  {pages} {024077} (\bibinfo {year} {2020})},\ \Eprint
  {http://arxiv.org/abs/2001.09082} {arXiv:2001.09082 [gr-qc]} \BibitemShut
  {NoStop}%
\bibitem [{\citenamefont {Gamba}\ \emph {et~al.}(2021)\citenamefont {Gamba},
  \citenamefont {Breschi}, \citenamefont {Carullo}, \citenamefont {Rettegno},
  \citenamefont {Albanesi}, \citenamefont {Bernuzzi},\ and\ \citenamefont
  {Nagar}}]{Gamba:2021gap}%
  \BibitemOpen
  \bibfield  {author} {\bibinfo {author} {\bibfnamefont {R.}~\bibnamefont
  {Gamba}}, \bibinfo {author} {\bibfnamefont {M.}~\bibnamefont {Breschi}},
  \bibinfo {author} {\bibfnamefont {G.}~\bibnamefont {Carullo}}, \bibinfo
  {author} {\bibfnamefont {P.}~\bibnamefont {Rettegno}}, \bibinfo {author}
  {\bibfnamefont {S.}~\bibnamefont {Albanesi}}, \bibinfo {author}
  {\bibfnamefont {S.}~\bibnamefont {Bernuzzi}}, \ and\ \bibinfo {author}
  {\bibfnamefont {A.}~\bibnamefont {Nagar}},\ }\href@noop {} {\bibfield
  {journal} {\bibinfo  {journal} {Submitted to Nature Astronomy}\ } (\bibinfo
  {year} {2021})},\ \Eprint {http://arxiv.org/abs/2106.05575} {arXiv:2106.05575
  [gr-qc]} \BibitemShut {NoStop}%
\bibitem [{\citenamefont {Gold}\ and\ \citenamefont
  {Br{\"u}gmann}(2013)}]{Gold:2012tk}%
  \BibitemOpen
  \bibfield  {author} {\bibinfo {author} {\bibfnamefont {R.}~\bibnamefont
  {Gold}}\ and\ \bibinfo {author} {\bibfnamefont {B.}~\bibnamefont
  {Br{\"u}gmann}},\ }\href {\doibase 10.1103/PhysRevD.88.064051} {\bibfield
  {journal} {\bibinfo  {journal} {Phys. Rev.}\ }\textbf {\bibinfo {volume}
  {D88}},\ \bibinfo {pages} {064051} (\bibinfo {year} {2013})},\ \Eprint
  {http://arxiv.org/abs/1209.4085} {arXiv:1209.4085 [gr-qc]} \BibitemShut
  {NoStop}%
\bibitem [{\citenamefont {Barausse}\ \emph
  {et~al.}(2012{\natexlab{b}})\citenamefont {Barausse}, \citenamefont
  {Buonanno},\ and\ \citenamefont {Le~Tiec}}]{Barausse:2011dq}%
  \BibitemOpen
  \bibfield  {author} {\bibinfo {author} {\bibfnamefont {E.}~\bibnamefont
  {Barausse}}, \bibinfo {author} {\bibfnamefont {A.}~\bibnamefont {Buonanno}},
  \ and\ \bibinfo {author} {\bibfnamefont {A.}~\bibnamefont {Le~Tiec}},\ }\href
  {\doibase 10.1103/PhysRevD.85.064010} {\bibfield  {journal} {\bibinfo
  {journal} {Phys.Rev.}\ }\textbf {\bibinfo {volume} {D85}},\ \bibinfo {pages}
  {064010} (\bibinfo {year} {2012}{\natexlab{b}})},\ \Eprint
  {http://arxiv.org/abs/1111.5610} {arXiv:1111.5610 [gr-qc]} \BibitemShut
  {NoStop}%
\bibitem [{\citenamefont {Antonelli}\ \emph {et~al.}(2020)\citenamefont
  {Antonelli}, \citenamefont {van~de Meent}, \citenamefont {Buonanno},
  \citenamefont {Steinhoff},\ and\ \citenamefont {Vines}}]{Antonelli:2019fmq}%
  \BibitemOpen
  \bibfield  {author} {\bibinfo {author} {\bibfnamefont {A.}~\bibnamefont
  {Antonelli}}, \bibinfo {author} {\bibfnamefont {M.}~\bibnamefont {van~de
  Meent}}, \bibinfo {author} {\bibfnamefont {A.}~\bibnamefont {Buonanno}},
  \bibinfo {author} {\bibfnamefont {J.}~\bibnamefont {Steinhoff}}, \ and\
  \bibinfo {author} {\bibfnamefont {J.}~\bibnamefont {Vines}},\ }\href
  {\doibase 10.1103/PhysRevD.101.024024} {\bibfield  {journal} {\bibinfo
  {journal} {Phys. Rev.}\ }\textbf {\bibinfo {volume} {D101}},\ \bibinfo
  {pages} {024024} (\bibinfo {year} {2020})},\ \Eprint
  {http://arxiv.org/abs/1907.11597} {arXiv:1907.11597 [gr-qc]} \BibitemShut
  {NoStop}%
\bibitem [{\citenamefont {Mishra}\ \emph {et~al.}(2015)\citenamefont {Mishra},
  \citenamefont {Arun},\ and\ \citenamefont {Iyer}}]{Mishra:2015bqa}%
  \BibitemOpen
  \bibfield  {author} {\bibinfo {author} {\bibfnamefont {C.~K.}\ \bibnamefont
  {Mishra}}, \bibinfo {author} {\bibfnamefont {K.~G.}\ \bibnamefont {Arun}}, \
  and\ \bibinfo {author} {\bibfnamefont {B.~R.}\ \bibnamefont {Iyer}},\ }\href
  {\doibase 10.1103/PhysRevD.91.084040} {\bibfield  {journal} {\bibinfo
  {journal} {Phys. Rev.}\ }\textbf {\bibinfo {volume} {D91}},\ \bibinfo {pages}
  {084040} (\bibinfo {year} {2015})},\ \Eprint
  {http://arxiv.org/abs/1501.07096} {arXiv:1501.07096 [gr-qc]} \BibitemShut
  {NoStop}%
\bibitem [{\citenamefont {Schmidt}\ \emph {et~al.}(2021)\citenamefont
  {Schmidt}, \citenamefont {Breschi}, \citenamefont {Gamba}, \citenamefont
  {Pagano}, \citenamefont {Rettegno}, \citenamefont {Riemenschneider},
  \citenamefont {Bernuzzi}, \citenamefont {Nagar},\ and\ \citenamefont
  {Del~Pozzo}}]{Schmidt:2020yuu}%
  \BibitemOpen
  \bibfield  {author} {\bibinfo {author} {\bibfnamefont {S.}~\bibnamefont
  {Schmidt}}, \bibinfo {author} {\bibfnamefont {M.}~\bibnamefont {Breschi}},
  \bibinfo {author} {\bibfnamefont {R.}~\bibnamefont {Gamba}}, \bibinfo
  {author} {\bibfnamefont {G.}~\bibnamefont {Pagano}}, \bibinfo {author}
  {\bibfnamefont {P.}~\bibnamefont {Rettegno}}, \bibinfo {author}
  {\bibfnamefont {G.}~\bibnamefont {Riemenschneider}}, \bibinfo {author}
  {\bibfnamefont {S.}~\bibnamefont {Bernuzzi}}, \bibinfo {author}
  {\bibfnamefont {A.}~\bibnamefont {Nagar}}, \ and\ \bibinfo {author}
  {\bibfnamefont {W.}~\bibnamefont {Del~Pozzo}},\ }\href {\doibase
  10.1103/PhysRevD.103.043020} {\bibfield  {journal} {\bibinfo  {journal}
  {Phys. Rev. D}\ }\textbf {\bibinfo {volume} {103}},\ \bibinfo {pages}
  {043020} (\bibinfo {year} {2021})},\ \Eprint
  {http://arxiv.org/abs/2011.01958} {arXiv:2011.01958 [gr-qc]} \BibitemShut
  {NoStop}%
\bibitem [{\citenamefont {Katz}\ \emph {et~al.}(2021)\citenamefont {Katz},
  \citenamefont {Chua}, \citenamefont {Speri}, \citenamefont {Warburton},\ and\
  \citenamefont {Hughes}}]{Katz:2021yft}%
  \BibitemOpen
  \bibfield  {author} {\bibinfo {author} {\bibfnamefont {M.~L.}\ \bibnamefont
  {Katz}}, \bibinfo {author} {\bibfnamefont {A.~J.~K.}\ \bibnamefont {Chua}},
  \bibinfo {author} {\bibfnamefont {L.}~\bibnamefont {Speri}}, \bibinfo
  {author} {\bibfnamefont {N.}~\bibnamefont {Warburton}}, \ and\ \bibinfo
  {author} {\bibfnamefont {S.~A.}\ \bibnamefont {Hughes}},\ }\href@noop {} {\
  (\bibinfo {year} {2021})},\ \Eprint {http://arxiv.org/abs/2104.04582}
  {arXiv:2104.04582 [gr-qc]} \BibitemShut {NoStop}%
\bibitem [{\citenamefont {Hughes}\ \emph {et~al.}(2021)\citenamefont {Hughes},
  \citenamefont {Warburton}, \citenamefont {Khanna}, \citenamefont {Chua},\
  and\ \citenamefont {Katz}}]{Hughes:2021exa}%
  \BibitemOpen
  \bibfield  {author} {\bibinfo {author} {\bibfnamefont {S.~A.}\ \bibnamefont
  {Hughes}}, \bibinfo {author} {\bibfnamefont {N.}~\bibnamefont {Warburton}},
  \bibinfo {author} {\bibfnamefont {G.}~\bibnamefont {Khanna}}, \bibinfo
  {author} {\bibfnamefont {A.~J.~K.}\ \bibnamefont {Chua}}, \ and\ \bibinfo
  {author} {\bibfnamefont {M.~L.}\ \bibnamefont {Katz}},\ }\href@noop {} {\
  (\bibinfo {year} {2021})},\ \Eprint {http://arxiv.org/abs/2102.02713}
  {arXiv:2102.02713 [gr-qc]} \BibitemShut {NoStop}%
\bibitem [{\citenamefont {Bardeen}\ \emph {et~al.}(1972)\citenamefont
  {Bardeen}, \citenamefont {Press},\ and\ \citenamefont
  {Teukolsky}}]{Bardeen:1972fi}%
  \BibitemOpen
  \bibfield  {author} {\bibinfo {author} {\bibfnamefont {J.~M.}\ \bibnamefont
  {Bardeen}}, \bibinfo {author} {\bibfnamefont {W.~H.}\ \bibnamefont {Press}},
  \ and\ \bibinfo {author} {\bibfnamefont {S.~A.}\ \bibnamefont {Teukolsky}},\
  }\href {\doibase 10.1086/151796} {\bibfield  {journal} {\bibinfo  {journal}
  {Astrophys. J.}\ }\textbf {\bibinfo {volume} {178}},\ \bibinfo {pages} {347}
  (\bibinfo {year} {1972})}\BibitemShut {NoStop}%
\end{thebibliography}%

\end{document}